\documentclass[a4paper,11pt]{report}

\usepackage[a4paper, total={6.8in, 8.9in}, bottom=1.5in]{geometry}
\usepackage{cite}
\usepackage[toc,page]{appendix}
\usepackage{amsmath}
\usepackage[british]{babel}
\usepackage{kpfonts}
\usepackage[utf8]{inputenc}
\usepackage{layout}
\usepackage{color} 

\usepackage{multicol} 
\usepackage{epsfig}
\usepackage{cancel}
\usepackage{physics}

\usepackage{caption}
\usepackage{epsf}
\usepackage{fancyhdr}
\usepackage{feynmf} 
\usepackage{frontespizio}
\usepackage{rotating}
\usepackage{subfigure}
\usepackage{pstricks}
\usepackage{lettrine}
\usepackage{bbold}
\usepackage{calligra}
\usepackage{tikz}

\usepackage{subfigure}
\usepackage{mathrsfs}
\usepackage{curve2e}
\usepackage{setspace}
\usepackage{indentfirst}
\usepackage{emptypage}
\usepackage{relsize}
\usepackage{slashed}
\usepackage{mathrsfs}
\usepackage{stackengine}
\usepackage{calc}
\usepackage{hyperref}
\hypersetup{
	colorlinks,
	citecolor=green,
	filecolor=black,
	linkcolor=blue,
	urlcolor=black
}
\usepackage{bbold}

\let\a=\alpha   \let\b=\beta   \let\g=\gamma   \let\d=\delta
\let\e=\epsilon         
    \let\k=\kappa  \let\l=\lambda  \let\m=\mu
\let\n=\nu      \let\x=\xi     \let\p=\pi      \let\r=\rho
\let\s=\sigma        \let\f=\phi
     \let\y=\psi    
  \let\D=\Delta   \let\L=\Lambda
     \let\P=\Pi

\newcommand{\beq}{\begin{equation}}
\newcommand{\eeq}{\end{equation}}
\newcommand{\beqa}{\begin{eqnarray}}
\newcommand{\eeqa}{\end{eqnarray}}

\newcommand{\tphi}{\tilde{\phi}}
\newcommand{\pd}{\partial}

\newcommand{\eq}[1]{Eq.~(\ref{#1})}
\newcommand{\sdfrac}[2]{\mbox{\small$\displaystyle\frac{#1}{#2}$}}

\newcommand{\nn}{\nonumber \\}
\renewcommand{\Re}{\textrm{Re}}

\newcommand{\sm}{\mathcal{S}}
\newcommand{\na}{\nabla}
\newcommand{\vf}{\phi}
\newcommand{\bes}{\begin{subequations}}
\newcommand{\ees}{\end{subequations}}
\newcommand{\figref}[1]{Fig.~\ref{#1}}			
\newcommand{\secref}[1]{Section~\ref{#1}}		
\newcommand{\appref}[1]{Appendix~\ref{#1}}		
\newcommand{\app}[4]{F_{\!#1}\!
	\left(\left.\substack{\Scale[1]{ #2} \\[1.5ex] \Scale[1]{#3}}\right| #4 \right) }
\def\de{\delta} 

\def\nbar{\bar{\nabla}}

\def\app{{}_{(4)}}
\def\apd{{}_{(d)}}
\def\ape{{}_{(e)}}
\def\rgdo{\rgp \rge}
\def\rgpb{\sqrt{ \app \tilde g }}
\def\rgdb{\sqrt{ \apd \bar g }}

\def\rge{\sqrt{ \ape g }}
\def\rgp{\sqrt{ \app g }}

\newcommand{\del}{\delta}
\def\D{\mathcal{D}}

\def\rgd{\sqrt{ \apd g }}

\newcommand{\cS}{\mathcal{S}}
\newcommand{\mO}{\mathcal{O}}
\def\rg{\sqrt{g}}

\def\rgb{\sqrt{ \bar g }}

\newcommand{\be}{\begin{equation}}
\newcommand{\ee}{\end{equation}}
\newcommand{\bal}{\begin{aligned}}
\newcommand{\eal}{\end{aligned}}
\def\nbox#1#2{\vcenter{\hrule \hbox{\vrule height#2in
			\kern#1in \vrule} \hrule}}
\def\sq{\,\raise.5pt\hbox{$\nbox{.09}{.09}$}\,}
\def\sqb{\,\raise.5pt\hbox{$\overline{\nbox{.09}{.09}}$}\,}
\def\bpde{\bar \partial}
\def\bnabla{\bar \nabla}
\def\Box{\sq}
\numberwithin{equation}{section}

\usepackage{accents}

\newcommand{\pa}{\partial}	
\newcommand{\dfun}[2]{ \frac{\delta #1}{\delta #2}}
\def\lq{\left[}
\def\rq{\right]}
\def\lt{\left(}
\def\rt{\right)}

\usepackage{pifont}
\newcommand{\cmark}{\ding{51}}%
\newcommand{\xmark}{\ding{55}}%
\newcommand{\HRule}{\rule{\linewidth}{0.5mm}}
\def\sq{\,\raise.5pt\hbox{$\nbox{.09}{.09}$}\,}
\def\sqb{\,\raise.5pt\hbox{$\overline{\nbox{.09}{.09}}$}\,}
\begin{document}
\thispagestyle{empty}
\begin{titlepage}
	\begin{center}
		{\includegraphics[scale=0.12]{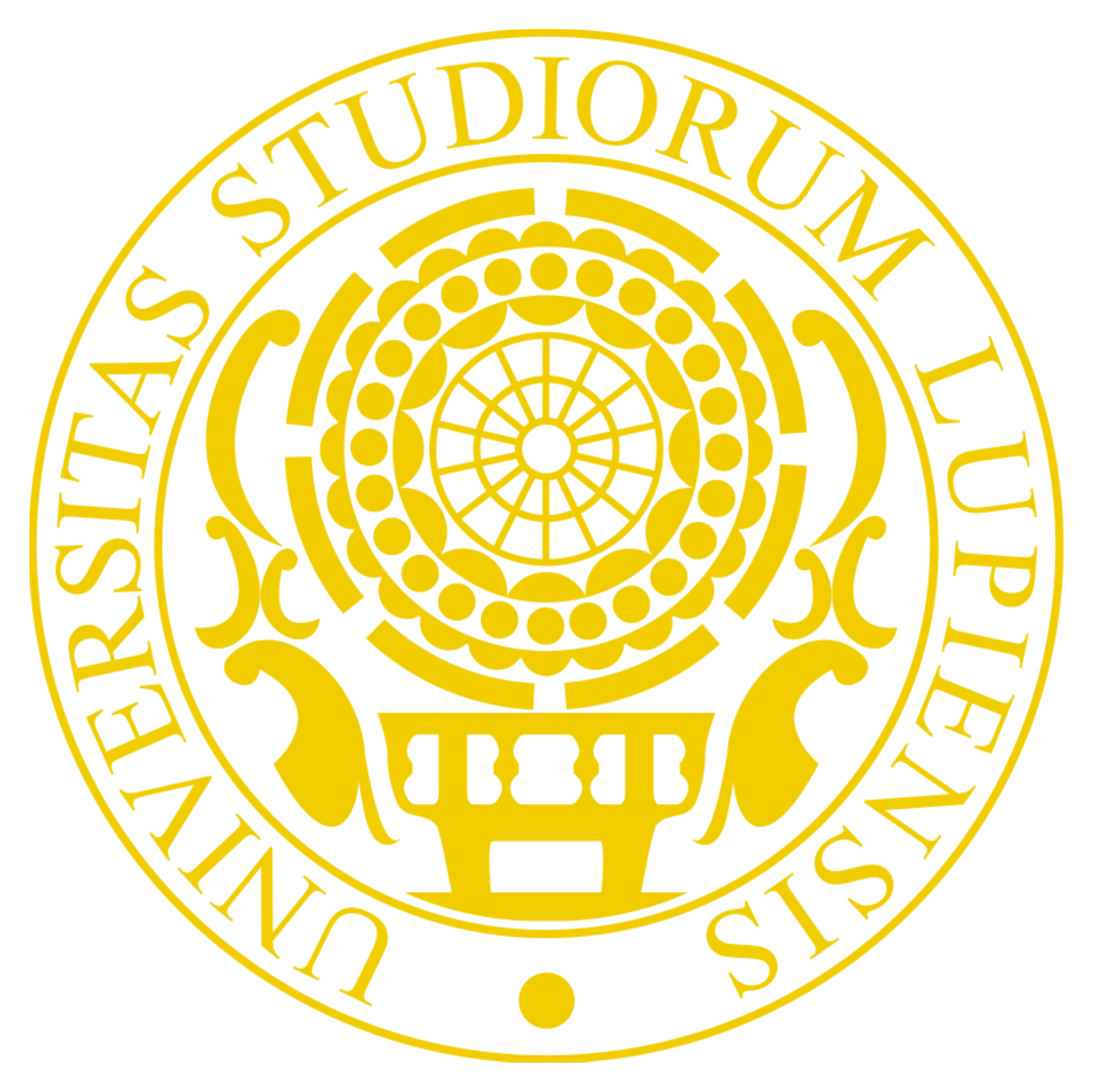}}\\[0.3cm]
		\scalebox{1.2}{\bf{Universit\`a del Salento}}\\[0.1cm]
		\HRule\\[0.3cm]
		\scalebox{1.5}{ Dipartimento di Matematica e Fisica ``Ennio De Giorgi''}	\\[2ex]
		\scalebox{1}{Corso di Dottorato di Ricerca in Fisica e Nanoscienze}\\[5cm]		
		{\Large { \bf{Four point functions in momentum space and topological terms in gravity }}}
		\\[4cm]
		\begin{minipage}{0.4\textwidth}
			\begin{flushleft}\large
				\textit{Supervisor}\\
				Prof.~Claudio Corian\`o
			\end{flushleft}
		\end{minipage}
		\begin{minipage}{0.4\textwidth}
			\begin{flushright} \large
				\textit{Candidate}\\
				Dimosthenis Theofilopoulos
			\end{flushright}
		\end{minipage}\\[3cm]
	\HRule\\
		Tesi di Dottorato in Fisica e Nanoscienze - XXXIV ciclo\\
		Anno Accademico 2021-2022
		\vfill
	\end{center}
\end{titlepage}

\begin{flushright}
	\thispagestyle{empty}
	\textit{Little by little one travels far.\\
J.R.R Tolkien}
\end{flushright}

\chapter*{Aknowledgements}
I would like to thank my supervisor Prof. Claudio Corian\'o for the interaction we had over the last three and half years. Particularly, I would like to thank him for the discussions we shared, the insights he gave me about Physics and the guidance as his student. Also, I am deeply thankful for his attention and his constant presence to hear and suggest solution to my problems even if they weren't related to my research.
Another person that I would like to thank is Prof. Nikos Irges. As a student of him, in my master degree we developed a respectful relation and he was ever present also during my PhD studies with support and advice. I would like also to thank him for his hospitality during our collaboration in NTUA. These two persons have played a great role in my aging as a Physicist.\\
The next person that I would like to thank is Dr. Matteo Maria Maglio, with who we collaborated constantly over the last years. I would like to thanks him for his advice, his help with dealing with out of research topics in Italy and also for the moments, filled with coffee that we shared. Next, is Dr. Fotis Koutroulis which I met in NTUA years ago and we developed a relationship that went beyond the margins of Physics. The discussions we had and the blackboards that we wrote together are moments that I really enjoyed.\\
Another person that I would like to thank is Dr. Konstantinos Rigatos, with whom we glued from the first moment we met and we have developed a deep friendship. Our endless talks about everything in life, from Physics to music and politics, were great and helpful the last three years. \\
Lastly, as this section can go on and on, I would like to thank my family for their endless support in all ways during this journey that started years ago. Also I am deeply thankful for the help that Dottoressa Maria Tsigdinoy and her family have given to me.
A special mention to some people that I hope they already know my feelings for them. These people are Panagiotis, Savvas, Vaggelis, Nikos, mia amica Elena, Pavlina, Vanessa, Natassa, Fotis, Anna, Apostolis, Michalis, Giannis, Mitsos, Athina and many more that I have probably forgotten. 
\begin{flushright}
\textit{Amicitia fortior. Fatum nos iunget}
\end{flushright}
\chapter*{List of publications}
This thesis discusses the following research papers
\begin{itemize}
\item Topological Corrections and Conformal Backreaction in the Einstein Gauss-Bonnet/Weyl Theories of Gravity at $D=4$,\\Corian\`o, Claudio and Maglio, Matteo Maria and Theofilopoulos, Dimosthenis\\ arXiv:2203.04213 [hep-th]
\item The Conformal Anomaly Action to Fourth Order (4T) in $d=4$ in Momentum Space\\
Corian\`o, Claudio and Maglio, Matteo Maria and Theofilopoulos, Dimosthenis\\Mar 25, 2021, Published in : Eur.Phys.J.C 81 (2021) 6, 740  \\ \href{https://arxiv.org/abs/2103.13957}{\bfseries arXiv:2103.13957}
    
    \item The conformal $N$-point scalar correlator in coordinate space\\
    Irges, Nikos and Koutroulis, Fotis and Theofilopoulos, Dimosthenis\\Jan 20, 2020, arXiv:2001.07171\\
    \href{https://arxiv.org/abs/2001.07171}{\bfseries arXiv:2001.07171}
    \item Four-Point Functions in Momentum Space: Conformal Ward Identities in the Scalar/Tensor case\\
    Corian\`o, Claudio and Maglio, Matteo Maria and Theofilopoulos, Dimosthenis \\Dec 4, 2019, Published in: Eur.Phys.J.C 80 (2020) 6, 540 \\ \href{https://arxiv.org/abs/1912.01907}{\bfseries arXiv:1912.01907 }
   \end{itemize}
   and on the following proceedings
   \begin{itemize}
        \item Exact Correlators from Conformal Ward Identities in Momentum Space and Perturbative Realizations\\Corian\`o, Claudio and Maglio, Matteo Maria and Tatullo, Alessandro and Theofilopoulos, Dimosthenis\\ Apr 30, 2019, Published in: PoS CORFU2018 (2019) 072\\
    \href{https://arxiv.org/abs/1904.13174}{\bfseries arXiv:1904.13174  }
   \end{itemize}
   The following papers and proceedings were published during my PhD program but are not discussed in this thesis:
   \begin{itemize}
       \item An axion-like particle from an $SO(10)$ seesaw with $U (1)_X$\\
    Corian\`o, Claudio and Frampton, Paul H. and Tatullo, Alessandro and Theofilopoulos, Dimosthenis
    \\Jun 13, 2019, Published in: Phys.Lett.B 802 (2020) 135273,\\ \href{https://arxiv.org/abs/1906.05810}{\bfseries arXiv:1906.05810 }
    \item Dark Matter with Light and Ultralight St\"uckelberg Axions\\
    Corian\`o, Claudio and Maglio, Matteo Maria and Tatullo, Alessandro and Theofilopoulos, Dimosthenis\\May 5, 2020,Published in: PoS CORFU2019 (2020) 080 \\\href{https://arxiv.org/abs/2005.02292}{\bfseries arXiv:2005.02292}
   \end{itemize}
   \newpage

\tableofcontents
\newpage

\chapter*{Prologue}
Conformal Field Theory (CFT) has  gathered great attention over the last 20 and more years and there have been major advancements in the understanding and the use of CFT. The impact of CFTs spans from AdS/CFT to string theory, the studies of fixed points in Quantum Field Theories (QFTs) and in condensed matter physics. Naturally, most of these studies have to do with conformal correlators and their solutions in $d>2$ dimensions. For textbooks and reviews, we recommend \cite{Poland:2018epd, DiFrancesco:1997nk, Fradkin:1996is,Blumenhagen:2009zz}. While in $d=2$ conformal symmetry is infinite dimensional, in $d>2$ dimensions the algebra of the conformal group is forms the group  $SO(d,2)$. There has been a lot of interest in conformal correlators from different research areas of Physics such as particle physics and cosmology. Recently, due to experimental result coming from condensed matter physics regarding topological materials such as Weyl or Dirac semimetals there has been great effort to connect with CFTs \cite{Chernodub:2013kya,Chernodub:2019tsx,Ambrus:2019khr,Ambrus:2019khr1,Arjona:2019lxz,Gooth:2017mbd}. This is because these material can present anomalies, chiral or conformal, and these  can be studied with CFTs.\\ Regarding holography,  there has been an explosive interest towards the Sachdev-Ye-Kitaev (SYK)  model and CFT in $d=1$ \cite{Maldacena:2016hyu,Rosenhaus:2018dtp,Polchinski:2016xgd, Gross:2017aos}. This model has interest properties  and exhibits conformal symmetry which is spontaneously broken. It has been completely solved and this marks it as an extremely interesting candidate for a concrete AdS/CFT example. The gravity dual part has not been yet found but it has opened various roads and advancements in matrix models etc \cite{Saad:2019lba,Stanford:2019vob}. SYK is a model that fall into the category of models with generalized conformal structure \cite{Taylor:2017dly}. \\
 Conformal symmetry greatly constrains the form of the correlators. Two and three-point functions,scalar or tensorial are completely fixed module some constants. For the study of higher point function new methods have been introduced. The operator product expansion (OPE) is a tool as we will see that connects/bootstraps n-point function with lower point functions through conformal blocks. The bootstrap method is now considered on the most important topics in CFT and is also studied numerically \cite{Poland:2018epd,El-Showk:2012cjh}. There is an alternative way to solve correlation functions coming from path integral. As we know, each continuous symmetry imposes constrains in correlation functions through the Ward Identities. Thus, one can solve the Conformal Ward Identities (CWIs) and reproduce the already know results in a self-consistent and autonomous way. The study of CWIs is one of the main topics of this thesis. 
\section*{CFT in momentum space}
Most of the mentioned topics have been established in coordinate space.
The analysis of the conformal constraints in CFT's in momentum space provides new insight into the structure of the corresponding correlators. The advantages of study in momentum space are many. It allows a direct comparison between  general CFT predictions and those derived within the traditional S-matrix approach - based on the study of scattering amplitudes - widely investigated in a perturbative context. \\
Up to 3-point functions, the conformal Ward identities (CWI) are sufficient to fix all the correlators in terms only of the conformal data, which amount to a set of constants. A similar analysis of higher point functions is far more demanding, since it requires the use of the operator product expansion and the study of conformal partial waves (conformal blocks) associated to a given CFT \cite{Poland:2018epd,Dolan:2000ut}.\\
By turning to momentum space, even the analysis of 3-point functions becomes nontrivial, and one has to proceed with a  reformulation of the action of the conformal generators in these new variables, which show the hypergeometric nature of the solutions of the CWIs 
\cite{Coriano:2013jba, Bzowski:2013sza, Bzowski:2019kwd, Bzowski:2018fql,Coriano:2018bsy1,Coriano:2018bbe,Maglio:2019grh}. These can all be 
reformulated as systems of partial differential equations (pde's), whose solutions are linear combinations of Appell functions ($F_4$), which are hypergeometric functions of 2 variables. \\
In tensor correlators, by appropriate shifts of the parameters of such solutions, it is possible to solve for all the form factors \cite{Coriano:2018bbe, Coriano:2018bsy1} of a given tensorial parameterization. Equivalently, one can map such solutions to parametric integrals (3K integrals) of Bessel functions \cite{Bzowski:2013sza}. \\
There are four fundamental solutions of a hypergeometric system of pde's generated by the CWIs of a scalar 3-point function, as discussed in \cite{Coriano:2013jba}. Any other solution, obtained by requiring specific symmetries of the correlation function, is built around such a basis \cite{Coriano:2018bsy1,Coriano:2018bbe}. This holds also for inhomogeneous systems, as illustrated for nontrivial correlators such as the $TJJ$, $TTT$, where the several form factors appearing in the tensor decomposition can all be determined explicitly in terms of few constants \cite{Bzowski:2013sza}
.\\
Regarding 4-point functions, CWIs cease to provide sufficient information for the complete identification of the corresponding correlators, and it is necessary to define a bootstrap program in momentum space which is consistent with the same CWIs, in analogy with coordinate space. Several  studies have widened the goal of this activity, addressing issues such as the use of conformal blocks/CP symmetric blocks (Polyakov blocks) \cite{Isono:2019ihz,Isono:2018rrb,Isono:2019wex} \cite{Chen:2019gka} , light-cone blocks \cite{Gillioz:2019lgs, Gillioz:2018mto,Gillioz:2018kwh}, analytic continuations to Lorentzian spacetimes \cite{Bautista:2019qxj} and spinning correlators. Also in another analysis the authors have explored the link to Witten diagrams within the AdS/CFT correspondence \cite{Anand:2019lkt,Albayrak:2019yve}. Furhermore, the extension of these investigations to de Sitter space has opened  new applications in cosmology \cite{Arkani-Hamed:2018kmz,Baumann:2019oyu,Arkani-Hamed:2017fdk,Benincasa:2019vqr,Benincasa:2018ssx} \cite{Kundu:2014gxa,Almeida:2017lrq} and in gravitational waves \cite{BeltranAlmeida:2019gku}. Finally, investigations of such correlators in Mellin space \cite{Penedones:2010ue,Fitzpatrick:2011ia} \cite{Gopakumar:2016cpb,Gopakumar:2016wkt} offer a new perspective on the bootstrap program both in flat and in curved space \cite{Sleight:2019mgd,Sleight:2019hfp} and connecting  momentum space and Mellin variables.  \\

\subsection*{CFTs and anomalies}
The second major advantage that CFTs in momentum space have to offer is the study of the conformal anomaly and interpretation of this phenomenon in a more physical way, respect to coordinate space.

 The correlators with the insertion of stress energy tensors play a special role in CFTs due to the presence of the conformal anomaly \cite{Coriano:2012wp}. Analysis of 4-point functions have been extended apart from  scalar correlators in flat \cite{Maglio:2019grh} \cite{Bzowski:2019kwd} and curved backgrounds \cite{Arkani-Hamed:2018kmz,Baumann:2019oyu} to also tensorial correlators \cite{mirko,tttt}
 CWIs of correlators with insertion of stress energy tensors have to reproduce the correct expression of the conformal anomaly which introduces significant complications. At the same time, in  coordinate space  
 the issues of the ultraviolet  behaviour at coincident spacetime points of the corresponding operators are not addressed. 
 In coordinate space, the problem has been investigated in few cases - for instance in the $TTT$ case - quite directly, by solving the CWIs separately in their homogenous and inhomogeneous (anomalous) forms, by adding to the homogeneous solution one extra contribution \cite{Osborn:1993cr}. \\
  Such additional contribution amounts to an ultralocal term in the corresponding correlation function, generated when all the coordinates of the operators coalesce \cite{Osborn:1993cr,Erdmenger:1996yc} and reproduced by a variation of the anomaly functional.\\
In this context, studies of such correlators in momentum space find significant guidance from free field theory realizations. For example, direct one-loop computations in classical conformal invariant theories 
(such as massless QED and QCD) indicate that the anomalous breaking of conformal symmetry is associated with the exchange of massless poles \cite{Giannotti:2008cv,Armillis:2009pq,Coriano:2018bsy1}. This special feature unifies both conformal and chiral anomalies, as found in supersymmetric studies \cite{Coriano:2014gja}, and it has been shown to be consistent with the solutions of the CWIs  of three point functions, such as the $TJJ$ \cite{Armillis:2009pq,Coriano:2018zdo} and the $TTT$ \cite{Coriano:2018bsy1}.
\section*{Organization of the thesis}
Here we will briefly expose the contents and the organization of the content of this thesis which  is divided in two parts. More details can be found in the starting of each chapter. In the first chapter, we will concentrate on CFT in coordinate space. As we have stated in the introduction we will lay the foundation of Conformal Field Theory in a brief manner but we will also demonstrate a method where by using the embedding formalism we can derive up to n-point scalar conformal correlators. This  section is based on \cite{Irges:2020lgp}.
Moving to the next chapter \ref{cftmom}, we proceed  with our analysis in momentum space and how the notions we have introduced in the previous chapter take form. In the third chapter, we illustrate the theory of the conformal anomalies and we proceed our analysis with the renormalization method through counterterms. We conclude the first part of the thesis by including some comments about the anomaly action, having prepared the ground to investigate four point correlation functions in part two.\\\\

The second part consists of the main research topic of this thesis. For a smoother progression, we will first present an overview of the four-point scalar correlator, based on \cite{Coriano:2020ees,Maglio:2019grh} in section \ref{OOOO}. After that we will move to tensorial correlators. The section \ref{tooo} is based on \cite{Coriano:2019nkw} and it will demonstrate the first attempt to study a conformal tensorial four-point correlator in momentum space. In this section,
we derive and analyze the conformal Ward identities (CWIs) of a tensor 4-point function of a generic CFT in momentum space. The correlator involves the stress-energy tensor $T$ and three scalar operators $O$ ($TOOO$).  We derive the structure of the corresponding CWIs in two different sets of variables, relevant for the analysis of the 1-to-3 (1 graviton $\to$ 3 scalars) and 2-to-2 (graviton + scalar $\to$ two scalars)  scattering processes. In both cases we discuss the structure of the equations and their possible behaviors in various asymptotic limits of the external invariants. A comparative analysis of the systems of equations for the $TOOO$ and those for the $OOOO$, both in the general (conformal) and dual-conformal/conformal (dcc) cases, is presented. We show that in all the cases the Lauricella functions are homogenous solutions of such systems of equations, also described as parametric 4K integrals of modified Bessel functions.\\
  Then, we will move to a much more complex four-point conformal correlator, the one made of four stress-energy tensors. The chapter \ref{tttt} is based on \cite{tttt}. We elaborate on the structure of the conformal anomaly effective action up to 4-th order, in an expansion in the gravitational fluctuations $(h)$ of the background metric, in the flat spacetime limit.  We discuss the renormalization of 4-point functions containing insertions of stress-energy tensors (4T), in conformal field theories in four spacetime dimensions with the goal of identifying the structure of  the anomaly action. We focus on a separation of the correlator into its transverse/traceless and longitudinal components, applied to the trace and conservation Ward identities (WI) in momentum space. These are sufficient  to identify, from their hierarchical structure, the anomaly contribution, without the need to proceed with a complete determination of all of its independent form factors. Renormalization induces sequential bilinear graviton-scalar mixings on single, double and multiple trace terms, corresponding to $R\square^{-1}$ interactions of the scalar curvature, with intermediate virtual massless exchanges. These dilaton-like terms couple to the conformal anomaly, as for the chiral anomalous WIs. We show that at 4T level a new traceless component appears after renormalization.
  Along with \cite{mirko}, this work was of the first to study this correlator in momentum space. More details will be given in the start of the corresponding section. Moreover, although chapters \ref{tooo} and \ref{renorm} deal with four-point functions in momentum space, each chapter has a different focus based on the structure of the correlator. Particularly, chapter \ref{tooo} studies particular solutions of the CWI in various limits while chapter \ref{renorm} focuses on the renormalization and the anomalous CWI identities of the correlator. In the final chapter based on \cite{Coriano:2022ftl}, we deepen our analysis on the topological terms that are involved in the renormalization and consequently corrections of gravitational theories and are connected with the analysis we have already illustrated in the the previous chapters. For that reason, in the end of each chapter we have added a section with conclusion regarding this particular chapter.
\part{Conformal Field Theory and the Conformal Anomaly}

\chapter{Conformal Field Theory in coordinate space}\label{CFTcord}
In the first chapter, we will concentrate on CFT in coordinate space. As we have stated in the introduction we will lay the foundation of Conformal Field Theory in a brief manner but we will also demonstrate a method where by using the embedding formalism we can derive up to n-point scalar conformal correlators. This  section is based on \cite{Irges:2020lgp}

 After a brief introduction to conformal transformations along with the conformal group and its representations we turn our focus in correlation function that exhibit conformal symmetry. We start the discussion  by giving an overview of the embedding formalism. Afterwards, we present a method where by using this formalism we can derive up to n-point scalar conformal correlators. This  section is based on \cite{Irges:2020lgp}. We end the section by giving a short recap of tensorial correlators and the Operator Product Expansion formalism.  In  section \ref{cwisection}, we provide a review of the Conformal Ward Identities (CWI) and by a simple example we show how they are used to derive the two and three point functions. Finally in section \ref{secstress}, we examine the consequences of conformal symmetry in the stress energy tensor and prepare the ground for the phenomenon of conformal anomaly.
\section{Conformal Transformations}
In this section we will provide a short review of conformal transformations. For more details one can look at \cite{Poland:2018epd, DiFrancesco:1997nk, Fradkin:1996is}. In a d-dimensional spacetime we define the conformal transformations as the transformations that leave invariant the metric $g_{\mu \nu}$ up to a local scale factor.
\begin{equation}
g_{\mu \nu}'(x')=\Lambda(x) g_{\mu \nu}(x) \, .
\end{equation}
For $\Lambda(x)=1$, we have isometries. In the flat space where $g_{\mu \nu}=\eta_{\mu \nu}$, the group of isometries is the Poincare group, a subgroup of the conformal group. The case where $\Lambda(x)=\mathrm{const.}$ corresponds to scale transformations/dilatations. \\
These transformations preserve the angle between intersecting curves. This is easy to check. Let's define the angle $\theta$ between two vector $x, y$ in a manifold $\mathcal{M}$ 
\begin{equation}
\mathrm{cos}\theta=\frac{x\cdot y}{\sqrt{x^2 y^2}}=\frac{g_{\m \n}x^\n y^\n}{\sqrt{g_{\k \l}x^\k x^\l g_{\r \s}x^\s y^\r}}.
\end{equation}
If we apply a conformal transformation, the new angle $\theta'$ is 
\begin{equation}
\mathrm{cos}\theta'=\frac{x\cdot y}{\sqrt{x^2 y^2}}=\frac{\L(x) g_{\m \n}x^\n y^\n}{\sqrt{\L(x)g_{\k \l}x^\k x^\l \L(x) g_{\r \s}y^\s y^\r}}=\mathrm{cos}\theta.
\end{equation}
An infinitesimal coordinate transformation is expressed as :
\begin{equation}
x^{\mu}\rightarrow x'^{\mu}=x^{\mu}+\epsilon^{\mu}(x)\, ,
\end{equation}
where $\epsilon^{\mu}(x)$ is very small. The metric is a (0,2) tensor and under coordinate transformations transforms as:
\begin{equation}
g_{\mu \nu}'(x')=\frac{\partial x^a}{\partial x'^{\mu}}\frac{\partial x^b}{\partial x'^{\nu}}g_{ab} \, .
\end{equation}
Under the previous infinitesimal transformation, we have:
\begin{equation}
\begin{split}
g_{\mu \nu}'&=(\delta^a{}_{\mu}-\partial_{\mu}\epsilon^a)(\delta^b{}_{\nu}-\partial_{\nu}\epsilon^b)g_{ab}\\&=g_{\mu \nu}-(\partial_{\mu}\epsilon_{\nu}+\partial_{\nu}\epsilon_{\mu})+\mathcal{O}(\epsilon^2) \, .
\end{split}
\end{equation}
With $\Lambda(x)\simeq 1- f(x)$, we get 
\begin{equation}\label{eq:f}
(\partial_{\mu}\epsilon_{\nu}+\partial_{\nu}\epsilon_{\mu})=f(x)g_{\mu\nu} \, .
\end{equation}
By taking the trace of the previous expression:
\begin{equation}
f(x)=\frac{2}{d}\partial_\rho \epsilon^{\rho} \, .
\end{equation}
For our purposes, we take $g_{\mu\nu}=\eta_{\mu \nu}=\mathrm{diag}(1,1,...,1)$. The treatment will be identical to Minkowski spacetime too. Once we take the derivative $\partial_{\rho}$ of \eqref{eq:f}, we have:
\begin{equation}
(\partial_{\rho}\partial_{\mu}\epsilon_{\nu}+\partial_{\rho}\partial_{\nu}\epsilon_{\mu})=(\partial_{\rho}f(x))g_{\mu\nu} \,  .
\end{equation}
Permuting $(\mu \leftrightarrow \rho)$ and $(\nu \leftrightarrow \rho)$, we get
\begin{equation}
2\partial_{\mu}\partial_{\nu}\epsilon_{\rho}=\eta_{\mu\rho}\partial_{\nu}f+\eta_{\nu \rho}\partial_{\mu}f-\eta_{\mu \nu}\partial_{\rho}f \, .
\end{equation}
Multiplying by $\eta^{\mu \nu}$ we get:
\begin{equation}
\begin{split}
2\eta^{\mu \nu}\partial_{\mu}\partial_{\nu}\epsilon_{\rho}&=\eta^{\mu \nu}\eta_{\mu\rho}\partial_{\nu}f+\eta^{\mu \nu}\eta_{\nu \rho}\partial_{\mu}f-\eta^{\mu \nu}\eta_{\mu \nu}\partial_{\rho}f\\
\Leftrightarrow 2\partial^2\epsilon_{\rho}&=(2-d)\partial_{\rho}f \, .
\end{split}
\end{equation}
Differentiating with $\partial_{\nu}$ and using $\partial^2$ of \eqref{eq:f}, we end up with:
\begin{equation}\label{eq:dim}
(2-d)\partial_{\mu}\partial_{\nu}f=\eta_{\mu \nu}\partial ^2 f\rightarrow(d-1)\partial^2f=0 \, .
\end{equation}
For the case $d=1$, we have no restrictions in $f$. The $d=2$ is special and we will not study this case in this thesis.\\
 For $d>2$, the above equation implies that $\partial_{\mu}\partial_{\nu}$ so $f$ is at most  linear in $x^{\mu}$.
\begin{equation}
f=A+B_{\mu}x^{\mu}\, ,
\end{equation}
with $A,B$ constants. So at the level of $\epsilon$, this implies that $\epsilon$ is at most quadratic in coordinates.
\begin{equation}
\epsilon_{\mu}=a_{\mu}+b_{\mu \nu}x^{\nu}+c_{\mu \nu \rho} x^{\nu}x^{\rho}\, ,
\end{equation}
with $c_{\mu \nu \rho}=c_{\mu \rho \nu} $. Plugging this in \eqref{eq:f} we get:
\begin{equation*}
\partial_{\mu}(a_{\nu}+b_{\mu \kappa}x^{\kappa}+c_{\mu \kappa \rho} x^{\kappa}x^{\rho})+\partial_{\nu}(a_{\mu}+b_{\mu \nu}x^{\nu}+c_{\mu \nu \rho} x^{\nu}x^{\rho})=\frac{2}{d}\partial_\lambda \epsilon_{\mu}\eta^{\mu \lambda}.
\end{equation*}
\newpage
Comparing the term of the right and left side of the equation we get:
\begin{itemize}
\item $a_{\mu}$ has no constraints and corresponds to infinitesimal translations.
\item $b_{\mu \nu}=\alpha \eta_{\mu \nu}+ m_{\mu \nu}$ with $m_{\mu \nu}=-m_{ \nu \mu}$. The trace part corresponds to an infinitesimal scale transformation $\alpha=\frac{1}{d}b^{\lambda}{}_{\lambda}$ and $m_{\mu \nu}$ corresponds to an infinitesimal Lorentz rotation.
\item $c_{\mu \rho \nu}=\eta_{\mu \rho}b_{\nu}+\eta_{\mu \nu}b_{\rho}-\eta_{\nu \rho}b_{\mu}$ with $b_{\mu}=\frac{1}{d}c^{\kappa}{}_{\kappa \mu}$ a constant vector.
\end{itemize}
The last transformation is called Special Conformal Transformation (SCT) and acts on coordinates as:
\begin{equation}
x^{\mu}\rightarrow x'^{\mu}=x^{\mu}+2(x \cdot b)x^{\mu}-b^{\mu}x^2 \, .
\end{equation}
\subsubsection{Analysis of the special conformal transformation}
We now that if we apply an infinitesimal special conformal transformation(SCT) the change in the coordinates is
\begin{equation}\label{infi}
\d x^{\m}= 2(x \cdot b)x^{\m}-x^2 b^{\m}.
\end{equation}
The first observation is that is is not linear in $x$. To find the finite form of the transformation we have to apply \eqref{infi} multiple times. The usual way is to integrate the infinitesimal form. The other way, and since we know that the transformations satisfy the conformal Killing equation, is to find the integral curve of the corresponding conformal Killing vector field( since they are equal.
In other words this mean to solve the differential equation
\begin{equation}\label{Dif}
\dot{x}^{\m}=\frac{d x^{\m}}{d t}=X^{\m}(x)=2(x \cdot b)x^{\m}-x^2 b^{\m}.
\end{equation}
 Instead of solving this non-linear ODE we make the substitute $y^{\m}=\frac{x^{\m}}{x^2}$, where y has the form of an inversion. We have
\begin{equation}
\dot{y}^{\m}=\frac{\dot{x}^{\m} x^2-2x^2 \dot{x}^{\m} }{(x^2)^2}=-b^{\m}.
\end{equation}
So instead of solving \eqref{Dif}, we solve $\dot{y}^{\m}=-b^{\m}$. The solution is clearly 
\begin{equation}
y^{\mu}(t)=y_0^{\m}-b^{\m}t.
\end{equation}
Now we substitute again we get in terms of $x^{\mu}$ and now we have 
\begin{equation}\label{FinitealmostSCT}
\frac{x^{\m}}{x^2}=\frac{x^{\m}_0}{x_0^2}-b^{\m}t.
\end{equation}
We almost have the finite SCT but the $1/x^2$ has to eliminated. There are two ways to do it:
\begin{itemize}
\item Squaring both sides of \eqref{FinitealmostSCT}, we get:
\begin{equation*}
1/x^2=\frac{1}{x_0^2}+b^2 t^2-\frac{2x_0\cdot b t}{x_0^2}=1-2 (x_0\cdot b) t+(2b t)^2 x^2.
\end{equation*}
Substituting in \eqref{FinitealmostSCT} we end up with the desired form 
\begin{equation}\label{SCT}
x'^{\m}(t)=\frac{x_0^{\m}-x_0^2 b^{\mu} t}{1-2 (x_0\cdot b) t+2(b t)^2 x_0^2} .
\end{equation}
We can see also that for $t=0$ we have $x'^{\m}(0)=x_0^{\mu}$ which is an important property that the integral curve has to satisfy.
\item We can also solve the equation $\frac{x^{\mu}}{x^2}=A^{\mu}$. The solution is $$ x^{\m}=A^{\m}/A^2 $$ For $A^{\mu}=\frac{x^{\m}_0}{x_0^2}-b^{\m}t$ we end up again with \eqref{SCT}.
\end{itemize}
To double check \eqref{SCT},  we can Taylor expand \eqref{SCT} around $b^{\m}\approx 0$ to get the infinitesimal transformation
\begin{equation}
x'^{\mu}(b^{\m}\approx 0)=x'^{\mu}(0)+b^{\nu}\frac{\partial x'^{\mu}}{\partial b^{\nu}}|_{b=0}+\mathcal{O}(b^2)=x^{\mu}+b^{\nu}\left(2x^{\mu}x_{\nu}-\delta^{\mu}_{\nu} x^2\right)=x_0^{\mu}+2(b\cdot x)x^{\mu}-b^{\mu}x^2.
\end{equation}
So we confirmed that the infinitesimal transformation is 
\begin{equation}
\d x^{\m}= 2(x \cdot b)x^{\m}-x^2 b^{\m}.
\end{equation}
If we introduce the inversion transformation $I$ such that
\begin{equation}
I: x^{\mu}\rightarrow x'^{\mu}=\frac{x^{\mu}}{x^2}\, ,
\end{equation}
with $I^2=1$ we can see that the SCT is equivalent with performing an inversion followed by a translation and then another inversion.\\\\
The finite versions of the above transformations are
\begin{itemize}
\item Translations: $x'^{\mu}=x^{\mu}+\alpha^{\mu}.$
\item Dilatations:  $x'^{\mu}=ax^{\mu}.$
\item Rotations: $x'^{\mu}=M^{\mu}_{\nu}x^{\nu}.$
\item SCT: $x'^{\mu}=\frac{x^{\mu}-b^{\mu}x^2}{1-b\cdot x +b^2x^2}.$
\end{itemize}

Finally, counting the generators  in d dimensions we have: $d$ generators from translations, $d$ generators from dilatations, $\frac{d(d-1)}{2}$ generators from rotations and $d$ generators from SCT. So, we have $\frac{(d+1)(d+2)}{2}$ generators.
\section{Conformal Group}
 The composition of conformal  transformations yields another conformal transformation and every conformal transformation has an inverse that is a conformal transformation too. The conformal transformations posses the structure of a group.
 
 To study the algebra of the group generators, we will  construct a representation of conformal generators that act on  scalar fields. 

Given a conformal transformation $x\rightarrow x'=x'(x)$ we define the action on fields $\Phi(x)$ as
\begin{equation}
\Phi(x')\equiv \Phi(x) \, .
\end{equation}
We can always write an infinitesimal coordinate transformation as
\begin{equation}
x'^{\mu}=x^{\mu}+ \omega_{\alpha} \frac{\delta x^{\mu}}{\delta \omega_{\alpha}} \, ,
\end{equation}
where $\omega_{\alpha}$ is very small. We define the generators $G_{\alpha}$ of such transformation as
\begin{equation}
\delta_{\omega} \Phi(x)=\Phi'(x)-\Phi(x)=-i\omega_{\alpha}G_{\alpha}\Phi(x) \, .
\end{equation}
Taylor expanding $\Phi'(x)$ we have
\begin{equation}
\Phi'(x)=\Phi(x-\omega_{\alpha} \frac{\delta x^{\mu}}{\delta \omega_{\alpha}})\simeq \Phi(x) -\omega_{\alpha} \frac{\delta x^{\mu}}{\delta \omega_{\alpha}}\partial_{\mu}\Phi(x) \, .
\end{equation}
In the end we have
\begin{equation}\label{gen}
iG_{\alpha}\Phi(x)= \frac{\delta x^{\mu}}{\delta \omega_{\alpha}}\partial_{\mu}\Phi(x) \, .
\end{equation}
We focus now on the conformal transformations. For translations, $x'^{\mu}=x^{\mu}+\omega^{\nu}\delta^{\mu}{}_{\nu}$. Thus, $\frac{\delta x^{\mu}}{\delta \omega_{\alpha}}=\delta^{\mu}{}_{\nu}$. Putting this in the equation \eqref{gen}, we find that the the generator for the translations is
\begin{equation}
P_{\nu}=-i\partial_{\nu} \, .
\end{equation}
An infinitesimal Lorentz transformation can be written as
\begin{equation}
\begin{split}
x'^{\mu}&=x^{\mu}+\omega^{\mu}{}_{\nu}x^{\nu}=x^{\mu}+\omega_{\rho \nu}\eta^{\rho \mu}x^{\nu}\\ &= x^{\mu}+\frac{1}{2}(\omega_{\rho \nu}\eta^{\rho \mu}x^{\nu}+\omega_{\nu \rho}\eta^{\nu \mu}x^{\rho}) \, ,
\end{split}
\end{equation}
where $\omega_{\mu \nu}=-\omega_{ \nu \mu}$. Using this relation we find that $$\frac{\delta x^{\mu}}{\delta \omega_{\rho \nu}}=\frac{1}{2}(\eta^{\rho \mu}x^{\nu}-\eta^{\nu \mu}x^{\rho}).$$
The generator that corresponds to Lorentz transformation is given by the expression
\begin{equation}
iG_{\rho \nu}\Phi(x)= \frac{1}{2}(\eta^{\rho \mu}x^{\nu}-\eta^{\nu \mu}x^{\rho})\partial_{\mu}=\frac{1}{2}(x^{\nu}\partial_{\rho}-x^{\rho}\partial_{\nu}) \, .
\end{equation}
Thus, we get the familiar expression
\begin{equation}\label{eq:genlambda}
L_{\mu \nu}=i(x_{\mu}\partial_{\nu}-x_{\nu}\partial_{\mu}) \, .
\end{equation}
For a dilatation, we write the infinitesimal transformation as
\begin{equation}
x'^{\mu}=x^{\mu}+\alpha x^{\mu}=x^{\mu}+\alpha \eta^{\mu\nu} x_{\nu} \, .
\end{equation}
We can easily see that the generator of dilatations is given by
\begin{equation}
D=-i x^{\mu} \partial_{\mu}  .
\end{equation}
We summarize the generators of the conformal algebra as differential operators acting on functions
\begin{itemize}
\item Translations: $P_{\mu}=-i\partial_{\mu}.$
\item Rotations: $L_{\mu \nu}=i(x_{\mu}\partial_{\nu}-x_{\nu}\partial_{\mu}).$
\item Dilatations: $D=-i x^{\mu} \partial_{\mu}.$
\item Special CT: $K_{\mu} -i(2x_{\mu}x^{\nu}\partial_{\nu} -x^2\partial_{\mu}).$
\end{itemize}
We can now obtain the conformal algebra
\begin{align*}
&[D,P_{\mu}]=iP_{\mu}.\\
&[D,K_{\mu}] =-iK_{\mu}.\\
&[K_{\mu},P_{\nu}]=2i(\eta_{\mu \nu}D-L_{\mu \nu}).\\
&[L_{\mu \nu},P_{\rho}]=-i(\eta_{\mu \rho}P_{\nu}-\eta_{\nu \rho} P_{\mu}).\\
&[L_{\mu \nu},K_{\rho}]=-i(\eta_{\mu \rho}K_{\nu}-\eta_{\nu \rho} K_{\mu}).\\
&[L_{\mu \nu},L_{\rho \sigma}]=-i(L_{\mu \rho}\eta_{\nu \sigma}-L_{\mu \sigma}\eta_{\nu \rho}-L_{\nu \rho}\eta_{\mu \sigma}+L_{\nu \sigma}\eta_{\mu \rho}).\\
&[D,L_{\mu \nu}]=0.\\
&[P_{\mu},P_{\nu}]=0.\\
&[K_{\mu},K_{\nu}]=0.\\
&[D,D]=0
\end{align*}

We define the above generators
$$J_{\mu \nu}=L_{\mu \nu} , \quad   J_{-1,\mu}=\frac{1}{2}(P_{\mu}-K_{\mu}).$$
$$J_{-1,0}=D, \quad   J_{0,\mu}=\frac{1}{2}(P_{\mu}+K_{\mu}).$$

and $J_{a,b}=-J_{b,a}$ with $a,b \in \{-1,0,1..,d \}$. The above generators satisfy the following algebra
\begin{equation}\label{lorentzgroup}
[J_{ab},J_{cd}]=i(\eta_{ad}J_{bc}+\eta_{bc}J_{ad}-\eta_{ac}J_{bd}-\eta_{bd}J_{ac}) \, ,
\end{equation}
with $\eta_{ab}=\mathrm{diag}(-1,1,...,1)$.

The above relations show the isomorphism between the conformal group in $d$ dimensions and the $ SO (d+1,1)$ or ($SO(d,2)$ in Minkowski space) which have $\frac{(d+1)(d+2)}{2}$ generators too. This fact is crucial for establishing the embedding space formalism. It is important to notice that the Poincare group together with translations form a subgroup of the full conformal group. This means that a theory invariant under rotations, translations and dilatations is not necessarily invariant under the special conformal transformations.
\subsection{Represantations of the Conformal Group}
We will demonstrate the act of the conformal group in operators.  In the Heisenberg picture the spacetime dependence of a multicomponent operator $\phi_{\alpha}(x)$ is given by
\begin{equation}\label{eq:trans}
\phi_{\alpha}(x)=e^{-iPx}\phi_{\alpha}(0)e^{+iPx} \, .
\end{equation}
By taking the derivative we obtain the action of the translation generator on the operator
\begin{equation}
[P_{\mu},\phi_{\alpha}(x)]=i\partial_{\mu}\phi_{\alpha}(x) \, .
\end{equation}
To find the action of the remaining operator, we focus in the stability group, the group that leaves the origin invariant. In case of the conformal group, it is spanned by Lorentz rotations, dilatations and special conformal transformations. We define the actions of these operators at the origin
\begin{align}
[D,\phi_{\alpha}(0)]&=i\Delta \phi_{\alpha}(0).\\
[L_{\mu \nu},\phi_{\alpha}(0)]&=i(S_{\mu \nu})^{\beta}_{\alpha} \phi_{\beta}(0).\\
[K_{\mu},\phi_{\alpha}(0)]&=0.
\end{align}
where $\Delta$ is the scaling dimension and $S_{\mu \nu}$ is a spin associated matrix which is zero for scalar fields. The above transformations are the definition for a primary operator of scaling dimension $\Delta$. A primary operator is an operator that is annihilated by a special conformal transformation at the origin.

With the use of \eqref{eq:trans} combined with the conformal algebra we are able to derive the action the action of the conformal generators on $\phi_{\alpha}(x)$. We will also use the Haussdorff formula
$$e^{-A}Be^{A}=B+[B,A]+\frac{1}{2!}[[B,A],A]+...$$
For the dilatation generator we get
\begin{equation}
\begin{split}
[D,\phi_{\alpha}(x)]&=De^{-iPx}\phi_{\alpha}(0)e^{iPx}-e^{-iPx}\phi_{\alpha}(0)e^{iPx}D\\
&=e^{-iPx}e^{iPx}De^{-iPx}\phi_{\alpha}(0)e^{iPx}-e^{-iPx}\phi_{\alpha}(0)e^{iPx}De^{-iPx}e^{iPx}\\
&=e^{-iPx}\hat{D}\phi_{\alpha}(0)e^{iPx}-e^{-iPx}\phi_{\alpha}(0)\hat{D}e^{iPx}\\
&=e^{-iPx}[\hat{D},\phi_{\alpha}(0)]e^{iPx} \, ,
\end{split}
\end{equation}
where we have defined $\hat{D}=e^{iPx}De^{-iPx}$. Using the Haussdorff formula we find
\begin{equation}
\hat{D}=D+ix^{\mu}[P_{\mu},D]+...=D+x^{\mu}P_{\mu} \, .
\end{equation}
Thus, we obtain
\begin{equation}
[D,\phi_{\alpha}(x)]=i(\Delta+x^{\mu}\partial_{\mu})\phi_{\alpha}(x) \, .
\end{equation}
For the generator of rotations we get
\begin{equation}
[L_{\mu \nu}, \phi_{\alpha}(x)]=e^{-iPx}[\hat{L_{\mu \nu}},\phi_{\alpha}(0)]e^{iPx} \, .
\end{equation}
Finally, 
\begin{equation}
[L_{\mu \nu}, \phi_{\alpha}(x)]=-i(x_{\mu}\partial_{\nu}-x_{\nu}\partial_{\mu})\phi_{\alpha}(x)+i(S_{\mu \nu})_{\alpha \beta}\phi^{\beta}(x) \, .
\end{equation}
Following the same procedure for the generator of special conformal transformations we find
\begin{equation}
[K_{\mu},\phi_{\alpha}(x)]=2ix_{\mu}\Delta\phi_{\alpha}(x)+i(2x_{\mu}x^{\nu}\partial_{\nu}-x^2\partial_{\mu})\phi_{\alpha}(x)+2ix^{\rho}(S_{\rho \mu})_{\alpha \beta}\phi_{\beta}(x) \, .
\end{equation}
A primary scalar field $\Phi(x)$ of scaling dimension $\Delta$ transforms as 
\begin{equation}
\Phi(x')\rightarrow \Phi'(x')=\left| \frac{\partial x'}{\partial x}\right|^{-\Delta /d} \Phi(x) \, .
\end{equation}
The Jacobian of the transformation is given by
\begin{equation}
\left| \frac{\partial x'}{\partial x}\right|=\Lambda(x)^{-d/2} \, .
\end{equation}
Thus, we have
\begin{equation}
\Phi'(x')=\Lambda (x)^{\Delta/2}\Phi(x) \, .
\end{equation}
Since  $\delta g_{\m \n}= f(x) \delta_{\m \n}$  the Jacobian of the conformal transformation is
\begin{equation}
J=\frac{\partial x'^\m}{\partial x^n}= \sqrt{f(x)} M^\m_\n=b(x) M^\m_\n,
\end{equation}
where $ M^\m_\n$ is an orthogonal matrix of the rotation $SO(D)$ group.
Consequently, an operator in an irreducible representation of  $SO(D)$ will transform as
\begin{equation}
\phi'(x')=b^{\Delta}R[M^\m_\n]\phi.
\end{equation}
For a scalar field $R[M^\m_\n]=1$ while for a spin 1 $R[M^\m_\n]=M^\m_\n$. A stress-energy tensor ,which has scaling dimensions $\Delta=d$, transforms as 
\begin{equation}
T'^{\m \n}=b^{d }M^\m_\a M^\n_\b T^{\m \n}
\end{equation}
Using the infinitesimal form of the matrices we arrive at 
\begin{equation}
	\delta T^{\mu\nu}(x)=-d\, f(x)\, T^{\mu\nu} -\epsilon\cdot \partial \,T^{\mu\nu}(x) +
	\frac{1}{2}\partial_{[\alpha }\epsilon_{\mu]}\,T^{\alpha\nu} +\frac{1}{2}\partial_{[\nu }\epsilon_{\alpha]}T^{\mu\alpha}.
\end{equation}
For a special conformal transformation  
\begin{equation}
	\epsilon_{\mu}(x)=b_\mu x^2 -2 x_\mu b\cdot x
\end{equation}
with a generic parameter $b_\mu$ and  $f(x)=\frac{2}{d}\partial_\rho \epsilon^{\rho}$ to obtain
\begin{equation}
	\delta T^{\mu\nu}(x)=-(b^\alpha x^2 -2 x^\alpha b\cdot x )\, \partial_\alpha  T^{\mu\nu}(x)   - d f(x) T^{\mu\nu}(x)+
	2(b^\mu x_\alpha- b_\alpha x^\mu)T^{\alpha\nu} + 2 (b^\nu x_\alpha -b_\alpha x^\nu)\, T^{\mu\alpha}(x).
\end{equation}
It is sufficient to differentiate this expression respect to $b_\kappa$ in order to derive the form of the special conformal transformation $K^\kappa$ on $T$ in its finite form 
\begin{align}
	\mathcal{K}^\kappa T^{\mu\nu}(x)&\equiv\delta_\kappa T^{\mu\nu}(x) =\frac{\partial}{\partial b_\kappa} (\delta T^{\mu\nu})\notag\\
	&= -(x^2 \partial^\kappa - 2 x^\kappa x\cdot \partial) T^{\mu\nu}(x) + 2\Delta_T x^\kappa T^{\mu\nu}(x) +
	2(\delta_{\mu\kappa}x_\alpha -\delta_{\alpha}^{\kappa}x_\mu) T^{\alpha\nu}(x) + 2 (\delta^\kappa_\nu x_{\alpha} -\delta_\alpha^\kappa x_\nu )T^{\mu\alpha}. 
	\label{ith1}
\end{align}
\section{Conformal correlation functions}
\subsection{The embedding formalism}
We will briefly review the embedding formalism that we are going to use to constrain the conformal correlation function. More details can be found in \cite{Weinberg:2010fx, Weinberg:2012mz, Poland:2018epd}. \\
We have seen that the algebra of the conformal group \eqref{lorentzgroup} forms the group $SO(d,2)$. Based on this, we can embed the four dimensional system in a flat spacetime of two dimensions higher with metric signature $(-,+,+,+,+,-)$. The coordinates are denoted as $y^A$, $A=\m,5,6$ and 
project back to the original 4d space by the null-cone condition $y^A y_A=0$ and the identification for the $4d$ coordinates
\begin{equation}
x^\m = \frac{y^\m}{y^+}\, , \hskip 1cm y^+ = y_5^+ + y^+_6.
\end{equation}
The conformal transformations in the projective space are
\begin{equation}
y^A\rightarrow\L^A_B x^B
\end{equation}
and satisfy the following conditions
\begin{equation}
L^A_C L^B_D \eta^{CD}=\eta^{AB}, \quad Det\L=1.
\end{equation}
The algebra of the generators satisfies the following commutations relation 
\begin{equation}
i[J^{KL},J^{MN}]=\eta^{ML}J^{KN}-\eta^{KM}J^{LN}-\eta^{NL}J^{KM}+\eta^{KN}J^{ML},
\end{equation}
where the conformal generators are
\begin{equation}
P^{\m}=J^{5\m}+J^{6\m}, \quad K^{\m}=-J^{5\m}+J^{6\m}, \quad D=J^{65}.
\end{equation}
Since in the projective space the conformal transformation are rotations and/or boosts along with the null cone conditions, the only non-zero invariants are the inner products of the coordinates
\begin{align}
y_i\cdot y_j&=y_{i,\m} y_j^\m+\frac{1}{2}(y_i^5-y_i^6)y_j^+\frac{1}{2}y_i^+(y_j^5-y_j^6)\notag \\  &
= -\frac{1}{2} (y_i^+ y_j^+) (x_i-x_j)^2 .
\end{align}

Also, it has been show \cite{Weinberg:2010fx} that the field ${\cal O}_q(x) = (y^+)^{\Delta_q} \Phi (x,y^+)$ depend only on $x$, have scaling dimension $\Delta_q$ and for a
conformally invariant correlator the $\Phi$'s must contribute to it terms proportional to the product of all possible inner products
$y_i\cdot y_j$.
\subsection{The conformal N-point scalar correlation function}\label{npoint}
In this section we are going to derive the conformal n-point scalar function using a general method that exploits the embedding space formalism\cite{Irges:2020lgp}. Based on the previous result, the 4-point function for example must be of the form
\begin{equation}
\langle {\cal O}_1(x_1){\cal O}_2(x_2){\cal O}_3(x_3){\cal O}_4(x_4) \rangle = \frac{\prod_{a=1,\cdots,4}\left(y^+_a\right)^{\Delta_a}}{\prod_{i,j} \left(y_i\cdot y_j\right)^{e_{ij}}}\, ,
\hskip .75cm 1,2,3 =i < j=2,\cdots,4
\end{equation}
Imposing the self consistency condition that the right hand side is $y_a^+$-independent and restricting to identical scalar operators,
we arrive at \begin{equation}\label{4cor}
\langle {\cal O}(x_1){\cal O}(x_2){\cal O}(x_3){\cal O}(x_4) \rangle = R_4\, g(u,v)\, ,\hskip 1cm R_4=\frac{1}{x_{12}^{2\D}x_{34}^{2\D}}\, ,
\end{equation}
where $x_{ij}=|x_i-x_j|$.The conformally invariant cross-ratios $u$ and $v$ are 
\begin{equation}
u= \frac{x_{12}x_{34}}{x_{13}x_{24}}, \hskip 1cm v= \frac{x_{23}x_{14}}{x_{13}x_{24}}\, .
\end{equation}
The function $g(u,v)$ remains unconstrained by the
conformal symmetry itself, but satisfies additional relations, obtained by the requirement that the correlator, in Euclidean space, is symmetric
under the interchange $x_i \leftrightarrow x_j$, symbolized by the notation $(i\, j)$. The action $(i\, j)$ on a function of coordinates 
induces an action denoted as $g_{ij}$. Invariance of the correlator under all possible such exchanges imposes the two exchange-symmetry constraints
\begin{equation}\label{4cross}
g(u,v) = g\left(\frac{u}{v}, \frac{1}{v}\right), \hskip 1cm g(u,v) = \left(\frac{u}{v}\right)^{2\D} g(v,u)\, .
\end{equation}
These are (the only) independent constraints on $g$. More constrains could come from the OPE technique.
Also, $R_4$ and $g(u,v)$ are both and separately conformally invariant, so that the correlator in \ref{4cor}
can be seen as being factorized in coordinate space, in the product of at least two invariant substructures.\\
This methodology can be straightforwardly generalized. For the correlator of $N$ scalar operators
\begin{equation}\label{Ncor}
\langle {\cal O}_1(x_1)\cdots {\cal O}_N(x_N) \rangle \sim \frac{\prod_{a=1,\cdots,N}\left(y^+_a\right)^{\Delta_a}}{\prod_{i,j} \left(y_i\cdot y_j\right)^{e_{ij}}}\, ,
\hskip .75cm 1,\cdots, N-1 =i < j
\end{equation}
 the conditions that restrict its form come from the requirement of its independence from the $y^+$'s, as before.
To extract these so we must decide which $e_{ij}$ to solve for.
Since we have $N$ equations, we have to pick $N$ exponents.
Any loss of generality involved in this choice will be lifted by the exchange-symmetry constraints.
A convenient choice is to solve for $e_{1i}, \, i=2,\cdots,N$ and $e_{23}$.
Defining the vectors $E=(e_{12},\cdots, e_{1N},e_{23})$, $D=(\Delta_1,\Delta_i - (\s^N_i + \r_2^i))$ the equation
to be solved for $E$ is ($T$ stands for transpose)
\begin{equation}\label{system}
M E^T = D^T\, .
\end{equation}
In the above we have defined the partial sums
\begin{align}
\s_i^N &= e_{i,i+1} + e_{i,i+2}+\cdots + e_{i,N} \nonumber\\
\r_2^i &= e_{2,i}+e_{3,i}+\cdots + e_{i-1,i}
\end{align}
 where $i,j=2,\cdots,N$ and $\s_i^j$ and $\r_i^j $ are non-zero only when $i<j$. The matrix $M$ is 
 \begin{equation}
 M = \begin{bmatrix}
1 & 1 & 1 & \cdots & 1 & 1 & 0  \\
1 & 0 & 0 & \cdots & 0 & 0 & 1  \\
0 & 1 & 0 & \cdots & 0 & 0 & 1  \\
0 & 0 & 1 & \cdots & 0 & 0 & 0  \\
. & . & . & \cdots & . & . & .  \\
. & . & . & \cdots & . &. & .  \\
. & . & . & \cdots & . & . & .  \\
0 & 0 & 0 & \cdots & 1 & 0 & 0  \\
0 & 0 & 0 & \cdots & 0 & 1 & 0  \\
\end{bmatrix}\, .
 \end{equation}
For $N\ge d+3$ correlators degeneracies originating from the necessary linear dependence of some of the $y_i$
may start to arise that must be dealt with \cite{Weinberg:2010fx, Weinberg:2012mz,Costa:2011dw}. They appear by making the above matrix have some linearly dependent rows and columns.
In this case these rows/columns must be moved to the right hand side of \eq{system}, into ${D}$. As a result,
some of the exponents in $E$ will not be independent. We will not complicate our analysis any further by such a possibility since apart from this technicality the
logic is the same as for the non-degenerate case $N < d+3$.
The easiest way to solve this system of equations is to discard the first row and last column, which leaves an
$N-1$ dimensional unit submatrix in $M$, trivially invertible. The solution is given though in terms of $e_{23}$ due to the
missing row and column. Fortunately we can solve for $e_{23}$ separately, by combining for example the sum of all $N-1$
equations with the constraint that comes from the observation that \eq{Ncor} must be invariant under the trivial rescaling $y_i\rightarrow \l y_i$.
The result is
\begin{equation}\label{e23}
2e_{23} = -\Delta_1 + \Delta_2 + \cdots + \Delta_N - 2(\r_2^4 + \cdots + \r_2^N)
\end{equation}
and then the $N-1$ dimensional system of equations collapses to 
\begin{equation}
2e_{1i} = 2\Delta_i -2 (\s^N_i + \r_2^i)\, ,
\end{equation}
where the only thing to remember is to substitute for $e_{23}$ from \eq{e23} when it appears in either $\s^N_i $ or $\r_2^i$
which occurs twice, once in $e_{12}$ and once in $e_{13}$.
It is illuminating to show the explicit form of the final solution:
\begin{align}
2e_{23}&= (-\Delta_1+\Delta_2+\cdots + \Delta_N) -2 (\r_2^4+\cdots+\r_2^N)\nonumber\\
2e_{12} &= 2\Delta_2 - (-\Delta_1+\Delta_2+\cdots + \Delta_N) - 2 {\hat \s}_2^N + 2(\r_2^4+\cdots+\r_2^N)\nonumber\\
2e_{13} &= 2\Delta_3 - (-\Delta_1+\Delta_2+\cdots + \Delta_N) - 2 {\s}_3^N + 2(\r_2^4+\cdots+\r_2^N)\nonumber\\
2e_{1i} &= 2\Delta_i - 2(\s_i^N + \r_2^i)\, ,\hskip .75cm i=4,\cdots, N \nonumber\\
\end{align}
where ${\hat \s}_2^N = \s_2^N-e_{23}$. Then 
\begin{equation}
\langle {\cal O}_1(x_1)\cdots {\cal O}_N(x_N) \rangle = \frac{1}{x_{12}^{2e_{12}}\cdots x_{1N}^{2e_{1N}} x_{23}^{2e_{23}}} \frac{1}{\prod_{ij}x_{ij}^{2e_{ij}}}
\end{equation}
 where in the product: $i=2,\cdots, N-1$ and $j=i+1,\cdots N$ and $ij\ne 23$. 
To prepare this expression for an exchange-symmetry analysis we first define
\begin{align}\label{exponents}
\Delta_{23} &= -\Delta_1+\Delta_2+\cdots + \Delta_N\nonumber\\
\Delta_{12} &= 2\Delta_2-(-\Delta_1+\Delta_2+\cdots + \Delta_N)\nonumber\\
\Delta_{13} &= 2\Delta_3-(-\Delta_1+\Delta_2+\cdots + \Delta_N)\nonumber\\
\Delta_{1i} &= 2\Delta_i\, , \hskip .75cm i=4,\cdots,N
\end{align}
and write the correlator as
\begin{align}\label{Ncor2}
\langle {\cal O}_1(x_1)\cdots {\cal O}_N(x_N) \rangle = R_N\, 
\frac{x_{23}^{2 (\r_2^4+\cdots+\r_2^N)} \prod_{i=4}^N x_{1i}^{2(\s_i^N + \r_2^i)} }{x_{12}^{- 2 {\hat \s}_2^N + 2(\r_2^4+\cdots+\r_2^N)}x_{13}^{- 2 {\s}_3^N + 2(\r_2^4+\cdots+\r_2^N)}   
\prod_{ij}x_{ij}^{2e_{ij}}}
\end{align}
with the same restrictions on the $i,j$ indices as above and
\begin{equation}
R_N = \frac{1}{x_{23}^{\Delta_{23}}\prod_{a=2}^{N} x_{1a}^{\Delta_{1a}}}\, .
\end{equation}
The geometric interpretation says that if we think of the correlator as a sort of a representation of a discrete metric on the points $\{x_1,\cdots,x_N\}$ then 
$R_N$ is its radial part, the rest is the angular part and rotations correspond to transformations that exchange $x_i\leftrightarrow x_j$. 
When the $N$ operators are distinct, the gauging of the exchange-symmetry group does not leave the triangle defined by any three points invariant (apart from the identity action)
and the radial part $R_N$ will contain a $123$ sector, corresponding to the triangle defined by $x_1,x_2,x_3$. For $N=3$ this is a conformally invariant structure.
When the operators in the correlator are identical, it may happen that a non-trivial (not an identity) combination of the exchange-symmetry group elements
leaves the $123$ triangle invariant and then the corresponding sector has no reason to appear in $R_N$. 
Instead, all information for structures built from triangles is contained in the angular part $f$. Such is the case of the $N=4$ correlator of identical scalars.

The statement of exchange-symmetry (in Euclidean $x$-space) is that 
\begin{equation}
g_{1a} \langle {\cal O}_{1}(x_1)\cdots {\cal O}_{N}(x_N) \rangle = \langle {\cal O}_{1}(x_1)\cdots {\cal O}_{N}(x_N) \rangle\, , \hskip 1cm a=2,\cdots,N
\end{equation}
since the $(1\, a)$ generate the permutation group $S_N$. We also define here the important quantity
\begin{equation}
J_a = R_N^{-1}\, (g_{1a} R_N)\, ,
\end{equation}
a sort of discrete version of a Jacobian, originating from the transformation induced by the $g_{1a}$.
The only ingredient we are missing are the conformally invariant cross-ratios. These can be straightforwardly obtained from
\eq{Ncor2} by collecting all the $x_{mn}$ under a fixed power $2e_{kl}$. This is not a unique decomposition of the correlator but is easy to generalize. Then, we obtain
the $2(N-3)$ conformally invariant, order two, cross-ratios
\begin{equation}\label{u2}
u_{2k} = \frac{x_{23}x_{1k}}{x_{13}x_{2k}}\, ,\hskip 1cm u_{3k} = \frac{x_{23}x_{1k}}{x_{12}x_{3k}}\, , \hskip .5 cm k=4,\cdots,N
\end{equation}
and the $\frac{1}{2}(N-3)(N-4)$, order three, cross-ratios\footnote{The existence of these has been noticed in \cite{Rosenhaus:2018zqn,Fortin:2019fvx} for $N=5$.}
\begin{equation}\label{u3}
u_{ji} = \frac{x_{23} x_{1i}x_{1j}}{x_{12}x_{13}x_{ji}}\, , \hskip 1cm 4\le j < i=5,\cdots,N
\end{equation}
These are ratios of 3-point functions but being dimensionless moduli, appear in the angular part of the correlator.
The counting is right, since $1+2+\cdots + (N-4) + 2(N-3) = \frac{1}{2}N(N-3)$.
It seems that cross-ratios of higher order do not form and any higher order cross-ratio can be expressed in terms of the order two
and the order three ratios, may it be of even or odd order. An interesting fact is that while for $N=4$ we see only order two cross-ratios
and for $N=6$ the order two are twice as many as the order three ones, for large $N$ the order three cross ratios start to dominate.
Note also the useful identities
\begin{equation}\label{permgroupu}
g_{23} u_{2k} = u_{3k} \, ,\hskip .5cm g_{jk} u_{ji} = u_{ki}\,\,\, (k\ne i), \hskip .5 cm g_{ik} u_{ji} = u_{jk}\,\,\, (k\ne j)
\end{equation}
which tell us that we can start from $u_{24}$ and $u_{45}$ and generate all other cross-ratios by acting on them with the elements of $S_N$.
The last step is to generalize in the expression for the correlator the part that depends on the unfixed exponents to a general
function of its conformally invariant cross-ratios, which we will refer to also as the conformal coordinates
\begin{equation}
\langle {\cal O}_{1}(x_1)\cdots {\cal O}_{N}(x_N) \rangle = R_N f_{1\cdots N}(u_{24},\cdots,u_{2N},u_{34},\cdots,u_{4N},\cdots,u_{N-1,N})\, .
\end{equation}
Now we are done, since one can use these expressions and obtain explicitly all $N$-correlators of scalar operators of scaling dimension $\Delta_i$ from \eq{Ncor2} and their 
$N-1$ cross symmetry constraints ($a=2,\cdots,N$):
\begin{align}\label{crossgen}
& f_{q_1\cdots q_N}(u_{24},\cdots,u_{2N},u_{34},\cdots,u_{3N},\cdots,u_{N-1,N}) = \nonumber\\
& J_a f_{g_{1a}[q_1\cdots q_N]}(g_{1a}u_{24},\cdots,g_{1a}u_{2N},g_{1a}u_{34},\cdots,g_{1a}u_{3N},\cdots,g_{1a}u_{N-1,N})\, .
\end{align}
Below we will focus on two examples, the  $N=4$ and $N=6$ correlators, that will illustrate in details the described process. In both cases, the correlators will be constructed 
with identical operators, in which case $\Delta_i=\Delta$ and $f_{q_1\cdots q_N}=f$.
\subsection{The $N=4$ correlator}

For $N=4$ there are two coordinates of the type \eq{u2}:
\begin{equation}
u_{24} = \frac{x_{23}x_{14}}{x_{13}x_{24}}\, , \hskip 1cm u_{34} = \frac{x_{23}x_{14}}{x_{12}x_{34}}
\end{equation}
and no coordinates of the type \eq{u3}. Also, for identical operators $\Delta_{12}=\Delta_{13}=0$ and $\Delta_{14}=\Delta_{23}=2\D$.
The correlator in this case is
\begin{equation}
\langle {\cal O}(x_1){\cal O}(x_2){\cal O}(x_3){\cal O}(x_4) \rangle = R_4 f(u_{24},u_{34})\, , \hskip .5cm R_4=\frac{1}{x_{14}^{2\D}x_{23}^{2\D}}
\end{equation}
The action of the generators of $S_4$ on the conformal coordinates are
\begin{align}
& g_{12} u_{24} = \frac{1}{u_{24}}\, , \hskip .75cm g_{12} u_{34} = \frac{u_{34}}{u_{{24}}}\nonumber\\
& g_{13} u_{24} = \frac{u_{24}}{u_{34}}\, , \hskip .75cm g_{13} u_{34} = \frac{1}{u_{{34}}}\nonumber\\
& g_{14} u_{24} = u_{34}\, , \hskip .85cm g_{14} u_{34} = u_{24}
\end{align}
and the three Jacobians are
\begin{equation}
J_2 = u_{24}^{2\D}\, , \hskip .75cm J_3= u_{34}^{2\D}\, , \hskip .75cm  J_4=1\, .
\end{equation}
These imply the three exchange-symmetry constraints
\begin{align}\label{4crosscov}
g_{12}:\hskip .25cm f(u_{24},u_{34}) &= u_{24}^{2\D} f\left(\frac{1}{u_{24}},\frac{u_{34}}{u_{24}}\right)\nonumber\\
g_{13}:\hskip .25cm f(u_{24},u_{34}) &= u_{34}^{2\D} f\left(\frac{u_{24}}{u_{34}},\frac{1}{u_{34}}\right)\nonumber\\
g_{14}:\hskip .25cm f(u_{24},u_{34}) &= f(u_{34},u_{24})
\end{align}
Some comments need to be made here.
One is related to the observation that in \eq{4cross} we presented only two exchange-symmetry constraints and here we just found three.
What happens is that out of the three covariant constraints in \eq{4crosscov} only two are independent,
as it is easy to check that $g_{12}g_{13}g_{12}\sim g_{14}$, where the $\sim$ sign indicates not a group theory relation between $S_N$ elements but an equivalence of their action on 
the correlator and the $u_{24}, u_{34}$. In other words, the transformation with the unit Jacobian in \eq{4crosscov} for example is not independent.
This seems to be a reflection of the fact that one can bring the four points $x_1,x_2,x_3,x_4$ on a plane by conformal transformations, thus the trivial Jacobian.
Furthermore, one can place these points on the corners of a tilted rectangle that the gauging of the exchange-symmetry group turns into a square.
As a result, the exchange symmetry effectively reduces to the dihedral group $D_4$
and if the freedom to choose which three points define the plane on which the fourth point is projected on is taken into account,
the symmetry reduces further to $D_3$, which is isomorphic to $S_3$. The latter is generated by $g_{12}$ and $g_{13}$ indeed.
Thus, the $g_{14}$ operation can not be independent.
The second comment is that according to our previous geometric arguments we expect to see no 3-point subsector in $R_4$ as
the information about the invariance of the triangles inside the parallelogram under rotations about its two diagonals, are contained in the action of the $g_{1a}$.
Indeed, we saw that $\Delta_{12}=\Delta_{13}=0$ and the radial part of the correlator $R_4$ contains only two disconnected $x_{ij}$'s
($x_{14}x_{23}$ in the $u_{24},u_{34}$ angular coordinates and $x_{12}x_{34}$ in the $u,v$ coordinates).
Lastly, the two independent constraints in \eq{4crosscov} are equivalent to the ones in \eq{4cross} by a coordinate change,
even though the trivial Jacobian transformation in the $(u_{24},u_{34})$ coordinates maps to a non-trivial one in the $(u,v)$ coordinates and vice versa.

\subsection{The $N=6$ correlator}

Next, we will concentrate on the $N=6$ scalar correlator.
The algorithm we described then yields via \eq{u2} and \eq{u3} the nine ($=3+3+2+1$) conformal coordinates
\begin{align}
& u_{24} = \frac{x_{14}x_{23}}{x_{13}x_{24}},\hskip .75cm u_{25} = \frac{x_{15}x_{23}}{x_{13}x_{25}},\hskip .75cm u_{26} = \frac{x_{16}x_{23}}{x_{13}x_{26}}\nonumber\\
& u_{34} = \frac{x_{14}x_{23}}{x_{12}x_{34}},\hskip .75cm u_{35} = \frac{x_{15}x_{23}}{x_{12}x_{35}},\hskip .75cm u_{36} = \frac{x_{16}x_{23}}{x_{12}x_{36}}\nonumber\\
& \hskip 3.15cm u_{45} = \frac{x_{14}x_{15}x_{23}}{x_{12}x_{13}x_{45}},\hskip .25cm u_{46} = \frac{x_{14}x_{16}x_{23}}{x_{12}x_{13}x_{46}}\nonumber\\
& \hskip 6.35cm u_{56} = \frac{x_{15}x_{16}x_{23}}{x_{12}x_{13}x_{56}}
\end{align}
in terms of which the 6-point correlator is
\begin{equation}
\langle {\cal O}_1(x_1)\cdots {\cal O}_6(x_6)\rangle = R_6 f_{q_1\cdots q_6}\left(u_{24},u_{25},u_{26},u_{34},u_{35},u_{36},u_{45},u_{46},u_{56}\right)
 \end{equation}
 with the radial prefactor
 \begin{equation}
  R_6 = \frac{1}{x_{12}^{\Delta_{12}}x_{13}^{\Delta_{13}}x_{14}^{\Delta_{14}}x_{15}^{\Delta_{15}}x_{16}^{\Delta_{16}}x_{23}^{\Delta_{23}}}\, .
 \end{equation}
 In the case of scalar operators of the same scaling dimensions $\Delta$ this reduces to
 \begin{equation}
 \langle {\cal O}(x_1)\cdots {\cal O}(x_6)\rangle = \left(\frac{x_{12}x_{13}}{x_{14}x_{15}x_{16}x_{23}^2}\right)^{2\D}\, f\left(u_{24},u_{25},u_{26},u_{34},u_{35},u_{36},u_{45},u_{46},u_{56}\right)\, .
\end{equation}
The five corresponding exchange-symmetry constraints are
\begin{align}
g_{12}: \,\, & f\left(u_{24},u_{25},u_{26},u_{34},u_{35},u_{36},u_{45},u_{46},u_{56}\right)=\nonumber\\
& (u_{24}u_{25}u_{26})^{2\D}
f\left(\frac{1}{u_{24}},\frac{1}{u_{25}},\frac{1}{u_{26}},\frac{u_{34}}{u_{24}},\frac{u_{35}}{u_{25}},\frac{u_{36}}{u_{26}},\frac{u_{45}}{u_{24}u_{25}},\frac{u_{46}}{u_{24}u_{26}},\frac{u_{56}}{u_{25}u_{26}}\right)
\end{align}
\begin{align}
g_{13}: \,\, & f\left(u_{24},u_{25},u_{26},u_{34},u_{35},u_{36},u_{45},u_{46},u_{56}\right)=\nonumber\\
& (u_{34}u_{35}u_{36})^{2\D}
f\left(\frac{u_{24}}{u_{34}},\frac{u_{25}}{u_{35}},\frac{u_{26}}{u_{36}},\frac{1}{u_{34}},\frac{1}{u_{35}},\frac{1}{u_{36}},\frac{u_{45}}{u_{34}u_{35}},\frac{u_{46}}{u_{34}u_{36}},\frac{u_{56}}{u_{35}u_{36}}\right)
\end{align}
\begin{align}
g_{14}: \,\, & f\left(u_{24},u_{25},u_{26},u_{34},u_{35},u_{36},u_{45},u_{46},u_{56}\right)=\nonumber\\
& \left(\frac{u_{45}u_{46}}{u_{24}u_{34}}\right)^{2\D}
f\left(u_{34},\frac{u_{25}u_{34}}{u_{45}},\frac{u_{26}u_{34}}{u_{46}},u_{24},\frac{u_{24}u_{35}}{u_{45}},\frac{u_{24}u_{36}}{u_{46}},\frac{u_{24}u_{34}}{u_{45}},\frac{u_{24}u_{34}}{u_{46}},\frac{u_{24}u_{34}u_{56}}{u_{45}u_{46}}\right)\nonumber\\
\end{align}
\begin{align}
g_{15}: \,\, & f\left(u_{24},u_{25},u_{26},u_{34},u_{35},u_{36},u_{45},u_{46},u_{56}\right)=\nonumber\\
& \left(\frac{u_{45}u_{56}}{u_{25}u_{34}}\right)^{2\D}
f\left(\frac{u_{24}u_{35}}{u_{45}},u_{35},\frac{u_{26}u_{35}}{u_{56}},\frac{u_{25}u_{34}}{u_{45}},u_{25},\frac{u_{25}u_{36}}{u_{56}},\frac{u_{25}u_{35}}{u_{45}},\frac{u_{25}u_{35}u_{46}}{u_{45}u_{56}},\frac{u_{25}u_{35}}{u_{56}}\right)\nonumber\\
\end{align}
\begin{align}
g_{16}: \,\, & f\left(u_{24},u_{25},u_{26},u_{34},u_{35},u_{36},u_{45},u_{46},u_{56}\right)=\nonumber\\
& \left(\frac{u_{46}u_{56}}{u_{26}u_{36}}\right)^{2\D}
f\left(\frac{u_{24}u_{36}}{u_{46}},\frac{u_{25}u_{36}}{u_{56}},u_{36},\frac{u_{26}u_{34}}{u_{46}},\frac{u_{26}u_{35}}{u_{56}},u_{26},\frac{u_{26}u_{36}u_{45}}{u_{46}u_{56}},\frac{u_{26}u_{36}}{u_{46}},\frac{u_{26}u_{36}}{u_{56}}\right)\nonumber\\
\end{align}
There is no trivial Jacobian, so we expect these constraints to be independent. 
To conclude this section, some more comments will be highlighted.

One could argue that most of our results could have been constructed in a 'bottom-up' spirit, like for example the structure of the 
$R_N$ prefactor and that of the cross ratios. Indeed, the former depends only on the choice of the $x_{ij}$ basis and the latter
can be given without having to solve the system \eq{system}. One could do this in our basis by noticing that 
each of the $u_{nm}$ is characterized by $x_{nm}$ (reflecting their independence in the basis) and the rest
of the coordinate differences in it just balance conformal invariance using the basis elements $x_{1a}$ and $x_{23}$. Like this,
all $u_{nm}$'s can be guessed and one could even live without the order 3 cross-ratios, by redefining for example
the $u_{ji}$ in \eq{u3} as $u_{ji}\to u_{ji}/(u_{2j}u_{3i})$, turning them into order 2 cross ratios.
What can not be guessed however without an explicit solution are the exponents in \eq{exponents}.
Another comment concerns the issue of the possible degeneracies when $N\ge d+3$. We have chosen not to address in detail the issue because it seems that
it may involve some points that need to be clarified. For example, in the simple case we are discussing where all operators are scalar, if we consider
the case where $d=3$ and $N=6$, one of the position vectors in the embedding space, say $y_6$, must be linearly dependent on
the other position vectors $y_{1,\cdots,5}$. This makes the 6-point function collapse into a 5-point function which is similar
to its OPE reduction but in a special gauge. This could be a property of the embedding space where due to the null condition,
linearly dependent vectors project on the same points in $x$-space.
\subsection{Tensorial Correlation functions}
In this section we will review the constraints imposed by conformal symmetry to a correlation function with field with spin. For more details on this topic, see \cite{Poland:2018epd,Osborn:1993cr,DiFrancesco:1997nk} \\
We will start with a correlator composed of two vector currents $J_{\mu}$. Due to Lorentz invariance we have
\begin{equation}
\langle J_{\mu}(x) J_{\nu}(y) \rangle =\frac{I_{\mu \nu(x-y)}}{\abs {x-y}^{2\Delta}}.
\end{equation}
We need to construct $I_{\mu \nu(x-y)}$ in a way that it satisfies the following condition. It must be symmetric to $\mu \leftrightarrow nu$  and of course be made of the quantities that enter the problem.
A general solution is
\begin{equation}
I_{\mu \nu(x-y)}=\eta_{\mu \nu}+\alpha \frac{(x-y)_\mu(x-y)_{\nu}}{(x-y)^{2}}.
\end{equation}
The constant $\alpha$ will be fixed by applying a special conformal transformation to the correlator. Using the transformation law for a vector
\begin{equation}
J'_{\mu}(x')=\frac{\partial x^{\kappa}}{\partial x'^{\mu}}J_{\nu}(x)
\end{equation}
along with the statement that the correlator should be invariant under this transformation $\langle J'_{\mu}(x') J'_{\nu}(y') \rangle=\langle J_{\mu}(x) J_{\nu}(y) \rangle$ we arrive at
\begin{equation}
I_{\mu \nu(x-y)}=\eta_{\mu \nu}-2 \frac{(x-y)_\mu(x-y)_{\nu}}{(x-y)^{2}}.
\end{equation}
Regarding the scaling dimension of the fields, we can use 
\begin{equation}
\frac{\partial}{\partial x_{\mu}}\langle J_{\mu}(x) J_{\nu}(y) \rangle=0
\end{equation}
for conserved currents. We find as it was expected $\Delta=d-1$. \\ The next interesting correlator that we can examine is the one of two stress-energy tensors $\langle T_{\mu \nu}T_{\rho \sigma}\rangle$.  Using the constructed tensor $I_{\mu \nu(x-y)}$ along with Lorentz invariance and the symmetries of the correlator we find 
\begin{equation}\label{tttrace}
\langle T_{\mu \nu}(x)T_{\rho \sigma}(y)\rangle= \frac{C}{\abs {x-y}^{2\Delta}}\left(\frac{1}{2}\left(I_{\mu \nu}(x-y)I_{\mu \nu}(x-y)+I_{\mu \rho}(x-y)I_{\mu \sigma} (x-y)\right)-\frac{2}{d}\eta_{\mu \nu} \eta_{ \rho \sigma}\right)
\end{equation}
For a conserved stress-energy tensor we have the condition
\begin{equation}
\frac{\partial}{\partial x_{\mu}}\langle T_{\mu \nu}(x)T_{\rho \sigma}(y)\rangle=0.
\end{equation}
We find that $\Delta=d$, which is a necessary condition for a spin-2 field to be conserved. The fact that the current/stress-energy tensor conservation led us to their scaling dimensions also implies that in a CFT, these dimensions are protected from quantum corrections. This does not hold, for example in case of scalar primary fields. Also, once can check that \ref{tttrace} is traceless. We can use the following method to construct 3-point functions of various fields.  A way to simplify this procedure is to identify the all possible tensor structures that will be involved in a correlator.
\section{The OPE formalism}
The operator product expansion (OPE) is used in QFT to write a product of two operators that are close to each other, as a product of local operators at the middle point. In CFTs we will see that the OPE acquires powerful properties thanks to the radial quantization. For more details on this topic, see \cite{Poland:2018epd, DiFrancesco:1997nk}

Lets consider a state where we have inserted two operators
\begin{equation}
\ket{\psi}=\phi_1(x)\phi_2(0)\ket{0}.
\end{equation}
Then we can expand this state in a basis of eigenstates of the dilatation operator $\ket{E_n}$. Thus,
\begin{equation}
\ket{\psi}=\sum_n c_n(x) \ket{E_n}
\end{equation}
But as we have seen through the state/operator correspondence, each $\ket{E_n}$  is linear combination of primaries and their derivatives/descendants. We can write
\begin{equation}
\phi_1(x)\phi_2(0)\ket{0}=\sum_{primaries}C_{\Delta}(x,\partial)\phi_{\Delta}(0)\ket{0}.
\end{equation}
This expression has algebraic origin since we have expanded a state in a complete basis. Practically, this means that in contrary with QFT the operators don't have to be close, just inside the sphere. Now, we analyse the functions $C_{\Delta}(x,\partial)$. For simplicity, we consider only one primary field $\phi_{\Delta}(0)$.
\begin{equation}\label{eq:lhs}
\phi_1(x)\phi_2(0)\ket{0}=\frac{\mathrm{const.}}{\abs{x}^k}(\phi_{\Delta}(0)+...)\ket{0},
\end{equation}
where the dots stand for descendants and other primaries. We act with the dilatation on the L.H.S of \eqref{eq:lhs}. We get
\begin{equation}
\begin{split}
D\phi_1(x)\phi_2(0)\ket{0}&=i(\Delta+x^{\mu}\partial_{\mu}\phi_1(x)\phi_2(0)\ket{0}+i\Delta_2\phi_1(x)\phi_2(0)\ket{0}\\&=i(\Delta_1+\Delta_2-k)\frac{\mathrm{const.}}{\abs{x}^k}(\phi_{\Delta}(0)+...)\ket{0},
\end{split}
\end{equation}
where in the last line we have used the R.H.S  of \eqref{eq:lhs}. Now, we act with the dilatation operator on the R.H.S  of \eqref{eq:lhs}.
\begin{equation}
D\frac{\mathrm{const.}}{\abs{x}^k}(\phi_{\Delta}(0)+...)\ket{0}=i\Delta\frac{\mathrm{const.}}{\abs{x}^k}(\phi_{\Delta}(0)+...)\ket{0}.
\end{equation}
Comparing the previous equations we arrive at
\begin{equation}
k=\Delta_1+\Delta_2-\Delta.
\end{equation}
Next, we focus on the descendant contribution, that is the next order term of $C_{\Delta}(x,\partial)$. We have
\begin{equation}\label{eq:opedesc}
\phi_1(x)\phi_2(0)\ket{0}=\frac{\mathrm{const.}}{\abs{x}^{\Delta_1+\Delta_2-\Delta
}}(\phi_{\Delta}(0)+cx^{\mu}\partial_{\mu}\phi_{\Delta}(0) +...)\ket{0}.
\end{equation}
As before we use the conformal symmetry to fix the constant c. Now, we will act with $K_{\mu}$ on both sides of \eqref{eq:opedesc}. This will allow us to use the definition of the primary operator, that is the annihilation by $K_{\mu}$. Moreover, we will use the transformation rule for a scalar field
\begin{equation}
[K_{\mu},\phi_{\alpha}(x)]=2ix_{\mu}\Delta\phi_{\alpha}(x)+i(2x_{\mu}x^{\nu}\partial_{\nu}-x^2\partial_{\mu})\phi_{\alpha}(x).
\end{equation}
Acting on the LHS of \eqref{eq:opedesc}. We get
\begin{equation}
\begin{split}
K_{\mu}\phi_1(x)\phi_2(0)\ket{0}&=2ix_{\mu}\Delta_1\phi_{\alpha}(x)+i(2x_{\mu}x^{\nu}\partial_{\nu}-x^2\partial_{\mu}))\phi_1(x)\phi_2(0)\ket{0}\\&=ix_{\mu}(\Delta_1+\Delta-\Delta_2)\frac{\mathrm{const.}}{\abs{x}^{\Delta_1+\Delta_2-\Delta
}}(\phi_{\Delta}(0)+...)\ket{0}.
\end{split}
\end{equation}
Acting on the RHS of \eqref{eq:opedesc} and using $[K_{\mu},P_{\nu}]\phi_{\Delta}(0)=K_{\mu}P_{\nu}\phi_{\Delta}(0)=2i\eta_{\mu \nu}\Delta\phi_{\Delta}(0)$ and $-i[P_{\mu},\phi_{\alpha}(x)]=\partial_{\mu}\phi_{\alpha}(x)$, we arrive at
\begin{equation}
\begin{split}
&K_{\mu}\left(\frac{\mathrm{const.}}{\abs{x}^{\Delta_1+\Delta_2-\Delta
}}(\phi_{\Delta}(0)+cx^{\mu}\partial_{\mu}\phi_{\Delta}(0) +...)\ket{0}\right)\\&=
\frac{\mathrm{const.}}{\abs{x}^{\Delta_1+\Delta_2-\Delta
}}(2ic\Delta x_{\mu}\phi_{\Delta}(0)+...).
\end{split}
\end{equation}
Comparing again the two sides of the equation, we fix the constant c
\begin{equation}
c=\frac{\Delta_1+\Delta_2-\Delta}{2\Delta}.
\end{equation}
Once again, the conformal invariance fixes the constant. Following this procedure, we can deduce that conformal invariance fully fixes the function $C_{\Delta}(x,\partial)$, up to an overall factor $\mathcal{C}_{12\Delta}$. It is important to notice that the function $C_{\Delta}(x,\partial)$ has dependence only on the scaling dimensions of the inserted fields and the dimension of the primary.

We consider now a three-point function of primaries and we take the OPE of the first two operators. We have
\begin{equation}
\langle \phi_1(x)\phi_2(0)\phi_{\Delta}(z)\rangle=\sum_{\mathrm{primaries} \Delta'} C_{12\Delta'}C_{\Delta'}\langle \phi_{\Delta'}(y)\phi_{\Delta} \rangle \vert_{y=0}.
\end{equation}
Considering the two-point function of the above equation, we have seen that for the two primaries to be correlated they must have the same dimensions. Moreover, we assume we have contribution from only one primary. Thus, it must have dimensions $\Delta$. The three-point function then, becomes
\begin{equation}\label{eq:three}
\langle \phi_1(x)\phi_2(0)\phi_{\Delta}(z)\rangle= C_{12\Delta}C_{\Delta} \langle \phi_{\Delta}(y)\phi_{\Delta} \rangle \vert_{y=0}.
\end{equation}
In the previous section, we have found the expressions for a conformal invariant two and three-point function. Thus we can substitute in the above expression. To determine $C_{\Delta}(x,\partial)$, we have to expand in powers of x the three-point function on the L.H.S of \ref{eq:three}. The coefficients $C_{12\Delta}$ are the same that appear in the three-point function and are called OPE coefficients.
\subsection{Conformal blocks}
 Equipped with the function $C_{\Delta}(x,\partial)$, with successive application of the OPE, we can fix the building blocks of any correlation function. Let's see how this is done for the four-point function. We apply the OPE two times
 \begin{equation}
 \begin{split}
  \langle \phi(x_1)\phi(x_2)\phi(x_3)\phi(x_4) \rangle&= \sum_{\Delta} c_{12\Delta}C_{\Delta}(x_{12},\partial_y) \langle \phi_{\Delta}(y)\phi(x_3)\phi(x_4)\rangle \\&=\sum_{\Delta} c_{12\Delta}c_{34\Delta}\Big[C_{\Delta}(x_{12},\partial_y)C_{\Delta}(x_{34},\partial_z)\langle \phi_{\Delta}(y)\phi_{\Delta}(z) \rangle\Big].
\end{split}
 \end{equation}
Since, $C_{\Delta}(x_{12},\partial_y),C_{\Delta}(x_{34},\partial_z)$ together with the two-point function are fixed by conformal invariance, the whole quantity in the brackets is fixed. Recalling the general form of the four-point function, we can define
\begin{equation}
\Big[C_{\Delta}(x_{12},\partial_y)C_{\Delta}(x_{34},\partial_z)\langle \phi_{\Delta}(y)\phi_{\Delta}(z) \rangle\Big]=\frac{G_{\Delta,l}(u,v)}{\abs{x_{12}}^{2\Delta}\abs{x_{34}}^{2\Delta}}=\mathcal{F}_{1234}^{\Delta}.
\end{equation}
The functions $G_{\Delta,l}(u,v)$ are called conformal blocks and they only depend on the dimensions of the primaries, their spin $l$ and the dimension $\Delta$ of the operator appearing during the OPE. In the end, the four-point function can be written as
\begin{equation}
\langle \phi(x_1)\phi(x_2)\phi(x_3)\phi(x_4) \rangle= \sum_{\Delta} c_{12\Delta}c_{34\Delta}\mathcal{F}_{1234}^{\Delta}.
\end{equation}
We have arrived at a remarkable conclusion.\\

\noindent\fbox{\parbox{\textwidth}{In a conformal field theory, the dimensions of the primaries along with the OPE coefficients and the structure of the OPE is enough to write any correlation function.}}
\newpage
\section{Conformal Ward Identities}\label{cwisection}
In this section we will review the Conformal Ward Identities (CWIs) and the constraints they apply on the correlation functions of a CFT.
Let's review the derivation of WIs in general. Consider a correlation function
\begin{equation}
\langle \Phi(x_1)....\Phi(x_n) \rangle=\int [d\Phi] \quad  \Phi(x_1)....\Phi(x_n) \quad \exp^{-S[\Phi]}
\end{equation}
We can write the following
\begin{align}
\langle \Phi(x_1)....\Phi(x_n) \rangle&=\int [d\Phi] \quad  \Phi(x_1)....\Phi(x_n) \quad \exp^{-S[\Phi]}\\&=\int [d\Phi'] \quad  \Phi'(x_1)....\Phi'(x_n) \quad \exp^{-S[\Phi']}\\&=\int [d\Phi] \quad  \Phi(x_1)....\Phi(x_n) \quad \exp^{-S'[\Phi]}\\&=\int [d\Phi] \quad  \Phi'(x_1)....\Phi'(x_n)\exp^{-S[\Phi']}\\&=\langle \Phi'(x_1)....\Phi'(x_n) \rangle
\end{align}
In the second line we have applied  a transformation on the fields and in the third line we used the fact that the action is invariant under the transformation and the integration measure is also invariant, if the theory is non-anomalous. Using 
\begin{equation}
\delta_{\omega} \Phi(x)=\Phi'(x)-\Phi(x)=-i\omega_{\alpha}G_{\alpha}\Phi(x) \, . \notag
\end{equation}
we arrive at 
\begin{equation}
\langle \left(G_{\alpha}\Phi(x_1)\right)\Phi(x_2)...\Phi(x_n)\rangle+\langle \Phi(x_1)\left(G_{\alpha}\Phi(x_2)\right)\Phi(x_2)...\Phi(x_n)\rangle+\langle \Phi(x_1)\Phi(x_2)...\left(G_{\alpha}\Phi(x_n)\right)\rangle=0
\end{equation}
We will apply this technique to conformal transformations
From the Ward identities corresponding to the Poincare group(Lorentz rotations and translations) we know that the two-point function must be of the form
\begin{equation}
\langle \phi_1(x_1) \phi(x_2) \rangle=f(|x_1-x_2|^2).
\end{equation}
\subsubsection{Dilatation Ward Identity}
First, we are going to apply the Dilatation Ward Identity
\begin{equation}\label{WardDil}
\langle \left(D\cdot \phi_1(x_1)\right)\phi(x_2) \rangle+\langle  \phi_2(x_1)\left(D\cdot\phi(x_2)\right \rangle=0.
\end{equation}
The action of the Dilatation operator on a scalar primary field is
\begin{equation}
D\cdot \phi_i(x)=(\Delta+x\cdot\partial)\phi_i(x).
\end{equation}
Then, we get from \eqref{WardDil}
\begin{equation}\label{WardDil1}
\left(\Delta_1+x_1\cdot\partial_{x_1}+\Delta_2+x_2\cdot\partial_{x_2}\right)f\Big((x_1-x_2)^2\Big)=0.
\end{equation}
We use the following rule
\begin{equation}
x_1^{\m}\frac{\partial }{\partial x_1^{\m}}=x_1^{\m} \frac{\partial (x_1-x_2)^2 }{\partial x_1^{\m}}\frac{\partial }{\partial (x_1-x_2)^2}=2(x_1^2-x_1 \cdot x_2)\frac{\partial }{\partial (x_1-x_2)^2}.
\end{equation}
Also, we get
\begin{equation}
x_2^{\m}\frac{\partial }{\partial x_2^{\m}}=2(x_2^2-x_1 \cdot x_2)\frac{\partial }{\partial (x_1-x_2)^2}.
\end{equation}
Plugging the above expression in \eqref{WardDil1} we end up with
\begin{equation}
\left[ \frac{\Delta_1+\Delta_2}{2}+(x_1-x_2)^2\frac{\partial }{\partial (x_1-x_2)^2}\right]f\Big((x_1-x_2)^2\Big)=0.
\end{equation}
From the above we deduce that the two-point function must be a function of order $-\frac{\Delta+\Delta_2}{2}$. Thus 
\begin{equation}\label{2ptDil}
\langle \phi_1(x_1) \phi(x_2) \rangle=\frac{C_{12}}{\Big((x_1-x_2)^2\Big)^{\frac{\Delta_1+\Delta_2}{2}}}.
\end{equation}
As expected this result is in total agreement with the one that somebody gets if he/she applies the scaling law of the fields that comes from dilatation along with Poincare symmetry.
\subsubsection{Special Conformal Ward Identity}
We will now impose the Special Conformal Ward Identity on \eqref{2ptDil}
\begin{equation}
\langle \left(\mathcal{K}_{\mu}\cdot \phi_1(x_1)\right)\phi(x_2) \rangle+\langle  \phi_2(x_1)\left(\mathcal{K}_{\mu}\cdot\phi(x_2)\right \rangle=0.
\end{equation}
The action of the operator $\mathcal{K}_{\mu}$ on a scalar field is
\begin{equation}
\mathcal{K}_{\mu}\cdot \phi_i(x)=\left[2\Delta_i x_{\mu}+2x_{\m}(x\cdot \partial)-x^2\partial_{\mu}\right]\phi_i(x).
\end{equation}
This means that we have to solve\label{SCWI2pt}
\begin{equation}
\left[2\Delta_1 x_{1\mu}+2x_{1\m}(x\cdot \partial)-x_1^2\partial_{1\mu}+2\Delta_2 x_{2\mu}+2x_{2\m}(x\cdot \partial)-x_2^2\partial_{2\mu}\right]\frac{C_{12}}{\Big((x_1-x_2)^2\Big)^{\frac{\Delta_1+\Delta_2}{2}}}=0.
\end{equation}
The various derivatives are
\begin{equation}
x_{1\m}\left(x_1^{\nu}\cdot \frac{\partial  }{\partial x_{1\m}}\right)\langle \phi_1(x_1) \phi(x_2) \rangle=-2\left(\frac{\Delta_1+\Delta_2}{2}\right)(x_1^2-x_1 \cdot x_2)\frac{C_{12}}{\Big((x_1-x_2)^2\Big)^{\frac{\Delta_1+\Delta_2-2}{2}}}.
\end{equation}
\begin{equation}
x_{2\m}\left(x_2^{\nu}\cdot \frac{\partial  }{\partial x_{1\m}}\right)\langle \phi_1(x_1) \phi(x_2) \rangle=-2\left(\frac{\Delta_1+\Delta_2}{2}\right)(x_2^2-x_1 \cdot x_2)\frac{C_{12}}{\Big((x_1-x_2)^2\Big)^{\frac{\Delta_1+\Delta_2-2}{2}}}.
\end{equation}
\begin{equation}
x_1^2\partial_{1\mu}\langle \phi_1(x_1) \phi(x_2) \rangle=-2\left(\frac{\Delta_1+\Delta_2}{2}\right)x_1^2(x_{1\m}-x_{2\m})\frac{C_{12}}{\Big((x_1-x_2)^2\Big)^{\frac{\Delta_1+\Delta_2-2}{2}}}.
\end{equation}
\begin{equation}
x_2^2\partial_{2\mu}\langle \phi_1(x_1) \phi(x_2) \rangle=-2\left(\frac{\Delta_1+\Delta_2}{2}\right)x_2^2(x_{2\m}-x_{1\m})\frac{C_{12}}{\Big((x_1-x_2)^2\Big)^{\frac{\Delta_1+\Delta_2-2}{2}}}
\end{equation}
Plugging the above expressions in \eqref{SCWI2pt} and after some manipulations we arrive at
\begin{equation}
\begin{split}
&\left[2\Delta_1 x_{1\mu}+2\Delta_1 x_{1\mu}+\frac{\Delta_1+\Delta_2}{(x_1-x_2)^2} \left((x_1-x_2)^2 \right) \left( x_{1\m}-x_{2\m} \right) \right]\frac{C_{12}}{\Big((x_1-x_2)^2\Big)^{\frac{\Delta_1+\Delta_2}{2}}}=0\\
&\Leftrightarrow \left[ (\Delta_1-\Delta_2) \left(x_{1\m}-x_{2\m}\right)\right]\frac{C_{12}}{\Big((x_1-x_2)^2\Big)^{\frac{\Delta_1+\Delta_2}{2}}}=0
\end{split}
\end{equation}
From the above, we see that for $C_{12}\neq 0$ we get $\Delta_1=\Delta_2$. Without surprises we arrive at the already familiar expression
\begin{align}
&\langle \phi_1(x_1) \phi(x_2) \rangle=\frac{C_{12}}{(x_1-x_2)^{2\D}},&\Delta_1=\Delta_2=\Delta, \\
 &\langle \phi_1(x_1) \phi(x_2) \rangle=0, &\Delta_1 \neq \Delta_2.
\end{align} 
\subsubsection{Three-point function}
The same method can also be applied to the three-point function. The three point function is
\begin{equation}
\langle \phi_1(x_1) \phi(x_2) \phi(x_3) \rangle=f\left(x_{12}^2,x_{23}^2,x_{13}^2\right).
\end{equation}
When we apply the partial derivative we have to be careful because
\begin{equation}
x_1^{\mu}\partial_{1\m}f=x_1\cdot \partial_1 x_{12}^2 \frac{\partial f}{\partial x_{12}}+x_1\cdot \partial_1 x_{13}^2 \frac{\partial f}{\partial x_{13}}.
\end{equation}
From the dilatation WI we get
\begin{equation}
a+b+c=\frac{\Delta_1+\Delta_2+\Delta_3}{2}.
\end{equation}
while from the special CWI
we have
\begin{align}\label{3ptcons}
a=&\frac{\Delta_1+\Delta_2-\Delta_3}{2}\\ 
b=&\frac{-\Delta_1+\Delta_2+\Delta_3}{2}\\
c=&\frac{\Delta_3+\Delta_1-\Delta_2}{2}
\end{align}
So we have
\begin{equation}
\langle \phi_1(x_1) \phi(x_2) \phi(x_3) \rangle=\frac{C_{123}}{x_{12}^{2a}x_{23}^{2b}x_{13}^{2b}},
\end{equation}
along with the relation \eqref{3ptcons}.

\section{The stress-energy tensor in Conformal Field Theories}\label{secstress}
For our discussion, we will consider infinitesimal transformations
\begin{align}\label{eq:inftrans}
x'^{\mu}&=x^{\mu}+\omega_a\frac{\delta x^{\mu}}{\delta \omega_a},\\
\phi'(x')&=\phi(x)+\omega_a \frac{\delta \mathcal{F}}{\delta \omega_a}.
\end{align}
We consider now the change of the action under these transformations. The Jacobian changes as follows
\begin{equation}
\frac{\partial x'^{\nu}}{\partial x^{\mu}}=\delta^{\nu}_{\mu}+\partial_{\mu}\omega_a\frac{\delta x^{\nu}}{\delta \omega_a}.
\end{equation}
Using the identity $\mathrm{det}(1+A)\approx 1+\mathrm{Tr}A$, we obtain
\begin{equation}
\left|\frac{\partial \textbf{x}'}{\partial \textbf{x}}\right|=1+\partial_{\mu}\omega_a\frac{\delta x^{\mu}}{\delta \omega_a}.
\end{equation}
Thus, the transformed action becomes
\begin{equation}
\begin{split}
S'&=\int d^d x\Big(1+\partial_{\mu}\omega_a\frac{\delta x^{\mu}}{\delta \omega_a}\Big)\\&\times \mathcal{L}\Big(\phi(x)+\omega_a \frac{\delta \mathcal{F}}{\delta \omega_a},[\delta^{\nu}_{\mu}+\partial_{\mu}\omega_a\frac{\delta x^{\nu}}{\delta \omega_a}](\partial_{\nu}\phi(x)+\partial_{\nu}\omega_a \frac{\delta \mathcal{F}}{\delta \omega_a})\Big) .
\end{split}
\end{equation}
We can now expand the Lagrangian and keep only terms of first derivatives of $\omega_a$. We get
\begin{equation}\label{eq:action}
\delta S=-\int d^dxj^{\mu}_a\omega_a,
\end{equation}
with the current associated to the infinitesimal transformation
\begin{equation}
j^{\mu}_a=\Big(\frac{\partial \mathcal{L}}{\partial(\partial_{\mu}\phi)}\partial_{\nu}\phi-\delta^{\mu}_{\nu}\mathcal{L}\Big)\frac{\delta x^{\nu}}{\delta \omega_a}-\frac{\partial \mathcal{L}}{\partial(\partial_{\mu}\phi)}\frac{\delta \mathcal{F}}{\delta \omega_a}.
\end{equation}
The Noether theorem states that to every continuous symmetry of the action, we may associate a current that is classically conserved. Assuming now that the transformation \eqref{eq:inftrans} is symmetry of the action, that means that it leaves the action invariant, and the fields satisfy the equations of motion then $\delta S=0$ for every variation $\omega_a$. We integrate by parts \eqref{eq:action} and  arrive at the conservation equation
\begin{equation}
\partial_{\mu}j^{\mu}_a=0.
\end{equation}
We can then define a conserved charges
\begin{equation}
Q_a=\int d^{d-1} xj^{0}_a.
\end{equation}
Then,
\begin{equation}
\partial_tQ_a=\int d^{d-1}\partial_t xj^{0}_a=\int d^{d-1}\partial_i xj^{i}_a=0.
\end{equation}
In the last step, we have assumed that the fields and thus $j^{i}_a$ vanish sufficiently fast at infinity. We can always redefine the given current as 
\begin{equation}
j^{\mu}_a\rightarrow j^{\mu}_a+\partial_{\nu}B^{\mu\nu},
\end{equation}
with $B^{\mu\nu}=-B^{\nu \mu}$. It is is to see that the redefined current is also conserved as
\begin{equation}
\partial_{\mu}j^{\mu}_a+\partial_{\mu}\partial_{\nu}B^{\mu\nu}=\partial_{\mu}j^{\mu}_a=0,
\end{equation}
as the term $\partial_{\mu}\partial_{\nu}B^{\mu\nu}$ is the product of derivatives, which are symmetric under $\mu \leftrightarrow \nu$, with the antisymmetric $B^{\mu\nu}$ and consequently it vanishes.
\subsubsection{Energy-momentum tensor}
Now we consider an infinitesimal translation $x^{\mu}\rightarrow x^{\mu}+\epsilon^{\mu}$. Then we have
\begin{equation*}
\frac{\delta x^{\mu}}{\delta x^{\nu}}=\delta^{\mu}_{\nu} , \qquad  \frac{\delta \mathcal{F}}{\delta \epsilon ^{\nu}}=0.
\end{equation*}
Using these relations, we get
\begin{equation}\label{energy}
T^{\mu \nu}_c=\frac{\partial \mathcal{L}}{\partial(\partial_{\mu}\phi)}\partial_{\nu}\phi-\eta^{\mu\nu}\mathcal{L},
\end{equation}
where $T^{\mu \nu}_c$ is the canonical energy momentum tensor. The conservation law becomes
\begin{equation}
\partial_{\mu}T^{\mu \nu}_c=0.
\end{equation}
The conserved charge is the momentum
\begin{equation}
P^{\nu}=\int d^{d-1}xT^{0 \nu}_c.
\end{equation}
For example, the energy is
\begin{equation}
P^{0}=\int d^{d-1}xT^{0 0}_c=\int d^{d-1}x\Big( \frac{\partial \mathcal{L}}{\partial \dot{\phi}}-\mathcal{L}\Big)=\int d^{d-1}x \mathcal{H}.
\end{equation}

It is found out that when the theory is Poincare invariant, we redefine a new symmetric tensor.
\begin{equation}
T^{\mu \nu}_c \rightarrow T^{\mu \nu},
\end{equation}
with $T^{\mu \nu}=T^{\nu \mu}$. This new tensor is called the Belifante tensor. It is based on the freedom we have to redefine the energy momentum tensor as
\begin{equation}
T^{\mu \nu}=T^{\mu \nu}_c+ \partial_{\rho} B^{\rho \mu \nu},
\end{equation}
with $B^{\rho \mu \nu}=-B^{ \mu \rho \nu}$. Of course this redefinition does not violate the conservation law. The tensor $B^{\rho \mu \nu}$ is then constructed in such a way that the antisymmetric part of the redefined $T^{\mu \nu}$ vanishes. For more details, see \cite{DiFrancesco:1997nk}.

We consider now an infinitesimal diffeomorphism
\begin{equation}
x'^{\mu}=x^{\mu}+\epsilon^{\mu}(x).
\end{equation}
From \eqref{eq:action}, we have
\begin{equation}\label{stresss}
\delta S=-\int d^d xT^{\mu \nu}\partial_\mu \epsilon_{\nu}=-\frac{1}{2}\int d^d xT^{\mu \nu}(\partial_\mu \epsilon_{\nu}+\partial_\nu \epsilon_{\mu}).
\end{equation}
\subsubsection{Conformal invariance and stress-energy tensor}
As we will now see the scale and the conformal symmetry has important implications for the stress-energy tensor. Suppose we have translation and Poincare invariance and we impose now conformal invariance. We have shown that for an infinitesimal conformal transformation
\begin{equation}
\partial_\mu \epsilon_{\nu}+\partial_\nu \epsilon_{\mu}=\eta_{\mu\nu}f(x).
\end{equation}
Thus, equation \eqref{stresss} becomes
\begin{equation}
\delta S=-\int d^d xT^{\mu}_{\mu} f(x).
\end{equation}
For a scale invariant action, we have $\delta S=0$ and $f(x)=a$. Thus, for scale invariant theories the stress-energy tensor is traceless, $T^{\mu}_{\mu}=0$. For special conformal transformation, the function $f(x)$ is not arbitrary. But if the stress-energy tensor is traceless, $T^{\mu}_{\mu}=0$ then we have once again $\delta S=0$. Then, the theory is conformally invariant. 
\bigskip

\noindent\fbox{\parbox{\textwidth}{A conformal field theory must have a conserved and symmetric stress energy tensor ($\partial_{\mu}T^{\mu \nu}=0,T^{\mu \nu}=T^{\nu \mu}$) that is also traceless$T^{\mu}_{\mu}=0$.}}

\subsubsection{The free boson}
As an example of the above, we will study the Euclidean action for a free boson. The action is
\begin{equation}
S[\phi]=\frac{1}{2}\int d^d x \partial_{\mu}\phi\partial^{\mu}\phi.
\end{equation}
From equation \eqref{energy}, we obtain the stress-energy tensor of the theory
\begin{equation}
T^{\mu \nu}=-\frac{1}{2}\eta^{\mu\nu}\partial_{\kappa}\phi\partial^{\kappa}\phi+\partial^{\nu}\phi\partial^{\mu}\phi.
\end{equation}
Of course, the tensor is already symmetric. It also conserved when we use the e.o.m $\partial_{\mu}\partial^{\mu}\phi=0$. The trace is
\begin{equation}
\begin{split}
T^{\mu}_{\mu}&=\eta^{\mu\nu}T^{\mu \nu}=-\frac{1}{2}d\partial_{\kappa}\phi\partial^{\kappa}\phi+\partial_{\kappa}\phi\partial^{\kappa}\phi\\&=-\frac{1}{2}(d-2)\partial_{\kappa}\phi\partial^{\kappa}\phi.
\end{split}
\end{equation}
We see that in $d=2$, the trace vanishes and thus the theory is conformally invariant.\\\\
 Another useful formula that we will use in extend in this thesis is the alternative definition
\begin{equation}
T_{\m \n}=\frac{2}{\sqrt{g}}\frac{\d S}{\d g^{\m \n}}
\end{equation}
We will see that when the the trace of the stress-energy tensor is not zero, we have a trace or conformal anomaly. 
\chapter{Conformal Field Theory in momentum space}\label{cftmom}
We turn our attention to momentum space and how the notions we have introduced in the previous chapter take form. We start by establishing the CWI in momentum space and define their action on scalar and spinorial correlators. The next step is to see in action how we can solve the constraints. We illustrate the procedure for the two-point function where by solving the CWI we derive its form. Then we move to the three-point function where illustrate the CWI in scalar form and then briefly review the hypergeometric character of this function. Regarding these more details can be found \cite{Coriano:2019sth,Coriano:2020ccb,Coriano:2020ees}.
\section{The dilatation and  the Special Conformal operator}
In this section we are going to derive the action of the Dilatation and the SC operators on fields living on momentum space. In coordinate space we have the following
\begin{align}
&[D,\Phi_{\Delta}(x)]=i\left(\Delta+x^{\m} \frac{\partial}{\partial x^{\m}}\right) \Phi_{\Delta}(x),\\&
\left[ K_{\rho},\Phi_{\Delta}(x) \right]=i\left(2\D x_{\rho}-2x^{\sigma} S_{\rho \sigma}+2x_{\rho}x^{\mu}\partial_{\m}-x^2\partial_{\rho} \right) \Phi_{\Delta}(x),
\end{align}
where $ S_{\rho \sigma}$ is the spin part of the generator. For a field with spin l it is
\begin{equation}
S_{\rho \sigma}\Phi_{\m_1...\m_l}=\sum_{i=1}^l \left(\eta_{\rho \m_i}\delta^{\tau}_{\sigma}-\eta_{\sigma \m_i}\delta^{\tau}_{\rho}\right)\Phi_{\m_1..\tau..\m_l}.
\end{equation}
We are going to present two equivalent ways to get the corresponding expressions in momentum space. The first is based on using the Fourier transform
\begin{equation}
f(x)=\int d^d x e^{i x\cdot p}f(p).
\end{equation}
In our case 
\begin{equation}
\begin{split}
&D\int d^d p e^{i x\cdot p} \Phi_{\Delta}(p)=i\left(\Delta+x^{\m} \frac{\partial}{\partial x^{\m}}\right)\int d^d p e^{i x\cdot p} \Phi_{\Delta}(p)=i\D+i x^{\m} \int d^d p (ip_{\m}) e^{i x\cdot p} \Phi_{\Delta}(p)\\&=
i\D+i \int d^d p \left(-i\frac{e^{i x\cdot p}}{\partial p^{\mu}}\right) (ip_{\m}) \Phi_{\Delta}(p)=i\D-i \int d^d p e^{i x\cdot p}\left(-i^2\frac{\partial (p_{\m} \Phi_{\Delta}(p))}{\partial p^{\mu}}\right)\\&=i(\Delta-d- p_{\mu} \frac{\partial}{\partial p_{\mu}})\int d^d p e^{i x\cdot p}\Phi_{\Delta}(p)).
\end{split}
\end{equation}
Thus the action of the dilatation generator in momentum space is
\begin{equation}
[D,\Phi_{\Delta}(p)]=i\left(\Delta-d- p_{\mu} \frac{\partial}{\partial p_{\mu}}\right) \Phi_{\Delta}(p).
\end{equation}
The second way is to  replace in the expression in coordinate space the following
\begin{eqnarray}
&x_{\mu} \rightarrow -i\frac{\partial}{\partial p_{\mu}}\\&
\partial_{\mu}\rightarrow -ip_{\mu}
\end{eqnarray}
Substituting in $$[D,\Phi_{\Delta}(x)]=i\left(\Delta+x^{\m} \frac{\partial}{\partial x^{\m}}\right) \Phi_{\Delta}(x)$$ we get
\begin{equation}\label{DileqMom}
\begin{split}
[D,\Phi_{\Delta}(p)]=i\left(\Delta -i\frac{\partial}{\partial p_{\mu}} \left( -ip_{\mu}\Phi_{\Delta}(p)\right) \right)=i\left(\Delta-d- p_{\mu} \frac{\partial}{\partial p_{\mu}}\right) \Phi_{\Delta}(p).
\end{split}
\end{equation}
The results are the same as expected. To derive the expression for the Special Conformal operator we are going to use the second method. We start from $$\left[ K_{\rho},\Phi_{\Delta}(x) \right]=i\left(2\Delta x_{\rho}-2x^{\sigma} S_{\rho \sigma}+2x_{\rho}x^{\mu}\partial_{\m}-x^2\partial_{\rho} \right) \Phi_{\Delta}(x).$$
\begin{itemize}
\item \begin{equation*}
i2\Delta x_{\rho}\rightarrow 2\Delta \frac{\partial}{\partial p_{\rho}}\Phi_{\Delta}(p)=2\Delta \partial_{\rho}\Phi_{\Delta}(p)
\end{equation*}
\item 
\begin{equation*}
-i 2x^{\sigma} S_{\rho \sigma}=- 2	\hat{S}_{\rho \sigma}\Phi_{\Delta,\m_1...\m_l}(p)=-2\Delta \sum_{i=1}^l \left(\eta_{\rho \m_i}\partial^{\tau}-\delta^{\tau}_{\rho}\partial_{\m_i}\right)\Phi_{\Delta,\m_1.\tau.\m_l}(p).
\end{equation*}
\item
\begin{equation*}
i 2x_{\rho}x^{\mu} \partial_{\m} \rightarrow 2i\left( -i\partial_{\rho}(-i \partial_{\mu}(-ip_{\mu}\Phi(p)))\right)=-2\left(d \partial_{\rho}+(p\cdot \partial)\partial_{\rho}+\partial_{\rho}\right)\Phi(p)
\end{equation*}
\item
\begin{equation*}
-ix^2 \partial_{\rho} \rightarrow \partial_{\mu}\partial^{\mu}(p_{\rho}\Phi(p))=2\partial_{\rho}\Phi_{\Delta}(p)+p_{\rho}\Box \Phi_{\Delta}(p)
\end{equation*}
\end{itemize}
Combining the above terms we arrive at
\begin{equation}\label{KeqMOM}
\left[ K_{\rho},\Phi_{\Delta}(p) \right]=\left((2\Delta-d)\partial_{\rho}- 2	\hat{S}_{\rho \sigma}-2(p\cdot \partial)\partial_{\rho}+p_{\rho}\Box \right) \Phi_{\Delta}(p).
\end{equation}
It is crucial to point that in  \eqref{KeqMOM} all the derivatives are respect to momenta.
Using the equations \eqref{DileqMom},\eqref{KeqMOM} we can now proceed to the analysis of correlation functions.
\section{Conformal Ward Identities in momentum space}
The analysis of the conformal n-point functions in momentum space is done by applying the Conformal Ward Identities(CWI) to the correlation function and demanding the result to be 0. Then one has to solve the differential equations that will arise. The Dilatation Ward Identity is expressed as
\begin{equation}\label{Dilinx}
D\langle \mathcal{O}_1(x_1)\mathcal{O}_2(x_2)...\mathcal{O}_n(x_n)\rangle=\left[\sum_{j=1}^n \Delta_j +\sum_{j=1}^n x_j^a\frac{\partial}{\partial x_j^a}\right]\langle \mathcal{O}_1(x_1)\mathcal{O}_1(x_2)...\mathcal{O}(x_n)\rangle.
\end{equation}
Using the translational invariance of the correlator, we can reduce \eqref{Dilinx} to
\begin{equation}
D'\langle\mathcal{O}_1(x_1-x_n)\mathcal{O}_2(x_2-x_n)...\mathcal{O}_n(0)\rangle=0,
\end{equation}
where $$D'=D-x_{n}^{\kappa}\frac{\partial}{\partial x_n^\kappa}=\sum_{j=1}^{n-1}(x_j-x_n)^{\kappa}\frac{\partial}{\partial x_j^\kappa}+\sum_{j=1}^n\Delta_j.$$
The next step is to Fourier transform this expression. First of all we should notice that the translation invariance means momentum conservation. This can be written as
\begin{equation}\label{fourier}
\langle \phi(x_1) \phi(x_2)..\phi(x_n) \rangle=\int d^d p_1 ..d^d p_n\delta\left(\sum_{i=1}^n p_i\right)e^{\sum_{i=1}^n p_i x_i} \langle \phi(p_1) \phi(p_2)..\phi(p_n) \rangle.
\end{equation}
Using the delta function we can remove one of the momenta. Using $\overline{p}_n=-\sum_{j=1}^{n-1}p_j$ we have
\begin{equation}
\langle \phi(x_1) \phi(x_2)..\phi(x_n) \rangle=\int d^d p_1 ..d^d p_{n-1}\delta\left(\sum_{i=1}^n p_i\right)e^{\sum_{i=1}^{n-1} p_i x_i+\overline{p}_n x_n} \langle \phi(p_1) \phi(p_2)..\overline{p}_n \rangle.
\end{equation}

Now we can Fourier transform the above equation and we arrive at
\begin{equation}\label{DWI}
(2\pi)^d \delta^d \left(\sum_j p_j \right)\left[\Delta_{\textrm{tot}}-(n-1)d-\sum_{i=1}^{n-1} p_i^{\lambda} \frac{d}{dp_i^{\lambda}}\right]  \langle \mathcal{O}_1(p_1)\mathcal{O}_2(p_2)...\mathcal{O}_n(\overline{p}_n)\rangle=0.
\end{equation}
For the special conformal operator we have to Fourier transform the following expression
\begin{equation}
	\sum_{j=1}^{n} \left(- x_j^2\frac{\partial}{\partial x_j^\kappa}+ 2 x_j^\kappa x_j^\alpha \frac{\partial}
	{\partial x_j^\alpha} +2 \Delta_j x_j^\kappa\right)\langle \phi(x_1) \phi(x_2)..\phi(x_n) \rangle=0
\end{equation}
Using \eqref{fourier} we arrive at
\begin{equation}
	\sum_{j=1}^n \int \P_j^n {d^d p_j}\left(p_j^\kappa \frac{\partial^2}{\partial p_j^\alpha \partial p_j^\kappa} -
	2 p_j^\alpha \frac{\partial^2}{\partial p_j^\alpha \partial p_j^\kappa}  -2 \Delta_j\frac{\partial}{\partial p_j^\kappa}\right) 
	e^{i\sum_{i=1}^np_i\cdot x_i} \delta^d(P)\langle \phi(p_1) \phi(p_2)..\phi(p_n) \rangle=0,
\end{equation}
where the action of the operator is only on the exponential. Now we use integration by parts so the operator will act on the correlator and on the delta function
\begin{equation}
	\sum_{j=1}^n \int \P_j^n {d^d p_j}\delta^d(P)e^{i\sum_{i=1}^np_i\cdot x_i} K \langle \phi(p_1) \phi(p_2)..\phi(p_n) \rangle+ \mathrm{\delta_{term}}=0,
\end{equation}
Focusing on the delta term and using distributional identities \cite{Coriano:2018bbe} we have
\begin{align}
	\delta'_\textrm{term}&= \int {\underline{d^d p}}\,e^{i\underline{p\cdot x}}
	\Bigg[ \frac{\partial}{\partial P^\alpha} \delta^d(P)\sum_{j=1}^n\left(p_j^\alpha\frac{\partial}{\partial p_j^\kappa}- p_j^\kappa\frac{\partial}{\partial p_j^\alpha}\right)\langle \phi(p_1) \phi(p_2)..\phi(p_n) \rangle \notag\\
	&\hspace{5cm}+ 2 \frac{\partial}{\partial P^\kappa} \delta^d(P)\left(\sum_{j=1}^n\left (\Delta_j -  p_j^\alpha 
	\frac{\partial}{\partial p_j^\alpha} \right)- (n-1) d \right)\langle \phi(p_1) \phi(p_2)..\phi(p_n) \rangle\Bigg].
\end{align}

The above term vanishes for the following reason. The first term is zero since the correlator possess rotational invariance. In coordinate space the rotational invariance can be written as 
\begin{equation}
	\sum_{j=1}^n L_{\mu\nu}(x_j) \langle \langle \phi(x_1) \phi(x_2)..\phi(x_n) \rangle =0, 
\end{equation}
with 
\begin{equation}
	L_{\mu\nu}(x)=i\left(x_\mu\partial_\nu - x_\nu\partial_\mu \right),
\end{equation}
In momentum space this translates as 
\begin{equation}
	\sum_{j=1}^n\left(p_j^\alpha\frac{\partial}{\partial p_j^\kappa}- p_j^\kappa\frac{\partial}{\partial p_j^\alpha}\right)\langle \phi(p_1) \phi(p_2)..\phi(p_n) \rangle=0.
\end{equation}
The second term is the dilatation ward identity which is zero for a conformal correlator. Consequently, the $\mathrm{\delta_{term}}$ vanishes. The final expression of the special conformal operator action on a scalar correlator is \begin{equation}\label{SCWIscalar}
\Sigma_{i=1}^{n-1} \Big[ 2(\Delta_i-d)\frac{\partial}{\partial p_{i,\kappa}}-2p_i^{\alpha}\frac{\partial}{\partial p_i^{\alpha}}\frac{\partial}{\partial p_i^{\kappa}}  +p_i^{\kappa}\frac{\partial}{\partial p_i^{\alpha}}\frac{\partial}{\partial p_{i,\alpha}} \Big]\langle \mathcal{O}_1(p_1)\mathcal{O}_2(p_2)...\mathcal{O}_n(\overline{p}_n)\rangle=0,
\end{equation}
If the correlator has spin fields we have to add the following contribution
 the spin contribution is
\begin{equation}\label{SCWIspin}
2\sum_{j=1}^{n-1}\sum_{h=1}^{r_j}\left[\delta^{\kappa\mu_{j_h}}\frac{\partial}{\partial p_j^{\alpha_{j_h}}}-\delta^{\kappa}_{\alpha_{j_h}}\frac{\partial}{\partial p_{j_{\mu_{j_h}}}}\right]\langle\mathcal{O}^{\mu_{1_1}...\mu_{1_{r_1}}(p_1)}...\mathcal{O}^{\mu_{j_1}...\alpha_{j_h}...\mu_{j_{rj}}}(p_j)...\mathcal{O}^{\mu_{n_1}...\mu_{n_{rn}}}(\bar{p_n})\rangle=0.
\end{equation}
Combing the above, the action of the Special Conformal Ward Identity is written as 
 \begin{equation}\label{SCWI}
 \begin{split}
&\Bigg\{\Sigma_{i=1}^{n-1} \Big[ 2(\Delta_i-d)\frac{\partial}{\partial p_{i,\kappa}}-2p_i^{\alpha}\frac{\partial}{\partial p_i^{\alpha}}\frac{\partial}{\partial p_i^{\kappa}}  +p_i^{\kappa}\frac{\partial}{\partial p_i^{\alpha}}\frac{\partial}{\partial p_{i,\alpha}} \Big]\nn \\&
+2\sum_{j=1}^{n-1}\sum_{h=1}^{r_j}\left[\delta^{\kappa\mu_{j_h}}\frac{\partial}{\partial p_j^{\alpha_{j_h}}}-\delta^{\kappa}_{\alpha_{j_h}}\frac{\partial}{\partial p_{j_{\mu_{j_h}}}}\right]\Bigg\}\langle\mathcal{O}^{\mu_{1_1}...\mu_{1_{r_1}}(p_1)}...\mathcal{O}^{\mu_{j_1}...\alpha_{j_h}...\mu_{j_{rj}}}(p_j)...\mathcal{O}^{\mu_{n_1}...\mu_{n_{rn}}}(\bar{p_n})\rangle=0,
\end{split}
\end{equation}

Some comments need  to be made before we proceed. Due to momentum conservation,  the action of any differential 
operator which is separable on each of the coordinate $x_i$ violates the Leibniz rule, if we want to differentiate only the independent momenta. This because of momentum conservation, which is a consequence of the translational invariance of the correlator. More details can be found in \cite{Coriano:2018bbe}. Also in equations \eqref{DWI},\eqref{SCWI} the sum runs over the first $n-1$ momenta.
\subsection{Two-point functions}
\subsubsection{Scalar two-point function}
We focus now on the two-point function of two scalars. It can be written as
\begin{equation}
\langle\mathcal{O}(p_1)\mathcal{O}(p_2)\rangle=(2\pi)^d\d (p_1+p_2) \langle\langle\mathcal{O}(p_1)\mathcal{O}(p_2)\rangle\rangle,
\end{equation}
but using the conservation of momentum, we end up with
\begin{equation} \label{2pt}
\langle \mathcal{O}(p)\mathcal{O}(-p) \rangle,
\end{equation}
where $p=\sqrt{p_1^2}$ is the magnitude of the momentum. Now we apply the CWI to \eqref{2pt}. We have to solve
\begin{align}
 &\Big[ 2(\Delta_1-d)\frac{\partial}{\partial p_{\m}}-2p^{\n}\frac{\partial}{\partial p^{\n}}\frac{\partial}{\partial p_{\m}}  +p^{\m}\frac{\partial}{\partial p^{\n}}\frac{\partial}{\partial p_{\n}} \Big]\langle \langle \mathcal{O}(p)\mathcal{O}(-p)\rangle \rangle=0,\\
 &\Big[\Delta_1+\Delta_2-d-p^{\m}\frac{\partial}{\partial p_{\m}}
 \Big]\langle \langle \mathcal{O}(p)\mathcal{O}(-p)\rangle \rangle=0
\end{align}
But since, due to Poincare invariance, the correlation function will depend only on the magnitude, we can simplify the above equations by using the following rules
\begin{equation}
\label{chain1}
\frac{\partial}{\partial p^{\m}}=\frac{\partial p}{\partial p^{\m}}\frac{\partial}{\partial p}=\frac{p_{\m}}{p}\frac{\partial}{\partial p}.
\end{equation}
Then $p^{\m}\frac{\partial}{\partial p^{\m}}=p\frac{\partial}{\partial p}$. Now the Dilatation Ward Identity is simplified. Moreover, we must rewrite terms like
\begin{equation}
\begin{split}
2p^{\n}\frac{\partial}{\partial p^{\n}}\frac{\partial}{\partial p_{\m}}=2p^{\n}\frac{\partial }{\partial p^{\n}}\left(\frac{p_{\m}}{p}\frac{\partial}{\partial p}\right)=2p^{\n}\frac{\partial }{\partial p^{\n}}\left(\frac{p_{\m}}{p}\right)\frac{\partial}{\partial p}+\frac{2p^{\n}p_{\m}}{p}\frac{p_{\n}}{p}\frac{\partial^2}{\partial p^2}.
\end{split}
\end{equation}
Using 
\begin{equation}
\frac{\partial }{\partial p^{\n}}\left(\frac{p_{\m}}{p}\right)=\frac{\d^{\m}_{\n}}{p}+p^{\m}\left(-\frac{1}{p^2}\right)\frac{p_{\n}}{p},
\end{equation}
we arrive at
\begin{equation}
2p^{\n}\frac{\partial}{\partial p^{\n}}\frac{\partial}{\partial p_{\m}}=-2p^{\m}\frac{\partial}{\partial p^2}.
\end{equation}
Finally, we see that in the Special Conformal Ward Identity, we get a term like
\begin{equation}
K^{\m}=p^{\m}\cdot K=p^{\m}\left[-\frac{d^2}{d p^2}+\frac{2\Delta_1-d-1}{p}\frac{d}{d p}\right]
\end{equation}

We will try to solve directly the second order differential equation coming from the Special Conformal Ward Identity
\begin{equation}
\left[\frac{d^2}{d p^2}+\frac{-2\Delta_1+d+1}{p}\frac{d}{d p}\right]\langle \langle \mathcal{O}(p)\mathcal{O}(-p)\rangle \rangle=0
\end{equation}
 For simplicity, we set
$$\langle \langle \mathcal{O}(p)\mathcal{O}(-p)\rangle \rangle=f(p),-2\Delta_1+d+1=A$$ and using $u(p)=f'(p)$ we get
\begin{equation}
\begin{split}
u'(p)+\frac{A}{p}u(p)=0\\
\Leftrightarrow u(p)=p^{-A}.
\end{split}
\end{equation}
Then, by integrating we arrive at
\begin{equation}
\langle \langle \mathcal{O}(p)\mathcal{O}(-p)\rangle \rangle=c+\frac{p^{-A+1}}{1-A}=c+\frac{p^{2\Delta_1-d}}{2\Delta_1-d},\; \; \; \Delta_1\neq d/2
\end{equation}
The two solutions agree up to a constant, but we have to be careful about $\Delta_1\neq d/2$.
 For $\Delta_1= d/2$, we have the equation
\begin{equation}
f''(p)+\frac{1}{p}f'(p)=0\Leftrightarrow u'(p)+\frac{1}{p}u(p)=0
\end{equation}
The solution is $u=\frac{1}{p}$ and by integrating we arrive at
\begin{equation}
\langle \langle \mathcal{O}(p)\mathcal{O}(-p)\rangle \rangle=c_0+c \textrm{ lnp},\; \; \; \Delta_1= d/2.
\end{equation}

Focusing on $\Delta_1\neq d/2$, we will try to solve the Dilation Ward Identity. We have found that
$$
\langle \langle \mathcal{O}(p)\mathcal{O}(-p)\rangle \rangle=c_0+c_1 p^{2\Delta_1-d}.
$$
The Dialtation Ward Identity can be written as
\begin{equation}
\left[d-\Delta_1-\Delta_2+{p}\frac{d}{d p}\right]\langle \langle \mathcal{O}(p)\mathcal{O}(-p)\rangle \rangle=0.
\end{equation}
Plugging our solution, we arrive at
\begin{equation}
c_0(d-\Delta_1-\Delta_2)+c_1\left(\Delta_1-\Delta_2)p^{2\Delta_1-d}\right)=0
\end{equation}
Thus, after ignoring the trivial solution $c_1=0$, we get
\begin{align*}
c_0=0,\\
\Delta_1=\Delta_2=\Delta.
\end{align*}
The fact that the scaling dimensions of the two operators are equal is expected, since we know that in position space the conformal symmetry imposes this condition on the two-point function. Thus our solutions are
\begin{align}
&\langle \langle \mathcal{O}(p)\mathcal{O}(-p)\rangle \rangle=c_1 p^{2\Delta-d},\; \; \; \Delta_1 \neq d/2\\
&\langle \langle \mathcal{O}(p)\mathcal{O}(-p)\rangle \rangle=c_0+c \textrm{ lnp,}\; \; \; \Delta_1= d/2.
\end{align}
If we redefine
\begin{equation}
C=c_{S 12} \,  \frac{\pi^{d/2}}{4^{\Delta_1 - d/2}} \frac{\Gamma(d/2 - \Delta_1)}{\Gamma(\Delta_1)},
\end{equation}
in terms of the new integration constant $c_{S 12}$, the two-point function takes the form
\begin{equation}
\label{TwoPointScalar2}
\langle \langle \mathcal{O}(p)\mathcal{O}(-p)\rangle \rangle =  G_S(p^2)=  \delta_{\Delta_1 \Delta_2}  \, c_{S 12} \,  \frac{\pi^{d/2}}{4^{\Delta_1 - d/2}} \frac{\Gamma(d/2 - \Delta_1)}{\Gamma(\Delta_1)} 
(p^2)^{\Delta_1 - d/2} \,,
\end{equation}
and after a Fourier transform in coordinate space it takes the familiar form
\begin{equation}
\langle \mathcal O_1(x_1) \mathcal O_2(x_2) \rangle \equiv \mathcal{F.T.}\left[ G_S(p^2) \right] =  \delta_{\Delta_1 \Delta_2} \,  c_{S 12} 
\frac{1}{(x_{12}^2)^{\Delta_1}} \,,
\end{equation}
where $x_{12} = x_1 - x_2$. 
The ratio of the two Gamma functions relating the two integration constants $C$ and $c_{S 12}$ correctly reproduces the ultraviolet singular behavior of the correlation function and plays a role in the discussion of the origin of the scale anomaly.

\subsubsection{Tensorial two-point functions}
Next, we turn our attention to two point function that include spin operators \cite{Bzowski:2013sza,Bzowski:2015pba}. The simplest case is the one with two vectors. Defining $G_V^{\alpha \beta}(p) \equiv \langle V_1^\alpha(p) V_2^\beta(-p) \rangle$. If the 
vector current is conserved, $\partial^\mu V_\mu = 0$, then the tensor structure of the two-point correlation function is entirely fixed as 
\begin{equation}
\label{TwoPointVector0}
G_V^{\alpha \beta}(p) =  \pi^{\alpha\beta}(p) \, f_V(p^2)\,, \quad   \text{where} \quad 
\pi^{\alpha\beta}(p) = \eta^{\alpha \beta} -\frac{p^\alpha p^\beta}{p^2}, 
\end{equation}
where $f_V$ is a function of the invariant square $p^2$ whose form, as in the scalar case, is determined by the conformal constraints. 
Following the same procedure as in the scalar case we find that
\begin{equation}
\label{TwoPointVector}
G_V^{\alpha \beta}(p) = \delta_{\Delta_1 \Delta_2}  \, c_{V 12}\, 
\frac{\pi^{d/2}}{4^{\Delta_1 - d/2}} \frac{\Gamma(d/2 - \Delta_1)}{\Gamma(\Delta_1)}\,
\left( \eta^{\alpha \beta} -\frac{p^\alpha p^\beta}{p^2} \right)\
(p^2)^{\Delta_1-d/2} \,,
\end{equation}
with $c_{V12}$ being an arbitrary constant. 
Applying \eqref{SCWI}    in \eqref{TwoPointVector} we find that the scale dimension is $\Delta_1 = d - 1$. 
Finally, we present the solution  of the conformal two-point function of two energy momentum tensor operators which are symmetric, conserved and traceless
\begin{equation}
\label{EMTconditions}
T_{\mu\nu} = T_{\nu\mu} \,,  \qquad \partial^{\mu} T_{\mu\nu} = 0 \,,  \qquad {T_{\mu}}^{\mu} = 0 \,.
\end{equation}
From  the conditions defined in \eqref{EMTconditions} we can  define the tensor structure of the correlation 
function $G^{\alpha\beta\mu\nu}_T(p) = \Pi_{d}^{\alpha\beta\mu\nu}(p) \, f_T(p^2)$ with
\begin{equation} 
\label{TT}
\Pi^{\alpha\beta\mu\nu}_{d}(p) = \frac{1}{2} \bigg[ \pi^{\alpha\mu}(p) \pi^{\beta\nu}(p) + \pi^{\alpha\nu}(p) \pi^{\beta\mu}(p) 
\bigg] 
- \frac{1}{d-1} \pi^{\alpha\beta}(p) \pi^{\mu\nu}(p) \,,
\end{equation}
By requiring the invariance under 
dilatations and special conformal transformations, we obtain
\begin{equation}
\label{TwoPointEmt}
G^{\alpha\beta\mu\nu}_T(p) = \delta_{\Delta_1 \Delta_2}  \, 
c_{T 12}\,\frac{\pi^{d/2}}{4^{\Delta_1 - d/2}} \frac{\Gamma(d/2 - \Delta_1)}{\Gamma(\Delta_1)}\, 
\Pi^{\alpha\beta\mu\nu}_{d}(p) \, (p^2)^{\Delta_1 - d/2} \,.
\end{equation}
As for the conserved vector currents, also for the energy momentum tensor the scaling dimension is fixed by \eqref{SCWI} and it is given by $\Delta_1 = d$. This particular value ensures that $\partial^\mu T_{\mu\nu}$ is also a quasi primary (vector) field.
\section{The three-point scalar function and the hypergeometric solution}
\label{fuchs}
In this section, we present the hypergeometric character of the CWIs. This approach was  presented in \cite{Coriano:2013jba, Coriano:2018bsy1, Coriano:2018bbe, Maglio:2019grh}. Moreover, another analysis performed in \cite{Bzowski:2013sza} has connected the solutions of such equations to 3K (or triple-K) integrals. The analysis of \cite{Coriano:2013jba} proved that the fundamental basis for the most general solutions of such equations is given by Appell functions $F_4$.
Here demonstrate the analysis of scalar three-point functions. The four-point scalar function case will be presented in the second part of the thesis. 
\section{CWI of the three-point scalar function}
Before presenting the hypergeometric character of the the three-pint function we are going to examine the scalar form of the CWI and introduce some useful notation.  As we have already seen due to Lorentz invariance, the correlator can be expressed as a function of  the magnitudes of the momenta, defined as ${p}_i=\sqrt{p_i^2}$
\begin{equation}
\langle{\mathcal{O}_1(p_1)\mathcal{O}_2(p_2)\mathcal{O}_3(\bar{p}_3)}\rangle=\Phi(p_1,p_2,p_3).
\end{equation} 
Using \ref{chain1}, the CWI can be written in scalar form. By using
\begin{equation}
\label{chainr}
\frac{\partial \Phi}{\partial p_i^\mu}=\frac{p_i^\mu}{  p_i}\frac{\partial\Phi}{\partial  p_i} 
-\frac{\bar{p}_3^\mu}{  p_3}\frac{\partial\Phi}{\partial   p_3}, \quad i=1,2,
\end{equation}
where $\bar{p}_3^\mu=-p_1^\mu-p_2^\mu$ and $p_3=\sqrt{(p_1+p_2)^2}$, the covariant differential operator is 
\begin{equation}
\sum_{i=1}^2\,p_i^\mu\,\frac{\partial }{{p_i}^\mu}\Phi(p_1,p_2,p_3)=\left(
{p}_1\frac{ \partial }{\partial   p_1} +   p_2\frac{ \partial }{\partial   p_2} +   p_3\frac{ \partial }{\partial   p_3}\right)\Phi(p_1,p_2,p_3).
\end{equation}
Therefore, the dilatation equation becomes 
\begin{equation}
\label{scale}
\left(\sum_{i=1}^3\Delta_i -2 d - \sum_{i=1}^3    p_i \frac{ \partial}{\partial   p_i}\right)\Phi(p_1,p_2,\bar{p}_3)=0.
\end{equation}
Using the same procedure, the special conformal Ward identities in $d$ dimensions become
\begin{align}
0&={K}_{scalar}^{\kappa}\Phi(p_1,p_2,p_3)=\bigg(p_1^\kappa\,K_1+p_2^\kappa\,K_2+\bar{p}_3^\kappa\,K_3\bigg)\Phi(p_1,p_2,p_3)\notag\\
&=p_1^\kappa\Big(K_1-K_3\Big)\Phi(p_1,p_2,p_3)+p_2^\kappa\Big(K_2-K_3\Big)\Phi(p_1,p_2,p_3),\label{SCWIs}
\end{align}
where we have used the conservation of the total momentum, with the $K_i$ operators defined as
\begin{equation}\label{operatorK}
{ K}_i\equiv \frac{\partial^2}{\partial    p_i \partial    p_i} 
+\frac{d + 1 - 2 \Delta_i}{   p_i}\frac{\partial}{\partial   p_i}, \quad i=1,2,3. 
\end{equation}
The equation \eqref{SCWIs} is satisfied if every coefficient of the independent four-momenta $p_1^\mu,\,p_2^\mu$ is equal to zero. This condition leads to the scalar form of the special conformal constraints 
\begin{equation}
\frac{\partial^2\Phi}{\partial   p_i\partial   p_i}+
\frac{1}{  p_i}\frac{\partial\Phi}{\partial  p_i}(d+1-2 \Delta_1)-
\frac{\partial^2\Phi}{\partial   p_3\partial   p_3} -
\frac{1}{  p_3}\frac{\partial\Phi}{\partial  p_3}(d +1 -2 \Delta_3)=0\qquad i=1,2,
\label{3k1}
\end{equation}
and defining 
\begin{equation}
\label{kij}
K_{ij}\equiv {K}_i-{K}_j,
\end{equation}
\eqref{3k1} can be written in the concise form 
\begin{equation}
\label{3k2}
K_{13}\,\Phi(p_1,p_2,p_3)=0, \qquad K_{23}\,\Phi(p_1,p_2,p_3)=0.
\end{equation}
In the derivation of \eqref{SCWIs} we used  the derivative of the dilatation WI 
\begin{equation}
  p_1\frac{\partial^2\Phi}{\partial   p_3\partial  p_1} 
+   p_2\frac{\partial^2\Phi}{\partial   p_3\partial  p_2}=(\Delta -2 d -1)
\frac{\partial\Phi}{\partial  p_3} -  p_3\frac{\partial^2\Phi}{\partial   p_3\partial   p_3}.
\end{equation}
\section{The hypergeometric character of the three-point scalar function}\label{OOOhyper}
The hypergeometric character of the CWIs emerges in various ways. One may proceed from \eqref{SCWI} and introduce the change of variables - as originally done in
\cite{Coriano:2013jba} -
\begin{eqnarray}
\frac{\partial}{\partial p_{1}^{\mu}}  &=   2 (p_{1\, \mu} + p_{2 \, \mu}) \frac{\partial}{\partial p_3^2} + \frac{2}{p_3^2}\left( 
(1- x) p_{1 \, \mu}  - x  \,  p_{2 \, \mu} \right) \frac{\partial}{\partial x} - 2  (p_{1\, \mu} + p_{2 \, \mu}) \frac{y}{p_3^2} 
\frac{\partial}{\partial y} \,, \nn \\
\frac{\partial}{\partial p_{2}^{\mu}}  &= 2 (p_{1\, \mu} + p_{2 \, \mu}) \frac{\partial}{\partial p_3^2}   -   2  (p_{1\, \mu} + 
p_{2 \, \mu}) \frac{x}{p_3^2} \frac{\partial}{\partial x}   + \frac{2}{p_3^2}\left( (1- y) p_{2 \, \mu}  - y  \,  p_{1 \, \mu} 
\right) \frac{\partial}{\partial y}. \, 
\end{eqnarray}
with $x=\frac{p_1^2}{p_3^2}$ and $y=\frac{p_2^2}{p_3^2}$. Replacing $p_3$ by $p_1$ will yield a different result but since this two choices are equivalent, the results are related by some known inversion formula of the hypergeometric function $F_4$.  

Consider the case of the scalar correlator 
$\Phi(p_1,p_2, p_3)$, which is simpler, defined by the two homogeneous conformal equations
\begin{equation}
K_{31}\Phi=0,  \qquad K_{21}\Phi=0,
\end{equation}
obtained by subtracting the relations in \eqref{3k2}, and combined with the scaling equation 
\begin{equation}
\sum_{i=1}^3 p_i\frac{\partial}{\partial p_i} \Phi=(\Delta_t-2 d) \Phi, 
\end{equation}
with $\Delta_t=\Delta_1+\Delta_2+\Delta_3$. As shown in \cite{Coriano:2013jba}, the ansatz for the solution can be taken of the form 
\begin{equation}
\label{ans}
\Phi(p_1,p_2,p_3)=p_1^{\Delta_t - 2 d} x^{a}y^{b} F(x,y).
\end{equation}
We require that $\Phi$ is homogeneous of degree $\Delta_t-2 d$ under a scale transformation, according to (\ref{scale}). In (\ref{ans}) this condition is taken into account by the factor $p_1^{\Delta - 2 d}$. This procedure will be used extensively in the search of hypergeometric solutions also of other correlators, even in four-point functions, as shown by us for the dual conformal/conformal (dcc) solutions that we will discuss below.

The use of the scale invariant variables $x$ and $y$, now defined as
\begin{equation}
x=\frac{p_2^2}{p_1^2},\qquad y=\frac{p_3^2}{p_1^2},
\end{equation} 
reduces the equations to a generalized hypergeometric form 
\begin{align}
&K_{21}\Phi(p_1,p_2,p_3) =\notag\\
&= 4 p_1^{\Delta -2d -2} x^a y^b
\left(  x(1-x)\frac{\partial }{\partial x \partial x}  + (A x + \gamma)\frac{\partial }{\partial x} -
2 x y \frac{\partial^2 }{\partial x \partial y}- y^2\frac{\partial^2 }{\partial y \partial y} + 
D y\frac{\partial }{\partial y} + \left(E +\frac{G}{x}\right)\right)F(x,y)=0,
\end{align}
with the parameters of the equations given by
\begin{align}
&A=D=\Delta_2 +\Delta_3 - 1 -2 a -2 b -\frac{3 d}{2} \qquad \gamma(a)=2 a +\frac{d}{2} -\Delta_2 + 1,
\notag\\
& G=\frac{a}{2}(d +2 a - 2 \Delta_2),
\notag\\
&E=-\frac{1}{4}(2 a + 2 b +2 d -\Delta_1 -\Delta_2 -\Delta_3)(2 a +2 b + d -\Delta_3 -\Delta_2 +\Delta_1).
\end{align}
Similar constraints are obtained from the equation $K_{31}\Phi=0$, with the obvious exchanges $(a,b,x,y)\to (b,a,y,x)$
\begin{align}
&K_{31}\Phi(p_1,p_2,p_3) =\notag\\
&= 4 p_1^{\Delta -2 d -2} x^a y^b
\left(  y(1-y)\frac{\partial }{\partial y \partial y}  + (A' y + \gamma')\frac{\partial }{\partial y} -
2 x y \frac{\partial^2 }{\partial x \partial y}- x^2\frac{\partial^2 }{\partial x \partial x} + 
D' x\frac{\partial }{\partial x} + \left(E' +\frac{G'}{y}\right)\right)F(x,y)=0,
\end{align}
with
\begin{equation}
\begin{aligned}
&A'=D'= A,   &\hspace{1cm}\gamma'(b)=2 b +\frac{d}{2} -\Delta_3 + 1,\\
& G'=\frac{b}{2}(d +2 b - 2 \Delta_3), & \hspace{1cm}E'= E.
\end{aligned}
\end{equation}
In the mathematical analysis of hypergeometric systems of Appell type, one encounters 4 possible values for the ``indices'' $a$ and $b$ of the ansatz that we have introduced above. As shown  in \cite{Coriano:2018bbe}, such values are exactly those that set the $1/x , 1/y$ terms of the equations, in the new $x,y$ variables to zero.  This gives 
\begin{equation}
\label{cond1}
a=0\equiv a_0 \qquad \textrm{or} \qquad a=\Delta_2 -\frac{d}{2}\equiv a_1.
\end{equation}
From the equation $K_{31}\Phi=0$ we obtain a similar condition for $b$ by setting $G'/y=0$, thereby fixing the two remaining indices
\begin{equation}
\label{cond2}
b=0\equiv b_0 \qquad \textrm{or} \qquad b=\Delta_3 -\frac{d}{2}\equiv b_1.
\end{equation}
The four independent solutions of the CWIs will all be characterized by the same 4 pairs of indices $(a_i,b_j)$ $(i,j=1,2)$.
Setting 
\begin{equation}
\alpha(a,b)= a + b + \frac{d}{2} -\frac{1}{2}(\Delta_2 +\Delta_3 -\Delta_1), \qquad \beta (a,b)=a +  b + d -\frac{1}{2}(\Delta_1 +\Delta_2 +\Delta_3),
\label{alphas}
\end{equation}
then
\begin{equation}
E=E'=-\alpha(a,b)\beta(a,b), \qquad A=D=A'=D'=-\left(\alpha(a,b) +\beta(a,b) +1\right).
\end{equation}
the solutions take the form 
\begin{align}
\label{F4def}
F_4(\alpha(a,b), \beta(a,b); \gamma(a), \gamma'(b); x, y) = \sum_{i = 0}^{\infty}\sum_{j = 0}^{\infty} \frac{(\alpha(a,b), {i+j}) \, 
	(\beta(a,b),{i+j})}{(\gamma(a),i) \, (\gamma'(b),j)} \frac{x^i}{i!} \frac{y^j}{j!},
\end{align}
where $(\alpha,i)=\Gamma(\alpha + i)/ \Gamma(\alpha)$ is the Pochhammer symbol. We will refer to $\alpha\ldots \gamma'$ as to the first,$\ldots$, fourth parameters of $F_4$.\\ 
This hypergeometric function is an Appell function \cite{APPELL,Bateman:100233,Bateman:1935,Slater:1966}. The 4 independent solutions are  of the form $x^a y^b F_4$, with
$a$ and $b$ fixed by (\ref{cond1}) and (\ref{cond2}). Specifically we have
\begin{equation}
\Phi(p_1,p_2,p_3)=p_1^{\Delta-2 d} \sum_{a,b} c(a,b,\vec{\Delta})\,x^a y^b \,F_4(\alpha(a,b), \beta(a,b); \gamma(a), \gamma'(b); x, y),
\label{compact}
\end{equation}
where the sum runs over the four values $a_i, b_i$ $i=0,1$ with arbitrary constants $c(a,b,\vec{\Delta})$, with $\vec{\Delta}=(\Delta_1,\Delta_2,\Delta_3)$. Equation \eqref{compact} is a  compact way to write down the solution but once these types of solutions of a homogeneous hypergeometric system are inserted into an inhomogeneous system of equations, the sum over $a$ and $b$ 
needs to be made explicit. For this, we define 
\begin{align} 
&\alpha_0\equiv \alpha(a_0,b_0)=\frac{d}{2}-\frac{\Delta_2 + \Delta_3 -\Delta_1}{2},\, & \beta_0\equiv \beta(b_0)=d-\frac{\Delta_1 + \Delta_2 +\Delta_3}{2},  \nn \\
&\gamma_0 \equiv \gamma(a_0) =\frac{d}{2} +1 -\Delta_2,\, &\gamma'_0\equiv \gamma(b_0) =\frac{d}{2} +1 -\Delta_3.
\end{align}
to be the 4 basic (fixed) hypergeometric parameters, and express all the remaining ones by shifts with respect to these. The 4 independent solutions can be re-expressed in terms of the parameters above as 
\begin{align}
S_1(\alpha_0, \beta_0; \gamma_0, \gamma'_0; x, y)&\equiv F_4(\alpha_0, \beta_0; \gamma_0, \gamma'_0; x, y) = \sum_{i = 0}^{\infty}\sum_{j = 0}^{\infty} \frac{(\alpha_0,i+j) \, 
(\beta_0,i+j)}{(\gamma_0,i )\, (\gamma'_0,j)} \frac{x^i}{i!} \frac{y^j}{j!}, \\
S_2(\alpha_0, \beta_0; \gamma_0, \gamma'_0; x, y) &= x^{1-\gamma_0} \, F_4(\alpha_0-\gamma_0+1, \beta_0-\gamma_0+1; 2-\gamma_0, \gamma'_0; x,y) \,,  \\
S_3(\alpha_0, \beta_0; \gamma_0, \gamma'_0; x, y) &= y^{1-\gamma'_0} \, F_4(\alpha_0-\gamma'_0+1,\beta_0-\gamma'_0+1;\gamma_0,2-\gamma'_0 ; x,y) \,,  \\
S_4(\alpha_0, \beta_0; \gamma_0, \gamma'_0; x, y) &= x^{1-\gamma_0} \, y^{1-\gamma'_0} \, 
F_4(\alpha_0-\gamma_0-\gamma'_0+2,\beta_0-\gamma_0-\gamma'_0+2;2-\gamma_0,2-\gamma'_0 ; x,y) \, .
\end{align}
A particular feature of the scalar case is that we can impose the complete symmetry of the correlator under the exchange of the 3 external momenta and scaling dimensions, as discussed in \cite{Coriano:2013jba}. This reduces the four
constants to just one.

\chapter{Conformal Anomalies and the anomaly action}\label{anomalyaction}
 In the last chapter \ref{anomalyaction} of the first part, we review the conformal anomaly by giving an example in a scalar theory and examining  the space time symmetries of the action. Then we focus on the counter terms needed for the process of renormalization of some correlation functions. We study the counterterms and give additional details regarding the Euler-Gauss-Bonnet counterterm.  More details regarding section \ref{counterr} can be found in \cite{tttt}. 
 \\\\
If a theory possesses a symmetry at classical level and due to quantum correction this symmetry is broken, then we call this phenomenon an anomaly. A well know example is the chiral anomaly. The chiral symmetry of the classical theory is broken when we include quantum correction and particularly the so called triangle Feynman diagrams. This symmetry breaking can also be seen from the non invariance of the integral measure of the partition function. 
Another known anomaly, which is of great research interest and a main topic of this thesis is the conformal anomaly. There are  classical theories that have scale invariance such as massless Quantum Electrodynamics (QED) and the $\phi^4$ theory in four dimensions. This symmetry of the action is broken when we proceed with renormalization. This happens due to the fact that we have to introduce some energy scale during this process. As a result, the stress-energy tensor acquires a non vanishing  trace. This trace anomaly can be generated by a purely gravitational, since the matter is integrated out, action.  This action provides a semi-classical description of the interaction of the gravitational field with ordinary matter.  This action is not uniquely defined but defined modulo traceless contributions.

These actions may differ  by the number of asymptotic degrees of freedom introduced in the action itself. By ``asymptotic'' we mean fields which are part of the effective action, but not of the original theory.  This is the case for local anomaly actions constructed by the inclusion of a dilaton using the Weyl gauging procedure, as we will present in a following section.\\ 
All the local forms of such actions introduce one extra degree of freedom, in the form of a dilaton field. 
This dilaton is introduced as a Goldstone mode that couples to the divergence of the conformal current $(J_c)$, which is not conserved at quantum level. This coupling has been worked out in \cite{Coriano:2013nja,Coriano:2013xua} and is at most quartic in $d=4$. 
In the case of supersymmetric theories the dilaton field turns into a multiplet with a dilaton, an axion and an axino and can be described by a St\"uckelberg-like Lagrangian \cite{Coriano:2010ws}. 

There are various ways to compute the conformal anomaly.
For example one can proceed with perturbative computations in momentum space, which allow to identify on a diagrammatic basis some components of the anomaly functional. 
In this approach one can consider correlators with several insertions of stress energy tensors and/or currents, which are sensitive to different components of the anomaly functional. For instance, the trace of the $TJJ$ accounts for the $F^2$ part of the anomaly, where $F$ is the field strength of the gauge field coupled to the $J$ current, but it is insensitive to other components, which come from gravity. For this reason a complete picture of the trace anomaly requires other, more complex correlators containing multiple insertions of stress energy tensors.\\
An advantage of this approach is to bring us close to the core of an interacting theory, with the emergence of dynamical degrees of freedom. Ultimately, quantum field theory is a theory of massive and massless particles propagating in space time either as real or virtual 
 states. Such effective interactions can be worked out directly in a traditional Feynman expansion. 

An alternative approach is to start from the vacuum persistence amplitude, which takes to the DeWitt-Schwinger or heat-kernel method. One important feature of the method is to provide an expression for the anomaly action which allows to identify 
the general structure of the anomaly functional. This generality is the main advantage of this method  compared to the standard perturbative approach, and for that we are going 
to briefly present it in this section. 

\section{The anomaly action}
The conformal anomaly effective action has an important role in the identification of the impact of conformal symmetry at a certain physical scale, after having integrated out a certain number of degrees of freedom in the functional integral. For a given Lagrangian CFT, $\mathcal{S}(g)$ ( see also \cite{Cappelli:1988vw,Cappelli:2001pz,Erdmenger:1998xv,Asorey:2003uf,Asorey:2006rm,Buchbinder:1992rb, Hamada}) is a functional of the external metric  $g$, and its expansion in the fluctuations around a given background $\bar{g}$ ($g= \bar{g} + \delta g)$ allows to define the $n$-point correlation functions of stress-energy tensors for any $n$.  In general it is generated by integrating out some conformal matter in the path integral in arbitrary external metric. The simplest example at $d=4$ is provided by scalar, fermionic and spin-$1$ free field theories coupled to gravity  \cite{Coriano:2018bbe,Coriano:2018bsy1,Serino:2020pyu}.

Integrating out the conformal matter in the path integral allows to identify its back-reaction on 
gravity, following an approach that has some similarity with Sakharov's theory of induced gravity, where integration over ordinary matter, for a generic metric $g$, is expected to generate terms of the form  \cite{Visser:2002ew}
\begin{equation} 
\label{st}
\sm(g)\sim \int d^4 x \sqrt{g}\left( \Lambda + c_1(g) R + c_2 R^2\right).
\end{equation}
These terms correspond to a cosmological constant, the Einstein-Hilbert action and to generic higher derivative invariant terms. The integration over conformal matter does not introduce any scale, if the result of the integration turns out to be finite, and the structure of \eqref{st} simplifies. $\mathcal{S}(g)$ is, in this case, non-local and Weyl invariant. 

The process of renormalization, as $d\to 4$ induces a non-vanishing variation of $\mathcal{S}(g)$ under a Weyl rescaling of the metric. This can be attributed to the need of including a dimensional constant $\mu$ in order to balance the mass-dimensions of the counterterms that are need in order to make the integration on the quantum degrees of freedom, finite.
This appears as a balancing factor $\mu^{-\epsilon}$ - with $\epsilon=d-4$ - in the structure of the counterterms, which are expressed in terms of Weyl-invariant operators in $d=4$. 
The renormalized action then acquires a $\log k^2/\mu^2$ dependence, where $k$ is a generic momentum, determining the breaking of dilatation invariance. 

 In an ordinary field theory the relation between the partition function and the functional of all the connected correlators is  given by the functional relation
\begin{equation}
\label{defg}
e^{-\mathcal{S}(g)}=Z(g) \leftrightarrow \mathcal{S}(g)=-\log Z(g).
\end{equation}
 $Z(g)$ can be thought of as related to a functional integral in which we integrate the action of a generic CFT  over a field $(\phi)$  in a background metric $g_{\mu\nu}$, 
\begin{equation} 
\label{induced}
Z(g)=N \int D\phi e^{-S_0(g,\phi)},   \qquad Z(\eta)=1.
\end{equation}
Its logarithm, $\mathcal{S}(g)$, is our definition of the effective action, while $S_0(g,\phi)$ is the classical action. As usual $Z(g)$  will contain both connected and disconnected graphs, while $\mathcal{S}(g)$ collects only connected graphs. One can verify that this 
collection corresponds also to 1PI (1 particle irreducible) graphs only in the case of free field theories embedded in external (classical) gravity.  

The emergence of bilinear mixing on the external graviton lines, as we are going to realize at the end of our analysis, should then be interpreted as a dynamical response 
of the theory, induced by the process of renormalization, with the generation of an intermediate dynamical degree of freedom propagating with the $1/\square$ operator. For this reason, the presence of such terms does not invalidate the 1PI nature of this functional.

\subsection{The scalar case}
  We may assume that $S(\phi,g)$ describes a free scalar field $\phi$ in a generic background. The action, in  this case, is given by 
\begin{equation}
\label{phi}
S_0=\frac{1}{2}\int\, d^dx\,\sqrt{-g}\left[g^{\mu\nu}\nabla_\mu\phi\nabla_\nu\phi-\chi\, R\,\phi^2\right],
\end{equation}
where   $\chi(d)=\frac{1}{4}\frac{(d-2)}{(d-1)}$ is the conformal coupling. This particular choice of $\chi(d)$ guarantees the conformal invariance of this action in $d$ dimensions and generates a term of improvement for the stress-energy tensor in the flat limit, which becomes symmetric and traceless in this limit. For a more detailed perturbative analysis of this term when the Standard Model is coupled to gravity, we refer to
 \cite{Coriano:2011zk,Coriano:2011ti}. One can verify explicitly that the ordinary counterterms which renormalize the Lagrangian of the Standard Model, renormalize also insertions of the stress energy tensor of the theory, only if the Higgs sector is conformally coupled.\\
The stress energy tensor,  takes the form
  \begin{equation}
 \label{defT}
T^{\mu\nu}_{scalar}
\equiv\frac{2}{\sqrt{g}}\frac{\delta S_0}{\delta g_{\mu\nu}}\nonumber \\
=\nabla^\mu \phi \, \nabla^\nu\phi - \frac{1}{2} \, g^{\mu\nu}\,g^{\alpha\beta}\,\nabla_\alpha \phi \, \nabla_\beta \phi
+ \chi \bigg[g^{\mu\nu} \Box - \nabla^\mu\,\nabla^\nu + \frac{1}{2}\,g^{\mu\nu}\,R - R^{\mu\nu} \bigg]\, \phi^2 .
\end{equation}
A conformal free field theory realization stays conformal at quantum level, except for the appearance  of an anomaly in even dimensions. For an interacting theory, on the other hand, anomalous dimensions appear as soon as we switch-on an interaction in \eqref{phi}.\\
In the thesis, $n$-point correlation functions are defined as
\begin{equation}
\label{exps1}
\langle T^{\mu_1\nu_1}(x_1)\ldots T^{\mu_n\nu_n}(x_n)\rangle \equiv\frac{2}{\sqrt{g_1}}\ldots \frac{2}{\sqrt{g_n}}\frac{\delta^n \sm[g]}{\delta g_{\mu_1\nu_1}(x_1)\delta g_{\mu_2\nu_2}(x_2)\ldots \delta g_{\mu_n\nu_n}(x_n)} ,
\end{equation}
with $\sqrt{g_1}\equiv \sqrt{|\textrm{det} \, g_{{\mu_1 \nu_1}}(x_1)} $ and so on. \\
The action $\sm$ collects all the connected contributions of the correlation functions in the expansion with respect to the metric fluctuations, and may as well be expressed in a covariant expansion as 
\begin{equation}
\label{exps2}
\sm(g)=\sm(\bar{g})+\sum_{n=1}^\infty \frac{1}{2^n n!} \int d^d x_1\ldots d^d x_n \sqrt{g_1}\ldots \sqrt{g_n}\,\langle T^{\mu_1\nu_1}\ldots \,T^{\mu_n\nu_n}\rangle_{\bar{g}}\delta g_{\mu_1\nu_1}(x_1)\ldots \delta g_{\mu_n\nu_n}(x_n).
\end{equation}
 For a scalar theory in a flat background, in terms of Feynman diagrams
\begin{align}
\label{figg}
\sm(g)=& \sum_n \quad\raisebox{-6.9ex}{{\includegraphics[width=0.2\linewidth]{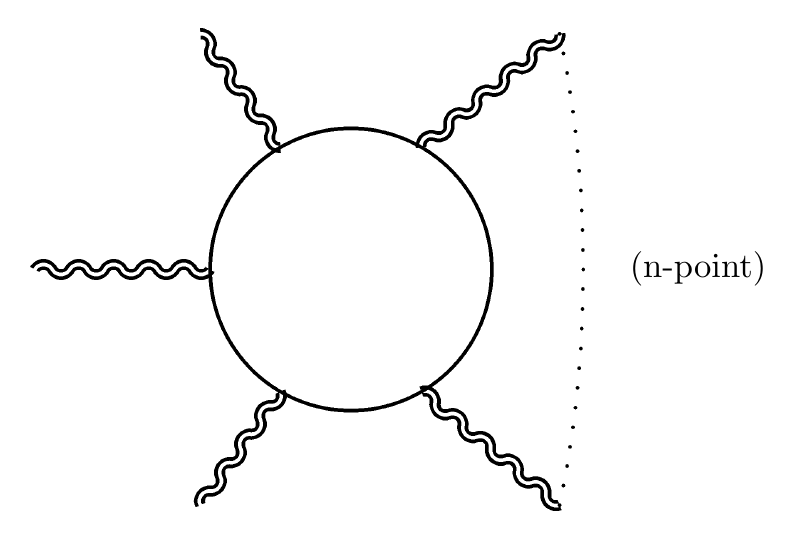}}}\,,
\end{align}
and defines the bare action in $d$ spacetime dimensions. As $d\to 4$ it will develop singularities in the form of single poles in $1/\epsilon$ that will be removed by the action of the counterterms.  
 The external weavy lines represent gravitational fluctuations, in terms of contributions that are classified as tadpoles, two-point , three-point  and $n$-point correlation functions of stress energy tensors. Tadpoles are removed in dimensional regularization (DR) in flat space, and the sum in \eqref{figg} starts from two-point  functions. The anomaly contributions start from three-point  functions.\\ 
Indeed, in the case of the scalar free field theory presented above, the topological contributions coming from the 4-$T$ are summarized by the following vertices
\begin{align}
\label{lighter}
\mathcal{S}_4(g)=\quad\raisebox{-4.9ex}{{\includegraphics[width=0.20\linewidth]{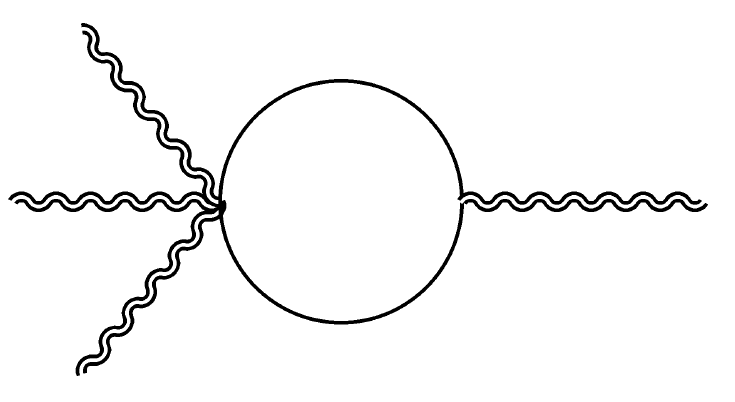}}}+\quad\raisebox{-4.5ex}{{\includegraphics[width=0.15\linewidth]{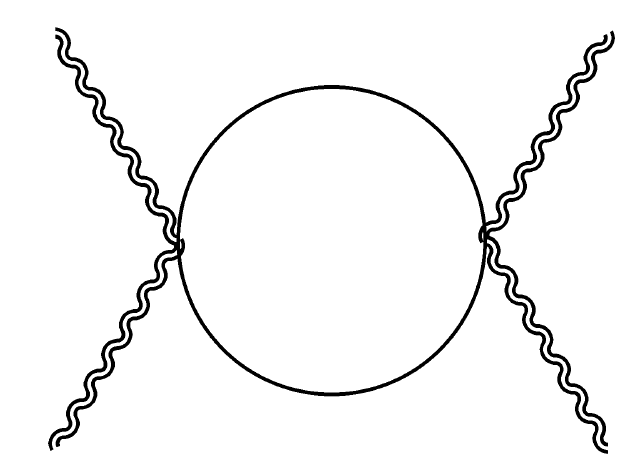}}}+\quad\raisebox{-3.5ex}{{\includegraphics[width=0.15\linewidth]{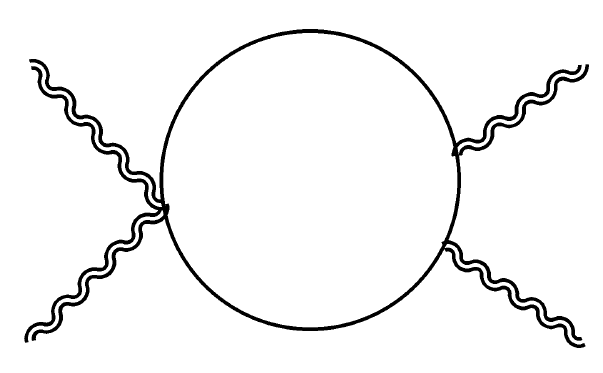}}}+\quad\raisebox{-6ex}{{\includegraphics[width=0.2\linewidth]{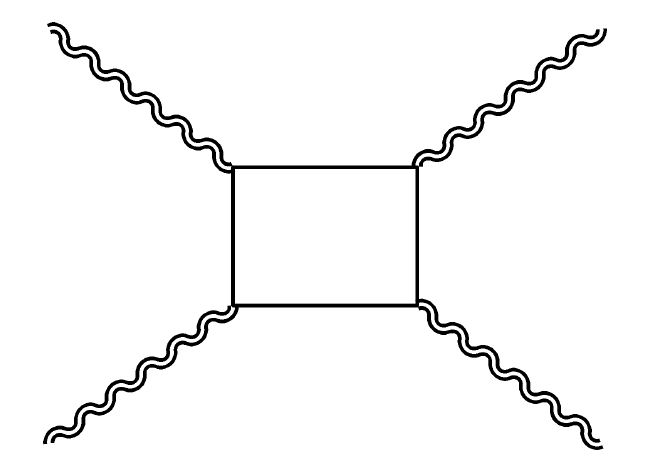}}}\nonumber \\
\end{align}
which can be directly computed in perturbation theory \cite{mirko}. 
 A counterterm action will be included for the process of renormalization. For example, in the action \eqref{phi} the counterterm action will be determined by two gravitational counterterms, which will be defined below \eqref{counter}, and the renormalized partition functional will be defined as 
\begin{equation}
Z[g]=\mathcal{N}\int D\phi e^{i(S_0(g,\phi) + S_{ct}(g))}=\mathcal{N}e^{i S_{ct}(g)}\int D\phi e^{i(S_0(g,\phi)}, 
\end{equation} 
where 
$\mathcal{N}$ is a normalization constant. 
  The role of the two counterterms is to remove the $1/\epsilon$ singularities present in the bare effective action $\sm_B$
\begin{equation}
\sm_B(g,d)=-\log\left(\int D\Phi e^{-S(\Phi,g)}\right) +\log\mathcal{N},
\end{equation}
and allow to define the regularized effective action in the form 
\begin{equation}
\label{rena}
\mathcal{S}_R(g,d)=\mathcal{S}_B(g,d) +  b' \frac{1}{\epsilon}V_E(g,d) + b \frac{1}{\epsilon}V_{C^2}(g,d).
\end{equation}
 $V_E$ and $V_{C^2}$ are related to the Euler density $E$ and to the Weyl tensor squared $C^2$, respectively, 
defined by the expressions
\begin{align}
\label{ffr}
V_{C^2}(g, d)\equiv & \mu^{\varepsilon}\int\,d^dx\,\sqrt{-g}\, C^2, \notag \\
V_{E}(g,d)\equiv &\mu^{\varepsilon} \int\,d^dx\,\sqrt{-g}\,E , 
\end{align}
where $\mu$ is a renormalization scale while $\varepsilon=d-4$. The counterterm vertices will be simply obtained by multiple differentiations of the two expressions above. We will start our analysis of the terms from the next section. 

For the case of the  scalar theory, this implies that the loop contributions are renormalized just by using the vertices obtained by the functional differentiation of $S_{ct}$ with respect to the metric. Specific choices of different backgrounds allows connect with cosmological scenarios. For instance, if $\bar{g}$ is chosen to be Weyl-flat, giving the opportunity for studying the conformal back reaction in De Sitter cosmology.

\subsection{The effective action and space time symmetries}
The analysis until this section is mainly focused in flat space time. Especially the derivation of the CWI can be done if we start from the effective action in a  curved background and take the flat limit. These results will be of course equivalent to the corresponding results that one can obtain from a conformal field theory. This is rather obvious since a CFT doesnt require a particular curved background to be formulated.
The infinitesimal Weyl variation of the metric  is 
  \begin{equation}
\delta_\sigma g_{\mu\nu}= 2 \sigma g_{\mu\nu}, 
\end{equation}
and its variation under diffeomorphisms ($x^\mu\to x^\mu+ \epsilon(x)$)  is
\begin{equation} 
\delta_\epsilon g_{\mu\nu}=-\nabla_{\mu}\epsilon_{\nu}- \nabla_{\nu}\epsilon_{\mu}.
\end{equation}
The variation of the action with respect to $\sigma$ is given by 
\begin{equation}
\label{anomx}
\frac{\delta \sm}{\delta \sigma(x)}=\sqrt{g} \,g_{\mu\nu}\,\langle T^{\mu\nu}\rangle, 
\qquad\textrm{where}\qquad  \langle T^{\mu\nu}\rangle = \frac{2}{\sqrt{g}}\frac{\delta \sm}{\delta g_{\mu\nu}},
\end{equation} 
Thus Weyl and diffeomorphism invariance of $\sm$ mean
\begin{equation} 
\label{eww1}
\delta_\sigma \sm=0, \qquad \delta_\epsilon \sm=0,
\end{equation}
are then summarised by the relations 
 \begin{equation}
\label{comby1}
\langle T^\mu_\mu\rangle=0, \qquad \textrm{and}\qquad  \nabla_\mu\langle T^{\mu\nu}\rangle=0.
\end{equation}
 By taking the functional derivative  of $\sm[g]$ with respect to the metric background, we can derive the trace and the conformal WI. In the presence of an anomaly, the equation take the form

\begin{equation}
\label{plus1}
\delta_\sigma \sm=\int d^4 x\sqrt{g} \bar{\mathcal{A}}(x), \qquad \langle T^\mu_\mu\rangle =\bar{\mathcal{A}}(x),
\end{equation}
which violates Weyl invariance.  $\sqrt{g} \bar{\mathcal{A}}(x)$ is the anomaly functional,that will be discussed later. Now, taking functional derivatives of the anomaly equation will produce the anomalous WI that play an important role in this thesis.\\


It is certainly interesting to examine the implications of  spacetime symmetries which that the effective action invariant. Two example are provided by metrics which allow Killing (KV) or conformal Killing vectors (CKV). The requirement that the action $\sm$ is invariant under such symmetries allows to define some conserved currents. These are constructed by contracting the stress energy tensor derived from $\sm$ with such 4-vector fields. 
The relations that one derives are valid at quantum level, since the stress energy tensor derived by varying $\sm$ is a quantum average. In particular, the use of conformal Killing vectors allows to define a conformal current $J_c$ which plays a key role in the derivation of the CWIs and of the corresponding CWIs in the flat spacetime limit.\\
In Minkowski space, scale invariance of a certain theory implies that the corresponding stress energy tensor has zero trace, and one can then define the conserved current
\begin{equation} 
J^\mu=x_\nu T^{\mu\nu}, \qquad \partial \cdot J=T^\mu_\mu=0.
\end{equation} 
At quantum level, in a curved background, we can generalize this approach by defining similar currents 
\begin{equation} 
\label{confcur}
\langle J^\mu\rangle =\epsilon_\nu^{(K)} \langle T^{\mu\nu}\rangle, 
\end{equation}
where $\epsilon_\mu^{(K)}(x)$ is a Killing or a conformal Killing vector field of the metric $g$. The proof of their conservation at quantum level follows closely the derivation at classical level. For instance, if we assume that $g$ allows vector isometries $\epsilon_\mu^{(K)}(x)$, which for changes of coordinates $x\to  x + \epsilon^{(K)}_\mu$ leave the metric invariant 
\begin{equation}
\delta_\sigma g_{\mu\nu}=-(\nabla_\mu \epsilon^{(K)}_\nu + \nabla_\nu \epsilon^{(K)}_\mu)=0,
\end{equation}
then the requirement of diffeomorphism invariance of the effective action $\sm[g]$ in that metric implies that the quantum average of $T^{\mu\nu}$  is conserved
\begin{equation} 
\nabla_\mu \langle T^{\mu\nu}\rangle=0.
\end{equation}
Combining this condition with the requirement that $\epsilon^{(K)}_\mu$ are Killing vectors,
then $J$ in \eqref{confcur} is conserved at quantum level
\begin{equation}
\label{iso}
\nabla\cdot \langle J\rangle =0. 
\end{equation}
A conformal current $J_c$ can be defined analogously to \eqref{confcur} by assuming that
the background metric $g$ allows conformal Killing vectors. 
In this more general case, we recall that the CKVs are solution of the equation
\begin{equation}
(ds')^2 = e^{2 \sigma(x)}(ds)^2 \qquad \leftrightarrow\qquad  \nabla_\mu \epsilon_\nu^{(K)} + \nabla_\nu \epsilon_\mu^{(K)} = 2 \sigma \delta_{\mu\nu}\qquad \sigma=\frac{1}{4}\nabla\cdot\epsilon^{(K)}.
\end{equation}
We remark that if we introduce a conformal current, now  using the CKVs of the background metric as in \eqref{confcur} in order to define $J_c$, if conditions \eqref{comby1} are respected by $\sm$, then $J_c$ is conserved as in the isometric case \eqref{iso}. \\
On the other hand, anomalous CWIs are generated if we allow a Weyl-variant term in $\sm$, which takes place in $d=4$, after renormalization, as in the case of an anomaly action. \\
In this case the anomaly induces a non-zero trace, and modifies the semi-classical condition \eqref{iso} into the new form
\begin{eqnarray}
\nabla\cdot \langle J_c\rangle =\frac{1}{4}\nabla \cdot \epsilon^{(K)} \langle T^\mu_\mu\rangle  +\epsilon^{(K)}_\nu \nabla_{\mu} \langle T^{\mu\nu}\rangle=\frac{1}{4}\nabla \cdot \epsilon^{(K)} \langle T^\mu_\mu\rangle .
\end{eqnarray}
This relation can still be used for the derivation of the special CWIs of $n$-point functions, as shown in \cite{Coriano:2017mux,tttt}.
Notice that $\sigma(x)$ is, at the beginning, a generic scalar function, which in a Taylor expansion around a given point $x^\mu$ is characterized by an infinite and arbitrary number of constants. Their number gets drastically reduced if we require that the spacetime manifold with metric $g$ allows a tangent space at each of its points, endowed with a flat conformal symmetry. 

In an equivalent, more general definition, one introduces an infinite dimensional abelian Weyl group $\mathcal{G}$ of transformations, with generators $J_x$ acting on the space of fields 
\begin{equation}
\label{jx}
J_x=-2 g^{\mu\nu}\frac{\delta}{\delta g^{\mu\nu}(x)} - \Delta_f  \Phi_f(x),
\end{equation}
summed over the fields $\Phi_f$, where $\Delta_f$ denote their scaling dimensions.
 The finite action of an element of $\mathcal{G}$ can be expressed as
 \begin{equation}
 e^{\sigma(x)\cdot J_x},\qquad \textrm{where} \qquad {\sigma(x)\cdot J_x}\equiv \int d^4 x \sigma(x) J_x,
 \end{equation}
with Weyl invariance of a generic action action expressed in the form 
\begin{equation} J_x\mathcal{S}_0[g,\Phi_f]=0. 
\end{equation}
In the case of the anomaly action $\Gamma[g]$, the expression \eqref{jx} includes only the variation of the metric, since all the remaining fields $\Phi_f$ have been integrated out, 
\begin{equation}
J_x \sm[g]=0.
\end{equation}
In flat space, the conformal Killing equation identify CKVs $\epsilon^\mu$ which are at most quadratic in $x$, are expressed in terms of the fifteen parameters $(a^\mu,\omega^{\mu\nu}, \lambda_s, b^\mu)$ of the conformal group, indicated as $K^\mu(x)$ 
\begin{equation}
\label{Kil}
\epsilon^{\mu}(x)\vline_{flat}\to K^\mu(x)= a^\alpha + \omega^{\mu\nu} x^\nu +\lambda_s x^\mu + b^\mu x^2 -2 x^\mu b\cdot x.
\end{equation}
Using such CKVs, the derivation of the special CWIs, following the approach of \cite{Coriano:2017mux}, can be performed directly in $d=4$, and takes to anomalous special CWIs. 
\section{A first look on the counterterms}\label{counterr}
Around a flat metric background, if we use a  mass-independent regularization scheme, the structure of the counterterms is polynomial in momentum space. \\
As we have seen,  the breaking of Weyl invariance  due to the presence of an anomaly produces
\begin{equation}
 \delta_\sigma \Gamma=\frac{1}{(4 \pi)^2}\int d^4 x \sqrt{g}\left(c_1 R_{\mu\nu\rho\sigma}R^{\mu\nu\rho\sigma} + c_2 R_{\mu\nu}R^{\mu\nu} +c_3 R^2\right).
\end{equation}
The exact form  is constrained by the Wess-Zumino consistency condition 
\begin{equation}
\label{WZcons}
\left[\delta_{\sigma_1},\delta_{\sigma_2}\right]\Gamma=0.
\end{equation}
Applying this condition we arrive at
\begin{equation}
\label{anom1}
\delta_\sigma \Gamma=\frac{1}{(4\pi)^2}\int d^4 x\sqrt{g}\sigma\left( b_1 C^{(4)}_{\mu\nu\rho\sigma}C^{(4)\mu\nu\rho\sigma} + b_2 E_{4} +b_3 \square R\right),
\end{equation}
where
\begin{align}
\label{fourd}
E_4&\equiv R_{\mu\nu\alpha\beta}R^{\mu\nu\alpha\beta}-4R_{\mu\nu}R^{\mu\nu}+R^2  ,\\
( C^{(4)})^2&\equiv R_{\mu\nu\alpha\beta}R^{\mu\nu\alpha\beta}-2R_{\mu\nu}R^{\mu\nu}+\frac{1}{3}R^2, \label{fourd2}
\end{align}
which are the Euler-Gauss-Bonnet (GB) invariant and the square of the Weyl conformal tensor, respectively, in $d=4$. 
In dimensional regularization, we need  an expansion of $d$ around 4 for a proper definition of  $C^2$. The dependence of $C^2$ on $d$ allows to introduce   
both a $(C^{(d)})^2$,  and a $(C^{(4)})^2$ operator, both defined in $d$ dimensions. Consequently the contraction of the indices in both cases are all performed in $d$ spacetime dimensions while the  coefficients in their definition  are {\em parametrically} dependent on $d$ 
\begin{equation}
C^{(d) \alpha\beta\gamma\delta}C^{(d)}_{\alpha\beta\gamma\delta}
=
R^{\alpha\beta\gamma\delta}R_{\alpha\beta\gamma\delta} -\frac{4}{d-2}R^{\alpha\beta}R_{\alpha\beta}+\frac{2}{(d-2)(d-1)}R^2,
\end{equation}
with
\begin{equation}
C^{(d)}_{\alpha\beta\gamma\delta} = R_{\alpha\beta\gamma\delta} -
\frac{2}{d-2}( g_{\alpha\gamma} \, R_{\delta\beta} + g_{\alpha\delta} \, R_{\gamma\beta}
- g_{\beta\gamma} \, R_{\delta\alpha} - g_{\beta\delta} \, R_{\gamma\alpha} ) +
\frac{2}{(d-1)(d-2)} \, ( g_{\alpha\gamma} \, g_{\delta\beta} + g_{\alpha\delta} \, g_{\gamma\beta}) R\, .
\end{equation}
We prefer to separate the parametric dependence from the range of variability of the tensor indices of this operator, since different choices of their range (and of the parametric dependence) induce finite renormalization of the effective action and change the anomaly by local terms. Here the term ``local'' refers to 
terms which are obtained from the Weyl variation of a local action. The remaining contributions, usually termed ``non-local'', refer to operators appearing in the anomaly functional which can be derived by varying a non-local action. An example of such action is the Riegert action.
In general one could define 
\begin{align}
& C^{(d)}_{\lambda\mu\nu\rho}\qquad     \lambda,\mu,\nu,\rho=0,1\ldots d-1,\notag \\
& C^{(4)}_{\lambda\mu\nu\rho}\qquad     \lambda,\mu,\nu,\rho=0,1\ldots d-1,\notag \\
& C^{(4) r}_{\lambda\mu\nu\rho},\qquad  C^{(d) r}_{\lambda\mu\nu\rho}  \qquad \lambda,\mu,\nu,\rho=0,1,2,3 \qquad  \textrm{restricted forms},
\end{align}
as possible operators appearing in the counterterm action that renormalizes the effective action. After differentiation of these expression with respect to the metric, only a contraction of the result with the metric itself will set a difference between these different definitions. If the indices of the differentiation are left open, there will be no extra dependence on $d$ which is generated in the various cases. \\
The operators above  have a {\em parametric} dependence on the dimension {\em and} an index variability, and both  can be 4 or $d$. 

The Weyl-scaling of the metric
\begin{equation} 
\label{GAUGED} g_{\mu \nu} = e^{2 \phi} \bar g_{\mu \nu}. 
\end{equation}

gives
\begin{equation}  
C{}_{\lambda\mu\nu\rho}=e^{2 \sigma} \bar{C}_{\lambda\mu\nu\rho},\qquad C^{\lambda\mu\nu\rho}=e^{-6  \sigma} \bar{C}^{ \lambda\mu\nu\rho},
\end{equation}
and corresponds to Weyl covariance. 
By contraction, one gets 
\begin{align}
C^2\equiv g^{\alpha\mu}g^{\beta\nu}C^{\lambda}_{\gamma\alpha\beta}C^{\gamma}_{\lambda\mu\nu}=& e^{-4\sigma(x)}\bar{C}^{\lambda}_{\gamma\alpha\beta}\bar{C}^{\gamma}_{\lambda\mu\nu}=e^{-4\sigma(x)}\bar{C}^2.  
\end{align}
Moreover, this tensor carries the symmetries of the Riemann tensor along with the corresponding Bianchi identity.
\begin{equation}
C_{\mu\nu\rho\sigma}=C_{\rho\sigma\mu\nu},
\end{equation}
\begin{equation}
C_{\mu\nu\rho\sigma}=-C_{\nu\mu\rho\sigma}=-C_{\mu\nu\sigma\rho},
\end{equation}
\begin{equation}
C_{\mu\nu\rho\sigma}+C_{\mu\sigma\nu\rho}+C_{\mu\rho\sigma\nu}=0.
\end{equation}
The choice of the Weyl counterterm is affected by the prescription dependence.  One could choose a counterterm of the form 
\begin{equation} 
\label{choice1}
\frac{1}{\epsilon}V_{C^2}=\frac{\mu^{-\epsilon}}{\epsilon}\int d^d x\sqrt{g} (C^{(d)})^2,
\end{equation}
 or 
\begin{equation} 
\frac{1}{\epsilon}\tilde{V}_{C^2}=\frac{\mu^{-\epsilon}}{\epsilon}\int d^d x\sqrt{g} (C^{(4)})^2.
\end{equation}
Let's examine the differences between the two choices. Setting $d=4-\epsilon$ in  \eqref{choice1}, we get
\begin{equation}
C^{(d)\alpha\beta\gamma\de}C_{(d)\alpha\beta\gamma\de}=R^{\alpha\beta\gamma\de}R_{\alpha\beta\gamma\de}-2R^{\alpha\beta}R_{\alpha\beta}+\frac{1}{3}R^2+\epsilon\lt -R^{\alpha\beta}R_{\alpha\beta}+\frac{5}{18}R^2 \rt,
\end{equation}
which can be written as 
\begin{equation}
\label{interm}
(C^{(4-\epsilon)})^2=(C^{(4)})^2 +\epsilon\left( - (R_{\mu\nu})^2 + \frac{5}{18}R^2\right).
\end{equation}
Using the scaling property of the integration measure 
\begin{equation} 
g=\textrm{det} g_{\alpha\beta} =\epsilon^{\mu_1\ldots \mu_d}g_{0 \mu_1}\ldots g_{0\mu_d}\to e^{2 d \sigma} g ,
\end{equation}
 one obtains 
\begin{equation} 
\frac{\delta}{\delta \sigma(x)}\int d^d x \sqrt{-g}(C^{4)})^2(x)=\epsilon \sqrt{g}(C^{(4)})^2,
\end{equation}
as well as 
\begin{equation} 
\frac{\delta}{\delta \sigma(x)}\int d^d x \sqrt{-g}(C^{d)})^2(x)=\epsilon \sqrt{g}(C^{(d)})^2.
\end{equation}
The use of \eqref{interm} before the Weyl variation in $\sigma$, produces  a different result, since the operation of expanding in $d$ and the variation do not commute. Looking separately to the variation of the second  term of \eqref{interm} we have
\begin{equation} 
\frac{\delta}{\delta \sigma(x)}\int d^d x \sqrt{-g}\left( - R_{\mu\nu}^2 + \frac{5}{18}R^2\right)=
-\frac{2}{3}\epsilon \sqrt{g} \Box R.
\end{equation}
Thus we obtain
\begin{equation}
\label{C2 piu boxR}
\frac{1}{d-4}\frac{\de}{\de\sigma(y)}\int d^dx \rg (C^{(4-\epsilon)})^2=\rg\lt (C^{(4)})^2 -\frac{2}{3}\Box R\rt.
\end{equation}
 The renormalization of the entire functional is then obtained by the addition of the counterterm action
\begin{align}
\label{counter}
\sm_{ct}&=-\frac{\mu^{-\varepsilon}}{\varepsilon}\,\int\,d^dx\,\sqrt{-g}\left(b\,(C^{(d)})^2+b'\,E\right),
\end{align}
corresponding to the Weyl tensor squared and the Euler density in $d=4$ as in \eqref{fourd} and \eqref{fourd2}. Here, $\mu$ is a renormalization scale.While the Gauss-Bonnet term does not carry any parametric dependence in $d$   its tensorial structure is expanded in $d$ dimensions, according to the regularization. The scaling of $E$ is more involved 
\begin{equation} 
\label{GaussB}
\rg E=\sqrt{\bar g} e^{(d-4)\phi}\biggl \{ \bar E+(d-3)\bar\nabla_\mu \bar J_1^\mu+(d-3)(d-4) \bar K  \biggl \} ,
\end{equation}
where
\begin{equation} 
\bar J_1^\mu=8\bar R^{\mu\nu}\bar\nabla_\nu\phi-4\bar R\bar \nabla^\mu\phi+4(d-2)(\bar\nabla^\mu\phi\bar \Box \phi-\bar \nabla^\mu\bar\nabla^\nu\phi\bar \nabla_\nu\phi+\bar\nabla^\mu\phi\bar\nabla_\lambda\phi\bar\nabla^\lambda\phi),
\end{equation}

\begin{equation} 
\bar K=4\bar R^{\mu\nu}\bar\nabla_\mu\phi\bar\nabla_\nu\phi-2\bar R\bar\nabla_\lambda\phi\bar\nabla^\lambda\phi+4(d-2)\bar\Box\phi\bar\nabla_\lambda\phi\bar\nabla^\lambda\phi+(d-1)(d-2)(\bnabla_\lambda \phi\bnabla^\lambda \phi)^2.
\end{equation}
and all the barred operators are computed respect to the fiducial metric $\bar{g}_{\mu\nu}$.\\
The Weyl tensor can be used to build a conformally invariant action at $d=4$ in the form
\begin{equation}
S_G=\alpha\int d^4 x \rg (C^{(4)})^2,
\label{azione gravitazionale conforme di weyl}
\end{equation}
which is the action of Weyl gravity. In the context of anomaly actions, the derivative of $V_{C^2}$ (i.e. $V'_{C^2}$) with respect to the dimension $d$ will be part of the renormalized action and the approach can be modified by the inclusion of a finite renormalization. 
For the remaining section we will  choose $(C^{(d)})^2$ as the operator appearing in the counterterm action, and we will be using the variation
\begin{equation}
\label{oneq}
\frac{\delta}{(d-4)\delta \sigma (x) }\int d^d x \sqrt{-g} (C^{(d)})^2 =\sqrt{-g} (C^{(d)})^2,
\end{equation}
which differs from 
\begin{equation}
\label{twoq}
\frac{\delta}{(d-4)\delta \sigma (x) }\int d^d x \sqrt{-g} (C^{(4)})^2 =\sqrt{-g}\Bigg( (C^{(4)})^2
-\frac{2}{3}\square R\Bigg) ,
\end{equation}
introduced in some of the literature on the conformal anomaly. 
We remark that the number of Weyl invariants operators depends on the dimension. The case of $d=6$ is discussed in several works \cite{Bastianelli:2000hi,Ferreira:2015lna,Ferreira:2017wqz}.\\
One can derive an anomaly action of the Wess-Zumino form starting from those invariants, using the Weyl gauging approach \cite{Iorio:1996ad,Codello:2012sn}. An example, in  $d=6$, can be found in \cite{Coriano:2013nja}.  \\
The anomaly action that one introduces via the regularization process, is derived using counterterms which are defined  for any even dimension ($2 k$), and analytically continued around the integer value $k$.
 
 In generic  spacetime dimensions, the structure of the counterterm Lagrangian is modified accordingly, with the Euler/ GB  density formula
\begin{align}
\label{Ed}
E_d = \frac{ 1}{2^{d/2}}
\delta_{\mu_1 \cdots \mu_d}^{\nu_1 \cdots \nu_d}
{R^{\mu_1 \mu_2}}_{\nu_1 \nu_2} \cdots
{R^{\mu_{d-1} \mu_d}}_{\nu_{d-1} \nu_d} \ ,
\end{align}
which, for $d=4$, is quadratic in the curvatures. The topological nature 
of $V_E$ is evident only if $d=2k$, but the regularization procedure will modify the effective action with the inclusion of a finite contribution ($V'_E$) which is metric-dependent, produced by the integration of the Euler density  in $d=2 k +\epsilon$ dimensions.
 \subsection{ Gauss-Bonnet term and finite renormalization}
 The following section is based on \cite{tttt}. We will address the nature of the finite renormalization induced by the Gauss-Bonnet counterterm and examine this counterterm in more detail.
\subsubsection{The finite renormalization induced by $V_E$} 
It is quite obvious that the inclusion of $V_E$ induces a finite renormalization of the effective action in $d=4$. This can be simply shown by noticing that both $V_E$ and $V_{C^2}$ manifest an explicit dependence on $\varepsilon$, i.e. 
\begin{equation}
V_{E/C^2}\equiv V_{E/C^2}(d),
\end{equation} 
and the counterterm contributions can be expanded around $d=4$. 
For this purpose, given a scalar functional  $f(d)$, it will be convenient to denote its Taylor expansion around $d=4$ in the form 
\begin{equation} 
f(d)=\left[f\right] + \varepsilon \left[f'\right] \qquad \left[f\right]\equiv f(4), \left[f'\right] \equiv 
f'(4) 
\end{equation}
with coefficients which are square bracketed once they are computed at $d=4$. The expansion of $V_{E/C^2}(d)$, using these notations, takes the form 

\begin{equation} 
\label{expand}
\frac{1}{\varepsilon}V_{E/C^2}(d)=\frac{\mu^{-\varepsilon}}{\varepsilon}\left( \left[V_{E/C^2}\right] + \varepsilon 
\left[V_{E/C^2}'\right] +O(\varepsilon^2) \right),
\end{equation}
where the first correction to the residue at the pole in $\varepsilon$ comes from the derivative respect to the dimension $d$. 
Notice that $\left[V_E\right]$ is a topological term and it is therefore metric-independent. Its value is related to the global topology of the spacetime and it is therefore a pure number. 
Then, it is clear that the $1/\varepsilon$ term will not contribute to the renormalization of the bare 4-T vertex, since each counterterm vertex  is generated by functional differentiation of \eqref{expand} with respect to the background metric. \\
The only contribution of the $V_E$ as $\varepsilon\to 0$  is related to $\left[V'_E\right]$, and it is indeed finite, as a 0/0 contribution in $\varepsilon$. Therefore, the inclusion of $V_E$, will induce only a finite renormalization of the bare vertex, and henceforth of the entire effective action, since this result remains valid to all orders.\\
Finally, we can relate $\left[V'_E\right]$ to the anomaly by the equation 
\begin{equation}
\sqrt{g} E(x)=2 g_{\alpha\beta}\left[ V'_E\right]^{\alpha\beta}
\end{equation} 
where the indices of $\left[ V'_E\right]^{\alpha\beta}$ run in four dimensions. 
\subsubsection{Open indices in $V_E$}
We can investigate this point in more detail.\\
 The extraction of a counterterm vertex, as mentioned above, requires a functional differentiation of \eqref{expand}. 
It will appear in the $\varepsilon\to 0$ limit, in the form
\begin{align}
\label{ff}
\left[ \frac{\delta}{\delta g_{\mu_1\nu_1} (x_1)} \frac{\delta}{\delta g_{\mu_2\nu_2} (x_2)}\ldots  \frac{\delta}{\delta g_{\mu_n\nu_n} (x_n)}\frac{\mu^{-\varepsilon}}{\varepsilon}V_{E}(d)\right]= &
\frac{\delta}{\delta g_{\mu_2\nu_2} (x_2)}\ldots  \frac{\delta}{\delta g_{\mu_n\nu_n} (x_n)}\left[ V_{E}'\right]\notag \\
&\equiv \left[ V_{E}'\right]^{\mu_1\nu_1\ldots  \mu_n\nu_n}(x_1,\ldots x_n).\notag \\
\end{align} 
The rhs of the this expression is clearly regular as $\varepsilon\to 0$, and defines a finite renormalization of the corresponding $n$-graviton vertex. Obviously, the same is not true for the $V_{C^2}$ counterterm since $\left[V_{C^2}\right]$ will be metric-dependent. In this second case the analogous of \eqref{ff} is 
\begin{align}
\label{open}
\left[ \frac{\delta}{\delta g_{\mu_1\nu_1} (x_1)} \frac{\delta}{\delta g_{\mu_2\nu_2} (x_2)}\ldots  \frac{\delta}{\delta g_{\mu_n\nu_n} (x_n)}\frac{\mu^{-\varepsilon}}{\varepsilon}V_{C^2}(d)\right]= &
\frac{1}{\varepsilon} \left[V_{C^2}\right]^{\mu_1\nu_1\ldots  \mu_n\nu_n}(x_1,\ldots x_n)+ \left[V_{C^2}'\right]^{\mu_1\nu_1\ldots  \mu_n\nu_n}(x_1,\ldots x_n).\notag \\
\end{align} 
The second, finite term of this expression  ($V'_{C^2}$), will appear in all the renormalized 
anomalous WI that we will present below, when we will perform the  $\varepsilon\to 0$ limit of those eqs. in a flat background. \\
One important comment concerns possible ambiguities arising from the differentiation of 
terms such as $\left[V'_E\right]$ with respect to the metric $g_{\mu\nu}$. We mention that such ambiguities arise only in the presence of contractions with a metric, and not otherwise. We will be dealing with this second case, in the subsection below.\\
As in any practical application of DR, once the renormalization of a vertex is completed, a tensor structure with open indices, which is generated by the procedure, is automatically dimensionally reduced to the $d=4$ subspace, giving the final, finite expression of such vertex. \\
Indeed, expanding around $d=4$, \eqref{expand} can be rewritten in the form 
\begin{align}
 \frac{\delta}{\delta g_{\mu_1\nu_1} (x_1)} \frac{\delta}{\delta g_{\mu_2\nu_2} (x_2)}\ldots  \frac{\delta}{\delta g_{\mu_n\nu_n} (x_n)}V_{E}(d)= &
\left[V_{E}\right]^{\mu_1\nu_1\ldots  \mu_n\nu_n}(x_1,\ldots x_n)+ \varepsilon\left[V_{E}'\right]^{\mu_1\nu_1\ldots  \mu_n\nu_n}(x_1,\ldots x_n),\notag \\
\end{align}  
which in the $d\to 4$ limit vanishes, since the two terms on the rhs vanish separately. 
Notice that the first term $\left[V_{E}\right]^{\mu_1\nu_1\ldots  \mu_n\nu_n}$ is topological and is simply obtained by the differentiation of $V_E$ in $d=4$. As far as we leave all the indices open, then the following relation holds 
\begin{align}
\label{frr}
\lim_{d\to 4} \left(\frac{\delta}{\delta g_{\mu_1\nu_1} (x_1)} \frac{\delta}{\delta g_{\mu_2\nu_2} (x_2)}\ldots  \frac{\delta}{\delta g_{\mu_n\nu_n} (x_n)}V_{E}(d)\right)=&
 \frac{\delta}{\delta g_{\mu_1\nu_1} (x_1)} \frac{\delta}{\delta g_{\mu_2\nu_2} (x_2)}\ldots  \frac{\delta}{\delta g_{\mu_n\nu_n} (x_n)} \lim_{d\to 4} V_{E}(d) \notag \\
=& 0
\end{align} 
 In other words, the operation of functional differentiation with open indices and limit to $d=4$ commute. The topological nature of this part of the 3-point vertex counterterm was originally noted in \cite{Osborn:1993cr} (see also \cite{Zwiebach:1985uq}).
The explicit check of this topological relation requires the dimensional reduction of the $\delta^{d}_{\mu\nu}$ to $\delta^{(4)}_{\mu\nu}$, using the $n$-$p$ basis (discussed in Sec. \ref{nnp}) to take into account momentum degeneracies in specific dimensions, and indeed it has been verified for $n=3$ and $n=4$ \cite{Bzowski:2015pba,Serino:2020pyu}, i.e. for three- and four-point functions. We stress once again that \eqref{frr} holds as far as we do not perform any contraction of the tensor indices with the external metric. \\ 
In the presence of metric contractions, these results can be re-addressed using a Weyl variation. We are going to illustrate this point in some detail below, defining a straightforward procedure in order to handle this second case correctly. \\
Contributions proportional to $V'_E$ will be present once we contract the WIs with the $\Sigma$ projector defined below in \eqref{Lproj}, using the longitudinal-trace/ transverse-traceless decomposition of the correlators. 
One finds, by a direct computation, the emergence of virtual scalar exchanges (or mixing terms) in the decomposed WI, which signal the breaking of the conformal symmetry induced by the conformal anomaly. Such terms are easy to derive simply due to the presence of a combined $\pi^{\mu\nu}\delta^{\alpha\beta}$ projector in $\Sigma$, in each of the external legs of the gravitational vertex. This defines a coupling of a scalar pole to the anomaly functional. 
\subsubsection{Closed indices}
The analysis of relations involving the topological counterterm $V_E$ in the presence of contraction with the metric cannot be performed as above, but, as we have mentioned, 
can be addressed correctly by relating the contraction to a Weyl variation. A crucial relation is
\begin{equation}
\label{form}
\frac{\delta}{(d-4)\delta \sigma (x) }\int d^d x \sqrt{-g} E_4 =\sqrt{-g} E_4.
\end{equation} 
For this purpose consider the Weyl scaling relations 
\begin{equation}
\sqrt{g}=e^{d \sigma.}\sqrt{\bar{g}}\qquad E=e^{-4 \sigma}\bar{E},
\end{equation}
from which we derive the constraints 
 \begin{equation}
 \label{third}
 \frac{\delta}{\delta \sigma(x_3)} \frac{\delta}{\delta \sigma(x_2)} \frac{\delta}{\delta \sigma(x_1)}V_E=\varepsilon^3 \sqrt{g}E \delta^d(x_1-x_2)\delta^d(x_2-x_3)
\end{equation}
which can be generalized to any multiple derivative 
\begin{equation} 
\label{generalder}
\frac{\delta}{\delta \sigma(x_n)} \ldots\frac{\delta}{\delta \sigma(x_2)} \frac{\delta}{\delta \sigma(x_1)}V_E=\varepsilon^n \sqrt{g}E \delta^d(x_1-x_2)\delta^d(x_2-x_3)\ldots \delta^d(x_{n-1}-x_n).
\end{equation}
If we use 
\begin{equation}
\label{var}
2 g_{\alpha\beta} \frac{\delta}{\delta g_{\alpha\beta}}=\frac{\delta}{\delta \sigma(x)}
\end{equation} 
and
\begin{equation}
\frac{\delta}{\delta \sigma(x_2)}g^{\mu\nu}(x_1)=-2 g^{\mu\nu}\delta^4(x_2-x_1) 
\end{equation}
we derive several relations. For instance, expanding the left-hand side of \eqref{generalder} for $n=2$ we obtain  
\begin{align}
\frac{\delta}{\delta\sigma(x_2)}\frac{\delta}{\delta\sigma(x_1)}V_E=& 4 g_{\mu\nu} V_E^{\mu\nu}(x_1)
\delta^d(x_1-x_2) + 4 g_{\alpha\beta}(x_2)g_{\mu\nu}(x_1) V_E^{\alpha\beta\mu\nu}(x_2,x_1)\notag \\
=&\varepsilon^2{\sqrt{g(x_1)}}E(x_1)\delta^d(x_1-x_2).
\end{align}
The expression above can be evaluated in the flat limit both in 4 and $d$ dimensions. Consider, for instance, the flat limit with $d$ generic. We use the fact that $E$ and $V_E^{\mu\nu}$ vanish in the flat limit to obtain the constraint
\begin{align}
\label{double}
\delta^{(d)}_{\alpha\beta}\delta^{(d)}_{\mu\nu}V_E^{\alpha\beta\mu\nu}=0, 
\end{align}
which remains valid for $d=4$ since the equation is analytic in $d$, and in particular in $d=4$. 
Similarly, one can derive relations for the traces of higher point functions, which are discussed in the \appref{TraceRelations}. 
\subsubsection{Open and closed indices}
We can generalize the method and perform combined variations respect to $\sigma$ and 
to the metric in order to derive some additional relations. 
For this purpose we consider the expression 

\begin{equation}
J^{\mu_2\nu_2\mu_3\nu_3\mu_4\nu_4}\equiv
\frac{\delta}{\delta \sigma(x_1)}\frac{\delta}{\delta g_{\mu_2\nu_2}(x_2)}\frac{\delta}{\delta g_{\mu_3\nu_3}(x_3)}\frac{\delta}{\delta g_{\mu_4\nu_4}(x_4)}V_{E}.
\end{equation}
We can use the commutation relation
\begin{equation}
\left[ \frac{\delta }{\delta \sigma(x_1)},\frac{\delta}{\delta g_{\mu_2\nu_2}(x_2)}\right] V_E=
-2 V_E^{\mu_2\nu_2}(x_2)\delta^d(x_1-x_2)
\end{equation}
 to rearrange the differentiations in the form 
\begin{align}
J^{\mu_2\nu_2\mu_3\nu_3\mu_4\nu_4}=&
\frac{\delta}{\delta g_{\mu_2\nu_2}(x_2)}\frac{\delta}{\delta g_{\mu_3\nu_3}(x_3)}\frac{\delta}{\delta g_{\mu_4\nu_4}(x_4)}\frac{\delta}{\delta \sigma(x_1)}V_{E} \notag \\- 
& 2 V_E^{\mu_2\nu_2\mu_3\nu_3\mu_4\nu_4}(x_2,x_3,x_4) \left(\delta^d(x_1-x_2) +
\delta^d(x_1-x_3) +\delta^d(x_1-x_4)\right)\notag \\
\end{align} 
and use \eqref{form} to derive the constraint

\begin{align}
J^{\mu_2\nu_2\mu_3\nu_3\mu_4\nu_4}=& (d-4)\left(\sqrt{g} E\right)^{\mu_2\nu_2\mu_3\nu_3\mu_4\nu_4}(x_2,x_3,x_4) \notag \\- 
& 2 V_E^{\mu_2\nu_2\mu_3\nu_3\mu_4\nu_4}(x_2,x_3,x_4) \left(\delta^d(x_1-x_2) +
\delta^d(x_1-x_3)+ \delta^d(x_1-x_4)\right),\notag \\
\end{align}
 which is \eqref{mom} in coordinate space. The second term in the expression above, related to the pinched contributions $\delta (x_1-x_i)\, (i=2,3,4)$ is also d-dimensional and needs to be expanded around $d=4$. The expansion is performed as in \eqref{expand}  
 \begin{equation} 
V_E^{\mu_2\nu_2\mu_3\nu_3\mu_4\nu_4}= \left[V_E^{\mu_2\nu_2\mu_3\nu_3\mu_4\nu_4}\right] + \varepsilon \left[{V'_E}^{\mu_2\nu_2\mu_3\nu_3\mu_4\nu_4}\right]. 
 \end{equation}
Notice that the first contribution vanishes for being purely topological, while the second, transformed to momentum space, will appear in all the renormalized anomalous CWIs 
that we will discuss in the sections below 
\begin{equation}
(d-4)\left[{V'_E}^{\mu_2\nu_2\mu_3\nu_3\mu_4\nu_4}\right]\to
(d-4)\left[{V'}_{E}^{\m_2\n_2\dots\mu_4\nu_4}(p_1+p_2,p_3,p_4)\right]. 
\end{equation} 
It accounts for a finite $0/0$ contribution in the expansion of the $1/\varepsilon \, V_E(d)$ counterterm in the variable $\varepsilon$.  

For a generic $nT$ correlator, the only counterterm needed for its renormalization, is obtained by the inclusion of a classical gravitational vertex generated by the differentiation of \eqref{counter} $n$ times. The coefficients $b$ and $b'$ will be fixed by the field content, number of scalars, fermions and spin-$1$ field, of the quantum corrections.\\

\subsection{The inclusion of the anomaly}
To study the anomaly contribution to each correlation function, we start from the 1-point function. In a generic background $g$, the renormalized 1-point function is decomposed as 
\begin{equation}
\langle T^{\mu\nu}\rangle_{R}=\frac{2}{\sqrt{g}}\frac{\delta \sm_{R}}{\delta g_{\mu\nu}} =\langle T^{\mu\nu} \rangle_A  + \langle \overline{T}^{\mu\nu}\rangle_f,
\end{equation}
with
\begin{equation}
g^{\mu\nu}\frac{\delta \Gamma}{\delta g^{\mu\nu}} = g^{\mu\nu}\frac{\delta \sm_A}{\delta g^{\mu\nu}}\equiv \frac{\sqrt{g}}{2} g_{\mu\nu} \langle T^{\mu\nu} \rangle_A ,
\end{equation}
being the trace anomaly equation, and $\langle \overline{T}^{\mu\nu}\rangle_f$ is the Weyl-invariant (traceless) term. We will omit below the subscript $R$ from the quantum average $\langle T\rangle_R$.

According to \eqref{anom1}, these scaling violations can be written  in the form
\begin{equation}
\label{anomeq}
\langle{T^{\mu}_{\ \ \mu}(x)}\rangle=\mathcal{A}(x),
\end{equation} 
- having dropped the suffix {\em Ren} from the renormalized stress energy tensor - 
where the finite terms on the right hand side of this equation denote the anomaly contribution
with
\begin{equation}
\mathcal{A}(x)=\sqrt{-g(x)}\bigg[b\,C^2(x)+b'E(x)\bigg],
\end{equation}
being the anomaly functional. 
For $n$-point functions the trace anomaly, as well as all the other CWIs, are far more involved, and take a hierarchical structure. \\

For all the other WIs, in DR the structure of the equations can be analyzed in two different frameworks. \\
In one of them, we are allowed to investigate the correlators directly in $d$ spacetime dimensions, deriving ordinary (anomaly-free) CWIs, which are then modified by the inclusion of the 4-dimensional counterterm as $d\to 4$. In this limit, the conformal constraints become anomalous and the hierarchical equations are modified by the presence of extra terms which are anomaly-related. \\
Alternatively, one can proceed by working out the equations directly in d=4, with the inclusion of the contributions coming from the anomaly functional.
\part{Four-point functions in momentum space}

\chapter{A brief review of the scalar four-point correlators}\label{OOOO}

In this section which is heavily based \cite{Coriano:2020ees,Maglio:2019grh} we are going to present some properties of the four-point scalar function. It is well known that four-point and higher functions are not completely constrained by the the conformal symmetry. As we have already shown in \ref{CFTcord}, for the case of the four-point correlator one can identify the invariant cross sections
\begin{equation}
\label{uv}
u(x_i)=\frac{x_{12}^2 x_{34}^2}{x_{13}^2 x_{24}^2}, \qquad v(x_i)=\frac{x_{23}^2 x_{41}^2}{x_{13}^2 x_{24}^2},
\end{equation}
The general solution can be written in the form 
\begin{equation}
\label{general}
\langle \mO_1(x_1)\mO_2(x_2)\mO_3(x_3)\mO_4(x_4)\rangle=  \frac{h(u(x_i),v(x_i))}{\left(x_{12}^2\right)^\frac{\Delta_1 + \Delta_2}{2}\left(x_{3 4}^2\right)^\frac{\Delta_3 + \Delta_4}{2}},
\end{equation}
where $h(u(x_i),v(x_i))$ remains unspecified. 
As we have seen, in momentum space
 the correlator depends on six invariants that we will normalize as
$p_i=|\sqrt{p_{i}\,{\hspace{-0.09cm}}^2}|$, $i=1,\dots,4$, representing the magnitudes of the momenta, and $s=|\sqrt{(p_1+p_2)^2}|$, $t=|\sqrt{(p_2+p_3)^2}|$ the two Mandelstam invariants redefined by a square root, for which 
\begin{equation}
\langle O(p_1)\,O(p_2)\,O(p_3)\,O(\bar{p}_4)\rangle=\Phi(p_1,p_2,p_3,p_4,s,t).\label{invariant}
\end{equation}
This correlation function
 has to verify the dilatation Ward identity
\begin{equation}
\left[\sum_{i=1}^4\Delta_i-3d-\sum_{i = 1}^3p_i^{\mu}\frac{\partial}{\partial p_i^\mu}\right]\Phi(p_1,p_2,p_3,p_4,s,t)=0,
\end{equation}
and the special conformal Ward identities
\begin{equation}
\sum_{i=1}^3\left[2(\Delta_i-d)\frac{\partial}{\partial p_{i\,\k}}-2p_i^\alpha\frac{\partial^2}{\partial p_i^\alpha\partial p_i^\kappa}+p_i^\kappa\frac{\partial^2}{\partial p_i^\alpha\partial p_{i\,\alpha}}\right]\Phi(p_1,p_2,p_3,p_4,s,t)=0.
\end{equation}
One can split these equations in terms of the invariants of the four-point function written in \eqref{invariant}, by using the chain rules
\begin{align}
\frac{\partial}{\partial p_{1\,\mu}}&=\frac{p_1^\mu}{p_1}\frac{\partial}{\partial p_1}-\frac{\bar{p}_4^\mu}{p_4}\frac{\partial}{\partial p_4}+\frac{p_1^\mu+p_2^\mu}{s}\frac{\partial}{\partial s},\\
\frac{\partial}{\partial p_{2\,\mu}}&=\frac{p_2^\mu}{p_2}\frac{\partial}{\partial p_2}-\frac{\bar{p}_4^\mu}{p_4}\frac{\partial}{\partial p_4}+\frac{p_1^\mu+p_2^\mu}{s}\frac{\partial}{\partial s}+\frac{p_2^\mu+p_3^\mu}{t}\frac{\partial}{\partial t},\\
\frac{\partial}{\partial p_{3\,\mu}}&=\frac{p_3^\mu}{p_3}\frac{\partial}{\partial p_3}-\frac{\bar{p}_4^\mu}{p_4}\frac{\partial}{\partial p_4}+\frac{p_2^\mu+p_3^\mu}{t}\frac{\partial}{\partial t},
\end{align}
where $\bar{p}_4^\mu=-p_1^\mu-p_2^\mu-p_3^\mu$. From this prescription the dilatation WI becomes
\begin{align}
\bigg[(\Delta_t-3d)-\sum_{i=1}^4p_i\frac{\partial}{\partial p_i}-s\frac{\partial}{\partial s}-t\frac{\partial}{\partial t}\bigg]\Phi(p_1,p_2,p_3,p_4,s,t)=0,\label{Dilatation4ch}
\end{align}
with $\Delta_t=\sum_{i=1} ^4\Delta_i$ is the total scaling,
and the special CWIs can be written as
\begin{equation}
\sum_{i=1}^3\ p_i^\kappa\, C_i=0\label{primary},
\end{equation}
where the coefficients $C_i$ are differential equations of the second order with respect to the six invariants previously defined. Being $p_1^\k,\ p_2^\k,\ p_3^\k$, in \eqref{primary} independent variables, 
we derive three scalar second order equations for each of the three $C_i$, which must vanish independently.\\
At this stage the procedure in order to simplify the corresponding equations is similar to the one described in \cite{Coriano:2018bsy1,Coriano:2018bbe}. Thus equation are rewritten in the form
\begin{align}
C_1&=\bigg\{\frac{\partial^2}{\partial p_1^2}+\frac{(d-2\Delta_1+1)}{p_1}\frac{\partial}{\partial p_1}-\frac{\partial^2}{\partial p_4^2}-\frac{(d-2\Delta_4+1)}{p_4}\frac{\partial}{\partial p_4}\notag\\[1.5ex]
&\qquad+\frac{1}{s}\frac{\partial}{\partial s}\left(p_1\frac{\partial}{\partial p_1}+p_2\frac{\partial}{\partial p_2}-p_3\frac{\partial}{\partial p_3}-p_4\frac{\partial}{\partial p_4}\right)+\frac{(\Delta_3+\Delta_4-\Delta_1-\Delta_2)}{s}\frac{\partial}{\partial s}\notag\\[1.5ex]
&\qquad+\frac{(p_2^2-p_3^2)}{st}\frac{\partial^2}{\partial s\partial t}\bigg\}\,\Phi(p_1,p_2,p_3,p_4,s,t)=0\label{C1ch},
\end{align}
for $C_1$ and similarly for the other coefficients. \\
It has been shown that these equations involve a differential operator which is of hypergeometric type
combined with other terms.
\section{The dcc solutions} 
 We consider possible solutions of the conformal constraints \eqref{C1ch} which are built around specific dual conformal ans\"atze \cite{Maglio:2019grh}. It is noteworthy that the property of an object to be conformally and dual conformally invariant is also known as conformal Yangian (CY) invariance.

The number of independent equations in \eqref{primary}, by using the ansatz that we are going to present below, will then reduce from three down to two. We choose  the ansatz
		\begin{equation}
		\Phi(p_i,s,t)=\big(s^2t^2\big)^{n_s}\,F(x,y)\label{ansatz},
		\end{equation}
		where $n_s$ is a coefficient (scaling factor of the ansatz) that we will fix below by the dilatation WI, and the variables $x$ and $y$ are defined by the quartic ratios
		\begin{equation}
		x=\frac{p_1^2\,p_3^2}{s^2\,t^2},\qquad y=\frac{p_2^2\,p_4^2}{s^2\,t^2}.
		\end{equation}
		By inserting the ansatz \eqref{ansatz} the dilatation Ward identities, and turning to the new variables $x$ and $y$, we obtain from \eqref{Dilatation4ch} the condition
		\begin{equation}
		\label{dil1}
		\bigg[(\Delta_t-3d)-\sum_{i=1}^4p_i\frac{\partial}{\partial p_i}-s\frac{\partial}{\partial s}-t\frac{\partial}{\partial t}\bigg]\big(s^2t^2\big)^{c}\,F(x,y)=\big(s^2t^2\big)^{n_s}\big[(\Delta_t-3d)-4 n_s\big] \,F(x,y)=0,
		\end{equation}
		which determines $n_s=(\Delta_t-3d)/4$, giving
		\begin{equation}
		\Phi(p_i,s,t)=\big(s^2t^2\big)^{(\Delta_t-3d)/4}\,F(x,y).\label{ansatz2}
		\end{equation}
		The functional form of $F(x,y)$ will then be furtherly constrained.

\subsection{Equal scaling solutions}

	We start investigating the solutions of these equations by assuming that the scaling dimensions of all the scalar operators are  equal $\Delta_i=\Delta$, $i=1,\dots,4$.
	The special conformal Ward identities, re-expressed in terms of $x$ and $y$, are given in the form
	\begin{equation}
		\left\{\begin{aligned}
		&\bigg[y(1-y)\partial_{yy} -2x\,y\,\partial_{xy}-x^2\partial_{xx}-(1-2n_s)x\,\partial_x+\left(1-\Delta+\frac{d}{2}-y(1-2n_s)\right)\,\partial_y-n_s^2\bigg] F(x,y)=0\\[2ex]
		&\bigg[x(1-x)\partial_{xx} -2x\,y\,\partial_{xy}-y^2\partial_{yy}-(1-2n_s)y\,\partial_y+\left(1-\Delta+\frac{d}{2}-x(1-2n_s)\right)\,\partial_x-n_s^2\bigg] F(x,y)=0,
		\end{aligned}\right.\label{EqualScal},
	\end{equation}
	where we recall that $n_s$ is the scaling of the correlation function under dilatations, now given by
	\begin{equation}
	\label{scaled}
	n_s=\Delta-\frac{3d}{4},
	\end{equation}
	since $\Delta_t=4\, \Delta$.  Eqs. \eqref{EqualScal} correspond to  a hypergeometric system and its solutions can be expressed as linear combinations of four Appell functions $F_4$ of two variables $x$ and $y$, as in the case of three-point  functions. The general solution of such system is expressed as
\begin{align}
\Phi(p_i,s,t) &=\big(s^2t^2\big)^{(\Delta_t-3d)/4}\, \sum_{a,b} c(a,b,\Delta) x^a y^b F_4\left(\a(a,b),\b(a,b),\g(a),\g'(b);x, y\right).\label{solutionEq}
\end{align}
Notice that the solution is similar to that of the three-point  functions given by \eqref{compact}.\\
The general solution \eqref{solutionEq} has been written as a linear superposition of these, with independent constants $c(a,b)$ labeled by the exponents $a,b$ 
\begin{align}
&a=0,\,\Delta-\frac{d}{2},&b=0,\,\Delta-\frac{d}{2},\label{FuchsianPoint}
\end{align}
which fix the dependence of the $F_4$  
\begin{align}
\label{s2}
&\a(a,b)=\frac{3}{4}d-\Delta+a+b,&\b(a,b)=\frac{3}{4}d-\Delta+a+b,\notag\\
&\Gamma(a)=\frac{d}{2}-\Delta+1+2a,&\g'(b)=\frac{d}{2}-\Delta+1+2b.
\end{align}

\subsection{More general conditions}
The solution that we have identified in the equal scaling case can be extended by relaxing the conditions on the scaling dimensions, in the form
\begin{equation}
\Delta_1=\Delta_3=\Delta_x,\qquad \Delta_2=\Delta_4=\Delta_y.
\end{equation}
 In this case the CWIs give the system of equations
\begin{equation}
\left\{\begin{aligned}
&\bigg[y(1-y)\partial_{yy} -2x\,y\,\partial_{xy}-x^2\partial_{xx}-(1-2n_s)x\,\partial_x+\left(1-\Delta_y+\frac{d}{2}-y(1-2n_s)\right)\,\partial_y-n_s^2\bigg] F(x,y)=0\\[2ex]
&\bigg[x(1-x)\partial_{xx} -2x\,y\,\partial_{xy}-y^2\partial_{yy}-(1-2n_s)y\,\partial_y+\left(1-\Delta_x+\frac{d}{2}-x(1-2n_s)\right)\,\partial_x-n_s^2\bigg] F(x,y)=0
\end{aligned}\right.,
\end{equation}
where now $n_s$ is defined as 
\begin{equation}
n_s=\frac{\Delta_x}{2}+\frac{\Delta_y}{2}-\frac{3}{4}d,
\end{equation}
whose solutions are expressed as
\begin{equation}
\label{ress}
\Phi(p_i,s,t)=\big(s^2t^2\big)^{(\Delta_t-3d)/4}\,\sum_{a,b} c(a,b,\vec\Delta_t) x^a y^bF_4\left(\a(a,b),\b(a,b),\g(a),\g'(b);x,y\right),
\end{equation}
with $\vec\Delta_t=(\Delta_x,\Delta_y,\Delta_x,\Delta_y)$, $\Delta_t=2 \Delta_x + 2 \Delta_y$ and the Fuchsian points are fixed by the conditions
\begin{align}
&a=0,\,\Delta_x-\frac{d}{2},&b=0,\,\Delta_y-\frac{d}{2},\notag\\
&\a(a,b)=\frac{3}{4}d-\frac{\Delta_x}{2}-\frac{\Delta_y}{2}+a+b,&\b(a,b)=\frac{3}{4}d-\frac{\Delta_x}{2}-\frac{\Delta_y}{2}+a+b,\notag\\
&\g(a)=\frac{d}{2}-\Delta_x+1+2a,&\g'(b)=\frac{d}{2}-\Delta_y+1+2b.
\end{align}
The domain of convergence of such solutions
 for $F_4$ is bounded by the relation 
\begin{equation}
\sqrt{x}+\sqrt{y}< 1, 
\end{equation}
which is satisfied in a significant kinematic region, and in particular at large energy and momentum transfers. Notice that the analytic continuation of \eqref{ress} in the physical region can be simply obtained by sending $t^2\to -t^2$ (with $t^2<0$) and leaving all the other invariants untouched. In this case we get
\begin{equation}
 \sqrt{p_1^2 p_3^2}  +\sqrt{p_2^2 p_4^2} < \sqrt{- s^2 t^2}.
\end{equation}
At large energy and momentum transfers the correlator exhibits a power-like behavior of the form 
\begin{equation}
\Phi(p_i,s,t)\sim \frac{1}{(- s^2 t^2)^{(3 d - \Delta_t)/4}}. 
\end{equation}
Given the connection between the function $F_4$ and the triple-K integrals, we will reformulate this solution in terms of such integrals. They play a key role in the solution of the CWIs for tensor correlators, as discussed in \cite{Bzowski:2013sza}, for three-point  functions.

The link between $3$- and four-point functions outlined in the previous section allows to re-express the solutions in terms of a class of parametric integrals of 3 Bessel functions, as done in the case of the  scalar and tensor correlators  \cite{Bzowski:2013sza}, with the due modifications.
We consider the case of the solutions characterized by $\Delta_1=\Delta_2=\Delta_3=\Delta_4=\Delta$ or $\Delta_1=\Delta_3=\Delta_x$ and\  $\Delta_2=\Delta_4=\Delta_y$. We will show that the solution can be written in terms of triple-K integrals which are connected to the Appell function $F_4$ by the relation 
\begin{align}
& \int_0^\infty d x \: x^{\alpha - 1} K_\lambda(a x) K_\mu(b x) K_\nu(c x) =\frac{2^{\alpha - 4}}{c^\alpha} \bigg[ B(\lambda, \mu) + B(\lambda, -\mu) + B(-\lambda, \mu) + B(-\lambda, -\mu) \bigg], \label{3K}
\end{align}
where
\begin{align}
B(\lambda, \mu) & = \left( \frac{a}{c} \right)^\lambda \left( \frac{b}{c} \right)^\mu \Gamma \left( \frac{\alpha + \lambda + \mu - \nu}{2} \right) \Gamma \left( \frac{\alpha + \lambda + \mu + \nu}{2} \right) \Gamma(-\lambda) \Gamma(-\mu) \times \notag\\
& \qquad \times F_4 \left( \frac{\alpha + \lambda + \mu - \nu}{2}, \frac{\alpha + \lambda + \mu + \nu}{2}; \lambda + 1, \mu + 1; \frac{a^2}{c^2}, \frac{b^2}{c^2} \right), \label{3Kplus}
\end{align}
valid for
\begin{equation}
\Re\, \alpha > | \Re\, \lambda | + | \Re \,\mu | + | \Re\,\nu |, \qquad \Re\,(a + b + c) > 0, \nn
\end{equation}
with the Bessel functions $K_\nu$ satisfying the equations 
\begin{equation}
\begin{split}
\frac{\partial}{\partial p}\big[p^\b\,K_\b(p\,x)\big]&=-x\,p^\b\,K_{\b-1}(p x),\\
K_{\b+1}(x)&=K_{\b-1}(x)+\frac{2\b}{x}K_{\b}(x). 
\end{split}\label{der}
\end{equation}
In particular  the solution can be written as
\begin{equation}
I_{\a\{\b_1,\b_2,\b_3\}}(p_1\,p_3; p_2\,p_4;s\,t)=\int_0^\infty\,dx\,x^\a\,(p_1\,p_3)^{\b_1}\,(p_2\,p_4)^{\b_2}\,(s\,t)^{\b_3}\,K_{\b_1}(p_1\,p_3\,x)\,K_{\b_2}(p_2\,p_4\,x)\,K_{\b_3}(s\,t\,x).\label{trekappa}
\end{equation}
  Using various relations, the dilatation WI's \eqref{Dilatation4ch} become
\begin{equation}
(\Delta_t-3d) I_{\a\{\b_1,\b_2,\b_3\}}+2p_1^2p_3^2\ I_{\a+1\{\b_1-1,\b_2,\b_3\}}+2p_2^2p_4^2\ I_{\a+1\{\b_1,\b_2-1,\b_3\}}+2s^2t^2\ I_{\a+1\{\b_1,\b_2,\b_3-1\}}=0,
\end{equation}
where the arguments of the $I_{\a\{\b_1\b_2\b_3\}}$ function, written explicitly in \eqref{trekappa}, have been omitted for simplicity. The $I$ integrals satisfy the differential equations
\begin{align}
\frac{1}{s}\frac{\partial}{\partial s}\left(p_1\frac{\partial}{\partial p_1}+p_2\frac{\partial}{\partial p_2}-p_3\frac{\partial}{\partial p_3}-p_4\frac{\partial}{\partial p_4}\right)I_{\a\{\b_1,\b_2,\b_3\}}&=0,\\
\frac{1}{t}\frac{\partial}{\partial t}\left(p_1\frac{\partial}{\partial p_1}+p_4\frac{\partial}{\partial p_4}-p_2\frac{\partial}{\partial p_2}-p_3\frac{\partial}{\partial p_3}\right)I_{\a\{\b_1,\b_2,\b_3\}}&=0,
\end{align} 
which can be checked using the relations given in the same appendix, and we finally obtain
\begin{equation}
(\Delta_t-3d+2\a+2-2\b_t) I_{\a\{\b_1,\b_2,\b_3\}}=0,
\end{equation}
where $\b_t=\b_1+\b_2+\b_3$. In order to satisfy this equation the $\a$ parameter has to be equal to a particular value given by 
\begin{equation}
\tilde{\a}\equiv \frac{3}{2}d+\b_t-1-\frac{\Delta_t}{2}.
\end{equation}
 In the particular case of $\Delta_i=\Delta$, the special conformal Ward identities are given by
\begin{equation}
\left\{\begin{aligned}&\\[-1.2ex]
&\bigg [\frac{\partial^2}{\partial p_1^2}+\frac{(d-2\D+1)}{p_1}\frac{\partial}{\partial p_1}-\frac{\partial^2}{\partial p_3^2}-\frac{(d-2\D+1)}{p_3}\frac{\partial}{\partial p_3}+\frac{(p_1^2-p_3^2)}{st}\frac{\partial^2}{\partial s\partial t}\bigg ]\,I_{\tilde\a\{\b_1,\b_2,\b_3\}}=0\\[1ex]
&\bigg[\frac{\partial^2}{\partial p_2^2}+\frac{(d-2\D+1)}{p_2}\frac{\partial}{\partial p_2}-\frac{\partial^2}{\partial p_4^2}-\frac{(d-2\D+1)}{p_4}\frac{\partial}{\partial p_4}+\frac{(p_2^2-p_4^2)}{st}\frac{\partial^2}{\partial s\partial t}\bigg ]\,I_{\tilde\a\{\b_1,\b_2,\b_3\}}=0\\[1ex]
&\bigg[\frac{\partial^2}{\partial p_3^2}+\frac{(d-2\D+1)}{p_3}\frac{\partial}{\partial p_3}-\frac{\partial^2}{\partial p_4^2}-\frac{(d-2\D+1)}{p_4}\frac{\partial}{\partial p_4}+\frac{(p_2^2-p_1^2)}{st}\frac{\partial^2}{\partial s\partial t}\bigg ]\,I_{\tilde\a\{\b_1,\b_2,\b_3\}}=0\\[-0.7ex]
\label{neweq}
\end{aligned}\right.
\end{equation}
and using the properties of Bessel functions they can be rewritten in a simpler form. The first equation, for instance, can be written as
\begin{equation}
\label{oone}
(p_1^2-p_3^2)\bigg( (d-2\D+2\b_1)\,I_{\tilde\a+1\{\b_1-1,\b_2,\b_3\}}-2\b_3\,I_{\tilde\a+1\{\b_1,\b_2,\b_3-1\}} \bigg)=0,
\end{equation}
which is identically satisfied if the conditions 
\begin{equation}
\b_1=\Delta-\frac{d}{2},\qquad\b_3=0
\end{equation}
hold. In the same way we find that the second equation takes the form
\begin{equation}
\label{otwo}
(p_2^2-p_4^2)\bigg((d-2\D+2\b_2)\,I_{\tilde\a+1\{\b_1,\b_2-1,\b_3\}}-2\b_3\,I_{\tilde\a+1\{\b_1,\b_2,\b_3-1\}}\bigg)=0,
\end{equation}
and it is satisfied if 
\begin{equation}
\b_2=\Delta-\frac{d}{2},\qquad \b_3=0.
\end{equation}
One can check that the third equation 
\begin{equation}
p_2^2(d-2\D+2\b_2)\,I_{\tilde\a+1\{\b_1,\b_2-1,\b_3\}}-p_1^2(d-2\D+2\b_1)\,I_{\tilde\a+1\{\b_1-1,\b_2,\b_3\}}-2(p_2^2-p_1^2)\b_3\,I_{\tilde\a+1\{\b_1,\b_2,\b_3-1\}}=0,
\end{equation}
generates the same conditions given by \eqref{oone} and \eqref{otwo}.
 The solution for the four-point function, in this particular case, can be written as
\begin{equation}
\langle{O(p_1)\,O(p_2)\,O(p_3)\,O(\bar{p}_4)}\rangle=\,\bar{\a} \,I_{\frac{d}{2}-1\left\{\Delta-\frac{d}{2},\Delta-\frac{d}{2},0\right\}}(p_1\, p_3;p_2\,p_4; s\,t),
\end{equation}
where $\bar{\a}$ is an undetermined constant. 

In the case $\Delta_1=\Delta_3=\Delta_x$ and $\Delta_2=\Delta_4=\Delta_y$, the special CWIs can be written as
\begin{equation}
\left\{\begin{aligned}&\\[-1.9ex]
&\bigg [\frac{\partial^2}{\partial p_1^2}+\frac{(d-2\Delta_x+1)}{p_1}\frac{\partial}{\partial p_1}-\frac{\partial^2}{\partial p_3^2}-\frac{(d-2\Delta_x+1)}{p_3}\frac{\partial}{\partial p_3}+\frac{(p_1^2-p_3^2)}{st}\frac{\partial^2}{\partial s\partial t}\bigg ]\,I_{\tilde\a\{\b_1,\b_2,\b_3\}}=0\\[1ex]
&\bigg[\frac{\partial^2}{\partial p_2^2}+\frac{(d-2\Delta_y+1)}{p_2}\frac{\partial}{\partial p_2}-\frac{\partial^2}{\partial p_4^2}-\frac{(d-2\Delta_y+1)}{p_4}\frac{\partial}{\partial p_4}+\frac{(p_2^2-p_4^2)}{st}\frac{\partial^2}{\partial s\partial t}\bigg ]\,I_{\tilde\a\{\b_1,\b_2,\b_3\}}=0\\[1ex]
&\bigg[\frac{\partial^2}{\partial p_3^2}+\frac{(d-2\Delta_x+1)}{p_3}\frac{\partial}{\partial p_3}-\frac{\partial^2}{\partial p_4^2}-\frac{(d-2\Delta_y+1)}{p_4}\frac{\partial}{\partial p_4}+\frac{(p_2^2-p_1^2)}{st}\frac{\partial^2}{\partial s\partial t}\bigg ]\,I_{\tilde\a\{\b_1,\b_2,\b_3\}}=0\\[1.3ex]
\end{aligned}\right.
\end{equation}
whose solution is
\begin{align}
\langle{O(p_1)\,O(p_2)\,O(p_3)\,O(\bar{p}_4)}\rangle&=\,\bar{\bar{\a}} \,I_{\frac{d}{2}-1\left\{\Delta_x-\frac{d}{2},\Delta_y-\frac{d}{2},0\right\}}(p_1\, p_3;p_2\,p_4; s\,t),
\end{align}
which takes a form which is typical of the three-point function.
\section{Symmetric solutions as \texorpdfstring{$F_4$}{F4} hypergeometrics or triple-K integrals. The equal scalings case}
The derivation of symmetric expressions of such correlators requires and can be obtained by resorting to the formalism of the triple-K integrals \cite{Bzowski:2015yxv}. \\ 
A solution which is symmetric with respect to all the permutation of the momenta $p_i$, expressed in terms of 3 of the four constants $c(a,b)$ can be expressed in the form
\begin{align}
\langle{O(p_1)O(p_2)O(p_3)O(p_4)}\rangle&=\notag\\
&\hspace{-2cm}=\sum_{a,b}c(a,b)\Bigg[\,(s^2\,t^2)^{\Delta-\frac{3}{4}d}\left(\frac{p_1^2p_3^2}{s^2t^2}\right)^a\left(\frac{p_2^2p_4^2}{s^2t^2}\right)^bF_4\left(\a(a,b),\b(a,b),\g(a),\g'(b),\frac{p_1^2p_3^2}{s^2t^2},\frac{p_2^2p_4^2}{s^2t^2}\right)\notag\\
&\hspace{-1.5cm}+\,(s^2\,u^2)^{\Delta-\frac{3}{4}d}\,\left(\frac{p_2^2p_3^2}{s^2u^2}\right)^{a}\left(\frac{p_1^2p_4^2}{s^2u^2}\right)^{b}F_4\left(\a(a,b),\b(a,b),\g(a),\g'(b),\frac{p_2^2p_3^2}{s^2u^2},\frac{p_1^2p_4^2}{s^2u^2}\right)\notag\\
&\hspace{-1.5cm}+\,(t^2\,u^2)^{\Delta-\frac{3}{4}d}\,\left(\frac{p_1^2p_2^2}{t^2u^2}\right)^{a}\,\left(\frac{p_3^2p_4^2}{t^2u^2}\right)^{b}\,F_4\left(\a(a,b),\b(a,b),\g(a),\g'(b),\frac{p_1^2p_2^2}{t^2u^2},\frac{p_3^2p_4^2}{t^2u^2}\right)
\Bigg],
\label{fform1}
\end{align}
where the four coefficients $c(a,b)$'s given in \eqref{fform1} are reduced to three by the constraint 
\begin{align}
c\left(0,\Delta-\frac{d}{2}\right)=c\left(\Delta-\frac{d}{2},0\right).
\end{align}

The final symmetric result is given by
\begin{align}
\langle{O(p_1)O(p_2)O(p_3)O(p_4)}\rangle&= C\bigg[\,I_{\frac{d}{2}-1\{\Delta-\frac{d}{2},\Delta-\frac{d}{2},0\}}(p_1\,p_3,p_2\,p_4,s\,t)\notag\\
&\hspace{-1.5cm}+\, \,I_{\frac{d}{2}-1\{\Delta-\frac{d}{2},\Delta-\frac{d}{2},0\}}(p_2\,p_3,p_1\,p_4,s\,u)+\, \,I_{\frac{d}{2}-1\{\Delta-\frac{d}{2},\Delta-\frac{d}{2},0\}}(p_1\,p_2,p_3\,p_4,t\,u)\bigg],
\end{align}
written in terms of only one arbitrary constant overall, $C$. We can use the relation between the triple-K integrals and the $F_4$ written in \eqref{3K} and \eqref{3Kplus}, to re-express the final symmetric solution, originally given in \eqref{fform1}, in terms of a single constant in the form
\begin{align}
\langle{O(p_1)O(p_2)O(p_3)O(p_4)}\rangle&=2^{\frac{d}{2}-4}\ \ C\,\sum_{\l,\m=0,\Delta-\frac{d}{2}}\x(\l,\m)\bigg[\big(s^2\,t^2\big)^{\Delta-\frac{3}{4}d}\left(\frac{p_1^2 p_3^2}{s^2 t^2}\right)^\l\left(\frac{p_2^2p_4^2}{s^2t^2}\right)^\m\nonumber\\
&\hspace{-3cm}\times\,F_4\left(\frac{3}{4}d-\Delta+\l+\m,\frac{3}{4}d-\Delta+\l+\m,1-\Delta+\frac{d}{2}+\l,1-\Delta+\frac{d}{2}+\m,\frac{p_1^2 p_3^2}{s^2 t^2},\frac{p_2^2 p_4^2}{s^2 t^2}\right)\notag\\
&+\big(s^2\,u^2\big)^{\Delta-\frac{3}{4}d}\left(\frac{p_2^2 p_3^2}{s^2 u^2}\right)^\l\left(\frac{p_1^2p_4^2}{s^2u^2}\right)^\m\notag\\
&\hspace{-3cm}\times\,F_4\left(\frac{3}{4}d-\Delta+\l+\m,\frac{3}{4}d-\Delta+\l+\m,1-\Delta+\frac{d}{2}+\l,1-\Delta+\frac{d}{2}+\m,\frac{p_2^2 p_3^2}{s^2 u^2},\frac{p_1^2 p_4^2}{s^2 u^2}\right)\notag\\
&+\big(t^2\,u^2\big)^{\Delta-\frac{3}{4}d}\left(\frac{p_1^2 p_2^2}{t^2 u^2}\right)^\l\left(\frac{p_3^2p_4^2}{t^2u^2}\right)^\m\notag\\
&\hspace{-3cm}\times\,F_4\left(\frac{3}{4}d-\Delta+\l+\m,\frac{3}{4}d-\Delta+\l+\m,1-\Delta+\frac{d}{2}+\l,1-\Delta+\frac{d}{2}+\m,\frac{p_1^2 p_2^2}{t^2 u^2},\frac{p_3^2 p_4^2}{t^2 u^2}\right)\bigg].\label{finalSol}
\end{align}
where the coefficients $\x(\l,\m)$ are explicitly given by
\begin{equation}
\begin{split}
\x\left(0,0\right)&=\left[\Gamma\left(\frac{3}{4}d-\Delta\right)\right]^2\left[\Gamma\left(\Delta-\frac{d}{2}\right)\right]^2,\\
\x\left(0,\Delta-\frac{d}{2}\right)&=\x\left(\Delta-\frac{d}{2},0\right)=\left[\Gamma\left(\frac{d}{4}\right)\right]^2\Gamma\left(\Delta-\frac{d}{2}\right)\Gamma\left(\frac{d}{2}-\Delta\right),\\
\x\left(\Delta-\frac{d}{2},\Delta-\frac{d}{2}\right)&=\left[\Gamma\left(\Delta-\frac{d}{4}\right)\right]^2\left[\Gamma\left(\frac{d}{2}-\Delta\right)\right]^2.
\end{split}\label{xicoef}
\end{equation}
The solution found in \eqref{finalSol} is explicitly symmetric under all the possible permutations of the momenta and it is fixed up to one undetermined constant $C$. Equation \eqref{finalSol} gives the final expression of the solution obtained from the dual conformal ansatz \eqref{ansatz2}.
\chapter{The $TOOO$ correlator: CWIs in the scalar/tensor case}\label{tooo}
The investigation of the CWIs in momentum space that we are going to present is based on the reconstruction method of a tensor correlator starting from its transverse/traceless (tt) sector, formulated for 3-point functions  \cite{Bzowski:2013sza,Bzowski:2018fql,Bzowski:2015yxv}, that here we are going to extend to 4-point functions.  \\
In particular, in \cite{Bzowski:2013sza} a complete approach for the analysis of 3-point functions, up to the $TTT$ case, with three stress-energy tensors, valid for tensor correlators, has been formulated. The reconstruction of the entire correlator from its $tt$ projection involves the identification of a minimal set of form factors in this sector, and it is accompanied by a set of technical steps for re-assembling it in a systematic way. \\
This approach allows to identify primary and secondary CWIs of a tensorial 3-point function, with the former corresponding to second order partial differential equations (pde's) which can be solved independently in terms of a set of arbitrary constants.\\
Primary CWIs are equations involving only form factors of the tt sector and generate, for tensor correlators, inhomogeneous systems of pde's of hypergeometric type. Secondary CWIs, on the other end, connect the same form factors to 3-point functions via the corresponding canonical WI's, which impose extra constraints on the constants appearing in the solution of the primary equations. \\
The results that we present extend a previous analysis devoted to the scalar case, involving the $OOOO$ correlator  \cite{Maglio:2019grh}. We will re-investigate the scalar case, by taking a closer look at the structure of the equations and at their asymptotic behaviour. We will remark few additional properties of such correlators and highlight some properties of the asymptotic solutions of such equations, which have not been addressed before. 
This will allow us to gain a more general perspective both on the scalar and the tensor cases, especially in view of possible future extensions of our work to correlators of higher rank.\\
Scalar correlators are characterised only by primary CWIs and are therefore simpler to handle, differently from the $TOOO$ case where both primary and secondary equations are present.\\
In the scalar case the analysis of the conformal constraints will be performed by focusing on a special class of solutions of such equations which are conformal and dual conformal at the same time, derived in \cite{Maglio:2019grh}. These are obtained by imposing a specific condition on the scaling dimensions of the scalar operators, which allow to reduce the CWIs to a hypergeometric system, as in the case of 3-point and 4-point functions. More details can be found in \ref{OOOhyper} and \ref{OOOO}.

As we move from 3- to 4-point functions, all the equations, primary and secondary, are expressed in terms of 6 invariants, which are the external invariant masses $p_i^2$ and the two Mandelstam invariants $s$ and $t$. As far as we keep the external lines off-shell, and stay away from kinematical points where an invariant is exactly zero, the equations are 
well-defined and it is possible to investigate their structure.  As we are going to show, the selection of a set of specific invariants, compared to others, is particularly beneficial if we intend to uncover the symmetries of the equations and their redundancies under the permutations of the external momenta. \\
A crucial goal of our study is the identification of the asymptotic behaviour of the solutions of such constraints in specific kinematical limits. This may allow, in the near future, to relate results from ordinary 
perturbation theory - in ordinary Lagrangian realizations, at one loop level - to those 
derived from CFT's in the same limits. For instance, in \cite{Coriano:2018bbe,Coriano:2018bsy1} it is shown how to match the general solutions of the CWIs for the $TTT$ and $TJJ$ correlators, to free field theories with a specific content of fermions, scalars 
and spin 1 fields. The matching allows to re-express the solutions of such equations in terms of simple one-loop master integrals in full generality, for any CFT.  

\subsection{The search for asymptotic solutions}
For this reason, the search for asymptotic solutions of the CWIs, which acquire a simpler form in such limits, is particularly interesting. It may allow to establish a link with the classical factorization theorems proven in gauge theory amplitudes \cite{Sterman:2014nua}, especially if such CFT methods can be extended to multi-point functions.\\
We will investigate the structure of the equations in two specific limits. The first case that we will address will be the $1\to 3$, where the graviton line of the T is assumed to acquire a large invariant mass $(p_1^2)$ and decays into three scalar lines with small invariants 
$(p_2^2,p_3^2,p_4^2)$, while the remaining invariants $s$ and $t$  are large. We are going to derive some approximate asymptotic solutions of the equations which are separable in the $(p_2^2,p_3^2,p_4^2)$ and $(s,t,u)$ dependence. A similar analysis will be presented in the $2\to 2$ process, where one of the lines of the scalar operators is selected in the initial state together with the graviton line and the remaining scalar lines are in the final state.\\ Our work is organized as follows. \\
After a brief discussion of the conformal and canonical WI's in momentum space, we investigate the structure of the tt sector of the $TOOO$, identifying the symmetry constraints under the permutation of the momenta of the single form factor appearing in this correlator. \\
We then turn to a derivation of the primary and secondary CWIs of this correlator, written in a form which will be useful for the derivation of their asymptotic limits. We describe the orbits of such equations under the symmetry permutations, which allows to identify a subset of independent equations. \\
The analysis is repeated from scratch in the $2\to 2$ case and it is followed by a discussion of the asymptotic limits of such equations, after a brief overview of the approach in the scalar case. \\
We start from the scalar case, discuss the system of scalar equations and discuss its reduction to the dcc case, which can be solved exactly. The asymptotic behaviour of the 
dcc solutions provides an example and a guidance for a more general analysis first of the scalar case, and then of the tensor case, the $TOOO$. 
\section{Ward identities for the $TOOO$ in coordinate and in momentum space} 
 In this subsection, we use the formalism developed in chapter \ref{cftmom} to formulate the CWI our correlator 
\begin{equation}
\Gamma^{\mu\nu}(x_1,x_2,x_3,x_4)\equiv \langle T^{\mu\nu}(x_1)O(x_2)O(x_3)O(x_4) \rangle.
\end{equation}
Using  \eqref{SCWIs} we have 
\begin{align}
{\cal K}^\kappa \Gamma^{\mu\nu}(x_1,x_2,x_3,x_4) 
&=  {\cal K}^{ \kappa}_{scalar}(x_i) \Gamma^{\mu\nu}(x_1,x_2,x_3,x_4) 
 + 2 \left(  \delta^{\mu\kappa} x_{1\rho} - \delta_{\rho}^{\kappa }x_1^\mu  \right)\Gamma^{\rho \nu}(x_1,x_2,x_3,x_4)\nn\\ 
& + 2 \left(  \delta^{\nu\kappa} x_{1\rho} - \delta_{\rho}^{\kappa }x_1^\nu  \right)\Gamma^{\mu\rho}(x_1,x_2,x_3,x_4)\nonumber
=0,
\end{align}
 where the first contribution denotes the scalar part and the last two contributions the spin part, which are trivially absent in the case of a scalar correlator. \\
 The action of $\mathcal{K}^\kappa$ can be summarized by the expression 

 \begin{align}
&\sum_{j=1}^{3}\left[2(\Delta_j-d)\sdfrac{\partial}{\partial p_j^\k}-2p_j^\a\sdfrac{\partial}{\partial p_j^\a}\sdfrac{\partial}{\partial p_j^\k}+(p_j)_\k\sdfrac{\partial}{\partial p_j^\a}\sdfrac{\partial}{\partial p_{j\a}}\right]\langle{{T^{\mu_1\nu_1}(p_1)\,O(p_2)\,O( p_3)}O(\bar{p_4}}\rangle\notag \\
&\qquad + {\cal K}^\kappa_{spin}\langle{T^{\mu_1\nu_1}(p_1)\,O(p_2)\,O( p_3)}O(\bar{p_4}\rangle=0,\notag\\
\label{SCWTJJ}
\end{align}
where we have defined the spin part of ${\cal K}$ in momentum space as

\begin{align}
{\cal K}^\kappa_{spin}\langle{{T^{\mu_1\nu_1}(p_1)\,O(p_2)\,O( p_3)}O(\bar{p_4}}\rangle
&\equiv 4\left(\d^{\k(\mu_1}\sdfrac{\partial}{\partial p_1^{\a_1}}-\delta^{\k}_{\alpha_1}\delta^{\l(\mu_1}\sdfrac{\partial}{\partial p_1^\l}\right)\langle{{T^{\nu_1)\alpha_1}(p_1)\,O(p_2)\,O p_3)}O(\bar{p}_4)}\rangle\nn
\end{align}
(symmetrization is normalized with an overall factor $1/2$). \\
In the previous expression we have taken $p_4$ as a dependent momentum ($p_4\to \overline{p}_4$), which requires an implicit differentiation if we take $p_1,p_2$ and $p_3$ as independent momenta. The equations can be projected onto the three independent 
momenta, giving scalar equations which can be re-expressed in terms of all the scalar invariants parameterizing the form factors. The hypergeometric character of the 3-point functions, as well as for 4-point functions (for the dcc solutions), emerges after such reduction of the equations to a scalar form \cite{Coriano:2013jba,Bzowski:2013sza,Coriano:2018bsy1,Coriano:2018bbe}. \\ 
\subsection{Conservation and Trace Ward Identities}
The next objective is to derive the the trace and conservation Ward Identities. The starting point is 

\begin{equation}
Z[\phi_0,g^{\mu \nu}]=\int  \mathcal{D} \Phi \mathrm{exp}\big(-S_{CFT}[\phi,g^{\mu \nu}]-\sum_i \int d^d x \sqrt{g}\phi_0^{j}O_j \big),
\end{equation}
dependent on the background metric $g_{\mu \nu}$ and the classical source $\phi_0(x)$ coupled to the scalar operator $O(x)$, with the 1-point functions given by 

\begin{align}
&\langle T^{\mu \nu}(x) \rangle=\frac{2}{\sqrt{g(x)}} \frac{\delta Z}{\delta g_{\mu \nu}},\\  
&\langle O_j (x) \rangle =-\frac{1}{\sqrt{g(x)}}\frac{\delta Z}{\delta \phi_0^j(x)}.
\end{align}
In our case, in order to avoid some bulky notation, we consider only one type of scalar operator, with a unique scaling dimension $\Delta$. We will present the derivations of all the conformal and canonical WI's in this specific case. In section \ref{gensection} we will then provide  the expression of the same equations for general distinct $\Delta_i$'s, which can be obtained by a very similar procedure, as in the equal scaling case. \\
To get the transverse and trace Ward Identities, we require that the generating functional Z is invariant under diffeomorphisms and Weyl transformations respectively, which gives
\begin{align}
\nabla_{\nu} \langle T^{\mu \nu} (x_1) \rangle +\partial^{\mu}\phi_0 \cdot \langle O(x_1) \rangle =0,\\
g_{\mu \nu}\langle T^{\mu \nu} (x_1) \rangle+(d-\Delta)\phi_0 \langle O(x_1) \rangle =0.
\end{align}
The WI's for the  $\langle T^{\mu_1 \nu_1}(\textbf{p}_1)O(\textbf{p}_2)O(\textbf{p}_3)O(\textbf{p}_4)\rangle $ can be derived by taking three variations of the above identities with respect to the source $\phi_0$ of the scalar operator. At the end, by imposing the flat limit $g_{\mu \nu}=\delta_{\mu \nu}, \nabla_{\nu}=\partial_{\nu}$, turning off the sources ($\phi_0=0$) and using the definitions
\begin{align}
&\langle T^{\mu \nu} (x_1) O(x_2)O(x_3)O(x_4)\rangle=\frac{-2}{\sqrt{g(x_1)}\dots\sqrt{g(x_4)}}\frac{\delta^4 Z}{\delta g_{\mu \nu}(x_1)\delta \phi_0(x_2)\delta \phi_0(x_3)\delta \phi_0(x_4)},\\
&\langle O(x_1)O(x_2)O(x_3) \rangle=\frac{-1}{\sqrt{g(x_1)}\sqrt{g(x_2)}\sqrt{g(x_3)}}\frac{\delta^3Z}{\delta \phi_0(x_3)\delta \phi_0(x_2)\delta \phi_0(x_1)},
\end{align}
 the conservation WI gives the constraint
\begin{equation}\label{TransCord}
\begin{split}
\partial_{\nu}\langle T^{\mu \nu} (x_1)O(x_2)O(x_3)O(x_4) \rangle&=\partial^{\mu}\delta^{(d)}(x_1-x_2)\langle O(x_1)O(x_3)O(x_4) \rangle
\\
&\hspace{-2cm}+\partial^{\mu}\delta^{(d)}(x_1-x_3)\langle O(x_1)O(x_2)O(x_4) \rangle+\partial^{\mu}\delta^{(d)}(x_1-x_4)\langle O(x_1)O(x_2)O(x_3) \rangle,
\end{split}
\end{equation}
while the trace WI gives
\begin{align}\label{TraceCoord}
\delta_{\mu \nu} \langle T^{\mu \nu} (x_1)O(x_2)O(x_3))(x_4) \rangle&=(d-\Delta)\delta^{(d)}(x_1-x_2)\langle O(x_1)O(x_3)O(x_4) \rangle \notag\\
&\hspace{-3cm}+(d-\Delta)\delta^{(d)}(x_1-x_3)\langle O(x_1)O(x_2)O(x_4) \rangle +(d-\Delta)\delta^{(d)}(x_1-x_4)\langle O(x_1)O(x_2)O(x_3) \rangle.
\end{align}
The expressions of \eqref{TransCord} and \eqref{TraceCoord} in momentum space can be obtained  by a Fourier transform and are explicitly given by
\begin{equation}\label{TransMom}
\begin{split}
\delta^{(d)}\left(\sum_{i=1}^{4}\textbf{p}_i \right) p_{1\nu}\langle T^{\mu \nu} (\textbf{p}_1)O(\textbf{p}_2)O(\textbf{p}_3)O(\textbf{p}_4) \rangle=&-\Bigg( p_2^{\mu}\langle O(\textbf{p}_1+\textbf{p}_2)O(\textbf{p}_3)O(\textbf{p}_4) \rangle\\
&\hspace{-4cm}+p_3^{\mu}\langle O(\textbf{p}_1+\textbf{p}_3)O(\textbf{p}_2)O(\textbf{p}_4) \rangle+p_4^{\mu}\langle O(\textbf{p}_1+\textbf{p}_4)O(\textbf{p}_2)O(\textbf{p}_3) \rangle\Bigg)\delta^{(d)} \left(\sum_{i=1}^{4}\textbf{p}_i \right),
\end{split}
\end{equation}
and
\begin{equation}\label{TraceMom}
\begin{split}
\delta^{(d)}\left(\sum_{i=1}^{4}\textbf{p}_i \right)\delta_{\mu \nu}\langle T^{\mu \nu} (\textbf{p}_1)O(\textbf{p}_2)O(\textbf{p}_3)O(\textbf{p}_4) \rangle=&(d-\Delta)\Bigg( \langle O(\textbf{p}_1+\textbf{p}_2)O(\textbf{p}_3)O(\textbf{p}_4) \rangle\\
&\hspace{-3cm}+\langle O(\textbf{p}_1+\textbf{p}_3)O(\textbf{p}_2)O(\textbf{p}_4) \rangle +\langle O(\textbf{p}_1+\textbf{p}_4)O(\textbf{p}_2)O(\textbf{p}_3) \rangle\Bigg)\delta^{(d)} \left(\sum_{i=1}^{4}\textbf{p}_i \right),
\end{split}
\end{equation}
where on the right hand side of the equations appear only scalar 3-point functions.
 We will insert a bar over a momentum variable to indicate that it is treated as a dependent one. In the following we are going to make two separate choices of dependent momenta, respectively $\bar{p}_1$ and $\bar{p}_4$. If we choose $\bar{p}_1$ as the dependent momentum, the WI's take the form 
\begin{subequations}
	\begin{align}
	\bar{p}_{\mu_1}\,\langle{T^{\mu_1\nu_1}(\mathbf{\bar{p}}_1)\,O(\mathbf{p}_2)\,O(\mathbf{p}_3)\,O(\mathbf{p}_4)}\rangle&=-p_2^{\nu_1}\langle{O(\mathbf{p}_3+\mathbf{p}_4)\,O(\mathbf{p}_3)\,O(\mathbf{p}_4)}\rangle\notag\\[1.3ex]
	&\hspace{-3cm}-p_3^{\nu_1}\langle{O(\mathbf{p}_2+\mathbf{p}_4)\,O(\mathbf{p}_2)\,O(\mathbf{p}_4)}\rangle-p_4^{\nu_1}\langle{O(\mathbf{p}_2+\mathbf{p}_3)\,O(\mathbf{p}_2)\,O(\mathbf{p}_3)}\rangle,\\[2ex]
	\delta_{\mu_1\nu_1}\,\langle{T^{\mu_1\nu_1}(\mathbf{\bar{p}}_1)\,O(\mathbf{p}_2)\,O(\mathbf{p}_3)\,O(\mathbf{p}_4)}\rangle&=(d-\Delta)\Big[\langle{O(\mathbf{p}_3+\mathbf{p}_4)\,O(\mathbf{p}_3)\,O(\mathbf{p}_4)}\rangle\notag\\[1.3ex]
	&\hspace{-3cm}+\langle{O(\mathbf{p}_2+\mathbf{p}_4)\,O(\mathbf{p}_2)\,O(\mathbf{p}_4)}\rangle+\langle{O(\mathbf{p}_2+\mathbf{p}_3)\,O(\mathbf{p}_2)\,O(\mathbf{p}_3)}\rangle\Big],
	\end{align}\label{ConsWI}
\end{subequations}
and for $\bar{p}_4$
\begin{subequations}
	\begin{align}
	p_{1,\mu_1}\langle T^{\mu_1 \nu_1}(\textbf{p}_1)O(\textbf{p}_2)O(\textbf{p}_3)O(\mathbf{\bar{p}_4})\rangle=&+p_1^{\nu_1}\langle{ O(\textbf{p}_2)O(\textbf{p}_3)O(\mathbf{p_1+\bar{p}_4})}\rangle\notag \\&-p_2^{\nu_1}\Big(\langle O(\textbf{p}_1+\textbf{p}_2)O(\textbf{p}_3)O(\mathbf{\bar{p}_4})\rangle-\langle O(\textbf{p}_2)O(\textbf{p}_3)O(\mathbf{p_1+\bar{p}_4}) \rangle\Big)\notag \\&-p_3^{\nu_1}\Big(\langle O(\textbf{p}_2)O(\mathbf{p_1+p_3})O(\mathbf{\bar{p}_4})\rangle-\langle O(\textbf{p}_2)O(\textbf{p}_3)O(\mathbf{p_1+\bar{p}_4}) \rangle\Big),\\
	\delta_{\mu_1\nu_1}\langle T^{\mu_1 \nu_1}(\textbf{p}_1)O(\textbf{p}_2)O(\textbf{p}_3)O(\mathbf{\bar{p}_4})\rangle=&(d-\Delta)\Big[\langle O(\textbf{p}_1+\textbf{p}_2)O(\textbf{p}_3)O(\mathbf{\bar{p}_4}) \rangle+\langle O(\textbf{p}_1+\textbf{p}_3)O(\textbf{p}_2)O(\mathbf{\bar{p}_4}) \rangle \notag \\&+\langle O(\textbf{p}_1+\mathbf{\bar{p}_4})O(\textbf{p}_2)O(\textbf{p}_3) \rangle\Big].
	\end{align}\label{ConsWI22}
\end{subequations} 
The left hand sides of these equations will be related to the form factor identified from the $tt$ sector.
\section{The reconstruction method from 3- to 4-point functions and the $TOOO$}
Following \cite{Bzowski:2013sza}, we consider the four point correlation function formed by a stress-energy tensor $T^{\mu\nu}$ and three scalar operators $O(p_i)$ of the same kind and with the same scaling dimensions. We define
\begin{align}
p_i=\sqrt{\mathbf{p}_i^2},\quad s=\sqrt{(\mathbf{\bar{p}}_1+\mathbf{p}_2)^2}=\sqrt{(\mathbf{p}_3+\mathbf{p}_4)^2},\quad t=\sqrt{(\mathbf{p}_2+\mathbf{p}_3)^2},\quad u=\sqrt{(\mathbf{p}_2+\mathbf{p}_4)^2},
\end{align}
and introduce the $tt$ $(\Pi)$ and local $(\Sigma)$ projectors 
\begin{align}
\Pi^{\mu\nu}_{\alpha\beta}(\mathbf{p})&=\frac{1}{2}\left(\pi^{\mu}_{\alpha}(\mathbf{p})\pi^{\nu}_{\beta}(\mathbf{p})+\pi^{\mu}_{\beta}(\mathbf{p})\pi^{\nu}_{\alpha}(\mathbf{p})\right)-\frac{1}{d-1}\pi^{\mu\nu}(\mathbf{p})\pi_{\alpha\beta}(\mathbf{p})\\
\Sigma^{\mu\nu}_{\alpha\beta}(\mathbf{p})&=\delta^{(\mu}_{\alpha}\delta^{\nu)}_{\beta}-\Pi^{\mu\nu}_{\alpha\beta}(\mathbf{p})=\frac{1}{p^2}\left[2\,p_{\scriptsize{\raisebox{-0.9ex}{$(\beta$}}}\,\delta^{(\mu}_{\scriptsize{\raisebox{-0.7ex}{$\alpha)$}}} p^{\scriptsize{\raisebox{0.5ex}{$\nu)$}}}-\frac{p_\alpha p_\beta}{(d-1)}\left(\delta^{\mu\nu}+(d-2)\frac{p^\mu p^\nu}{p^2}\right)\right]+\frac{1}{d-1}\pi^{\mu\nu}(\mathbf{p})\delta_{\alpha\beta}\label{Sigma}.
\end{align}
The stress-energy tensor is decomposed in its transverse traceless $(tt)$ and local parts in the form 

\begin{equation}
\label{loca1}
T^{\mu\nu}=t^{\mu\nu} + t_{loc}^{\mu\nu}
\end{equation}
with 
\begin{align}
\label{loca2}
t_{loc}^{\mu\nu}(p)&=\frac{p^{\mu}}{p^2}Q^\nu + \frac{p^{\nu}}{p^2}Q^\mu -
\frac{p^\mu p^\nu}{p^4} Q +\frac{\pi^{\m\nu}}{d-1}(T - \frac{Q}{p^2})\nn
&=\Sigma^{\mu\nu}_{\alpha\beta} T^{\alpha\beta}
\end{align}
and 
\begin{equation}
Q^\mu=p_\nu T^{\mu\nu},\qquad T=\delta_{\mu\nu}T^{\mu\nu}, \qquad Q= p_\nu p_\mu T^{\mu\nu}.
\end{equation}

One can consider the decomposition of the $\langle{TOOO}\rangle$ correlation function as
\begin{align}
\langle{T^{\mu_1\nu_1}(\mathbf{\bar{p}}_1)O(\mathbf{p}_2)O(\mathbf{p}_3)O(\mathbf{p}_4)}\rangle&=\langle{t^{\mu_1\nu_1}(\mathbf{\bar{p}}_1)O(\mathbf{p}_2)O(\mathbf{p}_3)O(\mathbf{p}_4)}\rangle+\langle{t_{loc}^{\mu_1\nu_1}(\mathbf{\bar{p}}_1)O(\mathbf{p}_2)O(\mathbf{p}_3)O(\mathbf{p}_4)}\rangle\label{decomptooo}
\end{align}
where in bold we refer to vectors in the Euclidean $\mathbb{R}^d$ space, and we are considering $\bar{p}_1^\mu=-p_2^\mu-p_3^\mu-p_4^\mu$ from momentum conservation. \\
The first term on the right hand side of \eqref{decomptooo} is the $tt$ part of the correlation function, and the second represents the local ($loc$) part. The method consists in expanding the $tt$ sector into a minimal number of form factors, fixed by the symmetry of the correlator \cite{Bzowski:2013sza}. 
In our case the $tt$ and local parts take the form
\begin{align}
\langle{t^{\mu_1\nu_1}(\mathbf{\bar{p}}_1)O(\mathbf{p}_2)O(\mathbf{p}_3)O(\mathbf{p}_4)}\rangle&=\Pi^{\mu_1\nu_1}_{\alpha_1\beta_1}(\mathbf{\bar{p}}_1)\left[A\,p_2^{\alpha_1}p_2^{\beta_1}+A(p_2\leftrightarrow p_3)\,p_3^{\alpha_1}p_3^{\beta_1}+A(p_2\leftrightarrow p_4)\,p_4^{\alpha_1}p_4^{\beta_1}\right]\label{transvtrace}\\
\langle{t_{loc}^{\mu_1\nu_1}(\mathbf{\bar{p}}_1)O(\mathbf{p}_2)O(\mathbf{p}_3)O(\mathbf{p}_4)}\rangle&=\Sigma^{\mu_1\nu_1}_{\alpha_1\beta_1}(\mathbf{\bar{p}}_1)\langle{T^{\alpha_1\beta_1}(\mathbf{\bar{p}}_1)O(\mathbf{p}_2)O(\mathbf{p}_3)O(\mathbf{p}_4)}\rangle\label{local}.
\end{align}
From these expressions, one can observe that the local term is constrained by the conservation WI's \eqref{ConsWI}, which project on 3-point functions of the form $OOO$, as we have discussed in the previous section. Using \eqref{Sigma} and \eqref{ConsWI} in \eqref{local}, one can explicitly write the local term in the form
\begin{align}
\langle{t_{loc}^{\mu_1\nu_1}(\mathbf{\bar{p}}_1)O(\mathbf{p}_2)O(\mathbf{p}_3)O(\mathbf{p}_4)}\rangle&=\frac{2}{\bar{p}_1^2}\Big[-\bar{p}_1^{(\mu_1}p_2^{\nu_1)}\langle O(\mathbf{p}_3+\mathbf{p}_4)O(\mathbf{p}_3)O(\mathbf{p}_4)\rangle\notag\\
&\hspace{-4cm}-\bar{p}_1^{(\mu_1}p_3^{\nu_1)}\langle{O(\mathbf{p}_2+\mathbf{p}_4)O(\mathbf{p}_2)O(\mathbf{p}_4)}\rangle-\bar{p}_1^{(\mu_1}p_4^{\nu_1)}\langle{O(\mathbf{p}_2+\mathbf{p}_3)O(\mathbf{p}_2)O(\mathbf{p}_3)}\rangle\Big]+\frac{(d-\Delta)}{d-1}\pi^{\mu_1\nu_1}(\mathbf{\bar{p}}_1)\notag\\
&\hspace{-4.7cm} \times\Big[\langle{O(\mathbf{p}_2+\mathbf{p}_4)O(\mathbf{p}_2)O(\mathbf{p}_4)}\rangle+\langle{O(\mathbf{p}_2+\mathbf{p}_3)O(\mathbf{p}_2)O(\mathbf{p}_3)}\rangle+\langle{O(\mathbf{p}_3+\mathbf{p}_4)O(\mathbf{p}_3)O(\mathbf{p}_4)}\rangle\Big]\notag\\
&\hspace{-4cm}-\frac{1}{(d-1)}\left(\delta^{\mu_1\nu_1}+(d-2)\frac{\bar{p}_1^{\mu_1}\bar{p}_1^{\nu_1}}{\bar{p}_1^2}\right)\left[-\frac{\mathbf{\bar{p}}_1\cdot\mathbf{p}_2}{\bar{p}_1^2}\langle{O(\mathbf{p}_3+\mathbf{p}_4)O(\mathbf{p}_3)O(\mathbf{p}_4)}\rangle
\right.\notag\\
&\hspace{-3cm}\left.-\frac{\mathbf{\bar{p}}_1\cdot\mathbf{p}_3}{\bar{p}_1^2}\langle{O(\mathbf{p}_2+\mathbf{p}_4)O(\mathbf{p}_2)O(\mathbf{p}_4)}\rangle-\frac{\mathbf{\bar{p}}_1\cdot\mathbf{p}_4}{\bar{p}_1^2}\langle{O(\mathbf{p}_2+\mathbf{p}_3)O(\mathbf{p}_2)O(\mathbf{p}_3)}\rangle\right].
\end{align}
The scalar 3-point function appearing on the right hand side is exactly known. In this way, the task of finding the structure of the entire $\langle{TOOO}\rangle$ has been reduced to the identification of only its $tt$ part. In particular, as we are going to show, all the WI's will constrain a single form factor.\\
The parameterization of this form factor (A), eventually, can be chosen according to 
the type of amplitude that one intends to consider, in order to facilitate the analysis. \\
For instance, in the case in which one in interested in a comparison between the conformal prediction and a free field theory realization  - for example in a 1 (graviton) $\to$ three (scalars) process - then it is convenient to adopt the parameterization $A\equiv A(p_2,p_3,p_4,s,t,u)$  and derive the equations 
using such variables. This choice is the one which respects the symmetries of the process, since the three scalars can be treated equally, and it allows to discuss more easily its asymptotic behaviour. Notice that in this case, momentum $p_1$ is treated as a dependent one $(\overline{p}_1)$ and needs to be differentiated implicitly in the corresponding equations. 
\section{Conformal Ward Identities in the $1\to 3$ formulation }
Using the $1\to 3$ symmetric formulation and the parameterization presented in \eqref{local}, the $A$ form factor exhibits the following symmetries 
\begin{align}
A(p_3\leftrightarrow p_4)&=A(p_2,p_3,p_4,s,t,u)\\
A(p_2\to p_4\to p_3\to p_2)&=A(p_2\leftrightarrow p_4)\\
A(p_2\to p_3\to p_4\to p_2)&=A(p_2\leftrightarrow p_3),
\end{align}
which can be written in the form
\begin{align}
A(p_2,p_4,p_3,s,u,t)&=A(p_2,p_3,p_4,s,t,u)\\
A(p_4,p_2,p_3,t,u,s)&=A(p_4,p_3,p_2,t,s,u)\\
A(p_3,p_4,p_2,u,s,t)&=A(p_3,p_2,p_4,u,t,s).
\end{align}

In order to extract some information on A$(p_2,p_3,p_4,s,t,u)$, we turn to the dilatation and the special conformal WI's which it has to satisfy. In the case of scalars of equal scaling $\Delta$ these take the simplified forms 
\begin{align}
0&=\left[(3\Delta-2d)-\sum_{j=2}^{4}\,p_j^\mu\,\frac{\partial}{\partial p_j^\mu}
\right]\,\langle{T^{\mu_1\nu_1}(\mathbf{\bar{p}}_1)O(\mathbf{p}_2)O(\mathbf{p}_3)O(\mathbf{p}_4)}\rangle,\\
0&=\sum_{j=2}^4\left[2(\Delta-d)\frac{\partial}{\partial p_{j\,\kappa}}-2p_j^\alpha\,\frac{\partial^2}{\partial p_j^\alpha\,\partial\,p_{j\,\kappa}}+p_j^\kappa\,\frac{\partial^2}{\partial\,p_j^\alpha\,\partial\,p_{j\,\alpha}}\right]\,\langle{T^{\mu_1\nu_1}(\mathbf{\bar{p}}_1)O(\mathbf{p}_2)O(\mathbf{p}_3)O(\mathbf{p}_4)}\rangle,
\end{align}
where, as already mentioned, the momentum $p_1^\mu$ is taken as the dependent one. \\
As discussed in \cite{Coriano:2018bbe}, one of the external coordinates of the correlator can be set to vanish by translational symmetry, and its corresponding momentum, after Fourier transform, has to be taken as dependent on the other. For instance, in this case,  for convenience, we have chosen the coordinate of the stress-energy tensor to vanish ($x_1=0$), and taken its momentum as the dependent one ($p_1\to \overline{p}_1$). This implies that the spin part of the special conformal transformation will not act on the stress-energy tensor, and the action of this generator is  reduced to a pure scalar.\\
The differentiation is performed only respect to the independent momenta, using the chain rule while differentiating $\overline{p}_1$. This choice is optimal if we intend to derive symmetric equations for the $TOOO$, in which we treat the three scalar operators equally, as is the case if we intend to investigate this correlator in a $1\to 3$ kinematical configuration. In section \ref{asymtreat} we will reverse this choice, by taking one of the scalar momenta ($p_4$) as the dependent one, which is equivalent to choosing $x_4=0$ in coordinate space. In this second case the special conformal generator will act with its spin part on the indices of the stress-energy tensor as well, being the momentum $p_1$ one of the independent momenta.
\\
The procedure that we will apply in this case follows quite closely the approach implemented for 3-point functions, developed in \cite{Bzowski:2013sza}. Both equations are projected onto the transverse traceless sector using the $\Pi$ projector, whose action is endomorphic on this sector \cite{Bzowski:2013sza}. A more detailed discussion of this point can be found in \cite{Coriano:2018bbe}. \\
Henceforth, by applying $\Pi^{\rho_1\sigma_1}_{\mu_1\nu_1}(\mathbf{\bar{p}}_1)$ on the left of the dilatation and special conformal generators, we find 
\begin{equation}
\Pi^{\rho_1\sigma_1}_{\mu_1\nu_1}(\mathbf{\bar{p}}_1)\,\hat{D}\,\langle{t^{\mu_1\nu_1}(\mathbf{\bar{p}}_1)O(\mathbf{p}_2)O(\mathbf{p}_3)O(\mathbf{p}_4)}\rangle=0\label{Kdil}
\end{equation}
for the dilatation WI, and
\begin{align}
0&=\Pi^{\rho_1\sigma_1}_{\mu_1\nu_1}(\mathbf{\bar{p}}_1)\,\mathcal{K}^\kappa\,\left[\langle{t^{\mu_1\nu_1}(\mathbf{\bar{p}}_1)O(\mathbf{p}_2)O(\mathbf{p}_3)O(\mathbf{p}_4)}\rangle+\langle{t_{loc}^{\mu_1\nu_1}(\mathbf{\bar{p}}_1)O(\mathbf{p}_2)O(\mathbf{p}_3)O(\mathbf{p}_4)}\rangle\right]\notag\\
&=\Pi^{\rho_1\sigma_1}_{\mu_1\nu_1}(\mathbf{\bar{p}}_1)\,\mathcal{K}^\kappa\,\langle{t^{\mu_1\nu_1}(\mathbf{\bar{p}}_1)O(\mathbf{p}_2)O(\mathbf{p}_3)O(\mathbf{p}_4)}\rangle+\Pi^{\rho_1\sigma_1}_{\mu_1\nu_1}\left[\frac{4d}{\bar{p}_1^2}\bar{p}_{1\,\beta}\delta^{\mu_1\kappa}\langle{T^{\nu_1\beta}(\mathbf{\bar{p}}_1)O(\mathbf{p}_2)O(\mathbf{p}_3)O(\mathbf{p}_4)}\rangle\right]\label{prim}
\end{align}
for the conformal WI,
where we have used the relation
\begin{equation}
\Pi^{\rho_1\sigma_1}_{\mu_1\nu_1}\,{\cal K}^\kappa\,\langle{t_{loc}^{\mu_1\nu_1}(\mathbf{\bar{p}}_1)O(\mathbf{p}_2)O(\mathbf{p}_3)O(\mathbf{p}_4)}\rangle=\Pi^{\rho_1\sigma_1}_{\mu_1\nu_1}\left[\frac{4d}{\bar{p}_1^2}\,\delta^{\mu_1\kappa}\,\bar{p}_{1\,\beta}\,\langle{T^{\nu_1\beta}(\mathbf{\bar{p}}_1)O(\mathbf{p}_2)O(\mathbf{p}_3)O(\mathbf{p}_4)}\rangle\right].
\end{equation}
 The first term in \eqref{prim} can be explicitly written as
\begin{align}
\Pi^{\rho_1\sigma_1}_{\mu_1\nu_1}(\mathbf{\bar{p}}_1)\,\mathcal{K}^\kappa\,\langle{t^{\mu_1\nu_1}(\mathbf{\bar{p}}_1)O(\mathbf{p}_2)O(\mathbf{p}_3)O(\mathbf{p}_4)}\rangle&=\notag\\
&\hspace{-7.3cm}=\Pi^{\rho_1\sigma_1}_{\mu_1\nu_1}(\mathbf{\bar{p}}_1)\Bigg\{p_2^\kappa\Bigg[C_{11}\,p_2^{\mu_1} p_2^{\nu_1}+C_{12}\,p_3^{\mu_1} p_3^{\nu_1}+C_{13}\,p_4^{\mu_1} p_4^{\nu_1}\Bigg]+p_3^\kappa\Bigg[C_{21}\,p_2^{\mu_1} p_2^{\nu_1}+C_{22}\,p_3^{\mu_1} p_3^{\nu_1}+C_{23}\,p_4^{\mu_1} p_4^{\nu_1}\Bigg]\notag\\
&\hspace{-4cm}+p_4^\kappa\Bigg[C_{31}\,p_2^{\mu_1} p_2^{\nu_1}+C_{32}\,p_3^{\mu_1} p_3^{\nu_1}+C_{33}\,p_4^{\mu_1} p_4^{\nu_1}\Bigg]+\delta^{\mu_1\kappa}\Bigg[C_{41}\,p_2^{\nu_1}+C_{42}\,p_3^{\nu_1}+C_{43}\,p_4^{\nu_1}\Bigg]\Bigg\}\label{Ktt}
\end{align}
where we have used the chain rules
\begin{align}
\frac{\partial}{\partial p_2^\mu}&=\frac{p_2^\mu}{p_2}\frac{\partial}{\partial p_2}+\frac{p_2^{\mu}+p_3^{\mu}}{t}\frac{\partial}{\partial t}+\frac{p_2^\mu+p_4^\mu}{u}\frac{\partial}{\partial u}\\
\frac{\partial}{\partial p_3^\mu}&=\frac{p_3^\mu}{p_3}\frac{\partial}{\partial p_3}+\frac{p_3^{\mu}+p_4^{\mu}}{s}\frac{\partial}{\partial s}+\frac{p_2^\mu+p_3^\mu}{t}\frac{\partial}{\partial t}\\
\frac{\partial}{\partial p_4^\mu}&=\frac{p_4^\mu}{p_4}\frac{\partial}{\partial p_4}+\frac{p_2^{\mu}+p_4^{\mu}}{s}\frac{\partial}{\partial s}+\frac{p_2^\mu+p_4^\mu}{u}\frac{\partial}{\partial u}
\end{align}
in order to write the covariant derivatives in terms of scalar derivatives involving the invariants parameterizing $A$. The coefficients $C_{ij}$ in \eqref{Ktt} are linear combinations of differential operators acting on $A$.
The dilatation WI \eqref{Kdil} can be written in scalar form as
\begin{align}
0&=\Pi^{\rho_1\sigma_1}_{\mu_1\nu_1}(\mathbf{\bar{p}}_1)\,\hat{D}\,\langle{t^{\mu_1\nu_1}(\mathbf{\bar{p}}_1)O(\mathbf{p}_2)O(\mathbf{p}_3)O(\mathbf{p}_4)}\rangle\notag\\
&=\Pi^{\rho_1\sigma_1}_{\mu_1\nu_1}(\mathbf{\bar{p}}_1)\Bigg\{D_1\,p_2^{\mu_1} p_2^{\nu_1}+D_2\,p_3^{\mu_1} p_3^{\nu_1}+D_3\,p_4^{\mu_1} p_4^{\nu_1}\Bigg]\Bigg\}
\end{align}
where $D_i$ are terms involving scalar derivatives acting on the form factor $A(p_2,p_3,p_4,s,t,u)$. The previous equation is satisfied if all the $D_i$ vanish independently, giving a dilatation constraint on $A(p_2,p_3,p_4,s,t,u)$ of the form
\begin{align}
D_1=0\,\implies \left[\sum_{j=2}^4\,p_j\frac{\partial}{\partial p_j}+s\frac{\partial}{\partial s}+t\frac{\partial}{\partial t}+u\frac{\partial}{\partial u}\right]A(p_i,s,t,u)=(\Delta_t-3d-2)\,A(p_i,s,t,u),
\end{align} 
where $\Delta_t=\sum_{j=1}^4\Delta_j=d+3\Delta$, since $\Delta_1=d$ for the stress-energy tensor, and we have set  $\Delta_2=\Delta_3=\Delta_4=\Delta$. From the other conditions $D_i=0$, with $i=2,3$, we generate the same constraint as from $D_1$, modulo some permutations 
involving $(p_2\leftrightarrow p_3)$ and $(p_2\leftrightarrow p_4)$ respectively. 

\subsection{Primary Conformal Ward Identities}
From the expressions of \eqref{prim} and \eqref{Ktt}, after some lengthy algebraic manipulations, we derive the primary constraints as
\begin{align}
&C_{11}=0,&C_{12}=0,&C_{13}=0,\notag\\
&C_{21}=0,&C_{22}=0,&C_{23}=0,\notag\\
&C_{31}=0,&C_{32}=0,&C_{33}=0,
\end{align}
which are explicitly given in \ref{AppendixTOOO}.

\begin{figure}[h]
    \centering
\raisebox{-0.5\height}{\includegraphics[scale=0.8]{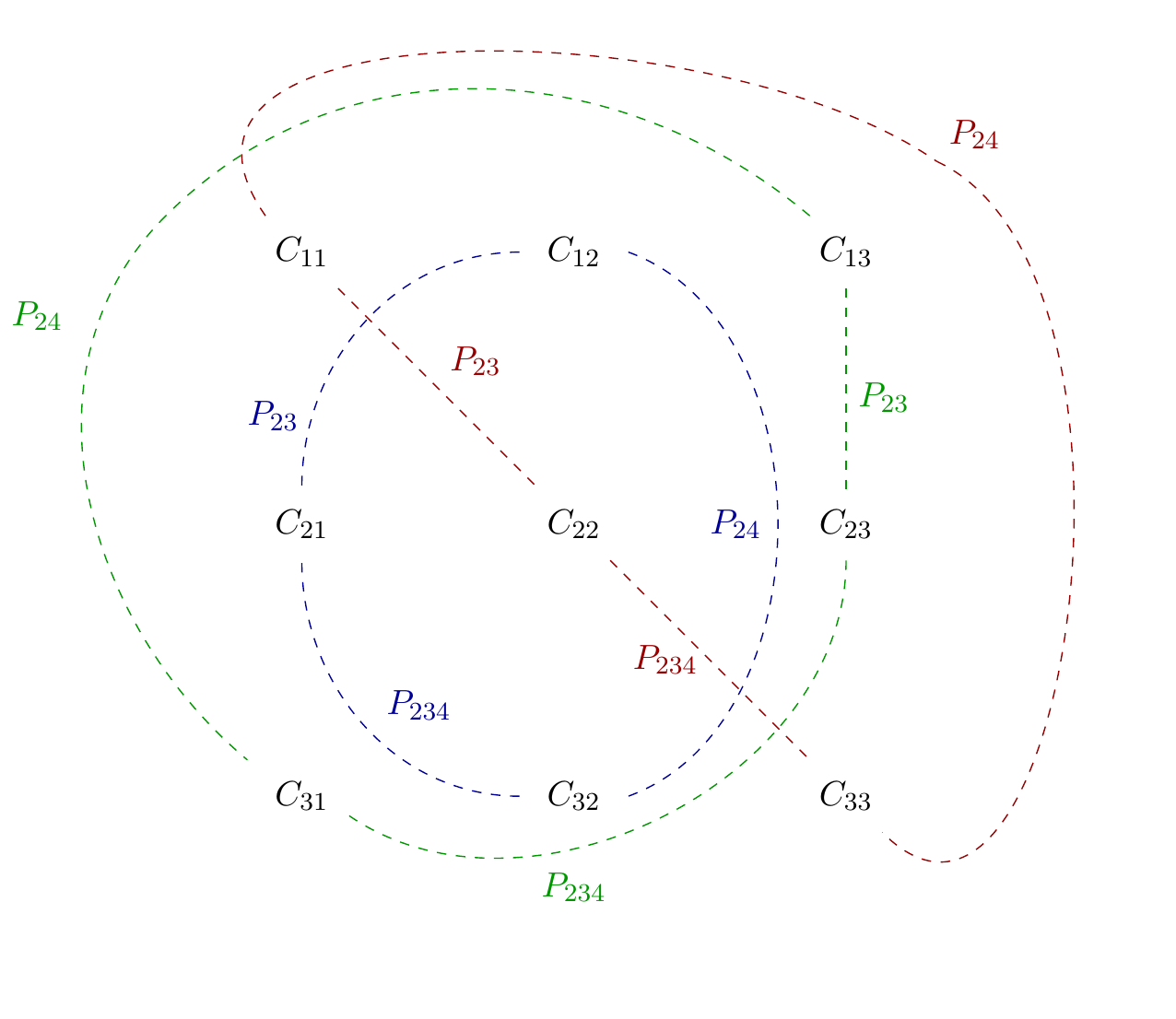}}
    \caption{Orbits of the primary CWIs of the $TOOO$ under $P_{23}$ and $P_{24}$}
    \label{tikzz}
\end{figure}

One can easily reorganize these equations by introducing the operators
\begin{align}
\bar{K}(p_2,p_3,p_4,s,t,u)&\equiv K_2+\frac{p_3^2-p_4^2}{s\,t}\frac{\partial}{\partial s \partial t}-\frac{p_3^2-p_4^2}{s\,u}\frac{\partial}{\partial s \partial u}+\frac{1}{t}\frac{\partial}{\partial t}\left(p_2\frac{\partial}{\partial p_2}+p_3\frac{\partial}{\partial p_3}-p_4\frac{\partial}{\partial p_4}\right)\notag\\
&\hspace{-3cm}+(d-\Delta)\left(\frac{1}{t}\frac{\partial}{\partial t}+\frac{1}{u}\frac{\partial}{\partial u}\right)+\frac{1}{u}\frac{\partial}{\partial u}\left(p_2\frac{\partial}{\partial p_2}-p_3\frac{\partial}{\partial p_3}+p_4\frac{\partial}{\partial p_4}\right)+\frac{2p_2^2+p_3^2+p_4^2-s^2-t^2-u^2}{t\,u}\frac{\partial}{\partial t\partial u}
\end{align}
and
\begin{equation}
L(s,t)\equiv\frac{2}{s}\frac{\partial}{\partial s}-\frac{2}{t}\frac{\partial}{\partial t}
\end{equation}
with
\begin{align}
L(s,t)=-L(t,s),
\end{align}
obtaining
\begin{align}
C_{11}&=\bar{K}(p_2,p_3,p_4,s,t,u)\,A(p_2,p_3,p_4,s,t,u)\notag\\
C_{12}&=\bar{K}(p_2,p_3,p_4,s,t,u)\,A(p_3,p_2,p_4,u,t,s)+L(t,u)\bigg(A(p_2,p_3,p_4,s,t,u)+A(p_3,p_2,p_4,u,t,s)\bigg)\notag\\
C_{13}&=\bar{K}(p_2,p_3,p_4,s,t,u)\,A(p_4,p_3,p_2,t,s,u)-L(t,u)\bigg(A(p_2,p_3,p_4,s,t,u)+A(p_4,p_3,p_2,t,s,u)\bigg)\notag\\[2ex]
C_{21}&=\bar{K}(p_3,p_2,p_4,u,t,s)\,A(p_2,p_3,p_4,s,t,u)-L(s,t)\bigg(A(p_2,p_3,p_4,s,t,u)+A(p_3,p_2,p_4,u,t,s)\bigg)\notag\\
C_{22}&=\bar{K}(p_3,p_2,p_4,u,t,s)\,A(p_3,p_2,p_4,u,t,s)\notag\\
C_{23}&=\bar{K}(p_3,p_2,p_4,u,t,s)\,A(p_4,p_3,p_2,t,s,u)+L(s,t)\bigg(A(p_4,p_3,p_2,t,s,u)+A(p_3,p_2,p_4,u,t,s)\bigg)\notag
\end{align}
\begin{align}
C_{31}&=\bar{K}(p_4,p_3,p_2,t,s,u)\,A(p_2,p_3,p_4,s,t,u)-L(s,u)\bigg(A(p_2,p_3,p_4,s,t,u)+A(p_4,p_3,p_2,t,s,u)\bigg)\notag\\
C_{32}&=\bar{K}(p_4,p_3,p_2,u,t,s)\,A(p_3,p_2,p_4,u,t,s)+L(s,u)\bigg(A(p_3,p_2,p_4,u,t,s)+A(p_4,p_3,p_2,t,s,u)\bigg)\notag\\
C_{33}&=\bar{K}(p_4,p_3,p_2,u,t,s)\,A(p_4,p_3,p_2,t,s,u).\notag
\end{align}
We illustrate in \eqref{tikzz} pictorially the action of the permutation operators $P_{ij}$, acting on the two momenta $p_i^\mu$ and $p_j^\mu$, on the various $C_{ij}$ presented above. The functional dependence of the form factor $A(p_2,p_3, p_4,s,t,u)$ will vary accordingly. \\
The orbits connect the various coefficients $C_{ij}$ which can be reached by the action of the various permutations. We start with $P_{23}$, $P_{24}$ and their product $P_{234}$.
 The orbits describe equivalent equations and we are allowed to choose any of the equations labeled by coefficients $C_{ij}$ belonging to separate orbits. Since there are three independent orbits under this subgroup, we start by selecting only three primary conformal WI's which are not related by the action of such permutations

\begin{figure}[h]
    \centering
\raisebox{-0.5\height}{\includegraphics[scale=0.8]{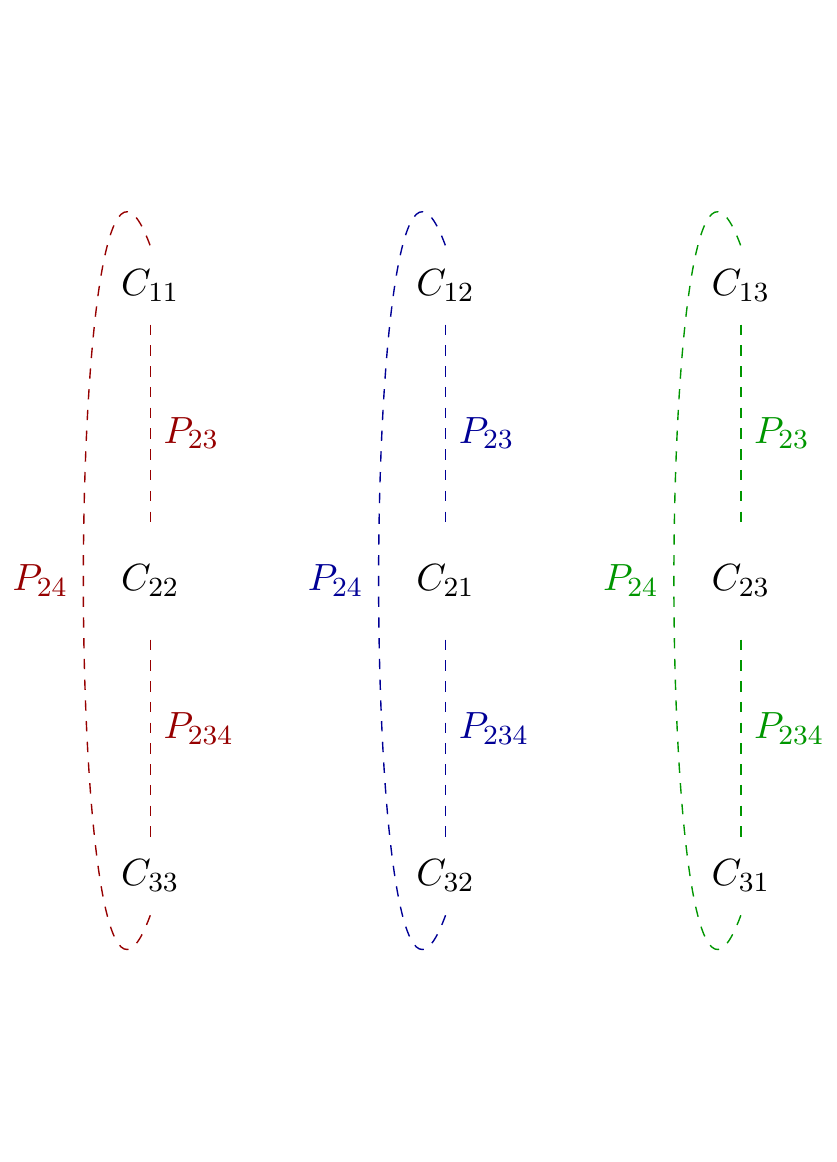}}
    \caption{Equivalent pictorial representation of the orbits as in Fig. \ref{tikzz}. }
    \label{}
\end{figure}
The equations that we are going to choose are
\begin{align}
C_{11}&=\bar{K}(p_2,p_3,p_4,s,t,u)\,A(p_2,p_3,p_4,s,t,u)\notag\\
C_{12}&=\bar{K}(p_2,p_3,p_4,s,t,u)\,A(p_3,p_2,p_4,u,t,s)+L(t,u)\bigg(A(p_2,p_3,p_4,s,t,u)+A(p_3,p_2,p_4,u,t,s)\bigg)\notag\\
C_{13}&=\bar{K}(p_2,p_3,p_4,s,t,u)\,A(p_4,p_3,p_2,t,s,u)-L(t,u)\bigg(A(p_2,p_3,p_4,s,t,u)+A(p_4,p_3,p_2,t,s,u)\bigg).
\end{align}
At this stage we include $P_{34}$, under whose action $C_{11}$ is mapped to itself, while $C_{12}\leftrightarrow C_{13}$. The mapping is illustrated below showing that the independent equations are only two.

\begin{figure}[h]
    \centering
\raisebox{-0.5\height}{\includegraphics[scale=1.2]{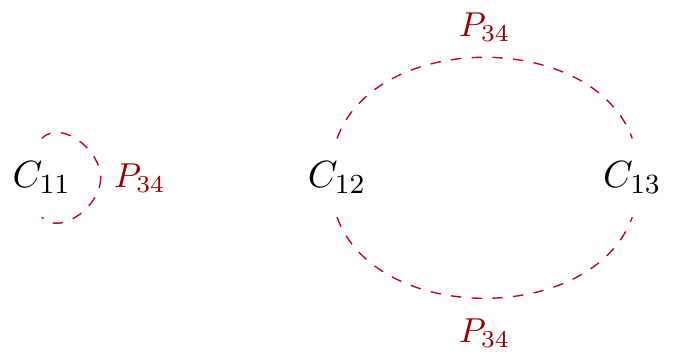}}
    \caption{Orbits of the CWIs for the $TOOO$ under $P_{34}$.}
    \label{three}
\end{figure}
 We take $C_{11}$ and $C_{12}$ as the independent ones, and all the other equations are obtained by acting on these two with a generic permutation of $(p_2,p_3,p_4)$. Therefore we have to solve only the two equations
\begin{equation}
\left\{
\begin{split}
0=&\bar{K}(p_2,p_3,p_4,s,t,u)\,A(p_2,p_3,p_4,s,t,u)\\
0=&\bar{K}(p_2,p_3,p_4,s,t,u)\,A(p_3,p_2,p_4,u,t,s)+L(t,u)\bigg(A(p_2,p_3,p_4,s,t,u)+A(p_3,p_2,p_4,u,t,s)\bigg)
\end{split}\right.
\end{equation}
as representatives of the set of the CWI, after taking into account all the symmetry properties of $A$. They can equivalently be set into the form
\begin{equation}
\left\{
\begin{split}
0&=\bar{K}(p_2,p_3,p_4,s,t,u)\,A(p_2,p_3,p_4,s,t,u)\\
0&=\bar{K}(p_3,p_2,p_4,u,t,s)\,A(p_2,p_3,p_4,s,t,u)-L(s,t)\bigg(A(p_2,p_3,p_4,s,t,u)+A(p_3,p_2,p_4,u,t,s)\bigg)
\end{split}\right.
\end{equation}
using the symmetry of the correlator. 
\subsection{Secondary Conformal Ward Identities}
\begin{figure}[h]
    \centering
\raisebox{-0.5\height}{\includegraphics[scale=1.2]{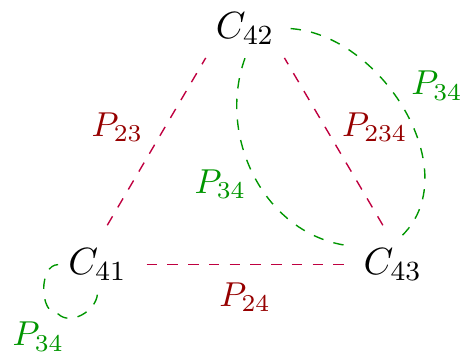}}
    \caption{Orbits of the secondary CWIs under permutations.}
    \label{four}
\end{figure}

The secondary CWIs for the correlator are first order differential equations derived from the coefficients $C_{4i}$, $i=1,\,2,\,3$ in \eqref{Ktt} together with Eq. \eqref{prim}. Such coefficients take the forms
\begin{align}
C_{41}&=\left[\frac{2(p_3^2-p_2^2-t^2)}{t}\frac{\partial}{\partial t}+\frac{2(p_4^2-p_2^2-u^2)}{u}\frac{\partial}{\partial u}-4\,p_2\,\frac{\partial}{\partial p_2}+\frac{2}{\bar{p}_1^2}\bigg(\frac{d(d-2)(p_2^2-s^2)-2s^2}{(d-1)}\bigg)\right.\notag\\
&\hspace{1cm}\left.+\frac{4\Delta(d-1)-2d^2-3(d-2)}{(d-1)}-\frac{(d-2)}{(d-1)}\frac{\big(p_2^2-s^2\big)^2}{\bar{p}_1^4}\right]\,A(p_2,p_3,p_4,s,t,u)
\notag\\
&-\left[\frac{2}{\bar{p}_1^2}\bigg(\frac{d(p_3^2-u^2)+2u^2}{(d-1)}\bigg)+\frac{(d-2)}{(d-1)}+\frac{(d-2)}{(d-1)}\frac{(p_3^2-u^2)^2}{\bar{p}_1^4}\right]\,A(p_3,p_2,p_4,u,t,s)\notag\\
&-\left[\frac{2}{\bar{p}_1^2}\bigg(\frac{d(p_4^2-t^2)+2t^2}{(d-1)}\bigg)+\frac{(d-2)}{(d-1)}+\frac{(d-2)}{(d-1)}\frac{(p_4^2-t^2)^2}{\bar{p}_1^4}\right]\,A(p_4,p_3,p_2,t,s,u)\,,
\end{align}

\begin{align}
C_{42}&=\left[\frac{2(p_2^2-p_3^2-t^2)}{t}\frac{\partial}{\partial t}+\frac{2(p_4^2-p_3^2-s^2)}{s}\frac{\partial}{\partial s}-4\,p_3\,\frac{\partial}{\partial p_3}+\frac{2}{\bar{p}_1^2}\bigg(\frac{d(d-2)(p_3^2-u^2)-2u^2}{(d-1)}\bigg)\right.\notag\\
&\hspace{1cm}\left.+\frac{4\Delta(d-1)-2d^2-3(d-2)}{(d-1)}-\frac{(d-2)}{(d-1)}\frac{\big(p_3^2-u^2\big)^2}{\bar{p}_1^4}\right]\,A(p_3,p_2,p_4,u,t,s)
\notag\\
&-\left[\frac{2}{\bar{p}_1^2}\bigg(\frac{d(p_2^2-s^2)+2s^2}{(d-1)}\bigg)+\frac{(d-2)}{(d-1)}+\frac{(d-2)}{(d-1)}\frac{(p_2^2-s^2)^2}{\bar{p}_1^4}\right]\,A(p_2,p_3,p_4,s,t,u)\notag\\
&-\left[\frac{2}{\bar{p}_1^2}\bigg(\frac{d(p_4^2-t^2)+2t^2}{(d-1)}\bigg)+\frac{(d-2)}{(d-1)}+\frac{(d-2)}{(d-1)}\frac{(p_4^2-t^2)^2}{\bar{p}_1^4}\right]\,A(p_4,p_3,p_2,t,s,u)\notag\\
\end{align}
and 
\begin{align}
C_{43}&=\left[\frac{2(p_3^2-p_4^2-s^2)}{s}\frac{\partial}{\partial s}+\frac{2(p_2^2-p_4^2-u^2)}{u}\frac{\partial}{\partial u}-4\,p_4\,\frac{\partial}{\partial p_4}+\frac{2}{\bar{p}_1^2}\bigg(\frac{d(d-2)(p_4^2-t^2)-2t^2}{(d-1)}\bigg)\right.\notag\\
&\hspace{1cm}\left.+\frac{4\Delta(d-1)-2d^2-3(d-2)}{(d-1)}-\frac{(d-2)}{(d-1)}\frac{\big(p_4^2-t^2\big)^2}{\bar{p}_1^4}\right]\,A(p_4,p_3,p_2,t,s,u)
\notag\\
&-\left[\frac{2}{\bar{p}_1^2}\bigg(\frac{d(p_3^2-u^2)+2u^2}{(d-1)}\bigg)+\frac{(d-2)}{(d-1)}+\frac{(d-2)}{(d-1)}\frac{(p_3^2-u^2)^2}{\bar{p}_1^4}\right]\,A(p_3,p_2,p_4,u,t,s)\notag\\
&-\left[\frac{2}{\bar{p}_1^2}\bigg(\frac{d(p_2^2-s^2)+2s^2}{(d-1)}\bigg)+\frac{(d-2)}{(d-1)}+\frac{(d-2)}{(d-1)}\frac{(p_2^2-s^2)^2}{\bar{p}_1^4}\right]\,A(p_2,p_3,p_4,s,t,u)\,,
\end{align}
where here $p_1^2$ is treated as a dependent variable, that is: $\bar{p}_1^2=s^2+t^2+u^2-p_2^2-p_3^2-p_4^2$. 
The actions of the operators enforcing the momentum permutations and the orbits of the $C_{ij}$ are illustrated in 
Fig. \eqref{four}, where a given equation is connected by a link if there is a permutation of the momenta which relates it to a different one. \\
It is clear from the figure that each single vertex of the triangle is mapped into itself under a permutation 
acting on the opposite edge, showing that there is only one independent secondary CWI. In particular, we choose as the independent one $C_{41}$, which can be re-expressed in the form

\begin{align}
&L_1\,A(p_2,p_3,p_4,s,t,u)\notag\\
&\qquad-L_2\,A(p_3,p_2,p_4,u,t,s)-L_2(p_3\leftrightarrow p_4)\,A(p_4,p_3,p_2,t,s,u)=-\frac{4d}{\bar{p}_1^2}\langle{O(\mathbf{p}_3+\mathbf{p}_4)O(\mathbf{p}_3)O(\mathbf{p}_4)}\rangle,\label{secondary}
\end{align}
where 
\begin{align}
L_1&=\left[\frac{2(p_3^2-p_2^2-t^2)}{t}\frac{\partial}{\partial t}+\frac{2(p_4^2-p_2^2-u^2)}{u}\frac{\partial}{\partial u}-4\,p_2\,\frac{\partial}{\partial p_2}+\frac{2}{\bar{p}_1^2}\bigg(\frac{d(d-2)(p_2^2-s^2)-2s^2}{(d-1)}\bigg)\right.\notag\\
&\hspace{1cm}\left.+\frac{4\Delta(d-1)-2d^2-3(d-2)}{(d-1)}-\frac{(d-2)}{(d-1)}\frac{\big(p_2^2-s^2\big)^2}{\bar{p}_1^4}\right]\\
L_2&=\left[\frac{2}{\bar{p}_1^2}\bigg(\frac{d(p_3^2-u^2)+2u^2}{(d-1)}\bigg)+\frac{(d-2)}{(d-1)}+\frac{(d-2)}{(d-1)}\frac{(p_3^2-u^2)^2}{\bar{p}_1^4}\right].
\end{align}
From \eqref{secondary}, one can check that the symmetry $p_3\leftrightarrow p_4$ is explicitly manifest.\\
In general, the role of the secondary WI's is to reduce the parameters of the solutions of the primary ones. For instance, in the case of 3-point functions, such solutions are determined from the primary equations modulo few constants, which are then fixed by the secondary ones. In that case, the right hand side of the secondary equations will be proportional to 2-point functions.\\
 The constraints on the primary solutions are obtained by taking special limits on the left hand side of the equations, in order to send two external coordinates into coalescence. This is obtained, for 3-point functions,  by taking two of the external invariant masses large and of unit ratio - $p_3^2/p_2^2\to 1$, for instance - which reduces the correlator to a 2-point function. \\
For 4-point functions this limit is far more involved, and we will be able to say little about it, the crucial point being that the primary solutions should contain arbitrary function(s), in this case a single function, as expected from the analysis in coordinate space, which are not identified in our formulation.\\
For this reason, we will try to discuss the asymptotic limit only of the primary solutions, where it is possible to underscore some specific behaviors of such solutions just by examining the structure of the equations. 
\section{The decomposition of the $TOOO$ in the $2\rightarrow2$ formulation}
\label{asymtreat}
In this section we will reconsider the  $TOOO$ correlator  $\langle T^{\mu_1 \nu_1}(\mathbf{p_1})O(\mathbf{p_2})O(\mathbf{p_3})O(\mathbf{\bar{p}_4})\rangle $ but with a different choice of the dependent momentum compared to the $1\to 3$ case,  which is suitable for the study of a $2\to 2$ process.\\
 We choose $p_4^{\mu}$ as the dependent momentum, $\bar{p_4}^{\mu}=-p_1^{\mu}-p_2^{\mu}-p_3^{\mu}$. Moreover, also the Mandelstam invariant $u^2$, will be taken as dependent variable  $\tilde{u}^2=-s^2-t^2+\sum p_i^2$.\\
We rewrite the decomposition \eqref{decomptooo} in the forms

\begin{equation}\label{tensordecomp}
\langle T^{\mu_1 \nu_1}(\mathbf{p_1})O(\mathbf{p_2})O(\mathbf{p_3})O(\mathbf{\bar{p}_4})\rangle=\langle t^{\mu_1 \nu_1}(\mathbf{p_1})O(\mathbf{p_2})O(\mathbf{p_3})O(\mathbf{\bar{p}_4})\rangle +
\langle t^{\mu_1 \nu_1}_{\textrm{loc}}(\mathbf{p_1})O(\mathbf{p_2})O(\mathbf{p_3})O(\mathbf{\bar{p}_4})\rangle,
\end{equation} 
which is symmetric in $p_2^{\mu},p_3^{\mu},p_4^{\mu}$. Furthermore, we require  the parameterization in the $tt$ sector to be symmetric under the exchange of the indices of the stress-energy tensor $\mu_1 \leftrightarrow \nu_1$. 
The $tt$ component of the $TOOO$ can then be parameterized as
\begin{equation}
\langle t^{\mu_1 \nu_1}(\mathbf{p_1})O(\mathbf{p_2})O(\mathbf{p_3})O(\mathbf{\bar{p}_4})\rangle=\Pi^{\mu_1 \nu_1}_{\alpha_1 \beta_1}(\mathbf{p_1})X^{\alpha_1 \beta_1},
\end{equation}
where $X^{\alpha_1 \beta_1}$ is a general rank-2  tensor built out momenta and Kronecker's delta's. There are two equivalent decompositions of such $tt$ term, that we will present below, but only one of them allows to obtain simplified expressions of the primary and secondary CWIs, which will turn very useful for our analysis.

\subsection{First decomposition}
First, we are going to derive the decomposition of the correlator by choosing as independent  momenta $p_2^{\mu},p_3^{\mu}$.  Terms that include $p_1^{\mu}$ will be eliminated by the transverse-traceless projector $\Pi^{\mu_1 \nu_1}_{\alpha_1 \beta_1}(\mathbf{p_1})$ and therefore will be omitted. We obtain the parameterization
\begin{equation}
X^{\alpha_1 \beta_1}=C^{\prime}(p_1,p_2,p_3,p_4,s,t)p_2^{\alpha_1}p_2^{\beta_1}+C^{\prime \prime}(p_1,p_2,p_3,p_4,s,t)p_3^{\alpha_1}p_3^{\beta_1}+C(p_1,p_2,p_3,p_4,s,t) p_2^{\alpha_1} p_3^{\beta_1},
\end{equation}
expressed in temrs of form factors $C, C^\prime, C^{\prime\prime}$.
Now, by imposing all the possible permutations (6 in total) of the momenta $p_2^{\mu},p_3^{\mu},p_4^{\mu}$, we derive the constraints
\begin{align*}
&C^{\prime}(p_1,p_2,p_3,p_4,s,t)=\frac{1}{2}\Big(C(p_1,p_2,p_3,p_4,s,t)+C(p_1,p_4,p_3,p_2,t,s)\Big),\\
&C^{\prime \prime}(p_1,p_2,p_4,p_3,s,\tilde{u})=\frac{1}{2}\Big(C(p_1,p_2,p_3,p_4,s,t)+C(p_1,p_2,p_4,p_3,s,\tilde{u})\Big).
\end{align*}
Finally, such $tt$ component takes the form
\begin{equation}\label{first}
\begin{split}
\langle t^{\mu_1 \nu_1}(\mathbf{p_1})O(\mathbf{p_2})O(\mathbf{p_3})O(\mathbf{\bar{p}_4})\rangle&=\Pi^{\mu_1 \nu_1}_{\alpha_1 \beta_1}\Bigg[\frac{1}{2}\Big(C(p_1,p_2,p_3,p_4,s,t)+C(p_1,p_4,p_3,p_2,t,s)\Big)p_2^{\alpha_1}p_2^{\beta_1}\\&
+\frac{1}{2}\Big(C(p_1,p_2,p_3,p_4,s,t)+C(p_1,p_2,p_4,p_3,s,\tilde{u})\Big)p_3^{\alpha_1}p_3^{\beta_1}+C(p_1,p_2,p_3,p_4,s,t) p_2^{\alpha_1} p_3^{\beta_1}\Bigg],
\end{split}
\end{equation}
 expressed in terms of a single form factor which exhibits the following symmetries
\begin{equation}
\begin{split}
\label{cc1}
&C(p_1,p_2,p_3,p_4,s,t)=C(p_1,p_3,p_2,p_4,\tilde{u},t),\\
&C(p_1,p_2,p_4,p_3,s,\tilde{u})=C(p_1,p_4,p_2,p_3,\tilde{u},s)\\
&C(p_1,p_4,p_3,p_2,t,s)=C(p_1,p_3,p_4,p_2,\tilde{u},s).
\end{split}
\end{equation}

\subsection{Second decomposition}
The second decomposition is obtained by using all the available momenta. 
It takes the form
\begin{equation}
\langle t^{\mu_1 \nu_1}(\mathbf{p_1})O(\mathbf{p_2})O(\mathbf{p_3})O(\mathbf{\bar{p}_4})\rangle=\Pi^{\mu_1 \nu_1}_{\alpha_1 \beta_1}(\mathbf{p_1})\tilde{X}^{\alpha_1 \beta_1},
\end{equation}
where 
\begin{equation}
\tilde{X}^{\alpha_1 \beta_1}=F(p_1,p_2,p_3,p_4,s,t) p_2^{\alpha_1}p_3^{\beta_1}+F^{\prime}(p_1,p_2,p_3,p_4,s,t) p_2^{\alpha_1}p_4^{\beta_1}+F^{\prime \prime}(p_1,p_2,p_3,p_4,s,t) p_3^{\alpha_1}p_4^{\beta_1}.
\end{equation}
Taking into account all the possible permutations, we end up with the expression
\begin{equation}\label{second}
\tilde{X}^{\alpha_1 \beta_1}=F(p_1,p_4,p_3,p_2,t,s) p_3^{\alpha_1}p_4^{\beta_1}+F(p_1,p_2,p_4,p_3,s,\tilde{u}) p_2^{\alpha_1}p_4^{\beta_1}+F(p_1,p_2,p_3,p_4,s,t) p_2^{\alpha_1}p_3^{\beta_1}.
\end{equation}
Our form factor obeys the following symmetries:
\begin{equation}
\begin{split}
&F(p_1,p_3,p_4,p_2,\tilde{u},s)=F(p_1,p_4,p_3,p_2,t,s),\\
&F(p_1,p_4,p_2,p_3,\tilde{u},s)=F(p_1,p_2,p_4,p_3,s,\tilde{u}),\\
&F(p_1,p_3,p_2,p_4,\tilde{u},t)=F(p_1,p_2,p_3,p_4,s,t).
\end{split}
\end{equation}
Now, we can impose momentum conservation on the first two terms of \eqref{first}. Then comparing with \eqref{second} and using the symmetry properties of the previous form factor $C$ in \eqref{cc1}, we obtain
\begin{align}
F(p_1,p_2,p_3,p_4,s,t)=-\frac{1}{2}\Big(C(p_1,p_2,p_3,p_4,s,t)+C(p_1,p_2,p_4,p_3,s,\tilde{u})\Big),\notag\\
F(p_1,p_2,p_4,p_3,s,\tilde{u})=-\frac{1}{2}\Big(C(p_1,p_2,p_3,p_4,s,t)+C(p_1,p_4,p_3,p_2,t,s)\Big),\notag\\
F(p_1,p_4,p_3,p_2,t,s)=-\frac{1}{2}\Big(C(p_1,p_2,p_3,p_4,s,t)+C(p_1,p_2,p_4,p_3,s,\tilde{u})\Big).
\label{relat}
\end{align}
The form factors $F$ and $C$ are related proving the equivalence between the two parameterizations. However $F$ is the one which generates CWIs of a simpler structure.

\subsection{Dilatation Ward Identity in the $2\rightarrow2$ formulation }
In this section we will proceed with the study of the dilatation WI. Using the form factor $F$, the full correlator is given by \eqref{tensordecomp} and the
exact parameterization of its $tt$ sector takes the form
\begin{equation}\label{TransverseFull}
\begin{split}
\langle t^{\mu_1 \nu_1}(\mathbf{p_1})O(\mathbf{p_2})O(\mathbf{p_3})O(\mathbf{\bar{p}_4})\rangle=&\Pi^{\mu_1 \nu_1}_{\alpha_1 \beta_1}(\mathbf{p_1})\Big(F(p_1,p_4,p_3,p_2,t,s) p_3^{\alpha_1}p_4^{\beta_1}+F(p_1,p_2,p_4,p_3,s,\tilde{u}) p_2^{\alpha_1}p_4^{\beta_1}\\&+F(p_1,p_2,p_3,p_4,s,t) p_2^{\alpha_1}p_3^{\beta_1}\Big),
\end{split}
\end{equation}
while the longitudinal sector is extracted by a contraction with the longitudinal projector
\begin{equation}\label{LocalFull}
\langle t^{\mu_1 \nu_1}_{\textrm{loc}}(\mathbf{p_1})O(\mathbf{p_2})O(\mathbf{p_3})O(\mathbf{\bar{p}_4})\rangle=\Sigma^{\mu_1 \nu_1}_{\alpha_1 \beta_1}(\mathbf{p_1})  \langle T^{\alpha_1 \beta_1}(\mathbf{p_1})O(\mathbf{p_2})O(\mathbf{p_3})O(\mathbf{\bar{p}_4})\rangle
\end{equation}
as in our previous analysis of the equations for the $1\to 3$.
We can express the CWI' s in terms of  6 invariants of the four-point function ($\sqrt{ p_i^2}=p_i$, $s=\sqrt{(\mathbf{p_1+p_2})^2},t=\sqrt{(\mathbf{p_2+p_3})^2}$) by using the chain rules
\begin{align}
& \frac{\partial}{\partial p_{1 \mu}}=\frac{p_1^{\mu}}{p_1}\frac{\partial}{\partial p_1}-\frac{\bar{p_4}^{\mu}}{p_4}\frac{\partial}{\partial p_4}+\frac{p_1^{\mu}+p_2^{\mu}}{s}\frac{\partial}{\partial s},\\
&\frac{\partial}{\partial p_{2 \mu}}=\frac{p_2^{\mu}}{p_2}\frac{\partial}{\partial p_1}-\frac{\bar{p_4}^{\mu}}{p_4}\frac{\partial}{\partial p_4}+\frac{p_1^{\mu}+p_2^{\mu}}{s}\frac{\partial}{\partial s}+\frac{p_2^{\mu}+p_3^{\mu}}{t}\frac{\partial}{\partial t},\\
&\frac{\partial}{\partial p_{3 \mu}}=\frac{p_3^{\mu}}{p_3}\frac{\partial}{\partial p_3}-\frac{\bar{p_4}^{\mu}}{p_4}\frac{\partial}{\partial p_4}+\frac{p_2^{\mu}+p_3^{\mu}}{t}\frac{\partial}{\partial t}.
\label{chain}
\end{align}
Applying the dilatation WI to \eqref{tensordecomp} we obtain
\begin{equation}\label{fulldilatation}
\big[(\Delta_{\textrm{t}}-3d)-\sum_{i=1}^{3} p_i^{\lambda} \frac{\partial}{\partial p_i^{\lambda}}\big]\langle T^{\mu_1 \nu_1}(\mathbf{p_1})O(\mathbf{p_2})O(\mathbf{p_3})O(\mathbf{\bar{p}_4})\rangle=0,
\end{equation}
which can be projected using the $tt$ projector $\Pi^{\rho \sigma}_{\mu_1 \nu_1}(\mathbf{p_1})$ obtaining  

\begin{equation}
\Pi^{\rho \sigma}_{\mu_1 \nu_1}(\mathbf{p_1})\hat{D}\langle t^{\mu_1 \nu_1}(\mathbf{p_1})O(\mathbf{p_2})O(\mathbf{p_3})O(\mathbf{\bar{p}_4})\rangle=0.
\end{equation}
Using  \eqref{TransverseFull} and differentiating by the chain rule \eqref{chain}, we finally obtain the equation \begin{equation}\label{DWIv1}
\left[(2+3d-\Delta_{\textrm{t}})+\sum_{i=1}^{4} p_i \frac{\partial }{\partial p_i}+s \frac{\partial }{\partial s}+t \frac{\partial  }{\partial t}\right]F(p_1,p_2,p_3,p_4,s,t)=0.
\end{equation}
The same equation holds also for $F(p_1,p_2,p_4,p_3,s,\tilde{u}),$ and $F(p_1,p_4,p_3,p_2,t,s)$.  The 2 in the first term of the sum $(2 + 3 d\ldots$), defines the tensorial dimension of the form factor, and counts the number of momenta with which it appears in the parameterization.

\subsection{Special CWI for the \texorpdfstring{ $\langle T^{\mu_1 \nu_1}(\mathbf{p_1})O(\mathbf{p_2})O(\mathbf{p_3})O(\mathbf{\bar{p}_4})\rangle $}{} }\label{scwilocal}
In this section we repeat the analysis of the $1\to 3$ case, with the new parameterization of the correlator that we have just derived, by selecting $p_4$ as the dependent momentum.
The action of the special conformal generator, as before, will take the form
\begin{equation}\label{CWI0}
\begin{split}
0&=\mathcal{K}^{\kappa}\langle{ T^{\mu_1 \nu_1}(\mathbf{p_1})O(\mathbf{p_2})O(\mathbf{p_3})O(\mathbf{\bar{p}_4})}\rangle\\[1.2ex]
&=\mathcal{K}^{\kappa}\langle t^{\mu_1 \nu_1}(\mathbf{p_1})O(\mathbf{p_2})O(\mathbf{p_3})O(\bar{\mathbf{p_4}})\rangle +\mathcal{K}^{\kappa}
\langle{ t^{\mu_1 \nu_1}_{\textrm{loc}}(\mathbf{p_1})O(\mathbf{p_2})O(\mathbf{p_3})O(\bar{\mathbf{p_4}})}\rangle,
\end{split}
\end{equation}
We will focus now on the local part related to $t_{loc}$.\\
Using \eqref{LocalFull}  we now have to compute
\begin{equation}
\Pi^{\rho \sigma}_{\mu_1 \nu_1}(p_1)\mathcal{K}^{\kappa}\Sigma^{\mu_1 \nu_1}_{\alpha_1 \beta_1}  \langle  T^{\alpha_1 \beta_1}(\mathbf{p_1})O(\mathbf{p_2})O(\mathbf{p_3})O(\mathbf{\bar{p}_4})\rangle ,
\end{equation}
projected on the $tt$ sector. 
We split our results into the scalar and the spin part of $\mathcal{K}^{\kappa}$. Acting with the projection $\Pi$ we obtain
\begin{equation}
\begin{split}
&\Pi^{\rho \sigma}_{\mu_1 \nu_1}(p_1)\mathcal{K}^{\kappa}_{\mathrm{scalar}}  \langle  t^{\mu_1 \nu_1}_{\textrm{loc}}(\mathbf{p_1})O(\mathbf{p_2})O(\mathbf{p_3})O(\mathbf{\bar{p}_4})\rangle=\\
&=\Pi^{\rho \sigma}_{\mu_1 \nu_1}(p_1)\Bigg[ 4(\Delta_1-d+1)\frac{\delta^{\kappa \mu_1}\delta^{\nu_1}_{\alpha_1}p_{1,\beta_1}}{p_1^2}+4\frac{p_1^{\kappa}}{p_1^2}\delta^{\mu_1}_{\beta_1}\delta^{\nu_1}_{\alpha_1} \Bigg]\langle T^{\alpha_1 \beta_1}(\mathbf{p_1})O(\mathbf{p_2})O(\mathbf{p_3})O(\mathbf{\bar{p}_4})\rangle,
\end{split}
\end{equation}
and
\begin{equation}
\begin{split}
&\Pi^{\rho \sigma}_{\mu_1 \nu_1}(p_1)\mathcal{K}^{\kappa}_{\alpha,\mathrm{spin}}  \langle  t^{\mu_1 \nu_1}_{\textrm{loc}}(\mathbf{p_1})O(\mathbf{p_2})O(\mathbf{p_3})O(\mathbf{\bar{p}_4})\rangle=\\
&=\Pi^{\rho \sigma}_{\mu_1 \nu_1}(\textbf{p}_1)\Bigg[4(d-1)\frac{\delta^{\kappa \mu_1}\delta^{\nu_1}_{\alpha_1}p_{1,\beta_1}}{p_1^2}-4\frac{p_1^{\kappa}}{p_1^2}\delta^{\mu_1}_{\beta_1}\delta^{\nu_1}_{\alpha_1}\Bigg]\langle  T^{\alpha_1 \beta_1}(\mathbf{p_1})O(\mathbf{p_2})O(\mathbf{p_3})O(\mathbf{\bar{p}_4})\rangle. 
\end{split}
\end{equation}
In our results, we have ignored terms that include $p_1^{\mu_1},p_1^{\nu_1},\delta^{\mu_1 \nu_1}$, because we have the freedom to apply a transverse-traceless projector of the form $\Pi^{\rho \sigma}_{\mu_1 \nu_1}(\mathbf{p}_1)$ to \eqref{CWI0}, so these terms will vanish. Adding the scalar and the spin contributions (and using $\Delta_1=d$), we get
\begin{equation}\label{CWIlocal}
\Pi^{\rho \sigma}_{\mu_1 \nu_1}(p_1)\mathcal{K}^{\kappa}\langle  t^{\mu_1 \nu_1}_{\textrm{loc}}(\mathbf{p_1})O(\mathbf{p_2})O(\mathbf{p_3})O(\mathbf{\bar{p}_4})\rangle=\Pi^{\rho \sigma}_{\mu_1 \nu_1}(p_1)\left(\frac{4d}{p_1^2}\delta^{\kappa \mu_1}p_{1\,\alpha}\langle  T^{\alpha \nu_1}(\mathbf{p_1})O(\mathbf{p_2})O(\mathbf{p_3})O(\mathbf{\bar{p}_4})\rangle\right).
\end{equation}

Now, we will apply the $\mathcal{K}^{\k}$ operator on the $tt$ part, followed by contraction with the $\Pi$ projector. We obtain the tensor equation

\begin{equation}\label{fullform}
\begin{split}
&\Pi^{\rho \sigma}_{\mu_1 \nu_1}(p_1)\mathcal{K}^{\kappa} \langle t^{\mu_1 \nu_1}(\mathbf{p_1})O(\mathbf{p_2})O(\mathbf{p_3})O(\mathbf{\bar{p}_4})\rangle=\\
&\Pi^{\rho \sigma}_{\mu_1 \nu_1}(p_1)\Big[p_1^{\kappa}\big(\tilde{C}_{11}p_2^{\mu_1}p_3^{\nu_1}+\tilde{C}_{12}p_2^{\mu_1}p_4^{\nu_1}+\tilde{C}_{13}p_3^{\mu_1}p_4^{\mu_1}\big)+p_2^{\kappa}\big(\tilde{C}_{21}p_2^{\mu_1}p_3^{\nu_1}+\tilde{C}_{22}p_2^{\mu_1}p_4^{\nu_1}+\tilde{C}_{23}p_3^{\mu_1}p_4^{\mu_1}\big)\\
&+p_3^{\kappa}\big(\tilde{C}_{31}p_2^{\mu_1}p_3^{\nu_1}+\tilde{C}_{32}p_2^{\mu_1}p_4^{\nu_1}+\tilde{C}_{33}p_3^{\mu_1}p_4^{\mu_1}\big)+\delta^{\mu_1 \kappa}\big(\tilde{C}_{41}p_2^{\nu_1}+\tilde{C}_{42}p_3^{\nu_1}\big)+\delta^{\nu_1 \kappa}\big(\tilde{C}_{51}p_2^{\mu_1}+\tilde{C}_{52}p_3^{\mu_1}\big)\Big]
\end{split}
\end{equation}
which will allow us to extract the independent conformal constraints.
\subsection{Primary Conformal Ward Identities}\label{PCWI22}
\label{primd}
The factors $\tilde{C}_{1j},\tilde{C}_{2j},\tilde{C}_{3j}$ are second-order differential equations involving the form factor $F$ and its various permutations. We see from \eqref{CWI0} and \eqref{fullform} that the coefficients  of the four-momenta $p_1^{\kappa},p_2^{\kappa},p_3^{\kappa}$ are zero. This translates into the equations  
\begin{align}
&\tilde{C}_{11}=0,&\tilde{C}_{12}=0,&\tilde{C}_{13}=0,\notag\\
&\tilde{C}_{21}=0,&\tilde{C}_{22}=0,&\tilde{C}_{23}=0,\notag\\
&\tilde{C}_{31}=0,&\tilde{C}_{32}=0,&\tilde{C}_{33}=0.
\end{align}

These are the primary  CWIs that we have mentioned before. Below we present the explicit expressions involving the $F(p_1,p_2,p_3,p_4,s,t)$ form factor. The remaining ones, which are obtained just by permutations of the momenta, can be found in \ref{AppendixTOOO}. We obtain

\begin{align}
\label{c11}
\tilde{C}_{11}=&\Bigg[\frac{\partial^2}{\partial p_4^2}+\frac{d-2\Delta+1}{p_4}\frac{\partial}{\partial p_4}-\frac{\partial^2}{\partial p_1^2}-\frac{1-d}{p_1}\frac{\partial}{\partial p_1}+\frac{1}{s}\frac{\partial}{\partial s}\left(p_4\frac{\partial}{\partial p_4}+p_3\frac{\partial}{\partial p_3}-p_1\frac{\partial}{\partial p_1}-p_2\frac{\partial}{\partial p_2}\right)\notag\\&
+\frac{d-\Delta}{s}\frac{\partial}{\partial s}
+\frac{p_3^2-p_2^2}{st}\frac{\partial^2}{\partial s \partial t}\Bigg]F(p_1,p_2,p_3,p_4,s,t)+\frac{2}{s}\frac{\partial F(p_1,p_4,p_3,p_2,t,s)}{\partial s}-\frac{2}{s}\frac{\partial F(p_1,p_2,p_4,p_3,s,\tilde{u})}{\partial s}
\end{align}
\begin{align}\label{c21}
\tilde{C}_{21}=&\Bigg[\frac{\partial^2 }{\partial p_4^2}+\frac{d-2\Delta+1}{p_4}\frac{\partial }{\partial p_4}-\frac{\partial^2 }{\partial p_2^2}-\frac{d-2\Delta+1}{p_2}\frac{\partial }{\partial p_2}+\frac{1}{s}\frac{\partial}{\partial s}\left(p_3\frac{\partial }{\partial p_3}+p_4\frac{\partial }{\partial p_4}-p_1\frac{\partial }{\partial p_1}-p_2\frac{\partial }{\partial p_2}\right)
\notag\\&
+\frac{\Delta-d-2}{t}\frac{\partial}{\partial t}+\frac{d-\Delta}{s}\frac{\partial}{\partial s}+\frac{1}{t}\frac{\partial}{\partial t}\left( p_1 \frac{\partial }{\partial p_1}+p_4\frac{\partial }{\partial p_4}-p_2\frac{\partial }{\partial p_2}-p_3\frac{\partial }{\partial p_3}\right)
\notag
\\&+\frac{p_4^2-p_2^2}{st}\frac{\partial^2}{\partial s \partial t}\Bigg]F(p_1,p_2,p_3,p_4,s,t)+\frac{2}{s}\frac{\partial F(p_1,p_4,p_3,p_2,t,s)}{\partial s}-\frac{2}{s}\frac{\partial F(p_1,p_2,p_4,p_3,s,\tilde{u})}{\partial s},\nonumber \\
\end{align}

and finally
\begin{align}\label{c31}
\tilde{C}_{31}=&\Bigg[\frac{\partial^2 }{\partial p_4^2}+\frac{d-2\Delta+1}{p_4}\frac{\partial }{\partial p_4}-\frac{\partial^2 }{\partial p_3^2}-\frac{d-2\Delta+1}{p_3}\frac{\partial }{\partial p_3}+\frac{2}{s}\frac{\partial}{\partial s}+\frac{p_1^2-p_2^2}{st}\frac{\partial^2}{\partial s \partial t}+\frac{\Delta-d-2}{t}\frac{\partial}{\partial t}\notag\\&+\frac{1}{t}\frac{\partial}{\partial t}\left(p_1\frac{\partial }{\partial p_1}+p_4\frac{\partial }{\partial p_4}-p_2\frac{\partial }{\partial p_2}-p_3\frac{\partial }{\partial p_3}\right)
\Bigg]F(p_1,p_2,p_3,p_4,s,t)+\frac{2}{s}\frac{\partial F(p_1,p_4,p_3,p_2,t,s)}{\partial s}\notag
\\&-\frac{2}{s}\frac{\partial F(p_1,p_2,p_4,p_3,s,\tilde{u})}{\partial s}.
\end{align}

\subsection{Secondary Conformal Ward Identities}
Since our 4-point function is symmetric in $\mu_1\leftrightarrow \nu_1$, the terms proportional to $\delta^{\mu_1 \kappa}$ and $\delta^{\nu_1 \kappa}$ given by the coefficients $\tilde{C}_{41}$ and $\tilde{C}_{51}$ identify a single constraint, as well as $\tilde{C}_{42}$ and $\tilde{C}_{52}$, and are explicitly given by the factors $\tilde{C}_{4j}$. They take the form
\begin{equation}
\begin{split}
&\tilde{C}_{41}=\hat{G}\Big(F(p_1,p_2,p_3,p_4,s,t)-F(p_1,p_4,p_3,p_2,t,s)\Big)+\hat{A}F(p_1,p_2,p_4,p_3,s,\tilde{u}),
\end{split}
\end{equation}
where 
\begin{equation}
\begin{split}
\hat{G}&=\frac{d(s^2+t^2-p_4^2-p_2^2-2p_1^2)+2\Delta p_1^2}{p_1^2}-\frac{t^2+p_3^2-p_2^2}{t}\frac{\partial}{\partial t}+\frac{p_2^2+p_4^2-s^2-t^2}{p_1}\frac{\partial}{\partial p_1}
\\&-\frac{s^2+p_3^2-p_4^2}{s}\frac{\partial}{\partial s}-2p_3\frac{\partial}{\partial p_3},
\end{split}
\end{equation}
and
\begin{equation}
\begin{split}
\hat{A}&=\Bigg(\frac{d(s^2+p_4^2-p_2^2-t^2)}{p_1^2}\Bigg)+\frac{t^2+p_2^2-p_3^2}{t}\frac{\partial}{\partial t}+\frac{p_2^2+t^2-p_4^2-s^2}{p_1}\frac{\partial}{\partial p_1}
\\&+\frac{p_3^2-p_4^2-s^2}{s}\frac{\partial}{\partial s}+2 p_2\frac{\partial}{\partial p_2}-2p_4\frac{\partial}{\partial p_4}.
\end{split}
\end{equation}
Moreover, we obtain
\begin{equation}
\begin{split}
&\tilde{C}_{42}=\hat{M}\Big(F(p_1,p_2,p_3,p_4,s,t)-F(p_1,p_2,p_4,p_3,s,\tilde{u})\Big)+\hat{N}F(p_1,p_4,p_3,p_2,t,s),
\end{split}
\end{equation}
where
\begin{equation}
\begin{split}
\hat{M}=&\left(\frac{2\Delta p_1^2-d(p_1^2-p_2^2+s^2)}{p_1^2}\right)-\frac{p_1^2+p_2^2-s^2}{p_1}\frac{\partial}{\partial p_1}+\frac{p_3^2-p_2^2-t^2}{t}\frac{\partial}{\partial t}-2 p_2\frac{\partial}{p_2}
\end{split}
\end{equation}
and
\begin{equation}
\begin{split}
\hat{N}=&\left(\frac{d \left(p_1^2+p_2^2+2p_4^2-s^2-2 t^2\right)}{p_1^2}\right)+\frac{t^2+p_3^2-p_2^2}{t}\frac{\partial}{\partial t}-\frac{p_1^2+p_2^2+2p_4^2-s^2-2t^2}{p_1}\frac{\partial}{\partial p_1}\\&+\frac{2 (p_3^2-p_4^2)}{s}\frac{\partial}{\partial s}-2p_4\frac{\partial}{\partial p_4}+2p_3\frac{\partial}{\partial p_3}.
\end{split}
\end{equation}

Combining  \eqref{ConsWI22} along with  \eqref{CWIlocal} and \eqref{fullform} we obtain the equations
\begin{equation}\label{example}
\begin{split}
&\hat{G}\Big(F(p_1,p_2,p_3,p_4,s,t)-F(p_1,p_4,p_3,p_2,t,s)\Big)+\hat{A}F(p_1,p_2,p_4,p_3,s,\tilde{u})=\\
&\hspace{2cm}=\frac{4d}{p_1^2}\Big(\langle O(\mathbf{p_1+p_2})O(\mathbf{p_3})O(\mathbf{\bar{p}_4})\rangle-\langle O(\mathbf{p_2})O(\mathbf{p_3})O(\mathbf{p_2+p_3}) \rangle \Big),
\end{split}
\end{equation}
and
\begin{equation}
\begin{split}
&\hat{M}\Big(F(p_1,p_2,p_3,p_4,s,t)-F(p_1,p_2,p_4,p_3,s,\tilde{u})\Big)+\hat{N}F(p_1,p_4,p_3,p_2,t,s)=\\
&\hspace{2cm}=\frac{4d}{p_1^2}\big(\langle O(\mathbf{p_2})O(\mathbf{p_1+p_3})O(\mathbf{\bar{p}_4})\rangle-\langle O(\mathbf{p_2})O(\mathbf{p_3})O(\mathbf{p_2+p_3}) \rangle\big).
\end{split}
\end{equation}
These are the secondary WI's for the $TOOO$. The 3-point function on the right hand side of this equation is uniquely given by a combination of hypergeometric functions and will be discussed below. 
\section{Asymptotics for scalar and dual conformal/conformal 4-point functions}
\label{asimpt}
Our goal, from this section on, will be to identify some of the properties of these primary and secondary equations for the $TOOO$, and for this reason it will be compelling to consider first the $(OOOO)$ correlator, which is slightly simpler compared to the former. Both cases show some similarities, starting from the fact that they are both characterised by a single form factor. The structure of the equations is expected to be similar, and indeed in both cases we will be able to identify also a similar behaviour in the corresponding form factors, in some kinematical limits. \\
The $OOOO$, as shown recently \cite{Maglio:2019grh}, allows a specific class of solutions which are uniquely identified by enlarging the original conformal symmetry to include a dual conformal symmetry as well. Indeed, these special solutions are very useful 
for studying the hypergeometric structure of the CWIs in some asymptotic limits. As we are going to see, hypergeometric solutions of 4-point functions are very special, as one expects on generic grounds, and the general CWIs, even in the scalar case, are not described by hypergeometric systems related to $F_4$. The  only exact statement that can be made concerning the structure of such systems of equations, as we are going to show, will be that Lauricella functions - i.e. hypergeometric functions of three variables - are exact solutions of all these systems of equations and can be interpreted as homogeneous (i.e. particular) solutions of such CWIs for arbitrary scaling dimensions of the scalar operators. \\  
We start our discussion by recalling that for the $OOOO$, the two CWIs take the form 
(general scalar CWIs)
\cite{Maglio:2019grh}

\begin{align}
S_1=&=\bigg\{\frac{\partial^2}{\partial p_2^2}+\frac{(d-2\Delta_2+1)}{p_2}\frac{\partial}{\partial p_2}-\frac{\partial^2}{\partial p_4^2}-\frac{(d-2\Delta_4+1)}{p_4}\frac{\partial}{\partial p_4}\notag\\
&\qquad+\frac{1}{s}\frac{\partial}{\partial s}\left(p_1\frac{\partial}{\partial p_1}+p_2\frac{\partial}{\partial p_2}-p_3\frac{\partial}{\partial p_3}-p_4\frac{\partial}{\partial p_4}\right)+\frac{\Delta_{3412}}{s}\frac{\partial}{\partial s}\notag\\
&\qquad+\frac{1}{t}\frac{\partial}{\partial t}\left(p_2\frac{\partial}{\partial p_2}+p_3\frac{\partial}{\partial p_3}-p_1\frac{\partial}{\partial p_1}-p_4\frac{\partial}{\partial p_4}\right)+\frac{\Delta_{1423}}{t}\frac{\partial}{\partial t}\notag\\[1.2ex]
&\qquad+\frac{(p_2^2-p_4^2)}{st}\frac{\partial^2}{\partial s\partial t}\bigg\}\,\Phi(p_1,p_2,p_3,p_4,s,t)=0
\label{C2}
\end{align}

\begin{align}
S_2&=\bigg\{\frac{\partial^2}{\partial p_1^2}+\frac{(d-2\Delta_1+1)}{p_1}\frac{\partial}{\partial p_1}-\frac{\partial^2}{\partial p_3^2}-\frac{(d-2\Delta_3+1)}{p_3}\frac{\partial}{\partial p_3}\notag\\
&\qquad+\frac{1}{s}\frac{\partial}{\partial s}\left(p_1\frac{\partial}{\partial p_1}+p_2\frac{\partial}{\partial p_2}-p_3\frac{\partial}{\partial p_3}-p_4\frac{\partial}{\partial p_4}\right)+\frac{\Delta_{3412}}{s}\frac{\partial}{\partial s}\notag\\
&\qquad+\frac{1}{t}\frac{\partial}{\partial t}\left(p_1\frac{\partial}{\partial p_1}+p_4\frac{\partial}{\partial p_4}-p_2\frac{\partial}{\partial p_2}-p_3\frac{\partial}{\partial p_3}\right)+\frac{\Delta_{1423}}{t}\frac{\partial}{\partial t}\notag\\[1.2ex]
&\qquad+\frac{(p_1^2-p_3^2)}{st}\frac{\partial^2}{\partial s\partial t}\bigg\}\,\Phi(p_1,p_2,p_3,p_4,s,t)=0.\label{Eq2}
\end{align}

\begin{align}
S_3&=\bigg\{\frac{\partial^2}{\partial p_1^2}+\frac{(d-2\Delta_1+1)}{p_1}\frac{\partial}{\partial p_1}-\frac{\partial^2}{\partial p_4^2}-\frac{(d-2\Delta_4+1)}{p_4}\frac{\partial}{\partial p_4}\notag\\[1.5ex]
&\qquad+\frac{1}{s}\frac{\partial}{\partial s}\left(p_1\frac{\partial}{\partial p_1}+p_2\frac{\partial}{\partial p_2}-p_3\frac{\partial}{\partial p_3}-p_4\frac{\partial}{\partial p_4}\right)+\frac{\Delta_{3412}}{s}\frac{\partial}{\partial s}\notag\\[1.5ex]
&\qquad+\frac{(p_2^2-p_3^2)}{st}\frac{\partial^2}{\partial s\partial t}\bigg\}\,\Phi(p_1,p_2,p_3,p_4,s,t)=0\label{C1}
\end{align}
where 
 \begin{equation}
 \label{deltas}
 \Delta_{ijkl}=\Delta_i +\Delta_j-\Delta_k-\Delta_l
 \end{equation}
 is a specific combination of the scaling parameters of the primary scalar operators $(O)$, which plays a special role in the derivation of the dcc solutions.
In \cite{Maglio:2019grh} the discussion dealt with two possible cases for the $OOOO$ in which the scaling combinations in \eqref{deltas} vanish: 1) the equal scaling case with $\Delta_i=\Delta$ (i=1,2,3,4) and 
2) the case in which two operators are pairwise of equal scalings. In both cases, the solutions satisfy the condition of being conformal and dual conformal invariant.\\
 The vanishing of \eqref{deltas} is necessary in order to 
remove the $\partial/\partial s$ and $\partial/\partial t$ terms and reduce the three $S_i$'s to a  hypergeometric system of equations \eqref{diff}. Notice that differently from the case of 3-point functions, where a similar system has been identified \cite{Coriano:2013jba}, as shown in Eq. \eqref{ipergio}, the variables are quartic - rather than quadratic - ratios of the invariants. \\
In order to derive such a system, which is extracted from the $S_i$'s, we need a product ans\"atz based on a quartic pivot $(s^2 t^2)$ with variables \cite{Maglio:2019grh}\cite{Bzowski:2013sza}
 \begin{equation}
 x=\frac{p_1^2 p_3^2}{s^2 t^2}\qquad  y=\frac{p_2^2 p_4^2}{s^2 t^2},
 \label{xy}
 \end{equation}
and  observe that this choice sets automatically to zero the mixed derivative terms in $p_i$ and $s$ and $t$ in Eqs. \eqref{C2},\eqref{Eq2} and \eqref{C1}. The ans\"atz for the solution is based on the product of a function G(x,y) and of powers of x and y - given by \eqref{xy} - of the form 
\begin{equation} 
\Phi\sim x^a y^b G(x,y),
\end{equation}
for suitable $a$ and $b$, quite similarly to the case of a scalar 3-point function.
On any function $G(x,y)$, terms of the form

\begin{equation}
\left( p_1\frac{\partial}{\partial p_1}+p_2\frac{\partial}{\partial p_2}-p_3\frac{\partial}{\partial p_3}-p_4\frac{\partial}{\partial p_4}\right)G(x,y)=0
\label{GG}
\end{equation}
vanish, if we choose $x$ and $y$ as the quartic ratios \eqref{xy}.
If we use the definition of the $K_{ij}$ operators  \eqref{operatorK} and the ans\"atz based on $G(x,y)$ as defined above, the three equations take the form (intermediate scalar CWIs)
\begin{equation}
\left\{
\begin{split}
&\bigg( K_{2 4} +\frac{(p_2^2-p_4^2)}{st}\frac{\partial^2}{\partial s\partial t}\bigg)\Phi(p_1,p_2,p_3,p_4,s,t)=-\bigg(\frac{\Delta_{3412}}{s}\frac{\partial}{\partial s}
+\frac{\Delta_{1423}}{t}\frac{\partial}{\partial t} \bigg)\,\Phi(p_1,p_2,p_3,p_4,s,t)
 \\ 
&\bigg( K_{1 3} +\frac{(p_1^2-p_3^2)}{st}\frac{\partial^2}{\partial s\partial t}\bigg)\Phi(p_1,p_2,p_3,p_4,s,t)=-\bigg(\frac{\Delta_{3412}}{s}\frac{\partial}{\partial s}
+\frac{\Delta_{1423}}{t}\frac{\partial}{\partial t} \bigg)\,\Phi(p_1,p_2,p_3,p_4,s,t)\\ 
&\bigg( K_{1 4} +\frac{(p_2^2-p_3^2)}{st}\frac{\partial^2}{\partial s\partial t}\bigg)\Phi(p_1,p_2,p_3,p_4,s,t)=-\frac{\Delta_{3412}}{s}\frac{\partial}{\partial s} \,\Phi(p_1,p_2,p_3,p_4,s,t)\\
\label{CC4}
\end{split}
\right.\\
\end{equation}
where we have removed all the mixed derivative terms in $(s, p_i^2),(t, p_i^2)$, thanks to  \eqref{GG}. Explicit dcc solutions of this system of equations are obtained if $\Delta_{ijkl}=0$, and the operators 
$K_{24}$ and $K_{13} $ depend separately on a {\em single scaling variable}, that is if 
$\Delta_2=\Delta_4$ and $\Delta_1=\Delta_3$. Notice that this condition is compatible with the vanishing of $\Delta_{3412}$ and $\Delta_{1423}$ and takes to a hypergeometric system of equations, which are again solved in terms of hypergeometrics of the variables $x$ and $y$ given in \eqref{xy}. In this case we could rewrite the system in the form (reduced scalar CWIs)
\begin{equation}
\left\{
\begin{split}
&\bigg( K_{2 4}(\Delta_2) +\frac{(p_2^2-p_4^2)}{st}\frac{\partial^2}{\partial s\partial t}\bigg)\Phi(p_1,p_2,p_3,p_4,s,t)=0
\\ 
&\bigg( K_{1 3}(\Delta_1) +\frac{(p_1^2-p_3^2)}{st}\frac{\partial^2}{\partial s\partial t}\bigg)\Phi(p_1,p_2,p_3,p_4,s,t)=0 \\
&\bigg( K_{1 4} +\frac{(p_2^2-p_3^2)}{st}\frac{\partial^2}{\partial s\partial t}\bigg)\Phi(p_1,p_2,p_3,p_4,s,t)=0
\label{CCC4}
\end{split}
\right.\\
\end{equation}
where the $K_{ij}(\Delta_i)$ indicates that such operators depend on a single scaling constant. \\
It is important to observe that the system \eqref{CCC4} admits explicit dcc solutions which are expressed as 
hypergeometric functions, or, equivalently, as 3K integrals, but the entire set of dcc solutions is not just composed of these functions..\\ 
Dual conformal symmetry constrains a certain ans\"atz (the dual conformal ans\"atz) to be expressed only in terms of the two quartic ratios $x$ and $y$, via a function $G(x,y)$. Functions $G$ of such ratios will then necessarily satisfy the condition \eqref{GG}, and henceforth the reduced system \eqref{CC4}. \\
The solutions of the three constraints in \eqref{CC4} of the form $G(x,y)$, will then characterize the most general set of dcc solutions for scalar primary operators, of which special cases are those found in \cite{Maglio:2019grh} and reported below in Eq. \eqref{fform}.
The additional reduction of the system \eqref{CC4} to \eqref{CCC4} obviously, allows us to work with explicit expressions which are all related by analytic continuations and therefore describe a unique solution, as shown in \cite{Maglio:2019grh}. Therefore, they are optimal for the study of several kinematical limits of the scalar correlator, that we are now going to investigate.

\subsection{Limits for equal scalings and $\Delta_{ijkl}=0$}
As we have mentioned, the choice $\Delta_{ijkl}=0$ is what renders the system \eqref{CCC4} a variant of the ordinary hypergeometric system, which in general takes the form  
\begin{equation}
K_{13}\Phi=0\qquad K_{23}\Phi=0, \qquad \textrm{where} \qquad K_{ij}=K_i -K_j.
\label{ipergio}
\end{equation} 
and it is solved by quadratic - rather than quartic - ratios of invariants. 
Once this gets reduced to \eqref{CCC4},  as already mentioned, the complete ans\"atz for the general solution of such system is constructed by multiplying the function $G(x,y)$ by the pivot, raised to a power $n_s$, fixed by the dilatation WI  
\begin{align}
\bigg[(\Delta_t-3d)-\sum_{i=1}^4p_i\frac{\partial}{\partial p_i}-s\frac{\partial}{\partial s}-t\frac{\partial}{\partial t}\bigg]\Phi(p_1,p_2,p_3,p_4,s,t)=0,\label{Dilatation4}
\end{align}
with $\Delta_t$ denoting the total scaling. If we choose as a pivot $s^2 t^2$,
 the solution indeed will take the form
\begin{equation}
 \Phi(p_i,s,t)= (s^2 t^2)^{n_s}G(x,y) \qquad  n_s=\frac{\Delta_t -3 d}{4}.
 \label{ans2}
\end{equation}
Few additional comments are in order concerning the homogeneous case 
$(\Delta_{ijkl}=0)$ and the system \eqref{CCC4}. We remark that the third equation of such system is identically satisfied if the first and the second equations are, which is the case if an ans\"atz of type \eqref{ans2} is chosen. This is clearly consistent with the fact the four functionally independent solutions of an Appell system of equations (for $F_4$) is based only on two independent equations \eqref{diff}.
 
 The solution of the homogeneous system \eqref{CCC4}, as already mentioned, can be written in terms of 4 Appell functions $F_4$ of the $x$ and $y$ ratios given in \eqref{xy} 
 \cite{Maglio:2019grh}
\begin{align}
\langle{O(p_1)O(p_2)O(p_3)O(p_4)}\rangle&=2^{\frac{d}{2}-4}\ \ C\,\sum_{\l,\m=0,\Delta-\frac{d}{2}}\x(\l,\m)\bigg[\big(s^2\,t^2\big)^{\Delta-\frac{3}{4}d}\left(\frac{p_1^2 p_3^2}{s^2 t^2}\right)^\l\left(\frac{p_2^2p_4^2}{s^2t^2}\right)^\m\nonumber\\
&\hspace{-3cm}\times\,F_4\left(\frac{3}{4}d-\Delta+\l+\m,\frac{3}{4}d-\Delta+\l+\m,1-\Delta+\frac{d}{2}+\l,1-\Delta+\frac{d}{2}+\m,\frac{p_1^2 p_3^2}{s^2 t^2},\frac{p_2^2 p_4^2}{s^2 t^2}\right)\notag\\
&+\big(s^2\,u^2\big)^{\Delta-\frac{3}{4}d}\left(\frac{p_2^2 p_3^2}{s^2 u^2}\right)^\l\left(\frac{p_1^2p_4^2}{s^2u^2}\right)^\m\notag\\
&\hspace{-3cm}\times\,F_4\left(\frac{3}{4}d-\Delta+\l+\m,\frac{3}{4}d-\Delta+\l+\m,1-\Delta+\frac{d}{2}+\l,1-\Delta+\frac{d}{2}+\m,\frac{p_2^2 p_3^2}{s^2 u^2},\frac{p_1^2 p_4^2}{s^2 u^2}\right)\notag\\
&+\big(t^2\,u^2\big)^{\Delta-\frac{3}{4}d}\left(\frac{p_1^2 p_2^2}{t^2 u^2}\right)^\l\left(\frac{p_3^2p_4^2}{t^2u^2}\right)^\m\notag\\
&\hspace{-3cm}\times\,F_4\left(\frac{3}{4}d-\Delta+\l+\m,\frac{3}{4}d-\Delta+\l+\m,1-\Delta+\frac{d}{2}+\l,1-\Delta+\frac{d}{2}+\m,\frac{p_1^2 p_2^2}{t^2 u^2},\frac{p_3^2 p_4^2}{t^2 u^2}\right)\bigg],\label{fform}
\end{align}
where the coefficients $\x(\l,\m)$ are explicitly given by
\begin{equation}
\begin{split}
\x\left(0,0\right)&=\left[\Gamma\left(\frac{3}{4}d-\Delta\right)\right]^2\left[\Gamma\left(\Delta-\frac{d}{2}\right)\right]^2\\
\x\left(0,\Delta-\frac{d}{2}\right)&=\x\left(\Delta-\frac{d}{2},0\right)=\left[\Gamma\left(\frac{d}{4}\right)\right]^2\Gamma\left(\Delta-\frac{d}{2}\right)\Gamma\left(\frac{d}{2}-\Delta\right)\\
\x\left(\Delta-\frac{d}{2},\Delta-\frac{d}{2}\right)&=\left[\Gamma\left(\Delta-\frac{d}{4}\right)\right]^2\left[\Gamma\left(\frac{d}{2}-\Delta\right)\right]^2,
\end{split}\label{xicoef2}
\end{equation}
which is explicitly symmetric under all the possible permutations of the momenta and it is fixed up to one undetermined constant $C$.\\
As shown in \cite{Maglio:2019grh}, \eqref{fform} can be re-expressed in the form  
\begin{align}
&I_{\frac{d}{2}-1\{\Delta-\frac{d}{2},\Delta-\frac{d}{2},0\}}(p_1p_3,p_2 p_4,s t)=\notag\\
&\qquad=\,(p_1p_3)^{\Delta-\frac{d}{2}}(p_2p_4)^{\Delta-\frac{d}{2}}\int_{0}^\infty\,dx\,x^{\frac{d}{2}-1}\,K_{\Delta-\frac{d}{2}}(p_1p_3\,x)\,K_{\Delta-\frac{d}{2}}(p_2 p_4\,x)\,K_{0}(st\,x).\label{Sol}
\end{align}
i.e. as a 3K integrals of quadratic $(p_1 p_3, s t, p_2 p_4)$ variables, which are solutions of a system of the form \eqref{diff} with quartic ratios $x, y$.

\subsection{Comparison between the general, the intermediate and the reduced systems}

To address the asymptotic behaviour of this solution of the general system of Eqs. 
\eqref{C2}\eqref{Eq2}\eqref{C1} (the $S_i$ constraints) and compare it with the intermediate 
\eqref{CC4} and the reduced \eqref{CCC4} ones, we clearly need to perform a special asymptotic limit. We can reasonably assume that at large $s$ and $t$ the general solution of the $S_{i}'s$ equations decays as $\sim 1/(s t)^\alpha$, with $\alpha > 0$. \\
 Both for the $S_i$ and for the intermediate system \eqref{CC4}, the action of the derivative operators $(1/s) \partial/\partial s$ and $(1/t) \partial/\partial t$ is suppressed by two additional powers of the kinematic invariant $s$ and $t$ and can reasonably be set to zero asymptotically.\\
 If we neglect such contributions, the equations in \eqref{CC4} turn again into a homogeneous system  \eqref{fform} which, however, is not hypergeometric any longer, nor the third equation is dependent from the previous two, as found in the $\Delta_{ijkl}=0$ case for the reduced system \eqref{CCC4}.  Although the three systems, general intermediate and reduced, look pretty similar in such limit, we can only safely state that their solutions have to share the same asymptotic behaviour. This is fixed by the scaling power 
 $n_s=\Delta_t -3/4 d $, extracted from the dilatation WI in the form  
\begin{equation}
\Phi(p_1,p_2,p_3,p_4)\sim\frac{1}{(s^2 t^2)^{- n_s}} + O(1/(s^2 t^2))
\end{equation}
which requires that $n_s$ be negative. \\
In the two sections below we will try to characterize the behaviour of the dcc solution of \eqref{CCC4} in various limits before coming back again to the three systems of equations, discussing some approximate factorised solutions of such equations.

\subsection{IR and equal mass limits of the dcc solutions}
The analysis of the infrared or soft limits at small $s$ and $t$ of the dual conformal solution, with $\Delta_{ijkl}=0$, for $\Delta_i=\Delta, i=1,2,3,4$ can be discussed using a second version of the solution given by 
\eqref{fform}, but completely equivalent to it, obtained by a sequence of analytic continuations  \cite{Maglio:2019grh}

\begin{equation}
\resizebox{0.6\hsize}{!}{$
\begin{aligned}
\Phi&=C_1\bigg\{\left(p_1^2\,p_3^2\right)^{\Delta-\frac{3}{4}d}\bigg[F_4\left(\frac{d}{4}\,,\,\frac{3}{4}d-\Delta\,,\,1\,,\,\frac{d}{2}-\Delta+1\,;\frac{s^2t^2}{p_1^2p_3^2}\,,\,\frac{p_2^2p_4^2}{p_1^2p_3^2}\right)\\[1.2ex]
&\hspace{-0.3cm}+\tau_1\left(\frac{p_2^2p_4^2}{p_1^2p_3^2}\right)^{\Delta-\frac{d}{2}} F_4\left(\Delta-\frac{d}{4}\,,\,\frac{d}{4}\, ,\,1\,,\,1-\frac{d}{2}+\Delta\,;\frac{s^2t^2}{p_1^2p_3^2}\,,\,\frac{p_2^2p_4^2}{p_1^2p_3^2}\right)\bigg]\\[1.2ex]
&+\left(p_2^2\,p_3^2\right)^{\Delta-\frac{3}{4}d}\bigg[F_4\left(\frac{d}{4}\,,\,\frac{3}{4}d-\Delta\, ,\,1\,,\,\frac{d}{2}-\Delta+1\,;\frac{s^2u^2}{p_2^2p_3^2}\,,\,\frac{p_1^2p_4^2}{p_2^2p_3^2}\right)\\[1.2ex]
&\hspace{-0.3cm}+\tau_1 \left(\frac{p_1^2p_4^2}{p_2^2p_3^2}\right)^{\Delta-\frac{d}{2}} F_4\left(\Delta-\frac{d}{4}\,,\,\frac{d}{4}\, ,\,1\,,\,1-\frac{d}{2}+\Delta\,;\frac{s^2u^2}{p_2^2p_3^2}\,,\,\frac{p_1^2p_4^2}{p_2^2p_3^2}\right)\bigg]\\[1.2ex]
&+\left(p_1^2\,p_2^2\right)^{\Delta-\frac{3}{4}d}\bigg[F_4\left(\frac{d}{4}\,,\,\frac{3}{4}d-\Delta\, ,\,1\,,\,\frac{d}{2}-\Delta+1\,;\frac{u^2t^2}{p_1^2p_2^2}\,,\,\frac{p_3^2p_4^2}{p_1^2p_2^2}\right)\\[1.2ex]
&\hspace{-0.3cm}+\tau_1 \left(\frac{p_3^2 p_4^2}{p_1^2 p_2^2}\right)^{\Delta -\frac{d}{2}}F_4\left(\Delta-\frac{d}{4}\,,\,\frac{d}{4}\, ,\,1\,,\,1-\frac{d}{2}+\Delta\,;\frac{u^2t^2}{p_1^2p_2^2}\,,\,\frac{p_3^2p_4^2}{p_1^2p_2^2}\right)\bigg]\bigg\}.
\end{aligned}$}
\label{twoo}
\end{equation}

\begin{equation}
\tau_1=\frac{\Gamma\left(\Delta-\frac{d}{4}\right)\Gamma\left(1+\Delta-\frac{3}{4}d\right)\Gamma\left(1-\Delta+\frac{d}{2}\right)}{\Gamma\left(\Delta-\frac{3}{4}d\right)\Gamma\left(1-\Delta+\frac{3}{4}d\right)\Gamma\left(1+\Delta-\frac{d}{2}\right)}
\end{equation}
From this expression, we can keep $t^2$ fixed and of order $O(p_i^2)$ and send $s^2\to 0 $ to derive the soft behaviour 
\begin{equation}
\Phi\sim (p_1^2 p_3^2)^{\Delta - \frac{3}{4}d} + O(s^2/p_i^2)
\end{equation}
if the external mass invariants $p_i^2$ are kept larger than the invariant $s^2$. 
\subsection{Equal mass limit with $p_i^2=M^2 >\, s^2,t^2$ }
The equal mass limit is obtained by taking $p_i^2=M^2$ for all the external invariants. In this case, using the relation between $F_4$ and the Gauss hypergeometric $F_{2 1}(a,b,c,x)$
\begin{equation}
F_4(\alpha,\beta,\gamma,\gamma',x,y)=\sum_{m=0}^{\infty}\frac{(\alpha)_m(\beta)_m}
{(\gamma)_m m!}F_{21}(\alpha + m,\beta +m,\gamma',y)x^m
\end{equation}
and 

\begin{equation} 
F_{21}(a,b,c,1)=\frac{\Gamma(c)\Gamma(c-a-b)}{\Gamma(c-a)\Gamma(c-b)},
\end{equation}
from \eqref{twoo} we then obtain the simplified expression 
\begin{equation}
\begin{split}
&\Phi=M^{4 \Delta-3d }\sum_{m=0}^\infty \frac{1}{m!} \Bigg(
\frac{\Gamma \left(\frac{d}{2}-\Delta +1\right) \Gamma
   \left(-\frac{d}{2}-2 m+1\right) \Gamma
   \left(\frac{d}{4}+m\right) \Gamma \left(\frac{3
   d}{4}-\Delta +m\right)}{ \Gamma
   \left(\frac{d}{4}\right) \Gamma (m+1) \Gamma
   \left(\frac{3 d}{4}-\Delta \right) \Gamma
   \left(-\frac{d}{4}-m+1\right) \Gamma
   \left(\frac{d}{4}-\Delta -m+1\right)} \nonumber \\
&\frac{\Gamma \left(\frac{d}{2}-\Delta +1\right) \Gamma
   \left(-\frac{3 d}{4}+\Delta +1\right) \Gamma
   \left(-\frac{d}{2}-2 m+1\right) \Gamma
   \left(\frac{d}{4}+m\right) \Gamma
   \left(-\frac{d}{4}+\Delta +m\right)}{\Gamma
   \left(\frac{d}{4}\right) \Gamma (m+1) \Gamma
   \left(\frac{3 d}{4}-\Delta +1\right) \Gamma
   \left(\delta -\frac{3 d}{4}\right) \Gamma
   \left(-\frac{d}{4}-m+1\right) \Gamma \left(-\frac{3
   d}{4}+\Delta -m+1\right)} \Bigg)\times \nn\\
 &  \times \Bigg(\left(\frac{s^2 t^2}{M^4}\right)^m + \left(\frac{s^2 u^2}{M^4}\right)^m + 
 \left(\frac{u^2 t^2}{M^4}\right)^m\Bigg)\\
   \end{split}
\end{equation}
which in $d=4$ becomes
\begin{equation}
\begin{split}
&\Phi=M^{4 \Delta-3d}\sum_{m=0}^\infty \frac{1 }{m!}  \left(\frac{\Gamma (-2 m-1) \Gamma (-\Delta
   +m+3)}{\Gamma (-m) \Gamma (-\Delta
   -m+2)}+\frac{\Gamma (3-\Delta ) \Gamma (\Delta -2)
   \Gamma (-2 m-1) \Gamma (\Delta +m-1)}{ \Gamma
   (4-\Delta ) \Gamma (\Delta -3) \Gamma (-m) \Gamma
   (\Delta -m-2)}\right)\times \\ \nonumber 
&\qquad \qquad    \times \Bigg(\left(\frac{s^2 t^2}{M^4}\right)^m + \left(\frac{s^2 u^2}{M^4}\right)^m + \left(\frac{u^2 t^2}{M^4}\right)^m\Bigg)\\
\end{split}
\end{equation}
and is convergent as far as $M^2 \gg s^2, t^2 $.  Explicitly (in $d=4$)

\begin{equation} 
\Phi =M^{4 \Delta-3d}\left[d_0 + d_1 (\frac{s^2 t^2}{M^4} + \frac{s^2 u^2}{M^4} + \frac{u^2 t^2}{M^4})+\ldots \right],
\end{equation}
where
\begin{equation}
d_0=\frac{1}{2} \left(\frac{\Gamma (\Delta -1)}{\Gamma
   (\Delta -2)}-\frac{\Gamma (3-\Delta )}{\Gamma
   (2-\Delta )}\right)
\end{equation}
and 
\begin{equation}
\begin{split}
& d_1=\frac{1}{12} \left(\frac{\Gamma (4-\Delta )}{\Gamma
   (1-\Delta )}-\frac{\Gamma (\Delta )}{\Gamma (\Delta
   -3)}\right).
\end{split}
\end{equation}
\subsection{The equal mass limit with $s^2, t^2 > M^2$} 
A similar limit can be performed starting from \eqref{fform}. We can take $s^2, t^2, u^2 > M^2$, which in \eqref{fform} takes to a univariate expression of $F_4$, $F_4(a,b,c,c';x,x)$. It can be expressed as a single series in $x$ using the relation 

\begin{equation}
F_4(a,b,c,c';x,x)= {}_4 F_{3}\left(a,b,\frac{c + c'}{2},\frac{c + c'-1}{2}| \,c,c', c + c'-1; 4 x\right)  
\end{equation}
due to Burchnall, as reported in \cite{Vidunas1}. \\
Setting $x_1=M^4/(s^2 t^2)\sim M^4/(s^2 u^2)\sim M^4/(u^2 t^2)$ and choosing, for instance, a scaling dimension of a scalar operator 
$\phi^2$ (with $\Delta=d-2$), in $d=4$ one obtains a simple expression for $\Phi$

\begin{equation}
\Phi=C\left(\frac{\log ^2\left(\frac{M^4}{s^2 t^2}\right)}{s^2
   t^2}+\frac{\log ^2\left(\frac{M^4}{s^2
   u^2}\right)}{s^2 u^2}+\frac{\log
   ^2\left(\frac{M^4}{t^2 u^2}\right)}{t^2
   u^2}+\frac{\pi ^2}{3 s^2 t^2}+\frac{\pi ^2}{3 s^2
   u^2}+\frac{\pi ^2}{3 t^2 u^2}\right) +O(x_1^2)
\end{equation}
and in $d=3$
\begin{equation}
\begin{split}
&\Phi=C\left(\frac{\pi  \Gamma \left(\frac{1}{4}\right)^2}{M^4
   \sqrt{s} \sqrt{t}}+\frac{\pi  \Gamma
   \left(\frac{1}{4}\right)^2}{M^4 \sqrt{s}
   \sqrt{u}}+\frac{\pi  \Gamma
   \left(\frac{1}{4}\right)^2}{M^4 \sqrt{t}
   \sqrt{u}}-\frac{4 \pi  \Gamma
   \left(\frac{3}{4}\right)^2}{M^2 s^{3/2}
   t^{3/2}}-\frac{4 \pi  \Gamma
   \left(\frac{3}{4}\right)^2}{M^2 s^{3/2}
   u^{3/2}}\right.\nn \\
 &\qquad \qquad  \left. -\frac{4 \pi  \Gamma
   \left(\frac{3}{4}\right)^2}{M^2 t^{3/2}
   u^{3/2}}+\frac{\pi  \Gamma
   \left(\frac{1}{4}\right)^2}{4 s^{5/2}
   t^{5/2}}+\frac{\pi  \Gamma
   \left(\frac{1}{4}\right)^2}{4 s^{5/2}
   u^{5/2}}+\frac{\pi  \Gamma
   \left(\frac{1}{4}\right)^2}{4 t^{5/2} u^{5/2}}\right) +O(x_1^2)
\end{split}
\end{equation} 
\section{Large $s$ and $t$ limits and the Lauricella system}
We have already mentioned that the system of Eqs. \eqref{C2}, \eqref{Eq2} \eqref{C1} reduces to \eqref{CC4} if we choose a combination of invariants given by \eqref{xy}. 
The system \eqref{CC4} turns into hypergeometric if $\Delta_{ijkl}=0$, with only two independent equations, as pointed out above. However, for a generic $\Delta_{ijkl}$ it is possible to uncover an approximate hypergeometric structure in the equations only in the large $s$ and $t$ limit, if we neglect the coupling between the $s, t$ and $p_i^2$ invariants. At the same time
we could assume that $\Delta_{ijkl}\ll1$, which allows to drop the $1/s\partial/\partial_s$ and  $1/t\partial/\partial_t$ terms in the differential operator. This approximate factorization has been discussed in \cite{Maglio:2019grh}, where it has been shown to take to a hypergeometric system of Lauricella type in three variables (see section \eqref{lauri}). This asymptotic analysis is based on the ans\"atz

\begin{equation}
\Phi(p_1^2,p_2^2,p_3^2,p_4^2,s, t)\sim \phi(p_1^2,p_2^2,p_3^2,p_4^2)\chi(s,t)
\label{factor}
\end{equation}
and invokes the separability of the asymptotic system \eqref{CC4} 
\begin{equation}
\label{lau}
\begin{split}
&K_{2 4} \phi=0 \\
& K_{1 3}\phi=0 \\
& K_{1 4}\phi=0\\ 
&\frac{1}{st}\frac{\partial^2}{\partial s\partial t}\,\chi(s,t)=0.\\
\end{split}
\end{equation}
The Lauricella system corresponds to the first three equations of \eqref{lau}. 
Lauricella systems have recently appeared also in CFT in coordinate space \cite{Chen:2019gka}.
One can easily realize that they characterize a homogenous solution in the variables $p_i^2$ of the entire (complete) system \eqref{C2}, \eqref{Eq2}\eqref{C1} as well as of \eqref{CC4}. They are exact solutions of such systems, before an asymptotic limit. Notice that logarithmic terms of the form  $f(1/(s^2 t^2)) \log^k(s^2/t^2)$ (k>0) are also compatible with the asymptotic structure of such systems, which are generically expected for scattering at fixed angle in perturbation theory (see \cite{Cheng:1982xq} for an example). 
 
\subsection{The general primary CWIs in the $2\to 2$  and $1\to 3$ formulations for the $TOOO$ and asymptotics} 
\label{gensection}
In order to get further insight into the structure of the CWIs for the $TOOO$, we proceed with a rearrangement of their expressions in order to reduce them to homogeneous equations, following the same approach of section \ref{asimpt}, adopted in the scalar case. We will proceed by generalizing the CWIs for different scalar coefficients from the equal scaling case presented in section \ref{primd}.
For simplicity we set 
\begin{equation}
F\equiv F(p_1,p_2,p_3,p_4,s, t), \,\,\, F(p_2\leftrightarrow p_4)\equiv F(p_1,p_2,p_4,p_3, t ,s),
\,\,\, F(p_3\leftrightarrow p_4)\equiv F(p_1,p_4,p_2,p_3, s,\tilde{u}).
\end{equation}
If we allow for different scaling $\Delta_i$, with $\Delta_1=d$ for the stress-energy tensor, then the equations given in \eqref{c11}-\eqref{c31} can be generalized as follows
\begin{equation}
\begin{split}
 \tilde{C}_{1 1} -\tilde{C}_{2 1}\to B_1=& \Bigg( K_{21} +\frac{\Delta_{1423}+2}{t}\frac{\partial}{\partial t} -\frac{1}{t}\frac{\partial}{\partial t}\left( p_1\frac{\partial}{\partial p_1}+
p_4\frac{\partial}{\partial p_4} - p_2\frac{\partial}{\partial p_2} - p_3\frac{\partial}{\partial p_3}\right) \\&+ \frac{p_3^2-p_4^2}{s t}\frac{\partial^2}{\partial s \partial t}\Bigg)F(1,2,3,4)=0.
\end{split}
\end{equation}

The other homogeneous equations for $F(p_1,p_2,p_3,p_4,s, t)$ are similarly derived in the form
\begin{align}
 \tilde{C}_{1 1} -\tilde{C}_{3 1}\to B_2=& \Bigg( K_{31} +\frac{\Delta_{1423}+2}{t}\frac{\partial}{\partial t}+\frac{\Delta_{1234}-2}{s}\frac{\partial}{\partial s} -\frac{1}{t}\frac{\partial}{\partial t}\left( p_1\frac{\partial}{\partial p_1}+
p_4\frac{\partial}{\partial p_4} - p_2\frac{\partial}{\partial p_2} - p_3\frac{\partial}{\partial p_3}\right)\nonumber\\
&+\frac{1}{s}\frac{\partial}{\partial s}\left( p_3\frac{\partial}{\partial p_3}+
p_4\frac{\partial}{\partial p_4} - p_1\frac{\partial}{\partial p_1} - p_2\frac{\partial}{\partial p_2}\right) + \frac{p_3^2-p_1^2}{s t}\frac{\partial^2}{\partial s \partial t}\Bigg)F(1,2,3,4)=0, 
\end{align}
and
\begin{align}
\tilde{C}_{2 1} -\tilde{C}_{3 1}\to B_3=& \Bigg( K_{32} +\frac{\Delta_{1234}-2}{s}\frac{\partial}{\partial s} +\frac{1}{s}\frac{\partial}{\partial s}\left( p_3\frac{\partial}{\partial p_3}+
p_4\frac{\partial}{\partial p_4} - p_1\frac{\partial}{\partial p_1} - p_2\frac{\partial}{\partial p_2}\right) \\& +\frac{p_4^2-p_1^2}{s t}\frac{\partial^2}{\partial s \partial t}\Bigg)F(1,2,3,4)=0.
\end{align}
One can show that $B_1,B_2,B_3$ are not independent, in fact
\begin{equation}
B_1+B_3=B_2,
\end{equation}
indicating that there are only two independent homogeneous equations involving the $F$ form factor. \\
Finally, one has to consider the system of three differential equations, composed of ($B_1,B_2)$ together with the analogous of $\tilde{C}_{21}$, given in \eqref{c21}, now for different $\Delta_i$'s, which can be written as
\begin{equation}
\left\{
\begin{split}
&\Bigg( K_{31} -\frac{1}{t}\frac{\partial}{\partial t}\left( p_1\frac{\partial}{\partial p_1}+
p_4\frac{\partial}{\partial p_4} - p_2\frac{\partial}{\partial p_2} - p_3\frac{\partial}{\partial p_3}\right)+\frac{1}{s}\frac{\partial}{\partial s}\left( p_3\frac{\partial}{\partial p_3}+
p_4\frac{\partial}{\partial p_4} - p_1\frac{\partial}{\partial p_1} - p_2\frac{\partial}{\partial p_2}\right)  \notag \\& + \frac{p_3^2-p_1^2}{s t}\frac{\partial^2}{\partial s \partial t}\Bigg)F=-\left(\frac{\Delta_{1423}+2}{t}\frac{\partial}{\partial t}+\frac{\Delta_{1234}-2}{s}\frac{\partial}{\partial s}\right)F\\[2ex]
&\Bigg[K_{42}+\frac{1}{s}\frac{\partial}{\partial s}\left( p_3\frac{\partial}{\partial p_3}+
p_4\frac{\partial}{\partial p_4} - p_1\frac{\partial}{\partial p_1} - p_2\frac{\partial}{\partial p_2}\right)+\frac{1}{t}\frac{\partial}{\partial t}\left( p_1\frac{\partial}{\partial p_1}+
p_4\frac{\partial}{\partial p_4} - p_2\frac{\partial}{\partial p_2} - p_3\frac{\partial}{\partial p_3}\right)\notag \\&+\frac{p_4^2-p_2^2}{st}\frac{\partial^2}{\partial s \partial t}\Bigg]F 
=\left(\frac{\Delta_{1423}+2}{t}\frac{\partial}{\partial t}-\frac{\Delta_{1234}}{s}\frac{\partial}{\partial s}\right)F-\frac{2}{s}\frac{\partial }{\partial s}\big(F(p_2\leftrightarrow p_4)-F(p_3\leftrightarrow p_4)\big)\\[2ex]
&\left( K_{21}  -\frac{1}{t}\frac{\partial}{\partial t}\left( p_1\frac{\partial}{\partial p_1}+
p_4\frac{\partial}{\partial p_4} - p_2\frac{\partial}{\partial p_2} - p_3\frac{\partial}{\partial p_3}\right) + \frac{p_3^2-p_4^2}{s t}\frac{\partial^2}{\partial s \partial t}\right)F=-\frac{\Delta_{1423}+2}{t}\frac{\partial}{\partial t}F
\end{split}
\right.\label{SystemOfHypergeometic}
\end{equation}
We will try to extract some information about the structure of such equations by discussing some possible limits.\\
In the characterization of the nature of the system we begin by considering the case in which all the scalings are different and work our way starting from the left hand side of 
\eqref{SystemOfHypergeometic}. We have different options. For instance, if we are looking for factorised solutions such as those discussed in the scalar case, of the form \eqref{factor}, 
then we could consider the asymptotic limit  $s,t\to\infty$ and identify the Lauricella component  of such solutions, since the equations above turn homogenous, and the left hand side, exactly as in \eqref{lau}, reduces to a Lauricella system of hypergeometrics \eqref{lauri}. This holds independently of the values of the scalings $\Delta_i$.
On the other hand, it is possible to identify, at least asymptotically, some hypergeometric solutions, different from the Lauricella's, but we need to constrain the scaling dimensions in such a way that the operators $K_{31}$ and $K_{42}$ are each characterised by a single conformal scaling  ($\Delta_4=\Delta_2$ and $\Delta_3=\Delta_1$). As discussed in the previous sections, we can choose as variable the scale invariant ratios \eqref{xy} for $x$ and $y$ in the ans\"atz for the solution, reobtaining the same left hand side of \eqref{CC4}. This approximate solutions would again be quite similar to those discussed in the scalar case.  
However, in the general case, as we have already mentioned, even for large $s$ and $t$, when we keep the scalings generic, one can show that the left hand side of such system of equations is not of hypergeometric form, and the explicit form of such solutions is unknown.

\begin{figure}[t]
	\centering
	\raisebox{-0.5\height}{\includegraphics[scale=0.8]{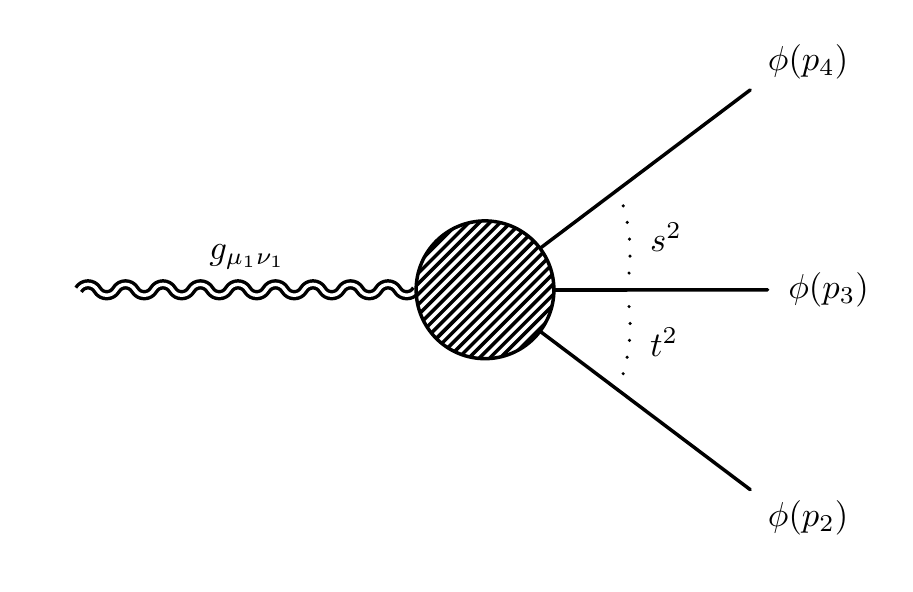}} \hspace{2cm}
	
	\caption{The $TOOO$ in a kinematical region in which can be described as a 
		$1\to 3$ process. }
	\label{nm1}
\end{figure} 
\subsection{The $1\to 3$ case} 
It is possible to perform other limits on the same form factor of the $TOOO$ in order to simplify the primary CWIs presented in the previous sections. We are going to focus our discussion on the $1\to 3$ formulation, which is symmetric in the momenta of the three scalar operators and provides a clear separation of the parametric dependence of the correlator in terms of a function of the external invariants $p_i^2$ times a function of $s,t$ and $u$, in analogy with the discussion presented in the $2\to 2$ case. \\
This kinematic choice is illustrated in Fig. 1.\\
In order to proceed with the investigation of this limit, it is convenient to perform an analytic continuation of the CWIs to the Minkowski region from their Euclidean definition, and take  all the invariants $t^2$ and $u^2$ and $s^2$ to be positive.  The kinematical region of interest, in this case, is delimited by the conditions
\begin{align}
& (p_2 + p_3)^2\leq t^2 \leq \left( p_1 - p_4\right)^2\nonumber \\
& (p_3 + p_4)^2\leq s^2 \leq \left( p_1 - p_2\right)^2 \nonumber\\
& (p_2+p_4)^2 \leq u^2 \leq \left(p_1-p_3\right)^2, 
\end{align}
with the usual relation
\begin{align}
& s^2 + t^2 + u^2= p_1^2 + p_2^2 + p_3^2 + p_4^2. 
\end{align}
We will be performing the large $p_1$ limit, where the invariant mass of the 
virtual graviton line gets asymptotically large, and assume that the invariants $s^2\sim t^2\sim u^2\sim p_1^2$ grow large with $p_1^2$. In this limit the primary CWIs simplify, and the equations become approximately separable in their dependence on the external $p_i^2 (i=2,3,4)$ and the remaining $(s, t, u)$ invariants. For this reason we choose asymptotic solutions of the form 
\begin{equation}
A(p_2,p_3,p_4,s,t,u)\sim \Phi(p_2,p_3,p_4)\chi(s,t,u).
\end{equation}
We study now the form of the $\chi(s,t,u)$. The corresponding equations for the $(s,t,u)$ invariants, from the primary conformal WI's, take the form
\begin{equation}
\frac{\partial^2 \chi}{\partial s\partial t } =0, \qquad  \frac{\partial^2 \chi}{\partial s \partial u } =0, \qquad 
\frac{\partial^2 \chi}{\partial t \partial u } =0,\label{mixed0}
\end{equation}
with the additional constraint imposed by the dilatation Ward identity. In particular, in this limit, the dilatation WI for the $(s,t,u)$ invariants takes the form
\begin{align}
\left[s\,\frac{\partial}{\partial s}+t\,\frac{\partial}{\partial t}+u\,\frac{\partial}{\partial u}\right]\chi(s,t,u)=0.\label{dilstu}
\end{align}
Notice that the remaining contribution to the dilatation WI is satisfied separately by the scale invariant condition on $\Phi(p_2,p_3,p_4)$
\begin{equation}
\left[p_2\,\frac{\partial}{\partial p_2}+p_3\,\frac{\partial}{\partial p_3}+p_4\,\frac{\partial}{\partial p_4}\right]\Phi(p_2,p_3,p_4)=(\Delta_t -3 d -2)\Phi(p_2,p_3,p_4),
\end{equation}
which takes to generalized hypergeometric $F_4$ solutions, functions of the ratios $p_2^2/p_4^2$ and $p_3^2/p_4^2$, as given in \eqref{compact}. The choice of the pivot ($p_4$ in this case) is arbitrary.\\
By differentiating \eqref{dilstu} with respect $s$ and using \eqref{mixed0}, one finds another constraint. Similar constraints are obtained by repeating the procedure with respect to $t$ and $u$. The resulting three equations obtained in this manner can be written in the form
\begin{align}
\left[s\frac{\partial^2}{\partial \,s^2}+\,\frac{\partial}{\partial s}\right]\chi(s,t,u)&=0\notag\\
\left[\,t\,\frac{\partial^2}{\partial\,t^2}+\,\frac{\partial}{\partial t}\right]\chi(s,t,u)&=0\notag\\
\left[u\frac{\partial^2}{\partial u^2}+\frac{\partial}{\partial u}\right]\chi(s,t,u)&=0,
\end{align}
giving the solution for $\chi$ of the form
\begin{align}
\chi(s,t,u)=c_1\log(s)+c_2\log(t)+c_3\log(u)+c_4,
\end{align}
where $c_1,\,c_2,\,c_3,\,c_4$ are undetermined constants. Imposing the dilatation WI on this solution we find some relations between the undetermined coefficients, with the solution rewritten in the form
\begin{equation}
\chi(s,t,u)=c_1\log\left(\frac{s}{t}\right)+c_2\log\left(\frac{u}{t}\right)+c_4.
\end{equation}
Finally, we also impose the symmetry constraint on the form factor $A$ of the $TOOO$
\begin{equation}
A(p_2,p_4,p_3,s,u,t)=A(p_2,p_3,p_4,s,t,u)
\end{equation}
which implies that 
\begin{align}
\phi(p_2,p_3,p_4)\chi(s,t,u)=\phi(p_2,p_4,p_3)\chi(s,u,t),
\end{align}
and recalling that the $\phi(p_2,p_3,p_4)$ is symmetric under the permutation of $\{p_2,p_3,p_4\}$, we obtain the condition $\chi(s,t,u)=\chi(s,u,t)$ or
\begin{align}
\big(2c_2+c_1\big)\log\left(\frac{u}{t}\right)=0.
\end{align}
Therefore the $\chi(s,t,u)$ function acquires the final form
\begin{equation}
\chi(s,t,u)=c_1\log\left(\frac{u\,t}{s^2}\right)+c_4.
\end{equation}
As we have seen from the last and the previous cursory analysis of such systems, it is possible to identify an approximate behaviour of such solutions, in one specific asymptotic limit in which 
the invariant $s$ and $t$ get large and of the same size. \\
In this approximate analysis the only exact statement that one can make is that Lauricella functions are indeed special 
solutions of such equations, and correspond to particular solutions of such inhomogenous systems. \\
We have been careful to rewrite all the CWIs for generic scalings $\Delta_i$, in such as way that the left hand sides of thse systems carry a close resemblance to those of 3-point functions, except for an extra term proportional to a double derivative in $s$ and $t$,  
$\sim 1/(s t)\partial^2/(\partial s\partial t)$, which is new for 4-point functions and absent in 3-point functions. \\
As we have stressed in the previous sections in the case of dcc solutions, this term does preserve the hypergeometric structure of the corresponding equations, although such solutions have little in common with those derived for genuine 3-point functions, for being quartic -rather than quadratic - ratios of momenta.  \\
The discovery of such solutions may not be accidental in the context of CFT's, since in ordinary perturbation theory similar dependences have been uncovered in the analysis of ladder diagrams \cite{Usyukina:1992jd}. However, one can easily check, following the discussion in \cite{Maglio:2019grh}, that box-like master integrals with propagators raised to generic powers, cannot be special cases of such dcc solutions, except for the ordinary box diagram. On general grounds, one expects that the simplified CWIs, which
are found in the scalar case for the dcc solutions, are related to an underlying Yangian symmetry \cite{Loebbert:2016cdm}, which is manifesting here in a bosonic, non supersymmetric, context.  
In the $TOOO$ such a symmetry, differently from the scalar case, is violated by the presence of a single stress-energy tensor. It could be restored in tensor correlators characterised by a single primary operator, such as the $JJJJ$ or the $TTTT$. 
\section{Conclusions}
The investigation of the CWIs of four point functions of a generic CFT in momentum 
space in $d >2$ is a new challenging domain of research,  with the possibility of establishing a direct connection with the analysis of scattering amplitudes in Lagrangian field theories. As in the case of lower point functions, 
one could envision several areas where such studies could find direct physical applications, from cosmology to condensed matter theory \cite{Chernodub:2017jcp,Chernodub:2019tsx}, due to the interplay, in the latter case, of quantum anomalies in transport phenomena.     
These studies need to be accompanied by investigations of the operator product expansion in the same variables, in order to develop a bootstrap program, as in coordinate space. \\
Obviously, while in coordinate space the operatorial expansion is well-behaved at separate spacetime points, in momentum space we gather information on such operators from all the spacetime regions, including those in which the external coordinates of a correlator coalesce. This makes the analysis in momentum space more demanding, and we have to worry about anomalies and address the issue of how to regulate a given theory. \\
It is then natural to advance our knowledge in this area starting from the analysis of simpler correlation functions, the scalar and the tensor/scalar cases being the first on the list.\\
For this reason we have derived the CWIs for the $TOOO$ and discussed their relation to those obtained in the case of 4 scalars. In both cases we have discussed their expressions in various limits, showing the hypergeometric character of the asymptotic solutions, if certain constraints on the scaling dimensions are respected. \\
While, obviously, we 
do not expect that a given correlator can be uniquely identified by these equations, neverthless they constrain quite significantly the structure of the possible solutions. \\
As mentioned, in our analysis we have concentrated on the structure of the equations in several kinematical limits, in order to gather some information about the behaviour of the corresponding solutions. 
In such limits, the differential operators take a simplified but a still nontrivial form. \\
The comparison between the $TOOO$ and scalar cases, allows to uncover some common features of the systems of equations that they need to satisfy.  In this context, of particular significance are those solutions which are dual conformal and conformal at the same time (or dcc solutions), which take a unique expression. Several different ans\"atze  take to the same hypergeometric form of such solutions, which are related by analytic continuations, and, as we have shown, turn useful for their study in specific kinematical limits. For such a 
reason they play a strategic role, since they can be used to investigate the behaviour of scalar 4-point functions in a rather direct way and allow to underscore some similarities between the CWIs both in the tensor and in the scalar contexts. \\
Specific features of such dcc solutions, extracted in several asymptotic limits, are expected to provide some indication on the behaviour of the more general (and unknown) solutions of the equations satisfied by scalar operators - the $OOOO$ for instance - for generic scaling dimensions of the primaries $O$. Both correlators are characterised by a single form factors, allowing particular solutions of Lauricella type. This suggests the presence of a more general underlying hypergeometric structure in such systems of equations. It could be of interest to investigate from a purely mathematical point of view the structure such equations in order to classify the structure of such solutions.\\
Our investigations can be extended in several directions, for instance to the study of the renormalization of the corresponding form factors, which requires a separate investigation, as in the case of 3-point functions \cite{Bzowski:2015pba}. There are also other and quite direct implications of our results and equations for the analysis of the decomposition of such correlators in terms of CP-symmetric (Polyakov) blocks. Indeed the CWIs that we have derived can be applied to constrain the block decomposition \cite{Isono:2018rrb}. 
\vspace{1cm}
\chapter{The $\langle TTTT \rangle $ correlator: Renormalization and the Conformal Anomaly Action  to Fourth Order  in $d=4$}\label{ttttchap}
\section{Introduction}
In this chapter, we will demonstrate an analysis of the $\langle TTTT \rangle $ correlator in momentum space. It is based on the work \cite{tttt}. 
As we move from 3- to 4-point functions, the solutions of the CWIs are affected by the appearance of arbitrary functions, which is a general feature of CFT,  and this approach is not of immediate help, in the sense that even if such bilinear mixings would be found in the perturbative description of a certain correlator, they would not find an immediate counterpart in the general solution of the conformal constraints. \\
Obviously, we would like to provide a proof of such behaviour with no reference to free field theory. 
We recall that the CWIs impose, at any $n$, hierarchical relations connecting $n$ to $n-1$ point functions, which we are going to investigate in great detail in momentum space. While these relations are not sufficient, for a generic CFT, to reconstruct the expression of $\sm(g)$ beyond $O(\delta g^3)$ (i.e. $n=3$), they nevertheless  constrain significantly its structure after renormalization.\\
 As we are going to show, the CWIs predict a well-defined structure for the anomaly contribution to $\sm_A$ at the level of the 4T, once we assume that their renormalization proceeds via the counterterm action \eqref{counter}. As already mentioned, this analysis does not require a complete classification of all the form factors which appear in the correlator, but just a careful study of the structure of the CWIs satisfied by it and by the 3T.\\
 
 \subsection{Bilinear mixings}
Bilinear mixings are, even in ordinary field theory, the signature that the functional expansion of the effective action is taking place in a nontrivial vacuum, as for the Higgs mechanism. \\
 In an ordinary gauge theory such mixings are removed by a suitable gauge choice, such as the `t Hooft or the unitary gauge. In the conformal case, a massless pole represents a virtual (nonlocal) interaction, which is directly coupled to the anomaly and defines a skeleton expansion which can be taken as a possible definition the anomaly action.  \\
 As we are going to elaborate in a following section, such terms correspond to an expansion of the same action - in coordinate space and in the flat limit- in the dimensionless variable $R\,\square^{-1}$. Here, $R$ is the scalar curvature, which has very often appeared in the analysis of nonlocal cosmologies, such as in 
 $f(R\,\square^{-1})$ models \cite{Deser:2007jk}. In this case, the goal has been that of explaining the late-time dark energy dominance of our universe. \\
These analysis have been limited in the past to 3-point functions \cite{Bzowski:2020kfw,Coriano:2019nkw,Maglio:2019grh}. Four-point functions have received attention only more recently \cite{Coriano:2019nkw,Serino:2020pyu}.
For 3-point functions, the analysis of conformal and non-conformal correlators in realistic 
theories such as QED and QCD, at one-loop, has covered the $TJJ$ and the $TTT$ (3T)\cite{Giannotti:2008cv,Coriano:2018bsy1,Coriano:2018bbe,Armillis:2009pq}.\\
 In this chapter we are going to show that this feature is generic. No complete solution of the CWIs at quartic level ($O(h^4)$ in flat space, with $ h$ the metric fluctuation) is needed in order to extract such behaviour, at least in flat space, once the contribution of the anomaly is correctly taken into account.  \\
Notice that the advantage of solving explicitly the CWIs, as in the case of conformal 3-point functions,  beside its indisputable value,  is that it gives the opportunity to apply and verify the renormalization procedure on the solution, using the known gravitational counterterms, as given in Eq. \eqref{counter} below. 
However, for the rest, the method becomes increasingly prohibitive if the goal is to infer the 
 all-order behaviour of the anomaly action.\\
  The purpose of the chpater is to show that a certain amount of information is hidden in the structure of the WIs, if we assume that they can be renormalized using the standard (known) counterterms typical of the anomaly functional. We work our way from this assumption backward, in order to characterize the structure of the CWIs and identify the implications for the structure of the anomaly action to a given order in $h^{\mu\nu}$ $(h)$, the fluctuation of the metric around its flat spacetime limit. \\
Our approach is, in a way, quite direct, and can be extended to n-point functions and to any spacetime dimension, following the same strategy. Since we  intend to characterize the method in its simplest formulation, we will focus in $d=4$ spacetime dimensions, and we will rely on an expansion of the anomaly functional around a flat background in Dimensional Regularization (DR).
\subsection{Nonlocal interactions in the 3T}
\begin{figure}[t]
	\centering
	{\includegraphics[scale=0.6]{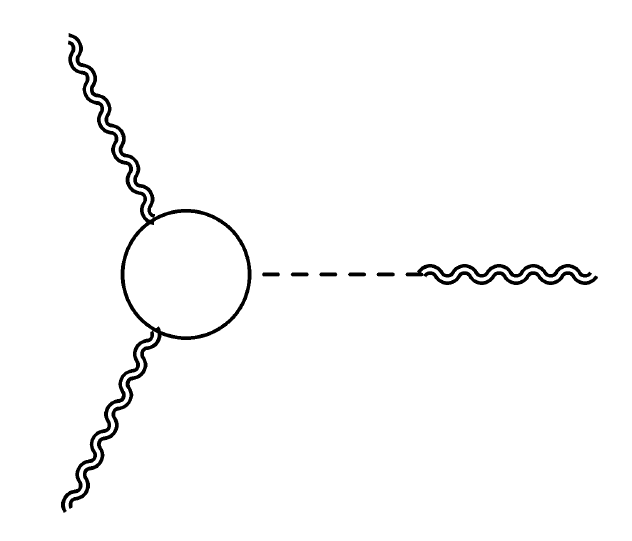}} \hspace{0.2cm}
	{\includegraphics[scale=0.6]{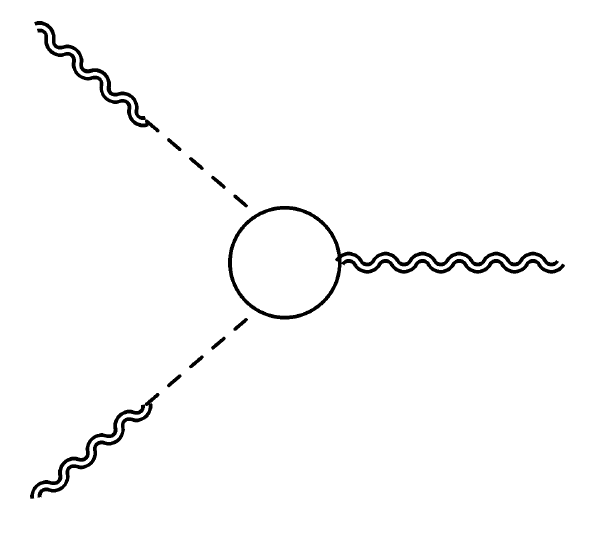}}\hspace{0.2cm}
	{\includegraphics[scale=0.6]{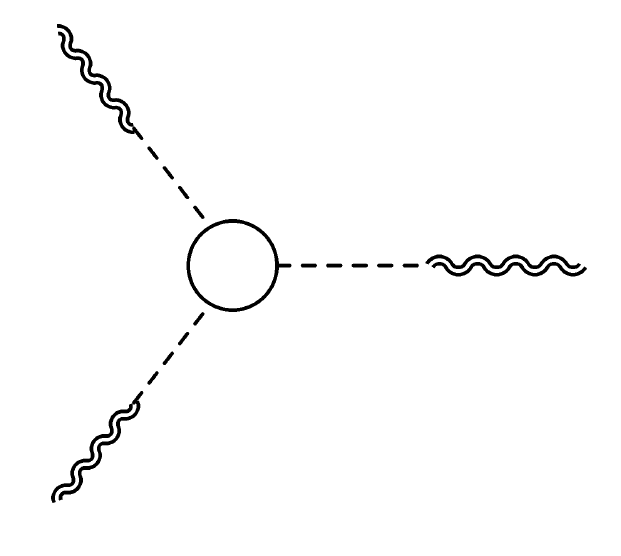}}
	\caption{Expansion of the anomaly contributions to the renormalized vertex for the 3T.\label{FeynTTTX}}
\end{figure}
In the case of the 3T all the contributions to the trace anomaly of the renormalized vertex  can be summarised by the three diagrams shown in \figref{FeynTTTX} and will be discussed below. \\ 
In this figure, the dashed lines indicate the inclusion of an operator
\begin{equation}
\hat{\pi}^{\mu\nu} =g^{\mu\nu}p^2  - p^\mu p^\nu
\end{equation}
on the external lines, which projects on the subspace transverse to a given momentum $p$. The projector is accompanied by a pole. Together,  $\hat{\pi}$ and the pole define the ordinary transverse projector 
\begin{equation}
\label{pole}
\pi^{\mu\nu}=\hat{\pi}^{\mu\nu}\frac{1}{p^2} 
\end{equation}
inserted on all the external lines in one, two and three copies. 
Poles in separate variables $1/p_i^2$,  $1/(p_i^2 p_j^2) (i\neq j)$ and $1/(p_1^2 p_2^2 p_3^2)$, are connected to separate external graviton lines, and each momentum invariant appears as a single pole. \\
This result had been obtained in \cite{Coriano:2018bsy1} by performing a complete perturbative analysis of the same vertex using free field theory realizations. In this case, the inclusion of 3 sectors, a scalar, a  fermion and a spin 1, allows to obtain the most general expression of this correlator, and its renormalization, performed by the addition of the general counterterm \eqref{counter}, has been verified.   \\
We will show that the Weyl-variant contribution to $\sm_A$  can be identified directly from the CWIs, under the assumption that  $\sm_{ct}$, as defined in \eqref{counter}, is all that is needed in order to proceed with the renormalization of a 3- or a 4-point function of correlators of stress energy tensors in $d=4$, in the flat spacetime limit.\\
If we move to coordinate space from momentum space 
and attach the metric fluctuations $h^{\mu\nu}$ on the external lines, it is easy to show that the inclusion of a projector of the form \eqref{pole} induces a bilinear mixing between the graviton and a virtual scalar propagating in the intermediate virtual (dashed) line. The interaction takes the form  
\begin{equation}
\label{coupling}
\frac{1}{p^2} \hat\pi^{\mu\nu} \leftrightarrow R^{(1)}\frac{1}{\square}
\end{equation}
In the 3T case, as one can immediately figure out from \figref{FeynTTTX}, such expressions can be easily traced back to nonlocal terms in the anomaly action $\sm_A$ of the form
\begin{equation}
\mathcal{S}_{A}\sim \int d^4 x\, d^4 y R^{(1)}(x)  \left(\frac{1}{\square}\right)(x,y) \left( b'\, E_4^{(2)}(y) + b\, (C^2)^{(2)}(y)\right),
\end{equation}
where the labels $(1)$, $(2)$ refer to the first and second variations of the invariants $R$, the scalar curvature, $C^2$ and $E_4$.
\begin{figure}[t]
	\centering
	{\includegraphics[scale=0.6]{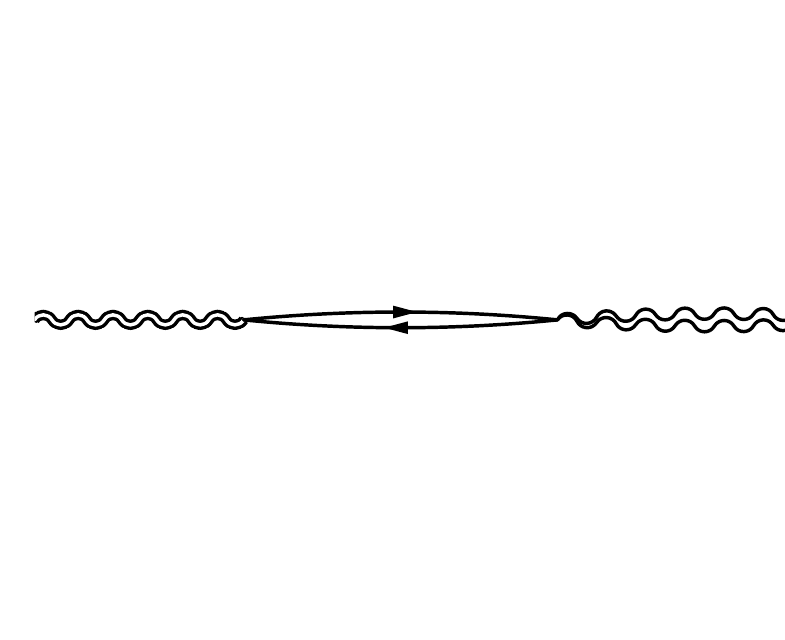}} \hspace{1cm}
	\raisebox{-1.5ex}{\includegraphics[scale=0.6]{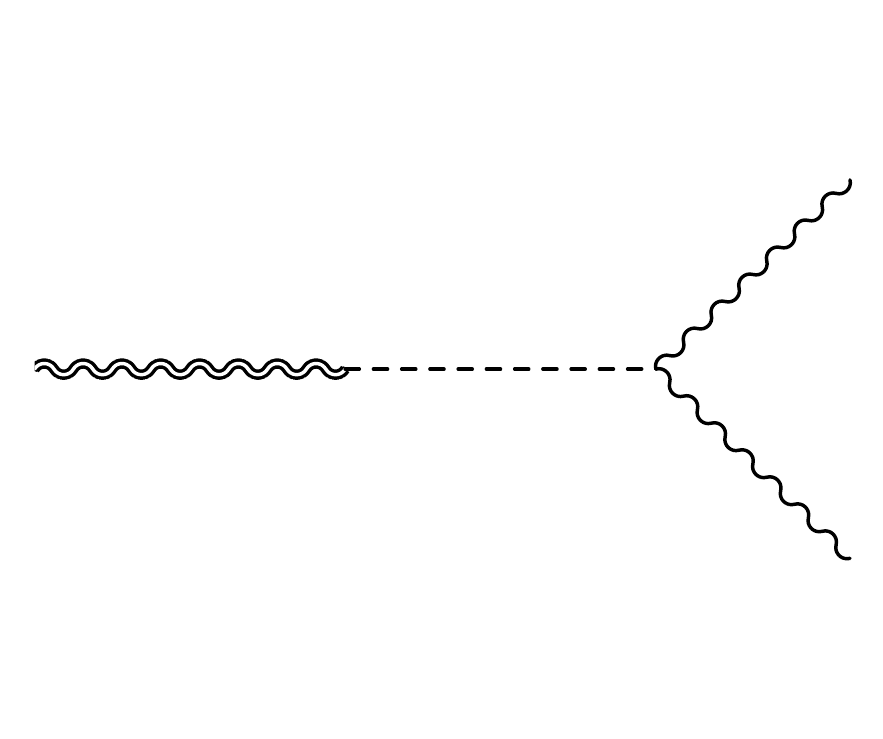}} 
	\caption{Expansion of the anomaly contributions to the renormalized vertex for the TJJ. Left: The collinear region in momentum space responsible for the origin of the pole. Right: Its interpretation as a scalar exchange. \label{FeynTJJ}}
\end{figure}
Analogously, in the case of the TJJ correlator in QED, the anomaly contribution, extracted by a complete perturbative analysis,  is shown in \figref{FeynTJJ} and takes the form
\begin{equation}
\mathcal{S}_{A}\sim \beta(e) \int d^4 x\, d^4 y \,R^{(1)}(x)  \left(\frac{1}{\square}\right)(x,y) F^{\mu\nu} F_{\mu\nu}(y),
\end{equation}
with $F^{\mu\nu}$ being the QED field strength and $\beta(e)$ the corresponding beta-function of the gauge coupling. Notice that in both examples the intermediate propagating virtual state is directly coupled to the anomaly. \\
This characterization of the anomaly action $\sm_A$ at cubic order, in each of the two cases, has been obtained by a complete analysis of the conformal constraints, using the explicit expression of their solutions, which for 3 point functions depends only on three constant for the 3T and on two for the TJJ. In both cases, the solutions, which are given by hypergeometric functions $F_4$, can be mapped to  general free field theory realizations, characterized by the inclusion at 1-loop level of an arbitrary number of scalars, fermions and spin-1 fields in the Feynman expansion. The map between the two approaches is an exact one in $d=3,5$.\\
In \cite{Coriano:2017mux} it has been shown that such nonzero trace contributions are automatically generated by a nonlocal conformal anomaly action of Riegert type \cite{Riegert:1984kt} expanded to first order in $\delta g$ and to second order in the external gauge field $A_{\mu}$, respectively. \\
As we move from 3- to 4-point functions, the solutions of the CWIs are affected by the appearance of arbitrary functions, which is a general feature of CFT,  and this approach is not of immediate help, in the sense that even if such bilinear mixings would be found in the perturbative description of a certain correlator, they would not find an immediate counterpart in the general solution of the conformal constraints. \\
We recall that the CWIs impose, at any $n$, hierarchical relations connecting $n$ to $n-1$ point functions, which we are going to investigate in great detail in momentum space. While these relations are not sufficient, for a generic CFT, to reconstruct the expression of $\sm(g)$ beyond $O(\delta g^3)$ (i.e. $n=3$), they nevertheless  constrain significantly its structure after renormalization.\\
The CWIs predict a well-defined structure for the anomaly contribution to $\sm_A$ at the level of the 4T, once we assume that their renormalization proceeds via the counterterm action \eqref{counter} as we will demonstrate. This analysis does not require a complete classification of all the form factors which appear in the correlator, but just a careful study of the structure of the CWIs satisfied by it and by the 3T.\\
For this reason, we are going to illustrate how this simplified procedure works first in the case of the 3T, before moving to the 4T in this new approach, showing how the structures of $\sm_A$ is constrained by the renormalized CWIs to assume a form very similar to the 3T case. \\
 As in previous studies in momentum space, we will rely on decompositions of the 3T and 4T  into a transverse traceless, a longitudinal, and a trace part, extending the approach formulated in \cite{Bzowski:2013sza} for 3-point functions to 4-point functions.

\subsection{Contents of the chapter}
This chapter is organised as follows. We will define our conventions and characterise the main features of the anomaly action  highlighting all the simplifications that are present when we consider the flat spacetime limit in the definition of the correlation functions. A complementary and more detailed  analysis of \ref{renorm} can be found in chapter \ref{anomalyaction}.There the discussion, is carried out in full generality in regard to the choice of the background metric, and provides a wider perspective on the structure of the effective action $(\mathcal{S})$ for arbitrary metrics, and on the increased complexity of the CWIs around such backgrounds. \\
It is well known that in more general backgrounds, for instance for Weyl flat/conformally flat metrics, the structure of the quantum average of the energy momentum tensor (its one-point function), is affected by tadpole contributions which are fixed by the anomaly, and need to be included in the analysis, and call for a generalization of our method. \\
Then we continue with a review of the structure of the CWIs in the longitudinal/transverse separation introduced in \cite{Bzowski:2013sza} and developed further in \cite{Bzowski:2018fql,Bzowski:2015pba,Bzowski:2015yxv}, followed by the analysis of the anomalous CWIs. In \secref{heres2} a new derivation, performed directly in $d=4$, of the special CWIs is presented for the $4T$. The approach  extends the one developed in \cite{Coriano:2017mux} for the 3T to the new case.\\
The renormalization of the 4T and its longitudinal/transverse decomposition is worked out afterwards, where, just for completeness, we also classify the singular form factors which are affected by renormalization.  \\
The structure of the anomaly action is slowly built starting from the 2-point function, that we review. The vanishing of the anomaly contribution to the anomaly action $\mathcal{S}$  
at $O(h^2)$ is illustrated in great detail. Then we derive the structure of the local (i.e. $t_{loc}$ or longitudinal stress energy tensor) contributions to the CWI in the renormalization procedure, and separate the anomaly contributions from the unknown but finite, renormalized  parts. This is the content of \secref{fourp}. \\
Even if the finite local terms are not explicitly given, we show that they are not necessary in order to identify the structure of the anomaly action at the level of the 4T. \\
The result is obtained  by working directly with the longitudinal components $t_{loc}$ of the counterterms to the same 4T correlator, combined with their trace WIs. We demonstrate that a structure with bilinear mixings appears quite naturally from the decomposition, together with an extra, trace-free contribution. 
This extra term, absent in the 3-point function, appears for the first time at the level of the 
4T.
\section{Renormalization in momentum space}\label{renorm}

For a generic $nT$ correlator, the only counterterm needed for its renormalization, is obtained by the inclusion of a classical gravitational vertex generated by the differentiation of \eqref{counter} $n$ times.\\
The renormalized effective action $\sm_R$ is then defined by the sum of the two terms
\begin{equation}
\sm_R(g)=\sm(g)+ \sm_{ct}(g)
\end{equation}
\begin{align}
\label{cct1}
\sm_{ct}=\raisebox{-0.8ex}{\includegraphics[width=0.15\linewidth]{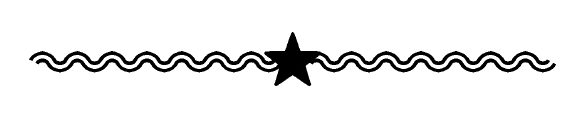}}+\raisebox{-5ex}{\includegraphics[width=0.15\linewidth]{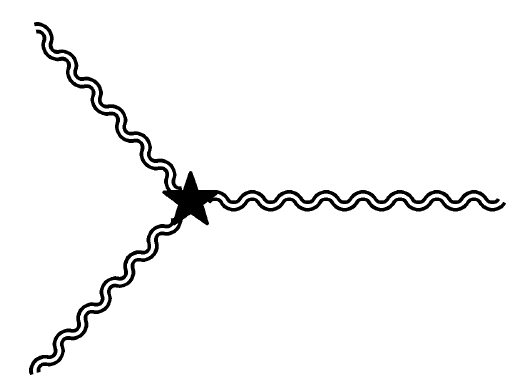}}+\raisebox{-6ex}{\includegraphics[width=0.15\linewidth]{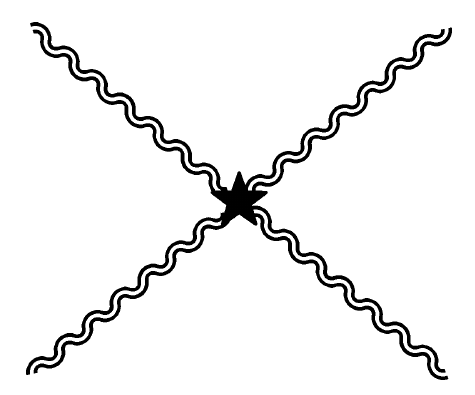}}+...
\end{align}
with $\sm_{ct}$ shown in \eqref{cct1}. Both terms of $\sm_R(g)$ are expanded in the metric fluctuations as in \eqref{cct1}. If we resort to a path integral definition of a certain CFT,  it is clear that any renormalized correlation function appearing in the expansion of $\sm_R(g)$ would be expressed in terms of a bare contribution accompanied by a counterterm vertex.\\
The correlation functions extracted by the renormalized action can be expressed as the sum of a finite $(f)$  correlator and of an anomaly term ($anomaly$) in the form
\begin{align}
\label{cct}
 \langle{T^{\mu_1\nu_1}T^{\mu_2\nu_2}\ldots T^{\mu_n\nu_n}}\rangle_{Ren}=
&\bigg[\langle{T^{\mu_1\nu_1}T^{\mu_2\nu_2}\ldots T^{\mu_n\nu_n}}\rangle_{bare}+\langle{T^{\mu_1\nu_1}T^{\mu_2\nu_2}\ldots T^{\mu_n\nu_n}}\rangle_{count}\bigg]_{d\to4}=\notag\\
&=\langle{T^{\mu_1\nu_1}T^{\mu_2\nu_2}\ldots T^{\mu_n\nu_n}}\rangle^{(d=4)}_{f}+\langle{T^{\mu_1\nu_1}T^{\mu_2\nu_2}\ldots T^{\mu_n\nu_n}}\rangle^{(d=4)}_{anomaly}
\end{align} 
The renormalized correlator shown above satisfies anomalous CWIs. \\
To characterize the anomaly contribution to each correlation function, we start from the 1-point function. In a generic background $g$, the 1-point function is decomposed as 
\begin{equation}
\label{decomponepoint}
\langle T^{\mu\nu}\rangle_{Ren}=\frac{2}{\sqrt{g}}\frac{\delta \sm_{Ren}}{\delta g_{\mu\nu}} =\langle T^{\mu\nu} \rangle_A  + \langle \overline{T}^{\mu\nu}\rangle_f
\end{equation}
with
\begin{equation}
g^{\mu\nu}\frac{\delta \sm}{\delta g^{\mu\nu}} = g^{\mu\nu}\frac{\delta \sm_A}{\delta g^{\mu\nu}}\equiv \frac{\sqrt{g}}{2} g_{\mu\nu} \langle T^{\mu\nu} \rangle_A \end{equation}
being the trace anomaly equation, and $\langle \overline{T}^{\mu\nu}\rangle_f$ is the Weyl-invariant (traceless) term.
\\
Following the discussion in \eqref{anom1}, these scaling violations may be written for the 1-point function in the form
\begin{equation}
\label{anomeq1}
\langle{T^{\mu}_{\ \ \mu}(x)}\rangle=\mathcal{A}(x)
\end{equation} 
- having dropped the suffix {\em Ren} from the renormalized stress energy tensor - 
where the finite terms on the right hand side of this equation denote the anomaly contribution
with
\begin{equation}
\label{AF}
\mathcal{A}(x)=\sqrt{-g(x)}\bigg[b\,C^2(x)+b'E(x)\bigg]
\end{equation}
being the anomaly functional. We will be needing several differentiation of this functional, 
evaluated in the flat limit. This procedure generates expressions which are  polynomial in the momenta, that can be found in the \appref{Mvc}.
 In general, one also finds additional dimension-4 local invariants $\mathcal{L}_i$, if there are couplings to other background fields, as for instance in the QED and QCD cases, with coefficients related to the $\beta$ functions of the corresponding gauge couplings. \\
For n-point functions the trace anomaly, as well as all the other CWIs, are far more involved, and take a hierarchical structure. \\
For all the other WIs, in DR the structure of the equations can be analyzed in two different frameworks. \\
In one of them, we are allowed to investigate the correlators directly in $d$ spacetime dimensions, deriving ordinary (anomaly-free) CWIs, which are then modified by the inclusion of the 4-dimensional counterterm as $d\to 4$. In this limit, the conformal constraints become anomalous and the hierarchical equations are modified by the presence of extra terms which are anomaly-related. \\
Alternatively, it is possible to circumvent this limiting procedure by working out the equations directly in $d=4$, with the inclusion of the contributions coming from the anomaly functional, as we are going to show below. This second approach has been formulated in \cite{Coriano:2017mux} and will be extended to the 4-point function in \secref{heres}. \\  
We recall that the counterterm vertex for the nT correlator, in DR, in momentum space takes the form
\begin{equation}
\langle{T^{\mu_1\nu_1}(p_1)\dots T^{\mu_n\nu_n}(\bar{p}_n)}\rangle_{count}=-\frac{\mu^{-\varepsilon}}{\varepsilon}\bigg(b\,V_{C^2}^{\mu_1\nu_1\dots\mu_n\nu_n}(p_1,\dots,\bar{p}_n)+b'\,V_{E}^{\mu_1\nu_1\dots\mu_n\nu_n}(p_1,\dots,\bar{p}_n)\bigg)
\label{nTcount},
\end{equation}
 where  
\begin{align}
V_{C^2}^{\m_1\n_1\dots\mu_n\nu_n}(p_1,\dots,\bar{p}_n)
&\equiv 2^n\big[\sqrt{-g}\,C^2\big]^{\m_1\n_1\dots\m_n\n_n}(p_1,\dots,\bar{p}_n)\notag\\
&=2^n\int\,d^dx_1\,\dots\,d^dx_n\,d^dx\,\bigg(\sdfrac{\d^n(\sqrt{-g}C^2)(x)}{\d g_{\m_1\n_1}(x_1)\dots\d g_{\m_n\n_n}(x_n)}\bigg)_{g=\delta}\,e^{-i(p_1\,x_1+\dots+p_nx_n)}\notag\\[2ex]
\end{align}
and
\begin{align}
V_{E}^{\m_1\n_1\dots\mu_n\nu_n}(p_1,\dots,\bar{p}_n)&\equiv 2^n\big[\sqrt{-g}\,E\big]^{\m_1\n_1\dots\m_n\n_n}(p_1,\dots,\bar{p}_n)\notag\\
&=2^n\int\,d^dx_1\,\dots\,d^dx_n\,d^dx\,\bigg(\sdfrac{\d^n(\sqrt{-g}E)(x)}{\d g_{\m_1\n_1}(x_1)\dots\d g_{\mu_n\nu_n}(x_n)}\bigg)_{g=\delta}\,e^{-i(p_1\,x_1+\dots+p_nx_n)}.
\end{align}
are the expressions of the two contributions present in \eqref{nTcount} in momentum space. 
One can also verify the following trace relations 
\begin{align}
\label{expabove}
&\delta_{\mu_1\nu_1}\,V_{C^2}^{\m_1\n_1\dots\mu_n\nu_n}(p_1,\dots,p_n)=2^{n-1}(d-4)\,\left[\sqrt{-g}C^2\right]^{\mu_2\nu_2\dots\mu_n\nu_n}(p_2,\dots,p_n)\notag\\
&-2\bigg[V_{C^2}^{\m_2\n_2\dots\mu_n\nu_n}(p_1+p_2,p_3,\dots,p_n)+V_{C^2}^{\m_2\n_2\dots\mu_n\nu_n}(p_2,p_1+p_3,\dots,p_n)+\dots+V_{C^2}^{\m_2\n_2\dots\mu_n\nu_n}(p_2,p_3,\dots,p_1+p_n)\bigg]
\end{align}
\begin{align}
\label{mom}
&\delta_{\mu_1\nu_1}\,V_{E}^{\m_1\n_1\dots\mu_n\nu_n}(p_1,\dots,p_n)=2^{n-1}(d-4)\,\left[\sqrt{-g}E\right]^{\mu_2\nu_2\dots\mu_n\nu_n}(p_2,\dots,p_n)\notag\\
&-2\bigg[V_{E}^{\m_2\n_2\dots\mu_n\nu_n}(p_1+p_2,p_3,\dots,p_n)+V_{E}^{\m_2\n_2\dots\mu_n\nu_n}(p_2,p_1+p_3,\dots,p_n)+\dots+V_{E}^{\m_2\n_2\dots\mu_n\nu_n}(p_2,p_3,\dots,p_1+p_n)\bigg]
\end{align}
that hold in general $d$ dimensions. \\
Obviously, the effective action that results from the renormalization can be clearly separated in terms of two contributions, as evident from \eqref{decomponepoint}, 
\begin{equation} 
\sm_R(g)=\sm_{A}(g) +{\sm}_f(g),
\end{equation}
corresponding to an anomaly part $\sm_{A}[g]$ and to a finite, Weyl-invariant term, which can be expanded in terms of fluctuations over a background $\bar{g}$ as for the entire effective action $\sm(g)$
\begin{equation}
\label{exps3}
\sm_f(g)=\sm(\bar{g})+\sum_{n=1}^\infty \frac{1}{2^n n!} \int d^d x_1\ldots d^d x_n \sqrt{g_1}\ldots \sqrt{g_n}\,\langle T^{\mu_1\nu_1}\ldots \,T^{\mu_n\nu_n}\rangle_{f }\delta g_{\mu_1\nu_1}(x_1)\ldots \delta g_{\mu_n\nu_n}(x_n).
\end{equation}
This functional collects finite correlators \eqref{cct} in $d=4$.
A similar expansion holds also for $\sm_{A}$, the anomaly part.\\ 
The anomaly effective action $\sm_A$ that results from this analysis in momentum space is a rational function of the external momenta, characterised by well-defined tensor structures, and it is free of logarithmic terms, as shown in direct perturbative studies of the TJJ and 3T \cite{Giannotti:2008cv,Armillis:2009pq,Armillis:2010qk}. \\
\section{Conservation Ward identities}
The anomaly action $\sm_A(g)$ is constrained by a hierarchical set of equations which can be derived by the symmetries of the general effective action $\sm$. We proceed assuming that the correlation functions can be derived by varying the path integral definition of $\sm(g)$ \eqref{defg} \eqref{induced} as in \eqref{exps1}.\\ 
Starting from the covariant definition of the stress-energy tensor, expressed in terms of a fundamental action as in \eqref{defT}, but for the rest, generic
\begin{equation}
\langle{T^{\mu\nu}(x)}\rangle_g=\frac{2}{\sqrt{-g(x)}}\frac{\delta S(g)}{\delta g_{\mu\nu}(x)},
\end{equation}
this 1-point function satisfies the fundamental Ward identity of covariant conservation in an arbitrary background $g$
\begin{equation}
\label{vat}
^{(g)}\nabla_\mu\langle T^{\mu\nu}(x)\rangle_g=0, \qquad i.e. \qquad \delta_\epsilon \sm(g)=0
\end{equation}
as a consequence of the invariance of $\sm(g)$ under diffeomorphisms.
Here $^{(g)}\nabla_\mu$ denotes the covariant derivative in the general background metric $g_{\mu\nu}(x)$. It can be expressed in the form 
\begin{equation}
\partial_\nu\left(\frac{\delta S(g)}{\delta g_{\mu\nu}(x)}\right)+\Gamma^\mu_{\nu\lambda}\left(\frac{\delta S(g)}{\delta g_{\lambda\nu}(x)}\right)=0,
\end{equation}
where $\Gamma^\mu_{\lambda\nu}$ is the Christoffel connection for the general background metric $g_{\mu\nu(x)}$.\\
Our definitions and conventions are summarised in an \appref{Definitions}.
In order to derive the conservation WIs for higher point correlation functions, one has to consider additional variations with respect to the metric of \eqref{vat} and then move to flat space, obtaining 
\begin{align}
&\partial_{\nu_1}\langle T^{\mu_1\nu_1}(x_1)T^{\mu_2\nu_2}(x_2)T^{\mu_3\nu_3}(x_3)T^{\mu_4\nu_4}(x_4)\rangle=\notag\\
=&-\left[2\left(\frac{\delta\Gamma^{\mu_1}_{\lambda\nu_1}(x_1)}{\delta  g_{\mu_2\nu_2}(x_2)}\right)_{g=\delta}\langle T^{\lambda\nu_1}(x_1)T^{\mu_3\nu_3}(x_3)T^{\mu_4\nu_4}(x_4)\rangle+(23)+(24)\right]\notag\\
& -\left[4\left(\frac{\delta^2\Gamma^{\mu_1}_{\lambda\nu_1}(x_1)}{\delta  g_{\mu_2\nu_2}(x_2)\delta  g_{\mu_3\nu_3}(x_3)}\right)_{g=\delta}\langle {T^{\lambda\nu_1}(x_1)T^{\mu_4\nu_4}(x_4)}\rangle+(24)+(34)\right],\label{transverseX}
\end{align}
where 
\begin{align}
\left(\frac{\delta\Gamma^{\mu_1}_{\lambda\nu_1}(x_1)}{\delta  g_{\mu_i\nu_i}(x_i)}\right)_{g=\delta}&=\frac{1}{2}\left(\delta^{\mu_1(\mu_i}\delta^{\nu_i)}_{\nu_1}\,\partial_\lambda\delta_{x_1x_i}+\delta^{\mu_1(\mu_i}\delta^{\nu_i)}_{\lambda}\,\partial_{\nu_1}\delta_{x_1x_i}-\delta^{(\mu_i}_\lambda\delta^{\nu_i)}_{\nu_1}\,\partial^{\mu_1}\delta_{x_1x_i}\right)\\
\left(\frac{\delta^2\Gamma^{\mu_1}_{\lambda\nu_1}(x_1)}{\delta  g_{\mu_i\nu_i}(x_i)\delta  g_{\mu_j\nu_j}(x_j)}\right)_{g=\delta}&=\notag\\
&\hspace{-3cm}=-\frac{\delta_{x_1x_i}}{2}\delta^{\mu_1(\mu_i}\delta^{\nu_i)\epsilon}\left(\delta^{(\mu_j}_\epsilon\delta^{\nu_j)}_{\nu_1}\,\partial_\lambda\delta_{x_1x_j}+\delta^{(\mu_j}_\epsilon\delta^{\nu_j)}_{\lambda}\,\partial_{\nu_1}\delta_{x_1x_j}-\delta^{(\mu_j}_\lambda\delta^{\nu_j)}_{\nu_1}\,\partial_\epsilon\delta_{x_1x_j}\right)+(ij),
\end{align}
are the first and second functional derivatives of the connection, in the flat limit.
We have explicitly indicated the symmetrization with respect to the relevant indices using the permutation $(ij)\equiv (i\leftrightarrow j)$. We have defined $ \delta^{(\mu_i}_\lambda\delta^{\nu_i)}_{\nu_1}\equiv1/2(\delta^{\mu_i}_\lambda\delta^{\nu_i}_{\nu_1}+\delta^{\nu_i}_\lambda\delta^{\mu_i}_{\nu_1})$, and introduced a simplified notation for the Dirac delta $\delta_{x_ix_j}\equiv \delta(x_i-x_j)$. All the derivative (e.g. $\partial_\lambda$) are taken with respect to the coordinate $x_1$  $(e.g. \partial/\partial x_1^\lambda)$. \\
We Fourier transform to momentum space with the convention 
\begin{align}
\langle{T^{\mu_1\nu_1}(p_1)T^{\mu_2\nu_2}(p_2)T^{\mu_3\nu_3}(p_3)T^{\mu_4\nu_4}( \overline{p}_4)}\rangle
&=\int d^4 x_1 d^4 x_2 d^4 x_3 e^{-i( p_1\cdot x_1 +p_2\cdot x_2 + p_3\cdot x_3)}\nonumber \\
& \times \langle{T^{\mu_1\nu_1}(x_1)T^{\mu_2\nu_2}(x_2)T^{\mu_3\nu_3}(x_3)T^{\mu_4\nu_4}(0)}\rangle.
\end{align}

Here we have used the translational invariance of the correlator in flat space, which allows to use momentum conservation to express one of the momenta (in our convention $p_4$) as combination of the remaining ones $\bar{p}_4=-p_1-p_2-p_3$. Details on the elimination of one of the momenta in the derivation of the CWIs and on the modification of the Leibnitz rule in the differentiation of such correlators in momentum space can be found in \cite{Coriano:2018bbe}. \\
The conservation Ward Identity \eqref{transverseX} in flat spacetime may be Fourier transformed, giving the CWIs in momentum space
\begin{align}
&p_{1\nu_1}\langle{T^{\mu_1\nu_1}(p_1)T^{\mu_2\nu_2}(p_2)T^{\mu_3\nu_3}(p_3)T^{\mu_4\nu_4}(\bar p_4)}\rangle=\notag\\
&=\Big[4\, \mathcal{B}^{\mu_1\hspace{0.4cm}\mu_2\nu_2\mu_3\nu_3}_{\hspace{0.3cm}\lambda\nu_1}(p_2,p_3)\langle{T^{\lambda\nu_1}(p_1+p_2+p_3)T^{\mu_4\nu_4}(\bar p_4)}\rangle+(34)+(2 4)\Big]\notag\\
&\hspace{0.5cm}+\Big[2 \, \mathcal{C}^{\mu_1\hspace{0.4cm}\mu_2\nu_2}_{\hspace{0.3cm}\lambda\nu_1}(p_2)\langle{T^{\lambda\nu_1}(p_1+p_2)T^{\mu_3\nu_3}(p_3)T^{\mu_4\nu_4}(\bar p_4)}\rangle+(2 3)+(2 4)\Big],\label{transverseP}
\end{align}
where we have defined
\begin{align}
\label{BB}
\mathcal{B}^{\mu_1\hspace{0.4cm}\mu_2\nu_2\mu_3\nu_3}_{\hspace{0.3cm}\lambda\nu_1}(p_2,p_3)&\equiv -\frac{1}{2}\delta^{\mu_1(\mu_2}{\delta^{\nu_2)\epsilon}}\left(\delta_\epsilon^{(\mu_3}\delta^{\nu_3)}_{\nu_1}\,p_{3\,\lambda}+\delta_\epsilon^{(\mu_3}\delta^{\nu_3)}_{\lambda}\,p_{3\,\nu_1}-\delta_\lambda^{(\mu_3}\delta^{\nu_3)}_{\nu_1}\,p_{3\,\epsilon}\right)+(23)\\[2ex]
\label{CC2}
\mathcal{C}^{\mu_1\hspace{0.4cm}\mu_2\nu_2}_{\hspace{0.3cm}\lambda\nu_1}(p_2)&\equiv \frac{1}{2}\left(\delta^{\mu_1(\mu_2}\delta^{\nu_2)}_{\nu_1}\,p_{2\,\lambda}+\delta^{\mu_1(\mu_2}\delta^{\nu_2)}_{\lambda}p_{2\,\nu_1}-\delta^{(\mu_2}_{\lambda}\delta^{\nu_2)}_{\nu_1}p_2^{\mu_1}\right),
\end{align}
related to the second and first functional derivatives of the Christoffel, connection respectively. 

\subsection{Conservation WI's for the counterterms} 
To illustrate the conservation WI in detail, we turn to the expression of the counterterm action \eqref{counter}, which generates counterterm vertices of the form
\begin{align}
&\langle{T^{\mu_1\nu_1}(p_1)T^{\mu_2\nu_2}(p_2)T^{\mu_3\nu_3}(p_3)T^{\mu_4\nu_4}(\bar{p}_4)}_{count}=\notag\\
&\qquad=-\frac{\mu^{-\varepsilon}}{\varepsilon}\bigg(b\,V_{C^2}^{\mu_1\nu_1\mu_2\nu_2\mu_3\nu_3\mu_4\nu_4}(p_1,p_2,p_3,\bar{p}_4)+b'\,V_{E}^{\mu_1\nu_1\mu_2\nu_2\mu_3\nu_3\mu_4\nu_4}(p_1,p_2,p_3,\bar{p}_4)\bigg),\label{TTTTcount}
\end{align}
where on the rhs of the expression above we have introduced the counterterm vertices (with $P= p_1 +\ldots p_4$)
\begin{align}
&V_{C^2}^{\m_1\n_1\m_2\n_2\m_3\n_3\mu_4\nu_4}(p_1,p_2,p_3,\bar{p}_4)
\,\delta^4(P)\equiv 16\, \delta^4(P)\,\big[\sqrt{-g}\,C^2\big]^{\m_1\n_1\m_2\n_2\m_3\n_3\mu_4\nu_4}(p_1,p_2,p_3,\bar{p}_4)\notag\\
=&16\int\,d^dx_1\,\,d^dx_2\,d^dx_3\,d^dx_4\,d^dx\,\bigg(\sdfrac{\d^4(\sqrt{-g}C^2)(x)}{\d g_{\m_1\n_1}(x_1)\d g_{\m_2\n_2}(x_2)\d g_{\m_3\n_3}(x_3)\d g_{\m_4\n_4}(x_4)}\bigg)_{g=\delta}\,e^{-i(p_1\,x_1+p_2\,x_2+p_3\,x_3+p_4x_4)},\notag\\[2ex]
&V_{E}^{\m_1\n_1\m_2\n_2\m_3\n_3\mu_4\nu_4}(p_1,p_2,p_3,\bar{p}_4)\delta^4(P)\equiv 16\,\,\delta^4(P)\big[\sqrt{-g}\,E\big]^{\m_1\n_1\m_2\n_2\m_3\n_3\m_4\n_4}(p_1,p_2,p_3,\bar{p}_4)\notag\\
=&16\int\,d^dx_1\,\,d^dx_2\,d^dx_3\,d^dx_4\,d^dx\,\bigg(\sdfrac{\d^4(\sqrt{-g}E)(x)}{\d g_{\m_1\n_1}(x_1)\d g_{\m_2\n_2}(x_2)\d g_{\m_3\n_3}(x_3)\d g_{\mu_4\nu_4}(x_4)}\bigg)_{g=\delta}\,e^{-i(p_1\,x_1+p_2\,x_2+p_3\,x_3+p_4x_4)},\label{count}
\end{align}
evaluated in the flat spacetime limit.
These vertices share some properties when contracted with flat metric tensors and the external momenta as we have already seen. In particular, from \eqref{expabove} and \eqref{mom}, when $n=4$ and in $d$ dimensions we have
\begin{align}
\delta_{\mu_1\nu_1}\,V_{C^2}^{\m_1\n_1\m_2\n_2\m_3\n_3\mu_4\nu_4}(p_1,p_2,p_3,\bar{p}_4)&=8(d-4)\,\left[\sqrt{-g}C^2\right]^{\mu_2\nu_2\mu_3\nu_3\mu_4\nu_4}(p_2,p_3,\bar{p}_4)\notag\\
&\hspace{-5.5cm}-2V_{C^2}^{\m_2\n_2\m_3\n_3\mu_4\nu_4}(p_1+p_2,p_3,\bar{p}_4)-2V_{C^2}^{\m_2\n_2\m_3\n_3\mu_4\nu_4}(p_2,p_1+p_3,\bar{p}_4)-2V_{C^2}^{\m_2\n_2\mu_3\nu_3\m_4\n_4}(p_2,p_3,p_1+\bar{p}_4),\\[2ex]
\delta_{\mu_1\nu_1}\,V_{E}^{\m_1\n_1\m_2\n_2\m_3\n_3\mu_4\nu_4}(p_1,p_2,p_3,\bar{p}_4)&=8(d-4)\,\left[\sqrt{-g}E\right]^{\mu_2\nu_2\mu_3\nu_3\mu_4\nu_4}(p_2,p_3,\bar{p}_4)\notag\\
&\hspace{-5.5cm}-2V_{E}^{\m_2\n_2\m_3\n_3\mu_4\nu_4}(p_1+p_2,p_3,\bar{p}_4)-2V_{E}^{\m_2\n_2\m_3\n_3\mu_4\nu_4}(p_2,p_1+p_3,\bar{p}_4)-2V_{E}^{\m_2\n_2\mu_3\nu_3\m_4\n_4}(p_2,p_3,p_1+\bar{p}_4),
\end{align}
which play a key role in the renormalization procedure.
Furthermore, the contraction of these vertices with the external momenta generates conservation WIs in $d$ dimensions, similar to \eqref{transverseP}, 
\begin{align}
&p_{1\,\nu_1}\,V_{C^2}^{\m_1\n_1\m_2\n_2\m_3\n_3\mu_4\nu_4}(p_1,p_2,p_3,\bar{p}_4)=\notag\\
&=\Big[4\, \mathcal{B}^{\mu_1\hspace{0.4cm}\mu_2\nu_2\mu_3\nu_3}_{\hspace{0.3cm}\lambda\nu_1}(p_2,p_3)V_{C^2}^{\lambda\n_1\mu_4\nu_4}(p_1+p_2+p_3,\bar{p}_4)+(34)+ (24)\Big]\notag\\
&\hspace{0.5cm}+\Big[2 \, \mathcal{C}^{\mu_1\hspace{0.4cm}\mu_2\nu_2}_{\hspace{0.3cm}\lambda\nu_1}(p_2)V_{C^2}^{\lambda\n_1\m_3\n_3\mu_4\nu_4}(p_1+p_2,p_3,\bar{p}_4)+(2 3)+(24)\Big]\\[2ex]
&p_{1\,\nu_1}\,V_{E}^{\m_1\n_1\m_2\n_2\m_3\n_3\mu_4\nu_4}(p_1,p_2,p_3,\bar{p}_4)=\notag\\
&=\Big[4\, \mathcal{B}^{\mu_1\hspace{0.4cm}\mu_2\nu_2\mu_3\nu_3}_{\hspace{0.3cm}\lambda\nu_1}(p_2,p_3)V_{E}^{\lambda\n_1\mu_4\nu_4}(p_1+p_2+p_3,\bar{p}_4)+(34)+ (24)\Big]\notag\\
&\hspace{0.5cm}+\Big[2 \, \mathcal{C}^{\mu_1\hspace{0.4cm}\mu_2\nu_2}_{\hspace{0.3cm}\lambda\nu_1}(p_2)V_{E}^{\lambda\n_1\m_3\n_3\mu_4\nu_4}(p_1+p_2,p_3,\bar{p}_4)+(2 3)+(24)\Big],
\end{align} 
where  $\mathcal{C}$ and $\mathcal{B}$ are given in \eqref{BB} and \eqref{CC2}.
These equations can be generalized to the case of $n$-point functions. 
\subsection{Conformal Ward Identities}
Turning to the ordinary (i.e. non anomalous) trace and conformal WIs, these can be obtained directly in flat space using the expression of operators of the dilatation and special conformal transformations.
The dilatation WI's for the 4T can be easily constructed from the condition of Weyl invariance of the effective action $\sm$, or equivalently, in the ordinary operatorial approach (see \cite{Coriano:2018bbe}) 

\begin{equation}
	\left[4d+\sum_{j=1}^4\,x_j^\a\sdfrac{\partial}{\partial x_j^\a}\right]\langle{T^{\mu_1\nu_1}(x_1)T^{\mu_2\nu_2}(x_2)T^{\mu_3\nu_3}(x_3)T^{\mu_4\nu_4}(x_4)}\rangle=0,
\end{equation}
where we have used the explicit expression of the scaling dimension of the stress energy tensor $\Delta_T=d$. Analogously, the special conformal WIs, corresponding to special conformal transformations, can be derived in the operatorial approach, applied to an ordinary CFT in flat space, 
relying on the change of $T^{\mu\nu}$ under a special conformal transformation, with a generic parameter  
 $b_\mu$, and  $\sigma=-2 b\cdot x$ 
\begin{equation}
\delta T^{\mu\nu}(x)=-(b^\alpha x^2 -2 x^\alpha b\cdot x )\, \partial_\alpha  T^{\mu\nu}(x)   - \Delta_T \sigma T^{\mu\nu}(x)+
2(b_\mu x_\alpha- b_\alpha x_\mu)T^{\alpha\nu} + 2 (b_\nu x_\alpha -b_\alpha x_\nu)\, T^{\mu\alpha}(x).
\end{equation}
 The action of the special conformal operator $\mathcal{K}^\kappa$ on $T$ in its finite form is obtained differentiating respect to the parameter $b_\kappa$ 
\begin{align}
\mathcal{K}^\kappa T^{\mu\nu}(x)&\equiv &\delta_\kappa T^{\mu\nu}(x) =\frac{\partial}{\partial b^\kappa} (\delta T^{\mu\nu})\nonumber \\
&= -(x^2 \partial_\kappa - 2 x_\kappa x\cdot \partial) T^{\mu\nu}(x) + 2\Delta_T x_\kappa T^{\mu\nu}(x) +
2(\delta_{\mu\kappa}x_\alpha -\delta_{\alpha \kappa}x_\mu) T^{\alpha\nu}(x) \nonumber \\ 
& + 2 (\delta_{\kappa\nu} x_{\alpha} -\delta_{\alpha \kappa} x_\nu )T^{\mu\alpha}.
\label{ith}
\end{align}
By using the Leibniz rule for the variation on correlation functions of multiple T's, it can be distributed over the entire correlator as 
\begin{equation}
\label{com}
\mathcal{K}^\kappa\langle{T^{\mu_1\nu_1}(x_1)T^{\mu_2\nu_2}(x_2)T^{\mu_3\nu_3}(x_3)T^{\mu_4\nu_4}(x_4)}\rangle=
\sum_{i=1}^4 \langle T(x_1)\ldots \delta_\kappa T(x_i)\ldots T(x_4) \rangle =0
\end{equation}
which takes the form
\begin{align}
	0=&\sum_{j=1}^n\left(2d\,x_j^k+2x_j^\k\,x_j^\a\sdfrac{\partial}{\partial x_j\a}-x_j^2\sdfrac{\partial}{\partial x_{jk}}\right)\,\langle{T^{\mu_1\nu_1}(x_1)T^{\mu_2\nu_2}(x_2)T^{\mu_3\nu_3}(x_3)T^{\mu_4\nu_4}(x_4)}\rangle\notag\\
	&+2\sum_{j=1}^4\left[(x_j)_{\a_{j}}\d^{\k\m_{j}}-x_j^{\m_{j}}\d^\k_{\a_{j}}\right]\,\langle{T^{\mu_1\nu_1}(x_1)\dots T^{\alpha_j\nu_j}(x_j) \dots T^{\mu_4\nu_4}(x_4)}\rangle\notag\\
	&+2\sum_{j=1}^4\left[(x_j)_{\a_{j}}\d^{\k\nu_{j}}-x_j^{\nu_{j}}\d^\k_{\a_{j}}\right]\,\langle{T^{\mu_1\nu_1}(x_1)\dots T^{\mu_j\alpha_j}(x_j) \dots T^{\mu_4\nu_4}(x_4)}\rangle,
\end{align}
where $\k$ is now a free Lorentz index. Notice that in order to be allowed to use an operatorial approach, one needs to rely on correlators defined via direct insertions of T's. Such correlation functions, are, in general,  different from the definition given above in \eqref{exps1} due to possible contact terms and the presence of nonvanishing tadpoles, not contemplated in \eqref{com}.\\
For this reason the CWIs derived by this operatorial method and by the functional method that we will present below, are naive expressions which are perfectly well-defined and equivalent, only in the presence of a suitable regularization scheme and of a flat background. In DR, which is well-defined in a flat spacetime, the vanishing of the 1-point function and the inclusion of vertex counterterms shows that we don't need to worry about such issues.  \\
Notice that these constraints are directly written in momentum space as 
\begin{align}
	0=&D\,\langle{T^{\mu_1\nu_1}(p_1)T^{\mu_2\nu_2}(p_2)T^{\mu_3\nu_3}(p_3)T^{\mu_4\nu_4}(\bar{p}_4)}\rangle\notag\\
	=&\left(d-\sum_{j=1}^3\,p_j^\a\sdfrac{\partial}{\partial p_j^\a}\right)\langle{T^{\mu_1\nu_1}(p_1)T^{\mu_2\nu_2}(p_2)T^{\mu_3\nu_3}({p}_3)T^{\mu_4\nu_4}(\bar{p}_4)}\rangle\label{Dequ}
\end{align}
and
	
\begin{align}	
	0=&\mathcal{K}^\kappa\langle{T^{\mu_1\nu_1}(p_1)T^{\mu_2\nu_2}(p_2)T^{\mu_3\nu_3}(p_3)T^{\mu_4\nu_4}(\bar{p}_4)}\rangle\notag\\
	=&\sum_{j=1}^{3}\left((p_j)^\kappa\frac{\partial}{\partial p_j^\alpha}\frac{\partial}{\partial p_{j\,\alpha}}-2p_j^\alpha\frac{\partial}{\partial p_j^\alpha}\frac{\partial}{\partial p_{j\,\kappa}}\right)\langle{T^{\mu_1\nu_1}(p_1)T^{\mu_2\nu_2}(p_2)T^{\mu_3\nu_3}(p_3)T^{\mu_4\nu_4}(\bar{p}_4)}\rangle\notag\\
	&\hspace{1cm}+4\left(\delta^{\kappa(\mu_1}\frac{\partial}{\partial p_1^{\alpha_1}}-\delta^\kappa_{\alpha_1}\Delta_\l^{(\m_1}\frac{\partial}{\partial p_{1\,\l}}\right)\langle{ T^{\nu_1)\alpha_1}(p_1)T^{\mu_2\nu_2}(p_2)T^{\mu_3\nu_3}(p_3)T^{\mu_4\nu_4}(\bar{p}_4)}\rangle\notag\\
	&\hspace{1cm}+4\left(\delta^{\kappa(\m_2}\frac{\partial}{\partial p_2^{\a_2}}-\delta^\kappa_{\a_2}\d^{(\m_2}_\l\frac{\partial}{\partial p_{2\,\l}}\right)\langle{ T^{\nu_2)\a_2}(p_2)T^{\mu_1\nu_1}(p_1)T^{\mu_3\nu_3}(p_3)T^{\mu_4\nu_4}(\bar{p}_4)}\rangle\notag\\
	&\hspace{1cm}+4\left(\delta^{\kappa(\m_3}\frac{\partial}{\partial p_3^{\a_3}}-\delta^\kappa_{\a_3}\d^{(\m_3}_\l\frac{\partial}{\partial p_{3\,\l}}\right)\langle{ T^{\nu_3)\a_3}(p_3)T^{\mu_1\nu_1}(p_1)T^{\mu_2\nu_2}(p_2)T^{\mu_4\nu_4}(\bar{p}_4)}\rangle\label{Keq},
\end{align}
in terms of a dilatation operator $D$
and a special conformal transformation operator $K^\kappa$. The action of the differential operators on the momenta is implicit on the 4th momentum, as discussed in the case of 3-point functions in previous works \cite{Coriano:2013jba, Bzowski:2013sza, Coriano:2018bbe}, with a modification of the Leibnitz rule.

\section{Trace and conformal anomalous Ward identities} 
The CWIs become anomalous as we move from $d$ spacetime dimensions to 4.
In $d$ dimensions the conformal symmetry of the correlator 4T is preserved and this property is reflected in the trace identity
\begin{equation}
\langle{T^{\mu}_{\ \ \mu}(x)}\rangle_g=g_{\mu\nu}\langle{T^{\mu\nu}(x)}\rangle_g=0,
\end{equation}
which generates, as we have already mentioned, a hierarchy of equations by functional differentiation of this result respect to the background metric $g$. 
Equivalently, the same equations can be derived from the condition of Weyl invariance of the effective action.\\
Following the same procedure as for the conservation WIs, we may derive the trace Ward identities for the four-point function 4T, in general $d$ dimensions, as
\begin{align}
\delta_{\mu_1\nu_1}\,\langle{T^{\mu_1\nu_1}(x_1)T^{\mu_2\nu_2}(x_2)T^{\mu_3\nu_3}(x_3)T^{\mu_4\nu_4}(x_4)}\rangle&=\notag\\
&\hspace{-3cm}=-2\Big[\delta_{x_1x_2}\langle{T^{\mu_2\nu_2}(x_1)T^{\mu_3\nu_3}(x_3)T^{\mu_4\nu_4}(x_4)}\rangle+(23)+(24)\Big],\label{traceD}
\end{align} 
that may be written in momentum space, after a Fourier transform, as
\begin{align}
&\delta_{\mu_1\nu_1}\langle{T^{\mu_1\nu_1}(p_1)T^{\mu_2\nu_2}(p_2)T^{\mu_3\nu_3}(p_3)T^{\mu_4\nu_4}(\bar p_4)}\rangle=-2\langle{T^{\mu_2\nu_2}(p_1+p_2)T^{\mu_3\nu_3}(p_3)T^{\mu_4\nu_4}(\bar p_4)}\rangle\notag\\[1.3ex]
&\hspace{2cm}-2\langle{T^{\mu_2\nu_2}(p_2)T^{\mu_3\nu_3}(p_1+p_3)T^{\mu_4\nu_4}(\bar p_4)}\rangle-2\langle{T^{\mu_2\nu_2}(p_2)T^{\mu_3\nu_3}(p_3)T^{\mu_4\nu_4}(p_1+\bar p_4)}\rangle.\label{traceP}
\end{align}
We have omitted an overall $\delta$ function, having replaced $p_4$ with $\bar{p}_4$.\\
In $d=4$  the equations need to be renormalized, by adding local covariant counterterms which will be generated from the action \eqref{counter}.\\
 If general covariance is respected by this procedure, the conservation WIs remain valid for the renormalized effective action and for its variations. This is reflected on the hierarchical structure of  the equations, which remain identical to the bare (naive) case. \\
Trace identities of the correlation functions involving at least three stress energy tensor operators are instead affected by the anomaly, due to the scaling violations induced by the regularization/renormalization procedure. 
In $d=4$ the corresponding anomalous Ward identities for the trace can be obtained by a functional variation of the equation \eqref{anomeq1}  with respect to the background metric. \\
In this case \eqref{traceD} is characterised by new contributions on its rhs, coming from the anomaly $\mathcal{A}(x)$
\begin{align}
&\delta_{\mu_1\nu_1}\langle{T^{\mu_1\nu_1}(p_1)T^{\mu_2\nu_2}(p_2)T^{\mu_3\nu_3}(p_3)T^{\mu_4\nu_4}(\bar p_4)}\rangle=-2\langle{T^{\mu_2\nu_2}(p_1+p_2)T^{\mu_3\nu_3}(p_3)T^{\mu_4\nu_4}(\bar p_4)}\rangle\notag\\[1.3ex]
&\hspace{2cm}-2\langle{T^{\mu_2\nu_2}(p_2)T^{\mu_3\nu_3}(p_1+p_3)T^{\mu_4\nu_4}(\bar p_4)}\rangle-2\langle{T^{\mu_2\nu_2}(p_2)T^{\mu_3\nu_3}(p_3)T^{\mu_4\nu_4}(p_1+\bar p_4)}\rangle\notag\\
&\hspace{2cm}+8\mathcal{A}^{\mu_2\nu_2\mu_3\nu_3\mu_4\nu_4}(p_2,p_3,\bar p_4),
\end{align}
where the trace anomaly functional $\mathcal{A}$ is given in Eq. \eqref{AF}, and the Fourier transform of its variation in the flat spacetime limit takes the form
\begin{align}
\mathcal{A}^{\mu_2\nu_2\mu_3\nu_3\mu_4\nu_4}(p_2,p_3,\bar p_4)\delta^4(P)=\int d^4 x_1d^4 x_2 d^4 x_3 d^4 x_4\,\frac{\delta^3\mathcal{A}(x_1)}{\delta g_{\mu_2\nu_2}(x_2)\delta g_{\mu_3\nu_3}(x_3)\delta g_{\mu_4\nu_4}(x_4)}\Bigg|_{flat, d=4}e^{i(p_1x_1+\dots+p_4x_4)}.
\end{align}
In the following we will also be using the simpler general definition 
\begin{equation}
\mathcal{A}^{\mu_1\nu_1\dots\mu_n\nu_n}(p_1,\dots,p_n)\equiv\bigg[\sqrt{-g}\bigg(b\,C^2+b'\,E\bigg)\bigg]^{\mu_1\nu_1\dots\mu_n\nu_n}(p_1,\dots,p_n)\label{defVarTrace}
\end{equation}
to denote the anomaly contributions to 3- and 4-point functions. Notice that these are genuine 4-dimensional terms, left over by the procedure of renormalization.

\section{The anomalous CWIs using conformal Killing vectors}
\label{heres}
The expressions of the anomalous conformal WIs can be derived in an alternative way following the formulation of  \cite{Coriano:2017mux}, that here we are going to extend to the 4-point function case.\\
The derivation of such identities relies uniquely on the effective action and can be obtained as follows. We illustrate it first in the TT case, and then move to the 4T.\\
We start from the conservation of the conformal current as derived in \eqref{iso}
\begin{equation}
\int d^d x \sqrt{g}\, \nabla^\alpha \left( \epsilon_\alpha \frac{2}{\sqrt{g}}\frac{\delta \sm}{\delta g_{\mu\alpha}}\right)=\int d^4 x \sqrt{g}\, \nabla_\mu\langle{\epsilon_\alpha T^{\alpha \nu}}\rangle =0.  
\end{equation}
In the TT case the derivation of the special CWIs is simplified, since there is no trace anomaly if the counterterm action is defined as in \eqref{counter}, a point that we will address in \secref{renren}. We rely on the fact that the conservation of the conformal current $J^\mu_{(K)}$ implies the conservation equation 
\begin{align}
0=\int\,d^dx\,\sqrt{-g}\, \,\nabla_\mu\,\langle{J^\mu_{(K)}(x)\,T^{\mu_1\nu_1}(x_1)}\rangle.
\end{align}
By making explicit the expression $J^\mu(x)=K_\nu(x)\,T^{\mu\nu}(x)$, with $\epsilon \to K$ in the flat limit,  the previous relation takes the form
 \begin{align}
0=\int\,d^dx\,\bigg(\partial_\mu K_\nu\,\langle{T^{\mu\nu}(x)\,T^{\mu_1\nu_1}(x_1)}\rangle+ K_\nu\,\partial_\mu\,\langle{T^{\mu\nu}(x)\,T^{\mu_1\nu_1}(x_1)}\rangle\bigg).\label{cons}
 \end{align}
We recall that $K_\nu$ satisfies the conformal Killing equation in flat space
\begin{align}
\label{flatc}
\partial_\mu K_\nu+\partial_\nu K_\mu=\frac{2}{d}\delta_{\mu\nu}\,\left(\partial\cdot K\right),
\end{align}
and by using this equation \eqref{cons} can be re-written in the form
\begin{align}
	0=\int\,d^dx\,\bigg(K_\nu\partial_\mu\,\langle{T^{\mu\nu}(x)\,T^{\mu_1\nu_1}(x_1)}\rangle+\frac{1}{d}\big(\partial\cdot K\big)\,\langle{T(x)\,T^{\mu_1\nu_1}(x_1)}\rangle\bigg).\label{newcons}
\end{align}
We can use in  this previous expression the conservation and trace Ward identities for the two-point function $\langle{TT}\rangle$, that in the flat spacetime limit are explicitly given by
\begin{align}
\partial_\mu\langle{T^{\mu\nu}(x)T^{\mu_1\nu_1}(x_1)}\rangle&=\bigg(\delta^{(\mu_1}_\mu\delta^{\nu_1)}_\lambda\partial^\nu\delta(x-x_1)-2\delta^{\nu(\mu_1}\delta^{\nu_1)}_\mu\partial_\lambda\delta(x-x_1)\bigg)\langle{T^{\lambda\mu}(x)}\rangle,\label{consTT}\\
\delta_{\mu\nu}\langle{T^{\mu\nu}(x)T^{\mu_1\nu_1}(x_1)}\rangle&\equiv\langle{T(x)T^{\mu_1\nu_1}(x_1)}\rangle=-2\delta(x-x_1)\langle{T^{\mu_1\nu_1}(x)}\rangle\label{traceTT}
\end{align}
and the explicit expression of the Killing vector $K^{(C)}_\nu$ for the special conformal transformations 
\begin{equation}
\begin{split}
K^{(C)\,\kappa}_\mu&=2x^\kappa\,x_\mu-x^2\delta^\kappa_\mu\\
\partial\cdot  K^{(C)\,\kappa}&=2d\,x^\kappa
\end{split}\label{spc}
\end{equation}
where $\kappa=1,\dots,d$. By using \eqref{spc} in the integral \eqref{newcons}, we can rewrite that expression as
\begin{align}
	0=\int\,d^dx\,\bigg[\big(2x^\kappa\,x_\nu-x^2\delta^\kappa_\nu\big)\partial_\mu\,\langle{T^{\mu\nu}(x)\,T^{\mu_1\nu_1}(x_1)}\rangle+2\,x^\kappa\,\langle{T(x)\,T^{\mu_1\nu_1}(x_1)}\rangle\bigg],
\end{align}
A final integrating by parts finally gives the relations 
\begin{align}
&\left(2d\,x_1^\kappa+2x_1^\kappa\,x^{\mu}_1\frac{\partial}{\partial x_1^\mu}+x_1^2\frac{\partial}{\partial x_{1\kappa}}\right)\langle{T^{\mu_1\nu_1}(x_1)}\rangle\notag\\
&\quad+2\bigg(x_{1\lambda}\,\delta^{\mu_1\kappa}-x_1^{\mu_1}\delta^\kappa_\lambda\bigg)\langle{T^{\lambda\nu_1}(x_1)}\rangle+2\bigg(x_{1\lambda}\,\delta^{\nu_1\kappa}-x_1^{\nu_1}\delta^\kappa_\lambda\bigg)\langle{T^{\mu_1\lambda}(x_1)}\rangle=0
\end{align}
that are  the special CWIs for the 1-point function $\langle{T^{\mu_1\nu_1}(x_1)}\rangle$. 
\section{ 4-point functions}\label{heres2}
The derivation above can be extended to n-point functions, starting from the identity 
\begin{equation} 
\int d^d x \sqrt{g} \nabla_\alpha(x) \langle J^\alpha_c(x)T^{\mu_1\nu_1}(x_1)\ldots T^{\mu_n\nu_n}(x_n)
\label{div}
\rangle=0.
\end{equation}
We have used the conservation of the conformal current in d dimensions under variations of the metric, induced by the conformal Killing vectors.\\
In absence of an anomaly, the conservation of the current $J^\mu_c$ follows from the conservation of the stress energy tensor plus the zero trace condition. 
As in the example illustrated above, we consider \eqref{div} in the flat limit 
\begin{align}
\int dx^d \,\partial_\nu\bigg[K_\mu(x)\langle{T^{\mu\nu}(x)T^{\mu_1\nu_1}(x_1)\dots T^{\mu_4\nu_4}(x_4)}\rangle\bigg]=0,\label{Killing1}
\end{align}
where we are assuming that the surface terms vanish, due to the fast fall-off behaviour of the correlation function at infinity. Expanding \eqref{Killing1} we obtain an expression similar to \eqref{newcons}

\begin{align}
0=\int d^dx\left\{K_{\mu}(x)\partial_\nu\langle{T^{\mu\nu}(x)T^{\mu_1\nu_1}(x_1)\dots T^{\mu_4\nu_4}(x_4)}\rangle+\frac{1}{d}\big(\partial\cdot K\big)\delta_{\mu\nu}\langle{T^{\mu\nu}(x)T^{\mu_1\nu_1}(x_1)\dots T^{\mu_4\nu_4}(x_4)}\rangle
\right\}.\label{killing01}
\end{align}
Starting  from this expression, the dilatation CWI is obtained by the choice of the CKV characterising the dilatations  
\begin{equation}
K^{(D)}_\mu(x)=x_\mu,\qquad\partial\cdot K^{(D)}=d
\end{equation}
and \eqref{killing01} becomes
\begin{align}
0=\int d^d x\bigg\{x_\mu\,\partial_\nu\langle{T^{\mu\nu}(x)T^{\mu_1\nu_1}(x_1)\dots T^{\mu_4\nu_4}(x_4)}\rangle+\delta_{\mu\nu}\langle{T^{\mu\nu}(x)T^{\mu_1\nu_1}(x_1)\dots T^{\mu_4\nu_4}(x_4)}\rangle\label{anomDil}
\bigg\}.
\end{align}
At this stage, we use the conservation and trace Ward identities in $d=4$ for the $4$-point function written as
\begin{align}
&\partial_\nu\langle{T^{\mu\nu}(x)T^{\mu_1\nu_1}(x_1)\dots T^{\mu_4\nu_4}(x_4)}\rangle=\notag\\
=&-8\bigg\{\left[\Gamma^{\mu}_{\nu\lambda}(x)\right]^{\mu_1\nu_1\mu_2\nu_2\mu_3\nu_3}(x_1,x_2,x_3)\langle{T^{\lambda\nu}(x)T^{\mu_4\nu_4}(x_4)}\rangle+(14)+(24)+(34)\bigg\}\notag\\
&-4\bigg\{\left[\Gamma^{\mu}_{\nu\lambda}(x)\right]^{\mu_1\nu_1\mu_2\nu_2}(x_1,x_2)\langle{T^{\lambda\nu}(x)T^{\mu_3\nu_3}(x_3)T^{\mu_4\nu_4}(x_4)}\rangle+(13)+(23)+(14)+(24)+(34)\bigg\}\notag\\
&-2\bigg\{\left[\Gamma^{\mu}_{\nu\lambda}(x)\right]^{\mu_1\nu_1}(x_1)\langle{T^{\lambda\nu}(x)T^{\mu_2\nu_2}(x_2)T^{\mu_3\nu_3}(x_3)T^{\mu_4\nu_4}(x_4)}\rangle+(12)+(13)+(14)\bigg\}\label{5ptcons}
\end{align}
and
\begin{align}
\delta_{\mu\nu}\langle{T^{\mu\nu}(x)T^{\mu_1\nu_1}(x_1)\dots T^{\mu_4\nu_4}(x_4)}\rangle&=-2\bigg\{\delta_{xx_1}\langle{T^{\mu_1\nu_1}(x)T^{\mu_2\nu_2}(x_2)\dots T^{\mu_4\nu_4}(x_4)}\rangle+(12)+(13)+(14)\bigg\}\notag\\
&\hspace{2cm}+2^4\big[\mathcal{A}(x)\big]^{\mu_1\nu_1\dots\mu_4\nu_4}(x_1,\dots,x_4).\label{5pttrace}
\end{align}
to finally derive the dilatation WI from \eqref{anomDil} in the form
\begin{align}
\left(4d+\sum_{j=1}^4\,x_j^\alpha\frac{\partial}{\partial x_j^\alpha}\right)\langle{T^{\mu_1\nu_1}(x_1)\dots T^{\mu_4\nu_4}(x_4)}\rangle=2^4\int dx \big[\mathcal{A}(x)\big]^{\mu_1\nu_1\dots\mu_4\nu_4}(x_1,\dots,x_4)\label{DilAnom}
\end{align}
where $d=4$. It is worth mentioning that \eqref{DilAnom} is valid in any even spacetime dimension if we take into account the particular structure of the trace anomaly in that particular dimension.

The special CWIs correspond to the $d$ special conformal Killing vectors in flat space given in \eqref{spc}, as in the $TT$ case. Also in this case we derive the identity 
\begin{align}
0=&\int d^d x\bigg\{\big(2x^\kappa\,x_\mu-x^2\delta^\kappa_\mu\big)\,\partial_\nu\langle{T^{\mu\nu}(x)T^{\mu_1\nu_1}(x_1)\dots T^{\mu_4\nu_4}(x_4)}\rangle\notag\\
&\hspace{4cm}+2x^\kappa\delta_{\mu\nu}\langle{T^{\mu\nu}(x)T^{\mu_1\nu_1}(x_1)\dots T^{\mu_4\nu_4}(x_4)}\rangle\label{anomSpecWI}
\bigg\}.
\end{align}
By using the relations \eqref{5ptcons} and \eqref{5pttrace} and performing the integration over $x$ explicitly in the equation above,  the anomalous special CWIs for the $4$-point function take the form
\begin{align}
&\sum_{j=1}^4\left[2x_j^\kappa\left(d+x_j^\alpha\frac{\partial}{\partial x_j^\alpha}\right)-x_j^2\,\delta^{\kappa\alpha}\frac{\partial}{\partial x_j^\alpha}\right]\langle{T^{\mu_1\nu_1}(x_1)\dots T^{\mu_4\nu_4}(x_4)}\rangle\notag\\
&+2\sum_{j=1}^4\left(\delta^{\kappa\mu_j}x_{j\,\alpha}-\delta^\kappa_\alpha x_j^{\mu_j}\right)\langle{T^{\mu_1\nu_1}(x_1)\dots T^{\nu_j\alpha}(x_j)\dots T^{\mu_4\nu_4}(x_4)}\rangle\notag\\
&+2\sum_{j=1}^4\left(\delta^{\kappa\nu_j}x_{j\,\alpha}-\delta^\kappa_\alpha x_j^{\nu_j}\right)\langle{T^{\mu_1\nu_1}(x_1)\dots T^{\mu_j\alpha}(x_j)\dots T^{\mu_4\nu_4}(x_4)}\rangle=\notag\\
&=2^5\,\int dx\,x^\kappa\big[\mathcal{A}(x)\big]^{\mu_1\nu_1\dots\mu_4\nu_4}(x_1,\dots,x_4),
\end{align}
where the presence of the anomaly term comes from the inclusion of the trace  WI, exactly as in the TT case. \\
At this stage, these equations can be transformed to momentum space, giving the final expressions of the CWIs in the form 
\begin{align}
\left(d-\sum_{j=1}^3\,p_j^\alpha\frac{\partial}{\partial p_j^\alpha}\right)\langle{T^{\mu_1\nu_1}(p_1)T^{\mu_2\nu_2}(p_2)T^{\mu_3\nu_3}(p_3)T^{\mu_4\nu_4}(\bar{p}_4)}\rangle=2^4\,\mathcal{A}^{\mu_1\nu_1\mu_2\nu_2\mu_3\nu_3\mu_4\nu_4}(p_1,p_2,p_3,\bar{p}_4)
\end{align} 
for the dilatation, and 
\begin{align}
&\sum_{j=1}^3\left(p_j^\kappa\frac{\partial^2}{\partial p_j^\alpha \partial p_{j\alpha}}-2p_j^\alpha\frac{\partial}{\partial p_j^\alpha\partial p_{j\kappa}}\right)\langle{T^{\mu_1\nu_1}(p_1)T^{\mu_2\nu_2}(p_2)T^{\mu_3\nu_3}(p_3) T^{\mu_4\nu_4}(\bar{p}_4)}\rangle\notag\\
&+2\sum_{j=1}^3\left(\delta^{\kappa\mu_j}\frac{\partial}{\partial p_{j\,\alpha}}-\delta^\kappa_\alpha \frac{\partial}{\partial p_j^{\mu_j}}\right)\langle{T^{\mu_1\nu_1}(p_1)\dots T^{\nu_j\alpha}(p_j)\dots T^{\mu_4\nu_4}(\bar{p}_4)}\rangle\notag\\
&+2\sum_{j=1}^3\left(\delta^{\kappa\nu_j}\frac{\partial}{\partial p_{j\,\alpha}}-\delta^\kappa_\alpha \frac{\partial}{\partial p_j^{\nu_j}}\right)\langle{T^{\mu_1\nu_1}(p_1)\dots T^{\mu_j\alpha}(p_j)\dots T^{\mu_4\nu_4}(\bar{p}_4)}\rangle\notag\\
&=-2^5\,\left[\frac{\partial}{\partial p_{4\kappa}}\,\mathcal{A}^{\mu_1\nu_1\mu_2\nu_2\mu_3\nu_3\mu_4\nu_4}(p_1,p_2,p_3,p_4)\right]_{\bar{p_4}=-p_1-p_2-p_3},
\end{align}
for the special CWIs, having used the definition \eqref{defVarTrace}. \\
When this procedure is applied to the $n$-point function, one finds that the anomalous CWIs are written as
\begin{align}
\left(d-\sum_{j=1}^{n-1}\,p_j^\alpha\frac{\partial}{\partial p_j^\alpha}\right)\langle{T^{\mu_1\nu_1}(p_1)\dots T^{\mu_n\nu_n}(\bar{p}_n)}\rangle=2^n\,\mathcal{A}^{\mu_1\nu_1\dots\mu_n\nu_n}(p_1,\dots,\bar{p}_n)
\end{align}
for the dilatation and 
	\begin{align}
		&\sum_{j=1}^{n-1}\left(p_j^\kappa\frac{\partial^2}{\partial p_j^\alpha \partial p_{j\alpha}}-2p_j^\alpha\frac{\partial}{\partial p_j^\alpha\partial p_{j\kappa}}\right)\langle{T^{\mu_1\nu_1}(p_1)\dots T^{\mu_n\nu_n}(\bar{p}_n)}\rangle\notag\\
		&+2\sum_{j=1}^{n-1}\left(\delta^{\kappa\mu_j}\frac{\partial}{\partial p_{j\,\alpha}}-\delta^\kappa_\alpha \frac{\partial}{\partial p_j^{\mu_j}}\right)\langle{T^{\mu_1\nu_1}(p_1)\dots T^{\nu_j\alpha}(p_j)\dots T^{\mu_n\nu_n}(\bar{p}_n)}\rangle\notag\\
		&+2\sum_{j=1}^{n-1}\left(\delta^{\kappa\nu_j}\frac{\partial}{\partial p_{j\,\alpha}}-\delta^\kappa_\alpha \frac{\partial}{\partial p_j^{\nu_j}}\right)\ \langle T^{\mu_1\nu_1}(p_1)\dots T^{\mu_j\alpha}(p_j)\dots T^{\mu_n\nu_n}(\bar{p}_n)\rangle\notag\\
		&=-2^{n+1}\left[\frac{\partial}{\partial p_{n\kappa}}\,\mathcal{A}^{\mu_1\nu_1\dots\mu_n\nu_n}(p_1,\dots,p_n)\right]_{p_n=\bar{p}_n},
	\end{align}
for the special conformal Ward identities, where $\bar{p}_n=-\sum_{i=1}^{n-1}p_i$ and we have used the definition \eqref{defVarTrace}. 

\section{Decomposition of the 4T} 
The analysis in momentum space allows to identify the contributions generated by the breaking of the conformal symmetry, after renormalization, in a direct manner. For this purpose we will be using the longitudinal transverse L/T decomposition of the correlator presented in \cite{Bzowski:2013sza} for 3-point functions, extending it to the 4T. This procedure has been investigated in detail for 3-point functions in \cite{Coriano:2018bsy1} in the context of a perturbative approach \cite{Coriano:2018zdo}. The perturbative analysis in free field theory shows how renormalization acts on the two L/T subspaces, forcing the emergence of a trace in the longitudinal sector.   \\
Due to the constraint imposed by conformal symmetry (i.e. their CWIs), the correlation functions can be decomposed into a transverse-traceless and a semilocal part. The term semilocal refers to contributions which are obtained from the conservation and trace Ward identities. Of an external off- shell graviton only its spin-2 component will couple to transverse-traceless part. \\
In general, by decomposing the gravitational fluctuations into their transverse-traceless and spin-1 and spin-0 components one finds an interesting separation of the anomaly effective action which can be useful also in a phenomenological context. We will address this point in a forthcoming paper.  \\
The split of the energy momentum operator in terms of a transverse traceless ($tt$) part and of a longitudinal (local) part \cite{Bzowski:2013sza} is defined in the form
\begin{equation}
T^{\mu_i\nu_i}(p_i)\equiv t^{\mu_i\nu_i}(p_i)+t_{loc}^{\mu_i\nu_i}(p_i)\label{decT}
\end{equation}
with
\begin{align}
\label{loct}
t^{\mu_i\nu_i}(p_i)&=\Pi^{\mu_i\nu_i}_{\alpha_i\beta_i}(p)\,T^{\alpha_i \beta_i}(p_i)\\
t_{loc}^{\mu_i\nu_i}(p_i)&=\Sigma^{\mu_i\nu_i}_{\alpha_i\beta_i}(p)\,T^{\alpha_i \beta_i}(p_i).
\end{align}
We have introduced the transverse-traceless ($\Pi$), transverse-trace $(\tau)$ and 
longitudinal ($\mathcal{I}$) projectors, given respectively by 
\begin{align}
\label{prozero}
\pi^{\mu}_{\alpha} & = \delta^{\mu}_{\alpha} - \frac{p^{\mu} p_{\alpha}}{p^2},  \qquad \tilde{\pi}^{\mu}_{\alpha} =\frac{1}{d-1}\pi^{\mu}_{\alpha} \\\
\Pi^{\mu \nu}_{\alpha \beta} & = \frac{1}{2} \left( \pi^{\mu}_{\alpha} \pi^{\nu}_{\beta} + \pi^{\mu}_{\beta} \pi^{\nu}_{\alpha} \right) - \frac{1}{d - 1} \pi^{\mu \nu}\pi_{\alpha \beta}\label{TTproj}, 
\end{align}
\begin{align}
\mathcal{J}^{\mu\nu}_{\alpha\beta}&=\frac{1}{p^2}p_{\beta}\left( p^{\mu}\delta^{\nu}_{\alpha} +p^{\nu}\delta^{\mu}_{\alpha} -
\frac{p_{\alpha}}{d-1}( \delta^{\mu\nu} +(d-2)\frac{p^\mu p^\nu}{p^2})    \right)\\
\mathcal{I}^{\mu\nu}_{\alpha\beta}&=\frac{1}{2}\left(\mathcal{J}^{\mu\nu}_{\alpha\beta} +\mathcal{J}^{\mu\nu}_{\beta\alpha}\right) \qquad \tau^{\mu\nu}_{\alpha\beta} =\tilde{\pi}^{\mu \nu}\delta_{\alpha \beta}\\
\mathcal{I}^{\mu\nu}_{\alpha}&=\frac{1}{p^2}\left( p^{\mu}\delta^{\nu}_{\alpha} +
p^{\nu}\delta^{\mu}_{\alpha} -
\frac{p_{\alpha}}{d-1}( \delta^{\mu\nu} +(d-2)\frac{p^\mu p^\nu}{p^2}  \right)\label{proone}\\
\mathcal{I}^{\mu\nu}_{\alpha\beta}&=\frac{1}{2}\left(p_\beta \mathcal{I}^{\mu\nu}_{\alpha}
+ p_\alpha \mathcal{I}^{\mu\nu}_{\beta}\right)\label{protwo}
\end{align}
with 
\begin{align}
\delta^{(\mu}_\alpha\delta^{\nu)}_{\beta}&=\Pi^{\mu \nu}_{\alpha \beta} +\Sigma^{\mu\nu}_{\alpha\beta} \\
\Sigma^{\mu_i\nu_i}_{\alpha_i\beta_i}&\equiv\mathcal{I}^{\mu_i\nu_i}_{\alpha_i\beta_i} +\tau^{\mu_i\nu_i}_{\alpha_i\beta_i}\notag\\
&=\frac{1}{p_i^2}\Big[2\delta^{(\nu_i}_{(\alpha_i}p_i^{\mu_i)}p_{i\,\beta_i)}-\frac{p_{i\alpha_i}p_{i\beta_i}}{(d-1)}\left(\delta^{\mu_i\nu_i}+(d-2)\frac{p_i^{\mu_i}p_i^{\nu_i}}{p_i^2}\right)\Big]+\frac{1}{(d-1)}\pi^{\mu_i\nu_i}(p_i)\delta_{\alpha_i\beta_i}\label{Lproj}.
\end{align}
Notice that we have combined together the operators $\mathcal{I}$ and $\tau$ into a projector 
$\Sigma$ which defines the local components of a given tensor $T$, according to \eqref{loct}, which are proportional both to a given momentum $p$ (the longitudinal contribution) and to the trace parts. Both $\Pi$ and $\tau$ are transverse by construction, while $\mathcal{I}$ is longitudinal and of zero trace.

The projectors induce a decomposition respect to a specific momentum $p_i$.
By using \eqref{decT}, the entire correlator is written as
\begin{align}
&\langle{T^{\mu_1\nu_1}(p_1)T^{\mu_2\nu_2}(p_2)T^{\mu_3\nu_3}(p_3)T^{\mu_4\nu_4}(\bar{p}_4)}\rangle=\notag\\
&=\langle{t^{\mu_1\nu_1}(p_1)t^{\mu_2\nu_2}(p_2)t^{\mu_3\nu_3}(p_3)t^{\mu_4\nu_4}(\bar{p}_4)}\rangle+\langle{T^{\mu_1\nu_1}(p_1)T^{\mu_2\nu_2}(p_2)T^{\mu_3\nu_3}(p_3)T^{\mu_4\nu_4}(\bar{p}_4)}\rangle_{loc}\label{decTTTT}
\end{align} 
where the first contribution is the transverse-traceless part which satisfies  by construction the conditions
\begin{equation}
\begin{split}
p_{i\,\mu_i}\langle{t^{\mu_1\nu_1}(p_1)t^{\mu_2\nu_2}(p_2)t^{\mu_3\nu_3}(p_3)t^{\mu_4\nu_4}(\bar{p}_4)}\rangle&=0,\qquad i=1,2,3,4\,,\\
\delta_{\mu_i\nu_i}\langle{t^{\mu_1\nu_1}(p_1)t^{\mu_2\nu_2}(p_2)t^{\mu_3\nu_3}(p_3)t^{\mu_4\nu_4}(\bar{p}_4)}\rangle&=0,\qquad i=1,2,3,4.
\end{split}\label{ttproperties}
\end{equation}
It is clear now that only the second term in \eqref{decTTTT} contributes entirely to the conservation WIs. Thus, the proper new information on the form factors of the 4-point function is entirely encoded in its transverse-traceless ($tt$) part, since the remaining longitudinal + trace contributions,  corresponding to the local term, are related to lower point functions. 

\section{Projecting the Conformal Ward Identities}
The action of $D$ and $\mathcal{K}$ on the 4T in \eqref{Dequ} and \eqref{Keq}, after the projection on the transverse traceless component, simplifies.
We start by considering at the dilatation operator $D$ which has the property of leaving unchanged the two subspaces identified by the $\Pi$ and $\Sigma$ projectors. 
These properties can be summarized by the relations
\begin{equation}
\begin{split}
		&\Sigma^{\mu_i\nu_i}_{\rho_i\sigma_i}(p_i)\,D\,\langle{t^{\mu_1\nu_1}(p_1)t^{\mu_2\nu_2}(p_2)t^{\mu_3\nu_3}(p_3)t^{\mu_4\nu_4}(p_4)}\rangle=0,\\
		&\Pi^{\mu_i\nu_i}_{\rho_i\sigma_i}(p_i)\,D\,\langle{T^{\mu_1\nu_1}(p_1)T^{\mu_2\nu_2}(p_2)T^{\mu_3\nu_3}(p_3)T^{\mu_4\nu_4}(p_4)}\rangle_{loc}=0,
\end{split}	\qquad\qquad  i=1,\dots,4.\label{projD}
\end{equation}
having used the properties of orthogonality and idempotence of the projectors. \\
For this reason, if one wants to project the dilatation equation \eqref{Dequ}, by using the transverse-traceless and longitudinal projectors, there are only two ways of doing it. These are the cases where we have either four $\Pi$'s or four $\Sigma$'s, due to the relations in \eqref{projD}. Therefore, the only relevant projected dilatation WIs are
\begin{align}
	\Pi^{\mu_1\nu_1}_{\alpha_1\beta_1}(p_1)\Pi^{\mu_2\nu_2}_{\alpha_2\beta_2}(p_2)\Pi^{\mu_3\nu_3}_{\alpha_3\beta_3}(p_3)\Pi^{\mu_4\nu_4}_{\alpha_4\beta_4}(\bar p_4)\,D\,\langle{T^{\alpha_1\beta_1}(p_1)T^{\alpha_2\beta_2}(p_2)T^{\alpha_3\beta_3}(p_3)T^{\alpha_4\beta_4}(\bar{p}_4)}\rangle&=0\\
	\Sigma^{\mu_1\nu_1}_{\alpha_1\beta_1}(p_1)\Sigma^{\mu_2\nu_2}_{\alpha_2\beta_2}(p_2)\Sigma^{\mu_3\nu_3}_{\alpha_3\beta_3}(p_3)\Sigma^{\mu_4\nu_4}_{\alpha_4\beta_4}(\bar p_4)\,D\,\langle{T^{\alpha_1\beta_1}(p_1)T^{\alpha_2\beta_2}(p_2)T^{\alpha_3\beta_3}(p_3)T^{\alpha_4\beta_4}(\bar{p}_4)}\rangle&=0
\end{align}
that can be simplified as
\begin{align}
	\Pi^{\mu_1\nu_1}_{\alpha_1\beta_1}(p_1)\Pi^{\mu_2\nu_2}_{\alpha_2\beta_2}(p_2)\Pi^{\mu_3\nu_3}_{\alpha_3\beta_3}(p_3)\Pi^{\mu_4\nu_4}_{\alpha_4\beta_4}(\bar p_4)\,D\,\langle{t^{\alpha_1\beta_1}(p_1)t^{\alpha_2\beta_2}(p_2)t^{\alpha_3\beta_3}(p_3)t^{\alpha_4\beta_4}(\bar{p}_4)}\rangle&=0,\label{Dtttt}\\[2ex]
	\Sigma^{\mu_1\nu_1}_{\alpha_1\beta_1}(p_1)\Sigma^{\mu_2\nu_2}_{\alpha_2\beta_2}(p_2)\Sigma^{\mu_3\nu_3}_{\alpha_3\beta_3}(p_3)\Sigma^{\mu_4\nu_4}_{\alpha_4\beta_4}(\bar p_4)\,D\,\langle{T^{\alpha_1\beta_1}(p_1)T^{\alpha_2\beta_2}(p_2)T^{\alpha_3\beta_3}(p_3)T^{\alpha_4\beta_4}(\bar{p}_4)}\rangle_{loc}&=0\label{DTTTTloc},
\end{align}
once we insert the decomposition of the 4T and use Eqs. \eqref{projD}. It is worth mentioning that \eqref{DTTTTloc} does not impose any additional constraints on the $4$-point function. This because the longitudinal part of the correlator is explicitly given in terms of lower point functions, and \eqref{DTTTTloc} is related to the dilatation WIs of the $3$- and $2$-point functions. The constraints on the 4-point function will be derived from \eqref{Dtttt}, which is related to the transverse traceless part of the correlator. 

Turning our attention towards the special CWIs, we observe that the action of the $\mathcal{K}^\kappa$ operator on the transverse-traceless part gives a result that it is still transverse and traceless
\begin{equation}
	\begin{split}
		&p_{i\mu_i}\,\mathcal{K}^\kappa\langle{t^{\mu_1\nu_1}(p_1)t^{\mu_2\nu_2}(p_2)t^{\mu_3\nu_3}(p_3)t^{\mu_4\nu_4}(p_4)}\rangle=0,\\
		&\delta_{\mu_i\nu_i}\,\mathcal{K}^\kappa\langle{t^{\mu_1\nu_1}(p_1)t^{\mu_2\nu_2}(p_2)t^{\mu_3\nu_3}(p_3)t^{\mu_4\nu_4}(p_4)}\rangle=0
	\end{split}
	\qquad i=1,\dots,4 \label{propKTTT}
\end{equation}
This can be shown exactly as in  the case of the 3T discussed in previous works. 
This property of $\mathcal{K}^\kappa$ allows us to identify the  relevant subspace where the special CWIs act. \\
As already pointed out, the transverse-traceless part of the 4T is the only part of the  correlator which is really unknown. Indeed, the longitudinal components can be expressed in terms of 2- and 3-point functions, by using the conservation and trace WIs.  Therefore, the  special CWIs will be constraining that unknown part, which can be parametrized in terms of a certain number of independent for factors, as in the 3T case. \\
By using the properties of the projectors $\Sigma$ and $\Pi$, from \eqref{propKTTT} one derives the relation
\begin{equation}
	\Sigma^{\rho_i\sigma_i}_{\mu_i\nu_i}(p_i)K^\kappa\langle{t^{\mu_1\nu_1}(p_1)t^{\mu_2\nu_2}(p_2)t^{\mu_3\nu_3}(p_3)t^{\mu_4\nu_4}(p_4)}\rangle=0, \qquad i=1,\dots,4,
\end{equation}
since the action of $K^\kappa$, as just mentioned, is endomorphic on the transverse traceless sector. 
With this result in mind, using the projectors $\Pi$ and $\Sigma$ we project \eqref{Keq} into all the possible subspaces and observe that when at least one $\Sigma$ is present, the equations reduce to the form
\begin{align}
	0&=\Sigma^{\rho_i\sigma_i}_{\mu_i\nu_i}(p_i)\mathcal{K}^\kappa\langle{T^{\mu_1\nu_1}(p_1)T^{\mu_2\nu_2}(p_2)T^{\mu_3\nu_3}(p_3)T^{\mu_4\nu_4}(p_4)}\rangle\notag\\
	&=\Sigma^{\rho_i\sigma_i}_{\mu_i\nu_i}(p_i)\mathcal{K}^\kappa\bigg[\langle{t^{\mu_1\nu_1}(p_1)t^{\mu_2\nu_2}(p_2)t^{\mu_3\nu_3}(p_3)t^{\mu_4\nu_4}(p_4)}\rangle+\langle{T^{\mu_1\nu_1}(p_1)T^{\mu_2\nu_2}(p_2)T^{\mu_3\nu_3}(p_3)T^{\mu_4\nu_4}(p_4)}\rangle_{loc}\bigg]\notag\\
	&=\Sigma^{\rho_i\sigma_i}_{\mu_i\nu_i}(p_i)\mathcal{K}^\kappa\bigg(\langle{T^{\mu_1\nu_1}(p_1)T^{\mu_2\nu_2}(p_2)T^{\mu_3\nu_3}(p_3)T^{\mu_4\nu_4}(p_4)}\rangle_{loc}\bigg), \qquad i=1,\dots,4.
\end{align}
The equations that result
\begin{equation}
\label{ccc}
	\Sigma^{\rho_i\sigma_i}_{\mu_i\nu_i}(p_i)\mathcal{K}^\kappa\bigg(\langle{T^{\mu_1\nu_1}(p_1)T^{\mu_2\nu_2}(p_2)T^{\mu_3\nu_3}(p_3)T^{\mu_4\nu_4}(p_4)}\rangle_{loc}\bigg)=0, \qquad  i=1,\dots,4
\end{equation}
will involve only 3- and 2-point functions, via the canonical Ward identities. For this reason, being the structure of the equations hierarchical, if we have already solved for the correlators of lower orders, no new constraint will be induced by \eqref{ccc}.\\
 The only  significant constraint will be derived when acting with 4 $\Pi$ projectors. For this reason we are interested in studying the equation
\begin{equation}
\label{ddd}
\Pi^{\mu_1\nu_1}_{\alpha_1\beta_1}(p_1)\Pi^{\mu_2\nu_2}_{\alpha_2\beta_2}(p_2)\Pi^{\mu_3\nu_3}_{\alpha_3\beta_3}(p_3)\Pi^{\mu_4\nu_4}_{\alpha_4\beta_4}(\bar p_4)\,\mathcal{K}^\kappa\,\langle{T^{\alpha_1\beta_1}(p_1)T^{\alpha_2\beta_2}(p_2)T^{\alpha_3\beta_3}(p_3)T^{\alpha_4\beta_4}(\bar{p}_4)}\rangle=0.
\end{equation}
If we insert the decomposition \eqref{decTTTT} into the equation above, one can prove that terms containing two or more $t_{loc}$ operators will vanish when projected on the transverse traceless component, for instance
\begin{align}
	\Pi^{\mu_1\nu_1}_{\alpha_1\beta_1}(p_1)\Pi^{\mu_2\nu_2}_{\alpha_2\beta_2}(p_2)\Pi^{\mu_3\nu_3}_{\alpha_3\beta_3}(p_3)\Pi^{\mu_4\nu_4}_{\alpha_4\beta_4}(\bar p_4)\,\mathcal{K}^{\kappa}\,\langle{t^{\alpha_1\beta_1}_{loc}(p_1)t^{\alpha_2\beta_2}_{loc}(p_2)T^{\alpha_3\beta_3}(p_3)T^{\alpha_4\beta_4}(\bar{p}_4)}\rangle&=0,
\end{align}
where the definitions \eqref{TTproj} and \eqref{Lproj} have been used. 

Proceeding with the reconstruction program, we need the action of the special conformal transformations on the correlators with a single $t_{loc}$, after projecting on the transverse traceless sector. After a lengthy but straightforward calculations we obtain
\begin{align}
	&\Pi^{\mu_1\nu_1}_{\alpha_1\beta_1}(p_1)\Pi^{\mu_2\nu_2}_{\alpha_2\beta_2}(p_2)\Pi^{\mu_3\nu_3}_{\alpha_3\beta_3}(p_3)\Pi^{\mu_4\nu_4}_{\alpha_4\beta_4}(\bar p_4)\,\,\mathcal{K}^{\kappa}\langle{T^{\alpha_1\beta_1}(p_1)T^{\alpha_2\beta_2}(p_2)T^{\alpha_3\beta_3}(p_3)t_{loc}^{\alpha_4\beta_4}(\bar{p}_4)}\rangle\notag\\
	&=4\Pi^{\mu_1\nu_1}_{\alpha_1\beta_1}(p_1)\Pi^{\mu_2\nu_2}_{\alpha_2\beta_2}(p_2)\Pi^{\mu_3\nu_3}_{\alpha_3\beta_3}(p_3)\Pi^{\mu_4\nu_4}_{\alpha_4\beta_4}(\bar p_4)\notag\\
	&\bigg\{\bigg[(d-1)\delta^{\kappa\alpha_4}+L_1^{\kappa\alpha_4}+L_2^{\kappa\alpha_4}+L_3^{\kappa\alpha_4}
	\bigg]\left(\frac{p_{4\rho_4}}{p_4^2}\langle{T^{\alpha_1\beta_1}(p_1)T^{\alpha_2\beta_2}(p_2)T^{\alpha_3\beta_3}(p_3)T^{\rho_4\beta_4}(\bar{p}_4)}\rangle\right)\notag\\
	&\hspace{1cm}+\left(S^{\kappa\alpha_4}\right)^{\alpha_1}_{\hspace{2ex}\rho_1}\left(\frac{p_{4\rho_4}}{p_4^2}\langle{T^{\rho_1\beta_1}(p_1)T^{\alpha_2\beta_2}(p_2)T^{\alpha_3\beta_3}(p_3)T^{\rho_4\beta_4}(\bar{p}_4)}\rangle\right)\notag\\
	&\hspace{1cm}+\left(S^{\kappa\alpha_4}\right)^{\alpha_2}_{\hspace{2ex}\rho_2}\left(\frac{p_{4\rho_4}}{p_4^2}\langle{T^{\alpha_1\beta_1}(p_1)T^{\rho_2\beta_2}(p_2)T^{\alpha_3\beta_3}(p_3)T^{\rho_4\beta_4}(\bar{p}_4)}\rangle\right)\notag\\
	&\hspace{1cm}+\left(S^{\kappa\alpha_4}\right)^{\alpha_3}_{\hspace{2ex}\rho_3}\left(\frac{p_{4\rho_4}}{p_4^2}\langle{T^{\alpha_1\beta_1}(p_1)T^{\alpha_2\beta_2}(p_2)T^{\rho_3\beta_3}(p_3)T^{\rho_4\beta_4}(\bar{p}_4)}\rangle\right)\bigg\},
\end{align}
where 
\begin{align}
	L_i^{\kappa\alpha_4}\equiv p_i^{\alpha_4}\frac{\partial}{\partial p_{i\,\kappa}}-p_i^\kappa\frac{\partial}{\partial p_{i\,\alpha_4}},\qquad i=1,2,3
\end{align}
are the $SO(4)$ (Lorentz) generators 
and $S_{\mu\nu}$ is the spin part, for which
\begin{align}
	\big(S^{\mu\nu}\big)^{\rho\sigma}=\delta^{\mu\rho}\delta^{\nu\sigma}-\delta^{\mu\sigma}\delta^{\nu\rho}.
\end{align}
By using the Lorentz Ward identities we obtain the expression
\begin{align}
	&\Pi^{\mu_1\nu_1}_{\alpha_1\beta_1}(p_1)\Pi^{\mu_2\nu_2}_{\alpha_2\beta_2}(p_2)\Pi^{\mu_3\nu_3}_{\alpha_3\beta_3}(p_3)\Pi^{\mu_4\nu_4}_{\alpha_4\beta_4}(\bar p_4)\,\,\mathcal{K}^{\kappa}\langle{T^{\alpha_1\beta_1}(p_1)T^{\alpha_2\beta_2}(p_2)T^{\alpha_3\beta_3}(p_3)t_{loc}^{\alpha_4\beta_4}(\bar{p}_4)}\rangle\notag\\
	&=\Pi^{\mu_1\nu_1}_{\alpha_1\beta_1}(p_1)\Pi^{\mu_2\nu_2}_{\alpha_2\beta_2}(p_2)\Pi^{\mu_3\nu_3}_{\alpha_3\beta_3}(p_3)\Pi^{\mu_4\nu_4}_{\alpha_4\beta_4}(\bar p_4)\left(\frac{4\,d}{p_4^2}\,\delta^{\alpha_4\kappa}\,p_{4\rho_4}\langle{T^{\alpha_1\beta_1}(p_1)T^{\alpha_2\beta_2}(p_2)T^{\alpha_3\beta_3}(p_3)T^{\rho_4\beta_4}(\bar{p}_4)}\rangle\right).
\end{align}
Analogous results are obtained for the other terms involving one $t_{loc}$ operator. \\
In summary, when we project \eqref{Keq} on the transverse traceless components we find
\begin{align}
	0=&\Pi^{\mu_1\nu_1}_{\alpha_1\beta_1}(p_1)\Pi^{\mu_2\nu_2}_{\alpha_2\beta_2}(p_2)\Pi^{\mu_3\nu_3}_{\alpha_3\beta_3}(p_3)\Pi^{\mu_4\nu_4}_{\alpha_4\beta_4}(\bar p_4)\,\,\mathcal{K}^{\kappa}\langle{T^{\alpha_1\beta_1}(p_1)T^{\alpha_2\beta_2}(p_2)T^{\alpha_3\beta_3}(p_3)T^{\alpha_4\beta_4}(\bar{p}_4)}\rangle\notag\\
	&=\Pi^{\mu_1\nu_1}_{\alpha_1\beta_1}(p_1)\Pi^{\mu_2\nu_2}_{\alpha_2\beta_2}(p_2)\Pi^{\mu_3\nu_3}_{\alpha_3\beta_3}(p_3)\Pi^{\mu_4\nu_4}_{\alpha_4\beta_4}(\bar p_4)\Bigg[\mathcal{K}^{\kappa}\langle{t^{\alpha_1\beta_1}(p_1)t^{\alpha_2\beta_2}(p_2)t^{\alpha_3\beta_3}(p_3)t^{\alpha_4\beta_4}(\bar{p}_4)}\rangle\notag\\
	&\hspace{2cm}+\frac{4\,d}{p_1^2}\,\delta^{\alpha_1\kappa}\,p_{1\rho_1}\langle{T^{\rho_1\beta_1}(p_1)T^{\alpha_2\beta_2}(p_2)T^{\alpha_3\beta_3}(p_3)T^{\alpha_4\beta_4}(\bar{p}_4)}\rangle\notag\\
	&\hspace{2cm}+\frac{4\,d}{p_2^2}\,\delta^{\alpha_2\kappa}\,p_{2\rho_2}\langle{T^{\alpha_1\beta_1}(p_1)T^{\rho_2\beta_2}(p_2)T^{\alpha_3\beta_3}(p_3)T^{\alpha_4\beta_4}(\bar{p}_4)}\rangle\notag\\
	&\hspace{2cm}+\frac{4\,d}{p_3^2}\,\delta^{\alpha_3\kappa}\,p_{3\rho_3}\langle{T^{\alpha_1\beta_1}(p_1)T^{\alpha_2\beta_2}(p_2)T^{\rho_3\beta_3}(p_3)T^{\alpha_4\beta_4}(\bar{p}_4)}\rangle\notag\\
	&\hspace{2cm}+\frac{4\,d}{p_4^2}\,\delta^{\alpha_4\kappa}\,p_{4\rho_4}\langle T^{\alpha_1\beta_1}(p_1)T^{\alpha_2\beta_2}(p_2)T^{\alpha_3\beta_3}(p_3)T^{\rho_4\beta_4}(\bar{p}_4)\rangle\Bigg]\label{KTTTTdec}.
\end{align}
It is worth mentioning that the last four terms in the previous equations are completely expressible in  terms of lower point functions via the longitudinal WIs. Then, \eqref{KTTTTdec} imposes some constraints on the transverse traceless part of the 4T and connects this part of the correlator with the lower point functions 3T and $TT$. 

\section{Identifying the divergent form factors and their reduction in $d=4$}
The general expression of the $tt$-contributions can be identified by imposing the transversality and trace-free \eqref{ttproperties} conditions on all the possible tensor structures which are allowed by the symmetries of the correlation function. In this section we are going first to proceed with the classification of the divergent ones which are present in the decomposition and are affected by the renormalization.  \\
We introduce a general decomposition of the counterterm for the 4T in general $d$ dimensions, in order to identify such form factors. Their number gets reduced once we move to $d=4$, due to the possibility of expressing the Kronecker $\delta^{\mu\nu}$ in terms of 3 of the 4 momenta of the correlator, and of a linearly independent 4-vector $n^\mu$. The latter is defined via a generalization of the external product by the $\epsilon^{\mu\nu\rho\sigma}$, as we are going to illustrate below (see \cite{Bzowski:2013sza} for the case $d=3$)

The decomposition can be generically written in the form
\begin{align}
&\langle{t^{\mu_1\nu_1}(p_1)t^{\mu_2\nu_2}(p_2)t^{\mu_3\nu_3}(p_3)t^{\mu_4\nu_4}(\bar{p}_4)}\rangle\equiv \Pi^{\mu_1\nu_1}_{\alpha_1\beta_1}(p_1)\Pi^{\mu_2\nu_2}_{\alpha_2\beta_2}(p_2)\Pi^{\mu_3\nu_3}_{\alpha_3\beta_3}(p_3)\Pi^{\mu_4\nu_4}_{\alpha_4\beta_4}(\bar{p}_4)\notag\\
&\times\Big\{\sum_{a,b\in\{2,3\}}\, \sum_{c,d\in\{3,4\}}\,\sum_{e,f\in\{4,1\}}\,\sum_{g,h\in\{1,2\}}A^{(8p)}_{a\,b\,c\,d\,e\,f\,g\,h}\ \ p_{a}^{\alpha_1}\,p_{b}^{\beta_1}\,p_{c}^{\alpha_2}\,p_{d}^{\beta_2}\,p_{e}^{\alpha_3}\,p_{f}^{\beta_3}\,p_{g}^{\alpha_4}\,p_{h}^{\beta_4}\notag\\
&\qquad+\Bigg[\delta^{\alpha_1\alpha_2}\sum_{a\in\{2,3\}}\sum_{b\in\{3,4\}}\sum_{c,d\in\{4,1\}}\sum_{e,f\in\{1,2\}} A^{(6p)}_{a\,b\,c\,d\,e\,f}\,p_{a}^{\beta_1}\,p_{b}^{\beta_2}\,p_{c}^{\alpha_3}\,p_{d}^{\beta_3}\,p_{e}^{\alpha_4}\,p_{f}^{\beta_4}+\text{(permutations)}\Bigg]\notag\\
&\qquad+\Bigg[\delta^{\alpha_1\alpha_2}\delta^{\beta_1\beta_2}\sum_{a,b\in\{4,1\}}\sum_{c,d\in\{1,2\}} A^{(I,4p)}_{a\,b\,c\,d}\,p_{a}^{\alpha_3}\,p_{b}^{\beta_3}\,p_{c}^{\alpha_4}\,p_{d}^{\beta_4}+\text{(permutations)}
\Bigg]\notag\\
&\qquad+\Bigg[\delta^{\alpha_1\alpha_2}\delta^{\beta_2\beta_3}\sum_{a\in\{2,3\}}\sum_{b\in\{4,1\}}\sum_{c,d\in\{1,2\}} A^{(II,4p)}_{a\,b\,c\,d}\,p_{a}^{\beta_1}\,p_{b}^{\alpha_3}\,p_{c}^{\alpha_4}\,p_{d}^{\beta_4}+\text{(permutations)}
\Bigg]\notag\\
&\qquad+\Bigg[\delta^{\alpha_1\alpha_2}\delta^{\alpha_3\alpha_4}\sum_{a\in\{2,3\}}\sum_{b\in\{3,4\}}\sum_{c\in\{4,1\}}\sum_{d\in\{1,2\}} A^{(III,4p)}_{a\,b\,c\,d}\,p_{a}^{\beta_1}\,p_{b}^{\beta_2}\,p_{c}^{\beta_3}\,p_{d}^{\beta_4}+\text{(permutations)}
\Bigg]\notag\\
&\qquad+\Bigg[\delta^{\alpha_1\beta_2}\delta^{\alpha_2\beta_3}\delta^{\alpha_3\beta_1}\sum_{a,b\in\{1,2\}} A^{(I,2p)}_{a\,b}\,p_{a}^{\alpha_4}\,p_{b}^{\beta_4}+\text{(permutations)}
\Bigg]\notag\\
&\qquad+\Bigg[\delta^{\alpha_1\beta_2}\delta^{\alpha_2\beta_1}\delta^{\alpha_3\alpha_4}\sum_{a\in\{4,1\}}\sum_{b\in\{1,2\}} A^{(II,2p)}_{a\,b}\,p_{a}^{\beta_3}\,p_{b}^{\beta_4}+\text{(permutations)}
\Bigg]\notag\\
&\qquad+\Bigg[\delta^{\alpha_1\beta_2}\delta^{\alpha_2\beta_4}\delta^{\alpha_4\beta_3}\sum_{a\in\{2,3\}}\sum_{b\in\{4,1\}} A^{(III,2p)}_{a\,b}\,p_{a}^{\beta_1}\,p_{b}^{\alpha_3}+\text{(permutations)}
\Bigg]\notag\\
&\qquad+\Bigg[\delta^{\alpha_1\beta_2}\delta^{\alpha_2\beta_3}\delta^{\alpha_3\beta_4}\delta^{\alpha_4\beta_1} A^{(I,0p)}+\text{(permutations)}\Bigg]+\Bigg[\delta^{\alpha_1\beta_2}\delta^{\alpha_2\beta_1}\delta^{\alpha_3\beta_4}\delta^{\alpha_4\beta_3} A^{(II,0p)}+\text{(permutations)}\Bigg]
\Bigg\}\label{tttt}
\end{align}  
in terms of form factors $A$. We have explicitly labeled the form factors with an index counting the number of momenta that each of them multiplies in the decomposition. This notation will be useful in our discussion below. 
The choice of the independent momenta of the expansion, similarly to the case of 3-point functions, can be different for each set of uncontracted tensor indices. We will choose  
\begin{align}
&(\alpha_1,\beta_1)\leftrightarrow p_2,p_3,\quad(\alpha_2,\beta_2)\leftrightarrow p_3,p_4\notag\\
&(\alpha_3,\beta_3)\leftrightarrow p_4,p_1,\quad(\alpha_4,\beta_4)\leftrightarrow p_1,p_2.
\end{align}
as basis of the expansion for each pair of indices shown above.
The linear dependence of $p_4$, which we will impose at a later stage, is not in contradiction with this choice, which allows to reduce the number of form factors, due to the presence of a single $tt$ projector for each external momentum. This strategy has been introduced in \cite{Bzowski:2013sza} for 3-point functions and it allows to reduce the number of form factors. These, in eq. \eqref{tttt} are functions of the six kinematic invariants
\begin{align}
p_i=\sqrt{p_i^2},\quad\,i=1,2,3,4,\qquad
s=\sqrt{(p_1+p_2)^2},\ t=\sqrt{(p_2+p_3)^2} 
\end{align}
or equivalently, in a completely symmetric formulation, they are functions of the six invariants
\begin{align}
s_{ij}=\sqrt{p_i\cdot p_j},\quad \,i\ne j\in\{1,2,3,4\}. 
\end{align}
As already mentioned, the local part of the 4T can be expressed entirely in terms of three- and two-point 
functions due to the transverse and trace WIs. The explicit form of the local contribution is indeed given by the expression
\begin{align}
&\langle{T^{\mu_1\nu_1}(p_1)T^{\mu_2\nu_2}(p_2)T^{\mu_3\nu_3}(p_3)T^{\mu_4\nu_4}(\bar{p}_4)}\rangle_{loc}=\notag\\ &=\Bigg[\langle{t_{loc}^{\mu_1\nu_1}(p_1)T^{\mu_2\nu_2}(p_2)T^{\mu_3\nu_3}(p_3)T^{\mu_4\nu_4}(\bar{p}_4)}\rangle+(1\,2)+(1\, 3)+(1\, 4)\Bigg]\notag\\
&-\Bigg[\langle{t_{loc}^{\mu_1\nu_1}(p_1)t_{loc}^{\mu_2\nu_2}(p_2)T^{\mu_3\nu_3}(p_3)T^{\mu_4\nu_4}(\bar{p}_4)}\rangle+(1\,3)+(1\,4)+(2\,3)+(2\,4)+(1\,3)(2\,4)\Bigg]\notag\\
&+\Bigg[\langle{t_{loc}^{\mu_1\nu_1}(p_1)t_{loc}^{\mu_2\nu_2}(p_2)t_{loc}^{\mu_3\nu_3}(p_3)T^{\mu_4\nu_4}(\bar{p}_4)}\rangle+(1\,4)+(2\,4)+(3\,4)\Bigg]\notag\\
&-\langle{t_{loc}^{\mu_1\nu_1}(p_1)t_{loc}^{\mu_2\nu_2}(p_2)t_{loc}^{\mu_3\nu_3}(p_3)t_{loc}^{\mu_4\nu_4}(\bar{p}_4)}\rangle\label{longTer},
\end{align}
where the insertion of $t_{loc}$ gives 
\begin{align}
&\langle{t_{loc}^{\mu_1\nu_1}(p_1)T^{\mu_2\nu_2}(p_2)T^{\mu_3\nu_3}(p_3)T^{\mu_4\nu_4}(\bar{p}_4)}\rangle\notag\\
&=\frac{1}{p_1^2}\Big[2\delta^{(\nu_1}_{(\alpha_1}p_1^{\mu_1)}p_{1\,\beta_1)}-\frac{p_{1\alpha_1}p_{1\beta_1}}{(d-1)}\left(\delta^{\mu_1\nu_1}+(d-2)\frac{p_1^{\mu_1}p_1^{\nu_1}}{p_1^2}\right)\Big]\langle{T^{\alpha_1\beta_1}(p_1)T^{\mu_2\nu_2}(p_2)T^{\mu_3\nu_3}(p_3)T^{\mu_4\nu_4}(\bar{p}_4)}\rangle\notag\\
&\qquad+\frac{1}{(d-1)}\pi^{\mu_1\nu_1}(p_1)\delta_{\alpha_1\beta_1}\langle{T^{\alpha_1\beta_1}(p_1)T^{\mu_2\nu_2}(p_2)T^{\mu_3\nu_3}(p_3)T^{\mu_4\nu_4}(\bar{p}_4)}\rangle\label{tloc1}.
\end{align}
Notice that the right-hand-side of \eqref{tloc1} is entirely expressed in terms of lower-point correlation functions, due to the WI's \eqref{transverseP} and \eqref{traceP}. Similar relations hold for all the other contributions contained in \eqref{longTer}.

\section{Divergences and Renormalization}
In order to investigate the implications of the CWIs on the anomaly contributions of the 4T, 
we turn to 4 spacetime dimensions and discuss the anomaly form of such equations. We start from the dilatation WIs.\\
The scale invariance of the correlator is expressed through the Dilatation Ward Identity \eqref{Dequ},
which in terms of the corresponding form factors takes to scalar equations of the form
\begin{align}
\left(d-n_p-\sum_{i=1}^3\,p_i^\mu\frac{\partial}{\partial p_i^\mu}\right)A^{(np)}_{a\dots}=0,\label{DA}
\end{align}
where $n_p$ is the number of momenta multiplying the form factors in the decomposition \eqref{tttt}. Eqs.  \eqref{DA} characterize the scaling behaviour of the form factors, and allow to identify quite easily those among all which will be manifestly divergent in the UV. For instance, the form factor  corresponding to eight momenta in \eqref{tttt} has degree $d-8$ and is finite in $d=4$. This simple dimensional counting can be done for all the form factors allowed by the symmetry of the correlator.\\
We have summarised the UV behaviour in the table below
\begin{center}
	\begin{tabular}{  c | c | c | c | c | c }
		 Form Factor &$A^{(8p)}$&$A^{(6p)}$&$A^{(4p)}$&$A^{(2p)}$&$A^{(0p)}$\\ 
		\hline Degree &$d-8$ & $d-6$ & $d-4$ & $d-2$ &$d$		
		\\ 	\hline
		UV divergent in $d=4$ & \xmark & \xmark & \cmark & \cmark &\cmark \\
	\end{tabular}
\end{center}

The expected form factors that will manifest divergences in $d=4$ are those of the form $A^{(4p)}$, $A^{(2p)}$ and $A^{(0p)}$ in \eqref{tttt}, which will show up as single poles in the regulator $\varepsilon$. The procedure of renormalization, obtained by the inclusion of the counterterm \eqref{counter}, will remove these divergences and will generate an anomaly. An explicit check of this cancellation is contained in \cite{Serino:2020pyu}, in the case of a conformally coupled 
free scalar theory.
\section{Explicit form of the divergences}
Being the anomaly generated by the renormalization procedure, it is possible to derive the structure of the anomaly contributions and the form of the anomalous CWIs' by applying the reconstruction procedure to the counterterms. \\
One can also work out the explicit structure of the counterterms for each of the divergent form factors $A^{(np)}$ identified above. Their renormalization is obviously guaranteed by the general counterterm Lagrangian \eqref{counter}. 
For example, considering the decomposition identified in \eqref{tttt}, the corresponding counterterms to the $A^{(np)}$ form factors, expanded in series of power in $\varepsilon=d+4$, can be determined in the form 
\begin{align}
A_{4411,\,count}^{(I,4p)}=&\frac{2}{\varepsilon}\,b'
+O(\varepsilon^0)\\
A_{2411,\,count}^{(II,4p)}=&\frac{4}{\varepsilon}\,b'+O(\varepsilon^0)\\
A_{2341,\,count}^{(III,4p)}=&-\frac{5}{8\,\varepsilon}\,\left(b-6\,b'\right)+O(\varepsilon^0)\\
A_{11,\,count}^{(I,2p)}=&-\frac{7}{16\,\varepsilon}\bigg[b\left(p_1^2-4p_2^2+p_3^2+3s^2+3t^2\right) - b' \left(p_1^2+4
p_2^2+p_3^2+4 p_4^2-s^2-t^2\right)\bigg]+O(\varepsilon^0)\\
A_{41\,,count}^{(II,2p)}=&-\frac{7}{8\,\varepsilon}\bigg[\frac{b}{6}\left(7p_1^2+7p_2^2+6p_3^2+6p_4^2-9s^2\right)
+b' \left(t^2-s^2\right)\bigg]+O(\varepsilon^0)\\
A_{34\,,count}^{(III,2p)}=&\frac{1}{2\,\varepsilon}\bigg[b\left(p_1^2-p_2^2-2p_3^2-6p_4^2-s^2\right)+ b' \left(3 p_1^2+p_2^2+2p_3^2-2 p_4^2+s^2\right)\bigg]+O(\varepsilon^0)\\
A^{(I,0p)}_{counter}=&-\frac{1}{\varepsilon}\Bigg\{\frac{b}{8}\bigg[12 p_1^4+p_1^2 \Big(2
(p_2^2+p_4^2)+8 p_3^2-10 \left(s^2+t^2\right)\Big)+12p_2^4+p_2^2 \Big(2 p_3^2+8 p_4^2-10 \left(s^2+t^2\right)\Big)\notag\\
&\hspace{1cm}+p_4^2 \Big(2 p_3^2-10\left(s^2+t^2\right)\Big)
+12 p_3^4-10\,p_3^2 \left(s^2+t^2\right)
+12p_4^4+10 s^4+4s^2 t^2+10t^4\bigg]\notag\\
&\hspace{0.8cm}+\frac{b'}{4}\bigg[-p_1^2
\left(p_2^2+p_4^2+3 \left(s^2+t^2\right)\right)+2
\left(p_1^2\right)^2-p_2^2 \left(p_3^2+3
\left(s^2+t^2\right)\right)+2
\left(p_2^2\right)^2\notag\\
&\hspace{1.5cm}-p_4^2 \left(p_3^2+3
\left(s^2+t^2\right)\right)+p_3^2 \left(2 p_3^2-3
\left(s^2+t^2\right)\right)+2 \left(p_4^2\right)^2+3
s^4+2 s^2 t^2+3 t^4\bigg]\Bigg\}+O(\varepsilon^0)\\
A^{(II,0p)}_{count}=&\frac{1}{24\varepsilon}\Bigg\{b \bigg[-\left(p_1^2+p_2^2\right) \left(p_3^2+p_4^2+9 s^2\right)+6 p_1^4+6 p_2^4-9 p_3^2 s^2+6 p_3^4-9 p_4^2 s^2+6 p_4^4+15 s^4\bigg]\notag\\
&\hspace{-0.5cm}-6 b' \bigg[p_1^2\,p_2^2+p_4^2\,p_3^2 -s^2 t^2-2 s^4-t^4-p_1^4-p_2^4-p_3^4-p_4^4+\left(p_1^2+p_2^2+p_3^2+p_4^2\right) \left(2 s^2+t^2\right)\bigg]\Bigg\}+O(\varepsilon^0),
\end{align}
where the finite contributions contain the scale dependence $\mu$. 
The scalar case, discussed in \cite{Serino:2020pyu}, can be obtained by assigning specific values to the $b$ and $b'$ coefficients of the counterterms \eqref{counter}, expressed in terms of a scalar field content.
\section{Simplifications in the $n$-$p$ basis} 
 \label{nnp}
Simplifications in the structure of the renormalized 4-point function are possible once we re-express all the contributions of the previous section in terms of the tensorial basis formed by the $n^\mu$ and $p_1,p_2$ and $p_3$ four-vectors
 \begin{equation}
n^{\mu}=\epsilon^{\m \a \b\g}p_{i,\a}p_{j,\b}p_{k,\g}, \quad i\neq j \neq k =1,2,3,4,
\end{equation}
with $(n^\mu, p_i,p_j, p_k)$ forming a tetrad that can be used as a basis of expansion in Minkowski space. The $n$-$p$ parametrization is also discussed in \appref{np}.
 We just recall that in the computation of the residue of the $\sim V_E$ counterterm, we 
need to parameterize the Kronecker $\delta_{\mu\nu}$ in this basis in the form 

\begin{equation}
\d^{(4)}_{\m \n}=\sum_{i,j}^4 p_i^{\m} p_j^{\n} (Z^{-1})_{j i},
\end{equation}
where $(Z^{-1})_{j i}$ is the inverse of the Gramm matrix, defined as $Z=[p_i\cdot p_j]_{i,j=1}^d$.\\
Using the expression of $\delta^{\mu\nu}$ in the $n$-$p$ basis, denoted as $\delta^{(4)}_{\mu\nu}$ 

\begin{align}\label{deltaepsilon}
\d^{(4)}_{\m \n}=&\frac{1}{n^2}\Bigg(2p_1^{(\mu } p_2^{\nu) } \left(p_3^2
   p_1\cdot p_2-p_1\cdot p_3 p_2\cdot p_3\right)+2p_1^{(\nu } p_3^{\mu )}
   \left(p_2^2 p_1\cdot p_3-p_1\cdot p_2 p_2\cdot p_3\right) \notag \\ 
 & \qquad \qquad  +2p_2^{(\nu } p_3^{\mu) }
   \left(p_1^2 p_2\cdot p_3-p_1\cdot p_2 p_1\cdot p_3\right)\notag +p_1^{\mu }
   p_1^{\nu } \left((p_2\cdot p_3)^2-p_2^2 p_3^2\right)\notag \\
   & \qquad \qquad +p_2^{\mu } p_2^{\nu }\left((p_1\cdot p_3)^2-p_1^2 p_3^2\right) +p_3^{\mu } p_3^{\nu }\left((p_1\cdot p_2)^2-p_1^2 p_2^2\right)+n^{\mu } n^{\nu } \Bigg)
   \end{align}
one derives several relations in $d=4$. We easily derive the constraint
\begin{align}\label{projtwon}
\Pi^{\m_i \n_i}_{\a_i \b_i}(p_i)n^{\a_i } n^{\b_i }=& -\Pi^{\m_i \n_i}_{\a_i \b_i}(p_i)\Bigg(2p_j^{(\b_i  } p_k^{\a_i) }
   \left(p_i^2 p_j\cdot p_k-p_i\cdot p_k p_i\cdot p_j\right)\notag \\
   & +p_j^{\a_i } p_j^{\b_i  }\left((p_i\cdot p_k)^2-p_i^2 p_k^2\right) +p_k^{\a_i } p_k^{\b_i  }\left((p_i\cdot p_j)^2-p_i^2 p_j^2\right) \Bigg).
\end{align}
while the relation $\d^{(4)}_{\a_i \b_i}\Pi(p_i)^{(4)\m_i \n_i}_{\a_i \b_i}=0$ is obviously satisfied in the new basis. Using these relations, it is possible to show the vanishing of 
$\left[ V^{\mu_1\nu_1\ldots \mu_n\nu_n}\right] $, which is obviously defined at $d=4$. 
An explicit check has been discussed in \cite{Serino:2020pyu} for $n=4$, using the $n$-$p$ decomposition. As we have elaborated above, this results holds in general, due to 
\eqref{frr}.  
\subsection{A simplified decomposition}
Using \eqref{deltaepsilon} and the fact that the transverse traceless sector is identified by a decomposition which is contracted with four $\Pi^{\m \n}_{\a \b}$ projectors, and for $i\neq j \neq k \neq l=1,2,3,4$ generates terms with different numbers of external momenta. We illustrate the procedure for a couple of cases. For instance, the form factor which in \eqref{tttt} is characterised by a a single Kronecker $\delta$ and 6 tensorial momenta, will be re-expressed in terms of tensorial expressions either with 8 momenta or of $n n$ terms times 6 momenta. 
At the end, the decomposition will consist of terms proportional to tensor structures which 
are independent of $n^\a_i$; those which are proportional to $n^\a_i n^\b_j$ and will include  a term proportional to $n^{\a_i}n^{\b_j}n^{\b_k}n^{\b_l}$. The last type of  terms do not appear in the case of the 3-point functions of stress energy tensors in $d=3$ \cite{Bzowski:2013sza}.
The new decomposition can be written as
\begin{align}\label{newDecomp}
&\langle{t^{\mu_1\nu_1}(p_1)t^{\mu_2\nu_2}(p_2)t^{\mu_3\nu_3}(p_3)t^{\mu_4\nu_4}(\bar{p}_4)}\rangle\equiv \Pi^{\mu_1\nu_1}_{\alpha_1\beta_1}(p_1)\Pi^{\mu_2\nu_2}_{\alpha_2\beta_2}(p_2)\Pi^{\mu_3\nu_3}_{\alpha_3\beta_3}(p_3)\Pi^{\mu_4\nu_4}_{\alpha_4\beta_4}(\bar{p}_4)\notag\\
&\times\Big\{\sum_{a,b\in\{2,3\}}\, \sum_{c,d\in\{3,4\}}\,\sum_{e,f\in\{4,1\}}\,\sum_{g,h\in\{1,2\}}B^{(8p)}_{a\,b\,c\,d\,e\,f\,g\,h}\ \ p_{a}^{\alpha_1}\,p_{b}^{\beta_1}\,p_{c}^{\alpha_2}\,p_{d}^{\beta_2}\,p_{e}^{\alpha_3}\,p_{f}^{\beta_3}\,p_{g}^{\alpha_4}\,p_{h}^{\beta_4}\notag\\
&\qquad+\Bigg[\frac{1}{n^2}\,n^{\alpha_1}n^{\alpha_2}\sum_{a\in\{2,3\}}\sum_{b\in\{3,4\}}\sum_{c,d\in\{4,1\}}\sum_{e,f\in\{1,2\}} B^{(6p)}_{a\,b\,c\,d\,e\,f}\,p_{a}^{\beta_1}\,p_{b}^{\beta_2}\,p_{c}^{\alpha_3}\,p_{d}^{\beta_3}\,p_{e}^{\alpha_4}\,p_{f}^{\beta_4}+\text{(permutations)}\Bigg]\notag\\
&\qquad+\Bigg[\frac{1}{n^4}\,n^{\alpha_1}n^{\alpha_2}n^{\alpha_3}n^{\alpha_4}\sum_{a\in\{2,3\}}\sum_{b\in\{3,4\}}\sum_{c\in\{4,1\}}\sum_{d\in\{1,2\}} B^{(4p)}_{a\,b\,c\,d}\,p_{a}^{\beta_1}\,p_{b}^{\beta_2}\,p_{c}^{\beta_3}\,p_{d}^{\beta_4}+\text{(permutations)}
\Bigg]
\end{align}
in terms of form factors $B$.  By power counting, we see that in \eqref{newDecomp}, the only form factors that will manifest divergences in $d=4$ are those of form $B^{(4p)}$. The $B$'s will be written in terms of the form factors $A$'s of the general decomposition, simplifying the divergent terms and leading a structure with only one type of form factor divergent, i. e. $B^{(4p)}$.
\section{Renormalization and anomaly in $d=2$ and $d=4$: the $TT$ case } 
\label{renren}
 It is clear, from the analysis presented in the previous sections, that $V_E$ induces only 
a finite renormalization on the $n$-point function. However, as we will be showing here using the example of the correlator $TT$ in $d=2$, it is consistent to introduce a topological counterterm to renormalize a correlator, even if this does not need one. \\
In $d=2$ the TT does not exhibit any singularity, since a $1/\varepsilon$ behaviour in a kinematical prefactor is accompanied by a tensor structure which vanishes as $d\to 2$. We describe the steps in this case, and move to the case of $d=4$ after that.\\
In $d=2$ the $TT$ does not exhibit any divergence. 
A one-loop calculation yields the correlator of two conserved traceless stress tensors as
\begin{align}
	\langle{T^{\mu_1\nu_1}(p)T^{\mu_2\nu_2}(-p)}\rangle= \frac{ c(d)}{(d-2)}\left(p^2\right)^{d/2}\Pi^{\mu_1\nu_1\mu_2\nu_2}_{(d)}(p)\label{TT2dim}
\end{align}
where $\Pi_{(d)}$ is the transverse traceless projector defined in general $d$ dimensions and the constant $c(d)$ is defined as
\begin{align}
	c(d)=4\,c_T \left(\frac{\pi}{4}\right)^{d/2}\frac{(d-1)\Gamma\left(2-\frac{d}{2}\right)}{\Gamma\left(d+2\right)}
\end{align}
with $c_T$ depending on the matter field realization of the conformal invariant action. It is worth noticing  that the constant $c(d)$ is finite for any $d>1$. 

In dimensional regularization \eqref{TT2dim} the correlator in $d=2(1+\varepsilon)$ takes the following form
\begin{align}
	\langle{T^{\mu_1\nu_1}(p)T^{\mu_2\nu_2}(-p)}\rangle= \frac{ c(2+2\varepsilon)}{2\varepsilon}\left(p^2\right)^{1+\varepsilon}\Pi^{\mu_1\nu_1\mu_2\nu_2}_{(d=2+2\varepsilon)}(p)\label{TTreg}
\end{align}
with the appearance of the UV divergence as $1/\varepsilon$ pole after the expansion around $d=2$.
However the limit $\varepsilon\to 0$ is ambiguous due to the fact that the transverse traceless projector in two dimensions goes to zero. This properties is the consequence of the tensor degeneracy when $n=d$ for an $n$-point function. The way out of this indeterminacy is to 
introduce a $n$-$p$ decomposition of the Kronecker $\delta_{\mu\nu}$  and perform the limit 
 to $d=2$ of the entire tensor structure in order to prove the finiteness of the result, as discussed in \cite{Deser:1993yx, Bzowski:2017poo}.\\
Indeed, for the case of the two point function in $d=2$, one could define the independent momentum $n^\mu$ using the Levi-Civita tensor $\epsilon^{\mu\nu}$ as
\begin{align}
	n^{\mu}=\epsilon^{\mu\nu}p_\nu,\label{indMom}
\end{align}
 orthogonal to the other momentum $p_\mu$. Having such two independent momenta $p_\mu,\,n_\mu$, then the metric is not an independent tensor and we can rewrite it as
\begin{equation}
	\delta^{\mu\nu}=\sum_{j,k=1}^2q_j^\mu\,q_k^\nu\big(Z^{-1}\big)_{kj},
\end{equation}
with $q_1^\mu=p^\mu$ and $q_2^\mu=n^\mu$, and $Z$ is the gram matrix, i. e. $Z_{ij}=(q_i\cdot q_j)^2_{i,j=1}$. The Gram matrix is trivially $Z=p^2 \mathbb 1$, because $n^2=p^2$. Then, the metric tensor can be written as
\begin{align}
	\delta^{\mu\nu}=\frac{1}{p^2}\left(p^\mu p^\nu+n^\mu n^\nu\right),
\end{align}
for which the transverse projector takes the form
\begin{equation}
	\pi^{\mu\nu}_{(d=2)}(p)\equiv\delta^{\mu\nu}-\frac{p^{\mu}p^{\nu}}{p^2}=\frac{n^\mu n^\nu}{p^2}
\end{equation}
and the transverse traceless projector in $d=2$ vanishes as
\begin{equation}
	\Pi^{\mu_1\nu_1\mu_2\nu_2}_{(d=2)}(p)=\pi^{\mu_1(\mu_2}\pi^{\nu_2)\nu_1}-\pi^{\mu_1\nu_1}\pi^{\mu_2\nu_2}=0.
\end{equation}
Starting from the transverse traceless projector defined in $d$ dimensions and taking into account this degeneracy in $d=2$, we write the projector around $d=2$ as 
\begin{equation}
	\Pi^{\mu_1\nu_1\mu_2\nu_2}_{d=2+2\varepsilon}(p)=\pi^{\mu_1(\mu_2}\pi^{\nu_2)\nu_1}-\frac{1}{1+2\varepsilon}\pi^{\mu_1\nu_1}\pi^{\mu_2\nu_2}=\frac{2\varepsilon}{1+2\varepsilon}\frac{n^{\mu_1}n^{\nu_1}n^{\mu_2}n^{\nu_2}}{p^4},
\label{ths}
\end{equation}
making the limit $\varepsilon\to0$ well defined. Notice that we are expanding the parametric 
dependence of the projector in $\varepsilon$ and then we are re-expressing the tensor structure in terms of the non-degenerate $n$-$p$ basis. \\
 Inserting \eqref{ths} in \eqref{TT2dim} one derives a finite result as $\varepsilon\to 0$ \begin{align}
	\langle{T^{\mu_1\nu_1}(p)T^{\mu_2\nu_2}(-p)}\rangle_{(d=2)}= \frac{ c(2+2\varepsilon)}{1+2\varepsilon}\left(p^2\right)^{\varepsilon-1}\,n^{\mu_1}n^{\nu_1}n^{\mu_2}n^{\nu_2}\,\,\overset{\varepsilon\to0}{=}\,\,c(2)\frac{n^{\mu_1}n^{\nu_1}n^{\mu_2}n^{\nu_2}}{p^2}
\end{align} 
which violates the tracelessness condition, making manifest the presence of the trace anomaly relation. The approach outlined above, is not relying on the introduction of any counterterm, but we are simply exploiting the degeneracy of the tensor structures in 
$d=2$ in order to to extract a renormalized expression of the correlation function. Notice, 
beside the generation of a trace anomaly in the renormalization process, also the absence of any scale dependence in the result. We are going to reproduce the same features of this result using a different approach, which will require the inclusion of a topological counterterm. 

\section{Dimensional Regularization}
This result can be obtained differently following the standard procedure of dimensional regularization scheme and renormalization. The same approach will be used in $d=4$.\\
 We start from the regulated expression of the $TT$ obtained by an expansion of \eqref{TTreg} in power of $\varepsilon$, that gives
\begin{align}
	\langle{T^{\mu_1\nu_1}(p)T^{\mu_2\nu_2}(-p)}\rangle_{Reg}=& \frac{ c(2)}{2\varepsilon}\left(p^2\right)\,\Pi^{\mu_1\nu_1\mu_2\nu_2}_{(2+2\varepsilon)}(p)+\frac{c(2)}{2}\Pi^{\mu_1\nu_1\mu_2\nu_2}_{(2)}\,p^2\log p^2+p^2c^\prime(2)\,\Pi^{\mu_1\nu_1\mu_2\nu_2}_{(2)}+O(\varepsilon),\label{RegTT}
\end{align}
In this case we are not making use of the tensor degeneracies in $d=2$ yet. We introduce the only possible counterterm action definable in $d=2$ as
\begin{align}
	S_{ct}=-\frac{1}{\varepsilon}\,\beta_c\,\int d^dx\, \sqrt{-g}\,\mu^{d-2}\,R
\end{align}
which will be included in the process of renormalization. By considering two functional variations of this action with respect to the metric, taking the flat space limit and after a Fourier transform we obtain the contribution
\begin{align}
	\langle{T^{\mu_1\nu_1}(p)T^{\mu_2\nu_2}(-p)}\rangle_{Count}=-\frac{\beta_c\,p^2\,\mu^{d-2}}{2\varepsilon}\,\left(\Pi^{\mu_1\nu_1\mu_2\nu_2}_{(d)}(p)-\frac{(d-2)}{(d-1)}\pi^{\mu_1\nu_1}(p)\pi^{\mu_2\nu_2}(p)\right), 
\end{align}
whose expansion in power of $\varepsilon$ is 
\begin{align}
	\langle{T^{\mu_1\nu_1}(p)T^{\mu_2\nu_2}(-p)}\rangle_{Count}=-\frac{\beta_c\,p^2}{2\varepsilon}\,\left(\Pi^{\mu_1\nu_1\mu_2\nu_2}_{(2+2\varepsilon)}(p)-\frac{2\varepsilon}{(1+2\varepsilon)}\pi^{\mu_1\nu_1}(p)\pi^{\mu_2\nu_2}(p)\right)\left(1+\varepsilon \log\mu^2\right).\label{CountTT}
\end{align}

Adding the counterterm contribution \eqref{CountTT} to the regularized correlator \eqref{RegTT}, and with the choice of $\beta_c=c(2)$ we remove the divergence. We obtain a finite renormalized result in the limit $\varepsilon\to0$ given by
\begin{align}
	\langle{T^{\mu_1\nu_1}(p)T^{\mu_2\nu_2}(-p)}\rangle_{(d=2)}^{Ren}=&\frac{c(2)}{2}\Pi^{\mu_1\nu_1\mu_2\nu_2}_{(2)}\,p^2\log\left( \frac{p^2}{\mu^2}\right)+p^2c^\prime(2)\,\Pi^{\mu_1\nu_1\mu_2\nu_2}_{(2)}+c(2)\,p^2\pi^{\mu_1\nu_1}(p)\pi^{\mu_2\nu_2}(p).\label{TTRen}
\end{align}
This result is not traceless, but is characterized by non zero anomalous trace 
\begin{equation}
	\langle{T^{\mu_1}_{\ \ \ \mu_1}(p)T^{\mu_2\nu_2}(-p)}\rangle_{(d=2)}^{Ren}=c(2)\,p^2\,\pi^{\mu_2\nu_2}(p)
\end{equation}
which coincides with the result of taking functional derivative of trace anomaly of the two point function in two dimensions
\begin{align}
	\mathcal{A}=c(2)\,\sqrt{-g}\,R.
\end{align}
While we have reproduced the correct structure of the anomaly, as in the previous section, 
we need to recover by this scheme the scale independence of the result in 
\eqref{TTRen}, which is not apparent from that equation. Notice that $c(2)$ identifies a topological contribution and therefore should be independent of any scale. \\
The apparent anomalous scale-dependence of the two point function is
\begin{align}
\label{rr}
	\mu\frac{\partial}{\partial\mu}	\langle{T^{\mu_1\nu_1}(p)T^{\mu_2\nu_2}(-p)}\rangle=-c(2)\,p^2\,\Pi^{\mu_1\nu_1\mu_2\nu_2}_{(2)}.
\end{align}
that shows an apparent contradiction. It is sufficient, at this stage, to use the degeneracy of the tensor structure $\Pi^{\mu_1\nu_1\mu_2\nu_2}_{(2)}$ to immediately realize that the 
the rhs of \eqref{rr} is zero. 
The use of the counterterm, even if it is zero in the dimensions on which we are going to project, of course, allows to obtain the correct expression of the renormalized correlator with iooen indices, while at the same time induces the correct expression of its anomaly.\\
 The advantage in following this procedure is evident especially for multi-point functions. 
\section{Longitudinal projectors in $d=4$}
To illustrate the emergence of longitudinal projectors in multi-point correlation functions in $d=4$, we start from the case of the $TT$, then move to the 3T and conclude our discussion with the 4T. 
The simplest context in which to discuss the renormalization of the $TT$ is in free field theory, and include three independent sectors with $n_S$ scalars, $n_F$ fermions and $n_G$ gauge fields. A direct perturbative computation gives 
\begin{align}
\langle{T^{\mu_1\nu_1}(p)T^{\mu_2\nu_2}(-p)}\rangle=&-\frac{\p^2\,p^4}{4(d-1)(d+1)}\,B_0(p^2)\,\Pi^{\mu_1\nu_1\mu_2\nu_2}(p)\Big[2(d-1)n_F+(2d^2-3d-8)n_G+n_S\Big]\notag\\
&+\frac{\p^2\,p^4\,n_G}{8(d-1)^2}(d-4)^2(d-2)\p^{\mu_1\nu_1}(p)\p^{\mu_2\nu_2}(p)\,B_0(p^2)\label{TTddim}
\end{align}
where $B_0(p^2)$ is the scalar 2-point function defined as
\begin{equation}
B_0(p^2)=\frac{1}{\pi^\frac{d}{2}}\,\int\,d^d\ell\,\frac{1}{\ell^2\,(\ell-p)^2}
\end{equation}
and \eqref{TTddim} shows the separation of the result into a transverse-traceless $(\Pi)$ and longitudinal part $(\pi^{\mu_1\nu_1})$. Around $d=4$, the projectors are expanded according to the relation 
\begin{equation}
\label{pexp}
\Pi^{\,\mu_1\nu_1\mu_2\nu_2}(p)=\Pi^{(4)\,\mu_1\nu_1\mu_2\nu_2}(p)-\frac{2}{9}\varepsilon\,\pi^{\mu_1\nu_1}(p)\,\pi^{\mu_2\nu_2}(p)+O(\varepsilon^2), 
\end{equation}
where the expansion is performed on the parametric dependence of the projector $\Pi(d)$. 
As usual in DR, the tensor indices are continued to $d$ dimensions and contracted with a d-dimensional Euclidean metric $(\delta^\mu_\mu=d)$.

Using \eqref{pexp} in \eqref{TTddim}, the latter takes the form
\begin{align}
\label{result}
\langle{T^{\mu_1\nu_1}(p)T^{\mu_2\nu_2}(-p)}\rangle&=-\frac{\p^2\,p^4}{4}\,\bigg(\frac{1}{\varepsilon}+\bar{B}_0(p^2)\bigg)\,\bigg(\Pi^{(4)\,\mu_1\nu_1\mu_2\nu_2}(p)-\frac{2}{9}\varepsilon\,\pi^{\mu_1\nu_1}(p)\,\pi^{\mu_2\nu_2}(p)+O(\varepsilon^2)\bigg)\notag\\
&\hspace{-2cm}\times\Bigg[\bigg(\frac{2}{5}+\frac{4}{25}\varepsilon+O(\varepsilon^2)\bigg)n_F+\bigg(\frac{4}{5}-\frac{22}{25}\varepsilon+O(\varepsilon^2)\bigg)n_G+\bigg(\frac{1}{15}+\frac{16}{225}\varepsilon+O(\varepsilon^2)\bigg)n_S\Bigg]\notag\\
&\hspace{-1cm}+\frac{\p^2\,p^4\,n_G}{8}\p^{\mu_1\nu_1}(p)\p^{\mu_2\nu_2}(p)\,\bigg(\frac{1}{\varepsilon}+\bar{B}_0(p^2)\bigg)\bigg[\frac{8}{9}\varepsilon^2+\frac{8}{27}\varepsilon^3+O(\varepsilon^4)\bigg]
\end{align}
where $\Pi^{ (4)\,\,\mu_1\nu_1\mu_2\nu_2}(p)$ is the transverse and traceless projector in $d=4$ and $\bar{B}_0(p^2)= 2 - \log(p^2)$ is the finite part in $d=4$ of the scalar integral in the $\overline{MS}$ scheme. The last term of \eqref{result}, generated by the addition of a non-conformal sector ($\sim n_G$), vanishes separately as $\varepsilon\to 0$. Finally, combining all the terms we obtain the regulated ($reg$) expression of the $TT$ around $d=4$ in the form

\begin{align}
\label{regular}
\langle{T^{\mu_1\nu_1}(p)T^{\mu_2\nu_2}(-p)}\rangle_{reg}&=-\frac{\p^2\,p^4}{60\,\varepsilon}\Pi^{(4)\,\mu_1\nu_1\mu_2\nu_2}(p)\left(6 n_F + 12 n_G + n_S\right)\notag\\
&\hspace{-3cm}+\frac{\p^2\,p^4}{270}\p^{\mu_1\nu_1}(p)\p^{\mu_2\nu_2}(p)\left(6 n_F + 12 n_G + n_S\right)-\frac{\p^2\,p^4}{300}\bar{B}_0(p^2)\Pi^{\mu_1\nu_1\mu_2\nu_2}(p)\left(30n_F+60n_G+5n_S\right)\notag\\
&\hspace{-3cm}-\frac{\p^2\,p^4}{900}\Pi^{\mu_1\nu_1\mu_2\nu_2}(p)\left(36n_F-198 n_G+16n_S\right)+O(\varepsilon)
\end{align}
The divergence in the previous expression can be removed through the one loop counterterm Lagrangian \eqref{nTcount}. In fact, the second functional derivative of $S_{count}$ with respect to the background metric gives 
\begin{align}
\langle{T^{\mu_1\nu_1}(p)T^{\mu_2\nu_2}(-p)}\rangle_{count}&\equiv -\sdfrac{\mu^{-\varepsilon}}{\varepsilon}\bigg(4b\,\big[\sqrt{-g}\,C^2\big]^{\m_1\n_1\m_2\n_2}(p,-p)\bigg)=-\frac{8(d-3)\,\mu^{-\varepsilon}\,b}{(d-2)\,\varepsilon}p^4\Pi^{(d)\,\mu_1\nu_1\mu_2\nu_2}(p)
\end{align}
having used the relation $V_{E}^{\m_1\n_1\m_2\n_2}(p,-p)=0$. In particular, expanding around $d=4$ and using again \eqref{pexp} we obtain
\begin{align}
&\langle{T^{\mu_1\nu_1}(p)T^{\mu_2\nu_2}(-p)}\rangle_{count}=\notag\\
&\hspace{1cm}-\frac{8\,b\,p^4}{\varepsilon}\bigg(\Pi^{(4)\,\mu_1\nu_1\mu_2\nu_2}(p)-\frac{2}{9}\varepsilon\,\pi^{\mu_1\nu_1}(p)\,\pi^{\mu_2\nu_2}(p)+O(\varepsilon^2)\bigg)\,\bigg(\frac{1}{2}-\frac{\varepsilon}{2}\left(\frac{1}{2}+\log\mu\right)+O(\varepsilon^2)\bigg)\notag\\
&\hspace{1cm}=-\sdfrac{4\,b}{\varepsilon}p^4\,\Pi^{(4)\,\mu_1\nu_1\mu_2\nu_2}(p)+4\,b\, p^4\bigg[\Pi^{(4)\,\mu_1\nu_1\mu_2\nu_2}(p)+\frac{2}{9}\p^{\mu_1\nu_1}(p)\p^{\mu_2\nu_2}(p)\bigg]+O(\varepsilon)
\end{align}
which cancels the divergence arising in the two point function, if one chooses the parameter $b$ as 
\begin{equation}
\label{choiceparm1}
b=-\frac{3\pi^2}{720}n_S-\frac{9\pi^2}{360}n_F-\frac{18\pi^2}{360}n_G,
\end{equation}
The renormalized 2-point function using  \eqref{choiceparm1} then takes the form  \begin{align}
\langle{T^{\mu_1\nu_1}(p)T^{\mu_2\nu_2}(-p)}\rangle_{Ren}&=\langle{T^{\mu_1\nu_1}(p)T^{\mu_2\nu_2}(-p)}\rangle+\langle{T^{\mu_1\nu_1}(p)T^{\mu_2\nu_2}(-p)}\rangle_{count}\notag\\
&=-\frac{\p^2\,p^4}{60}\bar{B}_0\left(\frac{p^2}{\mu^2}\right)\Pi^{\mu_1\nu_1\mu_2\nu_2}(p)\left(6n_F+12n_G+n_S\right)\notag\\
&\quad-\frac{\p^2\,p^4}{900}\Pi^{\mu_1\nu_1\mu_2\nu_2}(p)\big(126n_F-18n_G+31n_S\big).
\label{tren}
\end{align}

Notice that the final renormalized expression is transverse and traceless. Obviously, this result holds in the case in which we choose a counterterm in such a way that \ref{oneq} is satisfied.  If we had chosen the $C^2$ counterterm action to satisfy \ref{twoq}, we would have found the relation 
\begin{equation}
\delta_{\mu_1\nu_1}\langle T^{\mu_1\nu_1}(p_1)T^{\mu_1\nu_2}(-p_1)\rangle=
\mathcal{A}^{\mu_2\nu_2}(p_1),
\end{equation}
where on the right hand side we have the contribution of the $\square R$ term that can be removed by adding a local term $R^2$ in the effective action, obtaining a finite renormalization procedure. 
We will see that the same choice of parameters $b$ given in \eqref{choiceparm1} and for $b'$ as
\begin{equation}\label{choiceparm2}
	b'=\frac{\pi^2}{720}n_S+\frac{11\pi^2}{720}n_F+\frac{31\pi^2}{360}n_G.
\end{equation}
removes the divergences in the three point function, as we are going to discuss below. In fact, as we have seen, the expression of $b$ is related to the renormalization of the two point function, instead $b'$ is intrinsically related to the renormalization of the 3-point function.

It is worth mentioning that the renormalized result of the two point function \eqref{tren} does not contain any trace anomaly contributions but, due to the explicit presence of a $\mu$-dependence, it acquires anomalous dilatation Ward Identities written as
\begin{align}
\mu\frac{\partial}{\partial\mu}\langle{T^{\mu_1\nu_1}(p)T^{\mu_2\nu_2}(-p)}\rangle_{Ren}=-2\left[\frac{\pi^2\,p^4}{60}\,\Pi^{\mu_1\nu_1\mu_2\nu_2}(p)\left(6n_F+12n_G+n_S\right)\right].
\end{align}
\section{Three-point function}
We can now move to the 3-point function. 
In this case we start showing how the derivation of the anomaly part of the correlator can 
be worked out directly from the CWIs, in a simplified way.\\
As pointed out in \cite{Coriano:2018bsy1,Bzowski:2013sza, Bzowski:2017poo }, the 3T correlator can be separated into its transverse-traceless part and in its longitudinal one by using the projectors \eqref{prozero}\eqref{proone}\eqref{protwo} as
\begin{align}
&\langle{T^{\mu_1\nu_1}(p_1)T^{\mu_2\nu_2}(p_2)T^{\mu_3\nu_3}(\bar{p}_3)}\rangle=\langle{t^{\mu_1\nu_1}(p_1)t^{\mu_2\nu_2}(p_2)t^{\mu_3\nu_3}(\bar{p}_3)}\rangle\notag\\
&\hspace{1cm}+\bigg(\langle{t_{loc}^{\mu_1\nu_1}(p_1)T^{\mu_2\nu_2}(p_2)T^{\mu_3\nu_3}(\bar{p}_3)}\rangle+(\text{cyclic perm.})\bigg)\notag\\
&-\bigg(\langle{t_{loc}^{\mu_1\nu_1}(p_1)t_{loc}^{\mu_2\nu_2}(p_2)T^{\mu_3\nu_3}(\bar{p}_3)}\rangle+(\text{cyclic perm.})\bigg)+\langle{t_{loc}^{\mu_1\nu_1}(p_1)t_{loc}^{\mu_2\nu_2}(p_2)t_{loc}^{\mu_3\nu_3}(\bar{p}_3)}\rangle,\label{TTT}
\end{align}
where the transverse-traceless part is decomposed in terms of form factors and independent tensor structures as
\begin{align}
&\langle{t^{\mu_1\nu_1}(p_1)t^{\mu_2\nu_2}(p_2)t^{\mu_3\nu_3}(\bar{p}_3)}\rangle=\Pi^{\mu_1\nu_1}_{\a_1\b_1}(p_1)\Pi^{\mu_2\nu_2}_{\a_2\b_2}(p_2)\Pi^{\mu_3\nu_3}_{\a_3\b_3}(\bar{p}_3)\notag\\
&\times\Big\{A_1\,p_2^{\a_1} p_2^{\b_1} \bar{p}_3^{\a_2} p_3^{\b_2} p_1^{\a_3} p_1^{\b_3}+ A_2\,\d^{\b_1\b_2} p_2^{\a_1} p_3^{\a_2} p_1^{\a_3} p_1^{\b_3} 
+ A_2\,(p_1 \leftrightarrow p_3)\, \d^{\b_2\b_3}  p_3^{\a_2} p_1^{\a_3} p_2^{\a_1} p_2^{\b_1} \notag\\
&\hspace{0.8cm}+ A_2\,(p_2\leftrightarrow p_3)\, \d^{\b_3\b_1} p_1^{\a_3} p_2^{\a_1}  p_3^{\a_2} p_3^{\b_2}+ A_3\,\d^{\a_1\a_2} \d^{\b_1\b_2}  p_1^{\a_3} p_1^{\b_3} + A_3(p_1\leftrightarrow p_3)\,\d^{\a_2\a_3} \d^{\b_2\b_3}  p_2^{\a_1} p_2^{\b_1} \notag\\
&\hspace{1.2cm}
+ A_3(p_2\leftrightarrow p_3)\,\d^{\a_3\a_1} \d^{\b_3\b_1}  p_3^{\a_2} p_3^{\b_2} + A_4\,\d^{\a_1\a_3} \d^{\a_2\b_3}  p_2^{\b_1} p_3^{\b_2} + A_4(p_1\leftrightarrow p_3)\, \d^{\a_2\a_1} \d^{\a_3\b_1}  p_3^{\b_2} p_1^{\b_3} \notag\\
&\hspace{3.5cm}+ A_4(p_2\leftrightarrow p_3)\, \d^{\a_3\a_2} \d^{\a_1\b_2}  p_1^{\b_3} p_2^{\b_1} + A_5  \d^{\a_1\b_2}  \d^{\a_2\b_3}  \d^{\a_3\b_1}\Big\},\label{ttt}
\end{align}
while the longitudinal part is expressed in terms of lower point functions by using the CWIs, and takes the form
\begin{align}
p_{1\n_1}\langle{T^{\mu_1\nu_1}(p_1)\,T^{\m_2\n_2}(p_2)\,T^{\m_3\n_3}({p_3})}\rangle&=-p_2^{\m_1}\langle{T^{\m_2\n_2}(p_1+p_2)T^{\m_3\n_3}({p_3})}\rangle-{p_3^{\m_1}}\langle{T^{\m_2\n_2}(p_2)T^{\m_3\n_3}(p_1+{p_3})}\rangle\notag\\
&\hspace{-1.5cm}+p_{2\a}\left[\d^{\m_1\n_2}\langle{T^{\mu_2\a}(p_1+p_2)T^{\m_3\n_3}({p_3})}\rangle+\d^{\m_1\m_2}\langle{T^{\nu_2\a}(p_1+p_2)T^{\m_3\n_3}({p_3})}\rangle\right]\notag\\
&\hspace{-1.5cm}+{p_{3\a}}\left[\d^{\m_1\n_3}\langle{T^{\mu_3\a}(p_1+{p_3})T^{\m_2\n_2}(p_2)}\rangle+\d^{\m_1\m_3}\langle{T^{\nu_3\a}(p_1+{p_3})T^{\m_2\n_2}(p_2)}\rangle\right].
\label{long}
\end{align}

In $d=4$ this correlator manifests divergences in the forms of single poles in $1/\varepsilon$ ($\varepsilon=(4-d)$), as for any CFT affected by the trace anomaly. \\
These divergences are present in both the transverse-traceless and longitudinal parts. As discussed in detail in \cite{Coriano:2018bsy1}, the counterterm \eqref{nTcount}, for the $3$-point case, renormalizes the correlator \eqref{TTT} by canceling all the divergences with the same choice of the coefficients \eqref{choiceparm1} and \eqref{choiceparm2} but, at the same time, gives extra contributions in the final renormalized 3T from the local/longitudinal part. These extra contributions defines the anomalous part of the correlator. In the case of $n=4$ \eqref{cct} specializes in the obvious form
\begin{align}
&\bigg[\langle{T^{\mu_1\nu_1}T^{\mu_2\nu_2}T^{\mu_3\nu_3}}\rangle_{bare}+\langle{T^{\mu_1\nu_1}T^{\mu_2\nu_2}T^{\mu_3\nu_3}}\rangle_{count}\bigg]_{d\to4}=\notag\\
&=\langle{T^{\mu_1\nu_1}T^{\mu_2\nu_2}T^{\mu_3\nu_3}}\rangle^{(d=4)}_{f}+\langle{T^{\mu_1\nu_1}T^{\mu_2\nu_2}T^{\mu_3\nu_3}}\rangle^{(d=4)}_{anomaly}.
\end{align} 

 The anomaly part is given as
\begin{align}
&\langle{T^{\mu_1\nu_1}(p_1)T^{\mu_2\nu_2}(p_2)T^{\mu_3\nu_3}(p_3)}\rangle^{(d=4)}_{anomaly}=\notag\\
&=\left(\frac{4}{3}\,\p^{\mu_1\nu_1}(p_1)\,\mathcal{A}^{\mu_2\nu_2\mu_3\nu_3}(p_2,\bar{p}_3)+(\text{perm.})\right)\notag\\
&\hspace{0.5cm}-\left(\frac{4}{9}\,\p^{\mu_1\nu_1}(p_1)\,\p^{\mu_2\nu_2}(p_2)\,\Delta_{\a_2\b_2}\,\mathcal{A}^{\a_2\b_2\mu_3\nu_3}(p_2,\bar{p}_3)+(\text{perm.})\right)\notag\\
&\hspace{1cm}+\frac{4}{27}\,\p^{\mu_1\nu_1}(p_1)\,\p^{\mu_2\nu_2}(p_2)\,\p^{\mu_3\nu_3}(\bar{p}_3)\,\Delta_{\a_2\b_2}\Delta_{\a_3\b_3}\,\mathcal{A}^{\a_2\b_2\a_3\b_3}(p_2,\bar{p}_3),\label{3TAnomaly}
\end{align}
which is the expression depicted in \figref{FeynTTTX}. The equation above has indeed a clear and simple interpretation in terms of anomaly poles extracted from the $\pi^{\mu\nu}$ projectors attached to each of the external graviton legs. We can re-express it in the form
\begin{align}
\langle{T^{\mu_1\nu_1}(p_1)T^{\mu_2\nu_2}(p_2)T^{\mu_3\nu_3}(\bar{p}_3)}\rangle^{(4)}_{anomaly}&=\left(\frac{\p^{\mu_1\nu_1}(p_1)}{3}\langle{T(p_1)T^{\mu_2\nu_2}(p_2)T^{\mu_3\nu_3}(\bar{p}_3)}\rangle^{(4)}_{anomaly}+(\text{perm.})\right)\notag\\
&\hspace{-1cm}-\left(\frac{\p^{\mu_1\nu_1}(p_1)}{3}\frac{\p^{\mu_2\nu_2}(p_2)}{3}\langle{T(p_1)T(p_2)T^{\mu_3\nu_3}(\bar{p}_3)}\rangle^{(4)}_{anomaly}+(\text{perm.})\right)\notag\\
&\hspace{-1cm}+\frac{\p^{\mu_1\nu_1}(p_1)}{3}\frac{\p^{\mu_2\nu_2}(p_2)}{3}\frac{\p^{\mu_3\nu_3}(\bar{p}_3)}{3}\langle{T(p_1)T(p_2)T(\bar{p}_3)}\rangle^{(4)}_{anomaly},\label{3TAnomaly2}
\end{align}
from which it is clear that a contribution such as 
\begin{align}
\langle{T(p_1)T^{\mu_2\nu_2}(p_2)T^{\mu_3\nu_3}(\bar{p}_3)}\rangle^{(4)}_{anomaly}
&=\mathcal{A}^{\mu_2\nu_2\mu_3\nu_3}(p_2,\bar{p}_3)\notag\\
&=4b'[E]^{\mu_2\nu_2\mu_3\nu_3}(p_2,\bar{p}_3)+4b[C^2]^{\mu_2\nu_2\mu_3\nu_3}(p_2,\bar{p}_3)
\end{align} 
where we trace one of the three stress energy tensors, is obtained by differentiating the anomaly functional twice. Similar results hold for the other contributions with double
\begin{align}
\langle{T(p_1)T(p_2)T^{\mu_3\nu_3}(\bar{p}_3)}\rangle^{(4)}_{anomaly}&=\Delta_{\a_2\b_2}\,\mathcal{A}^{\a_2\b_2\mu_3\nu_3}(p_2,\bar{p}_3)\notag\\
&=\delta_{\alpha_2\beta_2}\left(4b'[E]^{\a_2\b_2\mu_3\nu_3}(p_2,\bar{p}_3)+4b[C^2]^{\a_2\b_2\mu_3\nu_3}(p_2,\bar{p}_3)\right)
\end{align}
and triple traces
\begin{align}
\langle{T(p_1)T(p_2)T(\bar{p}_3)}\rangle^{(4)}_{anomaly}&=\Delta_{\a_2\b_2}\Delta_{\a_3\b_3}\,\mathcal{A}^{\a_2\b_2\a_3\b_3}(p_2,\bar{p}_3)\notag\\
&=\Delta_{\a_2\b_2}\Delta_{\a_3\b_3}\left(4b'[E]^{\a_2\b_2\a_3\b_3}(p_2,\bar{p}_3)+4b[C^2]^{\a_2\b_2\a_3\b_3}(p_2,\bar{p}_3)\right).
\end{align}
It is clear from these expressions that all the possible anomaly contributions generated in the flat limit are associated with single, double and triple traces of the 3T, where on each external graviton leg we are allowing for a scalar exchange due to the presence of the transverse $\pi$  projector, as clear from \figref{FeynTTTX}. This result is strongly reminiscent of the emergence of an anomaly pole in the AVV chiral anomaly diagram for a $J_5 JJ $ correlator, with one axial-vector $(J_5)$ and two vector $(J)$ currents, that manifests a similar pattern.\\
 Indeed, an analogy with the behaviour of the 3T is present if we decompose the $AAA$ anomaly diagram as 
 \begin{equation}
AAA\to  1/3(AVV + VAV + VVA)
 \end{equation}
using its permutational symmetry on the axial-vector lines.\\
In both cases one encounters a scalar or a pseudoscalar mode, respectively, via a bilinear mixing term attached to the external graviton or gauge lines. This mode is directly coupled to the (chiral/conformal) anomaly. Therefore, we encounter a feature that unifies both the conformal and the chiral cases.\\
Notice that the spin-2 part of the gravitational fluctuations do not couple to such bilinear mixing term in the 3T case, which therefore mediates only spin-1 and spin-0 interactions. \\
 It is also clear that this massless scalar interaction, in the 3T, is not removed by the inclusion of other Weyl-invariant terms which are obviously present in the complete expression of the correlator, which are not identified by our method. \\
For this correlator the anomalous trace WI takes the form 
\begin{align}
&\delta_{\mu_1\nu_1}\langle{ T^{\mu_1\nu_1}(p_1)T^{\mu_2\nu_2}(p_2)T^{\mu_3\nu_3}(p_3)}\rangle\notag\\
&\hspace{2cm}=
4 \, \mathcal A^{\mu_2\nu_2\mu_3\nu_3}(p_2,p_3)
- 2 \, \langle{ T^{\mu_2\nu_2}(p_1+p_2)T^{\mu_3\nu_3}(p_3)}\rangle - 2 \, \langle{ T^{\mu_2\nu_2}(p_2)T^{\mu_3\nu_3}(p_1+p_3)}\rangle\nn \\
&\hspace{2cm}=
4 \, \bigg[ b\,\big[C^2\big]^{\mu_2\nu_2\mu_3\nu_3}(p_2,p_3)+ b'\, \big[E\big]^{\mu_2\nu_2\mu_3\nu_3}(p_2,p_3) \bigg]\nn \\
&\hspace{3cm}- 2 \, \langle{ T^{\mu_2\nu_2}(p_1+p_2)T^{\mu_3\nu_3}(p_3)}\rangle - 2 \, \langle{ T^{\mu_2\nu_2}(p_2)T^{\mu_3\nu_3}(p_1+p_3)}\rangle. \label{munu3PFanomaly}
\end{align}

\section{The four-point function}
\label{fourp}
We will now come to illustrate the reconstruction procedure for the renormalized 4T, showing how the separation of the vertex into a transverse-traceless part, a longitudinal one and an anomaly contribution takes place after renormalization. Clearly, by construction, the transverse traceless sector of the 4T is renormalized by adding the contribution coming from the counterterm 

\begin{align}
&\bigg[\langle{t^{\mu_1\nu_1}t^{\mu_2\nu_2}t^{\mu_3\nu_3}t^{\mu_4\nu_4}}\rangle_{bare}+\langle{t^{\mu_1\nu_1}t^{\mu_2\nu_2}t^{\mu_3\nu_3}t^{\mu_4\nu_4}}\rangle_{count}\bigg]_{d\to4}=\langle{t^{\mu_1\nu_1}t^{\mu_2\nu_2}t^{\mu_3\nu_3}t^{\mu_4\nu_4}}\rangle^{(d=4)}_{Ren}.
\end{align} 
If we consider a Lagrangian realization, the renormalization of this part and the corresponding form factors are ensured by the choice of the coefficients $b$ and $b'$ as in \eqref{choiceparm1} and \eqref{choiceparm2}, where $n_I$, $I=S,F,G$, are the number of scalar, fermion and gauge fields running into the virtual corrections of this correlator. For a general CFT, not directly related to a specific free field theory realization, the $b$ and $b'$ should be interpreted as fundamental constants of that theory, and are arbitrary.\\
  Now, we turn to the longitudinal part of the correlator, which is the most interesting component when the case $d=4$ is considered, due to the appearance of the anomaly. For instance, we study the bare part $\langle{t_{loc}TTT}\rangle$ in \eqref{longTer} that is explicitly written as
\begin{align}
&\langle{t_{loc}^{\mu_1\nu_1}(p_1)T^{\mu_2\nu_2}(p_2)T^{\mu_3\nu_3}(p_3)T^{\mu_4\nu_4}(p_4)}\rangle_{bare}=\notag\\
&=\mathcal{I}^{\mu_1\nu_1}_{\alpha_1}(p_1)\Bigg\{\left[4\, \mathcal{B}^{\alpha_1\hspace{0.4cm}\mu_2\nu_2\mu_3\nu_3}_{\hspace{0.3cm}\lambda\beta_1}(p_2,p_3)\langle{T^{\lambda\beta_1}(p_1+p_2+p_3)T^{\mu_4\nu_4}(\bar{p}_4)}\rangle+(34)+(24)\right]\notag\\
&\hspace{2cm}+\left[2 \, \mathcal{C}^{\alpha_1\hspace{0.4cm}\mu_2\nu_2}_{\hspace{0.3cm}\lambda\beta_1}(p_2)\langle{T^{\lambda\beta_1}(p_1+p_2)T^{\mu_3\nu_3}(p_3)T^{\mu_4\nu_4}(\bar p_4)}\rangle+(23)+ (24)\right]\Bigg\}\notag\\
&\hspace{2cm}-\frac{2}{(d-1)}\pi^{\mu_1\nu_1}(p_1)\bigg[\langle{T^{\mu_2\nu_2}(p_1+p_2)T^{\mu_3\nu_3}(p_3)T^{\mu_4\nu_4}(\bar p_4)}\rangle+(2 3)+(24)\bigg].\label{tTTT}
\end{align}
This contribution in $d=4$ manifests some divergences due to the presence of the $3$- and $2$- point functions on its rhs. \\
A similar equation holds for the counterterm in \eqref{TTTTcount}, which can be decomposed as well into the transverse-traceless part and the longitudinal one. The contribution that renormalises \eqref{tTTT} is
\begin{align}
&\langle{t_{loc}^{\mu_1\nu_1}(p_1)T^{\mu_2\nu_2}(p_2)T^{\mu_3\nu_3}(p_3)T^{\mu_4\nu_4}(\bar p_4)}\rangle_{count}=\notag\\
&=\mathcal{I}^{\mu_1\nu_1}_{\alpha_1}(p_1)\Bigg\{\left[4\, \mathcal{B}^{\alpha_1\hspace{0.4cm}\mu_2\nu_2\mu_3\nu_3}_{\hspace{0.3cm}\lambda\beta_1}(p_2,p_3)\langle{T^{\lambda\beta_1}(p_1+p_2+p_3)T^{\mu_4\nu_4}(\bar{p}_4)}\rangle_{count}+(34)+(24)\right]\notag\\
&\hspace{2cm}+\left[2 \, \mathcal{C}^{\alpha_1\hspace{0.4cm}\mu_2\nu_2}_{\hspace{0.3cm}\lambda\beta_1}(p_2)\langle{T^{\lambda\beta_1}(p_1+p_2)T^{\mu_3\nu_3}(p_3)T^{\mu_4\nu_4}(\bar p_4)}\rangle_{count}+(23)+ (24)\right]\Bigg\}\notag\\
&\hspace{2cm}-\frac{2}{(d-1)}\pi^{\mu_1\nu_1}(p_1)\bigg[\langle{T^{\mu_2\nu_2}(p_1+p_2)T^{\mu_3\nu_3}(p_3)T^{\mu_4\nu_4}(\bar p_4)}\rangle_{count}+(2 3) +(2 4)\bigg]\notag\\
&\quad+\frac{8\,(d-4)\,\mu^{-\varepsilon}}{(d-1)\varepsilon}\pi^{\mu_1\nu_1}(p_1)\Bigg\{b\left[\sqrt{-g}C^2\right]^{\mu_2\nu_2\mu_3\nu_3\mu_4\nu_4}(p_2,p_3,p_4)+b'\left[\sqrt{-g}E\right]^{\mu_2\nu_2\mu_3\nu_3\mu_4\nu_4}(p_2,p_3,p_4)
\Bigg\}\label{tloctttcount}
\end{align}
where we have taken into account the definition \eqref{nTcount}. It is worth mentioning that, as for the $1/\varepsilon\, V_E$ counterterm,  near $d=4$ one has to use the analogous expansion of the $d$-dimensional counterterms for $V_{C^2}$
\begin{align}
V_{C^2}^{\m_1\n_1\m_2\n_2\m_3\n_3\mu_4\nu_4}(p_1,p_2,p_3,\bar{p}_4)\simeq\bigg[ V_{C^2}^{\m_1\n_1\m_2\n_2\m_3\n_3\mu_4\nu_4}(p_1,p_2,p_3,\bar{p}_4)\bigg]_{d=4}+ \varepsilon {V'}_{C^2}^{\m_1\n_1\m_2\n_2\m_3\n_3\mu_4\nu_4}(p_1,p_2,p_3,\bar{p}_4)
\end{align}
and the expansion of the extra term present in \eqref{tloctttcount}, for instance, gives
\begin{align}
&\frac{8\,(d-4)\,\mu^{-\varepsilon}}{(d-1)\varepsilon}\pi^{\mu_1\nu_1}(p_1)\Bigg\{b\left[\sqrt{-g}C^2\right]^{\mu_2\nu_2\mu_3\nu_3\mu_4\nu_4}(p_2,p_3,p_4)+b'\left[\sqrt{-g}E\right]^{\mu_2\nu_2\mu_3\nu_3\mu_4\nu_4}(p_2,p_3,p_4)\Bigg\}\notag\\
&\simeq -\frac{8}{3}\pi^{\mu_1\nu_1}(p_1)\Bigg\{b\left[\sqrt{-g}C^2\right]^{\mu_2\nu_2\mu_3\nu_3\mu_4\nu_4}(p_2,p_3,p_4)+b'\left[\sqrt{-g}E\right]^{\mu_2\nu_2\mu_3\nu_3\mu_4\nu_4}(p_2,p_3,p_4)\Bigg\}+O(\varepsilon).
\end{align}
After adding the counterterms to the bare correlator and expanding around $d=4$ we obtain 
\begin{align}
&\bigg[\langle{t_{loc}^{\mu_1\nu_1}T^{\mu_2\nu_2}T^{\mu_3\nu_3}T^{\mu_4\nu_4}}\rangle_{bare}+\langle{t_{loc}^{\mu_1\nu_1}T^{\mu_2\nu_2}T^{\mu_3\nu_3}T^{\mu_4\nu_4}}\rangle_{count}\bigg]_{d\to4}=\notag\\
&=\langle{t_{loc}^{\mu_1\nu_1}T^{\mu_2\nu_2}T^{\mu_3\nu_3}T^{\mu_4\nu_4}}\rangle^{(d=4)}_{Ren}+\langle{t_{loc}^{\mu_1\nu_1}T^{\mu_2\nu_2}T^{\mu_3\nu_3}T^{\mu_4\nu_4}}\rangle^{(d=4)}_{anomaly}
\end{align}
with the inclusion of an extra anomalous contribution in the final expression. In particular, this contribution takes the explicit form
\begin{align}
&\langle{t_{loc}^{\mu_1\nu_1}T^{\mu_2\nu_2}T^{\mu_3\nu_3}T^{\mu_4\nu_4}}\rangle^{(d=4)}_{anomaly}=\frac{8\,\pi^{\mu_1\nu_1}(p_1)}{3}\,\mathcal{A}^{\mu_2\nu_2\mu_3\nu_3\mu_4\nu_4}(p_2,p_3,\bar p_4)\notag\\
&\hspace{0.5cm}+\mathcal{I}^{\mu_1\nu_1}_{\alpha_1}(p_1)\left[2 \, \mathcal{C}^{\alpha_1\hspace{0.4cm}\mu_2\nu_2}_{\hspace{0.3cm}\lambda\beta_1}(p_2)\langle{T^{\lambda\beta_1}(p_1+p_2)T^{\mu_3\nu_3}(p_3)T^{\mu_4\nu_4}(\bar p_4)}\rangle_{anomaly}+(23)+(24)\right]\notag\\
&\hspace{0.5cm}-\frac{2}{3}\pi^{\mu_1\nu_1}(p_1)\bigg[\langle{T^{\mu_2\nu_2}(p_1+p_2)T^{\mu_3\nu_3}(p_3)T^{\mu_4\nu_4}(\bar p_4)}\rangle_{anomaly}+(2 3)+(2 4)\bigg].\label{tlocTTTanom}
\end{align}
It is worth noticing that the anomaly of the $3$-point function \eqref{3TAnomaly} contributes to the anomaly part of the $4$-point function in \eqref{tlocTTTanom}, as expected. The same equation can be written in the simpler form 
\begin{align}
&\langle{t_{loc}^{\mu_1\nu_1}T^{\mu_2\nu_2}T^{\mu_3\nu_3}T^{\mu_4\nu_4}}\rangle^{(d=4)}_{anomaly}=\frac{\pi^{\mu_1\nu_1}(p_1)}{3}\,\langle{T(p_1)T^{\mu_2\nu_2}(p_2)T^{\mu_3\nu_3}(p_3)T^{\mu_4\nu_4}(\bar{p}_4)}\rangle_{anomaly}\notag\\
&\hspace{4cm}+\mathcal{I}^{\mu_1\nu_1}_{\alpha_1}(p_1)\,p_{1\beta_1}\,\langle{T^{\alpha_1\beta_1}(p_1)T^{\mu_2\nu_2}(p_2)T^{\mu_3\nu_3}(p_3)T^{\mu_4\nu_4}(\bar{p}_4)}\rangle_{anomaly}\label{tlocTTTanom2}
\end{align}
with 
\begin{align}
&\langle{T(p_1)T^{\mu_2\nu_2}(p_2)T^{\mu_3\nu_3}(p_3)T^{\mu_4\nu_4}(\bar{p}_4)}\rangle_{anomaly}\equiv\delta_{\mu_1\nu_1}\langle{T^{\mu_1\nu_1}(p_1)T^{\mu_2\nu_2}(p_2)T^{\mu_3\nu_3}(p_3)T^{\mu_4\nu_4}(\bar{p}_4)}\rangle_{anomaly}\notag\\
&\qquad=8\,\mathcal{A}^{\mu_2\nu_2\mu_3\nu_3\mu_4\nu_4}(p_2,p_3,\bar p_4)-2\langle{T^{\mu_2\nu_2}(p_1+p_2)T^{\mu_3\nu_3}(p_3)T^{\mu_4\nu_4}(\bar p_4)}\rangle_{anomaly}\notag\\
&\qquad\quad-2\langle{T^{\mu_2\nu_2}(p_2)T^{\mu_3\nu_3}(p_1+p_3)T^{\mu_4\nu_4}(\bar p_4)}\rangle_{anomaly}-2\langle{T^{\mu_2\nu_2}(p_2)T^{\mu_3\nu_3}(p_3)T^{\mu_4\nu_4}(p_1+\bar p_4)}\rangle_{anomaly},
\end{align}
and 
\begin{align}
&\,p_{1\beta_1}\,\langle{T^{\alpha_1\beta_1}(p_1)T^{\mu_2\nu_2}(p_2)T^{\mu_3\nu_3}(p_3)T^{\mu_4\nu_4}(\bar{p}_4)}\rangle_{anomaly}=\notag\\
&\qquad=2 \, \mathcal{C}^{\alpha_1\hspace{0.4cm}\mu_2\nu_2}_{\hspace{0.3cm}\lambda\beta_1}(p_2)\langle{T^{\lambda\beta_1}(p_1+p_2)T^{\mu_3\nu_3}(p_3)T^{\mu_4\nu_4}(\bar p_4)}\rangle_{anomaly}+(23)+(24),
\end{align} 
having taken into account the anomaly part of the $3$-point function defined in \eqref{3TAnomaly2}.\\
We proceed by studying the other local terms of the correlator that are responsible for the generation of extra contributions to its renormalized expression, thereby contributing to the anomaly in $d=4$. These are a term with two and three $t_{loc}$ contributions 
\begin{align}
&\bigg[\langle{t_{loc}^{\mu_1\nu_1}t_{loc}^{\mu_2\nu_2}T^{\mu_3\nu_3}T^{\mu_4\nu_4}}\rangle_{bare}+\langle{t_{loc}^{\mu_1\nu_1}t_{loc}^{\mu_2\nu_2}T^{\mu_3\nu_3}T^{\mu_4\nu_4}}\rangle_{count}\bigg]_{d\to4}=\notag\\
&=\langle{t_{loc}^{\mu_1\nu_1}t_{loc}^{\mu_2\nu_2}T^{\mu_3\nu_3}T^{\mu_4\nu_4}}\rangle^{(d=4)}_{Ren}+\langle{t_{loc}^{\mu_1\nu_1}t_{loc}^{\mu_2\nu_2}T^{\mu_3\nu_3}T^{\mu_4\nu_4}}\rangle^{(d=4)}_{anomaly}
\end{align}
where
\begin{align}
&\langle{t_{loc}^{\mu_1\nu_1}(p_1)t_{loc}^{\mu_2\nu_2}(p_2)T^{\mu_3\nu_3}(p_3)T^{\mu_4\nu_4}(p_4)}\rangle_{anomaly}^{(d=4)}\notag\\
&\quad=2\,\mathcal{I}^{\mu_2\nu_2}_{\alpha_2}(p_2)\,\mathcal{I}^{\mu_1\nu_1}_{\alpha_1}(p_1)\,p_{2\beta_2}\mathcal{C}^{\alpha_1\hspace{0.4cm}\alpha_2\beta_2}_{\hspace{0.3cm}\lambda\beta_1}(p_2)\,\langle{T^{\lambda\beta_1}(p_1+p_2)T^{\mu_3\nu_3}(p_3)T^{\mu_4\nu_4}(p_4)}\rangle_{anom}\notag\\
&+\bigg\{\frac{1}{3}\,\mathcal{I}^{\mu_1\nu_1}_{\alpha_1}(p_1)\,\pi^{\mu_2\nu_2}(p_2)\,\,p_{1\beta_1}\bigg[-2\,\langle{T^{\alpha_1\beta_1}(p_1+p_2)T^{\mu_3\nu_3}(p_3)T^{\mu_4\nu_4}(p_4)}\rangle_{anom}\notag\\
&\hspace{1cm}+8\,\mathcal{A}^{\alpha_1\beta_1\mu_3\nu_3\mu_4\nu_4}(p_1,p_3,p_4)\bigg]+(12)\bigg\}+\frac{8}{9}\pi^{\mu_1\nu_1}(p_1)\pi^{\mu_2\nu_2}(p_2)\bigg\{\delta_{\alpha_2\beta_2}\,\mathcal{A}^{\alpha_2\beta_2\mu_3\nu_3\mu_4\nu_4}(p_2,p_3,p_4)\notag\\
&\hspace{2cm}-\mathcal{A}^{\mu_3\nu_3\mu_4\nu_4}(p_3,p_4)-\mathcal{A}^{\mu_3\nu_3\mu_4\nu_4}(p_1+p_3,p_4)-\mathcal{A}^{\mu_3\nu_3\mu_4\nu_4}(p_3,p_1+p_4)\bigg\},
\end{align}
that can be re-expressed as
\begin{align}
&\langle{t_{loc}^{\mu_1\nu_1}(p_1)t_{loc}^{\mu_2\nu_2}(p_2)T^{\mu_3\nu_3}(p_3)T^{\mu_4\nu_4}(p_4)}\rangle_{anomaly}^{(d=4)}\notag\\
&\quad=\frac{\pi^{\mu_1\nu_1}(p_1)}{3}\frac{\pi^{\mu_2\nu_2}(p_2)}{3}\langle{T(p_1)T(p_2)T^{\mu_3\nu_3}(p_3)T^{\mu_4\nu_4}(\bar{p}_4)}\rangle_{anomaly}\notag\\
&\qquad+\,\mathcal{I}^{\mu_2\nu_2}_{\alpha_2}(p_2)\,\mathcal{I}^{\mu_1\nu_1}_{\alpha_1}(p_1)\,p_{2\beta_2}\,p_{1\beta_1}\langle{T^{\alpha_1\beta_1}(p_1)T^{\alpha_2\beta_2}(p_2)T^{\mu_3\nu_3}(p_3)T^{\mu_4\nu_4}(p_4)}\rangle_{anom}\notag\\
&\qquad+\mathcal{I}^{\mu_1\nu_1}_{\alpha_1}(p_1)\,\frac{\pi^{\mu_2\nu_2}(p_2)}{3}\,\,p_{1\beta_1}\langle{T^{\alpha_1\beta_1}(p_1)T(p_2)T^{\mu_3\nu_3}(p_3)T^{\mu_4\nu_4}(\bar{p}_4)}\rangle_{anomaly}\notag\\
&\qquad+\frac{\pi^{\mu_1\nu_1}(p_1)}{3}\,\mathcal{I}^{\mu_2\nu_2}_{\alpha_2}(p_2)\,\,p_{2\beta_2}\langle{T(p_1)T^{\alpha_2\beta_2}(p_2)T^{\mu_3\nu_3}(p_3)T^{\mu_4\nu_4}(\bar{p}_4)}\rangle_{anomaly}.
\end{align}
In a similar fashion, the terms involving three $t_{loc}$ will contribute as
\begin{align}
&\bigg[\langle{t_{loc}^{\mu_1\nu_1}t_{loc}^{\mu_2\nu_2}t_{loc}^{\mu_3\nu_3}T^{\mu_4\nu_4}}\rangle_{bare}+\langle{t_{loc}^{\mu_1\nu_1}t_{loc}^{\mu_2\nu_2}t_{loc}^{\mu_3\nu_3}T^{\mu_4\nu_4}}\rangle_{count}\bigg]_{d\to4}=\notag\\
&=\langle{t_{loc}^{\mu_1\nu_1}t_{loc}^{\mu_2\nu_2}t_{loc}^{\mu_3\nu_3}T^{\mu_4\nu_4}}\rangle^{(d=4)}_{Ren}+\langle{t_{loc}^{\mu_1\nu_1}t_{loc}^{\mu_2\nu_2}t_{loc}^{\mu_3\nu_3}T^{\mu_4\nu_4}}\rangle^{(d=4)}_{anomaly}
\end{align}
with 
\begin{align}
&\langle{t_{loc}^{\mu_1\nu_1}(p_1)t_{loc}^{\mu_2\nu_2}(p_2)t_{loc}^{\mu_3\nu_3}(p_3)T^{\mu_4\nu_4}(p_4)}\rangle_{anomaly}^{(d=4)}=\notag\\
&=\bigg\{\frac{8}{3}\,\mathcal{I}^{\mu_1\nu_1}_{\alpha_1}(p_1)\,\mathcal{I}^{\mu_2\nu_2}_{\alpha_2}(p_2)\pi^{\mu_3\nu_3}(p_3)\bigg[\,p_{2\beta_2}\,\mathcal{C}^{\alpha_1\hspace{0.4cm}\alpha_2\beta_2}_{\hspace{0.3cm}\lambda\beta_1}(p_2)\,\mathcal{A}^{\lambda\beta_1\mu_4\nu_4}(p_1+p_2,p_4)\bigg]+(13)+(23)\bigg\}\notag\\
&+\bigg\{\frac{8}{9}\,\mathcal{I}^{\mu_1\nu_1}_{\alpha_1}(p_1)\pi^{\mu_2\nu_2}(p_2)\,\pi^{\mu_3\nu_3}(p_3)\bigg[p_{1\beta_1}\,\delta_{\alpha_2\beta_2}\,\mathcal{A}^{\alpha_1\beta_1\alpha_2\beta_2\mu_4\nu_4}(p_1,p_2,p_4)-p_{1\beta_1}\,\mathcal{A}^{\alpha_1\beta_1\mu_4\nu_4}(p_1,p_4)\notag\\
&-p_{1\beta_1}\,\mathcal{A}^{\alpha_1\beta_1\mu_4\nu_4}(p_1+p_3,p_4)-p_{1\beta_1}\,\mathcal{A}^{\alpha_1\beta_1\mu_4\nu_4}(p_1,p_3+p_4)\bigg]+(12)+(13)\bigg\}\notag\\
&+\frac{8}{27}\,\pi^{\mu_1\nu_1}(p_1)\pi^{\mu_2\nu_2}(p_2)\,\pi^{\mu_3\nu_3}(p_3)\bigg\{\,\delta_{\alpha_2\beta_2}\,\delta_{\alpha_3\beta_3}\,\mathcal{A}^{\alpha_2\beta_2\alpha_3\beta_3\mu_4\nu_4}(p_2,p_3,p_4)\notag\\
&\hspace{1cm}-\delta_{\alpha_3\beta_3}\,\mathcal{A}^{\alpha_3\beta_3\mu_4\nu_4}(p_3,p_4)-\delta_{\alpha_3\beta_3}\,\mathcal{A}^{\alpha_3\beta_3\mu_4\nu_4}(p_1+p_3,p_4)-\delta_{\alpha_3\beta_3}\,\mathcal{A}^{\alpha_3\beta_3\mu_4\nu_4}(p_3,p_1+p_4)\bigg\}
\end{align}
that we rewrite in the form
\begin{align}
&\langle{t_{loc}^{\mu_1\nu_1}(p_1)t_{loc}^{\mu_2\nu_2}(p_2)t_{loc}^{\mu_3\nu_3}(p_3)T^{\mu_4\nu_4}(p_4)}\rangle_{anomaly}^{(d=4)}=\notag\\[1.5ex]
&=\frac{\pi^{\mu_1\nu_1}(p_1)}{3}\frac{\pi^{\mu_2\nu_2}(p_2)}{3}\,\frac{\pi^{\mu_3\nu_3}(p_3)}{3}\langle{T(p_1)T(p_2)T(p_3)T^{\mu_4\nu_4}(\bar{p}_4)}\rangle_{anomaly}\notag\\
&+\bigg\{\,\mathcal{I}^{\mu_1\nu_1}_{\alpha_1}(p_1)\,\mathcal{I}^{\mu_2\nu_2}_{\alpha_2}(p_2)\frac{\pi^{\mu_3\nu_3}(p_3)}{3}\langle{T^{\alpha_1\beta_1}(p_1)T^{\alpha_2\beta_2}(p_2)T(p_3)T^{\mu_4\nu_4}(\bar{p}_4)}\rangle_{anomaly}+(13)+(23)\bigg\}\notag\\
&+\bigg\{\,\mathcal{I}^{\mu_1\nu_1}_{\alpha_1}(p_1)\frac{\pi^{\mu_2\nu_2}(p_2)}{3}\,\frac{\pi^{\mu_3\nu_3}(p_3)}{3}\,p_{1\beta_1}\langle{T^{\alpha_1\beta_1}(p_1)T(p_2)T(p_3)T^{\mu_4\nu_4}(\bar{p}_4)}\rangle_{anomaly}+(12)+(13)\bigg\}.
\end{align}
Finally, the last term that involves four insertions of the operator $t_{loc}$, after renormalization takes the form
\begin{align}
&\bigg[\langle{t_{loc}^{\mu_1\nu_1}t_{loc}^{\mu_2\nu_2}t_{loc}^{\mu_3\nu_3}t_{loc}^{\mu_4\nu_4}}\rangle_{bare}+\langle{t_{loc}^{\mu_1\nu_1}t_{loc}^{\mu_2\nu_2}t_{loc}^{\mu_3\nu_3}t_{loc}^{\mu_4\nu_4}}\rangle_{count}\bigg]_{d\to4}=\notag\\
&=\langle{t_{loc}^{\mu_1\nu_1}t_{loc}^{\mu_2\nu_2}t_{loc}^{\mu_3\nu_3}t_{loc}^{\mu_4\nu_4}}\rangle^{(d=4)}_{Ren}+\langle{t_{loc}^{\mu_1\nu_1}t_{loc}^{\mu_2\nu_2}t_{loc}^{\mu_3\nu_3}t_{loc}^{\mu_4\nu_4}}\rangle^{(d=4)}_{anomaly},
\end{align}
generating an extra term that contributes to the anomaly part of the full correlator as
\begin{align}
&\langle{t_{loc}^{\mu_1\nu_1}t_{loc}^{\mu_2\nu_2}t_{loc}^{\mu_3\nu_3}t_{loc}^{\mu_4\nu_4}}\rangle^{(d=4)}_{anomaly}\notag\\
&\quad=\bigg\{\frac{8}{9}\,\mathcal{I}^{\mu_1\nu_1}_{\alpha_1}(p_1)\,\mathcal{I}^{\mu_2\nu_2}_{\alpha_2}(p_2)\,\pi^{\mu_3\nu_3}(p_3)\,\pi^{\mu_4\nu_4}(p_4)\,\bigg[\,p_{2\beta_2}\,\mathcal{C}^{\alpha_1\hspace{0.4cm}\alpha_2\beta_2}_{\hspace{0.3cm}\lambda\beta_1}(p_2)\,\delta_{\alpha_4\beta_4}\,\mathcal{A}^{\lambda\beta_1\alpha_4\beta_4}(p_1+p_2,p_4)\bigg]\notag\\
&\hspace{2cm}+(13)+(23)+(14)+(24)+(13)(24)\bigg\}\notag\\
&\quad+\bigg\{\frac{8}{27}\,\mathcal{I}^{\mu_1\nu_1}_{\alpha_1}(p_1)\,\pi^{\mu_2\nu_2}(p_2)\,\pi^{\mu_3\nu_3}(p_3)\,\pi^{\mu_4\nu_4}(p_4)\,\bigg[\,p_{1\beta_1}\,\delta_{\alpha_2\beta_2}\delta_{\alpha_3\beta_3}\,\mathcal{A}^{\alpha_1\beta_1\alpha_2\beta_2\alpha_3\beta_3}(p_1,p_2,p_3)\notag\\
&\hspace{1cm}-\,p_{1\beta_1}\,\delta_{\alpha_3\beta_3}\,\mathcal{A}^{\alpha_1\beta_1\alpha_3\beta_3}(p_1,p_3)-\,p_{1\beta_1}\,\delta_{\alpha_3\beta_3}\,\mathcal{A}^{\alpha_1\beta_1\alpha_3\beta_3}(p_1,p_3+p_4)\notag\\
&\hspace{3cm}-\,p_{1\beta_1}\,\delta_{\alpha_3\beta_3}\,\mathcal{A}^{\alpha_1\beta_1\alpha_3\beta_3}(p_1+p_4,p_3)\bigg]+(12)+(13)+(14)\bigg\}\notag\\
&\quad+\frac{8}{81}\,\pi^{\mu_1\nu_1}(p_1)\,\pi^{\mu_2\nu_2}(p_2)\,\pi^{\mu_3\nu_3}(p_3)\,\pi^{\mu_4\nu_4}(p_4)\notag\\
&\hspace{1cm}\times\bigg[\delta_{\alpha_2\beta_2}\delta_{\alpha_3\beta_3}\delta_{\alpha_4\beta_4}\mathcal{A}^{\alpha_2\beta_2\alpha_3\beta_3\alpha_4\beta_4}(p_2,p_3,p_4)-\delta_{\alpha_2\beta_2}\delta_{\alpha_4\beta_4}\mathcal{A}^{\alpha_2\beta_2\alpha_4\beta_4}(p_1+p_2,p_4)\notag\\
&\hspace{3cm}-\delta_{\alpha_2\beta_2}\delta_{\alpha_4\beta_4}\mathcal{A}^{\alpha_2\beta_2\alpha_4\beta_4}(p_2,p_1+p_4)-\delta_{\alpha_3\beta_3}\delta_{\alpha_4\beta_4}\mathcal{A}^{\alpha_3\beta_3\alpha_4\beta_4}(p_1+p_3,p_4)
\bigg].
\end{align}
It can be re-expressed as
\begin{align}
	&\langle{t_{loc}^{\mu_1\nu_1}t_{loc}^{\mu_2\nu_2}t_{loc}^{\mu_3\nu_3}t_{loc}^{\mu_4\nu_4}}\rangle^{(d=4)}_{anomaly}\notag\\
	&\quad=\frac{\pi^{\mu_1\nu_1}(p_1)}{3}\,\frac{\pi^{\mu_2\nu_2}(p_2)}{3}\,\frac{\pi^{\mu_3\nu_3}(p_3)}{3}\,\frac{\pi^{\mu_4\nu_4}(p_4)}{3}\langle{T(p_1)T(p_2)T(p_3)T(\bar{p}_4)}\rangle_{anomaly}\notag\\
	&\qquad+\bigg\{\,\mathcal{I}^{\mu_1\nu_1}_{\alpha_1}(p_1)\,\mathcal{I}^{\mu_2\nu_2}_{\alpha_2}(p_2)\,\frac{\pi^{\mu_3\nu_3}(p_3)}{3}\,\frac{\pi^{\mu_4\nu_4}(p_4)}{3}\,p_{1\beta_1}\,p_{2\beta_2}\langle{T^{\alpha_1\beta_1}(p_1)T^{\alpha_2\beta_2}(p_2)T(p_3)T(p_4)}\rangle_{anomaly}\notag\\
	&\hspace{2cm}+(13)+(23)+(14)+(24)+(13)(24)\bigg\}\notag\\
	&\quad+\bigg\{\,\mathcal{I}^{\mu_1\nu_1}_{\alpha_1}(p_1)\,\frac{\pi^{\mu_2\nu_2}(p_2)}{3}\,\frac{\pi^{\mu_3\nu_3}(p_3)}{3}\,\frac{\pi^{\mu_4\nu_4}(p_4)}{3}p_{1\beta_1}\langle{T^{\alpha_1\beta_1}(p_1)T(p_2)T(p_3)T(\bar{p}_4)}\rangle_{anomaly}\notag\\
	&\hspace{2cm}+(12)+(13)+(14)\bigg\}.
\end{align}
In summary we have obtained for the renormalized 4T the general expression
\begin{align}
&\bigg[\langle{T^{\mu_1\nu_1}T^{\mu_2\nu_2}T^{\mu_3\nu_3}T^{\mu_4\nu_4}}\rangle_{bare}+\langle{T^{\mu_1\nu_1}T^{\mu_2\nu_2}T^{\mu_3\nu_3}T^{\mu_4\nu_4}}\rangle_{count}\bigg]_{d\to4}=\notag\\
&=\langle{T^{\mu_1\nu_1}T^{\mu_2\nu_2}T^{\mu_3\nu_3}T^{\mu_4\nu_4}}\rangle^{(d=4)}_{Ren}+\langle{T^{\mu_1\nu_1}T^{\mu_2\nu_2}T^{\mu_3\nu_3}T^{\mu_4\nu_4}}\rangle^{(d=4)}_{anomaly}
\end{align} 
where the anomaly part can be identified using all the results presented above as
\begin{align}
&\langle{T^{\mu_1\nu_1}(p_1)T^{\mu_2\nu_2}(p_2)T^{\mu_3\nu_3}(p_3)T^{\mu_4\nu_4}(\bar{p}_4)}\rangle^{(d=4)}_{anomaly}=\notag\\ &\hspace{1cm}=\Bigg[\langle{t_{loc}^{\mu_1\nu_1}(p_1)T^{\mu_2\nu_2}(p_2)T^{\mu_3\nu_3}(p_3)T^{\mu_4\nu_4}(\bar{p}_4)}\rangle^{(d=4)}_{anomaly}+(1\,2)+(1\, 3)+(1\, 4)\Bigg]\notag\\
&-\Bigg[\langle{t_{loc}^{\mu_1\nu_1}(p_1)t_{loc}^{\mu_2\nu_2}(p_2)T^{\mu_3\nu_3}(p_3)T^{\mu_4\nu_4}(\bar{p}_4)}\rangle^{(d=4)}_{anomaly}+(1\,3)+(1\,4)+(2\,3)+(2\,4)+(1\,3)(2\,4)\Bigg]\notag\\
&\hspace{1cm}+\Bigg[\langle{t_{loc}^{\mu_1\nu_1}(p_1)t_{loc}^{\mu_2\nu_2}(p_2)t_{loc}^{\mu_3\nu_3}(p_3)T^{\mu_4\nu_4}(\bar{p}_4)}\rangle^{(d=4)}_{anomaly}+(1\,4)+(2\,4)+(3\,4)\Bigg]\notag\\
&\hspace{2cm}-\langle{t_{loc}^{\mu_1\nu_1}(p_1)t_{loc}^{\mu_2\nu_2}(p_2)t_{loc}^{\mu_3\nu_3}(p_3)t_{loc}^{\mu_4\nu_4}(\bar{p}_4)}\rangle^{(d=4)}_{anomaly}\label{TTTTanomaly}.
\end{align}
We have shown how the anomaly part of the 4T is extracted through the procedure of renormalization. It is clear from this procedure that such component is exactly the one predicted by the 4-dimensional reconstruction method, using the anomalous Ward identities. 

  \section{Summary}
In summary, we write the anomaly part of the correlator in the form
\begin{align}
&\langle{T^{\mu_1\nu_1}(p_1)T^{\mu_2\nu_2}(p_2)T^{\mu_3\nu_3}(p_3)T^{\mu_4\nu_4}(\bar{p}_4)}\rangle^{(d=4)}_{anomaly}=\notag\\
&\qquad=\langle{T^{\mu_1\nu_1}(p_1)T^{\mu_2\nu_2}(p_2)T^{\mu_3\nu_3}(p_3)T^{\mu_4\nu_4}(\bar{p}_4)}\rangle_{poles}+\langle{T^{\mu_1\nu_1}(p_1)T^{\mu_2\nu_2}(p_2)T^{\mu_3\nu_3}(p_3)T^{\mu_4\nu_4}(\bar{p}_4)}\rangle_{0-residue}
\end{align}
where the first contribution is anomalous (Weyl-variant) and the second one
is traceless
\begin{equation}
	\delta_{\mu_i\nu_i}\langle{T^{\mu_1\nu_1}(p_1)T^{\mu_2\nu_2}(p_2)T^{\mu_3\nu_3}(p_3)T^{\mu_4\nu_4}(\bar{p}_4)}\rangle_{0-residue}=0,\qquad i=1,2,3,4. 
\end{equation}
We call it the "zero residue" or the "zero trace" (0T) part, since the operation of tracing the anomalous part removes the anomaly pole in the bilinear mixing terms, leaving a residue which is proportional to the anomaly. This part carries no pole.\\
This contribution is explicitly given by the expression
\begin{align}
\label{Weyl}
&\langle{T^{\mu_1\nu_1}(p_1)T^{\mu_2\nu_2}(p_2)T^{\mu_3\nu_3}(p_3)T^{\mu_4\nu_4}(\bar{p}_4)}\rangle_{0-residue}=\notag\\
&=\mathcal{I}^{\mu_1\nu_1}_{\alpha_1}(p_1)\,p_{1\beta_1}\,\langle{T^{\alpha_1\beta_1}(p_1)T^{\mu_2\nu_2}(p_2)T^{\mu_3\nu_3}(p_3)T^{\mu_4\nu_4}(\bar{p}_4)}\rangle_{anomaly}+(perm.)\notag\\
&-\bigg\{\bigg[\,\mathcal{I}^{\mu_2\nu_2}_{\alpha_2}(p_2)\,\mathcal{I}^{\mu_1\nu_1}_{\alpha_1}(p_1)\,p_{2\beta_2}\,p_{1\beta_1}\langle{T^{\alpha_1\beta_1}(p_1)T^{\alpha_2\beta_2}(p_2)T^{\mu_3\nu_3}(p_3)T^{\mu_4\nu_4}(p_4)}\rangle_{anom}\notag\\
&\qquad+\mathcal{I}^{\mu_1\nu_1}_{\alpha_1}(p_1)\,\frac{\pi^{\mu_2\nu_2}(p_2)}{3}\,\,p_{1\beta_1}\langle{T^{\alpha_1\beta_1}(p_1)T(p_2)T^{\mu_3\nu_3}(p_3)T^{\mu_4\nu_4}(\bar{p}_4)}\rangle_{anomaly}\notag\\
&\qquad+\frac{\pi^{\mu_1\nu_1}(p_1)}{3}\,\mathcal{I}^{\mu_2\nu_2}_{\alpha_2}(p_2)\,\,p_{2\beta_2}\langle{T(p_1)T^{\alpha_2\beta_2}(p_2)T^{\mu_3\nu_3}(p_3)T^{\mu_4\nu_4}(\bar{p}_4)}\rangle_{anomaly}\bigg]+(perm.)\bigg\}\notag\\
&+\bigg\{\bigg[\mathcal{I}^{\mu_1\nu_1}_{\alpha_1}(p_1)\,\mathcal{I}^{\mu_2\nu_2}_{\alpha_2}(p_2)\frac{\pi^{\mu_3\nu_3}(p_3)}{3}\langle{T^{\alpha_1\beta_1}(p_1)T^{\alpha_2\beta_2}(p_2)T(p_3)T^{\mu_4\nu_4}(\bar{p}_4)}\rangle_{anomaly}+(13)+(23)\bigg]+(perm.)\notag
\end{align}
\begin{align}
&+\bigg[\,\mathcal{I}^{\mu_1\nu_1}_{\alpha_1}(p_1)\frac{\pi^{\mu_2\nu_2}(p_2)}{3}\,\frac{\pi^{\mu_3\nu_3}(p_3)}{3}\,p_{1\beta_1}\langle{T^{\alpha_1\beta_1}(p_1)T(p_2)T(p_3)T^{\mu_4\nu_4}(\bar{p}_4)}\rangle_{anomaly}+(12)+(13)\bigg]+(perm.)\bigg\}\notag\\
&\qquad-\bigg\{\,\mathcal{I}^{\mu_1\nu_1}_{\alpha_1}(p_1)\,\mathcal{I}^{\mu_2\nu_2}_{\alpha_2}(p_2)\,\frac{\pi^{\mu_3\nu_3}(p_3)}{3}\,\frac{\pi^{\mu_4\nu_4}(p_4)}{3}\,p_{1\beta_1}\,p_{2\beta_2}\langle{T^{\alpha_1\beta_1}(p_1)T^{\alpha_2\beta_2}(p_2)T(p_3)T(p_4)}\rangle_{anomaly}\notag\\
&\hspace{1.5cm}+(13)+(23)+(14)+(24)+(13)(24)\bigg\}\notag\\
&\quad-\bigg\{\,\mathcal{I}^{\mu_1\nu_1}_{\alpha_1}(p_1)\,\frac{\pi^{\mu_2\nu_2}(p_2)}{3}\,\frac{\pi^{\mu_3\nu_3}(p_3)}{3}\,\frac{\pi^{\mu_4\nu_4}(p_4)}{3}p_{1\beta_1}\langle{T^{\alpha_1\beta_1}(p_1)T(p_2)T(p_3)T(\bar{p}_4)}\rangle_{anomaly}\notag\\
&\hspace{2cm}+(12)+(13)+(14)\bigg\}.
\end{align}
On the other hand, the anomaly part is then explicitly given as
\begin{align}
&\langle{T^{\mu_1\nu_1}(p_1)T^{\mu_2\nu_2}(p_2)T^{\mu_3\nu_3}(p_3)T^{\mu_4\nu_4}(\bar{p}_4)}\rangle_{poles}=\notag\\
&=\frac{\pi^{\mu_1\nu_1}(p_1)}{3}\,\langle{T(p_1)T^{\mu_2\nu_2}(p_2)T^{\mu_3\nu_3}(p_3)T^{\mu_4\nu_4}(\bar{p}_4)}\rangle_{anomaly}+(perm.)\notag\\
&-\frac{\pi^{\mu_1\nu_1}(p_1)}{3}\frac{\pi^{\mu_2\nu_2}(p_2)}{3}\,\langle{T(p_1)T(p_2)T^{\mu_3\nu_3}(p_3)T^{\mu_4\nu_4}(\bar{p}_4)}\rangle_{anomaly}+(perm.)\notag\\
&+\frac{\pi^{\mu_1\nu_1}(p_1)}{3}\frac{\pi^{\mu_2\nu_2}(p_2)}{3}\frac{\pi^{\mu_3\nu_3}(p_2)}{3}\,\langle{T(p_1)T(p_2)T(p_3)T^{\mu_4\nu_4}(\bar{p}_4)}\rangle_{anomaly}+(perm.)\notag\\
&-\frac{\pi^{\mu_1\nu_1}(p_1)}{3}\frac{\pi^{\mu_2\nu_2}(p_2)}{3}\frac{\pi^{\mu_3\nu_3}(p_3)}{3}\frac{\pi^{\mu_4\nu_4}(p_4)}{3}\,\langle{T(p_1)T(p_2)T(p_3)T(\bar{p}_4)}\rangle_{anomaly}.
\end{align}
\begin{figure}[t]
 \centering
	\raisebox{-1.5ex}{\includegraphics[scale=0.5]{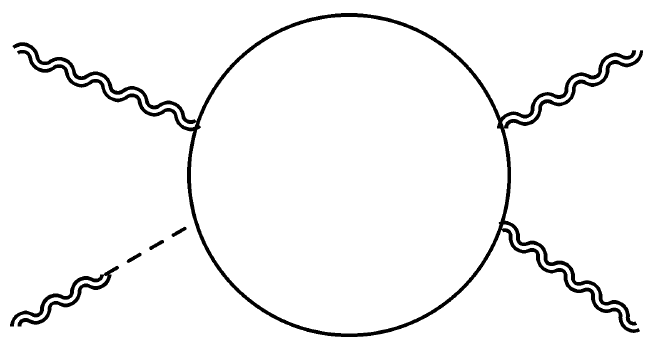}} 
	\raisebox{-1.5ex}{\includegraphics[scale=0.5]{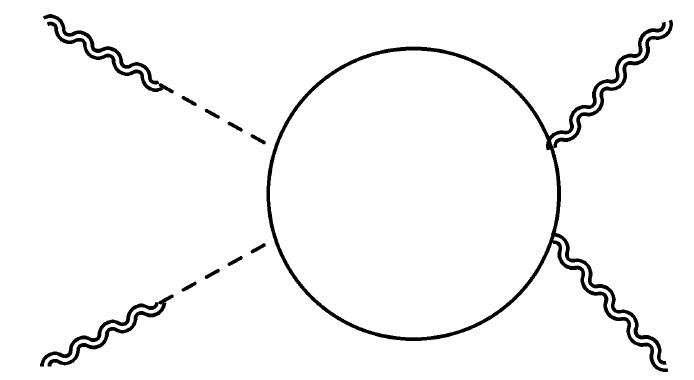}} 
	\raisebox{-1.5ex}{\includegraphics[scale=0.5]{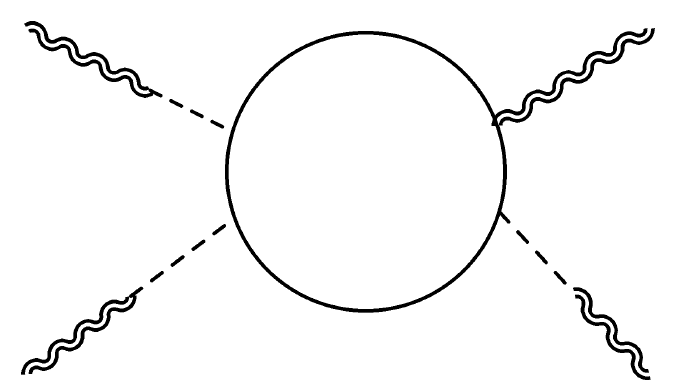}} 
	\raisebox{-1.5ex}{\includegraphics[scale=0.5]{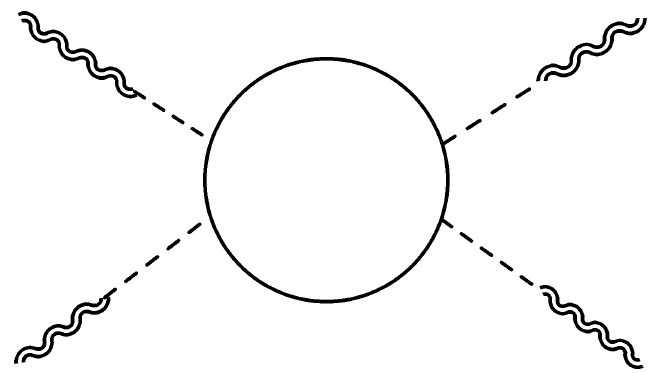}} 
	\caption{The Weyl-variant contributions from $\mathcal{S}_A$  to the renormalized vertex for the 4T with the corresponding bilinear mixings in $d=4$ \label{4T}}
\end{figure}
The picture that emerges from our analysis is shown in \figref{4T}, which generalizes the case of the 3T, with the inclusion of one extra bilinear graviton- spin-0 mixing term on the external legs. \\
However, we have shown though, that differently from the 3T, in the 4T there is an extra, Weyl-invariant  component in $\mathcal{S}_A$ that is also predicted by the reconstruction. It is also clear why this component is not present in the 3T. \\
Our result for the anomaly action, as predicted by the CWIs, can then be collected into the form 
\begin{align}
\label{res}
\mathcal{S}_A=&\int d^4 x_1 d^4 x_2 \langle T\cdot h(x_1)T\cdot h(x_2)\rangle +
\int d^4 x_1 d^4 x_2 d^4 x_3 \langle T\cdot h(x_1)T\cdot h(x_2)T\cdot h(x_3)\rangle_{pole}
\notag \\
&+ \int d^4 x_1 d^4 x_2 d^4 x_3 d^4 x_4 
\left( \langle T\cdot h(x_1)T\cdot h(x_2)T\cdot h(x_3)T\cdot h(x_4)\rangle_{pole} +\right.
\notag \\
&\qquad\qquad\qquad\hspace{2cm} \left. + \langle T\cdot h(x_1)T\cdot h(x_2)T\cdot h(x_3)T\cdot h(x_4)\rangle_{0T}\right),
\end{align}
where we have also included the (complete) $TT$, plus the extra traceless (0T) term appearing in the 4T, as identified in \eqref{Weyl}.\\
 It is quite clear from \eqref{res} that $\mathcal{S}_A$ can be organized at each order in the expansion in the gravitational fluctuations $h$, in terms of pole parts and of traceless contributions. In principle, all the traceless contributions can be omitted from the definition of $\mathcal{S}_A$ and the entire result can be expressed uniquely in terms of contributions affected by bilinear mixings (pole) terms, but this is matter of convention. The "true" and unique anomaly action can be defined only by the exact evalutation of the functional expansion which defines the complete effective action $\mathcal{S}(g)$ of which 
 $\mathcal{S}_A$ is part. \\
 Obviously, the key feature of $\mathcal{S}_A$ lays in its specific pole structure, which is characterised by the insertion of multiple $R\square^{-1}$ operators.
As we have  emphasized at various stages in this work, we don't allow massless tadpoles in our regularization scheme, and the linear terms in $h$, which otherwise would be present and dominant at the Planck scale, in a flat background are absent. These terms would be phenomenologically important if one were interested in extending our analysis to the cosmological case.\\
 For instance, one possible application would be to determine the contribution of the conformal anomaly both in the early and late stages of the cosmological evolution, addressing the issue of the dark energy dominance of its more recent epochs. This 
 can be performed using an extension of this procedure to more general backgrounds, starting from the Weyl-flat case.\\
The structure of the traceless contributions beyond the fourth order, induced by the CWIs at the level of the 5T and higher, can be worked out order by order, following the approach that we have illustrated, as we are going to describe in a related work.\\
Being the analysis formulated in the Euclidean case, it is clear that the extension of our result to Minkowski space requires an analytic continuation. It is quite obvious, though, that such continuation is trivial, since the residui at each pole are just polynomial in momentum space and do not involve any branch cut. The residui computed at each pole, for any nT, are just given by anomaly polynomials, obtained by functional differentiations of the anomaly functional to arbitrary high orders.\\
The logarithmic contributions which are part of the structure of the anomaly action, as shown in the case of the 3T, using the free field theory analysis, contribute only to the Weyl-invariant part of the effective action \cite{Coriano:2018bsy1}. \\
It is then clear that $\mathcal{S}_A$ predicts the emergence of intermediate massless states coupled to such anomaly polynomials. The entire anomaly action then appears to be quite simple, and can be identified in a very direct way. 
\chapter{Topological Terms in Gravitational Anomalies}
In this chapter we investigate the gravitational backreaction, generated by coupling a general conformal sector to external, classical gravity,  as described by a conformal anomaly effective action. We address the issues raised by the regularization of the topological Gauss-Bonnet and Weyl terms in these actions and the use of dimensional regularization (DR).    
We discuss both their local and nonlocal expressions, as possible IR and UV descriptions of conformal theories, below and above the conformal breaking scale. Our discussion overlaps with several recent studies of dilaton gravities - obtained via a certain singular limit of the Einstein-Gauss-Bonnet (EGB) theory - originally introduced as a way to bypass Lovelock's theorem.  
Nonlocal, purely gravitational realizations of such EGB theories, quadratic in the dilaton field, beside their local quartic forms, are possible by a finite renormalization of the Euler density.
Such nonlocal versions, which are deprived of any scale, can be expanded, at least around flat space,   
in terms of the combination $R \Box^{-1}$ times multiple variations of the anomaly functional, as pointed out in recent studies at $d=4$. Similar conclusions can be drawn for the proposed nonlocal EGB theory. The expansion emerges from previous investigations of the anomalous conformal Ward identities that constrain such theories around the flat spacetime limit in momentum space.
\section{Introduction} 
The search for corrections to general relativity (GR) and to its Einstein-Hilbert (EH) action by higher derivative terms, is characterized by a large number of both older and of more recent proposals. Their goal  is to address unsolved 
issues, such as the nature of dark energy and the mechanism of inflation of the early universe, in a 
more satisfactory way.\\
 From $f(R)$ theories to models incorporating a dilaton field (dilaton gravities), including Horndeski and Lovelock actions \cite{Charmousis:2014mia}, just to mention a few, important issues  need to be addressed both of phenomenological and of theoretical character.  
An important open question concerns the quantum consistency of these extensions, since the presence of higher order derivatives in the action leads, in general, to equations of motion of higher order. 
In quantum gravity, particular attention is paid to the stability and the consistency of such theories, by showing, for instance, the absence of tachyonic solutions as well as of ghosts and, eventually, addressing their renormalizability in a perturbative context. \\
Among these proposals, of particular interest are those extensions that lead to second-order equations of motion, even though they are generated by higher derivatives Lagrangians. Such Lagrangians may be introduced at classical level, or, alternatively, they may originate from the inclusion of quantum corrections, in models where gravity is still treated classically. Their structure depends on the specific type of matter sector that is integrated out of the quantum partition function. If the matter sector is conformal, we will refer to the ensuing semiclassical effective action as an action modified by a conformal backreaction.  The important role of conformal invariance and of anomalies in curved space, in the gravitational context, has been advocated for a long time in the physics of the the early universe \cite{Starobinsky:1980te}.
For example, it has been observed  that the embedding of flat spatial $R^3$
sections in de Sitter space induces a conformal invariant perturbation spectrum, with predictions for the shape of the non-Gaussian CMB bispectrum \cite{Antoniadis:2011ib}. 

\subsection{Higher derivative corrections}
The results of these analysis are, at any case, contained in classical Lagrangians which differ significantly respect to the simpler EH action density, by the inclusion of a larger number of invariants, quadratic, cubic and so on, in the curvatures. They are built out of the Riemann tensor and of its contractions, and their couplings are subject to distinct renormalization group evolutions with energy (see for instance \cite{Hamada:2014pba}). \\
In theories of induced gravity, the integration in the partition function of a matter sector can be sufficient, just by itself, to recover a EH action for gravity, accompanied by extra, higher derivative terms. Both the $R^2$ corrections and the EH term can be generated this way, realizing Sakharov's proposal of induced gravity \cite{Visser:2002ew}. In this case the spacetime is a Lorentzian manifold and the metric is essentially free, while its dynamics is entirely induced by the inclusion of quantum corrections due to a generic matter sector.  These induce an effective action of the form
\begin{equation}
\sm=\int d^4 x \sqrt{g}\left( c_1 \frac{M_P^2}{2}R + c_2 R^2 +\ldots   \right),
\end{equation}
where the ellipsis refer to extra contributions built out of higher order geometrical invariants.
In this approach, the entire gravitational theory can be viewed as the result of the quantum backreaction on a freely fluctuating metric, induced by the path integration over the matter sector. \\
One could also follow an intermediate path, by letting the quantum effects correct a tree-level action containing the standard EH term, or the EH plus the Weyl term $C^2$ or, alternatively,  just the Weyl term. In our discussion we will focus on the investigation of the quantum corrections to a classical gravity theory which can be of any of these types. 
The choice of a non-conformal background, such as the one given by Einstein gravity via the EH action, in principle can be also made conformal using the standard Weyl gauging procedure \cite{Codello:2012sn},\cite{Coriano:2013xua,Coriano:2013nja}, which modifies the theory by the inclusion of a dilaton field. Such dilaton 
theories, on physical grounds, are not of an easy interpretation, since the dilaton can be normalized to be of mass dimension one, only by the inclusion of a specific scale $(f)$. This can be interpreted as a conformal breaking scale, generating a local action that will be discussed in \secref{ef}.\\
The emergence of such a scale is an open and probably the most fundamental unsolved issue in conformal quantum theories of gravity, as pointed out by 't Hooft \cite{tHooft:2016uxd}. The Weyl gauging procedure that we will discuss in this work, which separates the conformal factor from a fiducial metric, leads to actions in which the Weyl symmetry is explicitly broken by the renormalization process.\\
 In general, from the Weyl-gauged action, the ordinary EH term is reobtained by assuming that the dilaton field will condense, causing a spontaneous breaking of the Weyl symmetry.
This possibility is an open one for theories affected by anomalies \cite{Armillis:2011hj},   
and makes the Weyl-gauged theory completely equivalent to the ordinary EH theory, with the Newton gravitational constant determined by the vev of the dilaton field. \\
 Also in this case, however, one falls short from proving that the dilaton will condense, unless either  a mechanism of ghost condensation or, alternatively, the generation of a nonperturbative extra potential for the dilaton, is, somehow, invoked. \\
 A similar behaviour is usually required in the case of non-abelian chiral gauge theories with the inclusion of a Stuckelberg field \cite{Coriano:2005js}, where the local shift symmetry of this field is expected to be broken by a periodic potential generated by instanton effects. Obviously, this raises the issue whether a similar phenomenon should be present for gravity, since boundary contributions in this theory are far less understood compared to those of ordinary gauge theories, as shown by recent progress in celestial amplitudes, and asymptotic flatness hides remarkable asymptotic symmetries.     

If the matter sector that is integrated out is conformally coupled to gravity, then this formulation adds to a tree-level Lagrangian, which may be conformal or not, the effect of conformal backreaction on the metric. The scale $f$ may be the result of the coupling of this sector to Einstein gravity, inducing corrections which take the form of an anomaly action. \\
This is a topic of special interest, since there are, obviously, strong reasons to believe that close to the Planck scale the dynamics of a matter sector may be conformal. 
 
We believe that such challenging issue motivate the analysis both of local and of nonlocal 
anomaly actions, under the assumption that both respond to different physical situations, covering the conformal dyamics around a scale $f$ or far above it. These may correspond to the cases   where $f\ll M_P$, with $M_P$ the Planck scale, for their local formulations while, in the nonlcal case they cover the UV limit, where the action can be expanded in the dimensionless variable 
$R\Box^{-1}$.
 \subsection{Content of our work}
In the next sections we are going to discuss the structure of the effective action in the most general setting, and in a quantum context, starting from the path integral formulation of this action and detailing all the steps which are relevant for its consistent definition. By considering the quantum case, obviously our discussion will be of wider scope compared to the one emerging from the purely classical analysis of Einstein-Gauss-Bonnet (EGB) theories discussed recently (see for instance \cite{Mann:1992ar}). For the rest, the two issues of 1) how to regulate and define a classical EGB theory and  of 2) how to correctly define the quantum effective action, after integrating out a conformal sector, run in close parallel, although the physical motivations for studying the two theories are very different. For this reason, in the next section we will address from scratch the issue of how to proceed with a consistent definition of a conformal backreaction, which includes, a least at formal level,  also the finite and Weyl-invariant contributions which are part of such actions, computed in a certain scheme. \\
Such Weyl-invariant terms are not uniquely identified just by the general analysis of the two counterterms, denoted by us as $V_E$ and $V_{C^2}$, which are used to regulate the quantum corrections of any effective action involving conformal sectors.\\
Our analysis, in this case, tries to identify all the terms present in the effective action which are missing in the context of anomaly-induced actions, and all the cutoffs that are present in the full renormalized action, that we will call $\sm_R$, starting from the basic path integral definition of the quantum effective action. Both for EGB theories as well as for $\sm_R$, the regularization procedures need to be defined by an appropriate embedding of the metric from 4 to $d$ dimensions. \\
\subsection{Evanescence}
An important issue for anomaly actions, as well as for the EGB theories introduced recently, is the fact that $V_E$ is an evanescent term at $d=4$. Several issues concerning the role of such topological terms have been addressed in the recent literature (see for instance \cite{Anastasiou:2020zwc,Matsumoto:2022fln}). 
\\
The Wess-Zumino (WZ) form of the topological actions, as discussed also in $d=2$ \cite{Mann:1992ar}, allows to remove the evanescence of the topological terms via a 0/0 procedure, which is common in the anomaly context. In this case, a topological term can be made finite by an infinite renormalization of its dimensionless coupling. 

In the case of the anomaly action in $d=4$, this approach is applied to both counterterms $V_E$ and $V_{C^2}$, in the form $V_{E/C^2}\to 1/(d-4)V_{E/C^2}$, as typical in DR. Notice that $V_E$ amounts to a finite renormalization, with $V_{C^2}$ being the only counterterm which is actually necessary in order to erase the pole singularities coming from a quantum conformal sector. \\
 As a last step, one needs to approach $d\to 4$ by dimensional reduction of the metric, with the results that a residual dilaton will be left in the spectrum. This is the point at which Weyl-invariant contributions are dropped in an apparently uncontrollable way. As we are going to show, this occurs for all the WZ actions defined in DR, for instance, by selecting a fiducial metric.  \\
One may argue that Riegert's scaling procedure  \cite{1984PhLB..134...56R}, which is applied strictly at $d=4$, is free of such ambiguities, although one may argue that a procedure based on DR should consistently move to $d=4$ from $d\neq 4$, and not just be limited to $d=4$. 
Mazur and Mottola \cite{Mazur:2001aa} have shown, however, that a finite extension at $O(\epsilon)$ in the definition of the topological term $V_E$, allows to recover the nonlocal action as in the 
$d=4$ case, applying a regularization in which the fields are, at the end, assumed not to depend on the extra coordinates of the embedding space.  The result in \cite{Mazur:2001aa} is, obviously quite significant, since it shows that a pure gravitational nonlocal theory can be consistently obtained in DR, following the canons of these prescriptions. \\
The effective action that results from this analysis has a clear $R\Box^{-1}$ structure around flat spacetime, and  provides an alternative formulation of EGB theories, as already pointed out in 
\cite{Coriano:2022knl}. \\
Our goal will be also to show, quite simply, that such nonlocal behaviour resurfaces from dilaton gravities once we include the conformal breaking scale into the theory and let this scale approach the Planck mass.    
\subsection{Regularizations}

The extraction of the conformal backreaction in specific metric backgrounds requires the identification of a regulated action containing, beside the tree-level classical gravity part, also the contribution coming from the finite (integral) terms 
\begin{equation}
V'_E=\frac{\partial V_E(g, d)}{\partial d}\mid_{d=4}, \qquad V'_{C^2}=\frac{\partial V_{C^2}(g,d)}{\partial d}\mid_{d=4},
\end{equation}
derived by a dimensional expansion of $V_E$ and $V_C^2$ around $d=4$. Both contributions 
are well-defined for non-singular metrics, if we assume asymptotic flatness, and the expansion in the spacetime dimension $d$ will clearly depend on the form of the metric chosen for the computation of such integrals. \\
The remaining part of the action is defined by the other finite contributions which result from the regularization procedure.  In general, these are not included in the explicit computation of the conformal backreaction, although they can be identified in specific backgrounds, such as in the quasi-flat and in the Weyl-flat cases. For example, if the coupling of the conformal sector to gravity is defined just by a free-field theory realization, they can be explicitly computed at one-loop. Examples are those discussed in \cite{Giannotti:2008cv,Coriano:2018bsy,Armillis:2009pq}, 
but similar computations, in principle, can be performed in De Sitter space.  \\
More recently, there has been a proposal \cite{Glavan:2019inb} in which the GB term has been singled out as a possible correction to Einstein gravity, in the same counterterm form discussed above, with $\alpha\to \alpha/(d-4)$. The claim has been that such  - purely classical - procedure generates second order equations of motion in a pure gravitational theory, evading, 
via this specific 0/0 limit, Lovelock's theorem \cite{Lovelock:1971yv}.\\
Soon after the apperance of \cite{Glavan:2019inb}, it was observed that the 0/0 limit on the $\alpha V_E(d)$ term, needed to be carefully re-examined, for being incorrect \cite{Gurses:2020ofy}. \\
 Indeed, the finite actions introduced as a follow up of \cite{Glavan:2019inb}, by introducing a suitable regularization, define various forms of dilaton gravities, and as such do not violate Lovelock's theorem after all, but can be classified as particular cases of Horndeski gravities \cite{Hennigar:2020lsl,Fernandes:2020nbq,Lu:2020iav}.  \\
In general, these analysis share similarities with traditional anomaly actions, since the ${\alpha}/{(d-4)}V_E$ counterterm and the corresponding regularization procedure have been part of the derivation of the latter  for a long time. These developments have drawn considerable attention \cite{Hennigar:2020lsl,Fernandes:2020nbq,Easson:2020mpq,Kobayashi:2020wqy,Konoplya:2020qqh,Bonifacio:2020vbk,Ai:2020peo,Wei:2020ght,Aoki:2020lig,Nojiri:2020tph,Konoplya:2020bxa,Guo:2020zmf,Fernandes:2020rpa,Casalino:2020kbt,Hegde:2020xlv,Ghosh:2020vpc,Doneva:2020ped,Zhang:2020qew,Konoplya:2020ibi,Singh:2020xju,Ghosh:2020syx,Konoplya:2020juj,Kumar:2020uyz,Zhang:2020qam,HosseiniMansoori:2020yfj,Wei:2020poh,Singh:2020nwo,Churilova:2020aca,Islam:2020xmy,Mishra:2020gce,Konoplya:2020cbv,Zhang:2020sjh,EslamPanah:2020hoj,Aragon:2020qdc,Aoki:2020iwm,Shu:2020cjw,Mahapatra:2020rds,Lu:2020iav,Gurses:2020ofy,Banerjee:2020dad,Ge:2020tid,Yang:2020jno,Lin:2020kqe,Yang:2020czk}. \\
As we are going to show,  such regulated actions, which modify the suggestion of  \cite{Glavan:2019inb},
are simply WZ actions, in a realization that can be defined as a local one, since it is explicitly characterised by the presence of a dilaton field and of quartic order in $\phi$ at $d=4$. \\
Topological terms are naturally included in an effective action once that a conformal sector is integrated out of the partition function and, as we have discussed above, they are naturally present in order to satisfy the WZ consistency condition. The 0/0 approach that we have illustrated can, obviously, be formulated for any dimension, since such topological terms are present in any even dimension and become part of the corresponding conformal anomaly action in that specific dimension. \\
The merit of \cite{Glavan:2019inb} has been that of drawing attention to the role of topological terms, which are evanescent in general, in the context of even a classical gravitational theory, following a procedure that mimics the ordinary dimensional renormalization methods of a quantum effective action.  \\

\section{The quantum effective action}
In this section we discuss the general structure of the quantum effective action, generated when a 
conformal sector is integrated out in the partition function, and characterize its Weyl-invariant contributions.    
 
The backreaction of a conformal sector on the gravitational metric can be discussed via the partition function $\mathcal{Z}_B(g)$,
identified by the bare functional (in the Euclidean case)
\begin{equation}
\label{partition}
\mathcal{Z}_B(g)=\mathcal{N}\int D\chi e^{-S_0(g,\chi)},
\end{equation} 
where $\mathcal{N}$ is a normalization constant. We have denoted by $\chi$, just as example, a conformal scalar. The logarithm of $\mathcal{Z}_B$ defines the effective action $\mathcal{S}(g)$, 

\begin{equation}
\label{defg}
e^{-\mathcal{S}_B(g)}=\mathcal{Z}_B(g) \leftrightarrow \mathcal{S}_B(g)=-\log\mathcal{Z}_B(g). 
\end{equation}
In our case, quantum matter fields are assumed to be in a conformal phase. In the case of a scalar field, for example, this can be described by the action 
\begin{align}
\label{phi}
S_0(g,\chi)=&\frac{1}{2}\int\, d^dx\,\sqrt{-g}\left[g^{\mu\nu}\nabla_\mu\chi\nabla_\nu\chi-c_0\, R\,\chi^2\right],
\end{align}
where we have included a conformal coupling $c_0(d)=\frac{1}{4}\frac{(d-2)}{(d-1)}$, and $R$ is the scalar curvature. In more general situations one can take into considerations other actions, with spin $1/2$ fermions, spin 1 ecc. \\
$\mathcal{S}_B(g)$ includes all the multiple insertions of the stress energy tensor 
  \begin{align}
 \label{defT}
T^{\mu\nu}_{scalar}
&\equiv\frac{2}{\sqrt{g}}\frac{\delta S_0}{\delta g_{\mu\nu}}\nonumber \\
=&\nabla^\mu \chi \, \nabla^\nu\chi - \frac{1}{2} \, g^{\mu\nu}\,g^{\alpha\beta}\,\nabla_\alpha \chi \, \nabla_\beta \chi
+ \chi \bigg[g^{\mu\nu} \Box - \nabla^\mu\,\nabla^\nu + \frac{1}{2}\,g^{\mu\nu}\,R - R^{\mu\nu} \bigg]\, \chi^2, 
\end{align}

and diagrammatically corresponds to the expression

\begin{align}
\label{figgx}
\sm(g)=& \sum_n \quad\raisebox{-6ex}{{\includegraphics[width=0.19\linewidth]{Figures/np}}} \,\scriptstyle \text{(n-point)}
\end{align}
which is expressed in  terms of stress energy tensor correlators $\langle T_1 T_2\ldots T_n\rangle$, with propagators and vertices that can be defined in any background, using \eqref{defT}.  The expansion can be constrained from the CWIs of the theory. \\
The simplest case that can be addressed is that of a flat background and, as shown in the figure above, can be computed by the ordinary Feynman expansion, order by order in $1/M_P^2$ in momentum space. The expansion accounts for the the metric fluctuations $h_{\mu\nu}$, with  $g_{\mu\nu}=\delta_{\mu\nu} + h_{\mu\nu}$, generated by the virtual corrections due to the scalar field in the loops. In principle, one can use any background and of particular interest is the case of a De Sitter metric.  

In general, the contributions of such diagrams are divergent as $d\to 4$ and  need to be renormalized. In turn, this can be performed by the addition of the two counterterms $V_E$ and $V_{C^2}$,
causing the violation of the conformal symmetry in the effective action, as we will be discussing next. \\
The entire set of correlation functions of stress energy tensors, to all orders in the fluctuations around certain metric background $\bar{g}$, is expressed in the form 
\begin{equation}
\label{exps2}
\sm(g)_B\equiv\sm(\bar{g})_B+\sum_{n=1}^\infty \frac{1}{2^n n!} \int d^d x_1\ldots d^d x_n \sqrt{g_1}\ldots \sqrt{g_n}\,\langle T^{\mu_1\nu_1}\ldots \,T^{\mu_n\nu_n}\rangle_{\bar{g} B}\delta g_{\mu_1\nu_1}(x_1)\ldots \delta g_{\mu_n\nu_n}(x_n),
\end{equation}
in terms of bare $(B)$ $nT$ correlators, with 
\begin{equation}
\label{exps1}
\langle T^{\mu_1\nu_1}(x_1)\ldots T^{\mu_n\nu_n}(x_n)\rangle_B \equiv\frac{2}{\sqrt{g_1}}\ldots \frac{2}{\sqrt{g_n}}\frac{\delta^n \sm_B(g)}{\delta g_{\mu_1\nu_1}(x_1)\delta g_{\mu_2\nu_2}(x_2)\ldots \delta g_{\mu_n\nu_n}(x_n)}, 
\end{equation}
where $\sqrt{g_1}\equiv \sqrt{|\textrm{det} \, g_{{\mu_1 \nu_1}}(x_1)} $ and so on. The renormalization of this functional expansion is rather involved in a general background, and can be best understood by borrowing the DR prescription around flat space. \\
In DR the divergences appear as single poles if we couple a conformal sector to gravity, and their renormalization, as already mentioned, is performed by expanding the counterterms around $d=4$. 
The two counterterms to be included are $V_E$ and $V_{C^2}$ that will be discussed next, giving a regularized effective action of the form 
\begin{equation}
\mathcal{Z}_R(g)_=\, \mathcal{N}\int D\Phi e^{-S_0(g,\Phi) + b' \frac{1}{\epsilon}V_E(g,d) + b \frac{1}{\epsilon}V_{C^2}(g,d)}.
\end{equation} 
where $\mathcal{N}$ is a normalization constant. 
Here, $b$ and $b'$ count the number of massless fields involved in the loop corrections. \\
The role of the two counterterms is to remove the $1/\epsilon$ singularities present in the bare effective action $\sm_B$
\begin{equation}
\sm_B(g,d)=-\log\left(\int D\Phi e^{-S(\Phi,g)}\right) +\log\mathcal{N},
\end{equation}
and allow to define the regularized effective action in the form 
\begin{equation}
\label{rena}
\mathcal{S}_R(g,d)=\mathcal{S}_B(g,d) +  b' \frac{1}{\epsilon}V_E(g,d) + b \frac{1}{\epsilon}V_{C^2}(g,d).
\end{equation}

\section{The counterterms }
 $V_E$ and $V_{C^2}$ are related to the Euler density $E$ and to the Weyl tensor squared $C^2$, respectively, 
defined by the expressions
\begin{align}
\label{ffr}
V_{C^2}(g, d)\equiv & \mu^{\varepsilon}\int\,d^dx\,\sqrt{-g}\, C^2, \notag \\
V_{E}(g,d)\equiv &\mu^{\varepsilon} \int\,d^dx\,\sqrt{-g}\,E , 
\end{align}
where $\mu$ is a renormalization scale while $\varepsilon=d-4$. The counterterm vertices will be simply obtained by multiple differentiations of the two expressions above. \\
We will omit $\mu$ from the expressions of such counterterms in most of our analysis, just for simplicity, by setting $\mu\to 1$, and we will reinsert it in \secref{WG}, when we move from a  naive regularization of the action to the application of a standard DR procedure.\\
The DR prescription is adopted in most of the results presented in the previous literature on the subject, sometimes ignoring the presence of cutoffs in the regularization process and neglecting Weyl-invariant terms. These terms may play a key 
role in generating consistent anomaly actions at all scales, the UV or the IR. 
Their rigorous definition, for a specific background metric, plays a central role in our analysis.\\
 It has long been known that the inclusion of $V_E$ induces a finite renormalization of the effective action, and, from this point of view, it does not seem that such couterterm  plays any role in this process. $V_E$ is necessary for the effective action to satisfy the Wess-Zumino (WZ) consistency condition. Indeed, at $d=4$, the integration of a conformal sector induces a renormalized effective action $\sm_R$ in \eqref{rena}, whose variation under an infinitesimal Weyl transformation of the metric 
 
\begin{equation}
\label{vars}
g_{\mu\nu}\to e^{2 \sigma(x)} g_{\mu\nu},\qquad 
\delta_\sigma g_{\mu\nu}= 2 \sigma g_{\mu\nu} 
\end{equation}
 ($\delta/\delta \sigma=2 g_{\mu\nu}\delta/\delta g_{\mu\nu}$) is equal to the conformal anomaly. Notice that, as far as we stay in $d$ dimensions, the $\delta \sigma$ variation 
 and the trace of the stress energy tensor of $\sm_R$ are identical, but things become more subtle as we take the $d\to 4$ limit of this action. \\
If we integrate out a conformal matter sector at quantum level, the gravitational action is modified only by contributions up to second order in the Riemann tensor
\begin{equation}
 \delta_\sigma \sm=\frac{1}{(4 \pi)^2}\int d^4 x \sqrt{g}\left(c_1 R_{\mu\nu\rho\sigma}R^{\mu\nu\rho\sigma} + c_2 R_{\mu\nu}R^{\mu\nu} +c_3 R^2 + c_4\square R\right), 
\label{var}
\end{equation}
which are constrained by the Wess-Zumino consistency condition 
\begin{equation}
\label{WZ}
\left[\delta_{\sigma_1},\delta_{\sigma_2}\right]\sm_R=0,
\end{equation}
 and the coefficients $c_i$ have to satisfy the relation $c_1+c_2 +3 c_3=0$, allowing to re-express \eqref{var} in the form 
 \begin{equation}
 \label{anof}
 \delta_\sigma \sm_R=  \frac{1}{(4 \pi)^2} \int d^4 x \sqrt{g}\delta\sigma(x) 
\left( a E + b C^2 + c\Box R\right). 
 \end{equation}
The coefficients $a,b,c$ are automatically fixed by the conformal sector that is integrated out, and the contributions $E$ and $C^2$ are both part of the variation of the renormalized effective action, generated by $V_E$ and $V_{C^2}$ contained in \eqref{rena}. 
Eq. \eqref{anof} is the usual expression of the conformal anomaly, generated by \eqref{rena}, with coefficients $a,b,c$ which are determined by the particle content of the theory: scalars, spin 1 vectors, and fermions $(n_s,n_V,n_f)$, that are integrated out in $d=4$. The last term ($\Box R$) is renormalization prescription dependent.\\ 
Since $\sm_B(g,d)$ is Weyl invariant, its Weyl variation according to  \eqref{vars} is zero 
and \eqref{anof} is entirely associated with the behaviour of $V_E$ and $V_{C^2}$ under the same transformations. There are some subtle regularization issues, on which we will come back in the next sections, that need to be readdressed once we perform the $d\to 4$ limit 
of $\sm_R(g,d)$.\\ 
In \eqref{anof} $C^2$  the Weyl tensor squared in $d=4$ 
\begin{align}
( C^{(4)})^2&\equiv R_{\mu\nu\alpha\beta}R^{\mu\nu\alpha\beta}-2R_{\mu\nu}R^{\mu\nu}+\frac{1}{3}R^2. \label{fourd2}
\end{align}
which is generalized to $d$ dimension by the expression
\begin{equation}\label{Geometry1}
C^{(d) \alpha\beta\gamma\delta}C^{(d)}_{\alpha\beta\gamma\delta}
=
R^{\alpha\beta\gamma\delta}R_{\alpha\beta\gamma\delta} -\frac{4}{d-2}R^{\alpha\beta}R_{\alpha\beta}+\frac{2}{(d-2)(d-1)}R^2,
\end{equation}
where
\begin{equation}
C^{(d)}_{\alpha\beta\gamma\delta} = R_{\alpha\beta\gamma\delta} -
\frac{1}{d-2}( g_{\alpha\gamma} \, R_{\delta\beta} + g_{\alpha\delta} \, R_{\gamma\beta}
- g_{\beta\gamma} \, R_{\delta\alpha} - g_{\beta\delta} \, R_{\gamma\alpha} ) +
\frac{1}{(d-1)(d-2)} \, ( g_{\alpha\gamma} \, g_{\delta\beta} - g_{\alpha\delta} \, g_{\gamma\beta}) R.\, 
\end{equation}
The four-dimensional expression is simply obtained by parametrically sending 
$d\to 4$ in the definiiton above. Eq. \eqref{anof} summarizes the usual $a-$ and $b-$ contributions of the conformal anomaly, the first being topological,  called the "nonlocal part", while the third term $\Box R$ is the local one, which can be attributed to  the variation of a $R^2$ term.\\
 The locality of the anomaly action, for the $E$ and $C^2$ contributions, is a result of the inclusion of one scalar field, the dilaton $\phi$, extracted from a fiducial metric as a conformal factor in the spectrum of the the theory. However, this is not the only option.\\
 The possibility of  performing a complete and consistent removal of $\phi$ in DR, and not just at $d=4$ as in Riegert's derivation \cite{1984PhLB..134...56R},  is associated with the freedom of performing an additional finite $O(\epsilon)$ renormalization   of  $V_E$,  as shown by Mazur and Mottola \cite{Mazur:2001aa} in their cohomological analysis of the effective action. \\
The nonlocal form of the action  that results in this second case, after dimensional reduction of the metric and of the dilaton field, coincides with Riegert's expression at $d=4$, and is completely consistent with DR. Notice, however, that all the approaches in the derivation of such actions differ in the power of the dilaton field and by the different inclusions of Weyl-invariant terms. This is a limitation that even an exact computation of the partition function cannot provide, due to the need of regulating the theory. Indeed, any regularization based on DR or on any other  procedure, necessarily has to reduce the metric from generic $d$ dimensions down to $d=4$, and will miss or include such terms according to a specific prescription. \\
We will clarify this point in the following sections, by carefully analising all the contributions to the effective action, separating the Weyl-invariant terms from the Weyl-variant ones, and addressing their regularization in specific schemes, all based on DR.\\
 Our results follow closely previous analysis of the anomaly-induced actions and overlap with such previous studies, but try to offer a unifying perspective on such actions. 
 Complete computations, which include also the Weyl-invariant contributions indicated as $\sm_f$ in our analysis, are possible only in specific backgrounds. One possibility is the case of a Weyl-flat metrics, that will be discussed elsewhere.\\
Possible shortcomings of either the local or non-local  forms of the quantum anomaly action are likely to be associated with such missing terms. This is an intrinsic limitation of DR, 
well-known in the case of perturbative gauge theories, where finite subtractions, handled according to the $MS$ (minimal subtractions) or to the $\overline{MS}$ (modified minimal subtraction) or the on-shell scheme, affect a final result.\\
In that context, the possibility of performing a computation up to higher peturbative orders of a given amplitudes, allows to reduce the impact of such dependences on the various scales (factorization, renormalization ecc) present in the theory. This is not deemed to be possible in gravity, given our limited knowledge of the theory at quantum level. 

\section{Topological terms and DR}

As just mentioned, one significant issue that we need to face in order to make sense of DR in field space, is to extend  
terms in the Lagrangian density to $d=p +\epsilon$ dimensions, from their original $p$-dimensional definition. 
In the case of topological terms, this is performed by first contracting all the indices in such a way to build the expressions of such contributions in terms of curvature scalars, before performing their reduction to $d=4$. For topological terms such as $E\equiv E_4$ or its 6-dimensional version $E_6$, which is cubic in the Riemann tensor, this scalar reduction is possible, due to the presence of two Levi-Civita antisymmetric tensors in their definitions, that can be reduced to Kronecker deltas. This allows to bypass the troublesome feature of regulating a single antisymmetric tensor in DR as, for example, in the chiral anomaly diagram.\\   
In the case of a curved background, we need a specific prescriptions for reducing the theory to $d=p$ from the original $d$ dimensions, and this procedure is not unique.

If we denote with $x^\mu$ the $p$ dimensional coordinates of the base manifold, and with $y^k$, \,  
$k=1,2,d-p$ the extra dimensional ones, all the invariants are embedded in the form 
\begin{equation}
(R_{\mu\nu\alpha\beta})^2 (x)\to (R_{M N A B})^2(x,y) \qquad \mu,\nu\ldots =1,2,3,4 \qquad M, N=1,\ldots d.
\end{equation}
The reduction to $d=4$ introduces, in general, a significant arbitrariness, due to the possibility of implementing various types of embeddings. This is a source of ambiguity of the corresponding effective action. For example, the extraction of a conformal factor either from the entire metric or, alternatively, only on a specific subspace of it, takes to different effective actions, which differ by Weyl-invariant terms. These two possibilities work as clear example of the difficulties encountered in DR and will be discussed next.\\
One way to recover the four-dimensional theory, for instance, is to require the independence of $d-$dimensional curvatures scalars from the extra dimensions using the conditions
 
 \begin{equation}
\label{dimred}
 R^2(x,y)\to R^2(x), \qquad  \partial_{y^k} R^2(x,y)=0,
\end{equation}

which is at the core of the regularization procedure. 
 In Minkowski space this procedure is unambiguous, since no extra degrees of freedom are induced by it, as known in ordinary gauge theories, but this is not the case for general backgrounds. \\
Perturbative computations around a flat spacetime proceed similarly to 
the non gravitational case, as shown in free field theory computations \cite{Giannotti:2008cv,Armillis:2009pq}, and in the analysis of the conformal Ward identities of 2- and 3-point functions \cite{Bzowski:2018fql,Coriano:2018bbe,Coriano:2018bsy, Bzowski:2013sza,Bzowski:2015pba}, but in more general backgrounds one needs particular care.

\subsection{Special features of the GB term} 
 Before turning to a general analysis of the quantum effective action, we briefly discuss some of the typical features of a GB $(V_E)$ term, which has always met a lot of 
 attention in several contexts, starting from string theory.  \\
Specific linear combinations of higher derivative invariants in the action may be deprived of double poles in their propagators.
In general, an expansion of the Riemann tensor in the fluctuations $h_{\mu\nu}$ around a flat Minkoswki vacuum, 
\begin{equation}
R_{\mu\nu\rho\sigma}=R^{(1)}_{\mu\nu\rho\sigma} + R^{(2)}_{\mu\nu\rho\sigma} +\ldots,
\end{equation}
shows immediately that at quadratic level the theory, arrested at the $"R^2"$ term, is plagued by propagating double poles, corresponding to the kinetic operator $\Box^2$, as one can easily derive from the action
\begin{equation}
\sm_2=\int d^d x \left( (R_{\mu\nu\rho\sigma})^2 + a (R_{\mu\nu})^2 + b R^2 \right),
\end{equation}
expanded to quadratic order in the fluctuations $h$. In the harmonic gauge ($\partial_\mu h^{\mu\nu}
=\frac{1}{2}\partial^{\nu}h$), with $h=h^{\mu}_\mu$, such double poles are associated with the action
\begin{equation}
\label{es1}
\sm_2^{(2)}=\frac{1}{4}\int d^d x \sqrt{g}\left(  (a+4) h^{\mu\nu}\Box^2 h_{\mu\nu} +(b-1) h\Box^2 h   \right), 
\end{equation}
and vanish if we choose for $\sm$ the Gauss-Bonnet (GB) combination 
\begin{equation}
\label{GB1}
 E=R^2 - 4 R^{\mu \nu} R_{\mu \nu} + R^{\mu \nu \rho \sigma} R_{\mu \nu \rho \sigma}. 
\end{equation}

Its contribution to the equation of motion of any action, either EH or Weyl, for instance, appears through its first functional derivative with respect to the metric 

\begin{equation}
{V}_E ^{\mu\nu}=\frac{\delta}{\delta g_{\mu\nu}} V_E,
\end{equation} 

explicitly given by the relation 
\begin{equation}
V_E^{\mu\nu}= 4R_{\mu\alpha\beta\sigma}R^{\;\,\alpha\beta\sigma}_\nu-8R_{\mu\alpha\nu\beta}R^{\alpha\beta}-8R_{\mu\alpha}R^{\;\,\alpha}_{\nu}+4RR_{\mu\nu}-g_{\mu\nu}{E},
\end{equation}
and it is characterised by a topological behaviour, for vanishing at $d=4$ in any metric. Notice that the variation of density $\sqrt{g} E$ with respect to the metric is a boundary distribution and not identically zero.\\
 The vanishing of $V_E^{\mu\nu}$, as shown by \eqref{es1}, is not expected to hold for a generic $d$. It is, instead, an obvious result at $d=4$, given the topological nature of the integral, for being proportional to the Euler number of the underlying spacetime manifold ($\mathcal{M}$)  
\begin{equation}
V_E=4 \pi \chi(\mathcal{M}). 
\label{top}
\end{equation}
$V_E$ shares properties similar to those of the EH action at $d=2$, 

\begin{equation}
\sm_{EH}(d)\equiv \int d^d x \sqrt{g} R,
\end{equation}
due to the topological nature of both functionals. \\
Therefore, all the functional derivatives of $V_E(4)$ or $\sm_{EH}(2)$ vanish for any 
metric $g_{\mu\nu}$ at $d=4$ and $d=2$ respectively,
\begin{equation}
\left(V_E(4)\right)^{\mu_1\nu_1\ldots \mu_n\nu_n}=0, \qquad \left(\sm_{EH}(2)\right)^{\mu_1\nu_1\ldots \mu_n\nu_n}=0, 
\end{equation}
while 
\begin{equation}
\label{ppf}
\left(V_E(d)\right)^{\mu_1\nu_1\ldots \mu_n\nu_n}\vert_{flat}=0, \qquad n=1,2.
\end{equation}
Eq. \eqref{ppf} shows that in an expansion around flat space, this term is responsible only for the generation of interacting vertices, from the trilinear level up. Therefore, the analysis of $V_E$ as a possible correction to the EH action, 

\begin{align}
\sm_{EGB}=&\sm_{EH}(d) +\alpha V_E(d)
= \frac{M_P^2}{2}\int d^d x \sqrt{g}\left(R + \frac{2}{M_P^2}\alpha E\right), 
\end{align}
giving equations of motion of the form 
\begin{equation}
G^{\mu\nu}  +\frac{2}{M_P^2}\alpha V_E^{\mu\nu}=0,\qquad G_{\mu\nu}=R_{\mu\nu} - \frac{1}{2}g_{\mu\nu} R, 
\end{equation}
is rather intricate.

As already mentioned, this point has recently come back to the attention of the 
cosmology/GR community, having been noticed 
\cite{Glavan:2019inb} that one could perform the tensorial limit  of the $d-$dimensional equation of motion down to $d=4$ together with a singular rescaling of the dimensionless coupling constant $\alpha$ in order to take care of the vanishing of $V_E^{\mu\nu}$. It has been suggested that such a trick generates a finite action that evades Lovelock's theorem at $d=4$. \\
We recall that Lovelock's theorem identifies the EH action - with the inclusion of a cosmological constant $\Lambda$ - as the unique, purely gravitational action, yielding second order equations of motion at $d=4$.  Obviously, the theorem does not contemplate in its assumptions the possibility of considering singular actions, regulated according to specific schemes, which is the case of 
\cite{Glavan:2019inb}. \\
A variant of this first proposal has been considered in \cite{Hennigar:2020lsl,Fernandes:2020nbq,Lu:2020iav}. The regulated action, with finite $0/0$ contributions to the gravitational equations of motion, is based on an approach that, as already mentioned, has been introduced in the past in $d=2$ for the EH action. The results of this procedure is to generate actions of Horndeski type, which are characterised by equations of motion of second order, but with the inclusion of a dilaton field. As such they correspond to dilaton gravities rather than to pure gravitational theories. 
In \cite{Coriano:2022knl} it has been pointed out that such regularization procedure identifies the regulated actions that ensues as a typical Wess-Zumino (WZ) action, limitedly to the topological contribution. The definition overlaps with the typical regularization of anomaly actions, here reproposed in a purely classical context. We are going to investigate these points in some detail, analysiing verly closely the way the regularization of such theories proceed and the addressing possible ambiguities.
 
 \section{The regularized quantum effective action in DR} 
The relevant expression for the analysis of the effective action, here defined as $\sm_{R}$, starts from its definition in $d$ dimensions, as given in \eqref{rena}. 
We implement DR on the counterterms, and use the analiticity of the two functionals $V$ respect to $d$, expanding their expressions around $d=4$, to obtain
\begin{equation} 
\label{expand1}
\frac{1}{\varepsilon}V_{E/C^2}(d)=\frac{1}{\varepsilon}\left( V_{E/C^2}(4) + \varepsilon 
V_{E/C^2}'(4) +O(\varepsilon^2) \right).
\end{equation}
The expansion above, which is standard in DR, at this stage is still rather formal, and we have not explicitly declared the form of the metric dependence on both members of the equation, since this requires a more careful analysis, that we will address next. 
This expansion plays an important role in our analysis of $\sm_R$, since the terms 
$V'(4)$ are those that appear as finite contributions in its final expression. \\
To clarify this point, we recall that $\sm_B$ is affected by single pole divergences in $1/\epsilon$ as we perform the $d\to 4$ limit. This can be isolated from its expression, using DR,  in the form 

\begin{equation}
\label{pone}
\sm_B(d)=\sm_f(d) +\frac{b}{\epsilon}V_{C^2}(4) + \frac{b'}{\epsilon}V_E(4),
\end{equation}
where $\sm_f$ is finite. In $d$ dimensions, $\sm_B$ is finite and Weyl-invariant, but as we isolate the singular contributions from $\sm_B$ and perform the $d\to 4$ limit, the Weyl variation needs to be carefully redefined.  \\
 In Eq. \eqref{expand1} we are expanding the residue at the $1/\epsilon$ pole for any background metric $g$ and the $O(\epsilon^0)$ terms $V_{E/C^2}(4)$ need also to be treated with care. In  particular, the topological nature of $V_E(4)$, as stated in \eqref{top},
with $\chi(\mathcal{M})$ the Euler number of the underlying space, will guarantee that such term will not contribute to the (infinite) renormalization of the bare quantum action $\sm_B$ at $d=4$ , for being independent of any metric variation. Indeed, $V'_E$, and its variants, defined via the WZ part of $\sm_R$ in \eqref{rena}, that we will investigate below, correspond to finite renormalizations of all the correlation functions of stress energy tensors, generated by the functional expansion of $\sm_B$ or of $\sm_f$, once we take the $d\to 4$ limit. 
    
The procedure of renormalization can then be summarised by the expression 
\begin{align}
\label{sum}
\sm_R(d)=&\Big(\sm_B(d)  +\frac{b}{\epsilon}V_{C^2}(4) + \frac{b'}{\epsilon}V_E(4)\Big) +
\frac{1}{\varepsilon}\left( b' V_{E}(4) + \varepsilon 
b' V_{E}'(4) +O(\varepsilon^2) \right) 
+ \frac{1}{\varepsilon}\left( b V_{C^2}(4) + \varepsilon 
b V_{C^2}'(4) +O(\varepsilon^2) \right),
\end{align}
and we can take separately the limit $\epsilon\to 0$ due to the absence of any singularity
\begin{equation}
\label{sf1}
\sm_f(4)=\lim_{d\to 4}\Big(\sm_B(d)  +\frac{b}{\epsilon}V_{C^2}(4) + \frac{b'}{\epsilon}V_E(4)\Big)
\end{equation}
In the expression of $\sm_R$ derived above, there is a cancellation between the $1/\epsilon$ contribution coming from $\sm_B$ (first bracket in \eqref{sum}) and those derived from the expansion of the counterterms $V_{E/C^2}$ (second and third bracket). 
The identification of such terms is quite involved, due to the need of computing propagators and vertices in a curved background. In few cases, they can be performed in DR using coordinate space methods. In the De Sitter case, for instance, such computations can be performed quite efficiently, especially for 1-point functions, such as the $\langle T_{\mu\nu}\rangle$, using a regularization by point-splitting and by other techniques.   

After the cancellation of the singular terms, we are left with the renormalized effective action
\begin{equation}
\label{simp}
\sm_R\equiv \sm_R(4)=\sm_f (4)  + V'_E(4) + V'_{C^2}(4). 
\end{equation}
Therefore, we can summarize the procedure in the $d\to 4$ limit by the expression

\begin{equation}
\label{ren1}
\sm_R(d)=\lim_{d\to 4}\left(\sm_B(d) +b'\frac{1}{\epsilon}V_E(d) + b\frac{1}{\epsilon}V_{C^2}\right)=\sm_f(4) +b'V'_E(4) + bV'_{C^2}(4),
\end{equation}
with 
\begin{equation}
\label{ps}
V'_{E/C^2}(4)=\frac{1}{\epsilon}\left(V_{E/C^2}(d)- V_{E/C^2}(4)\right).
\end{equation}
The anomaly-induced action which pertains to this regularization can then be defined in the form 
\begin{equation} 
\label{SA}
\sm_A=b' \,V'_{E}(4) + b \,V'_{C^2}(4).
\end{equation}
The expression above and \eqref{ps} are still rather formal, since they clearly depend on the way in which we extend the 4-dimensional metric to $d$ dimensions and then reduce it to four.
We are going to provide explicit examples of the expansion above, which is motivated by the fact that $V_E$ and $V_{C^2}$ are both analytic in $d$ but are not uniquely defined. Notice that both $V'_E$ and $V'_{C^2}$ take the form of local contributions only if we separate some degrees of freedom coming from the parameterization of the metric contained in their integrands. One example is the extraction from $g$ of a conformal factor. \\
The  nonlocal structure of such contributions will be apparent only if such extra field components are removed from $\sm_R$, although this is not always possible. This is the case of Wess-Zumino actions, which in $d$ dimensions contain up to $d$ dilaton fields and are commonly used as possible descriptions of anomaly-induced actions. In this case, the nonlocality of the action can be recovered only approximately, as we are going to show in 
\secref{WG}, since one is unable to solve explicitly for the dilaton field in terms of the fiducial metric.  
 \section{Conformal decompositions}  
In this section we describe the behaviour of $V_{E/C^2}$ and $\hat V_{E/C^2}$ under Weyl transformations, that will be essential in order to identify their dimensional expansions in $\epsilon$ around $d=4$.\\ 
For the moment we will adopt Greek indices $\mu,\nu$ etc., running from 1 to $d$.\footnote{We will later modify this notation, once we consider manifolds with the topology of a direct product, where the 
global indices will be uppercase latin ones $A,B,C$ etc, saving the $\mu,\nu$\ldots for the internal 4-dimensional manifold, and lowercase latin ones $a,b,c\ldots $  for the "external" manifold. Our spacetime will be chosen of the form $\mathcal{M}_d=\mathcal{M}_4\times \mathcal{M}_{e}$}.\\
 We choose a fiducial metric as in \eqref{mar}. 
The Christoffel symbol then transforms as
\begin{equation}
\label{dg}
\Gamma_{\mu\nu}^\lambda=\bar \Gamma_{\mu\nu}^\lambda+\de_\mu^\lambda\bar \nabla_\nu\phi+\de^\lambda_\nu\bar \nabla_\mu\phi-\bar g_{\mu\nu}\bar \nabla^\lambda\phi,
\end{equation}
and the curvature tensors as

 \begin{align}
 \frac{\delta}{\delta \phi}\int d^d x \sqrt{{g}}  {C}^2=& \frac{\delta}{\delta \phi}\int d^d x e^{2\epsilon \phi}\sqrt{\bar{g}}  \left(\bar{C}\right)^2\nn
 =&\epsilon \sqrt{g}{C}^2.
 \label{xxs}
\end{align}
Notice that at d=4
\begin{equation}
 \frac{\delta}{\delta \phi}\int d^4 x \sqrt{{g}}  {C}^2=\frac{\delta}{\delta \phi}\int d^4 x \sqrt{{\bar g}}  \bar{C}^2=0.
\end{equation}
Some relations concerning the variations of $C^2$ in $d=4$ and in general $d$ dimensions have been collected, for convenience, in an appendix. 
\section{ The Wess-Zumino action versus DR} 
The  WZ effective action provides a regularization of the quantum effective action $\sm_R$ that differs from standard DR by Weyl-invariant terms.\\
 Some ambiguities in the derivation of this part of the renormalized action can be noticed quite immediately. To illustrate this point, let's consider the conformal decomposition 
\begin{equation}
\label{mar}
g_{\mu\nu}=e^{2\phi(x)}\bar{g}_{\mu\nu} \qquad \bar{g}_{\mu\nu}= e^{-2\phi} g_{\mu\nu},
\end{equation}
expressed in terms of a fiducial metric $\bar{g}$ and a conformal factor $e^{2\phi(x)}$.
We recall that the regularization of the $V_E$ term 
 can be defined by a subtraction procedure of the form 
\begin{equation}
\label{WZ}
\mathcal{S}^{WZ}_E=\lim_{\epsilon\to 0}\frac{1}{\epsilon} \left(V_{E}(\bar{g} e^{2 \phi},d) -V_E(\bar{g},d)\right),
\end{equation}
and a similar one for $C^2$
\begin{equation}
\label{WZ}
\mathcal{S}^{WZ}_{C^2}=\lim_{\epsilon\to 0}\frac{1}{\epsilon} \left(V_{C^2}(\bar{g} e^{2 \phi},d) -V_{C^2}(\bar{g},d)\right),
\end{equation}
which is commonly used in the derivation of the  Wess-Zumino form (WZ) of the anomaly action
\footnote{From now on, for notational simplicity we will remove the $ \lim $ symbol in all the equations, which is a common praxis in DR.}  
\begin{equation}
\sm_{WZ}=\mathcal{S}^{WZ}_E +\mathcal{S}^{WZ}_{C^2}.
\end{equation}
The subtraction of the fiducial metric in the expressions above, in the term $V_E(\bar g,d)$, is performed in $d$ dimensions and clearly differs from the dimensional expansion introduced in \eqref{expand1}. $\sm_R$ inherits a different decomposition. Similarly to \eqref{rena}, we can write the relation
\begin{align}
\label{sumx}
\sm_R(d)=&\Big( \sm_B(d)  +\frac{b}{\epsilon}V_{C^2}(\bar g,d) +\frac{b'}{\epsilon}V_E(\bar g,d)\Big) +
\frac{1}{\epsilon}b'  V_{E}(g,d) \\
R^\lambda_{\mu\sigma\nu} =& \bar R^\lambda_{\mu\sigma\nu}+\de^\lambda_\nu\bar \Delta_{\mu\sigma}- \de^\lambda_\sigma\bar \Delta_{\mu\nu}+\bar g_{\mu\sigma}\bar \Delta^\lambda_\nu-\bar g_{\mu\nu}\bar \nabla^\lambda_\sigma +
(\de^\lambda_\nu\bar g_{\mu\sigma}- \de^\lambda_\sigma\bar g_{\mu\nu})\bar \nabla_\rho\phi\bar \nabla^\rho\phi, \\
\label{Rmunu underr finite}
R_{\mu\nu}=&\bar R_{\mu\nu}-(d-2)\bar \Delta_{\mu\nu}-\bar g_{\mu\nu}[\bar \square \phi+(d-2)\bar \nabla_\lambda\phi\bar \nabla^\lambda\phi], \\
\label{R under finite Weyl}
R=&e^{-2\phi}[\bar R-2(d-1)\bar \square \phi -(d-1)(d-2)\bar \nabla_\lambda\phi\bar \nabla^\lambda\phi], \end{align}
where we have defined the symmetric tensor $\bar \Delta_{\mu \nu}$ and its trace $\bar \Delta$ as
\begin{equation} \bar \Delta_{\mu \nu} = \bnabla_\mu \bnabla_\nu \phi - \bnabla_\mu \phi \bnabla_\nu \phi , \qquad \bar \Delta=\bar g^{\mu \nu} \bar \Delta_{\mu \nu} =\bar \square \phi - \bnabla_\lambda \phi \bnabla^\lambda \phi. \label{dd}\end{equation}
Notice that the symmetry of this tensor is ensured by the relation
\begin{equation} 
\bnabla_\mu \bnabla_\nu \phi = (\bpde_\mu \bpde_\nu - \bar \Gamma^\lambda_{\mu \nu} \bpde_\lambda) \phi= 
( \bpde_\nu \bpde_\mu - \bar\Gamma^\lambda_{\nu \mu} \bpde_\lambda )\phi = \bnabla_\nu \bnabla_\mu \phi . \end{equation}
In order to find $E$, we need the square of the curvatures
\begin{align} R_{\mu \nu \rho \sigma}^2 =& e^{-4\phi} \lt
\bar R_{\mu \nu \rho \sigma}^2 - 8 \bar R^{\mu \nu} \bar \Delta_{\mu \nu} - 4 \bar R\bnabla_\lambda \phi \bnabla^\lambda \phi + 4(d-2)\bar \Delta_{\mu \nu}^2 + 4\bar \Delta ^2  \right. \nn & \left.
+ 8(d-1)\bar \Delta \bnabla_\lambda \phi \bnabla^\lambda \phi  + 2d(d-1)(\bnabla_\lambda \phi \bnabla^\lambda \phi)^2 \rt 
\\
R_{\mu \nu}^2 =& e^{-4 \phi} \lq\bar R_{\mu \nu}^2 - 2(d-2) \bar R^{\mu \nu} \bar\Delta_{\mu \nu} 
- 2\bar R \bar\square \phi - 2(d-2) \bar R (\bnabla \phi)^2 + (d-2)^2 (\bnabla_\mu \bnabla_\nu \phi )^2 \right. \nn & \left.
- 2(d-2)^2 \bnabla_\mu \bnabla_ \nu \phi \bnabla^\mu \phi \bnabla^\nu \phi
+ (3d-4) (\bar\square \phi)^2  \right. \nn & \left. + 2(d-2)(2d-3)\bar\square \phi \bnabla_\lambda \phi \bnabla^\lambda \phi + (d-1)(d-2)^2 (\bnabla_\lambda \phi \bnabla^\lambda \phi)^2 \rq
\\
 R^2 =& e^{-4 \phi} \lq \bar R^2 - 4(d-1) \bar R \bar \square \phi - 2(d-1)(d-2)\bar R\bnabla_\lambda \phi \bnabla^\lambda \phi
+ 4 (d-1)^2 (\bar \square \phi)^2 \right. \nn & \left. + 4(d-1)^2 (d-2) \bar \square \phi \bnabla_\lambda \phi \bnabla^\lambda \phi
+ (d-1)^2 (d-2)^2 (\bnabla_\lambda \phi \bnabla^\lambda \phi)^2
 \rq .\end{align}
By using these relations, we can rewrite the rescaled $E$ as
\begin{align} E=& \bar E + 8 (d-3)\bar R^{\mu \nu} \bar\Delta_{\mu \nu} - 2(d-4)(d-3) \bar R (\bnabla \phi)^2 - 4 (d-3) \bar R \bar \square \phi  - 4(d-3)(d-2) (\bnabla_\mu \bnabla_\nu \phi)^2 \nn & + 8 (d-2)(d-1) \bnabla_\mu \bnabla_\nu \phi \bnabla^\mu \phi \bnabla^\nu \phi 
+ 4(d-3)(d-2)(\bar \square \phi)^2 + 4(d-3)^2 (d-2)\bar\square \phi (\bnabla \phi)^2 \nn & + (d-4)(d-3)(d-2)(d-1)(\bnabla \phi)^4 . \end{align}
Finally, by using the tensor relation for $\bar \Delta_{\mu \nu}$
$$\bar\Delta^2=(\bar\Box\phi)^2+(\bnabla_\lambda \phi \bnabla^\lambda \phi)^2-2\bnabla_\lambda \phi\bnabla^\lambda \phi\bar\Box\phi,$$
$$ \bar R_{\mu\nu}\bar \Delta^{\mu\nu}=\bar R_{\mu\nu}\bar\nabla^\mu\bar \nabla^\nu\phi-\bar R_{\mu\nu}\bar\nabla^\mu\phi\bar\nabla^\nu\phi,
$$
$$
\bar \Delta_{\mu\nu}^2=(\bar\nabla_\mu\bar\nabla_\nu\phi)^2+(\bnabla_\lambda \phi\bnabla^\lambda \phi)^2-2(\bar\nabla_\mu\bar\nabla_\nu\phi\bar\nabla^\mu\phi\bar\nabla^\nu\phi),
$$
we can collect the terms of the  GB density in the compact form 
\begin{equation}\label{form11}
\rg E=\sqrt{\bar g} e^{(d-4)\phi}\biggl \{ \bar E+(d-3)\bar\nabla_\mu \bar J^\mu(\bar{g},\phi) +(d-3)(d-4)\bar  K(\bar{g},\phi)  \biggl \},
\end{equation}
where we have defined
\begin{equation} \label{GBexJ}
\bar J^\mu(\bar{g},\phi)=8\bar R^{\mu\nu}\bar\nabla_\nu\phi-4\bar R\bar \nabla^\mu\phi+4(d-2)(\bar\nabla^\mu\phi\bar \Box \phi-\bar \nabla^\mu\bar\nabla^\nu\phi\bar \nabla_\nu\phi+\bar\nabla^\mu\phi\bar\nabla_\lambda\phi\bar\nabla^\lambda\phi),
\end{equation}
\begin{equation} \label{GBexK}
 \bar K(\bar{g},\phi)=4\bar R^{\mu\nu}\bar\nabla_\mu\phi\bar\nabla_\nu\phi-2\bar R\bar\nabla_\lambda\phi\bar\nabla^\lambda\phi+4(d-2)\bar\Box\phi\bar\nabla_\lambda\phi\bar\nabla^\lambda\phi+(d-1)(d-2)(\bnabla_\lambda \phi\bnabla^\lambda \phi)^2.
\end{equation}
The relation
\begin{equation} 
\label{ep2}
\frac{\delta}{\delta \phi}\int d^d y \sqrt{-g} E(y)=\epsilon \sqrt{g}E(x)
\end{equation} 
can be derived in two ways, either by scaling out the conformal factor from the background metric as specified above or, more simply, from a metric variation. In the latter case one gets
\begin{equation}
\label{ep3}
 \dfun{}{g_{\mu \nu}} \int d^d x \rg E_4 = \rg \lt \frac{1}{2} g^{\mu \nu} E_4 - 2 R^{\mu \alpha \beta \gamma}R^\nu_{\alpha \beta \gamma} + 4 R^{\mu \alpha} R^\nu_{\ \alpha} + 4 R^{\mu \alpha \nu \beta} R_{\alpha \beta} - 2 R R^{\mu \nu} \rt  \end{equation}
and \eqref{ep2} follows by a contraction with $2 g^{\mu\nu}$ 
\begin{equation}
\label{epx}
2 g_{\mu\nu}\frac{\delta}{\delta g_{\mu\nu}}\int d^d y \sqrt{-g} E(y)=\epsilon \sqrt{g}E(x).
\end{equation}

Alternatively, it can be written in the form  
\begin{equation}
\frac{\delta}{\delta \phi}V_E(g,d)=\epsilon \sqrt{g} E,
\end{equation} 
since the differentiation with respect to the conformal factor is equivalent to a metric differentiation plus a trace, at least in this case. This does not hold, in general, in the presence of a regularization.

The scaling relation \eqref{form11} allows to get some information on the behaviour of $V_E(g,d)$ expanded around $d=4$, as in \eqref{red1}. A differentiation of \eqref{form11} gives 

\begin{equation}
\frac{\delta}{\delta \phi}V_E(g,d)=\epsilon \sqrt{g}E + 
\int d^d x \sqrt{\bar g}e^{\epsilon \phi} \frac{\delta}{\delta\phi}\left(\bar E +\nabla_M J^M +\epsilon K(\bar g,\phi)\right).
\end{equation}
Notice that the second term has necessarily to be of $O(\epsilon^2)$ in such a way to satisfy the 
first relation in \eqref{ep2}, and this requires that 

\begin{align}
\int d^d x \sqrt{\bar g} \frac{\delta}{\delta\phi}\left(\bar E +\bar \nabla_M \bar J^M \right)=&0,\\
\int d^d x \sqrt{\bar g}\Bigg(\phi \frac{\delta}{\delta\phi}\left(\bar E +\bar \nabla_M\bar J^M  \right) + 
\frac{\delta}{\delta\phi} K(\bar g,\phi)\Bigg) =&O(\epsilon^2).
\end{align}
The first equation above is easy to prove, since $\bar E$ does not depend on the conformal factor and the integration is a boundary term if we pull the derivative w.r.t the conformal factor out of the integral. The second one can be verified by inspection, using some integration by parts and dropping some boundary terms.

A similar approach can be extended to $V_{C^2}$. Using

 \begin{align} 
 \dfun{}{g_{\mu \nu}} \int d^d x \rg \ C_{\alpha \beta \gamma \delta}^2 &= \rg \Big(
\f g^{\mu \nu} C_{\alpha \beta \gamma \delta}^2 - 2 R^{\mu \alpha \beta \gamma} R^\nu_{\ \alpha \beta \gamma} + 4 R^{\mu \alpha} R^\nu_{\ \alpha} \nn &- 4 \frac{d-4}{d-2}R^{\mu \alpha \nu \beta}R_{\alpha \beta} - \frac{4}{(d-2)(d-1)} RR^{\mu \nu} - 4 \frac{d-3}{d-2} \square R^{\mu \nu} \nn &+ 2 g^{\mu\nu} \frac{d-3}{(d-2)(d-1)} \square R + 2 \frac{d-3}{d-1} \nabla^\mu \nabla^\nu R \Big),  
\end{align} 
and after a direct computation, one gets

\begin{equation}
2 g_{\mu\nu}\dfun{}{g_{\mu \nu}} \int d^d x \rg C^2=\frac{\delta}{\delta \phi}\int d^d x \rg C^2=\epsilon \sqrt{g} C^2.
\end{equation}
Alternatively, by using the scaling relations above, one derives the condition \begin{align}
\sqrt{g}C^2=e^{\epsilon \phi}
- V_{E}(\bar g,d)   + 
\frac{1}{\epsilon}b \left( V_{C^2}(g,d) 
-V_{C^2}(\bar g,d)  \right) \nn
\end{align}
that we can rewrite in the form 
\begin{align}
\label{sumxp}
\sm_R(4)=&\tilde\sm_f(4) +\sm_{WZ},
\end{align}

with 
\begin{equation}
\label{sf2}
\tilde\sm_f(4)=\lim_{d\to 4} \Big(\sm_B(d)  +\frac{b}{\epsilon}V_{C^2}(\bar g,d) +\frac{b'}{\epsilon}V_E(\bar g,d)\Big), 
\end{equation}
and the DR and WZ procedures differ by Weyl-invariant terms. These are taken into account in $\sm_f(4)$ \eqref{sf1} and $\tilde \sm_f(4)$ in two different forms. To determine alla these cotributions explicitly we will be needing a careful implementation of DR. \\
The WZ action defined above, as already mentioned, is widely used in the context of the derivation of the anomaly actions in quantum gravity \cite{Coriano:2013nja,Coriano:2013xua}. In general, in these analysis, the attention goes either to the WZ part of $\sm_R$, denoted as $\sm_{WZ}$, or to $\sm_A$ defined in \eqref{SA}, and their completions in $\sm_R(4)$ are clearly different. These completions are $\tilde \sm_f (4)$ and $\sm_f(4)$, depending on whether we use either a standard DR or a WZ regularization of $\sm_R$.

\subsection{Detailing the anomaly-induced parts $\sm_A$ and $\sm_{WZ}$}
Then, to get more insight into the contributions in \eqref{WZ} we proceed with a rescaling of the metric integral defining $V_E$, isolating the conformal factor. For this we will be using 
\eqref{form11}. There, all the fields are defined on a $d$-dimensional spacetime manifold. The contractions of the capital letter indices, implicit and explicit, all run from 1 to $d$. 
$V_E$ can be expanded in $\epsilon$ in the form 
\begin{equation}
\label{red1}
V_E(g,d)=\int d^d x \sqrt{\bar{g}}\left(\bar E + \bar\nabla_M \bar J^M \right) +
\epsilon\int d^d x \sqrt{\bar g} \phi\left( \bar E + \bar\nabla_M \bar J^M \right) 
+\epsilon \int d^d x \sqrt{\bar g} K.
\end{equation}
The limiting value of $V_E(g,d)$, denoted as $V_E(\bar g, 4)$, can be immediately computed from 
\eqref{red1} without any consideration on DR in the form
\begin{equation}
 V_E(\bar g,4)=\int d^4 x \sqrt{\bar{g}}\left(\bar E + \bar\nabla_\mu \bar J^\mu \right) =\int d^4 x \sqrt{\bar{g}}\bar E,
\end{equation}
having dropped, in the last relation, a boundary term. 
This implies that in this limit, $V_E$ turns into a function only of the fiducial metric, without any conformal factor as $d\to 4$.
$V_E(\bar g,d)$ can be defined analogously 
\begin{equation}
 V_E(\bar g,d)=\int d^d x \sqrt{\bar{g}}\left(\bar E + \bar\nabla_\mu \bar J^\mu \right) =\int d^d x \sqrt{\bar{g}}\bar E.
\end{equation}
 The complete contribution to $\sm_A$ coming from the $V_E$ counterterm is then given by 
\begin{equation}
\label{like}
V'_E\equiv \frac{1}{\epsilon}\left( V_E(g,d)-V_E(\bar g,4)\right)=
\frac{\partial}{\partial d}V_E(\bar g, d)\mid_{d=4} +
\int d^d x \sqrt{\bar g} \phi\left( \bar E + \bar\nabla_M \bar J^M \right) 
+\int d^d x \sqrt{\bar g} K,
\end{equation}	
which is at variance respect to the renormalization used in the definition of the WZ effective action by the equation
\begin{equation}
\label{wzp}
\sm_E^{WZ}=\frac{1}{\epsilon}\left(V_E(g,d)-V_E(\bar g, d)\right)=V'_E- \frac{\partial}{\partial d}V_E(\bar g, d)\mid_{d=4},
\end{equation}
that is 
\begin{equation}
\label{wzp2}
\sm_E^{WZ}=\int d^d x \sqrt{\bar g} \phi\left( \bar E + \bar\nabla_M \bar J^M \right) 
+\int d^d x \sqrt{\bar g} K.
\end{equation}

A similar analysis, in the case of the quantum effective action, can be performed for $V_{C^2}$ with

\begin{equation}
\label{cc}
V_{C^2}(g,d)=\int d^d x \sqrt{\bar g} e^{\epsilon \phi} \bar C^2 =\int d^d x \sqrt{\bar g}  \bar C^2 
+\epsilon \int d^d x \sqrt{\bar g} \phi \bar C^2, 
\end{equation}
giving
\begin{equation}
 V_{C^2}(\bar g,4)=\int d^4 x \sqrt{\bar g}\bar C^2 
\end{equation}
and 
\begin{equation}
V'_{C^2}\equiv \frac{1}{\epsilon}\left( V_{C^2}(g,d)-V_{C^2}(\bar g,4)\right)=
\frac{\partial}{\partial d}V_{C^2}(\bar g, d)\mid_{d=4} + \int d^d x \sqrt{\bar g} \phi \bar C^2.
\end{equation}
giving,  similarly to \eqref{like},\eqref{wzp} and \eqref{wzp2},  
\begin{equation}
\sm^{WZ}_{C^2}=\int d^d x \sqrt{\bar g} \phi \bar C^2.
\end{equation}
It is then clear that $\sm_A$ and $\sm_{WZ}$ 
differ by Weyl-invariant contributions  
\begin{equation}
\sm_A = \sm_{WZ} +  \frac{\partial}{\partial d}V_E(\bar g, d)\mid_{d=4} + 
\frac{\partial}{\partial d}V_{C^2}(\bar g, d)\mid_{d=4}.
\end{equation}
The ambiguities in the definition of the Weyl-invariant terms is a natural result of the renormalization procedure, due to the prescriptions used in the regularization of the effective action. In general, the main difference between the different regularizations lays in the power of the dilaton field. In $\sm_A $ and $\sm_{WZ}$ $\phi$ the dependence is quartic, but the inclusion of a finite renormalizaton of the topological density makes it quadratic. This point will be addressed in \secref{mod}.

By the term "anomaly induced actions" we will refer to both the $\sm_A $ and $\sm_{WZ}$ contributions, which do not include the 
 finite term $\sm_f$ coming from the quantum corrections. The conformal backreaction is associated with $\sm_R$, and includes also $\sm_f$, as defined in \eqref{sf1}.\\
  In the case of the EGB theories discussed recently, 
$\sm_E^{WZ}$ is the only contribution derived above in $\sm_{WZ}$ that plays a role in those analysis. One notices the overlap betyween the regularization proposed in  \cite{Hennigar:2020lsl,Lu:2020iav,Fernandes:2020nbq} and the one discussed here and in previous works \cite{Mazur:2001aa}. Obviously, the latter are deprived of the quantum correction contained in $\sm_f$. 
 \subsection{Anomaly constraints} 

At this point, we are in condition to derive the anomaly constraints on $\mathcal{S}^{WZ}$ and on 
$\sm_A$, clarifying some of the intermediate steps in the derivations. We recall that the usual relation
\begin{equation}
2 g_{\mu\nu}\frac{\delta}{\delta g_{\mu\nu}}=\frac{\delta}{\delta\phi},
\end{equation}
holds only if a functional is a function of the entire metric $g_{\mu\nu}$. In general, for a regulated functional, separate function of $\bar{g}$ and $\phi$, say $F(\bar{g},\phi) $, this relation is not valid any longer. 
Given a functional $F(\bar{g})$, we use the relation $\bar{g}_{\mu\nu}=g_{\mu\nu}e^{-2 \phi}$, 
to derive the expressions
\begin{equation}
\label{rels}
\frac{\delta\bar{g}_{\mu\nu}(x) }{\delta g_{\alpha\beta}(y)}=\delta^{\alpha\beta}_{\mu\nu}e^{-2 \phi}\delta^d(x-y),
\qquad 
\frac{\delta F}{\delta g_{\alpha\beta}}= \frac{\delta F}{\delta \bar{g}_{\alpha \beta}}e^{-2 \phi},\qquad 
2 \bar g_{\alpha\beta}\frac{\del F}{\delta\bar g_{\alpha \beta} }=2  g_{\alpha\beta}\frac{\del F}{\delta g_{\alpha \beta} }.
\end{equation}
The regularization, by separating the dependence on the two components of $g$, leads to 
identities between the functional variations w.r.t. $\phi$ and $\bar{g}$ that, as we are going to show, are related by 
the anomaly. In EGB theories, the link is expressed only by the Euler density, but the constrain between the two variations is similar. This implies that the functional expansion of $\sm_R$ is far more involved, similar to \eqref{exps2}, but with the dilaton appearing on the external legs of the correlation functions. This requires a separate analysis of the anomalous CWIs, that we will discuss elsewhere. In general,  $\sm_R$ can be expressed either in terms of the couple $(\bar{g},\phi)$, but also in terms of $g$ and $\phi$, where the two variables shuould be treated as independent. We may easily derive several relations. \\ 
For instance, from $\sm_E^{WZ}$ we obtain
\begin{equation}
2 g_{\mu\nu}\frac{\delta \mathcal{S}_E^{WZ}}{\delta g_{\mu\nu}}=
\frac{1}{\epsilon}\left(2 g_{\mu\nu}\frac{\delta V_E(g,d)}{\delta g_{\mu\nu}}-
2 g_{\mu\nu}\frac{\delta V_E(\bar g,d)}{\delta g_{\mu\nu}}\right).
\end{equation}
Now using 
\begin{equation}
2 g_{\mu\nu}\frac{\delta}{\delta g_{\mu\nu}}V_E(\bar g,d)=2\bar g_{\mu\nu}\frac{\delta}{\delta \bar g_{\mu\nu}}V_E(\bar g,d)=\sqrt{\bar g}\bar E,
\end{equation}
together with \eqref{epx}, we obtain 
\begin{equation}
2g_{\mu\nu}\frac{\delta \mathcal{S}_E^{WZ}}{\delta g_{\mu\nu}}=\sqrt{g}E -\sqrt{\bar g }\bar E.
\end{equation}
On the other end, we have 

\begin{equation}
\label{ord}
\frac{\delta}{\delta \phi}V_E(g,d)=\epsilon \sqrt{g}E, \qquad \frac{\delta}{\delta \phi}V_E(\bar g,d)=0,
\end{equation}
giving
\begin{equation}
\label{equal}
\frac{\delta}{\delta \phi}\sm_E^{WZ}=\sqrt{g}E.
\end{equation}
Therefore, in the case of a GB theory one derives a constraint between the equation of motion of the dilaton and the trace of the stress energy tensor of the fiducial metric  
\begin{equation}
\label{break}
2g_{\mu\nu}\frac{\delta \mathcal{S}_E^{WZ}}{\delta g_{\mu\nu}}-\frac{\delta  \mathcal{S}_E^{WZ}}{\delta \phi}=-\sqrt{\bar g}\bar E.
\end{equation}

This relation shows that the Weyl variation and the trace/metric variation are not identical, if we perform a regularization.\\
A similar relation holds for $\mathcal{S}_{C^2}^{WZ}$

\begin{equation}
2g_{\mu\nu}\frac{\delta \mathcal{S}_{C^2}^{WZ}}{\delta g_{\mu\nu}}-\frac{\delta  \mathcal{S}_{C^2}^{WZ}}{\delta \phi}=-\sqrt{\bar g}\bar C^2.
\end{equation}
In summary, a WZ anomaly action, following the definitions above, will satisfy the anomaly condition 
\begin{equation}
\label{WZZ}
2g_{\mu\nu}\frac{\delta \mathcal{S}^{WZ}}{\delta g_{\mu\nu}}=b'\sqrt{g}E + b \sqrt{g} C^2 - 
\left(b'\sqrt{\bar g}\bar E + b \sqrt{\bar g} \bar C^2\right),
\end{equation}
and 
\begin{equation}
\label{WZZ1}
\frac{\delta \mathcal{S}^{WZ}}{\delta \phi}=b'\sqrt{g}E + b \sqrt{g} C^2 
\end{equation}
and, more generally, the condition 

\begin{equation}
\label{cons}
2g_{\mu\nu}\frac{\delta \mathcal{S}^{WZ}}{\delta g_{\mu\nu}}-\frac{\delta \mathcal{S}^{WZ}}{\delta \phi}= - 
\left(b'\sqrt{\bar g}\bar E + b \sqrt{\bar g} \bar C^2\right).
\end{equation}

It differs from the DR anomaly-induced action by terms 
\begin{equation}
\label{nv2}
\left(2g_{\mu\nu}\frac{\delta \mathcal{S}^{A}}{\delta g_{\mu\nu}}- 2g_{\mu\nu}\frac{\delta \mathcal{S}^{WZ}}{\delta g_{\mu\nu}}\right)= \frac{\partial}{\partial d} V_E(\bar g,d)\mid_{d=4} + \frac{\partial}{\partial d} V_{C^2}(\bar g,d)\mid_{d=4},
\end{equation}
that are Weyl invariant, since they do not depend on the conformal factor $\phi$. 

\subsection{$\sm_B(d)$ for $d\to 4$}
 $\sm_B(d)$ describes the finite quantum corrections that develop a singularity in the $d\to 4$ limit. Notice that this is a functional of the entire metric before the limit is taken, and it is Weyl-invariant. Therefore it satisfies the constraint
 \begin{equation}
 \label{ccd}
 2 g_{\mu\nu}\frac{\delta\sm_B }{\delta g_{\mu\nu}}=\frac{\delta \sm_B}{\delta\phi}=0.
 \end{equation}
 This relation implies that $\sm_B$ is only a functional of $\bar{g}$. 
 Using the relations  
 \begin{equation}
2 g_{\mu\nu}\frac{\delta V_{E/C^2}(g,4) }{\delta g_{\mu\nu}}=2 \bar g_{\mu\nu}\frac{\delta V_{E/C^2}(\bar g,4) }{\delta \bar g_{\mu\nu}}=0, 
\end{equation}
 which are valid at $d=4$, and using the definition in DR
 \begin{equation}
 \sm_f(d) =S_B(g,d) -b'\frac{1}{\epsilon}V_{E}(\bar g ,4)- b\frac{1}{\epsilon}V_{C^2}(\bar g ,4),
 \end{equation}
 one finds that $ \sm_f(d)$ is only a function of the fiducial metric, and its stress energy tensor has a vanishing trace. This property continues to hold in the  $d\to 4$ limit and henceforth 
\begin{equation}
\sm_f(4)\equiv \sm_f(\bar g).
 \end{equation}
This implies that $\sm_f(\bar g)$ does not contribute to the anomaly and we have consistently 

\begin{equation}
\label{course}
2 \bar g_{\mu\nu}\frac{\delta S_f(4)}{\delta \bar g_{\mu\nu}}=0, 
\qquad \frac{\delta \sm_f(4)}{\delta \phi}=0.
\end{equation}
In the WZ case, on the other end we define 
\begin{equation}
\tilde\sm_f(d) =S_B(g,d) +b'\frac{1}{\epsilon}V_{E}(\bar g ,d)+ b\frac{1}{\epsilon}V_{E}(\bar g ,d),
 \end{equation}
and in this case, the response of this functional metric and conformal variations are similar, 
but now 
\begin{equation}
\frac{\delta\tilde\sm_f}{\delta \phi}=0, 
\end{equation}
since the subtractions $V_{E/C^2}(\bar g ,d)$ do not depend on the conformal factor, as well as the bare action $\sm_B$. On the other hand we have 
\begin{equation}
2 \bar{g}_{\mu\nu}\frac{\delta V_{E}(\bar g ,d)}{\delta \bar g_{\mu\nu}}=\sqrt{\bar g} E
\end{equation}
and similarly for $V_{C^2}(\bar g ,d)$, thereby obtaining

\begin{equation}
\label{dd2}
2 \bar{g}_{\mu\nu}\frac{\delta \tilde\sm_f(\bar g ,d)}{\delta \bar g_{\mu\nu}}=
b'\sqrt{\bar g}\bar E  +b\sqrt{\bar g}\bar C^2 
\end{equation}

\subsection{The complete quantum action}
It is clear that the conformal backreaction 
is associated with the entire renormalized effective action $\sm_R$, rather than to the anomaly-induced actions $\sm_A$ or the WZ action $S^{WZ}$. The difference between $\sm_R$ and the previous two actions is, again, given by Weyl invariant terms. 
We recall that as far as we stay in $d$ dimensions, $S_{R}(g,d)$, defined by the sum of 
$\sm_B(g,d)$ and of the two counterterms $1/\epsilon V_E(g,d)$ and $1/\epsilon V_{C^2}$, under a Weyl variation behaves as
\begin{equation}
\frac{\delta}{\delta\phi}\sm_R(g,d)=\frac{\delta \sm_B(g,d)}{\delta\phi}  +\frac{\delta}{\delta\phi}\Bigg(\frac{1}{\epsilon}\left(b' V_E(g,d) + b V_{C^2}(g,d)\right) \Bigg).
\end{equation}
Using the invariance of $\sm_B$ \eqref{ccd}, we obtain
\begin{equation}
 \frac{\delta \sm_R(g,d)}{\delta\phi}=2 g_{\mu\nu}\frac{\delta \sm_R(g,d)}{\delta g_{\mu\nu}}=
 b'\sqrt{g}E + b\sqrt{g}C^2.
\end{equation}
This equation is modified by the separation into poles plus finite terms in DR in the form 
\begin{equation}
\frac{\delta \sm_R(g,4)}{\delta\phi}=\frac{\delta \sm_f(g,4)}{\delta\phi} +\frac{\delta \sm_A(g,4)}{\delta\phi}.
\end{equation}
In this case both a $\phi$- variation and trace-metric variation coincide and give 

\begin{equation}
\label{1p}
\frac{\delta \sm_A(4)}{\delta\phi}= b'\sqrt{g}E + b\sqrt{g}C^2
\end{equation} 

while 

\begin{equation}
\label{2p}
2 g_{\mu\nu}\frac{\delta \sm_A(4)}{\delta g_{\mu\nu}}=b'\sqrt{g}E + b \sqrt{g} C^2. 
\end{equation} 

The finite part of $\sm_R$, given by $\sm_f$, has no dependence on the conformal factor and its stress energy tensor has zero trace. Therefore the anomaly, in this case, is all generated by $\sm_A$
\begin{equation}
\label{canc}
\frac{\delta\sm_R}{\delta{\phi}}=2g_{\mu\nu}\frac{\delta \sm_R}{\delta g_{\mu\nu}}= 
 b'\sqrt{g}E + b\sqrt{g}C^2.
\end{equation}
 
 If we perform the renormalization of $\sm_B$ using \eqref{sumx} and \eqref{sumxp}, 
 then \eqref{canc} is obviously still valid. However both terms, $\tilde\sm_f$ and $\sm_{WZ}$  on the rhs of \eqref{sumxp} will contribute to the anomaly, with extra Weyl-invariant terms that carry opposite signs, as shown in \eqref{WZZ} and \eqref{dd2}, their sum reproducing again Eq. \eqref{canc}.
  
 \section{Expansions in DR}\label{WG}
In this section we turn to a discussion of the possible ways to reduce the $d$-dimensional integrals appearing in the counterterms, down to $d=4$, by selecting specific metrics. We start with the topological counterterm, by mentioning that 
from \eqref{form11}, using \eqref{GBexJ} and \eqref{GBexK} we obtain the relation
\begin{equation}
\bar \nabla_\mu\phi \bar J^\mu-\bar K=4\bar R^{\mu\nu}(\bnabla_\mu\phi\bnabla_\nu\phi)-2\bar R\bar \square \phi+2(\bnabla_\lambda \phi \bnabla^\lambda \phi )^2+4\bar\Box\phi\bnabla_\lambda \phi \bnabla^\lambda \phi ,
\end{equation}
written as a four-dimensional expression, but also valid  in the embedding space. As far as we do not specify the extra dimensional metric in some way, the Greek indices may be used, with no confusion, 
to describe the invariants over the entire embedding space.
 
After an integration by parts, we get the final form of the counterterm contibution, up to $O(\epsilon)$ terms, given  by
\begin{align} \label{WG in generic case}
\frac{1}{\epsilon}V_E(g,d) =& \frac{1}{\epsilon}\int d^dx \rgb \ \bar E+ \int d^dx \rgb\ \Big[\phi\bar E-(4\bar G^{\mu\nu}(\bar\nabla_\mu\phi\bar\nabla_\nu\phi) \nn & +2(\bnabla_\lambda \phi \bnabla^\lambda \phi )^2 +4\bar\Box\phi \bnabla_\lambda \phi \bnabla^\lambda \phi ) \Big].
\end{align}

This result holds in $d$ spacetime dimensions. 

If we intend to 
take the $d\to 4$ limit with more rigour, then we need to be more specific about the choice of the metric. In our case, for definiteness, we will consider a manifold of the form $\mathcal{M}_4\times \mathcal{M}_e$, split into a 4- and $(d-4)$-dimensional part.  
We will denote the $d$ dimensional indices as $M, N$, saving the Greek indices for the 4-dimensional part.  The $d$ dimensional metric is decomposed in the Weyl gauge and split into the direct sum of the metrics of the two submanifolds. ${}_e g_{m n }$ is the extra dimensional metric used for the regularization of the integral and $_4\tilde{g}_{\mu\nu}$ is its 4-dimensional part. The extra coordinates will be denoted as $y$. For example, we choose  
\begin{equation}
\apd g_{MN}(x,y)  =  e^{2 \phi(x)} \begin{bmatrix}  {}_4\tilde g_{\mu\nu}(x) &0\\
0 &  \ape  g_{mn}(y)
\end{bmatrix}  = e^{2 \phi(x)}\apd \bar g_{MN} .
\label{weylr}
\end{equation}
$_d\bar{g}_{\mu\nu}$ is the $d$-dimensional fiducial metric, from which we have extracted a conformal factor $\phi$, with the indices decomposed as $M=(\mu,m), \, (N=\nu,n)\ldots$, and so on.   
The original scaling relation \eqref{form11} in $d$ dimensions can be expressed, in this spacetime manifold, in the form
\begin{equation} \label{iter1a}
\int d^4x d^{d-4}y \rgd \ \apd E =  \int d^4x d^{d-4}y  \rgdb\ e^{(d-4)\phi} \lt \apd \bar E+(d-3)\bar\nabla_M \apd \bar J^M+(d-3)(d-4) \apd \bar K  \rt ,\end{equation}
where barred terms, including the covariant derivatives, are relative to the fiducial metric $\apd \bar g_{MN}$.
Since, with our choice, the dilaton does not appear in the extra dimensional part of the metric, the two blocks that make up the full $d$-dimensional metric $\apd \bar g_{MN}$ are only dependent on the coordinates of the submanifold which they belong to. The same is true for every curvature tensor, hence we have that ${}_4\tilde R_{\mu \nu \rho \sigma}$, the Riemann tensor of the base space, depends only on the $x$-coordinates, while ${}_e R_{abcd}$ depends only on the $y$-coordinates. Moreover, the connection has no mixed terms, hence the squared curvature tensors are decoupled
\begin{align} \apd \bar R^{ABCD} \apd \bar R_{ABCD}=& {}_4 \tilde R^{\mu \nu \rho \sigma} {}_4 \tilde  R_{\mu \nu \rho \sigma} + \ape R^{abcd} \ape R_{abcd}, \nn
\apd \bar R^{AB} \apd \bar R_{AB}=& {}_4 \tilde R^{\mu \nu} {}_4 \tilde R_{\mu \nu} + \ape R^{ab} \ape R_{ab}, \nn
\apd \bar R^2 =& {}_4 \tilde R^2 + \ape R^2 + 2 {}_4 \tilde R \ape R . \label{squaredcurvaturefact} \end{align}
The Gauss-Bonnet density
becomes 
\begin{equation} \apd \bar E = {}_4 \tilde E + \ape E + 2 {}_4 \tilde R  \ape R . \end{equation}
From the definitions \eqref{GBexJ} and \eqref{GBexK}, since $\phi(x)$ is a function only of the coordinates of the 4-dimensional subspace, we obtain
\begin{align} 
\apd \bar J^\mu =& {}_4 \tilde J^\mu - 4\ape R\ \tilde \nabla^\mu \phi , \qquad \ape \bar J^m = 0 ,\nn
\apd \bar K =& {}_4 \tilde K - 2\ape R\ \tilde \nabla^\lambda \phi  \tilde \nabla_\lambda \phi ,\end{align}
where $\tilde \nabla$ are covariant derivatives associated with the 4-dimensional metric ${}_4 \tilde g_{\mu \nu}$. \\
We may rewrite \eqref{iter1a} as 
\begin{align} \label{iter1b}
\int d^4x d^{d-4}y \rgd \ \apd E =&  \int d^4x d^{d-4}y  \rgdb\ e^{(d-4)\phi} \Big(
{}_4 \tilde E + (d-3)\tilde \nabla_\mu \,{}_4 \tilde J^\mu + (d-3)(d-4) {}_4 \tilde K  \nn & 
+ 2\, {}_4 \tilde R\ \ape R + \ape E-(d-3)\ape R [4\tilde \square \phi + 2(d-4)(\tilde \nabla_\lambda \phi \tilde \nabla^\lambda \phi)] \Big) . \end{align}
This equation is the starting point in order to proceed with a dimensional reduction of the fields in \eqref{dimred}.

\subsubsection{Weyl flat metric in $\mathcal{M}_e$ and DR}\label{example}
In dimensional reduction, as already pointed out, one usually assumes that all the $d$-dimensional fields do not depend on the coordinates of the extra dimensional manifold. The structure of the reduced theory carries symmetries which are decomposed with respect to the original ones. A classical example is that of $\mathcal{N}=1$ supersymmetric Yang-Mills theory in $d=10$, which turns into an $\mathcal{N}=4$ at $d=4$.\\
 In our case, with our metric choice, DR is implemented by assuming that the ${}_e g_{mn}$ metric becomes flat, in order to obtain a pure 4-dimensional integrand. In this case all the external curvatures $_e R$, $_e R_{mn}$ and so on, will obviously vanish.\\
We split the integration measure into $\rgd=\sqrt{{}_4 g} \rge$ and take the flat limit in ${}_e g$ 
($\sqrt{ {}_e{g}}\to 1$). We also reinsert the $\mu^{\epsilon}$ renormalization scale in the original defintion \eqref{ffr}, to obtain

\begin{align} \label{WG in generic case}
\frac{1}{d-4}V_E(g,d) =&\frac{1}{\epsilon} \left({L}{\mu}\right)^{\epsilon}\int d^4x \rg \  {}_4 \bar{E}+ \left({L}{\mu}\right)^{\epsilon}\int d^4x \rg\ \Big[\phi {}_4\bar{ E}-(4 {} G^{\mu\nu}(\bar\nabla_\mu\phi\bar\nabla_\nu\phi) \nn & +2(\nabla_\lambda \phi \nabla^\lambda \phi )^2 +4\Box\phi \nabla_\lambda \phi \nabla^\lambda \phi ) \Big],
\end{align}
where all the terms in the integrands are 4-dimensional and $L$ is a space cutoff in the $d-4$ extra dimensions. $L^{\epsilon}$ is the volume of the extra space. Taking the $\epsilon\to 0$ limit, and going back to the ordinary 4-$d$ notation, $g_{\mu\nu}=\bar g_{\mu\nu}e^{2 \phi}$ for the fiducial metric, we finally derive the expressions
\begin{align}
\label{rdef}
\hat{V}'_E( g, \phi)&\equiv \sm_E^{WZ}=\frac{1}{\epsilon}\left(V_E(g,d)-V_E(\bar g,d)\right)\nn
=&\int d^4x \rg \Big[\phi {}_4 E-(4 {} G^{\mu\nu}(\bar\nabla_\mu\phi\bar\nabla_\nu\phi)+2(\nabla_\lambda \phi \nabla^\lambda \phi )^2 +4\Box\phi \nabla_\lambda \phi \nabla^\lambda \phi ) \Big],
\end{align}
and 
\begin{equation}
V'_E=\hat{V}'_E(\bar g, \phi, d) +\log(L \mu)\int d^4x \sqrt{\bar g} \bar E.
\label{more}
\end{equation}
The dilaton appears with vertices up to order four. The result  reproduces in details the  analysis in 
\eqref{nv2}. In this case we identify the term 
\begin{equation}
\frac{\partial V_E(\bar g,d)}{\partial d}\mid_{d=4}=\log(L \mu)\int d^4x \sqrt{\bar g} \bar E
\end{equation}
which is the Weyl invariant mismatch between the regularization obtained via the WZ action and the one present in $\sm_R$ using standard DR. The example treated above defines a combined DR 
regularization for such types of actions.

A similar analysis can be performed for $V_{C^2}$. In this case, under a Weyl rescaling
\begin{equation}
V_{C^2}=\int d^4 x d^{d-4} y \rgd\ \apd C^2=\int d^4 x d^{d-4} y \rgdb\ e^{(d-4)\phi} \apd \bar C^2.
\end{equation}
As in the previous case, in the $ {}_{(d)}\bar g_{MN}$ metric, the squared curvatures are of the form \eqref{squaredcurvaturefact}, hence the $d$-dimensional Weyl tensor squared is expanded as
\begin{equation} \apd \bar C^2 = {}_4 \tilde C^2 + \ape C^2 + \frac{4}{(d-1)(d-2)} {}_4 \tilde R \ape R. \end{equation}
Then the countertem reads
\begin{equation} 
\label{cceq}
\int d^4 x d^{d-4} y \rgd\ \apd C^2=
\int d^4 x d^{d-4} y \rgdb\ e^{(d-4)\phi} \lt {}_4 \tilde C^2 + \ape C^2 + \frac{4}{(d-1)(d-2)} {}_4 \tilde R \ape R \rt . \end{equation}
Similarly we obtain
\begin{align} 
V_{C^2}=\int d^4 x d^{d-4} y \rgd\ \apd C^2 =&
\int d^4 x d^{d-4} y \sqrt{ {}_4 \tilde{g}}e^{(d-4)\phi} {}_4\tilde C^2 \nn 
=& L^{\epsilon} \int d^4 x  \sqrt{ {}_4 \tilde{g}}e^{(d-4)\phi} {}_4\tilde C^2.
\end{align}
Reinserting the $\mu$ dependence, the expansion in $\epsilon$ of this term generates two contributions
\begin{equation}
\frac{1}{d-4} V_{C^2}(d)=\frac{1}{\epsilon}\left(\mu L\right)^\epsilon \int d^4 x  \sqrt{ {g}}{} C^2 +O(\epsilon)
+ \int d^4 x  \sqrt{{g}}\phi C^2,
\end{equation}
the first of them relevant for the cancellation of the singular behaviour of $\sm_B(d)$ as $d$ goes to 4. Differently from the similar counterterm in $V_E$, this is necessary in order to regulate the divergences of $\sm_B(g,d)$ in the $d\to 4$ limit.   \\
Also in this case, by defining 
\begin{equation}
\hat{V}'_{C^2}\equiv S_{C^2}^{WZ}=\frac{1}{\epsilon}\left( V_{C^2}(g,d)-V_{C^2}(\bar g, d)\right),
\end{equation}
we have 
\begin{equation}
\hat{V}'_{C^2}(\bar g,d)=\int d^d x  \sqrt{{\bar g}}\phi \bar C^2
\end{equation}
and
\begin{equation}
V'_{C^2}(4)=\hat{V}'_{C^2}(\bar g,d) + \log( L \mu) \int d^4 x  \sqrt{{\bar g}}\bar C^2, 
\end{equation}
giving from \eqref{nv2} 
\begin{equation}
\left(2g_{\mu\nu}\frac{\delta \mathcal{S}^{A}}{\delta g_{\mu\nu}}- 2g_{\mu\nu}\frac{\delta \mathcal{S}^{WZ}}{\delta g_{\mu\nu}}\right)= \log( L \mu)\int d^4 x  \sqrt{\bar g}\left(b' \bar E +b\bar C^2\right).
 \end{equation}
The conformal backreaction, identified in $\sm_R$, can then be expressed in the final form
\begin{align}
\sm_R =&\sm_f + b'\int d^4x \rg \Big[\phi {}_4 E-(4 {} G^{\mu\nu}(\bar\nabla_\mu\phi\bar\nabla_\nu\phi)+2(\nabla_\lambda \phi \nabla^\lambda \phi )^2 +4\Box\phi \nabla_\lambda \phi \nabla^\lambda \phi ) \Big]\nn
& + b\int d^d x  \sqrt{{\bar g}}\phi \bar C^2 + \log( L \mu)\int d^4 x  \sqrt{\bar g}\left(b' \bar E +b\bar C^2\right),
\label{EE}
\end{align}
valid in DR, where the only missing term is $\sm_f$. Both $\sm_f$ and the log-contribution are Weyl-invariant (i.e. $\phi$-independent) terms which are part of the regulated action. If we limit our attention only to a GB theory, with a classical singular rescaling of the GB coupling, as in ordinary $d=2$ gravity \cite{Mann:1992ar},
then $\sm_f$ and the log terms are obviously absent, while at the same time we need to set $b=0$.

\subsection{Including the Einstein-Hilbert term: $EGBW_1$ and variants}
Extending $\sm_R$ in order to derive a Einsten GB/Weyl theory is quite straightforward, but it is not a unique procedure. At the same time, this theory can be accompanied by Lagriangian terms of various types. If one intends to adhere and modify the Lovelock classification at $d=4$, one can add the EH term. \\
We recall that the EH term may be expressed either in terms of the fiducial metric, as $\sm_{EH}(\bar g,4)$, where 
\begin{equation}
\sm_{EH}(\bar g,4)\equiv\int d^4 x \sqrt{\bar g}\left( M_P^2 \bar R + 2\Lambda\right),
\end{equation}
generating an action at $d=4$ of the form 
\begin{align}
\sm_{EGBW_1} \equiv \sm_{EH}(\bar g,4) +\sm_R(\bar g,\phi),
\end{align}
with
\begin{equation}
\sm_R(\bar g,\phi) = \sm_f(\bar g)  +\sm_A(\bar g,\phi), 
\end{equation}
and $\sm_A$ given by \eqref{EE},
or, alternatively, by promoting the entire EGBW theory to $d$-dimensions and performing the 
$d\to 4$ limit on all of its components. \\
In this second case, if we perform a Weyl trasformation also on the EH action, we derive the ordinary form of the dilaton gravity action
\begin{align} \int d^d x \rg ( M_P^2 R-2 \Lambda) =& 
\int d^dx \rgb \  e^{(d-2) \phi} \lt M_P^2 [_d\bar R - 2(d-1)\bar \square \phi - (d-1)(d-2) \bnabla_\lambda \phi \bnabla^\lambda \phi] - 2e^{2\phi} \Lambda \rt \nn &
=\int d^dx \rgb  e^{(d-2) \phi} \left(M_P^2[\bar R + (d-1)(d-2) \bnabla_\lambda \phi \bnabla^\lambda \phi] - 2e^{2 \phi}\Lambda \rt. \end{align}
Dimensional reduction of this action leads to the ordinary dilaton gravity $\sm_{EHd}$ in the Jordan (string) frame
\begin{equation}
\label{d1}
\sm_{EHd_1}(\bar g,\phi)=\int d^4x \rgb  e^{2 \phi} \left(M_P^2[\bar R + 6\bnabla_\lambda \phi \bnabla^\lambda \phi] - 2e^{2 \phi}\Lambda \rt. 
\end{equation}
Logarithmic, scale dependent terms are absent in this action, since we can smoothly take the $d\to 4$ limit in DR from $\sm_{EH}$ in $d$ dimensions, due to finiteness. 
 
We can add $\sm_R$ to this action, obtaining the corresponding EGBW action 
-denoted as  $EGBW_1$ -
\begin{align}
\sm_{EGBW_1} \equiv \sm_{EH{d_1}}(\bar g,\phi) +\sm_R(\bar g,\phi),
\end{align}
with $\sm_R(\bar g,\phi)$ given by \eqref{EE}.
We have also observed that there are variants of the 
theory in which the logarithmic  $\log(\mu L)$ scale dependent terms are absent from the counterterms. 
We work in the context of this variant, which corresponds to a redefiniton of the renormalized quantum effective action $\sm_R$ in the form
\begin{equation}
\tilde \sm_R=\sm_f(4) + \sm^{WZ},
\end{equation}
giving 
\begin{align}
\label{mod1}
\tilde \sm_{EGBW_1} \equiv \sm_{EH{d_1}}(\bar g,\phi) +\tilde\sm_R(\bar g,\phi).
\end{align}
This theory is defined according \eqref{EE} by the choice $\mu=1/L$, which removes the log terms in DR. In plane terms, in this case the subtraction is performed using the fiducial metric $\bar{g}$ in $d$ dimensions rather than at $d=4$. The need of performing an expansion in $\epsilon$ with dimensionless logs, justifies the presence of a second cutoff, since $\mu L$ needs to be dimensionless. The results is that extra Weyl invariant terms are present in the expansion an cannot be neglected. 

\subsection{A dilaton in $\mathcal{M}_e$ }\label{Dim red WG}
As already mentioned, one of the pressing issues in DR using higher dimensional metrics for the regularization of 
a certain action, is the ambiguity in the choice of the fiducial metric in the $d$-dimensional space. From this point of view, selecting a metric embedding such as the one discussed in the previous sections appears to be quite simple and appealing, but it is not unique. One criterion might be to ensure that fluctuations around flat spacetime, once the DR procedure is applied, do not include ghost-like degrees of freedom.   
We now bring an example that illustrates some of the problems encountered when the metric structure introduced above is modified by the inclusion of a dilator field only in the extradimensional manifold $\mathcal{M}_e$. Indeed, the expression of the effective action depends critically on the way the dilaton is extracted from the metric. 
We choose the metric of the form
\begin{equation} \label{lumetric}
\apd g_{MN}(x,y)=\begin{bmatrix}
{}_4 g_{\mu\nu}(x) &0\\
0 & e^{2\phi (x)} \ape g_{mn}(y)
\end{bmatrix} ,
\end{equation}

Now, consider the $d$-dimensional $V_E$ counterterm that we will compute by iterating the procedure \eqref{iter1b}. First, we factor the exponential in the $(d-4)$ metric
\begin{equation}
\apd g_{MN}(x,y)= e^{2 \phi(x)} \begin{bmatrix}
e^{-2\phi (x)} {}_4 g_{\mu\nu}(x) &0\\
0 &  \ape  g_{mn}(y)
\end{bmatrix} .
\end{equation}
In this way, the $d$-dimensional metric $\apd g_{MN}$ can be written as an overall exponential $e^{2\phi (x)}$ times a fiducial metric as in \eqref{weylr}, where now we define 
\begin{equation}
\label{gg}
{}_4 \tilde g_{\mu \nu} =  e^{-2 \phi} {}_4 g_{\mu \nu}.
\end{equation} 
Such rescaling affects $V_E$, since it corespond just to a Weyl transformation in $4$ dimensions, with a switch of the overall sign in front of $\phi$. Our goal now is to extract that factor from the metric..\\
For full conversion, we have to act with this reverse transformation on every single term of \eqref{iter1b}. \\
We start with the ${}_4 \tilde E$ term. We thus expand 
$\sqrt{{}_d g} =  \sqrt{{}_4 g}\rge$ and then use \eqref{form11} to get
\begin{align} 
\int d^4 x d^{d-4} y \rgdb\  e^{(d-4)\phi} \app \bar E =&
\int d^4 x d^{d-4} y \rgpb \rge\ e^{(d-4)\phi}  {}_4 \tilde E \nn =&
\int d^4 x d^{d-4} y \sqrt{_4 g} \rge\  e^{(d-4)\phi} \lt {}_4 E - (d-4)\nabla_\mu \phi \,\, {}_4 J^\mu_{-} \rt ,
\end{align}
where in the last equality we have integrated by parts.  The $(-)$ notation ${}_4 J^\mu_-$ is a remainder hat we have inverted the sign of $\phi$ inside the $J$ term 
\begin{equation}
{}_4 J^\mu_-=-8 {}_4 R^{\mu\nu}\nabla_\nu\phi+4 {}_4 R\nabla^\mu\phi+8(\nabla^\mu\phi \Box \phi- \nabla^\mu(\nabla^\nu\phi)\nabla_\nu\phi-\nabla^\mu\phi\nabla_\lambda\phi\nabla^\lambda\phi).
\end{equation}
Next, we deal with the second and the third term of \eqref{iter1b}. We integrate by parts the term $\tilde \nabla_\mu {}_4 \tilde J^{\mu}$ and expand using \eqref{GBexJ} and \eqref{GBexK} to obtain
\begin{align}
\Sigma&\equiv \int d^4 x d^{d-4}y \rgdb\ e^{(d-4)\phi}(d-3)(d-4)[-\tilde \nabla_\mu\phi {}_4 \tilde J^\mu+ {}_4 \tilde K] \nn
=&\int d^4 x d^{d-4} y\sqrt{{}_4\tilde g} \rge\ e^{(d-4)\phi}(d-3)(d-4)\biggl[- 4\lt {}_4 \tilde R^{\mu\nu}-\frac{1}{2}{}_4 \tilde g^{\mu\nu}{}_4 \tilde R\rt\tilde \nabla_\nu\phi\tilde \nabla_\mu\phi\nn
&+4(d-2)\tilde \nabla^\mu(\tilde \nabla^\nu\phi)\tilde \nabla_\nu\phi\tilde\nabla_\mu\phi+(d-2)(d-5)(\tilde\nabla^\lambda \phi \tilde \nabla_\lambda\phi)^2  \biggl)\biggl] .
\end{align}
At this point we must expand $\lt {}_4 \tilde R^{\mu\nu}-\frac{1}{2} {}_4 \tilde g^{\mu\nu} {}_4 \tilde R\rt$ using \eqref{Rmunu underr finite} and \eqref{R under finite Weyl}, taking care of the sign flip $\phi\rightarrow -\phi$, therefore obtaining 
$$\lt {}_4 \tilde R^{\mu\nu}-\frac{1}{2} {}_4 \tilde g^{\mu\nu} {}_4 \tilde R\rt\tilde \nabla_\nu\phi\tilde\nabla_\mu\phi={}_4 G^{\mu\nu}\nabla_\nu\phi\nabla_\mu\phi-(d-7)(\nabla_\lambda \phi \nabla^\lambda \phi)^2-3\Box\phi(\nabla_\lambda \phi \nabla^\lambda \phi).$$
After an integration by parts we finally get
\begin{align}
\Sigma=&\int d^4 x d^{d-4} y\sqrt{_4g} \rge e^{(d-4)\phi}(d-3)(d-4)\biggl[- 4{}_{(4)}G^{\mu\nu}\nabla_\nu\phi\nabla_\mu\phi+4(d-7)(\nabla\phi)^4+12\Box\phi(\nabla\phi)^2\nn
&-2(d-2)(d-4)(\nabla\phi)^4-2(d-2)(\nabla\phi)^2\Box\phi+4(d-2)(\nabla\phi)^4+(d-2)(d-5)(\nabla\phi)^4   \biggl)\biggl].\nn
\end{align}
At this point, only the second line of \eqref{iter1b} remains to be treated. We use \eqref{R under finite Weyl} to obtain 
\begin{align} && \int d^4 x d^{d-4}y\rgdb\ e^{(d-4) \phi} \ape R \ \app \tilde R 
=  \int d^4 x d^{d-4}y \rgdo \ e^{(d-6) \phi} \ape R\ \lt \app R + 6(d-7) \nabla_\lambda \phi \nabla^\lambda \phi \rt, \nn
\end{align}
where we have integrated by part in the last line, remembering that $\ape R$ does not depend on $x$. For the last term of \eqref{iter1b} we easily obtain 
\begin{align} 
&\int  d^4 x \, d^{d-4}y \rgpb \rge \ e^{(d-4) \phi} \ape R [4\tilde \square \phi + 2(d-4)(\tilde \nabla_\lambda \phi \tilde \nabla^\lambda \phi)]  \nn
=& \int d^4 x\, d^{d-4}y \rgdo\ e^{(d-6) \phi} \ape R [4\square \phi - (d-2)(\nabla_\lambda \phi \nabla^\lambda \phi)+ 2(d-4)(\nabla_\lambda \phi \nabla^\lambda \phi)]  \nn
=& -6(d-4)\int  d^4 x\, d^{d-4}y \sqrt{_4g}\sqrt{_eg} e^{(d-6) \phi} \ape R (\nabla_\lambda \phi \bar \nabla^\lambda \phi) . 
\end{align}
By adding all the terms,  the expanded counterterm takes the form
\begin{align} \label{Ved generic} V_E(d) =&
\int d^4 x\, d^{d-4}y \rgdo \ e^{(d-4)\phi} \lq _4E-4(d-4)(d-5) \app G^{\mu\nu}\nabla_\mu\phi\nabla_\nu\phi\right. \nn 
& -(d-4)(d-5)(d-6)[2(\nabla\phi)^2\Box\phi+(d-5)( \nabla_\lambda \phi \nabla^\lambda \phi)^2]+\ape E \nn  & \left. + 2e^{-2\phi}\ape R \lt  _4R + 3(d-10) \nabla_\lambda \phi \nabla^\lambda \phi \rt \rq .
\end{align}
Also in this case, to derive the renormalized EGB action, we consider a flat extra space ${}_4 g_{mn} = \de_{mn}$. With this choice, all the curvatures in $\mathcal{M}_e$ vanish. Finally, we drop the last line of \eqref{Ved generic} since $\ape R=0$ and obtain 
\begin{align}\label{dr in direct prod}
V_E(d) =& \int d^4 x \, d^{d-4}y \sqrt{4_g} \  e^{(d-4)\phi}\biggl[ {}_4 E-4(d-4)(d-5) {}_4 G^{\mu\nu}\nabla_\mu\phi\nabla_\nu\phi \nn
& -(d-4)(d-5)(d-6)[2(\nabla\phi)^2\Box\phi+(d-5)(\nabla\phi)^4] \biggl].
\end{align}

At this point, we need to expand \eqref{dr in direct prod} around $ d=4 $. As in the previous section, also in this case we need to intoduce a cutoff $L^{\epsilon}$ on the volume of $\mathcal{M}_e$, which will be removed in the process of dimensional reduction, since all the integrands are only supported on $\mathcal{M}_4$.

 Indeed we obtain 
\begin{align}\label{mixed pro E}
\frac{1}{\epsilon}V_E(d) =&\frac{1}{\epsilon}\left(\mu L\right)^{\epsilon}\int d^4 x  \sqrt{{}_{4}g}\,  {}_4 E +
\left(L\mu\right)^\epsilon \int d^4x \sqrt{{}_4 g}[\phi\, {}_{4} E+ 4{}_{4}G^{\mu\nu}\nabla_\mu\phi\nabla_\nu\phi) \nn &
-4\nabla_\lambda \phi\nabla^\lambda \phi \Box \phi+2(\nabla_\lambda \phi\nabla^\lambda \phi)^2],
\end{align}
from which we extract the finite contribution of $V_E$ to $\sm_R$, obtaining for this type of metric the following expression of the topological contribution to $\sm^{WZ}$
\begin{align}
\hat V'_E=&\int d^4x \sqrt{{}_4 g}[\phi\, {}_{4} E+ 4\, {}_{4}G^{\mu\nu}\nabla_\mu\phi\nabla_\nu\phi)  
-4\nabla_\lambda \phi\nabla^\lambda \phi \Box \phi+2(\nabla_\lambda \phi\nabla^\lambda \phi)^2].  
\end{align}
Also in this case, $\hat{V}'_E$ will differ from $V'_E$, by Weyl-invariant terms, exaclty as in \eqref{more}.
For the computation of the Weyl counterterm we use \eqref{cceq} and extract a conformal factor from the metric \eqref{gg}, obtaining 
\begin{align}
\int d^4x d^{d-4}y \rgdb \ e^{(d-4)\phi} \app\tilde C^2=\int d^dx \sqrt{{}_4 g} \ e^{(d-4)\phi} {}_4 C^2.
\end{align}
For the ${}_4\tilde R$ term, we resort again to \eqref{R under finite Weyl} obtaining
\begin{align}
\int d^4x d^{d-4}y \rgdb\ \ape R \app\ \tilde R =&
\int d^4x d^{d-4}y \rgdo\ e^{(d-6)\phi} \ape R  \lt \app R - 6 \square \phi - 6\nabla_\lambda \phi \nabla^\lambda \phi \rt \nn
=& \int d^4 x d^{d-4}y \rgdo\ e^{(d-6) \phi} \ape R \lt {}_4 R + 6(d-7) \nabla_\lambda \phi \nabla^\lambda \phi \rt.  \end{align}
The full $C^2$ counterterm thus reads
\begin{equation} \label{Vcd generic}
V_{C^2} = \int d^4x d^{d-4}y \sqrt{{}_4 g} e^{(d-4) \phi} \lq {}_4 C^2 + \ape C^2 + \frac{4}{(d-1)(d-2)} e^{-2 \phi} \ape R \lt _4R + 6 (d-7) \nabla_\lambda \phi \nabla^\lambda \phi \rt \rq . \end{equation}
Choosing a flat extra space $\ape g_{mn} = \de_{mn}$, \eqref{Vcd generic} reduces to the expression
\begin{equation} \label{Vcd flat}
V_{C^2} = \int d^4x d^{d-4}y \sqrt{{}_4 g}  \ e^{(d-4) \phi} \ {}_4 C^2 . \end{equation}
Also for this term, after factorizing out an overall volume on $\mathcal{M}_e$ we can reduce the integral to 4 dimensions. Expanding the exponential we get
\begin{equation} \label{VVcd flat}
\frac{1}{\epsilon} V_{C^2} = \frac{(L\mu)^\epsilon}{\epsilon}\int d^4x \sqrt{{}_4 g}\lt  {}_4 C^2+ \phi \, {}_4 C^2\rt +
L\mu)^\epsilon \int d^4x \phi \, {}_4 C^2,  \end{equation}
and the finite contribution coming from the Weyl counterterm, is here expressed only in terms of the fiducial metric ($g\to \bar g$)
\begin{equation}
\hat V'_{C^2} = \int d^4x\sqrt{\bar g} \phi \, {}_4 \bar C^2.
\end{equation}

To obtain the complete EGB action, we Weyl gauge the Einstein-Hilbert term by the same choice of metric. In fact, under a rescaling, as described in \eqref{R under finite Weyl}, we obtain
\begin{equation}
\int d^4 x  d^{d-4} y \rgd \ \lt \apd R-2\Lambda \rt=\int d^4 x d^{d-4}y \rgdo \ e^{(d-4)\phi} \lq {}_4 R-2e^{4\phi}\Lambda +(d-4)(d-5)(\nabla\phi)^2   \rq .
\end{equation}
From this expression we can factorize a volume $L^{\epsilon}$, if we take the extra dimensional metric flat ($g_{m n}(y)= \delta_{m n}$), giving a typical dilaton gravity action
\begin{equation}
\label{dgrav}
\sm_{EH} \to \sm_{EHd_2} =\left(L\mu\right)^\epsilon \int d^4 x e^{(d-4)\phi} \lq  R-2e^{4\phi}\Lambda +(d-4)(d-5)(\nabla\phi)^2   \rq,
\end{equation}
which at $d=4$ becomes
\begin{equation}
\label{re}
\sm_{EH} \to \sm_{EHd_2} =\int d^4 x  \lq  R-2e^{4\phi}\Lambda    \rq.
\end{equation}
There are some clear differences respect to $\sm_{EHd_1}$ in \eqref{d1}, since the dilaton kinetic term in \eqref{dgrav} is ghost-like at $O(\epsilon)$. Obviously, exactly at $d=4$, the instability is not present, but in \eqref{re} we miss a quadratic kinetic term for $\phi$ altogether. This clearly indicates that a consistent DR regularization should include metrics exhibiting a correct kinetic term for the dilaton, and this is clearly not always possible. At the same time, we need to exclude from the procedure those metrics which are not fully consistent, once we perform the flat spacetime limit at $d=4$.   

\subsection{Equations of motion for  $\widetilde{EGBW_1}$ }
 To derive the equations of motion we set $\kappa^2\equiv 1/(16 \pi G_N)=1$. 
We use the modified definition of $\tilde\sm_{EGBW_1}$ as given by \eqref{mod1}, which is deprived of logarithmic terms. As already remarked, this amounts to a finite renormalization of the effective action in field space.  
We obtain
\begin{equation}
\left( G^{\mu\nu}+ A^{\m \n}\right) -W^{\mu\nu}=\langle T^{\mu\nu}\rangle_f +  \frac{2}{\sqrt{g}}\lq  b' {\hat V'}_{E}^{\mu\nu}+b {\hat V'}_{C^2}^{\mu\nu}\rq 
\end{equation}
where $G^{\mu\nu}$ is the Einstein tensor, 
\begin{align}
A_{\m \n}&=\frac{2}{\rg}\frac{\de}{\de g^{\mu\nu}}\lq \int d^4 x \rg \  e^{2\phi} \lt[ R + 6\nabla_\lambda \phi \nabla^\lambda \phi ] -2e^{2\phi}\Lambda\rt\rq \nn &=
e^{2\phi}\left(-3g_{\m \n} \nabla_\l\phi\nabla^\l\phi+3\nabla_\m\phi\nabla^\m\phi+e^{2\phi}g_{\m \n} \Lambda\right),
\end{align}
and 
\begin{equation}\label{Wmn}
W_{\mu\nu}=\frac{2}{\rg}\frac{\de}{\de g^{\mu\nu}}\lq \int d^4x \sqrt{\bar g}\Big(\phi(\alpha\bar E+\alpha'\bar C^2)+\alpha[\bar G^{\mu\nu}(\bar\nabla_\mu\phi\bar\nabla_\nu\phi)+2(\bar \nabla\phi)^4+4\bar\Box\phi\bar\nabla\phi^2)]\Big)  \rq . \end{equation}
Explicitly
\begin{align}
&W_{\mu\nu}=2\bar R(\bar \nabla_\mu\bar \nabla_\nu \phi - \bar\nabla_\mu\phi \bar\nabla_\nu \phi) + 2\bar G_{\mu \nu}\lt(\bar\nabla\phi)^2-2\bar\Box\phi)\rt + 4\bar G_{\nu \alpha} (\bar\nabla^\alpha \bar\nabla_\mu \phi -\bar\nabla^\alpha \phi\bar \nabla_\mu \phi)\nn
& + 4\bar G_{\mu \alpha} \left(\bar\nabla^\alpha\bar \nabla_\nu \phi -\bar \nabla^\alpha \phi\bar \nabla_\nu \phi\right) + 4\bar R_{\mu \alpha \nu \beta}\left(\bar\nabla^\beta\bar \nabla^\alpha \phi - \bar\nabla^\alpha \phi\bar \nabla^\beta \phi\right)+ 4\bar\nabla_\alpha\bar \nabla_\nu \phi \left(\bar\nabla^\alpha \phi\bar \nabla_\mu \phi -\bar \nabla^\alpha\bar \nabla_\mu \phi \right)\nn
& +4 \bar \nabla_\alpha\bar \nabla_\mu \phi \bar\nabla^\alpha \phi\bar \nabla_\nu \phi - 4\bar\nabla_\mu \phi\bar \nabla_\nu \phi\lt (\bar\nabla\phi)^2+\bar\Box\phi)\rt+4\bar\Box\phi\bar \nabla_\nu \bar\nabla_\mu \phi - g_{\mu \nu} \biggl [2R\lt(\bar\Box\phi-\bar\nabla\phi)^2 )\rt\nn
& + 4 \bar G^{\alpha \beta} ( \bar \nabla_\beta \bar \nabla_\alpha \phi - \bar \nabla_\alpha \phi\bar  \nabla_\beta \phi ) + 2(\bar\Box\phi)^2 - ( \bar \nabla \phi)^4 + 2\bar \nabla_\beta\bar  \nabla_\alpha\phi(2\bar \nabla^\alpha \phi\bar  \nabla^\beta \phi - \bar \nabla^\beta\bar  \nabla^\alpha \phi )\biggl]+\nn
&+ \phi[\alpha H_{\mu\nu}+\alpha 'B_{\mu\nu}]+\alpha'\Big(-4R^{\m \n\a\b}\nabla_a\nabla_b\phi -4R^{\l(\m}\nabla_\l^{\n)}\phi+2R^{\m \n} \Box \phi+g^{\m \n} R^{\a \b} \nabla_a\nabla_b\phi \nn&-\frac{2}{3}R R^{\m \n} \phi+\frac{2}{3} R g^{\m \n}\Box \phi -\frac{2}{3}R \nabla^{\m}\nabla^{\n}\phi \Big),
\end{align}

where 
\begin{equation}
H_{\mu\nu}= 2\lq \bar R\bar R_{\mu\nu}-2\bar R_{\mu\alpha\nu\beta}\bar R^{\alpha\beta}+\bar R_{\mu\alpha\beta\kappa}\bar R_{\nu}{}^{\alpha\beta\kappa} -2\bar R_{\mu\alpha}\bar R^\alpha_\nu-\frac{1}{4}g_{\mu\nu}\bar E  \rq
\end{equation}
vanishes at $d=4$. Here, the renormalized quantum average of the stress energy tensor takes the form 
\begin{align}
\langle T^{\mu\nu}\rangle_R =&\langle T^{\mu\nu}\rangle_B + \lim_{d\rightarrow 4}\frac{2}{\sqrt{g}}\frac{1}{(d-4)}\lq  b' \hat V_{E}^{\mu\nu}+b \hat V_{C^2}^{\mu\nu}  \rq \nn
=&\langle T^{\mu\nu}\rangle_f + \frac{2}{\sqrt{g}}\lq  b' {\hat V'}_{E}^{\mu\nu}+b {\hat V'}_{C^2}^{\mu\nu} \rq
\end{align}
where  ${\hat V'}^{\mu\nu}\equiv\frac{\delta}{\delta g_{\mu\nu}} \hat V'$. The counterterms induce a cancellation of the singularities present in $\langle T^{\mu\nu}\rangle_B$, the stress energy tensor associated with $\sm_B$, as we perform the $d\to 4$ limit. The anomaly relation
\begin{equation}
g_{\mu\nu}\langle T^{\mu\nu}\rangle_R =\mathcal{A}(x),  \qquad \mathcal{A}(x)\equiv \sqrt{g}\left( b' E + b C^2\right),
\end{equation}
is generated by the $\hat V'$ counterterms since 
\begin{equation}
g_{\mu\nu}\langle T^{\mu\nu}\rangle_f=0
\end{equation}
and henceforth 
\begin{equation}
 \frac{2}{\sqrt{g}}g_{\mu\nu}\lq  b' {\hat V'}_{E}^{\mu\nu}+b{\hat V'}_{C^2}^{\mu\nu}\rq =\mathcal{A}(x). 
\end{equation}

\section{The quartic dilaton action and the conformal breaking scale}
\label{ef}
In this section we are going to address one aspect of the dilaton gravity action that we have indicated as $\tilde \sm_{EGBW_1}$, introduced in the previous sections.  
One of the most important issues 
concerns  the presence of constrains between the trace equation of motion of the fiducial metric and the equation of the conformal factor, as shown by \eqref{cons}. This relation is induced by the renormalization procedure and is obviously related to the anomaly, that breaks the residual invariance of the conformal decomposition \eqref{mar}. 
Notice that $\phi$ does not carry any dimension and it is clear that a correct normalization of this field requires the introduction of a scale $f$. Therefore, the selection of a given fiducial metric $\bar{g}$ is directly linked to the emergence of $f$, which breaks the conformal symmetry. To investigate this point, we proceed as follows.\\
Before expanding we send $\phi\rightarrow -\phi$, obtaining
\begin{align} \label{EGBW1}
&\tilde S_{EGBW_1} = \frac{1}{16\pi G}\int d^4 x \rg \  e^{-2\phi} \lt[ R + 6\nabla_\lambda \phi \nabla^\lambda \phi ] -2e^{-2\phi}\Lambda\rt  + \sm_f(4) \nn
&+\int d^4 x \rg \biggl[-\phi(b' E+ b C^2)-b'\left(4 G^{\mu\nu}(\nabla_\mu\phi\nabla_\nu\phi)+2(\nabla_\lambda \phi\nabla^\lambda \phi)^2-4\bar\Box\phi\nabla_\lambda \phi\nabla^\lambda \phi\right)\biggl].
\end{align}
We omit, for simplicity, the bar symbol on the gravitational metric.
It is quite straightforward to show that such a Lagrangian describe a spontaneously broken phase, due to the presence of a bilinear mixing between the scalar field $\phi$ and the metric. \\
To show this, it is convenient to introduce the field redefinition 
\begin{equation}
e^{-2 \phi}=1- \frac{\tilde{\phi}}{f} \qquad  \phi=-\frac{1}{2} \log(1-\frac{\tphi}{f})
\end{equation}  
for which the action is rewritten as 

\begin{align} \label{EGBW2}
&\tilde S_{EGBW_1} = \frac{M_P^2}{2}\int d^4 x \rg \left( R - \frac{1}{f}\tphi R + \frac{3}{2}\frac{1}{(1-\frac{\tphi}{f}) f^2}\partial_\lambda \tphi \partial^\lambda \tphi  -2(1-\frac{\tphi}{f})\Lambda \right)  + \sm_f(4) \nn
&+\int d^4 x \rg \Biggl[\frac{1}{2}\log(1-\frac{\tphi}{f})(b' E+ b C^2)- b' \left( G^{\mu\nu}\frac{1}{(1-\frac{\tphi}{f})^2 f^2}(\partial_\mu\tphi\partial_\nu\tphi)+\frac{1}{8(1-\frac{\tphi}{f})^4 f^4}(\partial_\lambda\tphi \partial^\lambda\tphi)^2\right.\nn
&\qquad \qquad \left. -\frac{1}{2(1-\frac{\tphi}{f})^3 f^3}\Box_0\tphi\partial_\lambda\tphi\partial^\lambda \tphi -\frac{1}{2(1-\frac{\tphi}{f})^4 f^4}\partial_\mu\tphi\partial^\mu\tphi \partial_\nu\tphi\partial^\nu\tphi +
 \frac{1}{2(1-\frac{\tphi}{f})^2 f^3} \Gamma^\lambda\partial_\lambda \tphi \partial_\sigma\tphi\partial^\sigma \tphi \right)\Bigg],
\end{align}

where $M_P^2=1/(8 \pi G_N)$ is the reduced Planck mass and $\Gamma^\lambda\equiv g^{\mu\nu}\Gamma^\lambda_{\mu\nu}\equiv-\frac{1}{\sqrt{g}}\partial_\mu(\sqrt{g}g^{\lambda \mu}) $. 
Notice that the coupling of the action can be organized in terms of 
interactions of increasing mass-dimensions in an expansion in $\tphi/f$. The presence of a bilinear mixing 
in the EH part of the effective action $(\sim M_P^2/f )\tphi R$ is indicating that we are describing a spontaneously broken phase.  A solution of the equations of motion can be obtained by setting 
 $\phi$ constant, and taking a flat fiducial metric $\bar{g}_{\mu\nu}=\delta_{\mu\nu}$, (i.e. a Weyl flat $g_{\mu\nu}$). In this case 
\begin{equation}
\phi=v, \qquad R_{\mu\nu}=\frac{1}{4}R g_{\mu\nu}, \qquad R=24 \lambda v^2.
\end{equation}
An alternative approach is to proceed by itroducing a different field redefinition of the form 
\begin{equation}
e^{-\phi}=\bar{\chi}(x),
\end{equation}
where $\bar{\chi}(x)$ can be related to a mass dimension-1 scalar as $\bar{\chi}(x)={\chi(x)}/({\sqrt{3}f})$
generating the coupling 
\begin{equation}
\mathcal{L} \supset \frac{M_P^2}{2 f^2}\sqrt{g}\left( \frac{1}{2}g^{\mu\nu}\partial_{\mu}\chi\partial_{\nu}\chi +\frac{1}{6}R \chi^2\ldots\right). 
\end{equation}
It is easy to realize that the $R\chi^2$ term carries the wrong sign, since for slowly varying curvature behaves essentially as a mass term 
with $m^2\sim R$. The presence either of mixing terms or of mass terms with the wrong sign are the signatures that the procedure of Weyl gauging generates a Lagrangian in a broken phase. \\
Concerning the asymptotic structure of $\tilde\sm_{EGBW_1}$, it is convenient to organize the terms appearing in it as an expansion in the two scales $1/f$ and $1/(f^n M_P^2)$, obtaining 
\begin{align}
\tilde\sm_{EGBW_1}=&\frac{M_P^2}{2}\int d^4 x \sqrt{g}\left( R -\frac{\tphi}{\bar{f}} R +
\frac{1}{2\bar{f}^2}(\partial_\mu\tphi)^2   + 2\frac{\tphi}{\bar{f}}\Lambda -2 \Lambda  + O(1/\bar{f}^2) \right.\nn
& \left. \qquad \qquad \qquad -\frac{\tphi}{\bar{f}M_P^2}(\alpha E + \alpha' C^2) +O(1/(\bar{f}^2 M_P^2) \right),
\end{align}

where we have redefined $\tphi\to \tphi/\sqrt{3}$ and $\bar{f}=\sqrt{3} f$.

At large $\bar{f}$, with $f \ll M_P$,  the dilaton field can be expressed in terms of the fiducial metric using the nonlocal relation 
\begin{equation}
\label{limit}
\tphi\sim \frac{1}{\Box}\left( - \bar{f}({R} + {\Lambda}) -\frac{\bar{f}}{M_P^2}(b'  E +b  C^2)\right),
\end{equation}
where the asymptotic expression of the field can be removed on-shell, via an auxiliary nonlocal interaction 
containing suppressed - by $\bar{f}/ M_P^2$ -  nonlocal couplings to the anomaly ($\frac{1}{\Box}(\alpha E +\alpha' C^2)$) and to the curvature ($\frac{1}{\Box}R$). If the conformal breaking scale $f$ grows 
towards $M_P$, the leading behaviour of the dilaton contribution is described by $\frac{1}{\Box}R$.
Similar couplings are found if one looks for a solution of the anomaly equation in four spacetime dimensions \cite{Riegert:1984kt}, although in this case, such behaviour is found simply as a result of the .  \\
We have seen that the structure of the effective action depends on the way we perform the $d\to 4$ limit and on the choice of the metric that is used for evaluating the finite contributions $V'_E$ and $V'_{C^2}$ or $\hat V'_E$ and $\hat V'_{C^2}$, resulting from the renormalization procedure.\\ 
Eq. \eqref{limit} indicates that after the spontaneous breaking of the conformal symmetry, the coupling of the dilaton to the anomaly is suppressed, compared to its coupling to the curvature or to the cosmological constant. One may observe that the enhanced coupling of such field to $R$, is a result of the Weyl gauging of the EH action, and it is not related to the inclusion of the quantum corrections. It shows up as a purely classical effect, which is expected to be present in any dilaton gravity model, given the generality of the procedure.

\subsection{Comments on \eqref{limit}}
It is important to pause for a moment and address an issue which is always present in these actions, concerning their local versus their nonlocal structure. It is expected that an anomaly action is nonlocal, since there are no local operators that allow to remove any anomaly in quantum field theory. For this reason, the terms $V'_E$ and $V'_{C^2}$ or $\hat V'_E$ and $\hat V'_{C^2}$ which result from the regularization procedure are local only apparently.\\
If we were able to solve for $\phi$ in terms of the background fiducial metric, then we would again derive a nonlocal action, expressed only in terms of the fiducial metric $\bar{g}$.
One can recover a nonlocal action only by removing the dilaton $\tphi$ from these expressions, generalizing the procedure 
that we have summarised in Eq. \eqref{limit}. 
Anomaly actions are all nonlocal, since the equation of motion of the dilaton field is sourced by the anomaly. \\
Another important point concerns the role of the regularization-dependent $\Box R$ term in the anomaly induced action, associated with the variation of local $R^2$ operators. As we 
are going to discuss in \secref{UV}, such terms may be introduced by a local renormalization of the topological density $\sqrt{g}E$, in order to generate a quadratic action in $\phi$. This procedure, obviously, remains valid for the classical EGB actions discussed in the recent literature, as illustrated in the Introduction. \\
It is clear that our redefinitions hide some important steps that are beyond our control, as $ f $ was not present in the original action and it has been introduced by hand through a dimensional redefinition of 
$\phi$. It is also obvious that a nonzero vev for the dilaton field can only be induced by a breaking of the Weyl simmetry \eqref{mar}.  In both cases these actions are missing a dynamical mechanism which may justify the origin of $f$.\\
In a classical EGB analysis, which does not contemplate the inclusion of any quantum correction, there is no obvious way to infer how such a scale should emerge. As we have emphasized, a classical renormalization of a singular action is equivalent to a phenomenological  parameterization of an effect which, in general, is controlled by some extra potential, but does not correspond to its fundamental and complete description.
Similar features have been discussed in Stuckelberg models \cite{Coriano:2005js} in the past in anomaly actions which emerge in the field theory limit of the Green-Schwarz mechanism of string theory. As we have already mentioned, in chiral anomaly models the solution of this conundrum forces us to advocate an extra potential whose origin, in that case, is attributed to the instanton sector of nonabelian gauge theories. Several analysis of this phenomenon have been performed in the past about this specific point. 
Here, essentially, we are after an analogous, but far more complex, gravitational one.
It is well known, however, that critical phenomena in classical gravity \cite{Gundlach:1999cu} are possible, and it is not excluded that the breaking of the $\sigma$-symmetry may be associated with the presence of classical unstable modes. We also remark that a parametric dependence in classical EGB models may be generated by the inclusion of finite boundary effects. The emergence of $f$ as a fundamental breaking scale in a scaleless GB theory could be communicated by the Einstein action, through the Planck scale, which is part of the EGB theory.

\section{Conformal constraints}
The contributions to the gravitational backreaction are due both to $\sm_f$, the Weyl-invariant part of the renormalized action $\sm_R$, and to $\sm_A$, as defined by \eqref{SA}.   
The computations of $\sm_B$, and of its finite part $\sm_f$, 
can be performed explicitly in particular backgrounds, using specific free field theory realizations. \\
Conformal Ward identities ( CWIs) constrain $\mathcal{S}_R$ in a significant way. Starting from the one-point function, under a Weyl rescaling \eqref{vars} and diffeomorhisms of the metric $x^{\mu}\to x^{\mu} + \epsilon(x)$ 
\begin{equation} 
 \delta_\epsilon g_{\mu\nu}=-\nabla_{\mu}\epsilon_{\nu}- \nabla_{\nu}\epsilon_{\mu}, 
\end{equation}
the responses of $\sm_B$ are assocated with the stress energy tensor  
 
\begin{equation}
\label{anomx}
\frac{\delta \sm_B}{\delta \sigma(x)}=\sqrt{g} \,g_{\mu\nu}\,\langle T^{\mu\nu}\rangle_B, 
\qquad
\end{equation} 
and its invariances are summarised by the relations 
\begin{equation} 
\label{eww}
\delta_\sigma \sm=0 \qquad \delta_\epsilon \sm=0,
\end{equation}
imposing trace and conservation conditions of the quantum averages of 
$T^{\mu\nu}$
 \begin{equation}
\label{comby}
\langle T^\mu_\mu\rangle_B=0 \qquad \qquad \nabla_\mu\langle T^{\mu\nu}\rangle_B=0.
\end{equation}
The conservation equation, in terms of the action takes the form 
\begin{equation}
\partial_\nu\left(\frac{\delta \sm_B(g)}{\delta g_{\mu\nu}(x)}\right)+\Gamma^\mu_{\nu\lambda}\left(\frac{\delta \sm_B(g)}{\delta g_{\lambda\nu}(x)}\right)=0\label{conserv},
\end{equation}
where $\Gamma^\mu_{\lambda\nu}$ is the Christoffel connection for the general background metric $g_{\mu\nu}$.\\
This equation is not affected by the process of renormalization and holds also for the renormalized action 
$S_R$
\begin{equation}
\partial_\nu\left(\frac{\delta \sm_R(g)}{\delta g_{\mu\nu}(x)}\right)+\Gamma^\mu_{\nu\lambda}\left(\frac{\delta \sm_R(g)}{\delta g_{\lambda\nu}(x)}\right)=0\label{conserv}.
\end{equation}
\subsection{Conformal constraints in a flat background}
In a flat background both equations become ordinary conservation equation for the quantum averaged stress energy tensor
\begin{equation}
\partial_\nu\left(\frac{\delta \sm_{R/B}(g)}{\delta g_{\mu\nu}(x)}\right)=0\label{conserv}.
\end{equation} 
 These are accompanied by special CWIs
\begin{align}
&\left(2d\,x_1^\kappa+2x_1^\kappa\,x^{\mu}_1\frac{\partial}{\partial x_1^\mu}+x_1^2\frac{\partial}{\partial x_{1\kappa}}\right)\langle T^{\mu_1\nu_1}(x_1)\rangle\notag\\
&\quad+2\bigg(x_{1\lambda}\,\delta^{\mu_1\kappa}-x_1^{\mu_1}\delta^\kappa_\lambda\bigg)\langle T^{\lambda\nu_1}(x_1)\rangle_B+2\bigg(x_{1\lambda}\,\delta^{\nu_1\kappa}-x_1^{\nu_1}\delta^\kappa_\lambda\bigg)\langle T^{\mu_1\lambda \rangle(x_1)}_B=0,
\end{align}
which becomes anomalous after renormalization in which $\sm_B \to \sm_R$. The two equations above, obviously, are regularization dependent, and need to be considered with care. In dimensional regularization, for instance, the computation of the 1-point function reduces to a combination of masselss tadpoles and vanishes identically before that any trace is taken. This is consistent from the point of view of the trace anomaly relation, since the anomaly functional $\mathcal{A}(x)$, that we will define below, vanishes in flat space. We need at least three derivatives in order for the anomaly contribution to appear in the special CWIs and trace WIs, which means that the first correlation function affected by a trace anomaly in their corresponding WIs is the $TTT$.  The dilatation WI, on the other end, in flat space shows up in the $TT$. Notice that the trace anomaly of the $TT$ can be removed by a finite renormalization of the counterterms $\sim \int d^d x \sqrt{g} R^2$.

Trace and CWI's can be derived from the equations above by functional differentiations of $\sm_R(g)$ with respect to the metric background. 
The anomalous trace WIs can be derived by allowing for an anomaly contribution on the rhs of the $\sigma$ variation in \eqref{eww}
\begin{equation}
\label{plus}
\delta_\sigma \sm_R=\int d^4 x\sqrt{g}\,\sigma\,\bar{\mathcal{A}}(x), \qquad \langle T^\mu_\mu\rangle_R =\bar{\mathcal{A}}(x),
\end{equation}
which violates Weyl invariance. Here, $\sqrt{g}\, \bar{\mathcal{A}}(x)$ is the anomaly functional 
with
\begin{equation}
\label{AF}
\mathcal{A}(x)=\sqrt{-g(x)}\bigg[b\,C^2(x)+b'E(x)\bigg].
\end{equation}

In a generic background $g$, following the renormalization procedure of \eqref{simp} \eqref{ren}, the 1-point function is decomposed as 
\begin{equation}
\label{decomp}
\langle T^{\mu\nu}\rangle_{R}=\frac{2}{\sqrt{g}}\frac{\delta \sm_{R}}{\delta g_{\mu\nu}} =\langle T^{\mu\nu} \rangle_A  + \langle \overline{T}^{\mu\nu}\rangle_f,
\end{equation}
with
\begin{equation}
g^{\mu\nu}\frac{\delta \sm_R}{\delta g^{\mu\nu}} = g^{\mu\nu}\frac{\delta \sm_A}{\delta g^{\mu\nu}}\equiv \frac{\sqrt{g}}{2} g_{\mu\nu} \langle T^{\mu\nu} \rangle_A \end{equation}
being the trace anomaly equation, and $\langle \overline{T}^{\mu\nu}\rangle_f$ is the traceless stress energy tensor derived from the Weyl-invariant functional $\sm_f$. Constraints on the entire expansions of 
$\sm_R$ or $\sm_B$, in the form of canonical WI, can be derived from the functional differentiation of 1-point functions in a rather direct way. For instance, 4-point functions are related to 3-point correlators by the hierarchical equations 

\begin{align}
&\partial_{\nu_1}\langle T^{\mu_1\nu_1}(x_1)T^{\mu_2\nu_2}(x_2)T^{\mu_3\nu_3}(x_3)T^{\mu_4\nu_4}(x_4)\rangle_R=\notag\\
=&-\left[2\left(\frac{\delta\Gamma^{\mu_1}_{\lambda\nu_1}(x_1)}{\delta  g_{\mu_2\nu_2}(x_2)}\right)_{g=\delta}\langle T^{\lambda\nu_1}(x_1)T^{\mu_3\nu_3}(x_3)T^{\mu_4\nu_4}(x_4)\rangle_R+(23)+(24)\right]\notag\\
& -\left[4\left(\frac{\delta^2\Gamma^{\mu_1}_{\lambda\nu_1}(x_1)}{\delta  g_{\mu_2\nu_2}(x_2)\delta  g_{\mu_3\nu_3}(x_3)}\right)_{g=\delta}\langle T^{\lambda\nu_1}(x_1)T^{\mu_4\nu_4}(x_4)\rangle_R+(24)+(34)\right],\label{transverseX}
\end{align}
where the correlators can either be the bare ones or the renormalized ones, the latter containing also 
the contributions coming from the functional differentiations of $\sm_A$, the anomaly part.  
In general, such equations can be expanded in any background, but they simplify significantly in 
a flat one, since the derivative of the connections take the form 
\begin{align}
\left(\frac{\delta\Gamma^{\mu_1}_{\lambda\nu_1}(x_1)}{\delta  g_{\mu_i\nu_i}(x_i)}\right)_{g=\delta}&=\frac{1}{2}\left(\delta^{\mu_1(\mu_i}\delta^{\nu_i)}_{\nu_1}\,\partial_\lambda\delta_{x_1x_i}+\delta^{\mu_1(\mu_i}\delta^{\nu_i)}_{\lambda}\,\partial_{\nu_1}\delta_{x_1x_i}-\delta^{(\mu_i}_\lambda\delta^{\nu_i)}_{\nu_1}\,\partial^{\mu_1}\delta_{x_1x_i}\right)\\[1.5ex]
\left(\frac{\delta^2\Gamma^{\mu_1}_{\lambda\nu_1}(x_1)}{\delta  g_{\mu_i\nu_i}(x_i)\delta  g_{\mu_j\nu_j}(x_j)}\right)_{g=\delta}&=\notag\\
&\hspace{-3cm}=-\frac{\delta_{x_1x_i}}{2}\delta^{\mu_1(\mu_i}\delta^{\nu_i)\epsilon}\left(\delta^{(\mu_j}_\epsilon\delta^{\nu_j)}_{\nu_1}\,\partial_\lambda\delta_{x_1x_j}+\delta^{(\mu_j}_\epsilon\delta^{\nu_j)}_{\lambda}\,\partial_{\nu_1}\delta_{x_1x_j}-\delta^{(\mu_j}_\lambda\delta^{\nu_j)}_{\nu_1}\,\partial_\epsilon\delta_{x_1x_j}\right)+(ij),
\end{align}
and in particular  in momentum space, where they become polynomial in the external momenta. \\
The inclusion of the backreaction in flat space is rather straightforward and can be addressed using the formalism developed in \cite{Coriano:2021nvn}. The two contributions indicated as $V'_E$ and ${V'}_{C^2}$ are free of ambiguities in a flat backgound.
By differentiation we get 
\begin{equation}
 \frac{1}{\epsilon}(V_E(d))^{\mu_1\nu_1\ldots \mu_n\nu_n}=\frac{1}{\epsilon}(V_E(4))^{\mu_1\nu_1\ldots \mu_n\nu_n} + (V'_E(4))^{\mu_1\nu_1\ldots \mu_n\nu_n}
\end{equation}
and collect the $O(\epsilon^0)$ terms that automatically appear in the expansion, once we perform the 
$g_{\mu\nu}\to\delta_{\mu\nu}$ limit. The approach does not need an explicit evaluation of $V_E(d)$, on the countrary of what required in the case of a general background metric $\bar{g}_{\mu\nu}$. The cancellations of the evanescent $0/0$ contributions are automatic, and can be checked in momentum space by a direct evaluation of $(V_E(4))^{\mu_1\nu_1\ldots \mu_n\nu_n}$. 
In this respect, the specification of a metric which separates the four-dimensional part from the extra-dimensional one will not be essential. \\
Terms of order $O(\epsilon^0)$ are generated only by the trace of the $d$-dimensional metric, using the condition $\delta^\mu_\mu= \epsilon + 4 $, in the flat limit.   
We can also readily move to momentum space by defining 

\begin{align}
V_{E}^{\m_1\n_1\dots\mu_n\nu_n}(p_1,\dots,\bar{p}_n)&\equiv 2^n\big[\sqrt{-g}\,E\big]^{\m_1\n_1\dots\m_n\n_n}(p_1,\dots,\bar{p}_n)\notag\\
&=2^n\int\,d^dx_1\,\dots\,d^dx_n\,d^dx\,\bigg(\sdfrac{\d^n(\sqrt{-g}E)(x)}{\d g_{\m_1\n_1}(x_1)\dots\d g_{\mu_n\nu_n}(x_n)}\bigg)_{g=\delta}\,e^{-i(p_1\,x_1+\dots+p_nx_n)}.
\end{align}
Then, the relation
\begin{equation}
\frac{\delta}{\delta g_{\mu_1\nu_1}}\ldots \frac{\delta}{\delta g_{\mu_n\nu_n}}\frac{\delta}{\delta \sigma}V_E=\epsilon (\sqrt{g}E)^{\mu_1\nu_1\ldots \mu_n\nu_n},
\end{equation}
in momentum space turns into the trace relation 
\begin{align}
\label{mom}
&\delta_{\mu_1\nu_1}\,V_{E}^{\m_1\n_1\dots\mu_n\nu_n}(p_1,\dots,p_n)=2^{n-1}(d-4)\,\left[\sqrt{-g}E\right]^{\mu_2\nu_2\dots\mu_n\nu_n}(p_2,\dots,p_n)\notag\\
&\qquad \qquad -2\bigg[V_{E}^{\m_2\n_2\dots\mu_n\nu_n}(p_1+p_2,p_3,\dots,p_n)+V_{E}^{\m_2\n_2\dots\mu_n\nu_n}(p_2,p_1+p_3,\dots,p_n)\notag \\
&\qquad \qquad \qquad +\dots+V_{E}^{\m_2\n_2\dots\mu_n\nu_n}(p_2,p_3,\dots,p_1+p_n)\bigg].
\end{align}
Of the two terms generated in the expression above, the second, proportional to $n-1$-point functions 
is obtained by the action of the functional derivative on the metric. \\
A simple example illustrating the result above is given  by the following identity
\begin{equation}
\frac{\delta}{\delta g_{\mu_2\nu_2}(x_2)}\left( 2 g_{\mu_1\nu_1}(x_1)\frac{\delta}{\delta g_{\mu_1\nu_1}(x_1)}V_E(x_1)\right)=
2 \left(\sqrt{g(x_1)}V_E\right)^{\mu_2\nu_2}\delta (x_2-x_1) + 2 g_{\mu_1\nu_1}(x_1) 
\left(\sqrt{g(x_1)}V_E\right)^{\mu_1\nu_1\mu_2\nu_2},
\end{equation}
where the last term on the rhs of this expression is further simplified by using the trace WIs. 
and expressed in terms of the one-point function. 

The correlation functions extracted by the renormalized action can be expressed as the sum of a finite $(f)$  correlator and of an anomaly term ({\em anomaly}) in the form
\begin{align}
\label{cct}
 \langle T^{\mu_1\nu_1}T^{\mu_2\nu_2}\ldots T^{\mu_n\nu_n}\rangle_{Ren} =
&\bigg[\langle T^{\mu_1\nu_1}T^{\mu_2\nu_2}\ldots T^{\mu_n\nu_n}\rangle_{bare}+\langle T^{\mu_1\nu_1}T^{\mu_2\nu_2}\ldots T^{\mu_n\nu_n}\rangle_{count}\bigg]_{d\to4}=\notag \\
&=\langle T^{\mu_1\nu_1}T^{\mu_2\nu_2}\ldots T^{\mu_n\nu_n}\rangle^{(d=4)}_{f}+\langle T^{\mu_1\nu_1}T^{\mu_2\nu_2}\ldots T^{\mu_n\nu_n}\rangle^{(d=4)}_{anomaly}
\end{align} 
This equation describes \eqref{simp} the correlators itroduced by the functional expansion of $\sm_R$, separated in terms of a Weyl-invariant $\sm_f$ and the anomaly part $\sm_A$. \\
Approaching the flat limit from a specific parameterization of a background metric, 
only requires some rearrangements of the functional expansion \eqref{exps2}. Indeed, if we 
parameterize the metric in terms of a conformal factor and of a fiducial background $\bar{g}_{\mu\nu}$
in $d$ dimensions, the flat limit is equivalently obtained by setting $\phi(x)\to 0$ and $\bar{g}_{\mu\nu}\to \delta_{\mu\nu}$. \\
Notice that the conformal factor, in this case, multiplies all the components of the metric, as described by Eq. \eqref{mar}. 
A dedicated comment deserves the case in which the extra dimensional space is characterised by a dimensionful constant. For instance, this could be the constant curvature of the extra space, as in Einstein spaces. In this case the inclusion of a dilaton component in the internal metric is necessary in order to preserve the scale invariance of the effective action in $d$ dimensions. We recall that scale invariance is 
preserved, in a theory which contain dimensionful constants, only if such constants are promoted to $x$-dependent fields. 
\section{Moving towards the UV: The reconstruction at $d=4$  for GB }\label{UV}
The reconstruction of the anomaly action in $d=4$ follows the standard procedure introduced long ago by Riegert, that we will review and extend to the GB case, in order to underline the difference between the various possible effective actions that may follow. 
Therefore, the regularization of the GB term can indeed generate regulated GB actions which can either take a local or a nonlocal form, depending on the way the conformal factor is treated in the regularization procedure. Both $V'_E$ and $V'_{C^2}$ are {\em local} expressions of the fiducial metric 
$\bar{g}_{\mu\nu}$ and of the field $\phi$. As already pointed out, their nonlocal structure will be apparent only if we are able to remove $\phi$, by re-expressing it in terms of the original metric $g_{\mu\nu}$, and this is not always possible. The case discussed by Riegert is one in which the conformal scaling relation \eqref{form11} is linear in $\phi$, and the dilaton can be removed by an integration procedure. This is a consequence of the fact that the rescaling is performed at $d=4$. 
Indeed, in this case the rescaling gives
\begin{equation}
\sqrt{g}\,\Big(E-\frac{2}{3}\,\Box R\Big)\,=\,
\sqrt{\bar{g}}\,\Big({\bar{E}}-\frac{2}{3}\,{\bar{\Box}} {\bar{R}}
+ 4{\bar{ \Delta_4}}\phi \Big)\,,
\label{119}
\end{equation}
where $\Delta_4$ is the fourth order self-adjoint 
operator, which is conformal invariant when it acts on a scalar function of vanishing scaling dimensions 
\begin{equation}
\Delta_4 = \na^2 + 2\,R^{\mu\nu}\na_\mu\na_\nu - \frac{2}{3}\,R{\Box}
+ \frac{1}{3}\,(\na^\mu R)\na_\mu\,.
\label{120}
\end{equation}
and satisfies the relation
\begin{equation}
\sqrt{-g}\,\D_4\chi=\sqrt{-\bar {g}}\,\bar{\D}_4 \chi,\label{point2}
\end{equation}
if $\chi$ is invariant (i.e. has scaling equal to zero) under a Weyl transformation. \\
Eq. \eqref{119} is crucial for the elimination of $\phi$ from the effective action.

Riegert's approach, obviously, can be modified by turning to $d$ dimensions, and the ambiguity in the derivation of the effective action, that may proceed either in d=4 or for $d=4 +\epsilon$, is an intrinsic part of the regularization. 
There are variants of $E$ that can be introduced in order to satisfy \eqref{ep2} and allow to 
eliminate $\phi$, quite closely to \eqref{119}, as we are going to discuss in the sections below. 
We can introduce for instance the modified and extended version of $E$ in the form 

\begin{equation}
\label{ext}
E^{ext}\equiv E +\frac{\epsilon}{2(d-1)^2}R^2,
\end{equation}
which is useful in order to investigate the contribution of the $V_E$ counterterm - and of its variants -  to the effective action. Notice that the two extra terms that appear on the rhs of \eqref{ext}, correspond to a boundary contribution $(\Box R)$, and to an $O(\epsilon)$ modification $(\sim \epsilon R^2)$ that vanish if we ensure either trivial boundary conditions on the metric, or we perform the $d\to 4$ limit. $E^{ext}$ plays a role in the identification of a form of the effective action which is quite close to Riegert's action. \\
The scaling relation \eqref{119} is rather unusual, in the sense that its metric variation links boundary terms in the two metrics $g_{\mu\nu}$ and $\bar{g}_{\mu\nu}$. One can show that in $d=4$, under a metric variation $\delta$

\begin{equation}
\label{intt}
\frac{1}{4}\delta (\sqrt{g} E)=\sqrt{g}\,\nabla_\sigma \delta X^\sigma,
\qquad 
\delta X^\sigma=\varepsilon^{\mu\nu\alpha \beta}\varepsilon^{\sigma\lambda\gamma\tau}
\delta \Gamma^\eta_{\nu\lambda}g_{\mu\eta}R_{\alpha \beta \gamma \tau},\qquad \varepsilon^{\mu\nu\alpha \beta}=\frac{\epsilon^{\mu\nu\alpha \beta}}{\sqrt{g}}
\end{equation}

\begin{equation}
\delta( \sqrt{g}\Box R)=\sqrt{g}\Box \delta\zeta,\qquad  \delta\zeta=-R^{\mu\nu}\delta g_{\mu\nu} +\nabla^\mu\nabla^\nu \delta g_{\mu\nu} 
-\Box(g^{\mu\nu}\delta g_{\mu\nu}).
\end{equation}

These relations follow after some integration by parts, having observed that the conformal factor varies like a scalar under the Weyl rescalings in two different frames $x$ and $x'$. This results from the fact that a fiducial metric transforms as an ordinary tensor in the two frames, hence 
\begin{equation}
g_{\mu\nu}(x)=\bar{g}_{\mu\nu}(x) e^{2 \phi(x)}  \qquad g'_{\mu\nu}(x')=\bar{g}'_{\mu\nu}(x') e^{2 \phi'(x')}
\qquad \phi'(x')=\phi(x).
 \end{equation} 
It is convenient to define 
\begin{equation}
\delta \Sigma^\sigma=\sqrt{g}g^{\sigma\beta}\partial_\beta \delta\zeta \qquad
\label{int1}
\end{equation} and vary 
both sides of \eqref{119} to obtain, using 
\begin{equation}
\delta_\phi \left(\sqrt{g}\Box R\right) = \epsilon \delta\phi \Box R +(d-6)\sqrt{g}\nabla^\lambda R \nabla_\lambda \delta\phi -2 \sqrt{g}R\nabla^2\delta \phi -2 (d-1)\sqrt{g}\nabla^4\delta \phi
\end{equation} 
the scaling relation at $d=4$
\begin{align}
& \delta_\phi\left(\frac{1}{4}\sqrt{-g}\left(E-\frac{2}{3}\square\, R\right)\right)=\sqrt{- g}\D_4\delta\phi, 
\label{pointd}
\end{align}

which simplifies in the form 

\begin{equation}
\partial_\sigma\left( \delta_\phi X^{\sigma} -\frac{1}{6}\delta_\phi\Sigma^\sigma\right)=\sqrt{- g}\D_4\delta\phi, 
\label{cc}
\end{equation}
if we use \eqref{intt} to relate it to a boundary contribution. Here we have used the general variation 
of $\delta X^\sigma$ specialised to changes in the dilaton field ($\delta_\phi$)
\begin{equation}
\delta_\phi X^\sigma=\varepsilon^{\mu\nu\alpha \beta}\varepsilon^{\sigma\lambda\gamma\tau}
\delta_\phi \Gamma^\eta_{\nu\lambda}g_{\mu\eta}R_{\alpha \beta \gamma \tau}.
\end{equation}
We have defined
\begin{equation}
\delta_\phi\Gamma_{\mu\nu}^\lambda=\de_\mu^\lambda \nabla_\nu\delta\phi+\de^\lambda_\nu \nabla_\mu\delta\phi- e^{-2 \phi} g_{\mu\nu}\nabla^\lambda\delta\phi,
\end{equation}
derived from \eqref{dg}, using $\bar\nabla_\mu\delta \phi=\nabla_\mu\delta\phi$ on scalars,
while and analogous variation $\delta\Sigma^\sigma$ in \eqref{int1} is specialised in the form 
\begin{equation}
\delta_\phi \Sigma^\sigma=\sqrt{g}g^{\sigma\beta}\partial_\beta \delta_\phi\zeta,  
\end{equation}
where
\begin{equation}
  \delta_\phi\zeta=-R^{\mu\nu}\delta_\phi g_{\mu\nu} +\nabla^\mu\nabla^\nu \delta_\phi g_{\mu\nu} 
-\Box(g^{\mu\nu}\delta_\phi g_{\mu\nu}),   \qquad  \textrm{with}\qquad   \delta_\phi g_{\mu\nu}=2g_{\mu\nu}\delta\phi.
\end{equation}
Eq. \eqref{cc}, integrated over spacetime, gives consistently 
\begin{equation}
\int d^4 x \partial_\sigma\left( \delta_\phi X^{\sigma} -\frac{1}{6}\delta_\phi\Sigma^\sigma\right)=0,
\end{equation}
if we assume asymptotic flatness, and therefore
\begin{equation}
\int d^4 x \sqrt{- g}\D_4\delta\phi=0,
\end{equation}
that follows from the self-adjointness of $\Delta_4$
\begin{equation}
\int d^4x\sqrt{-g}\,\y(\D_4\x)=\int d^4x\sqrt{-g}\,(\D_4\y)\x\label{point3},
\end{equation}
where $\x$ and $\psi$ are scalar fields of zero scaling dimensions. 

The scaling relation $\eqref{119}$ is strictly valid at $d=4$ and clearly is much simplified compared to 
\eqref{form11}, which is valid in $d$ dimensions
Clearly, Eq. \eqref{119} is not directly related to a DR procedure, but simply takes the expression of the anomaly as a given fundamental 4-dimensional result and integrates out the 
dilaton field from the scaling relation, to derive the nonlocal form of the action.  \\
We briefly review this point. It is convenient to redefine \eqref{119} in the form 
\begin{equation}
 J(x)=\bar{J}(x) + 4 \sqrt{g}\Delta_4\phi(x),\qquad     \bar J(x)\equiv \sqrt{\bar g}\left( \bar E-\frac{2}{3}\bar \Box \bar R\right), \qquad  J(x)\equiv \sqrt{ g}\left(  E-\frac{2}{3} \Box  R\right) 
 \end{equation}
\begin{equation}
(\sqrt{-g}\,\D_4)_xD_4(x,y)=\d^4(x,y).\label{point4}
\end{equation}

We invert \eqref{119} using the properties of the operator $\Delta_4$ to find the explicit form of the function $\phi(x)$, obtaining 
\begin{equation}
\label{onshell}
\phi(x)=\frac{1}{4}\int d^4y\,D_4(x,y)(J(y)- \bar{J}(y)).
\end{equation}
This sets $\phi$ on-shell. The derivation of $\sm_{WZ}$ requires the solution of the equation 
\begin{equation}
\frac{\delta \mathcal{S}_{WZ}^{(GB)}}{\delta \phi}=J,
\end{equation}
clearly identified in the form
\begin{equation}
\sm_{WZ}=\int d^4 x \sqrt{\bar g}\left(\bar J \phi + 2 \phi \Delta_4 \phi\right).
\end{equation}
At this stage it is just matter of inserting the on-shell expression of $\phi$ \eqref{onshell} into this equation to obtain 
the WZ action, in the form
\begin{equation}
\sm_{WZ}=\sm_{anom}(g)- \sm_{anom}(\bar g),
\label{WW}
\end{equation}
with 
\begin{equation}
\sm_{anom}(g)=\frac{1}{8}\int d^4 x d^4 y J(x) D_4(x,y) J(y),
\end{equation}
and a similar expression for $\sm_{anom}(\bar g)$. 
Using the explicit expression of $\phi$, and including the contributon from the rescaled $C^2$ term, we finally find the nonlocal and covariant anomaly effective action as
\begin{equation}
\mathcal{S}_{\rm anom}^{}(g) =\frac {1}{8}\!\int \!d^4x\sqrt{-g_x}\, \left(E - \frac{2}{3}\sq R\right)_{\!x} 
\int\! d^4x'\sqrt{-g_{x'}}\,D_4(x,x')\left[\frac{b'}{2}\, \left(E - \frac{2}{3}\sq R\right) +  b\,C^2\right]_{x'}.
\label{Snonl}
\end{equation}

\section{Modified Euler density and the nonlocal GB action} 
\label{mod}
Notice that if the rescaling is performed at $d=4$, and the extra field $\phi$ is reabsorbed into the definition 
of $g_{\mu\nu}$, giving a nonlocal action, then no scale of expansion is present in  $\sm_A(4)$. If we move away from 4 dimensions, and this is clearly allowed in DR, then it is obvious that extra components of the metric will be present in the expressions of $V'_E$ and $V'_{C^2}$, and the computation of the effective action will be affected by the choice of the fiducial metric over which we integrate in $d$ dimensions. \\
In the context of DR and, in particular, in the analysis of the effective actions, it is clear that variants of the topological terms are possible.\\
The functional differential form $V_E$, constructed out of $E_4\equiv E$ is not the only possible one. It is clearly exact since 
\begin{equation}
\label{abv}
\delta_\phi (\sqrt{g} E_4)=\epsilon \sqrt{g}E_4 \delta\phi,
\end{equation}
that can be verified directly by taking the trace of $V_E^{\mu\nu}$. Another way is to introduce separately the variations 
\begin{equation}
\delta_{\phi} \left(\sqrt{g}R^2\right)=\delta\phi \left(\epsilon\sqrt{g}R^2 -4 (d-1)\sqrt{g}\Box R\right),
\end{equation}
giving under integration 
\begin{equation}
\delta_\phi\int d^d x \sqrt{g}R^2=\epsilon\sqrt{g}R^2 -4 (d-1)\sqrt{g}\Box R.
\end{equation}
Similarly
\begin{equation}
\delta_\phi \left(\sqrt{g} (R_{\mu\nu\alpha\beta})^2\right) =\delta\phi\left(\epsilon \sqrt{g}  (R_{\mu\nu\alpha\beta})^2 
-8 \sqrt{g}\nabla_\mu\nabla_\nu R^{\mu\nu}
\right),
\end{equation}

\begin{equation}
\delta_\phi \left(\sqrt{g} (R_{\mu\nu})^2\right)=\epsilon \delta\phi \sqrt{g} (R_{\mu\nu})^2 -2 \sqrt{g}\Box R \delta\phi -2(d-2) \sqrt{g} \nabla_\mu\nabla_\nu R^{\mu\nu}\delta \phi,
\end{equation}
and combine them in the definition of $E_4$ using also the derivative Bianchi identity 
$\nabla_{\mu}R^\mu_\nu=\frac{1}{2}\nabla_{\nu} R$
to derive 
\begin{equation}
\label{abv}
\delta_\phi \int d^d x \sqrt{g} E_4=\epsilon \sqrt{g}E_4.  
\end{equation}

There are no boundary terms produced in the variation. 
Alternatively, from \eqref{GaussB} we obtain the two contributions 
\begin{equation}
\label{ebv1}
\frac{\delta}{\delta\phi} \int d^d x \sqrt{g}E_4=\epsilon \sqrt{g} E_4 + 
\int d^d x \sqrt{\bar g}e^{\epsilon\phi}\left( (d-3)\frac{\delta}{\delta \phi}
\bnabla_\mu \bar{J}^\mu +\epsilon(d-3) \frac{\delta}{\delta\phi}\bar{K}\right),
\end{equation}
and only the first term on the rhs of this equation contributes at $O(\epsilon)$. Consistency with \eqref{abv} 
induces the identities  
\begin{equation}
(d-3)\int d^d x \sqrt{\bar{g}}\left( \phi(x)\nbar_\mu\frac{\delta \bar J^\mu(x)}{\delta \phi(y)} +
\epsilon\frac{\delta\bar K(x)}{\delta\phi(y)}\right)=0,
\end{equation} 
where, we recall, $\bar{J}=\bar{J}(\bar{g},\phi)$ and $\bar{K}=\bar{K}(\bar{g},\phi)$.
Here we have used the boundary relation
\begin{equation}
\int d^d x \sqrt{\bar g}\frac{\delta}{\delta\phi}\bnabla_\mu \bar{J}^\mu=0.
\end{equation}
 Therefore \eqref{ebv1} implies that 
\begin{equation}
(d-3)\int d^d x \sqrt{g}e^{\epsilon\sigma}\left( \frac{\delta}{\delta \sigma}
\bnabla_\mu \bar{J}^\mu +\epsilon \frac{\delta}{\delta\sigma}\bar{K}\right)=O(\epsilon^2).
\end{equation}

Obviously, as we move away from $d=4$, modifications of such densities are possible.\\
 In general, we can modify such forms either by boundary terms, which play a role only if we include a spacetime boundary and/or by additional diffeomorphism invariant contributions of $O(\epsilon)$. \\
If we consider the extended expression of $E_4$ given by \eqref{ext},
in this case we define the counterterm
\begin{equation}
\tilde{V}_{E}=\int d^d x\sqrt{g} \left(E_4 + \epsilon\frac{R^2}{2 (d-1)^2}\right).
\end{equation}

Using the variations above, one obtains

\begin{equation}
\delta_\phi (\sqrt{g} E_{ext})=\delta\phi\epsilon\left(\sqrt{g}E_{ext}-\frac{2}{d-1}\sqrt{g}\Box R\right),
\end{equation} 
giving under integration 

\begin{equation}
\delta_\phi\int d^d x \sqrt{g} E_{ext}=\epsilon \sqrt{g}(E_{ext}- \frac{2}{d-1} \Box R). 
\end{equation}
Also in this case one needs to be careful about the $d\to 4$ limit since the metric is still $d$-dimensional 
and one has to proceed with an accurate definition of the corresponding invariants. One possibility is to perform a dimensional reduction as in \eqref{dimred}. As discussed in the previous sections, this introduces a cutofl 
$L^\epsilon$ that can be consistently removed as $\epsilon\to 0$. 

\subsection{The nonlocal EGB expansion} 
One of the standing issues concerning the consistency of the scaling approach introduced in \eqref{119} is that it is possible to make it consistent with DR, promoting to $d$ dimensions from $d=4$. This point can be addressed and solved by a redefinition of $S_{WZ}$
using $E_{ext}$ with $V_E\to \tilde{V}_E$, obtaining 
\begin{equation}
\tilde{V}_E=\int d^d x \sqrt{g}E_{ext}\, , 
 \end{equation}

\begin{equation}
\mathcal{S}^{(WZ)}_{GB} =\frac{\alpha}{\epsilon}\left(\tilde{V}_E(\bar{g}_{\mu\nu}e^{2\phi},d)- \tilde{V}_E(\bar{g}_{\mu\nu},d\right).
\label{inter}
\end{equation}

To derive its nonlocal expression, we can use the relation
\begin{equation}
\frac{\delta}{\delta\phi}\frac{1}{\epsilon}\tilde{V}_E(g_{\mu\nu},d)= \sqrt{g}\left(E-\frac{2}{3}\Box R +
\epsilon\frac{R^2}{2(d-1)^2}\right)
\end{equation}
in \eqref{inter}, to obtain  
 \begin{align}
 \frac{\delta \mathcal{S}^{(WZ)}_{GB}}{\delta\phi}=&\alpha\sqrt{g}\left(E-\frac{2}{3}\Box R \right)\nonumber \\
=&\alpha\sqrt{\bar g}\left(\bar E-\frac{2}{3}\bar \Box\bar R + 4 \bar\Delta_4 \phi\right),
\label{solve}
\end{align}
and henceforth
\begin{equation}
\mathcal{S}^{(WZ)}_{GB} = \alpha\int\,d^4x\,\sqrt{-\bar g}\,\left\{\left(\overline E - {2\over 3}
\bar{\Box} \overline R\right)\phi + 2\,\phi\bar\Delta_4\phi\right\},\,
\label{WZ2}
\end{equation}

As before, we can solve for $\phi$, deriving the regulated GB action
\begin{align}
 \mathcal{S}^{(WZ)}_{GB}& =
{\alpha\over 8} \int d^4x\,\sqrt{-g}\, \int d^4x'\,\sqrt{-g'}\,
\left(E_4 - {2\over 3} \Box R\right)_x\, \nonumber \\
&\qquad \times D_4(x,x')\left(E- {2\over 3} \Box R\right)_{x'},\,
\label{anomact}
\end{align}
that coincides with the result provided in \cite{Mazur:2001aa} by Mazur and Mottola. 

The nonlocal EGB action can be expanded, at least around a flat spacetime, in terms of the combination of the product of scalar curvature $R$ and the inverse of the D'Alembertian of flat space, i.e. of the variable $R\Box^{-1}$, which is dimensionless \cite{Coriano:2018bsy,Coriano:2017mux,Coriano:2021nvn}. This results both from perturbative computations performed around flat space and from studies of the hierarchical structure of the CWIs. The nonlocal action ca be derived by the standard approach. One considers the anomaly action written in the form  
\begin{align}
&\hspace{-1.5cm} \cS_{\rm anom}(g,\vf) \equiv -\sdfrac{1}{2} \int d^4x\,\sqrt{-g}\, \Big[ (\sq \vf)^2 - 2 \big(R^{\m\n} - \tfrac{1}{3} R g^{\m\n}\big)
(\na_\m\vf)(\na_\n \vf)\Big]\nn
& \hspace{1.5cm} +\, \sdfrac{1}{2}\,\int d^4x\,\sqrt{-g}\  \Big[\big(E - \tfrac{2}{3}\sq R\big)  \Big]\,\vf,
\label{Sanom}
\end{align}
that can be varied with respect to $\phi$, giving
\be
\sqrt{-g}\,\D_4\, \vf = \sqrt{-g}\left[\sdfrac{E}{2}- \sdfrac{\!\sq R\!}{3} \right] \label{phieom}.
\ee
The metric can be expanded perturbatively in the form 
\bes
\begin{align}
g_{\m\n} =& g_{\m\n}^{(0)} + g_{\m\n}^{(1)} + g_{\m\n}^{(2)} + \dots \equiv \eta_{\m\n} + h_{\m\n} + h_{\m\n}^{(2)} + \dots\\
\vf =& \vf^{(0)} +  \vf^{(1)} +  \vf^{(2)}  + \dots
\end{align}
\ees
The expansion above should be interpreted as a collection of terms generated by setting 
\begin{equation}
g_{\m\n} = \delta_{\mu\nu} + \kappa h_{\m\n} 
\end{equation}
having reinstated the coupling expansion $\kappa$, with $h$ of mass-dimension one,  
and collecting all the higher order terms in the functional expansion of \eqref{loc} of the order $h^2$, $h^3$ and so on. A similar expansion holds for $\vf$ if we redefine $ \vf^{(1)}=\kappa \bar\vf^{(1)},  \vf^{(2)}=\kappa^2 \bar \vf^{(2)}$ and so on. One obtains the relations
\bes
\begin{align}
&&\hspace{5cm}\sqb^2 \vf^{(0)} = 0 \label{eom0}\\
&&\hspace{-1.5cm}(\sqrt{-g} \D_4)^{(1)} \vf^{(0)} + \sqb^2 \vf^{(1)} = \left[\sqrt{-g}
\left( \sdfrac{E}{2}- \sdfrac{\!\sq R\!}{3} \right)\right]^{(1)}
= - \sdfrac{\!1\!}{3}\, \sqb R^{(1)} \label{eom1}\\
&&\hspace{-2cm}(\sqrt{-g} \D_4)^{(2)} \vf^{(0)} + (\sqrt{-g} \D_4)^{(1)} \vf^{(1)} + \sqb^2 \vf^{(2)} =
\left[\sqrt{-g}\left(\sdfrac{E}{2}- \sdfrac{\!\sq R\!}{3}  \right)\right]^{(2)} \nn
&&\hspace{5.5cm}= \sdfrac{1}{2}E^{(2)} - \sdfrac{1}{3}\, [\sqrt{-g}\sq R]^{(2)},  \label{eom2}
\end{align}
\ees
where $\sqb$ is the d'Alembert wave operator in flat Minkowski spacetime, and we have used the fact that $E$ and $C^2$ are second order in the fluctuations while the Ricci scalar $R$ starts at first order

\be
\vf^{(1)} = - \sdfrac{\!1 \!}{3\sqb}\, R^{(1)}
\label{vf1}
\ee
and the solution of (\ref{eom2}) is
\be
\vf^{(2)} = \sdfrac{1}{\sqb^2} \left\{ (\sqrt{-g} \D_4)^{(1)}\sdfrac{\!1 \!}{3\sqb} \, R^{(1)} 
+  \sdfrac{1}{2}E^{(2)} - \sdfrac{1}{3}\, [\sqrt{-g}\sq R]^{(2)} \right\}.
\label{vf2}
\ee
In this way we obtain the quadratic term
\be
\cS_{\rm anom}^{(2)} = - \sdfrac{1}{2} \,\int d^4x \, \vf^{(1)} \sqb^2 \vf^{(1)} + \sdfrac{1}{2}\,\int d^4x \, \left( - \sdfrac{2}{3} \sqb R^{(1)}\right) \vf^{(1)}
= \sdfrac{1}{18} \,\int d^4x \, \left(R^{(1)}\right)^2,
\label{Sanom2}
\ee
which is purely local, since all propagators cancel. 
The third order terms in the expansion of the anomaly action are
\begin{align}
\cS_{\rm anom}^{(3)} =&  - \sdfrac{1}{2} \int d^4x \, \left\{2\,\vf^{(1)} \sqb^2 \vf^{(2)} +\vf^{(1)} \big(\sqrt{-g} \D_4\big)^{(1)} \,\vf^{\!(1)} \right\}\nn
&\hspace{-1cm} + \sdfrac{1}{2} \int d^4x \left\{\left( - \sdfrac{2}{3} \sqb R^{(1)}\right) \vf^{(2)} + \left(E^{(2)} - \sdfrac{2}{3}\, \sqrt{-g}\sq R\right)^{\!(2)} \vf^{(1)} \right\}.
\label{Sanom3a}
 \end{align}
The remaining terms in (\ref{Sanom3a}) yield
\begin{align}
&\cS_{\rm anom}^{(3)} =- \sdfrac{1}{18} \int d^4x \, \left\{R^{(1)}\sdfrac{1}{\sqb} \big(\sqrt{-g} \D_4\big)^{\!(1)} \,\sdfrac{1}{\sqb} R^{(1)} \right\}
- \sdfrac{b'}{6} \int d^4x \left(E- \sdfrac{2}{3}\, \sqrt{-g}\sq R\right)^{\!(2)}\,\sdfrac{1}{\sqb} R^{(1)}.\nn
&\hspace{2cm}. 
\end{align}
In the variation of $\Delta_4$ it is convenient first to rewrite 
\begin{equation}
\Delta_4= \Box^2 +2 \nabla_\mu(R^{\mu\nu}\nabla^\nu) -\frac{2}{3}\nabla_\mu(R \nabla^\mu),
\end{equation}
having used the Leibnitz rule and the derivative Bianchi identity $\nabla_\mu R^{\mu\nu}=1/2 \nabla^{\nu} R$. An expansion of this operator to first order in $\delta g_{\mu\nu}$ gives

\be
\big(\sqrt{-g} \D_4\big)^{\!(1)} =  \big(\sqrt{-g} \sq^2\big)^{\!(1)} + 2\, \pa_{\m} \left(R^{\m\n} - \sdfrac{1}{3} \eta^{\m\n} R\right)^{\!(1)}\pa_{\n}.
\ee
An integration by parts gives 
\begin{align}
&&\hspace{-1.1cm}\cS_{\rm anom}^{(3)}\! =\!- \sdfrac{1}{18}\! \int\! d^4x \left\{\!R^{(1)}\!\sdfrac{1}{\sqb} \big(\sqrt{-g} \sq^2\big)^{\!(1)}\sdfrac{1}{\sqb} R^{(1)}\! \right\}
+ \sdfrac{1}{9}\! \int\! d^4x \left\{\!\pa_{\m} R^{(1)}\!\sdfrac{1}{\sqb} \! \left(\!R^{(1)\m\n}\! - \!\sdfrac{1}{3} \eta^{\m\n} R^{(1)}\!\right)\!
\sdfrac{1}{\sqb}\pa_{\n} R^{(1)}\!\right\}\hspace{-5mm}\nn 
&&\hspace{-8mm} - \sdfrac{1}{6}\! \int\! d^4x  E^{\!(2)} \sdfrac{1}{\sqb}R^{(1)}
+ \sdfrac{1}{9} \!\int\! d^4x\,  R^{(1)}  \sdfrac{1}{\sqb} \left(\sqrt{-g}\sq\right)^{\!(1)}R^{(1)}
+ \sdfrac{1}{9} \! \int\! d^4x\, R^{\!(2)}R^{(1)},
\label{Sanom3b}
\end{align}
which contains only single propagator poles. 
At this stage, using the covariant equation $\nabla_\mu\sqrt{g}=0$ on the tensor density $\sqrt{g}$ we rewrite 
\begin{equation}
\left(\sqrt{g}\Box^2\right)^{(1)}\equiv \delta\left(\sqrt{g}\Box^2\right)=\left(\frac{(\sqrt{g}\Box)^2}{\sqrt{g}}\right)^{(1)}\qquad \delta\left(\frac{1}{\sqrt{g}}
\right)(\sqrt{g}\Box)^2=-\delta(\sqrt{g})\Box^2,
\end{equation}
and
\begin{equation}
\left(\sqrt{g}\Box^2\right)^{(1)}=-\delta(\sqrt{g})\Box^2 +\frac{1}{\sqrt{g}}\delta(\sqrt{g}\Box)(\sqrt{g}\Box)
+\frac{1}{\sqrt{g}}(\sqrt{g}\Box)\left(\delta(\sqrt{g}\Box)\right),
\end{equation}
that in the flat limit becomes
\begin{equation}
\left(\sqrt{g}\Box^2\right)^{(1)}=-(\sqrt{g})^{(1)}\sqb^2 + (\sqrt{g}\Box)^{(1)}\sqb + \sqb(\sqrt{g}\Box)^{(1)},
\end{equation} 
obtaining finally
\begin{align}
&&\hspace{-5mm} \cS_{\rm anom}^{(3)} =
 \sdfrac{1}{9} \int\! d^4x \int\!d^4x'\!\int\!d^4x''\!\left\{\big(\pa_{\m} R^{(1)})_x\left(\sdfrac{1}{\sqb}\right)_{\!xx'}  
 \!\left(R^{(1)\m\n}\! - \!\sdfrac{1}{3} \eta^{\m\n} R^{(1)}\right)_{x'}\!
\left(\sdfrac{1}{\sqb}\right)_{\!x'x''}\!\big(\pa_{\n} R^{(1)})_{x''}\right\}\nn
&&\hspace{-6mm}- \sdfrac{1}{6}\! \int\! d^4x\! \int\!d^4x'\! \left(\, E^{\!(2)}\right)_{\!x}\! \left(\sdfrac{1}{\sqb}\right)_{\!xx'} \!R^{(1)}_{x'}
 + \sdfrac{1}{18} \! \int\! d^4x\, R^{(1)}\left(2\, R^{\!(2)} + (\sqrt{-g})^{(1)} R^{(1)}\right),
\label{S3anom3}
\end{align}
where the last term is purely local. This action describes graviton interactions up to trilinear fluctuations in the graviton field. \eqref{Sanom} can be expanded, with some extra effort, to quartic and higher orders, providing a consistent definition of the EGB theory, now in a completely nonlocal form and without a dilaton. The appearance of the Green's function of a quadratic $\Box_{-1}$ operator, once the expansion is performed around flat space has been shown in \cite{Giannotti:2008cv,Armillis:2009pq,Coriano:2018zdo} for 3-point functions in the $TJJ$ case and in \cite{Coriano:2017mux} for the 3-graviton vertex TTT.\\

\section{Conclusions}
In this chapter we have presented a discussion of the 
structure of the effective action and of its renormalization in DR in some detail, illustrating the main features of the procedure  that allow to identify its explicit expression. The conformal backreaction on a metric is characterised not just from the part of the action which is derived by the anomaly constraints, but also by important Weyl-invariant terms which should be explicitly computed in specific spacetimes.\\
 In the expansion around flat space, these are given by the set of correlation functions that have been discussed in the past in free field theory realizations, such as in the TT, TTT, and the 4T cases, but similar computations can be performed in other cases, such as in Weyl-flat backgrounds. The inclusion of these contributions would allow to obtain a consistent description of the backreaction, investigating its impact in a realistic cosmological context. 
\\
One of the objectives of this analysis was to establish a link between the anomalous actions in which the presence of the dilaton is manifest and those in which the dilaton can be removed.
The latter, originally derived by Riegert by a rescaling at $d=4$ of the metric, can be reconsidered in a complete DR scheme, by a redefinition of the topological density, as shown by Mazur and Mottola. 
We have pointed out that the former are expected to describe effective actions valid around a conformal breaking scale, a scale whose origin remains, at this time, not justified by the theory itself. The former are suitable for a description of the conformal anomaly action in the UV, and may play an important role in the analysis of the conformal backreaction in that domain.

We have also shown that the presence of a $R\Box^{-1}$ behaviour emerges from the quartic dilaton action derived in the "local" WZ approach, as we push the breaking scale up towards the Planck scale. This allows to establish a connection between the two formulations. Such behaviour shows that the description of anomaly actions can be realized without the inclusion of any scale, manifesting the nonlocal character of the action, reproducing, in the case of the conformal anomaly, a pattern already known for the chiral anomaly. \\
We have also shown how our results impact a class of theories, the EGB theories, in which a classical singular limit on the Gauss-Bonnet term is performed. These theories have recently received significant attention for defining local actions of dilaton gravities of Horndeski type which bypass Lovelock's theorem. We have shown that the relation between the equation of motion for the conformal factor and those of the metric, in theze theories, are constrained by the topological density. They can be classified as typical WZ actions. 
Also for such classical theories, a nonlocal formulations is possible. \\

\chapter*{Epilogue}
In this thesis we have studied  correlators in Conformal Field Theories. First, we demonstrated a method for deriving n-point scalar in coordinate space in section \ref{npoint}. Then we turned our attention to correlators in momentum space, which was the main focus of this thesis. Our goal, in both cases, was to illustrate  in a systematical way the methods one needs to study such correlators in general CFTs. In the two cases we studied in momentum space we focused on tensorial four-point correlators. 	While in coordinate space, these correlators have been studied  in great detail and by using various methods such as OPE and holography in momentum space we were among the first groups to deal with such correlators. We have used the tensor decomposition  along with the correlator invariance to find the tensor structure of each correlator. We derived the CWI and the trace and conservation WI for both cases and then for each correlator we focused our analysis in different aspects. For the $TOOO$, we focused on the structure of the CWI in two different scattering processes and then we studied various kinematical limits. Regarding the $TTTT$ correlator, due to the presence of the conformal anomaly in $d=4$, we analyzed the counterterms along with the WI that they have to satisfy and we continued with the renormalization procedure from which we derived the anomaly part of the correlator. In the final chapter, we have extended our analysis to different theories of gravities and turned our attention to conformal backreaction and topological terms. We have studied their implications with the goal to provide a clear explanation about anomalous actions with the presence of a dilaton and action in which the dilaton can be removed. \\
During the years of these studies there have been significant growing interest in CFT in momentum space for reason we have already discussed.  Some works have focused on the OPE in momentum space \cite{Gillioz:2019iye} while other have investigated  conformal/Polyakov blocks \cite{Albayrak:2018tam,Albayrak:2019yve}. Also, there have been works that exploit the formalism developed in momentum space to link  CFT with cosmology \cite{Arkani-Hamed:2018kmz,Baumann:2019oyu,Baumann:2020dch,Baumann:2021fxj,Sleight:2019mgd,Sleight:2019hfp}. In our work we based our analysis in CWI in flat space to study the correlators in interest. Moreover, there has been significant progress in the n-point scalar correlators \cite{Bzowski:2019kwd,Bzowski:2020kfw}.
Although, the analysis of CFT in momentum space is not developed as the one in coordinate space, there are many interesting directions one could take. Apart from the ones we mentioned above one could also investigate perturbative realizations of four-point or higher function. Another possible direction is further investigation of such correlators in curved spaces such as DeSitter and the study of anomalous actions in such spaces. More details for future direction are given in the conclusion of the previous chapters. \\
We hope that this thesis provides a structured and detailed study for the readers and that it will inspire them to delve deeper in this  promising and exciting research area of Theoretical Physics.

\begin{appendices}
\chapter{Appendix for the $\langle TOOO \rangle $ correlator}
\section{Primary Conformal Ward Identities }
\label{AppendixTOOO}
\subsection{Primary Conformal Ward Identities in \texorpdfstring{$\bar{p}_1$}{}
 }
Here we present the explicit expressions of the Primary Conformal Ward Identites in the case of $\bar{p}_1$ dependency. The $C_{ij}$ are given by
\begin{align}
C_{11}&=\Bigg[K_2+\frac{p_3^2-p_4^2}{s\,t}\frac{\partial}{\partial s \partial t}-\frac{p_3^2-p_4^2}{s\,u}\frac{\partial}{\partial s \partial u}+\frac{1}{t}\frac{\partial}{\partial t}\left(p_2\frac{\partial}{\partial p_2}+p_3\frac{\partial}{\partial p_3}-p_4\frac{\partial}{\partial p_4}\right)+(d-\Delta)\left(\frac{1}{t}\frac{\partial}{\partial t}+\frac{1}{u}\frac{\partial}{\partial u}\right)\notag\\
&+\frac{1}{u}\frac{\partial}{\partial u}\left(p_2\frac{\partial}{\partial p_2}-p_3\frac{\partial}{\partial p_3}+p_4\frac{\partial}{\partial p_4}\right)+\frac{2p_2^2+p_3^2+p_4^2-s^2-t^2-u^2}{t\,u}\frac{\partial}{\partial t\partial u}\Bigg]\,A(p_2,p_3,p_4,s,t,u)=0\\
C_{12}&=\Bigg[K_2+\frac{p_3^2-p_4^2}{s\,t}\frac{\partial}{\partial s \partial t}-\frac{p_3^2-p_4^2}{s\,u}\frac{\partial}{\partial s \partial u}+\frac{1}{t}\frac{\partial}{\partial t}\left(p_2\frac{\partial}{\partial p_2}+p_3\frac{\partial}{\partial p_3}-p_4\frac{\partial}{\partial p_4}\right)+(d-\Delta)\left(\frac{1}{t}\frac{\partial}{\partial t}+\frac{1}{u}\frac{\partial}{\partial u}\right)\notag\\
&+\frac{1}{u}\frac{\partial}{\partial u}\left(p_2\frac{\partial}{\partial p_2}-p_3\frac{\partial}{\partial p_3}+p_4\frac{\partial}{\partial p_4}\right)+\frac{2p_2^2+p_3^2+p_4^2-s^2-t^2-u^2}{t\,u}\frac{\partial}{\partial t\partial u}\Bigg]\,A(p_3,p_2,p_4,u,t,s)\notag\\
&\hspace{-0.5cm}+\frac{2}{t}\frac{\partial}{\partial t}\bigg(A(p_2,p_3,p_4,s,t,u)+A(p_3,p_2,p_4,u,t,s)\bigg)-\frac{2}{u}\frac{\partial}{\partial u}\bigg(A(p_2,p_3,p_4,s,t,u)+A(p_3,p_2,p_4,u,t,s)\bigg)
\end{align}
\begin{align}
C_{13}&=\Bigg[K_2+\frac{p_3^2-p_4^2}{s\,t}\frac{\partial}{\partial s \partial t}-\frac{p_3^2-p_4^2}{s\,u}\frac{\partial}{\partial s \partial u}+\frac{1}{t}\frac{\partial}{\partial t}\left(p_2\frac{\partial}{\partial p_2}+p_3\frac{\partial}{\partial p_3}-p_4\frac{\partial}{\partial p_4}\right)+(d-\Delta)\left(\frac{1}{t}\frac{\partial}{\partial t}+\frac{1}{u}\frac{\partial}{\partial u}\right)\notag\\
&+\frac{1}{u}\frac{\partial}{\partial u}\left(p_2\frac{\partial}{\partial p_2}-p_3\frac{\partial}{\partial p_3}+p_4\frac{\partial}{\partial p_4}\right)+\frac{2p_2^2+p_3^2+p_4^2-s^2-t^2-u^2}{t\,u}\frac{\partial}{\partial t\partial u}\Bigg]\,A(p_4,p_3,p_2,t,s,u)\notag\\
&\hspace{-0.5cm}-\frac{2}{t}\frac{\partial}{\partial t}\bigg(A(p_2,p_3,p_4,s,t,u)+A(p_4,p_3,p_2,t,s,u)\bigg)+\frac{2}{u}\frac{\partial}{\partial u}\bigg(A(p_2,p_3,p_4,s,t,u)+A(p_4,p_3,p_2,t,s,u)\bigg)
\end{align}
and 
\begin{align}
C_{21}&=\Bigg[K_3+\frac{p_2^2-p_4^2}{t\, u}\frac{\partial}{\partial t \partial u}-\frac{p_2^2-p_4^2}{s\,u}\frac{\partial}{\partial s \partial u}+\frac{1}{t}\frac{\partial}{\partial t}\left(p_2\frac{\partial}{\partial p_2}+p_3\frac{\partial}{\partial p_3}-p_4\frac{\partial}{\partial p_4}\right)+(d-\Delta)\left(\frac{1}{t}\frac{\partial}{\partial t}+\frac{1}{s}\frac{\partial}{\partial s}\right)\notag\\
&+\frac{1}{s}\frac{\partial}{\partial s}\left(p_3\frac{\partial}{\partial p_3}+p_4\frac{\partial}{\partial p_4}-p_2\frac{\partial}{\partial p_2}\right)+\frac{2p_3^2+p_2^2+p_4^2-s^2-t^2-u^2}{s\,t}\frac{\partial}{\partial s\partial t}\Bigg]\,A(p_2,p_3,p_4,s,t,u)\notag\\
&\hspace{-0.5cm}+\frac{2}{t}\frac{\partial}{\partial t}\bigg(A(p_2,p_3,p_4,s,t,u)+A(p_3,p_2,p_4,u,t,s)\bigg)-\frac{2}{s}\frac{\partial}{\partial s}\bigg(A(p_2,p_3,p_4,s,t,u)+A(p_3,p_2,p_4,u,t,s)\bigg)\\
C_{22}&=\Bigg[K_3+\frac{p_2^2-p_4^2}{t\, u}\frac{\partial}{\partial t \partial u}-\frac{p_2^2-p_4^2}{s\,u}\frac{\partial}{\partial s \partial u}+\frac{1}{t}\frac{\partial}{\partial t}\left(p_2\frac{\partial}{\partial p_2}+p_3\frac{\partial}{\partial p_3}-p_4\frac{\partial}{\partial p_4}\right)+(d-\Delta)\left(\frac{1}{t}\frac{\partial}{\partial t}+\frac{1}{s}\frac{\partial}{\partial s}\right)\notag\\
&+\frac{1}{s}\frac{\partial}{\partial s}\left(p_3\frac{\partial}{\partial p_3}+p_4\frac{\partial}{\partial p_4}-p_2\frac{\partial}{\partial p_2}\right)+\frac{2p_3^2+p_2^2+p_4^2-s^2-t^2-u^2}{s\,t}\frac{\partial}{\partial s\partial t}\Bigg]\,A(p_3,p_2,p_4,u,t,s)\\
C_{23}&=\Bigg[K_3+\frac{p_2^2-p_4^2}{t\, u}\frac{\partial}{\partial t \partial u}-\frac{p_2^2-p_4^2}{s\,u}\frac{\partial}{\partial s \partial u}+\frac{1}{t}\frac{\partial}{\partial t}\left(p_2\frac{\partial}{\partial p_2}+p_3\frac{\partial}{\partial p_3}-p_4\frac{\partial}{\partial p_4}\right)+(d-\Delta)\left(\frac{1}{t}\frac{\partial}{\partial t}+\frac{1}{s}\frac{\partial}{\partial s}\right)\notag\\
&+\frac{1}{s}\frac{\partial}{\partial s}\left(p_3\frac{\partial}{\partial p_3}+p_4\frac{\partial}{\partial p_4}-p_2\frac{\partial}{\partial p_2}\right)+\frac{2p_3^2+p_2^2+p_4^2-s^2-t^2-u^2}{s\,t}\frac{\partial}{\partial s\partial t}\Bigg]\,A(p_4,p_3,p_2,t,s,u)\notag\\
&\hspace{-0.5cm}+\frac{2}{s}\frac{\partial}{\partial s}\bigg(A(p_4,p_3,p_2,t,s,u)+A(p_3,p_2,p_4,u,t,s)\bigg)-\frac{2}{t}\frac{\partial}{\partial t}\bigg(A(p_4,p_3,p_2,t,s,u)+A(p_3,p_2,p_4,u,t,s)\bigg)
\end{align}
and finally 
\begin{align}
C_{31}&=\Bigg[K_4+\frac{p_2^2-p_3^2}{t\, u}\frac{\partial}{\partial t \partial u}-\frac{p_2^2-p_3^2}{s\,t}\frac{\partial}{\partial s \partial t}+\frac{1}{u}\frac{\partial}{\partial u}\left(p_2\frac{\partial}{\partial p_2}-p_3\frac{\partial}{\partial p_3}+p_4\frac{\partial}{\partial p_4}\right)+(d-\Delta)\left(\frac{1}{u}\frac{\partial}{\partial u}+\frac{1}{s}\frac{\partial}{\partial s}\right)\notag\\
&+\frac{1}{s}\frac{\partial}{\partial s}\left(p_3\frac{\partial}{\partial p_3}+p_4\frac{\partial}{\partial p_4}-p_2\frac{\partial}{\partial p_2}\right)+\frac{2p_4^2+p_2^2+p_3^2-s^2-t^2-u^2}{s\,u}\frac{\partial}{\partial s\partial u}\Bigg]\,A(p_2,p_3,p_4,s,t,u)\notag\\
&\hspace{-0.5cm}-\frac{2}{s}\frac{\partial}{\partial s}\bigg(A(p_2,p_3,p_4,s,t,u)+A(p_4,p_3,p_2,t,s,u)\bigg)+\frac{2}{u}\frac{\partial}{\partial u}\bigg(A(p_2,p_3,p_4,s,t,u)+A(p_4,p_3,p_2,t,s,u)\bigg)
\end{align}
\begin{align}
C_{32}&=\Bigg[K_4+\frac{p_2^2-p_3^2}{t\, u}\frac{\partial}{\partial t \partial u}-\frac{p_2^2-p_3^2}{s\,t}\frac{\partial}{\partial s \partial t}+\frac{1}{u}\frac{\partial}{\partial u}\left(p_2\frac{\partial}{\partial p_2}-p_3\frac{\partial}{\partial p_3}+p_4\frac{\partial}{\partial p_4}\right)+(d-\Delta)\left(\frac{1}{u}\frac{\partial}{\partial u}+\frac{1}{s}\frac{\partial}{\partial s}\right)\notag\\
&+\frac{1}{s}\frac{\partial}{\partial s}\left(p_3\frac{\partial}{\partial p_3}+p_4\frac{\partial}{\partial p_4}-p_2\frac{\partial}{\partial p_2}\right)+\frac{2p_4^2+p_2^2+p_3^2-s^2-t^2-u^2}{s\,u}\frac{\partial}{\partial s\partial u}\Bigg]\,A(p_3,p_2,p_4,u,t,s)\notag\\
&\hspace{-0.5cm}+\frac{2}{s}\frac{\partial}{\partial s}\bigg(A(p_3,p_2,p_4,u,t,s)+A(p_4,p_3,p_2,t,s,u)\bigg)-\frac{2}{u}\frac{\partial}{\partial u}\bigg(A(p_3,p_2,p_4,u,t,s)+A(p_4,p_3,p_2,t,s,u)\bigg)\notag\\
\end{align}
\begin{align}
C_{33}&=\Bigg[K_4+\frac{p_2^2-p_3^2}{t\, u}\frac{\partial}{\partial t \partial u}-\frac{p_2^2-p_3^2}{s\,t}\frac{\partial}{\partial s \partial t}+\frac{1}{u}\frac{\partial}{\partial u}\left(p_2\frac{\partial}{\partial p_2}-p_3\frac{\partial}{\partial p_3}+p_4\frac{\partial}{\partial p_4}\right)+(d-\Delta)\left(\frac{1}{u}\frac{\partial}{\partial u}+\frac{1}{s}\frac{\partial}{\partial s}\right)\notag\\
&+\frac{1}{s}\frac{\partial}{\partial s}\left(p_3\frac{\partial}{\partial p_3}+p_4\frac{\partial}{\partial p_4}-p_2\frac{\partial}{\partial p_2}\right)+\frac{2p_4^2+p_2^2+p_3^2-s^2-t^2-u^2}{s\,u}\frac{\partial}{\partial s\partial u}\Bigg]\,A(p_4,p_3,p_2,t,s,u).
\end{align}
\subsection{Primary Conformal Ward Identities in \texorpdfstring{$\bar{p}_4$}{}}
We present the remaining Primary CWIs of \secref{PCWI22}. We obtain
\begin{align}
\tilde{C}_{12}=&\Bigg[\frac{\partial^2}{\partial p_4^2}+\frac{d-2\Delta+1}{p_4}\frac{\partial}{\partial p_4}-\frac{\partial^2}{\partial p_1^2}-\frac{1-d}{p_1}\frac{\partial}{\partial p_1}+\frac{1}{s}\frac{\partial}{\partial s}\left(p_4\frac{\partial}{\partial p_4}+p_3\frac{\partial}{\partial p_3}-p_1\frac{\partial}{\partial p_1}-p_2\frac{\partial}{\partial p_2}\right)\notag\\&
+\frac{d-\Delta+2}{s}\frac{\partial}{\partial s}
+\frac{p_3^2-p_2^2}{st}\frac{\partial^2}{\partial s \partial t}\Bigg]F(p_1,p_4,p_3,p_2,t,s),\notag \\
\end{align}
\begin{align}
\tilde{C}_{13}=&\Bigg[\frac{\partial^2}{\partial p_4^2}+\frac{d-2\Delta+1}{p_4}\frac{\partial}{\partial p_4}-\frac{\partial^2}{\partial p_1^2}-\frac{1-d}{p_1}\frac{\partial}{\partial p_1}+\frac{1}{s}\frac{\partial}{\partial s}\left(p_4\frac{\partial}{\partial p_4}+p_3\frac{\partial}{\partial p_3}-p_1\frac{\partial}{\partial p_1}-p_2\frac{\partial}{\partial p_2}\right)\notag\\&
+\frac{d-\Delta-2}{s}\frac{\partial}{\partial s}
+\frac{p_3^2-p_2^2}{st}\frac{\partial^2}{\partial s \partial t}\Bigg]F(p_1,p_2,p_4,p_3,s,\tilde{u}),\\
\tilde{C}_{22}=&\Bigg[\frac{\partial^2 }{\partial p_4^2}+\frac{d-2\Delta+1}{p_4}\frac{\partial }{\partial p_4}-\frac{\partial^2 }{\partial p_2^2}-\frac{d-2\Delta+1}{p_2}\frac{\partial }{\partial p_2}+\frac{1}{s}\frac{\partial}{\partial s}\left(p_3\frac{\partial }{\partial p_3}+p_4\frac{\partial }{\partial p_4}-p_1\frac{\partial }{\partial p_1}-p_2\frac{\partial }{\partial p_2}\right)
\notag
\\&+\frac{\Delta-d}{t}\frac{\partial}{\partial t}+\frac{d-\Delta+2}{s}\frac{\partial}{\partial s}+\frac{1}{t}\frac{\partial}{\partial t}\left( p_1 \frac{\partial }{\partial p_1}+p_4\frac{\partial }{\partial p_4}-p_2\frac{\partial }{\partial p_2}-p_3\frac{\partial }{\partial p_3}\right)
\notag\\&+\frac{p_4^2-p_2^2}{st}\frac{\partial^2}{\partial s \partial t}\Bigg]F(p_1,p_4,p_3,p_2,t,s)+\frac{2}{t}\frac{\partial F(p_1,p_2,p_4,p_3,s,\tilde{u})}{\partial t}-\frac{2}{t}\frac{\partial F(p_1,p_2,p_3,p_4,s,t)}{\partial t},\notag\\
\end{align}
\begin{align}
\tilde{C}_{23}=&\Bigg[\frac{\partial^2 }{\partial p_4^2}+\frac{d-2\Delta+1}{p_4}\frac{\partial }{\partial p_4}-\frac{\partial^2 }{\partial p_2^2}-\frac{d-2\Delta+1}{p_2}\frac{\partial }{\partial p_2}+\frac{1}{s}\frac{\partial}{\partial s}\left(p_3\frac{\partial }{\partial p_3}+p_4\frac{\partial }{\partial p_4}-p_1\frac{\partial }{\partial p_1}-p_2\frac{\partial }{\partial p_2}\right)\notag
\\&+
\frac{\Delta-d+2}{t}\frac{\partial}{\partial t}+\frac{d-\Delta-2}{s}\frac{\partial}{\partial s}+\frac{1}{t}\frac{\partial}{\partial t}\left( p_1 \frac{\partial }{\partial p_1}+p_4\frac{\partial }{\partial p_4}-p_2\frac{\partial }{\partial p_2}-p_3\frac{\partial }{\partial p_3}\right)\notag\\&
+\frac{p_4^2-p_2^2}{st}\frac{\partial^2}{\partial s \partial t}\Bigg]F(p_1,p_2,p_4,p_3,s,\tilde{u}),\notag\\
\end{align}
\begin{align}
\tilde{C}_{32}=&\Bigg[\frac{\partial^2 }{\partial p_4^2}+\frac{d-2\Delta+1}{p_4}\frac{\partial }{\partial p_4}-\frac{\partial^2 }{\partial p_3^2}-\frac{d-2\Delta+1}{p_3}\frac{\partial }{\partial p_3}+\frac{p_1^2-p_2^2}{st}\frac{\partial^2}{\partial s \partial t}+\frac{\Delta-d+2}{t}\frac{\partial}{\partial t}\notag\\&+\frac{1}{t}\frac{\partial}{\partial t}\left(p_1\frac{\partial }{\partial p_1}+p_4\frac{\partial }{\partial p_4}-p_2\frac{\partial }{\partial p_2}-p_3\frac{\partial }{\partial p_3}\right)
\Bigg]F(p_1,p_4,p_3,p_2,t,s),\notag\\
\end{align}
\begin{align}
\tilde{C}_{33}=&\Bigg[\frac{\partial^2 }{\partial p_4^2}+\frac{d-2\Delta+1}{p_4}\frac{\partial }{\partial p_4}-\frac{\partial^2 }{\partial p_3^2}-\frac{d-2\Delta+1}{p_3}\frac{\partial }{\partial p_3}+\frac{p_1^2-p_2^2}{st}\frac{\partial^2}{\partial s \partial t}+\frac{\Delta-d}{t}\frac{\partial}{\partial t}\notag
\\&+\frac{1}{t}\frac{\partial}{\partial t}\left(p_1\frac{\partial }{\partial p_1}+p_4\frac{\partial }{\partial p_4}-p_2\frac{\partial }{\partial p_2}-p_3\frac{\partial }{\partial p_3}\right)-\frac{2}{s}\frac{\partial}{\partial s}
\Bigg]F(p_1,p_2,p_4,p_3,t,\tilde{u})\notag\\&
+\frac{2}{s}\frac{\partial F(p_1,p_2,p_3,p_4,s,t)}{\partial s}-\frac{2}{s}\frac{\partial F(p_1,p_4,p_3,p_2,t,s)}{\partial s}\notag\\
&+\frac{2}{t}\frac{\partial F(p_1,p_4,p_3,p_2,t,s)}{\partial t}-\frac{2}{t}\frac{\partial F(p_1,p_2,p_3,p_4,s,t)}{\partial t}.
\end{align}

\label{lauri}
\section{The hypergeometric system from \eqref{ipergio}}

Rewriting \eqref{ipergio}, with quadratic ratios $x=q_1^2/q_3^2, y=q_2^2/q_3^2$  
for the correlator $\Phi(q_1,q_2,q_3)=\langle OOO\rangle$  with scaling dimensions $\Delta_i=\Delta,\, i=1,2,3$
\begin{equation}
K_{13}\Phi=0\qquad K_{23}\Phi=0, \nonumber
\end{equation}
one obtains the system of equations
\begin{eqnarray}
\label{diff}
\begin{cases}
 \bigg[ x(1-x) \frac{\partial^2}{\partial x^2} - y^2 \frac{\partial^2}{\partial y^2} - 2 \, x \, y \frac{\partial^2}{\partial x \partial y} +  \left[ \gamma - (\alpha + \beta + 1) x \right] \frac{\partial}{\partial x}  
 - (\alpha + \beta + 1) y \frac{\partial}{\partial y}  - \alpha \, \beta \bigg] \Phi(x,y) = 0 \,,  \\
\bigg[ y(1-y) \frac{\partial^2}{\partial y^2} - x^2 \frac{\partial^2}{\partial x^2} - 2 \, x \, y \frac{\partial^2}{\partial x \partial y} +  \left[ \gamma' - (\alpha + \beta + 1) y \right] \frac{\partial}{\partial y}  
 - (\alpha + \beta + 1) x \frac{\partial}{\partial x}  - \alpha \, \beta \bigg] \Phi(x,y) = 0 \,, 
\end{cases} 
\end{eqnarray}
with parameters $\alpha(a,b),\beta(a,b),\gamma(a,b),\gamma'(a,b)$ given in \eqref{alphas} and \eqref{FuchsianPoint}, which are solved by ans\"atz of the form $x^a y^b G(x,y)$. $G$ is an 
Appell function of type $F_4(\alpha,\beta,\gamma,\gamma',x,y)$, given by 
\begin{equation} 
\label{f4}
F_4(\alpha,\beta,\gamma, \gamma';x,y)=\sum_{m, n=0}^{\infty}
\frac{(\alpha)_{m+n}(\beta)_{m+n}}{(\gamma)_{m}(\gamma')_n m! n!}x^m y^n.
\end{equation}
 Both in the case of quadratic or quartic \eqref{xy} ratios, in the variables $x$ and $y$,  the structure of \eqref{diff} is preserved, with appropriate values of the parameters $\alpha(a,b),\beta(a,b),\gamma(a,b),\gamma'(a,b)$ and indices $a,b$.

\subsection{The Lauricella system and 4K}
In the $s,t \to \infty$ limit the equations for the $K_{ij}$ operators, for arbitrary scalings $\Delta_i$, can be organized in the form

\begin{equation}
\textup{K}_{14}\phi=0,\qquad \textup{K}_{24}\phi=0,\qquad \textup{K}_{34}\phi=0\label{CWILaur}
\end{equation}
where 
\begin{align}
\textup{K}_i&=\frac{\partial^2}{\partial p_i^2}+\frac{(d-2\Delta_i+1)}{p_i}\frac{\partial}{\partial p_i},\qquad i=1,\dots,4\ ,\\
\textup{K}_{ij}&=\textup{K}_i-\textup{K}_j\ .
\end{align}
One can choose an arbitrary momentum as pivot in the ansatz for the solution of such system, for instance $(x,y,z,p_4^2)$, where 
\begin{equation}
x=\sdfrac{p_1^2}{p_4^2},\quad y=\sdfrac{p_2^2}{p_4^2},\quad z=\sdfrac{p_3^2}{p_4^2}
\end{equation}
are dimensionless quadratic ratios. The ans\"atz for the solution can be taken of the form
\begin{equation}
\phi(p_1,p_2,p_3,p_4)=(p_4^2)^{n_s}\,x^a\,y^b\,z^c\,F(x,y,z),
\end{equation}
satisfying the dilatation Ward identity with the condition
\begin{equation}
n_s=\frac{\Delta_t}{2}-\frac{3d}{2} \qquad \Delta_t=\sum_{i=1}^4 \Delta_i.
\end{equation}
With this ans\"atz the conformal Ward identities takes the form
\begin{align}
\textup{K}_{14}\phi=&4p_4^{\Delta_t-3d-2}\,x^a\,y^b\,z^c\,\bigg[(1-x)x\sdfrac{\partial^2}{\partial x^2}-2x\,y\sdfrac{\partial^2}{\partial x\partial y}-y^2\sdfrac{\partial^2}{\partial y^2}-2x\,z\sdfrac{\partial^2}{\partial x\partial z}-z^2\sdfrac{\partial^2}{\partial z^2}-2y\,z\sdfrac{\partial^2}{\partial y\partial z}\notag\\
&\hspace{2cm}+(Ax+\gamma)\sdfrac{\partial}{\partial x}+Ay\sdfrac{\partial}{\partial y}+Az\sdfrac{\partial }{\partial z}+\left(E+\sdfrac{G}{x}\right)\bigg]F(x,y,z)=0,
\end{align}
with
\begin{subequations}
	\begin{align}
	A&=\Delta_1+\Delta_2+\Delta_3-\sdfrac{5}{2}d-2(a+b+c)-1\\
	E&=-\sdfrac{1}{4}\big(3d-\Delta_t+2(a+b+c)\big)\big(2d+2\Delta_4-\Delta_t+2(a+b+c)\big)\\
	G&=\sdfrac{a}{2}\,\left(d-2\Delta_1+2a\right)\\
	\g&=\sdfrac{d}{2}-\Delta_1+2a+1.
	\end{align}
\end{subequations}
Similar constraints are obtained from the equations $\textup{K}_{24}\phi=0$ and $\textup{K}_{34}\phi=0$. \\
The reduction to the hypergeometric form requires that all the $1/x, 1/y$ and $1/z$ terms of the equations vanish. This implies that the Fuchsian points $a,b,c$ have  values

	\begin{align}\label{ind}
	a&=0,\,\Delta_1-\sdfrac{d}{2} \\
	b&=0,\,\Delta_2-\sdfrac{d}{2} \\
	c&=0,\,\Delta_3-\sdfrac{d}{2}
	\end{align}

and 
\begin{align}
\a(a,b,c)&=d+\Delta_4-\sdfrac{\Delta_t}{2}+a+b+c\notag\\
\b(a,b,c)&=\sdfrac{3d}{2}-\sdfrac{\Delta_t}{2}+a+b+c
\end{align}

\begin{equation}
\g(a)=\frac{d}{2}-\Delta_1+2a+1\,,\qquad\g'(b)=\frac{d}{2}-\Delta_2+2b+1\,,\qquad\g''(c)=\frac{d}{2}-\Delta_3+2c+1.
\end{equation}
With this redefinition of the coefficients, the equations are then expressed in the form
\begin{equation}
\resizebox{1\hsize}{!}{$
\left\{
\begin{matrix}
&x_j(1-x_j)\sdfrac{\partial^2F}{\partial x_j^2}+\hspace{-1cm}\sum\limits_{\substack{\hspace{1.3cm}s\ne j\ \text{for}\ r=j}}\hspace{-1.1cm}x_r\hspace{0.2cm}\sum x_s\hspace{0.5ex}\sdfrac{\partial^2F}{\partial x_r\partial x_s}+\left[\g_j-(\a+\b+1)x_j\right]\sdfrac{\partial F}{\partial x_j}-(\a+\b+1)\sum\limits_{k\ne j}\,x_k\sdfrac{\partial F}{\partial x_k}-\a\,\b\,F=0\\[3ex]
& (j=1,2,3)
\end{matrix}\right.\label{systemLauricella}$}
\end{equation}
where we have set $\g_1=\g$, $\ \g_2=\g'$ and $\g_3=\g''$ and $x_1=x$, $x_2=y$ and $x_3=z$. 
The system of equations admits as solutions hypergeometric functions of three variables, the Lauricella functions, of the form
\begin{equation}
F_C(\a,\b,\g,\g',\g'',x,y,z)=\sum\limits_{m_1,m_2,m_3}^\infty\,\frac{(\a)_{m_1+m_2+m_3}(\b)_{m_1+m_2+m_3}}{(\g)_{m_1}(\g')_{m_2}(\g'')_{m_3}m_1!\,m_2!\,m_3!}x^{m_1}y^{m_2}z^{m_3}.
\end{equation}
where the Pochhammer symbol $(\l)_{k}$ with an arbitrary $\l$ and $k$ a positive integer is  defined as \begin{equation}
(\alpha)_{k}=\frac{\Gamma(\alpha+k)}{\Gamma(\alpha)}=\alpha(\alpha+1)\dots(\alpha+k-1).\label{Pochh}
\end{equation}
 The convergence region of this series is defined by the condition
\begin{equation}
\left|\sqrt{x}\right|+\left|\sqrt{y}\right|+\left|\sqrt{z}\right|<1.
\end{equation}
The function $F_C$ is the generalization of the Appell $F_4$ for the case of three variables.
The system of equations \eqref{systemLauricella} admits 8 independent particular integrals (solutions) listed below. Finally, the solution for $\phi$  can be written as
\begin{equation}
\phi(p_i^2)=p_4^{\Delta_t-3d}\sum_{a,b} C_{i \,(a,b)}\ x^a y^b z^c F_C(\a(a,b),\b(a,b),\g(a,b),\g'(as,b),\g''(a,b),x,y,z)
\end{equation}
where $C_i$ are arbitrary constants. The sum runs over all the possible triple $(a,b,c)$ identified in \eqref{ind}. Introducing the 4K integral
\begin{equation}
I_{\a\{\b_1,\b_2,\b_3,\b_4\}}(p_1,p_2,p_3,p_4)=\int_0^\infty\,dx\,x^\a\,\prod_{i=1}^4(p_i)^{\b_i}\,K_{\b_i}(p_i\,x),
\end{equation}
the same solution can be re-expressed in the form 
\begin{align}
\phi(p_1,p_2,p_3,p_4)&=C\, I_{d-1\left\{\Delta_1-\frac{d}{2},\Delta_2-\frac{d}{2},\Delta_3-\frac{d}{2},\Delta_4-\frac{d}{2}\right\}}(p_1,p_2,p_3,p_4)\notag\\
&=\int_0^\infty\,dx\,x^{d-1}\,\prod_{i=1}^4(p_i)^{\Delta_i-\frac{d}{2}}\,K_{\Delta_i-\frac{d}{2}}(p_i\,x),
\label{4Kfin}
\end{align}
where $C$ is a undetermined constant. 
\chapter{Appendix for the  $\langle TTTTT \rangle $ correlator}
In this appendix, we provide useful details that complement the \ref{ttttchap} chapter. 
\label{AppendixTTTT}
\section{Contact terms}
\label{contact}
Contact terms generated by 4th variation of a fundamental action are generated in the form 
\begin{align}
&\left\langle\frac{\delta^4 S }{\delta g(d)\delta g(c)\delta g(b)\delta g(a)}\right\rangle =\frac{1}{2} \sqrt{\text{dg}(a)}
\left\langle \frac{\delta ^3 T(a)}{\delta g(b) \delta g(c) \text{$\delta
		$g}(d)}\right\rangle+\frac{1}{8} g(b) \sqrt{\text{dg}(c)} g(c) \delta (b,a) \delta (c,b) \left\langle
\frac{\text{$\delta $T}(a)}{\delta g(d)}\right\rangle \nonumber \\&
+\frac{1}{8} g(b)
\sqrt{\text{dg}(d)} g(d) \delta (b,a) \delta (d,b) \left\langle \frac{\text{$\delta
		$T}(a)}{\delta g(c)}\right\rangle\nonumber
+\frac{1}{8} g(c) \sqrt{\text{dg}(d)} g(d)
\delta (c,a) \delta (d,c) \left\langle \frac{\text{$\delta $T}(a)}{\text{$\delta
		$g}(b)}\right\rangle \nonumber \\&
+\frac{1}{8} \sqrt{\text{dg}(d)} g(d) \delta (b,a) \delta (c,b)
s(c,b) \delta (d,b) \langle T(a)\rangle
+\frac{1}{8} \sqrt{\text{dg}(c)} g(c) \delta (b,a)
\delta (c,b) \delta (d,b) s(d,b) \langle T(a)\rangle \nonumber \\&
+\frac{1}{8} g(b) \sqrt{\text{dg}(c)}
\delta (b,a) \delta (c,b) \delta (d,c) s(d,c) \langle T(a)\rangle
+\frac{1}{4}
\sqrt{\text{dg}(b)} g(b) \delta (b,a) \left\langle \frac{\delta ^2 T(a)}{\text{$\delta
		$g}(c) \delta g(d)}\right\rangle \nonumber \\&
	+\frac{1}{4} \sqrt{\text{dg}(c)} g(c) \delta
(c,a) \left\langle \frac{\delta ^2 T(a)}{\delta g(b) \text{$\delta
		$g}(d)}\right\rangle 
+\frac{1}{4} \sqrt{\text{dg}(d)} g(d) \delta (d,a) \left\langle
\frac{\delta ^2 T(a)}{\delta g(b) \delta g(c)}\right\rangle\nonumber \\&
+\frac{1}{16} g(b) g(c) \sqrt{\text{dg}(d)} g(d) \delta (b,a) \delta (c,b) \delta (d,c)
\langle T(a)\rangle  +\frac{1}{4} \sqrt{\text{dg}(b)} \delta (b,a) \delta (c,b) s(c,b)
\left\langle \frac{\text{$\delta $T}(a)}{\delta g(d)}\right\rangle \nonumber \\&
+\frac{1}{4}
\sqrt{\text{dg}(b)} \delta (b,a) \delta (d,b) s(d,b) \left\langle \frac{\text{$\delta
		$T}(a)}{\delta g(c)}\right\rangle
+\frac{1}{4} \sqrt{\text{dg}(c)} \delta (c,a)
\delta (d,c) s(d,c) \left\langle \frac{\text{$\delta $T}(a)}{\text{$\delta
		$g}(b)}\right\rangle\nonumber \\&
+\frac{1}{4} \sqrt{\text{dg}(b)} \delta (b,a) \delta (c,b) \delta
(d,b) \delta (d,c) \langle T(a)\rangle  s(d,c,b).
\end{align}
In the flat limit we obtain 
\begin{align}
\langle \frac{\delta^4 S}{\delta g(a)\delta g(b)\delta g(c)\delta g(d)} \rangle=&+\frac{1}{8} \delta ^{b}\delta ^{c} \delta (b,a) \delta (c,b) \left\langle \frac{\text{$\delta
		$T}(a)}{\delta g(d)}\right\rangle +\frac{1}{8} \delta ^{b}\delta ^{d} \delta (b,a) \delta
(d,b) \left\langle \frac{\text{$\delta $T}(a)}{\delta g(c)}\right\rangle
\nonumber \\&
+\frac{1}{8} \delta ^{d}\delta ^{c} \delta (c,a) \delta (d,c) \left\langle \frac{\text{$\delta
		$T}(a)}{\delta g(b)}\right\rangle
+\frac{1}{4} \delta (b,a) \delta (c,b) s(c,b)
\left\langle \frac{\text{$\delta $T}(a)}{\delta g(d)}\right\rangle\nonumber \\&
+\frac{1}{4}
\delta (b,a) \delta (d,b) s(d,b) \left\langle \frac{\text{$\delta $T}(a)}{\text{$\delta
		$g}(c)}\right\rangle +\frac{1}{4} \delta (c,a) \delta (d,c) s(d,c) \left\langle
\frac{\text{$\delta $T}(a)}{\delta g(b)}\right\rangle\nonumber \\& +\frac{1}{4} \delta ^d
\delta (d,a) \left\langle \frac{\delta ^2 T(a)}{\delta g(b) \text{$\delta
		$g}(c)}\right\rangle +\frac{1}{4} \delta ^c \delta (c,a) \left\langle \frac{\delta ^2
	T(a)}{\delta g(b) \delta g(d)}\right\rangle\nonumber \\&
+\frac{1}{4} \delta ^b
\delta (b,a) \left\langle \frac{\delta ^2 T(a)}{\delta g(c) \text{$\delta
		$g}(d)}\right\rangle +\frac{1}{2} \left\langle \frac{\delta ^3 T(a)}{\delta g(b)
	\delta g(c) \delta g(d)}\right\rangle.
\end{align}

\subsection{Construction of a new ($n$-$p$)  basis}
\label{np}
In $d=4$ we may construct a new orthogonal four-vector $n^\mu$ using the completely antisymmetric $\epsilon$ tensor and three of the four external momenta in the form 
\begin{equation}
n^{\mu}=\epsilon^{\m \a \b\g}p_{1,\a}p_{2,\b}p_{3,\g}.
\end{equation}
Discussions of this  basis can be found in \cite{Bzowski:2013sza,Serino:2020pyu,Bzowski:2017poo}.
We will refer to this basis as to the $n$-$p$ basis. Notice that this basis is the direct generalization of the orthogonalization procedure in $d=3$ of the usual external product of two 3-vectors $\mathbf{a},\mathbf{b}$ in Euclidean space $\mathbf{n}=\mathbf{a}\times \mathbf{b}$. In the process of renormalization, the $\delta^{\mu\nu}$ is taken as an independent symmetric tensor, along with the external momenta, which appears in the covariant expansion of the form factors for a generic $d$. After removing the singularities by the relevant counterterms, having performed all the contractions in $d$ dimensions and having obtained a finite expression, we can 
dimensionally reduced the indices of all the tensor components to $d=4$. This reduction allows us to use the $n$-$p$ basis as a basis of expansion. Notice that such simplifications are possible if we have in $d$ dimensions at least $d-1$ independent external momenta in a correlation function. For instance, in $d=3$, this simplification starts in the expansion of 3-point functions, and in $d=4$  from 4-point functions.

We are allowed to use this basis in order to expand $\delta^{\mu\nu}$ in the form 

\begin{equation}
\d^{\m \n}=\sum_{i,j}^4 p_i^{\m} p_j^{\n} (Z^{-1})_{j i},
\end{equation}
where $(Z^{-1})_{j i}$ is the inverse of the Gramm matrix, defined as $Z=[p_i\cdot p_j]_{i,j=1}^d$. In our case we have:
\begin{equation}
\begin{pmatrix}
p_1^2 & p_1\cdot p_2 &  p_1\cdot p_3 & 0\\
 p_2\cdot p_1 & p_2^2 & p_2\cdot p_3 & 0\\
 p_3 \cdot p_1 & p_3\cdot p_2 & p_3^2 &0\\
 0 & 0 &0 & n^2
\end{pmatrix}
\end{equation}
The zero entries in the matrix above come from the orthonormality relation between the vector $n^{\m}$ and the momenta.  We also have
\begin{equation}
n^2=-p_1^2 p_2^2 p_3^2+p_1^2 (p_2\cdot p_3)^2+p_2^2
   (p_1\cdot p_3)^2+p_3^2 (p_1\cdot p_2)^2-2 p_1\cdot p_2  p_1\cdot p_3
   p_2\cdot p_3
\end{equation}

This implies that when the transverse-traceless projector gets contracted with two vectors ${n^\mu}$, will generate a term whose tensorial structure involves two of the external momenta, such as ${p^\a_i}_j {p^\b_i}_k$ in the expression above, with the related 
scalar factors. 

\section{ Trace relations}\label{TraceRelations}
 An analysis similar to that discussed for \eqref{double} can be performed for the traces of the third functional derivatives of $V_E$ in $d$ dimensions. 
One finds, expanding the lhs of \eqref{third} 
\begin{align}
&8 g_{\mu_1\nu_1}(x_1)V_E^{\mu_1\nu_1}(x_1)\delta^d (x_3-x_1)\delta^d(x_2-x_1) +
8 g_{\mu_3\nu_3}(x_3)g_{\mu_1\nu_1}(x_1)\delta^d (x_1-x_2)V_E^{\mu_1\nu_1\mu_3\nu_3}(x_1,x_3) \notag \\
& +8 g_{\mu_1\nu_1}(x_1)g_{\mu_2\nu_2}(x_2)\delta^d (x_1-x_3)V_E^{\mu_1\nu_1\mu_2\nu_2}(x_1,x_2) + 
8 g_{\mu_1\nu_1}(x_1)g_{\mu_2\nu_2}(x_2) \delta^d(x_2-x_3)V_E^{\mu_1\nu_1\mu_2\nu_2}(x_1,x_2)\notag \\
& + 8g_{\mu_3\nu_3}(x_3)g_{\mu_2\nu_2}(x_2) g_{\mu_1\nu_1}(x_1)
V_E^{\mu_1\nu_1\mu_2\nu_2\mu_3\nu_3}(x_1,x_2,x_3) = \varepsilon^3{\sqrt{g(x_1)}}E(x_1)\delta^d(x_1-x_2)\delta^d(x_2-x_3).
\end{align}
By going to the flat limit and using \eqref{double} one obtains 
\begin{align}
\label{triple}
\delta^{d}_{\mu_1\nu_2}\delta^{d}_{\mu_2\nu_2}\delta^{d}_{\mu_3\nu_3}V_E^{\mu_1\nu_1\mu_2\nu_2\mu_3\nu_3}=0. 
\end{align}
Moving to the fourth derivative we have:
 \begin{equation}
 \label{fourth}
 \frac{\delta}{\delta \sigma(x_4)} \frac{\delta}{\delta \sigma(x_3)} \frac{\delta}{\delta \sigma(x_2)} \frac{\delta}{\delta \sigma(x_1)}V_E=\varepsilon^4 \sqrt{g}E \delta^d(x_1-x_2)\delta^d(x_3-x_4)\delta^d(x_2-x_3)
\end{equation}
Expanding the l.h.s of \eqref{fourth} we get
\begin{align}
&16 g_{\mu_1\nu_1}(x_1)V_E^{\mu_1\nu_1}(x_1)\delta^d (x_4-x_1)\delta^d (x_3-x_1)\delta^d(x_2-x_1) \notag\\
&+16 g_{\mu_4\nu_4}(x_4)g_{\mu_1\nu_1}(x_1)V_E^{\mu_1\nu_1\mu_4\nu_4}(x_1,x_4)\delta^d (x_1-x_2) \delta^d (x_1-x_3)\notag \\
& +16 g_{\mu_1\nu_1}(x_1)g_{\mu_3\nu_3}(x_3)V_E^{\mu_1\nu_1\mu_3\nu_3}(x_1,x_3)\big(\delta^d (x_1-x_4)\delta^d (x_1-x_2)+\delta^d (x_1-x_2)\delta^d (x_3-x_4)\big) + \notag \\&+
16 g_{\mu_1\nu_1}(x_1)g_{\mu_2\nu_2}(x_2) V_E^{\mu_1\nu_1\mu_2\nu_2}(x_1,x_2)\big(\delta^d(x_1-x_3)\delta^d(x_2-x_4)+\delta^d(x_1-x_3)\delta^d(x_1-x_4)\notag \\&+\delta^d(x_2-x_3)\delta^d(x_2-x_4)+\delta^d(x_2-x_3)\delta^d(x_1-x_4)\big)\notag\\
& + 16g_{\mu_4\nu_4}(x_4)g_{\mu_3\nu_3}(x_3) g_{\mu_1\nu_1}(x_1)
V_E^{\mu_1\nu_1\mu_2\nu_2\mu_3\nu_3}(x_1,x_3,x_4)\delta^d(x_1-x_2)\notag \\
& + 16g_{\mu_4\nu_4}(x_4)g_{\mu_2\nu_2}(x_2) g_{\mu_1\nu_1}(x_1)
V_E^{\mu_1\nu_1\mu_2\nu_2\mu_3\nu_3}(x_1,x_2,x_4)\big(\delta^d(x_1-x_3)+\delta^d(x_2-x_3)\big)\notag \\
& + 16g_{\mu_3\nu_3}(x_3)g_{\mu_2\nu_2}(x_2) g_{\mu_1\nu_1}(x_1)
V_E^{\mu_1\nu_1\mu_2\nu_2\mu_3\nu_3}(x_1,x_2,x_4)\big(\delta^d(x_1-x_4)+\delta^d(x_2-x_3)+\delta^d(x_3-x_4)\big)\notag \\& + 16g_{\mu_4\nu_4}(x_4)g_{\mu_3\nu_3}(x_3)g_{\mu_2\nu_2}(x_2) g_{\mu_1\nu_1}(x_1)
V_E^{\mu_1\nu_1\mu_2\nu_2\mu_3\nu_3 \mu_4\nu_4}(x_1,x_2,x_3,x_4) \notag\\
&= \varepsilon^4 \sqrt{g}E \delta^d(x_1-x_2)\delta^d(x_2-x_3)\delta^d(x_3-x_4).
\end{align}
By going to the flat limit and using \eqref{double} and \eqref{triple} one obtains 
\begin{align}
\label{quadro}
\delta^{d}_{\mu_1\nu_2}\delta^{d}_{\mu_2\nu_2}\delta^{d}_{\mu_3\nu_3}\delta^{d}_{\mu_4\nu_4}V_E^{\mu_1\nu_1\mu_2\nu_2\mu_3\nu_3\mu_4\nu_4}=0.
\end{align}
This analysis can be extended to any $n$, yielding zero for all the fully traced functional derivatives of $V_E$. Notice that in \eqref{generalder} we have used only Weyl variations, which combine changes in the metric accompanied by contractions, as clear from \eqref{var}.\\

\section{Definitions}\label{Definitions}
The Christoffel symbols are
\begin{equation}
\Gamma^\lambda_{\mu\nu}(x)=\frac{1}{2}g^{\lambda \kappa}(x)\big(\partial_\mu g_{\nu \kappa}+\partial_\nu g_{\mu \kappa}-\partial_\kappa g_{\mu \nu}\big).
\end{equation}
Our definition of the Riemann tensor is
\begin{align} \label{Tensors}
{R^\lambda}_{\mu\kappa\nu}
&=
\pd_\nu \Gamma^\lambda_{\mu\kappa} - \pd_\kappa \Gamma^\lambda_{\mu\nu}
+ \Gamma^\lambda_{\nu\eta}\Gamma^\eta_{\mu\kappa} - \Gamma^\lambda_{\kappa\eta}\Gamma^\eta_{\mu\nu}.
\end{align}
The Ricci tensor is defined by the contraction $R_{\mu\nu} = {R^{\lambda}}_{\mu\lambda\nu}$ 
and the scalar curvature by $R = g^{\mu\nu}R_{\mu\nu}$.\\
The traceless part of the Riemann tensor in $d$ dimension is the Weyl tensor, 

The functional variations with respect to the metric tensor are computed using the relations
\begin{align}\label{Tricks}
\delta \sqrt{-g} = -\frac{1}{2} \sqrt{-g}\, g_{\a\b}\,\delta g^{\a \b}\quad &
\delta \sqrt{-g} = \frac{1}{2} \sqrt{-g}\, g^{\a\b}\,\delta g_{\a \b}  \nonumber \\
\delta g_{\mu\nu} = - g_{\mu\a} g_{\nu\b}\, \delta g^{\a\b} \quad&
\delta g^{\mu\nu} = - g^{\mu\a} g^{\nu\b}\, \delta g_{\a\b}.\,
\end{align}
The following structure has been repeatedly used throughout the calculations
\begin{align}\label{Tricks2}
s^{\a\b\g\delta} \, \delta(z,x) &\equiv - \frac{\d g^{\a\b}(z)}{\d g_{\g\d}(x)} =
\frac{1}{2}\left[\delta^{\a\g}\delta^{\b\delta} + \delta^{\a\delta}\delta^{\b\g}\right]\delta(z,x)\, .
\end{align}
Another useful quantity is:
\begin{equation}
[\sqrt{-g}]^{\m_1\n_1 \m_2 \n_2 }=-\frac{1}{4} g^{\mu_1\nu_2} g^{\mu_2\nu_1}+\frac{1}{4} g^{\mu_1\nu_1} g^{\mu_2\nu_2}-\frac{1}{4} g^{\mu_1\mu_2} g^{\nu_1\nu_2}
\end{equation}
Regarding the derivatives of the Christoffel symbols we get, in the flat space limit:
\begin{align}
&[\Gamma^{\a}_{\b \chi}]^{\m_2 \n_2}(p_2)=-\frac{i}{2}\delta^{\a \kappa}\bigg(\delta^{(\m_2}_{\b}\delta^{\n_2)}_{\kappa}p_{2,\chi}+\delta^{(\m_2}_{\chi}\delta^{\n_2)}_{\kappa}p_{2,\beta}-\delta^{(\m_2}_{\b}\delta^{\n_2)}_{\chi}p_{2,\kappa}\bigg)=\delta^{\a \kappa}[\tilde{\Gamma}_{\kappa\b \chi}]^{\m_2 \n_2}(p_2),\\
&[\Gamma^{\a}_{\b \chi}]^{\m_2 \n_2 \m_3 \n_3}(p_2,p_3)=[g^{\a\kappa}]^{\m_2\n_2}[\tilde{\Gamma}_{\kappa \b \chi}]^{\m_3 \n_3}(p_3)+[g^{\a\kappa}]^{\m_3\n_3}[\tilde{\Gamma}_{\kappa\b \chi}]^{\m_2 \n_2}(p_2),\\
&[\Gamma^{\a}_{\b \chi}]^{\m_2 \n_2 \m_3 \n_3 \m_4 \n_4}(p_2,p_3,p_4)=[g^{\a \kappa}]^{\m_2\n_2 \m_4 \n_4}[\tilde{\Gamma}_{\kappa \b \chi}]^{\m_3 \n_3}(p_3)+[g^{\a \kappa}]^{\m_2\n_2 \m_3 \n_3}[\tilde{\Gamma}_{\kappa \b \chi}]^{\m_4 \n_4}(p_4)\notag\\&\hspace{5cm}+[g^{\a \kappa}]^{\m_3\n_3 \m_4 \n_4}[\tilde{\Gamma}_{\kappa \b \chi}]^{\m_2 \n_2}(p_2).
\end{align}
The above notation can be understood as follows:
\begin{equation}
[\Gamma^{\a}_{\b \chi}]^{\m_2 \n_2...\m_n \n_n}(p_2,\dots,p_n)=\int d^d x_1...d^d x_n e^{-i (x_1 p_1+\dots+x_n p_n)}\left(\sdfrac{\d^n \Gamma^{\a}_{\b \chi}(x_1)}{\d g_{\m_n \n_n}(x_n)\dots\d g_{\m_2 \n_2}(x_2)}\right)_{g=\delta}
\end{equation}
Also  we have defined (in the flat space-time limit)
\begin{align}
[g^{\a\kappa}]^{\m_2\n_2 \m_3 \n_3}=&\frac{1}{4} \d^{\alpha \n_3} \d^{\kappa \n_2} \d^{\m_2\m_3}+\frac{1}{4} \d^{\alpha \n_2} \d^{\kappa \n_3} \d^{\m_2\m_3}+\frac{1}{4} \d^{\alpha \n_2} \d^{\kappa \m_3} \d^{\m_2 \nu_3}+\frac{1}{4} \d^{\alpha \m_3} \d^{\kappa \n_2} \d^{\m_2\n_3}+\frac{1}{4} \d^{\alpha \n_3}
\d^{\kappa \m_2} \d^{\m_3\n_2}\notag\\&+\frac{1}{4} \d^{\alpha \m_2} \d^{\kappa \n_3} \d^{\mu_3\n_2}+\frac{1}{4} \d^{\alpha \m_3} \d^{\kappa \m_2} \d^{\n_2\n_3}+\frac{1}{4} \d^{\alpha
	\m_2} \d^{\kappa \m_3} \d^{\n_2\n_3}.
\end{align}
Using the above we can compute the derivatives of the Riemann, Ricci and the scalar curvature. Up to second derivative the expressions can be found at \cite{Coriano:2018bsy1}. For our convenience, we reproduce the formulae here.
In order to simplify the notation, we introduce the tensor components
\begin{align}
A^{\m_1\n_1\m\n}&\equiv\d^{\mu_1\nu_1}\d^{\m\n}-2\d^{\m(\m_1}\d^{\n_1)\n}\nn \\
B^{\m_1\n_1\m\n}&\equiv\d^{\mu_1\nu_1}\d^{\m\n}-\d^{\m(\m_1}\d^{\n_1)\n}\nn \\
C^{\m_1\n_1\m_2\n_2\m\n}&\equiv\d^{\m(\m_1}\d^{\n_1)(\m_2}\d^{\n_2)\n}+\d^{\m(\m_2}\d^{\n_2)(\m_1}\d^{\n_1)\n}\nn \\
\tilde{C}^{\m_1\n_1\m_2\n_2\m\n}&\equiv\d^{\m(\m_1}\d^{\n_1)(\m_2}\d^{\n_2)\n}\nn \\
D^{\m_1\n_1\m_2\n_2\m\n}&\equiv\d^{\m_1\n_1}\d^{\m(\m_2}\d^{\n_2)\n}+\d^{\m_2\n_2}\d^{\m(\m_1}\d^{\n_1)\n}\nn \\
E^{\m_1\n_1\m_2\n_2\m\n}&\equiv\d^{\m_1\n_1}B^{\m_2\n_2\m\n}+C^{\m_1\n_1\m_2\n_2\m\n},\nn \\
F^{\a_1\a_2\m\n}&\equiv\d^{\a_1[\m}\d^{\n]\a_2}\nn \\
\tilde{F}^{\a_1\a_2\m\n}&\equiv\d^{\a_1(\n}\d^{\m)\a_2}\nn \\
\tilde{F}^{\a_1\a_2}_{\m\n}&\equiv\d^{(\a_1}_{\n}\Delta_{\m}^{\a_2)}\nn \\
G^{\m_1\n_1\a_1\a_2\m\n}&\equiv\d^{\m[\n}\d^{\a_2](\m_1}\d^{\n_1)\a_1}+\d^{\a_1[\a_2}\d^{\n](\m_1}\d^{\n_1)\m}\nn \\
H^{\m_1\n_1\m_2\n_1\a_1\a_2\m\n}&\equiv A^{\m_1\n_1\m\a_1}\tilde{F}^{\m_2\n_2\n\a_2}-A^{\m_2\n_2\m\a_1}\tilde{F}^{\m_1\n_1\n\a_2}\nn \\
I^{\m_1\n_1\m_2\n_2\a_1\a_1\m\n}&\equiv\d^{\m_1\n_1}D^{\m\a_1\n\a_2\m_2\n_2}-\sdfrac{1}{2}\d^{\a_1\m}\d^{\a_2\n}A^{\m_1\n_1\m_2\n_2},
\end{align}
where we indicate with the circle brackets the symmetrization of the indices and with the square brackets the anti-symmetrization of the indices, as follows
\begin{align}
\d^{\m(\m_1}\d^{\n_1)\n}&\equiv\sdfrac{1}{2}\bigg(\d^{\m\m_1}\d^{\n_1\n}+\d^{\m\n_1}\d^{\m_1\n}\bigg)\nn
\d^{\m[\m_1}\d^{\n_1]\n}&\equiv\sdfrac{1}{2}\bigg(\d^{\m\m_1}\d^{\n_1\n}-\d^{\m\n_1}\d^{\m_1\n}\bigg).
\end{align}
The metric variation are consider in the flat space-time limit and the first variation of the square of the metric, Riemann, Ricci and the scalar curvature are given as
\begin{align}
\big[\sqrt{-g}\big]^{\m_i\n_i}&=\sdfrac{1}{2}\d^{\m_i\n_i}\nn \\
\big[R_{\m\a\n\b}\big]^{\m_i\n_i}(p_i)&=\frac{1}{2}\,\bigg(\Delta_\a^{(\m_i}\d^{\n_i)}_\b\,p_{i\m}\,p_{i\n}
+\Delta_\m^{(\m_i}\d^{\n_i)}_\n\,p_{i\a}\,p_{i\b}-\Delta_\m^{(\m_i}\d^{\n_i)}_\b\,p_{i\a}\,p_{i\n}-\Delta_\a^{(\m_i}\d^{\n_i)}_\n\,p_{i\m}\,p_{i\b}\bigg)\nn \\
\big[R^{\m\a\n\b}\big]^{\m_i\n_i}(p_i)&=\frac{1}{2}\,\bigg(\d^{\a(\m_i}\d^{\n_i)\b}\,p_i^\m\,p_i^\n
+\d^{\m(\m_i}\d^{\n_i)\n}\,p_i^\a\,p_i^\b-\d^{\m(\m_i}\d^{\n_i)\b}\,p_i^\a\,p_i^\n-\d^{\a(\m_i}\d^{\n_i)\n}\,p_i^\m\,p_i^\b\bigg)\nn \\
\big[R_{\m\n}\big]^{\m_i\n_i}(p_i)&=\frac{1}{2}\,\bigg(\Delta_\m^{(\m_i}\d^{\n_i)}_\n\,p_{i}^2
+\d^{\m_i\n_i}\,p_{i\m}\,p_{i\n}-p_i^{(\m_i}\d^{\n_i)}_\m\,p_{i\n}-p_i^{(\m_i}\d^{\n_i)}_\n\,p_{i\m}\bigg)\nn \\
\big[R^{\m\n}\big]^{\m_i\n_i}(p_i)&=\frac{1}{2}\,\bigg(\d^{\m(\m_i}\d^{\n_i)\n}\,p_{i}^2
+\d^{\m_i\n_i}\,p_i^\m\,p_i^\n-p_i^{(\m_i}\d^{\n_i)\m}\,p_i^\n-p_i^{(\m_i}\d^{\n_i)\n}\,p_i^\m\bigg)\nn \\
\big[R\big]^{\m_i\n_i}(p_i)&=\bigg(\d^{\m_i\n_i}\,p_{i}^2-p_i^{(\m_i}p_i^{\n_i)}\bigg)\nn \\
\big[\square R\big]^{\m_i\n_i}(p_i)&=p_i^2\,\bigg(p_i^{(\m_i}p_i^{\n_i)}-\d^{\m_i\n_i}\,p_{i}^2\bigg).
\end{align}
Their second variations can be calculated in the form
\begin{align}
\big[R^{\,\b}_{\ \ \n\r\s}\big]^{\m_1\n_1\m_2\n_2}(p_1,p_2)&=\Big[-\frac{1}{2}\,\tilde F^{\m_1\n_1\b\epsilon}p_{1\s}\big(\tilde F^{\m_2\n_2}_{\e\n}p_{2\r}+\tilde F^{\m_2\n_2}_{\e\r}p_{2\n}-\tilde F^{\m_2\n_2}_{\n\r}p_{2\e}\big)\notag\nn \\
&\hspace{-2.5cm}-\frac{1}{2}\big(\tilde C^{\m_1\n_1\m_2\n_2\b}_{\hspace{1.4cm}\r}\ p_{2\n}-\tilde F^{\m_1\n_1\b\e}\,\tilde F^{\m_2\n_2}_{\n\r}\,p_{2\e}\big)p_{2\s}\notag\nn \\
&\hspace{-2.5cm}-\frac{1}{4}\big(\tilde F^{\m_1\n_1}_{\a\n}p_{1\s}+\tilde F^{\m_1\n_1}_{\a\s}p_{1\n}-\tilde F^{\m_1\n_1}_{\s\n}p_{1\a}\big)\big(\tilde F^{\m_2\n_2\b\a}\,p_{2\r}+\tilde F^{\m_2\n_2\b}_{\qquad\r}\,p_{2}^\a-\tilde F^{\m_2\n_2\a}_{\qquad\r}\,p_{2}^{\b}\big)\Big]-(\s\leftrightarrow \r)\nn \\
\big[R_{\m\n\r\s}\big]^{\m_1\n_1\m_2\n_2}(p_1,p_2)&=\d^{(\m_1}_\m\d^{\n_1)}_\b\big[R^{\,\b}_{\ \ \n\r\s}\big]^{\m_2\n_2}(p_2)+\Delta_{\m\b}\big[R^{\,\b}_{\ \ \n\r\s}\big]^{\m_1\n_1\m_2\n_2}(p_1,p_2)\nn \\
\end{align}
\begin{align}
\big[R^{\m\n\r\s}\big]^{\m_1\n_1\m_2\n_2}(p_1,p_2)&=\d^{\a\n}\d^{\b\r}\d^{\s\g}\big[R^{\,\m}_{\ \ \a\b\g}\big]^{\m_1\n_1\m_2\n_2}(p_1,p_2)\notag\nn \\
&\hspace{-1.5cm}-\big(\d^{\a(\m_1}\d^{\n_1)\n}\d^{\b\r}\d^{\s\g}+\d^{\a\n}\d^{\b(\m_1}\d^{\n_1)\r}\d^{\s\g}+\d^{\a\n}\d^{\b\r}\d^{\s(\m_1}\d^{\n_1)\g}\big)\big[R^{\,\b}_{\ \ \n\r\s}\big]^{\m_2\n_2}(p_2)
\end{align}
\begin{align}
\big[R_{\n\s}\big]^{\m_1\n_1\m_2\n_2}(p_1,p_2)&=-\frac{1}{2}\tilde F^{\m_1\n_1\m_2\n_2}\left(p_{1\s}p_{2\n}-\sdfrac{1}{2}p_{1\n}p_{2\s}+p_{2\n}p_{2\s}\right)-\frac{1}{4}\d^{\m_2\n_2}\left(\tilde F^{\m_1\n_1}_{\a\n}\,p_{1\s}+\tilde F^{\m_1\n_1}_{\a\s}\,p_{1\n}\right)\,p_2^\a\notag\\
&\hspace{-1.5cm}+\frac{1}{2}\big(\tilde C^{\m_1\n_1\m_2\n_2\e}_{\hspace{1.4cm}\n}\,p_{2\s}+\tilde C^{\m_1\n_1\m_2\n_2\e}_{\hspace{1.4cm}\s}\,p_{2\n}\big)\,(p_1+p_2)_{\e}+\frac{1}{2}F^{\m_2\n_2}_{\a\s}\tilde F^{\m_1\n_1}_{\b\n}\,p_1^\a\,p_2^\b\notag\\
&\hspace{-1.5cm}-\frac{1}{2}\tilde F^{\m_2\n_2}_{\n\s}\,\tilde F^{\m_1\n_1\a\b}(p_1+p_2)_\a\,p_{2\b}-\frac{1}{2}\left(\tilde C^{\m_1\n_1\m_2\n_2}_{\hspace{1.25cm}\n\s}-\frac{1}{2}\d^{\m_2\n_2}\,\tilde F^{\m_1\n_1}_{\n\s}\right)\,p_1\cdot p_2\nn
\end{align}
\begin{align}
\big[R^{\n\s}\big]^{\m_1\n_1\m_2\n_2}(p_1,p_2)&=\d^{\n\a}\d^{\s\b}\big[R_{\a\b}\big]^{\m_1\n_1\m_2\n_2}(p_1,p_2)-\big(\d^{\n(\m_1}\d^{\n_1)\a}\d^{\s\b}+\d^{\n\a}\d^{\s(\m_1}\d^{\n_1)\b}\big)\big[R_{\a\b}\big]^{\m_2\n_2}(p_2)\\
\big[R\big]^{\m_1\n_1\m_2\n_2}(p_1,p_2)&=-\left(p_2^2+\sdfrac{1}{4}p_1\cdot p_2\right)\,\tilde F^{\m_1\n_1\m_2\n_2}+\frac{1}{4}\,A^{\m_1\n_1\m_2\n_2}\,p_1\cdot p_2\notag\\
&\hspace{-1.5cm}+\tilde C^{\m_1\n_1\m_2\n_2\a\b}\,(p_{1\a}+2p_{2\a})p_{2\b}-\d^{\m_2\n_2}\tilde F^{\m_1\n_1\a\b}\,(p_{1\a}+p_{2\a})p_{2\b}+\sdfrac{1}{2}\tilde C^{\m_2\n_2\m_1\n_1\a\b}\,p_{1\a}p_{2\b}
\end{align}
\begin{align}
\big[\square R\big]^{\m_1\n_1\m_2\n_2}(p_1,p_2)&=\tilde F^{\m_1\n_1\m_2\n_2}\,\bigg[p_2^2(p_1+p_2)^2+\sdfrac{3}{2}(p_2^2+p_1\cdot p_2)\bigg]+\sdfrac{1}{2}\d^{\m_1\n_1}\tilde F^{\m_2\n_2\a\b}(p_1\cdot p_2)\,p_{2\a}p_{2\b}\notag\\
&\hspace{-1.5cm}-\sdfrac{1}{2}\d^{\m_1\n_1}\d^{\m_2\n_2}(p_1\cdot p_2)\bigg[(p_1+p_2)^2-p_1\cdot p_2\bigg]+\d^{\m_2\n_2}F^{\m_1\n_1\a\b}p_{2\a}(p_1+p_2)_\b\bigg[(p_1+p_2)^2+p_2^2\bigg]\notag\\
&\hspace{-1.5cm}-\tilde F^{\m_2\n_2\a\b}p_{2\a}p_{2\b}\,\tilde F^{\m_1\n_1\gamma\d}p_{2\g}(p_1+p_2)_\d-(p_1+p_2)^2\,\tilde C^{\m_1\n_1\m_2\n_2\a\b}\bigg[2p_{2\a}p_{2\b}+p_{1\a}p_{2\b}+\sdfrac{1}{2}p_{2\a}p_{1\b}\bigg].
\end{align}

For the third derivatives the expressions can be written as
\begin{align}
&[R^{\a}_{\beta \chi \de}]^{\m_2 \n_2 \m_3 \n_3 \m_4 \n_4}(p_2,p_3,p_4)=i(p_2+p_3+p_4)_{\delta}[\Gamma^{\a}_{\b \chi}]^{\m_2 \n_2 \m_3 \n_3 \m_4 \n_4}(p_2,p_3,p_4)\notag\\&-i(p_2+p_3+p_4)_{\chi}[\Gamma^{\a}_{\b \delta}]^{\m_2 \n_2 \m_3 \n_3 \m_4 \n_4}(p_2,p_3,p_4)+[\Gamma^{\a}_{\delta \eta}]^{\m_2 \n_2  \m_4 \n_4}(p_2,p_4)[\Gamma^{\eta}_{\beta \chi}]^{ \m_3 \n_3 }(p_3)\notag\\&
+[\Gamma^{\a}_{\delta \eta}]^{\m_3 \n_3  }(p_3)[\Gamma^{\eta}_{\beta \chi}]^{ \m_2 \n_2 \m_4 \n_4 }(p_2,p_4)-[\Gamma^{\a}_{\chi \eta}]^{\m_2 \n_2  \m_3 \n_3}(p_2,p_3)[\Gamma^{\eta}_{\beta \delta}]^{ \m_4 \n_4 }(p_4)\notag\\&
-[\Gamma^{\a}_{\chi \eta}]^{\m_4 \n_4  }(p_4)[\Gamma^{\eta}_{\beta \delta}]^{ \m_2 \n_2 \m_3 \n_3 }(p_2,p_3)+\text{permutations}
\end{align}
 It follows that
\begin{align}
&[R_{\beta \delta}]^{\m_2 \n_2 \m_3 \n_3 \m_4 \n_4}(p_2,p_3,p_4)=[R^{\a}_{\beta \a\de}]^{\m_2 \n_2 \m_3 \n_3 \m_4 \n_4}(p_2,p_3,p_4)\\
&[R]^{\m_2 \n_2 \m_3 \n_3 \m_4 \n_4}(p_2,p_3,p_4)=\delta^{\b \d}[R_{\beta \delta}]^{\m_2 \n_2 \m_3 \n_3 \m_4 \n_4}(p_2,p_3,p_4).
\end{align}
\section{Metric variations of the anomaly}\label{Mvc}
From the anomaly functional
\begin{equation}
\mathcal{A}(x)=\sqrt{-g(x)}(b \,C^2(x)+b' E(x))=\sqrt{-g(x)} \chi(x)
\end{equation}
expanding around a flat space-time and transforming to momentum space we obtain \begin{equation}
[\mathcal{A}(p_1,p_2)]^{\m_1 \n_1 \m_2 \n_2}=\chi(p_1,p_2)^{\m_1 \n_1 \m_2 \n_2},
\end{equation}
where 
\begin{equation}
\chi(p_1,p_2)^{\m_1 \n_1 \m_2 \n_2}=b [C^2]^{\m_1\n_1\m_2\n_2}(p_1,p_2)+b' [E]^{\m_1\n_1\m_2\n_2}(p_1,p_2).
\end{equation}
The second-order derivatives of the Weyl tensor and the Euler density in flat space-time are
\begin{align}
\big[C^2\big]^{\m_1\n_1\m_2\n_2}(p_1,p_2)&=2\bigg([R_{abcd}]^{\m_1\n_1}(p_1)[R^{abcd}]^{\m_2\n_2}(p_2)-\sdfrac{4}{d-2}[R_{ab}]^{\m_1\n_1}(p_1)[R^{ab}]^{\m_2 \n_2}(p_2)\notag \\&+\sdfrac{2}{(d-2)(d-1)}[R]^{\m_1\n_1}(p_1)[R]^{\m_2\n_2 }(p_2)\bigg)
\end{align}
\begin{align}
\big[E\big]^{\m_1\n_1\m_2\n_2}(p_1,p_2)&=2\bigg([R_{abcd}]^{\m_1\n_1}(p_1)[R^{abcd}]^{\m_2\n_2}(p_2)-4[R_{ab}]^{\m_1\n_1}(p_1)[R^{ab}]^{\m_2 \n_2}(p_2)\notag \\&+[R]^{\m_1\n_1}(p_1)[R]^{\m_2\n_2 }(p_2)\bigg).
\end{align}
The third derivative of the anomaly is given by 
\begin{equation}
[\mathcal{A}(p_1,p_2,p_3)]^{\m_1 \n_1 \m_2 \n_2 \m_3 \n_3}=\big[\sqrt{-g(x)}\big]^{\m_3 \n_3}\chi(p_1,p_2)^{\m_1 \n_1 \m_2 \n_2}+\chi(p_1,p_2,p_3)^{\m_1 \n_1 \m_2 \n_2 \m_3 \n_3}+\text{permutations},
\end{equation}
where 
\begin{equation}
\chi(p_1,p_2,p_3)^{\m_1 \n_1 \m_2 \n_2 \m_3\n_3}=b [C^2]^{\m_1\n_1\m_2\n_2 \m_3\n_3}(p_1,p_2)+b' [E]^{\m_1\n_1\m_2\n_2 \m_3\n_3}(p_1,p_2).
\end{equation}
The derivatives of the relevant tensors are
\begin{align}
\big[C^2\big]^{\m_1\n_1\m_2\n_2\m_3\n_3}(p_1,p_2,p_3)&=\Bigg\{[R_{abcd}]^{\m_1\n_1\m_2\n_2}(p_1,p_2)[R^{abcd}]^{\m_3\n_3}(p_3)\notag\\
&\hspace{-3cm}+[R_{abcd}]^{\m_2\n_2}(p_2)[R^{abcd}]^{\m_1\n_1\m_3\n_3}(p_1,p_3)-\sdfrac{4}{d-2}[R_{ab}]^{\m_1\n_1\m_2\n_2}(p_1,p_2)[R^{ab}]^{\m_3\n_3}(p_3)\notag\\
&\hspace{-3cm}-\sdfrac{4}{d-2}[R_{ab}]^{\m_2\n_2}(p_2)[R^{ab}]^{\m_1\n_1\m_3\n_3}(p_1,p_3)+\sdfrac{2}{(d-2)(d-1)}[R]^{\m_1\n_1\m_2\n_2}(p_1,p_2)[R]^{\m_3\n_3}(p_3)\notag\\
&\hspace{-3cm}+\sdfrac{2}{(d-2)(d-1)}[R]^{\m_2\n_2}(p_2)[R]^{\m_1\n_1\m_3\n_3}(p_1,p_3)\Bigg\}+\text{permutations},
\end{align}
and 
\begin{align}
\big[E\big]^{\m_1\n_1\m_2\n_2\m_3\n_3}(p_1,p_2,p_3)&=\Bigg\{[R_{abcd}]^{\m_1\n_1\m_2\n_2}(p_1,p_2)[R^{abcd}]^{\m_3\n_3}(p_3)+[R_{abcd}]^{\m_2\n_2}(p_2)[R^{abcd}]^{\m_1\n_1\m_3\n_3}(p_1,p_3)\notag\\
&\hspace{-3.7cm}-4[R_{ab}]^{\m_1\n_1\m_2\n_2}(p_1,p_2)[R^{ab}]^{\m_3\n_3}(p_3)-4[R_{ab}]^{\m_2\n_2}(p_2)[R^{ab}]^{\m_1\n_1\m_3\n_3}(p_1,p_3)+[R]^{\m_1\n_1\m_2\n_2}(p_1,p_2)[R]^{\m_3\n_3}(p_3)\notag\\
&\hspace{-3cm}+[R]^{\m_2\n_2}(p_2)[R]^{\m_1\n_1\m_3\n_3}(p_1,p_3)\Bigg\}+\text{permutations}.
\end{align}
We also give the fourth-order derivatives of the counterterms. In compact notation they take the form

\begin{align}
\big[\sqrt{-g}\,C^2\big]^{\m_1\n_1\m_2\n_2\m_3\n_3\m_4\n_4}(p_1,p_2,p_3,p_4)&=\Bigg\{[R_{abcd}]^{\m_1\n_1}(p_1)[R^{abcd}]^{\m_2 \n_2 \m_3\n_3\m_4\n_4}(p_2,p_3,p_4)\notag\\&\hspace{-6.0cm}+[R_{abcd}]^{\m_2 \n_2 \m_3\n_3\m_4\n_4}(p_2,p_3,p_4)[R^{abcd}]^{\m_1 \n_1}(p_1)+[R_{abcd}]^{\m_1 \n_1 \m_2\n_2}(p_1,p_2)[R^{abcd}]^{\m_3 \n_3 \m_4 \n_4}(p_3,p_4)\notag\\&\hspace{-6.0cm}-\sdfrac{4}{d-2}[R_{ab}]^{\m_1 \n_1}(p_1)[R^{ab}]^{\m_2 \n_2 \m_3\n_3\m_4\n_4}(p_2,p_3,p_4)-\sdfrac{4}{d-2}[R_{ab}]^{\m_1 \n_1 \m_2\n_2}(p_1,p_2)[R^{ab}]^{\m_3\n_3\m_4\n_4}(p_3,p_4)\notag\\&\hspace{-6.0cm}-\sdfrac{4}{d-2}[R_{ab}]^{ \m_2 \n_2 \m_3 \n_3 \m_4 \n_4}(p_2,p_3,p_4)[R^{ab}]^{\m_1 \n_1 }(p_1)+\sdfrac{2}{(d-2)(d-1)}[R]^{\m_1\n_1}(p_1)[R]^{\m_2\n_2\m_3 \n_3 \m_4 \n_4}(p_2,p_3,p_4)\notag\\&\hspace{-6.0cm}+\sdfrac{2}{(d-2)(d-1)}[R]^{\m_1\n_1\m_2 \n_2 \m_4 \n_4}(p_1,p_2,p_4)[R]^{\m_3\n_3}(p_3)+\sdfrac{2}{(d-2)(d-1)}[R]^{\m_1\n_1\m_2 \n_2 }(p_1,p_2)[R]^{\m_3\n_3 \m_4 \n_4}(p_3,p_4)\Bigg\}\notag\\& \hspace{-6.0cm}+[\sqrt{-g}]^{\m_1\n_1}\big[C^2\big]^{\m_2\n_2\m_3\n_3 \m_4 \n_4}(p_2,p_3,p_4)+[\sqrt{-g}]^{\m_1\n_1 \m_2 \n_2 }\big[C^2\big]^{\m_3\n_3 \m_4 \n_4}(p_3,p_4)+\text{permutations},
\end{align}
and
\begin{align}
\big[\sqrt{-g}\,E\big]^{\m_1\n_1\m_2\n_2\m_3\n_3\m_4\n_4}(p_1,p_2,p_3,p_4)&=\Bigg\{[R_{abcd}]^{\m_1\n_1}(p_1)[R^{abcd}]^{\m_2 \n_2 \m_3\n_3\m_4\n_4}(p_2,p_3,p_4)\notag\\&\hspace{-6.0cm}+[R_{abcd}]^{\m_2 \n_2 \m_3\n_3\m_4\n_4}(p_2,p_3,p_4)[R^{abcd}]^{\m_1 \n_1}(p_1)+[R_{abcd}]^{\m_1 \n_1 \m_2\n_2}(p_1,p_2)[R^{abcd}]^{\m_3 \n_3 \m_4 \n_4}(p_3,p_4)\notag\\&\hspace{-6.0cm}-4[R_{ab}]^{\m_1 \n_1}(p_1)[R^{ab}]^{\m_2 \n_2 \m_3\n_3\m_4\n_4}(p_2,p_3,p_4)-4[R_{ab}]^{\m_1 \n_1 \m_2\n_2}(p_1,p_2)[R^{ab}]^{\m_3\n_3\m_4\n_4}(p_3,p_4)\notag\\&\hspace{-6.0cm}-4[R_{ab}]^{ \m_2 \n_2 \m_3 \n_3 \m_4 \n_4}(p_2,p_3,p_4)[R^{ab}]^{\m_1 \n_1 }(p_1)+[R]^{\m_1\n_1}(p_1)[R]^{\m_2\n_2\m_3 \n_3 \m_4 \n_4}(p_2,p_3,p_4)\notag\\&\hspace{-6.0cm}+[R]^{\m_1\n_1\m_2 \n_2 \m_4 \n_4}(p_1,p_2,p_4)[R]^{\m_3\n_3}(p_3)+[R]^{\m_1\n_1\m_2 \n_2 }(p_1,p_2)[R]^{\m_3\n_3 \m_4 \n_4}(p_3,p_4)\Bigg\}\notag\\&\hspace{-6.0cm}+[\sqrt{-g}]^{\m_1\n_1}\big[E\big]^{\m_2\n_2\m_3\n_3 \m_4 \n_4}(p_2,p_3,p_4)+[\sqrt{-g}]^{\m_1\n_1 \m_2 \n_2 }\big[E\big]^{\m_3\n_3 \m_4 \n_4}(p_3,p_4)+\text{permutations}.
\end{align}

\chapter{Functional relations and boundary terms in $V_{C^2}$}
We summarize the following expressions for the Weyl squared terms in 4 and in $d$ dimensions

\begin{align}
(C^{(4)})^2&=R^{\mu \nu \rho \sigma}R_{\mu \nu \rho \sigma}-2R^{\mu \nu }R_{\mu \nu}+R^2\notag \\ 
(C^{(d)})^2&=R^{\mu \nu \rho \sigma}R_{\mu \nu \rho \sigma}-\frac{4}{d-2}R^{\mu \nu }R_{\mu \nu}+\frac{2}{(d-2)(d-1)}R^2\notag \\ 
(C^{(d)})^2&=(C^{(4)})^2+\frac{d-4}{d-2}\left(2R^{\mu \nu }R_{\mu \nu}-\frac{d+1}{3(d-1)}R^2\right)\notag 
\end{align}
We have the following relation 
\begin{equation}
2g_{\mu \nu}\frac{\delta}{\sqrt{-g}\delta g_{\mu \nu}}\int d^d x  \sqrt{-g} (C^{(d)})^2=(d-4)(C^{(d)})^2.
\end{equation}
By using the relation between  $(C^{(4)})^2$ and $(C^{(d)})^2$ we can write 
\begin{equation}
2g_{\mu \nu}\frac{\delta}{\sqrt{-g}\delta g_{\mu \nu}}\int d^d x  \sqrt{-g} \left((C^{(4)})^2+\frac{d-4}{d-2}\left(\frac{d-4}{d-2}R^{\mu \nu }R_{\mu \nu}-\frac{d+1}{3(d-1)}R^2\right)\right)=(d-4)(C^{(d)})^2.
\end{equation}
By rearranging the terms above we get
\begin{align}
\label{var1}
&2g_{\mu \nu}\frac{\delta}{\sqrt{-g}\delta g_{\mu \nu}}\int d^d x  \sqrt{-g} (C^{(4)})^2 =(d-4)(C^{(d)})^2\nn \\
& -2g_{\mu \nu}\frac{\delta}{\sqrt{-g}\delta g_{\mu \nu}}\int d^d x \sqrt{-g}\frac{d-4}{d-2}\left(2R^{\mu \nu }R_{\mu \nu}-\frac{d+1}{3(d-1)}R^2\right).\nn
\end{align}
By a direct computation we obtain
\begin{align}
& 2g_{\mu \nu}\frac{\delta}{\sqrt{-g}\delta g_{\mu \nu}}\int d^d x \sqrt{-g}\frac{d-4}{d-2}\left(2R^{\mu \nu }R_{\mu \nu}-\frac{d+1}{3(d-1)}R^2\right)\nn \\
& =(d-4)\left\lbrace \frac{d-4}{d-2}\left(2R^{\mu \nu }R_{\mu \nu}-\frac{d+1}{3(d-1)}R^2\right)-\frac{2}{3}\Box R \right\rbrace.
\end{align}
Then we can substitute this expression in \eqref{var1} to obtain
\begin{align}\label{finalvar}
2g_{\mu \nu}\frac{\delta}{\sqrt{-g}\delta g_{\mu \nu}}\int d^d x  \sqrt{-g} (C^{(4)})^2&=(d-4)\left[(C^{(d)})^2+\frac{2}{3}\Box R \right] -\frac{(d-4)^2}{(d-2)}\left(2R^{\mu \nu }R_{\mu \nu}-\frac{d+1}{3(d-1)}R^2\right) \notag \\
&=(d-4)\left[(C^{(4)})^2+\frac{2}{3}\Box R+\frac{(d-4)}{(d-2)}\left(2R^{\mu \nu }R_{\mu \nu}-\frac{d+1}{3(d-1)}R^2\right) \right]\notag \\&-\frac{(d-4)^2}{(d-2)}\left(2R^{\mu \nu }R_{\mu \nu}-\frac{d+1}{3(d-1)}R^2\right)\notag\\
&=(d-4)\left[(C^{(4)})^2+\frac{2}{3}\Box R \right].
\end{align}
We obtain the same result by a direct computation
\begin{equation}
2g_{\mu \nu}\frac{\delta}{\sqrt{-g}\delta g_{\mu \nu}}\int d^d x  \sqrt{-g} (C^{(4)})^2=(d-4)\left[(C^{(4)})^2+\frac{2}{3}\Box R \right].
\end{equation}
\end{appendices}

\bibliography{bibliography.bib}

\bibliographystyle{h-physrev-mod1}
\end{document}